\documentclass[english,preprint, number]{elsarticle}
\usepackage[T1]{fontenc}
\usepackage[latin9]{inputenc}
\usepackage{url}
\usepackage{amsmath}
\usepackage{amssymb}
\usepackage{esint}

\makeatletter

\providecommand*{\perispomeni}{\char126}
\AtBeginDocument{\DeclareRobustCommand{\greektext}{%
  \fontencoding{LGR}\selectfont\def\encodingdefault{LGR}%
  \renewcommand{\~}{\perispomeni}%
}}
\DeclareRobustCommand{\textgreek}[1]{\leavevmode{\greektext #1}}
\DeclareFontEncoding{LGR}{}{}

\newcommand{\lyxmathsym}[1]{\ifmmode\begingroup\def\b@ld{bold}
  \text{\ifx\math@version\b@ld\bfseries\fi#1}\endgroup\else#1\fi}

\makeatother

\usepackage{babel}

\begin{document}

\title{Decoherence effects in Bose-Einstein condensate interferometry\\
 I. General Theory}

\author{B. J. Dalton}

\address{ARC Centre for Quantum-Atom Optics and Centre for Atom Optics and
Ultrafast Spectroscopy\\
Swinburne University of Technology\\
Melbourne, Victoria 3122, Australia\pagebreak{}}
\begin{abstract}
The present paper outlines a basic theoretical treatment of decoherence
and dephasing effects in interferometry based on single component
Bose-Einstein condensates in double potential wells, where two condensate
modes may be involved. Results for both two mode condensates and the
simpler single mode condensate case are presented. The approach involves
a hybrid phase space distribution functional method where the condensate
modes are described via a truncated Wigner representation, whilst
the basically unoccupied non-condensate modes are described via a
positive P representation..The Hamiltonian for the system is described
in terms of quantum field operators for the condensate and non-condensate
modes. The functional Fokker-Planck equation for the double phase
space distribution functional is derived. Equivalent Ito stochastic
equations for the condensate and non-condensate fields that replace
the field operators are obtained, and stochastic averages of products
of these fields give the quantum correlation functions that can be
used to interpret interferometry experiments. The stochastic field
equations are the sum of a deterministic term obtained from the drift
vector in the functional Fokker-Planck equation, and a noise field
whose stochastic properties are determined from the diffusion matrix
in the functional Fokker-Planck equation. The stochastic properties
of the noise field terms are similar to those for Gaussian-Markov
processes in that the stochastic averages of odd numbers of noise
fields are zero and those for even numbers of noise field terms are
the sums of products of stochastic averages associated with pairs
of noise fields. However each pair is represented by an element of
the diffusion matrix rather than products of the noise fields themselves,
as in the case of Gaussian-Markov processes. The treatment starts
from a generalised mean field theory for two condensate modes, where
generalised coupled Gross-Pitaevskii equations are obtained for the
modes and matrix mechanics equations are derived for the amplitudes
describing possible fragmentations of the condensate between the two
modes. These self-consistent sets of equations are derived via the
Dirac-Frenkel variational principle. Numerical studies for interferometry
experiments would involve using the solutions from the generalised
mean field theory in calculations for the stochastic fields from the
Ito stochastic field equations.\pagebreak{} 
\end{abstract}
\maketitle

\section{Introduction}

The creation of Bose-Einstein condensates (BEC) in cold atomic gases
has enabled the realisation of a controllable quantum system on a
macroscopic scale. With all bosons occupying the same single particle
state (or mode) the BEC exhibits coherence somewhat analogous to the
coherence for an idealised single mode laser and interference effects
were soon observed \citep{Andrews97a}, \citep{Hall98a}. Interferometry
using BECs was a natural outcome, and much research centres around
developing BEC interferometric systems, motivated not only by wishing
to study coherence, interference and entanglement in macroscopic systems
but also because of their potential applications for precision measurement,
including the development of BEC interferometry for measurements at
the Heisenberg limit \citep{Bouyer97a}, \citep{Dunningham02a}, \citep{Kasevich02a},
\citep{Dunningham04a}, \citep{Esteve08a}. Experiments demonstrating
precision beyond the standard quantum limit have recently been reported
\citep{Gross10a}, \citep{Riedel10a}. Reviews covering general aspects
of BEC interferometry include \citep{Bongs04a}, \citep{Dunningham05a},
\citep{Cronin09a}.

Interferometry with BECs is a quantum effect. In its simplest form
quantum interferometry essentially involves transitions between an
initial prepared state and a final measured state for the interferometer
system, where the overall transition probability amplitude for transitions
is split into two partial amplitudes associated with different intermediate
states, which are then recombined. The two amplitudes must remain
coherent but depend differently on the feature being measured. A variety
of such features can produce interferometric effects, ranging from
a transition frequency between states of interest to an asymmetry
in a trapping potential due to gravity effects. The partial amplitudes
for the differing intermediate states may result from various types
of time evolution, including free evolution stages and interaction
stages, where the system is subjected to external classical fields.
As the feature changes, constructive and destructive interference
between the partial amplitudes results, leading to the changes in
measurement probability for the final state.

In the case of interferometry with single atoms, the review by Cronin
\textit{et al }\citep{Cronin09a} outlines how \emph{Ramsey interferometry}
can be described in these terms. Here the interferometric system is
a two level atom with internal states $\left\vert a\right\rangle $,
$\left\vert b\right\rangle $, the first being the initial state and
the second is the final state. The feature that produces the interferometric
effect is the transition frequency $\omega_{ba}$ and the interferometer
is used to obtain a precise measurement of $\omega_{ba}$ - to use
for example in an atomic clock. The atoms are in a beam with a fixed
velocity and pass through two short interaction regions when a resonant
classical field of pulse area $\pi/2$ couples the two internal states,
turning each into different orthogonal linear superpositions of $\left\vert a\right\rangle $,
$\left\vert b\right\rangle $ - say $\left\vert a\right\rangle \rightarrow(\left\vert a\right\rangle +\left\vert b\right\rangle )/\sqrt{2}$
and $\left\vert b\right\rangle \rightarrow(\left\vert a\right\rangle -\left\vert b\right\rangle )/\sqrt{2}$.
Between the interaction regions the atoms undergo free evolution for
time $T$, with $\left\vert a\right\rangle \rightarrow\exp(i\omega_{a}T)\left\vert a\right\rangle $
and $\left\vert b\right\rangle \rightarrow\exp(i\omega_{b}T)\left\vert b\right\rangle $.
The states $\left\vert a\right\rangle $, $\left\vert b\right\rangle $
also act as two possible intermediate states for the process $a\rightarrow b$,
and there are two distinct pathways $a\rightarrow b\rightarrow b\rightarrow b$
and $a\rightarrow a\rightarrow a\rightarrow b$ whose partial amplitudes
interfere. In the first pathway the resonant classical field transition
$a\rightarrow b$ occurs in the first step, in the second it is in
the last step, and between the first and last steps free evolution
occurs in different states - $b$ for the first pathway and $a$ for
the second. The partial amplitudes are $(-1/\sqrt{2})\exp(i\omega_{b}T)(+1/\sqrt{2})$
for the first pathway and $(+1/\sqrt{2})\exp(i\omega_{a}T)(+1/\sqrt{2})$
for the second, giving a total amplitude proportional to $\sin(\omega_{ba}T/2)$
resulting from interference between the two partial amplitudes. This
produces oscillations in the measurement probability, enabling $\omega_{ba}$
to be determined. Single atom \emph{Mach-Zender interferometry} \citep{Hansel01a},
\citep{Sidorov06a} involving a \emph{double well} is another case
where a similar description applies. The initial state is the lowest
symmetric state $\left\vert S(0)\right\rangle $ for an atom in a
single well trap, the final state $\left\vert AS(T)\right\rangle $
is the lowest antisymmetric state in the same single well. The process
$\left\vert S(0)\right\rangle \rightarrow\left\vert AS(T)\right\rangle $
involves splitting the single well to a slightly asymmetric double
well and then recombining back to the single well during a time $T$.
The intermediate state can be chosen as two localised states \citep{Sidorov06a}
for the actual double well, one $\left\vert L(T/2)\right\rangle $
being localised in the left well the other $\left\vert R(T/2)\right\rangle $
in the right well. The two pathways whose transition amplitudes interfere
are $\left\vert S(0)\right\rangle \rightarrow\left\vert L(T/2)\right\rangle \rightarrow\left\vert AS(T)\right\rangle $
and $\left\vert S(0)\right\rangle \rightarrow\left\vert R(T/2)\right\rangle \rightarrow\left\vert AS(T)\right\rangle $,
the overall process being driven by non-adiabatic evolution during
the splitting and recombination stages. Asymmetry in the trapping
potential produces the interferometric effect. In the case of single
atom \emph{Bragg interferometry} \citep{Torii00a}, \citep{Bongs04a}
an atom in a zero momentum state is subjected to three Bragg pulses
with pulse areas $\pi/2$, $\pi$, $\pi/2$, where each pulse involves
counterpropagating photons of two slightly differing wave numbers
$k_{\lambda}$, $k_{\mu}$ \ A two-photon off-resonant Raman process
removes a photon from one of the laser beams in the Bragg pulse and
adds a photon to the other beam. The momentum difference changes the
atom's momentum from zero to $2\hbar k=k_{\lambda}+k_{\mu}$. For
a given $k$ the wave numbers $k_{\lambda}$, $k_{\mu}$ can be adjusted
to satisfy energy as well as momentum conservation. Bragg interferometry
can be described in terms of two momentum states $\left\vert p=0\right\rangle $
and $\left\vert p=2\hbar k\right\rangle $ for the atom. The $\pi/2$
pulses change each state.into linear combinations of these two states
- say $\left\vert 0\right\rangle \rightarrow(\left\vert 0\right\rangle -\exp(-i\phi)\left\vert 2\hbar k\right\rangle )/\sqrt{2}$
and $\left\vert 2\hbar k\right\rangle \rightarrow(\exp(+i\phi)\left\vert 0\right\rangle +\left\vert 2\hbar k\right\rangle )/\sqrt{2}$.
The $\pi$ pulse changes each momentum state.into the other state
- say $\left\vert 0\right\rangle \rightarrow-\exp(-i\phi)\left\vert 2\hbar k\right\rangle $
and $\left\vert 2\hbar k\right\rangle \rightarrow\exp(+i\phi)\left\vert 0\right\rangle $.
Here $\phi$ is a phase factor for the Bragg pulse involved. For an
overall process say $\left\vert 0\right\rangle \rightarrow\left\vert 0\right\rangle $
there are two pathways each with its own transition amplitude $\left\vert 0\right\rangle \rightarrow\left\vert 0\right\rangle \rightarrow\left\vert 2\hbar k\right\rangle \rightarrow\left\vert 0\right\rangle $
and $\left\vert 0\right\rangle \rightarrow\left\vert 2\hbar k\right\rangle \rightarrow\left\vert 0\right\rangle \rightarrow\left\vert 0\right\rangle $,
the successive steps involving the pulses $\pi/2$, $\pi$, $\pi/2$
respectively. If we choose $\phi=0$ in the first two steps and $\phi\neq0$
in the final $\pi/2$ step, the transition probability is given by
$(1+\cos\phi)/2$, giving interferometric effects as $\phi$ is changed.

Ramsey, Mach-Zender and Bragg interferometry \citep{Torii00a}, \citep{Shin04a},
\citep{Bongs04a}, \citep{Cronin09a} can also be carried out using
BECs rather than single atoms, and a generalised version of the above
approach could be used to describe these. Quantum interference in
double well BEC interferometry is discussed qualitatively in \citep{Dalton07a}
in terms of interfering transition amplitudes. However, since BECs
involve a large number $N$ of atoms rather than just one, there are
a number of additional complexities that need to be taken into account,
notably associated with the feature that macroscopic numbers of atoms
may occupy each single particle state. Firstly, large numbers of partial
transition amplitudes may now be involved in the overall process,
and evaluating all the partial transition amplitudes and then recombining
them becomes a formidable task. The analysis for the single atom case
establishes the general point that for interferometry to occur there
must be \emph{at least two} different single particle states (or \emph{modes})
that an atom can occupy - otherwise two or more pathways for the overall
process to occur would not be available. This suggests immediately
that interferometry using BECs must at least be based on a \emph{two-mode
theory}. For \emph{single component} BECs, the two single particle
states would be represented by two orthogonal, normalised spatial
mode functions $\phi_{1}{\small(\mathbf{r})}$, $\phi_{2}{\small(\mathbf{r})}$.
Time dependences are left implicit. For double well interferometry,
these could be either \emph{localised} in each of the two potential
wells or \emph{delocalised} symmetrically or antisymmetrically over
the two wells. For Bragg interferometry the two modes could be two
different momentum eigenfunctions. For \emph{two component} BECs,
with internal (hyperfine) states $\left\vert F\right\rangle $, $\left\vert G\right\rangle $
the two single particle states would be represented by $\phi_{F}{\small(\mathbf{r})}\left\vert F\right\rangle $,
$\phi_{G}{\small(\mathbf{r})}\left\vert G\right\rangle $, where the
associated \emph{normalised} spatial mode functions are $\phi_{F}{\small(\mathbf{r})}$,
$\phi_{G}{\small(\mathbf{r})}$. The significance of two-mode theories
for BECs is well recognised \citep{Leggett01a}, \citep{Molmer03a}
and points to the existence of \emph{Josephson effect} \citep{Gati07a}
physics in cold quantum gases. The idea of the BEC being equivalent
to a \emph{giant spin system}, with direct linkages to angular momentum
theory \citep{Rose57a}, spin squeezing \citep{Kitagawa93a} etc.
stems from two-mode theory, as will be outlined below. For the quantum
description of Ramsey, double well and Bragg interferometry with BECs
however, even if each atom is restricted to one of two single particle
states there are now $N+1$ distinct ways of dividing the atoms between
the two single particle states, corresponding to \emph{Fock states}
with occupancies given by $\frac{N}{2}-k$, $\frac{N}{2}+k$ with
$(k=-N/2,-N/2+1,..,N/2)$ in the two modes. The Fock states can describe
BECs that are \emph{fragmented}, with two modes having macroscopic
occupancy \citep{Leggett01a}, \citep{Pitaevskii03a}. Consequently,
in any overall process there are a great many pathways involved, so
the overall transition amplitude can contain many contributions. Having
more interfering pathways raises both the possibility of sharper interferometric
effects \citep{Dalton86a} but also the possibility that effects can
be degraded depending on how the phases and magnitudes of the.partial
amplitudes are related. These sorts of effects are also familiar from
multiple slit optical interference. Secondly, the need to consider
steps in the process where the intermediate states already have many
atoms occupying each of the two single particle states raises the
possibility of \emph{bosonic enhancement} of contributions to the
partial transition amplitude from the step involved. These sorts of
effects are familiar from the theory of lasers and in super-radiance.
The effects could occur because two-mode BECs are like a giant spin
system rather than a collection of independent atoms, and implies
that the simple analysis described above for single atoms is no longer
valid. However, a closer analysis (see \citep{Dalton10a}) suggests
that bosonic enhancement and super-radiance effects are not in fact
present. Thirdly, the evolution is not as simple as in the single
atom case, since collisions between the atoms need to be taken into
account. Even with only two single particle states allowed, \emph{dephasing}
between the contributing amplitudes can occur - which tend to degrade
interferometric features but which may also produce collapse and revival
effects \citep{Wright97a}, \citep{Dunningham04a}. Also, even if
the BEC is close to zero temperature, collisions could remove atoms
from the macroscopically occupied pair of single particle states and
deposit them into previously unoccupied higher energy \emph{thermal
states}. The unoccupied thermal states act as a kind of environment
(or reservoir) so the system defined in terms of the two macroscopically
occupied single particle states suffers \emph{decoherence}. Collective
phonon-like states of the BEC called \emph{Bogoliubov excitations}
\citep{Bogoliubov47a}, \citep{Leggett01a}, \citep{Pitaevskii03a}
can be created. These processes again generally result in degradation
of interferometric effects. However, some aspects of the interferometric
process will still be similar to the single atom interferometry case.
These include the presence of interaction regions in which the BEC
is subjected to external pulsed classical fields with pulse areas
$\pi/2$, $\pi$ that couple internal states, the effect of Bragg
pulses that change the momentum of each atom, the presence of asymmetries
in trapping potentials that confine the BEC, as well as periods of
free evolution of the BEC - though now of course collisions need to
be taken into account. The difference is though a more elaborate theory
is needed to treat quantum interferometry in BECs allowing for all
these effects.

Theories of BEC interferometry that take into account the many body
nature of the system are of various levels of sophistication \citep{Molmer03a}
depending on the range of effects taken into account. The Hamiltonian
is often expressed in terms of \emph{field operators}. For single
component BECs the field annihilation operator $\widehat{\Psi}({\small\mathbf{r}}\mathbf{)}$
destroys a bosonic atom at position $\mathbf{r}$, whilst for two
component BECs the field annihilation operator $\widehat{\Psi}_{a}({\small\mathbf{r}}\mathbf{)}$
$(a=F,G)$ destroys a bosonic atom with internal state $\left\vert a\right\rangle $
at position $\mathbf{r}$. Interferometry experiments are generally
interpreted in terms of \emph{quantum correlation functions}, which
are expectation values of products of field annihilation operators
with the associated field creation operators, and are related to bosonic
many atom position measurements \citep{Bach04a}. Actual measurements
of quantum correlation functions may be made via scattering a weak
probe beams (atoms, light) off the system, \citep{CohenTannoudji01a}.
If boson-boson interactions were absent and the BEC isolated from
the environment, idealised forms of the quantum correlation functions
would result, with clearly visible interferometric effects. Even where
external environmental effects are absent, the internal boson-boson
interactions can still result in dephasing (associated with interactions
within the condensate modes) and decoherence effects (associated with
interactions causing transitions from the condensate modes) that degrade
the interference pattern. Many treatments of BEC interferometry are
based on the simplest assumption, namely that during the interferometric
process the condensate is \emph{unfragmented}, with all bosons occupying
the \emph{same} single particle state $\left\vert \,\chi\right\rangle $.
This situation is a special case of a two mode theory, with the occupied
single particle state written as a linear superposition of the two
modes. For single component BEC interferometry linear combinations
of the form $\left\langle {\small\mathbf{r\,|\,\chi}}\right\rangle =\chi{\small(\mathbf{r})=\,}\alpha_{1}\phi_{1}{\small(\mathbf{r})}+\alpha_{2}\phi_{2}{\small(\mathbf{r})}$
are involved, for two component BEC interferometry superpositions
$\left\langle {\small\mathbf{r\,|\,\chi}}\right\rangle =\chi_{F}{\small(\mathbf{r})}\left\vert F\right\rangle +\chi_{G}{\small(\mathbf{r})}\left\vert G\right\rangle $
of the two internal states occur. Equations for the spatial wave functions
associated with these single particle states can be obtained using
variational principles \citep{Dirac30a}, \citep{Frenkel34a}. For
the single component case the well-known \emph{Gross-Pitaevskii equation}
\citep{Gross61a}, \citep{Pitaevskii61a} applies for the so-called
\emph{condensate wave function} $\chi{\small(\mathbf{r})}$, for the
two component case \emph{coupled Gross-Pitaevskii equations} \citep{Esry98a},
\citep{Blakie99a} apply for the condensate wave functions $\chi_{F}{\small(\mathbf{r})}$,
$\chi_{G}{\small(\mathbf{r})}$ associated with the two internal states.
The Gross-Pitaevskii equations are \emph{non-linear}, with collision
effects occuring via \emph{mean field} terms. Treatments of BEC interferometry
based on assuming the condensate is unfragmented include \citep{Ostrovskaya00a},
\citep{Ananikian06a}\emph{\ }for the single component case and \citep{Williams99a},
\citep{Anderson09a} for the two component case.

However, there are two distinct single particle states each boson
could occupy, and for $N$ bosons the $N+1$ dimensional state space
for two mode theories allows for more general quantum states that
are \emph{fragmented}, with macroscopic occupancy of two single particle
states. The basis states can be chosen as Fock states $\left\vert \,\frac{{\small N}}{{\small2}},k\right\rangle $
$(k=-N/2,..,N/2)$ in which $\frac{N}{2}-k$ bosons occupy one of
the two single particle states ($\phi_{1}{\small(\mathbf{r})}$ for
the single component case, $\phi_{F}{\small(\mathbf{r})}\left\vert F\right\rangle $
for the two component case) and $\frac{N}{2}+k$ bosons occupy the
other two single particle states ($\phi_{2}{\small(\mathbf{r})}$
for the single component case, $\phi_{G}{\small(\mathbf{r})}\left\vert G\right\rangle $
for the two component case). Each Fock state is a fragmented state,
with definite numbers $\frac{N}{2}\mp k$ of bosons respectively in
the two modes. In two mode theory the general quantum state of the
$N$ boson system is written as a superposition of the Fock states
with general amplitudes $b_{k}$. The unfragmented states are just
special cases called \emph{binomial states} since the amplitudes $b_{k}$
are determined from binomial coefficients. The Dirac-Frenkel variational
principle \citep{Dirac30a}, \citep{Frenkel34a} can be used to obtain
\emph{matrix mechanics} equations for the $N+1$ general amplitudes
$b_{k}$ and \emph{generalized Gross-Pitaevskii equations}.for the
two mode functions ($\phi_{1}{\small(\mathbf{r})}$, $\phi_{2}{\small(\mathbf{r})}$
for the single component case, $\phi_{F}{\small(\mathbf{r})}$, $\phi_{G}{\small(\mathbf{r})}$
for the two component case). The $N+1$ amplitude equations describe
the system evolution amongst the possible Fock states, and involve
Fock state Hamiltonian and rotation matrix elements which depend on
the two mode functions. The two coupled Gross-Pitaevskii equations
are again non-linear in the mode functions due to collision terms
- which occur via \emph{generalised} mean fields - and involve the
trap potential, with an additional intercomponent coupling term in
the two component case. They contain as coefficients one and two body
correlation functions that depend quadratically on the amplitudes,
and which reflect the relative importance of the different Fock states
during the interference process. The combined amplitude and mode equations
are \emph{self-consistent}, and are more general than the equations
for the unfragmented BEC case. It should be noted however that other
authors \citep{Castin98a}, \citep{Sorensen02a}, \citep{Gardiner07a}
define the condensate mode functions via a diferent approch, namely
in terms of the eigenfunctions of the first order quantum correlation
function that have macroscopic eigenvalues. This approach is discussed
below in Section \ref{Sect Phase Space Dist Fnal}. Two mode theories
similar to the present treatment have previously been developed for
single component BECs with two orthogonal spatial modes (such as in
double-well interferometry) \citep{Spekkens99a}, \citep{Menotti01a},
\citep{Dalton07a}, \citep{Streltsov06a}, \citep{Streltsov07a},
\citep{Alon08a}, \citep{Sakmann10a} and fragmentation effects shown
in \citep{Streltsov07a}, \citep{Alon08a}, \citep{Sakmann10a}. Two
mode theories for the two component case have been presented in \citep{Sinatra00a},
\citep{Li09a} and elsewhere \citet{Dalton10a}. Two mode theories
incorporate dephasing effects associated with transfers of bosons
between the two modes, but decoherence effects and Bogoliubov excitations
are outside the scope of the theory. Both the general two-mode theory
and the single mode theory are referred to as \emph{mean field theories},
since collisional effects occur via mean fields.

To allow for decoherence and Bogoliubov excitations the theory must
include large numbers of non-condensate modes, which are modes with
very small occupancy. Bogoliubov theory is perturbation theory in
the interaction between condensate and non-condensate modes, and treatments
of Bogoliubov excitations for BEC interferometry have been made \citep{Sorensen02a}
by adapting general BEC Bogoliubov theory \citep{Gardiner97a}, \citep{Castin98a},
\citep{Girardeau98a}, \citep{Gardiner07a}to treat two-component
BECs. Another approach that could be applied to BEC interferometry
is a master equation method \citep{Gardiner98a}, \citep{Gardiner00a},
in which a condensate density operator is defined and a master equation
is derived allowing for interactions with non-condensate modes, which
constitute a reservoir. The quantum state is now non-pure so a density
operator is needed to describe the system. The difficulty with this
method is that it is hard to evaluate the non-condensate contributions
to quantum correlation functions. A further approach could be based
on the Heisenberg equation method that have been applied in numerous
many-body theory problems. Heisenberg equations for field operators
and products of field operators are derived, and taking the expectation
values with the initial density operator results in a heirarchy of
coupled equations for quantum correlation functions. An ansatz (such
as assuming that a suitable high order correlation function factorises)
produces a truncated set of coupled equations from which correlation
functions of the required order can be calculated. The problem with
this method is that it is hard to confirm the validity of the ansatz.
In view of there being very large numbers of modes, phase space theories
have also been developed with the density operator represented by
a quasi-distribution functional in a phase space \citep{Steel98b}.
Quantum correlation functions are then expressed as functional integrals
in the phase space, involving products of the distribution functional
with the several field functions that replace the field operators.
The Liouville-von Neumann equation for the density operator is replaced
by a functional Fokker-Planck equation (FFPE) for the distribution
functional. Finally, the FFPE are finally replaced by coupled Ito
stochastic equations (c-number Langevin equations) for the field functions,
where the Ito equations contain deterministic and random noise terms
- identifiable from the FFPE. Stochastic averages of the field functions
then give the quantum correlation functions. Phase space distribution
functional treatments were originally developed to treat problems
in quantum optical physics \citep{Graham70a}, \citep{Graham70b},
\citep{Drummond87a}, \citep{Kennedy88a}, \citep{Gatti97a}, but
have since been adapted for BECs. There are several different phase
space theories that have or could be used to treat BEC interferometry,
depending on the nature of the distribution functional chosen to represent
the density operator. The positive P representation has been used
by \citep{Poulsen01a} to treat spin squeezing in two component BECs.
However, because most atoms will be in one or two highly occupied
modes and these bosons can be treated approximately in terms of mean
field theories, a more natural representation to use is the truncated
Wigner representation. Such theories have been developed \citep{Steel98b}
and applied to BEC interferometry \citep{Isella06a}, \citep{Scott09a}.
In the truncated Wigner FFPE there are no second order functional
derivatives, so there are no random noise terms in the Ito equations.
Quantum noise is embodied in the initial state, and Bogoliubov equations
are used to describe this state. Based on the truncated Wigner representation,
stochastic modifications of the Gross-Pitaevskii equation to allow
for the effects due to non-condensate modes have been derived for
the case where the condensate modes have macroscopic occupancy, and
these methods could be applied to BEC interferometry. These approaches
include the Projected Gross-Pitaevskii equation method \citep{Davis01a},
\citep{Blakie05a} and the Stochastic Gross-Pitaevskii equation theory
\citep{Gardiner02a}, \citep{Bradley05a}. A review of these methods
is given in \citep{Blakie08a}. In developing a quantum kinetic theory
of BECs, Gardiner and Zoller \citep{Gardiner98a}, \citep{Gardiner00a}
divided the field operator for the bosonic system as a sum of condensate
and non-condensate mode contributions. An alternative treatment also
based on distinguishing condensate and non-condensate modes is the
hybrid representation, with the highly occupied condensate modes described
via a truncated Wigner representation (since the bosons in condensate
modes behave like a classical mean field), whilst the basically unoccupied
non-condensate modes are described via a positive P representation
(these bosons should exhibit quantum effects). Such an approach has
been developed by \citep{Dalton07b}, \citep{Hoffmann08a}, \citep{Krachmalnicoff10a}
and in the present paper. Finally, a more elaborate phase space treatment
of BECs called the Gaussian quantum operator representation has been
formulated \citep{Corney03a} and could be applied to BEC interferometry.
Pairs of bosonic annihilation, creation operators as well as single
operators are represented by c-numbers in the phase space distribution
function. The approach is based on representing the density operator
via Gaussian rather than just coherent state projectors, as applies
for the simpler phase space theories.

As well as being suitable for studying macroscopic decoherence and
dephasing effects, interferometry with Bose-Einsten condensates is
closely linked to another fundamental feature of the quantum physics
in macroscopic systems - \emph{entanglement}. Entanglement is linked
to several important issues such as the EPR paradox, Bell inequalities
and Schrodinger cats. A number of papers have discussed entanglement
for two mode macroscopic systems, including \citep{Sorensen01a},
\citep{Simon02a}, \citep{Micheli03a}, \citep{Wiseman03a}, \citep{Hines03a},
\citep{Toth03a}, \citep{Esteve08a} and \citep{Goold09a}. Reviews
include \citep{Reid09a}, \citep{Amico09a}. Measures of entanglement
are more straightforward for bi-partite systems such as bosonic systems
based on two modes, where the two modes constitute the two subsystems.
The \emph{entropy of mode entanglement} is a useful measure, being
the difference in entropy between that for the original state and
that associated with the reduced density operator describing a sub-system,
and thus related to the change of \emph{quantum information}. The
connection to interferometry can be seen with a simple example \citep{Simon02a}.
If $\widehat{a}$, $\widehat{b}$ are the annihilation operators for
the modes $a$, $b$ then the pure quantum state for the $N$ boson
system given in terms of the corresponding creation operators and
the vacuum state $\left\vert 0\right\rangle $ as \begin{equation}
\left\vert \Phi\right\rangle _{E}=\frac{1}{\sqrt{N!}}\left(\frac{\widehat{a}^{\dagger}+\widehat{b}^{\dagger}}{\sqrt{2}}\right)^{N}\left\vert 0\right\rangle =\left(\frac{1}{\sqrt{2}}\right)^{N}\sum_{n=0}^{N}\sqrt{C_{n}^{N}}\,\left\vert n\right\rangle _{a}\left\vert N-n\right\rangle _{b}\label{Eq.PureEntangled}\end{equation}
is an \emph{entangled state}, being a quantum superposition of separable
states $\left\vert n\right\rangle _{a}\left\vert N-n\right\rangle _{b}$
in which there are $n$ bosons in mode $a$ and $N-n$ in mode $b$.
This state is a \emph{binomial state}, since its form is determined
by binomial coefficients $C_{n}^{N}$. The reduced density operator
for the mode $a$ subsystem is easily found to be\begin{equation}
\widehat{\rho}_{a}^{E}=\left(\frac{1}{2}\right)^{N}\sum_{n=0}^{N}C_{n}^{N}\,\left\vert n\right\rangle _{a}\left\langle n\right\vert _{a}\label{Eq.ReducedDensOprModeAEnt}\end{equation}
which is clearly a mixed state and the entropy of entanglement is
non-zero. Another pure state for the $N$ boson system is \begin{equation}
\left\vert \Phi\right\rangle _{NE}=\frac{1}{\sqrt{N!}}\left(\widehat{a}^{\dagger}\right)^{N}\left\vert 0\right\rangle =\left\vert N\right\rangle _{a}\left\vert 0\right\rangle _{b}\label{Eq.PureNonEntangled}\end{equation}
which is a \emph{non-entangled state}, being a separable product of
the states $\left\vert N\right\rangle _{a}$ and $\left\vert 0\right\rangle _{b}$.
The reduced density operator for the mode $a$ subsystem is easily
found to be\begin{equation}
\widehat{\rho}_{a}^{NE}=\,\left\vert N\right\rangle _{a}\left\langle N\right\vert _{a}\label{Eq.ReducedDensOprModeANonEnt}\end{equation}
which is clearly a pure state and the entropy of entanglement is zero.
If we now consider an interferometry experiment applied to each of
these two states, we will see that the entangled and non-entangled
states leads to differing interferometric effects. The experiment
involves applying a 50:50 beam--splitter process to each state and
then measuring the number of bosons in modes $a$, $b$. The beam
splitter process is associated with an evolution operator which transforms
the mode annihilation operators as $\widehat{a}\rightarrow(\widehat{a}+\widehat{b})/\sqrt{2}$,
$\widehat{b}\rightarrow(\widehat{a}-\widehat{b})/\sqrt{2}$. For single
component BEC in a double well with modes localised in each well,
such a process is associated with quantum tunneling through the potential
barrier during a period short enough that collisions can be ignored.
For two component BEC in a single well, the process is associated
with applying a two-photon classical field during a similar short
period. For the two initial states of interest the states change as
\begin{eqnarray}
\left\vert \Phi\right\rangle _{E} & \rightarrow & \frac{1}{\sqrt{N!}}\left(\widehat{a}^{\dagger}\right)^{N}\left\vert 0\right\rangle =\left\vert N\right\rangle _{a}\left\vert 0\right\rangle _{b}\nonumber \\
\left\vert \Phi\right\rangle _{NE} & \rightarrow & \frac{1}{\sqrt{N!}}\left(\frac{\widehat{a}^{\dagger}+\widehat{b}^{\dagger}}{\sqrt{2}}\right)^{N}\left\vert 0\right\rangle =\left(\frac{1}{\sqrt{2}}\right)^{N}\sum_{n=0}^{N}\sqrt{C_{n}^{N}}\,\left\vert n\right\rangle _{a}\left\vert N-n\right\rangle _{b}\label{Eq.PureStatesAfterBeamSplit}\end{eqnarray}
Measurements of the mean boson numbers in each mode give $\left\langle \widehat{a}^{\dagger}\widehat{a}\right\rangle =N$,
$\left\langle \widehat{b}^{\dagger}\widehat{b}\right\rangle =0$ for
the initially entangled state and $\left\langle \widehat{a}^{\dagger}\widehat{a}\right\rangle =N/2$,
$\left\langle \widehat{b}^{\dagger}\widehat{b}\right\rangle =N/2$
for the initially non-entangled state. Hence there is a difference
in the interferometric results for the two cases. More generally,
for an arbitrary \emph{mixed non-entangled} state for $N$ bosons
the density operator is of the form \begin{equation}
\widehat{\rho}^{NE}=\sum_{n=0}^{N}p_{n}\left\vert n\right\rangle _{a}\left\langle n\right\vert _{a}\otimes\left\vert N-n\right\rangle _{b}\left\langle N-n\right\vert _{b}\label{Eq.MixedNonEntangled}\end{equation}
and the reduced density operator for mode $a$ is \begin{equation}
\widehat{\rho}_{a}^{NE}=\sum_{n=0}^{N}p_{n}\left\vert n\right\rangle _{a}\left\langle n\right\vert _{a}\label{eq.ReducedDensOprModeAMixedNE}\end{equation}
As there is no entropy change between the original state and the state
for mode $a$, the entropy of entanglement is zero. In \citep{Simon02a}
it is shown that applying the beam-splitter process to this state
gives a new state where again $\left\langle \widehat{a}^{\dagger}\widehat{a}\right\rangle =N/2$,
$\left\langle \widehat{b}^{\dagger}\widehat{b}\right\rangle =N/2$.
Hence \emph{all} non-entangled states for for $N$ bosons give no
difference between the output measurements of boson numbers in the
two modes. This contrasts the situation for entangled states, as our
example has shown. Thus interferometry with BEC would be a possible
measurement system for demonstrating entanglement effects. 

In the present paper it will be assumed that the interferometry regime
is such that at most two condensate modes have a macroscopic occupancy.
The mean field theory treatment for this case is a time-dependent
version of the approach in an earlier two-mode theory paper \citep{Dalton07a}.
This approach leads to a set of self consistent equations for the
two mode functions and for the probability amplitudes for finding
the system in states with specific occupancies of the two modes. The
mode equations are generalised time-dependent Gross-Pitaevskii equations
involving non-linear mean field terms, and these equations include
coefficients that depend on the amplitudes. The amplitude equations
are matrix mechanics equations involving Hamiltonian and rotation
matrix elements, that depend on the mode functions and their spatial
and temporal derivatives. These self-consistent sets of equations
are derived via the Dirac-Frenkel variational principle. This generalised
mean field theory does allow for certain dephasing effects and for
transitions between the two condensate modes. Thermal and decoherence
effects are not included. For the purposes of the present paper it
will be assumed that the solutions to the generalised mean field two
mode theory have been obtained, and are available albeit in numerical
form for applications of the present theory. Numerical solutions of
equivalent equations have been published by Streltsov et al \citep{Streltsov07a},
\citep{Streltsov06a}, \citep{Alon08a}.

The present paper outlines a basic theoretical treatment of decoherence
and dephasing effects in interferometry based on single component
BECs in double potential wells, where we assume that only two condensate
modes could have macroscopic occupancy. Results for both two mode
condensates and the simpler single mode condensate case are presented.
The approach involves a hybrid phase space distribution functional
method where the condensate modes are described via a truncated Wigner
representation, whilst the basically unoccupied non-condensate modes
are described via a positive P representation \citep{Dalton07b},
\citep{Hoffmann08a}..The Hamiltonian for the system is described
in terms of quantum field operators for the condensate and non-condensate
modes. The functional Fokker-Planck equation for the double phase
space distribution functional is derived. Equivalent Ito stochastic
equations for the condensate and non-condensate fields that replace
the field operators are obtained and stochastic averages of products
of these fields give the quantum correlation functions that can be
used to interpret interferometry experiments. The treatment starts
from the generalised mean field theory for two condensate modes. Numerical
studies for interferometry experiments would involve using the solutions
from the generalised mean field theory in calculations for the stochastic
fields from the Ito stochastic field equations. 

Previous papers \citep{Graham70a}, \citep{Graham70b}, \citep{Drummond87a},
\citep{Gatti97a}, \citep{Steel98b} using distribution functional
and stochastic field approaches only contain brief explanations of
the method, so the present paper is aimed at a more complete exposition.
In Section \ref{Sect Ham & Field Oprs} the Hamiltonian for the single
component bosonic system is described in terms of field operators.
The field operators are the sum of condensate and non-condensate mode
contributions. The Hamiltonian is decomposed into contributions scaling
with decreasing powers of $\sqrt{N}$, and within the weak interaction
regime some terms are discarded, leaving a Hamiltonian which allows
for Bogoliubov excitations. Certain linear coupling terms involving
both condensate and non-condensate field operators are written in
a new form based aroud the condensate mode functions as obtained from
time-dependent Gross-Pitaevskii equations. In Section \ref{Sect Phase Space Dist Fnal}
phase space distribution functionals of a hybrid type are introduced
(Wigner for condensate fields, P+ for non-condensate fields) starting
with the characteristic functional, and quantum correlation functions
(symmetric ordering for condensate fields, normal ordering for non-condensate
fields) are expressed in terms of phase space functional integrals,
with field functions replacing the field operators and the distribution
functional replacing the density operator. The justification for these
phase space functional integral results is carefully outlined. Correspondence
rules and functional Fokker-Planck equations are obtained in Section
\ref{Sect Func Fokker Planck}, the key steps in the derivation of
the correspondence rules and functional Fokker-Planck equations being
explicitly covered. The derivation of the equivalent Ito stochastic
field equations is fully set out in Section \ref{Sect Ito Stochastic Eqns}.
Results for both two mode and single mode condensates are presented.
The single mode condensate results are compared with equations recently
presented in \citep{Krachmalnicoff10a}. The paper is summarised in
Section \ref{Sect Summary}.

Online supplementary material and a website version of this paper
\citep{Dalton10b} contains details for the derivations of results
in this paper which are too lengthy to present in the journal version.
Quantities involved in the two-mode theory equations are listed in
\ref{App: Quantities-1}. In Appendix B the key ideas of functional
calculus involving c-number functions are outlined. Results for quantum
correlation functions are derived in Appendix C. The derivation of
the correspondence rules and their application to deriving the functional
Fokker-Planck equation is given in Appendix D and Appendix E respectively.
The Ito stochastic equations details are in Appendix F. \pagebreak{}

\section{Hamiltonian and field operators}

\label{Sect Ham & Field Oprs}

In this section we describe the bosonic system in terms of field operators.
The field operators are written as the sum of two contributions, one
associated with the condensate modes, the other with the non-condensate
modes. The Hamiltonian is introduced for the single component bosonic
system within the zero range approximation for the boson-boson interactions.
The situation is restricted in this paper to the weak interaction
regime, and the Hamiltonian decomposed into contributions that scale
with decreasing powers of $\sqrt{N}$, where $N$ is the number of
bosons. After discarding the two smallest contibutions that scale
as $(\sqrt{N})^{-1}$ and $(\sqrt{N})^{-2}$, we are left with the
Bogoliubov Hamiltonian \citep{Gardiner97a}, \citep{Castin98a}, \citep{Girardeau98a},
\citep{Gardiner07a}. The condensate in this work dealing with applications
in double well interferometry is assumed to involve at most two modes,
and the Dirac-Frenkel principle \citep{Dirac30a}, \citep{Frenkel34a}
is used to obtain two coupled generalised Gross-Pitaevskii equations
for the two time-dependent mode functions. For the single condensate
mode situation the same approach gives the standard Gross-Pitaevskii
equation for the mode function. Our previous two mode theory \citep{Dalton07a}
yielded adiabatic mode functions, rather than the time-dependent modes
used here. Results from the Gross-Pitaevskki equations are then used
to simplify one of the terms in the Bogoliubov Hamiltonian, thereby
enabling functional Fokker-Planck equations to be derived.

\subsection{Field Operators for Condensate, Non-Condensate Modes}

For the application to double-well BEC interferometry most of the
bosons occupy one or maybe two modes, and that all the other modes
are essentially unoccupied. The two modes with macroscopic occupancy
will be referred to as the condensate modes, the remaining modes are
non-condensate modes. These physically based distinctions between
the two types of modes will be embodied in the theoretical treatment,
and it will be convenient to use two different phase space methods
for the condensate and non-condensate bosons. In the present paper
it is assumed that the interferometry regime is such that at most
two condensate modes have a macroscopic occupation.

The \emph{field operators} can be expanded in mode functions

\begin{eqnarray}
\widehat{\Psi}(\mathbf{r}) & = & \sum\limits _{k}\widehat{a}_{k}\phi_{k}(\mathbf{r})\label{Eq.FieldOpr1}\\
\widehat{\Psi}^{\dagger}(\mathbf{r}) & = & \sum\limits _{k}\phi_{k}^{\ast}(\mathbf{r})\widehat{a}_{k}^{\dagger}\label{Eq.FieldOpr2}\end{eqnarray}
where the mode functions are orthonormal\begin{equation}
\int d\mathbf{r}\phi_{i}^{\ast}(\mathbf{r})\phi_{j}(\mathbf{r})=\delta_{ij}\label{Eq.Orthog}\end{equation}
Throughout this paper both the mode functions and their accompanying
annihilation, creation operators are time dependent in general, but
for simplicity of notation the time dependence is usually left implicit.
Note however that the field operators $\widehat{\Psi}(\mathbf{r})$
and $\widehat{\Psi}^{\dagger}(\mathbf{r})$ are always time independent.

In the mode expansion we will assume that there is a cut-off at some
large mode number $K$ (momentum cut-off). This is to be consistent
with using the zero range approximation in the Hamiltonian. Accordingly
the completeness expression for the mode functions does not give the
ordinary delta function but a restricted delta function $\delta_{K}(\mathbf{r},\mathbf{r}^{\prime})$
which is no longer singular when $\mathbf{r}=\mathbf{r}^{\prime}$.
\begin{equation}
\sum\limits _{k}\phi_{k}(\mathbf{r})\phi_{k}^{\ast}(\mathbf{r}^{\prime})=\delta_{K}(\mathbf{r},\mathbf{r}^{\prime})\label{Eq.Completeness1}\end{equation}

Accordingly although the annihilation, creation operators satisfy
the standard bosonic commutation rules, the field operators satisfy
modified rules for which the non-zero results are\begin{eqnarray}
\lbrack\widehat{a}_{k},\widehat{a}_{l}^{\dagger}] & = & \delta_{kl}\nonumber \\
\lbrack\widehat{\Psi}(\mathbf{r}),\widehat{\Psi}^{\dagger}(\mathbf{r}^{\prime})] & = & \delta_{K}(\mathbf{r,r}^{\prime})\label{Eq.BoseCommutationRules}\end{eqnarray}
In obtaining these rules those for the annihilation, creation operators
are treated as fundamental and those for the field operators then
derived. If the cut-off is made very large then the restricted delta
function approaches the ordinary delta function. 

To exploit the distinction between condensate modes with a macroscopic
occupancy and non-condensate mode the field operator is written as
the sum of a \emph{condensate} term and a \emph{non-condensate} term.
In the \emph{two-mode approximation} it is assumed that there are
two condensate modes that may have macroscopic occupancy, in the standard
\emph{single mode approximation} only one.

For the \emph{two mode} case

\begin{eqnarray}
\widehat{\Psi}(\mathbf{r}) & = & \widehat{\Psi}_{C}(\mathbf{r})+\widehat{\Psi}_{NC}(\mathbf{r})\label{Eq.FieldOpr}\\
\widehat{\Psi}_{C}(\mathbf{r}) & = & \widehat{a}_{1}\phi_{1}(\mathbf{r})+\widehat{a}_{2}\phi_{2}(\mathbf{r})\label{Eq.CondOpr}\\
\widehat{\Psi}_{C}^{\dagger}(\mathbf{r}) & = & \phi_{1}^{\ast}(\mathbf{r})\widehat{a}_{1}^{\dagger}+\phi_{2}^{\ast}(\mathbf{r})\widehat{a}_{2}^{\dagger}\label{Eq.CondOpr2}\\
\widehat{\Psi}_{NC}(\mathbf{r}) & = & \sum\limits _{k\neq1,2}^{K}\widehat{a}_{k}\phi_{k}(\mathbf{r})\label{Eq.NonCondOpr}\\
\widehat{\Psi}_{NC}^{\dagger}(\mathbf{r}) & = & \sum\limits _{k\neq1,2}^{K}\phi_{k}^{\ast}(\mathbf{r})\widehat{a}_{k}^{\dagger}\label{Eq.NonCondOpr2}\end{eqnarray}
where the condensate is described via the two modes $\phi_{1}(\mathbf{r})$,
$\phi_{2}(\mathbf{r})$ and the non-condensate via the remaining modes
$\phi_{k}(\mathbf{r})$, which are cut off for momenta greater than
$K\sim\hbar/a$ where $a$ is the distance scale of the.short range
boson-boson interaction. In view of the orthogonality of the condensate
and non-condensate modes, the contributions to the field operator
commute. For the condensate and non-condensate field operator components
we have the following non-zero results\begin{eqnarray}
\lbrack\widehat{\Psi}_{C}(\mathbf{r}),\widehat{\Psi}_{NC}^{\dagger}(\mathbf{r})] & = & 0\nonumber \\
\lbrack\widehat{\Psi}_{C}(\mathbf{r}),\widehat{\Psi}_{C}^{\dagger}(\mathbf{r}^{\prime})] & = & \phi_{1}(\mathbf{r})\phi_{1}^{\ast}(\mathbf{r}^{\prime})+\phi_{2}(\mathbf{r})\phi_{2}^{\ast}(\mathbf{r}^{\prime})\nonumber \\
 & = & \delta_{C}(\mathbf{r,r}^{\prime})\label{Eq.RestrictedCondDeltaTwoMode}\\
\lbrack\widehat{\Psi}_{NC}(\mathbf{r}),\widehat{\Psi}_{NC}^{\dagger}(\mathbf{r}^{\prime})] & = & \sum\limits _{k\neq1,2}\phi_{k}(\mathbf{r})\phi_{k}^{\ast}(\mathbf{r}^{\prime})\nonumber \\
 & = & \delta_{NC}(\mathbf{r,r}^{\prime})\label{Eq.BoseCommutationRules2}\end{eqnarray}
The quantities $\delta_{C}(\mathbf{r,r}^{\prime})$ and $\delta_{NC}(\mathbf{r,r}^{\prime})$
act as restricted Dirac delta functions rather than ordinary delta
functions, in that for functions $\psi_{C}(\mathbf{r})$ and $\psi_{NC}(\mathbf{r})$
only involving condensate or non-consensate modes respectively (and
$\psi_{C}^{+}(\mathbf{r})$ and $\psi_{NC}^{+}(\mathbf{r})$ only
involving their complex conjugates), we have\begin{eqnarray}
\psi_{C}(\mathbf{r}) & = & \alpha_{1}\phi_{1}(\mathbf{r})+\alpha_{2}\phi_{2}(\mathbf{r})\qquad\psi_{C}^{+}(\mathbf{r})=\phi_{1}^{\ast}(\mathbf{r})\alpha_{1}^{+}+\phi_{2}^{\ast}(\mathbf{r})\alpha_{2}^{+}\label{eq:CondFieldFn}\\
\psi_{NC}(\mathbf{r}) & = & \sum\limits _{k\neq1,2}\alpha_{k}\phi_{k}(\mathbf{r})\qquad\psi_{NC}^{+}(\mathbf{r})=\sum\limits _{k\neq1,2}\phi_{k}^{\ast}(\mathbf{r})\alpha_{k}^{+}\label{eq:NonCondFieldFn}\\
\psi_{C}(\mathbf{r}) & = & \int d\mathbf{r}^{\prime}\mathbf{\,}\delta_{C}(\mathbf{r,r}^{\prime})\psi_{C}(\mathbf{r}^{\prime})\qquad\psi_{C}^{+}(\mathbf{r})=\int d\mathbf{r}^{\prime}\mathbf{\,}\psi_{C}^{+}(\mathbf{r}^{\prime})\delta_{C}(\mathbf{r}^{\prime}\mathbf{,r})\nonumber \\
\psi_{NC}(\mathbf{r}) & = & \int d\mathbf{r}^{\prime}\mathbf{\,}\delta_{NC}(\mathbf{r,r}^{\prime})\psi_{NC}(\mathbf{r}^{\prime})\,\,\,\,\,\,\psi(\mathbf{r})=\int d\mathbf{r}^{\prime}\mathbf{\,}\psi_{C}^{+}(\mathbf{r}^{\prime})\delta_{C}(\mathbf{r}^{\prime}\mathbf{,r})\label{Eq.RestrictedDeltaFns}\end{eqnarray}
Clearly\begin{equation}
\delta_{K}(\mathbf{r,r}^{\prime})=\delta_{C}(\mathbf{r,r}^{\prime})+\delta_{NC}(\mathbf{r,r}^{\prime})\label{Eq.Completeness2}\end{equation}
These features involving restricted delta functions will be useful
in deriving the functional Fokker-Planck equation.

In the \emph{single mode} case the condensate, non-condensate field
operators and restricted delta functions are now given by \begin{eqnarray}
\widehat{\Psi}_{C}(\mathbf{r}) & = & \widehat{a}_{1}\phi_{1}(\mathbf{r})\qquad\widehat{\Psi}_{C}^{\dagger}(\mathbf{r})=\phi_{1}^{\ast}(\mathbf{r})\widehat{a}_{1}^{\dagger}\label{Eq.CondFieldOprsSingleMode}\\
\widehat{\Psi}_{NC}(\mathbf{r}) & = & \sum\limits _{k\neq1}^{K}\widehat{a}_{k}\phi_{k}(\mathbf{r})\qquad\widehat{\Psi}_{NC}^{\dagger}(\mathbf{r})=\sum\limits _{k\neq1}^{K}\phi_{k}^{\ast}(\mathbf{r})\widehat{a}_{k}^{\dagger}\label{Eq.NonCondFieldOprsSingleMode}\\
\delta_{C}(\mathbf{r,r}^{\prime}) & = & \phi_{1}(\mathbf{r})\phi_{1}^{\ast}(\mathbf{r}^{\prime})\qquad\delta_{NC}(\mathbf{r,r}^{\prime})=\sum\limits _{k\neq1}\phi_{k}(\mathbf{r})\phi_{k}^{\ast}(\mathbf{r}^{\prime})\label{Eq.RestrictedDeltaFnsSingleMode}\end{eqnarray}
with the time dependences of the mode functions and annihilation,
creation operators left understood as usual. With obvious modifications,
(\ref{Eq.RestrictedCondDeltaTwoMode}) - (\ref{Eq.Completeness2})
also apply in the single mode case.

\subsection{Bogoliubov Hamiltonian}

The full \emph{Hamiltonian} in terms of field operators is given by\begin{eqnarray}
\widehat{H} & = & \int d\mathbf{r(}\frac{\hbar^{2}}{2m}\nabla\widehat{\Psi}(\mathbf{r})^{\dagger}\cdot\nabla\widehat{\Psi}(\mathbf{r})+\widehat{\Psi}(\mathbf{r})^{\dagger}V\widehat{\Psi}(\mathbf{r})+\frac{g}{2}\widehat{\Psi}(\mathbf{r})^{\dagger}\widehat{\Psi}(\mathbf{r})^{\dagger}\widehat{\Psi}(\mathbf{r})\widehat{\Psi}(\mathbf{r}))\nonumber \\
\label{Eq.HamiltFieldOprs}\\ & = & \widehat{K}+\widehat{V}+\widehat{U}\label{Eq.HamiltonianContns}\end{eqnarray}
the sum of a kinetic energy, trap potential energy and boson-boson
interaction energy terms. As usual the zero range approximation is
made with $g=4\pi\hbar^{2}a_{S}/m$, where $a_{S}$ is the s-wave
scattering length.

The condensate mode occupation is of order the total boson number
$N$. For bosons in a trap of frequency $\omega$, with harmonic oscillator
length scale $a_{0}=\sqrt{(\hbar/2m\omega)}$, the density is of order
$\rho=N/(a_{0})^{3}$. In the \emph{weak interaction regime} \citep{Castin98a}
we have $\rho(a_{S})^{3}\ll1$, or \begin{equation}
N(\frac{a_{S}}{a_{0}})^{3}\ll1\label{Eq.WeakIntnRegime}\end{equation}
For Rb$^{87}$ in a trap with $\omega=2\pi.58$ s$^{-1}$ we have
$a_{0}=1$ $\mu$m and $a_{S}=5$ nm, so that the weak interaction
regime applies for reasonably large boson numbers $N\ll10^{7}$. Also,
as has been shown \citep{Castin98a}, it is possible to consider a
situation for the weak interaction regime where $Na_{S}/a_{0}$ and
$k_{B}T/\hbar\omega$ are kept constant but with $N$ becoming very
large whilst $a_{S}$ remains finite so that \begin{equation}
gN=g_{N}\label{Eq.ConditionG}\end{equation}
where $g_{N}$ is constant. This can be achieved by decreasing the
trap frequency.

In the weak interaction regime and with $g=g_{N}/N$ it is convenient
to write the Hamiltonian as the sum of five terms in decreasing powers
of $\sqrt{N}$, based on using Eq.(\ref{Eq.FieldOpr}) and assuming
the condensate operators scale like $\sqrt{N}$. We can then express
the Hamiltonian in the form\begin{equation}
\widehat{H}=\widehat{H}_{1}+\widehat{H}_{2}+\widehat{H}_{3}+\widehat{H}_{4}+\widehat{H}_{5}\label{Eq.HamOrdersN}\end{equation}
where\begin{eqnarray}
\widehat{H}_{1} & = & \int d\mathbf{r}(\frac{\hbar^{2}}{2m}\nabla\widehat{\Psi}_{C}^{\dagger}(\mathbf{r})\cdot\nabla\widehat{\Psi}_{C}(\mathbf{r})+\widehat{\Psi}_{C}^{\dagger}(\mathbf{r})V\widehat{\Psi}_{C}(\mathbf{r})\nonumber \\
 &  & +\frac{g_{N}}{2N}\widehat{\Psi}_{C}^{\dagger}(\mathbf{r})\widehat{\Psi}_{C}^{\dagger}(\mathbf{r})\widehat{\Psi}_{C}(\mathbf{r})\widehat{\Psi}_{C}(\mathbf{r}))\label{Eq.H1}\end{eqnarray}
\begin{eqnarray}
\widehat{H}_{2} & = & \int d\mathbf{r(}\widehat{\Psi}_{NC}(\mathbf{r})^{\dagger}\,\left\{ \mathbf{-}\frac{\hbar^{2}}{2m}\nabla^{2}\widehat{\Psi}_{C}(\mathbf{r})+V\widehat{\Psi}_{C}(\mathbf{r})+\frac{g_{N}}{N}\widehat{\Psi}_{C}^{\dagger}(\mathbf{r)}\widehat{\Psi}_{C}(\mathbf{r})\widehat{\Psi}_{C}(\mathbf{r})\right\} \nonumber \\
 &  & +\left\{ -\frac{\hbar^{2}}{2m}\nabla^{2}\widehat{\Psi}_{C}^{\dagger}(\mathbf{r)}+\widehat{\Psi}_{C}(\mathbf{r})^{\dagger}V+\frac{g_{N}}{N}\widehat{\Psi}_{C}^{\dagger}(\mathbf{r})\widehat{\Psi}_{C}^{\dagger}(\mathbf{r})\widehat{\Psi}_{C}(\mathbf{r})\right\} \widehat{\Psi}_{NC}(\mathbf{r}))\label{Eq.H2}\end{eqnarray}
\begin{eqnarray}
\widehat{H}_{3} & = & \int d\mathbf{r}(\frac{\hbar^{2}}{2m}\nabla\widehat{\Psi}_{NC}^{\dagger}(\mathbf{r})\cdot\nabla\widehat{\Psi}_{NC}(\mathbf{r})+\widehat{\Psi}_{NC}^{\dagger}(\mathbf{r)}V\widehat{\Psi}_{NC}(\mathbf{r})\nonumber \\
 &  & +\frac{g_{N}}{2N}\left\{ \widehat{\Psi}_{NC}^{\dagger}(\mathbf{r})\widehat{\Psi}_{NC}^{\dagger}(\mathbf{r})\widehat{\Psi}_{C}(\mathbf{r})\widehat{\Psi}_{C}(\mathbf{r})+\widehat{\Psi}_{C}^{\dagger}(\mathbf{r})\widehat{\Psi}_{C}^{\dagger}(\mathbf{r})\widehat{\Psi}_{NC}(\mathbf{r})\widehat{\Psi}_{NC}(\mathbf{r})\right\} \nonumber \\
 &  & +\frac{g_{N}}{2N}\left\{ 4\widehat{\Psi}_{NC}^{\dagger}(\mathbf{r})\widehat{\Psi}_{C}^{\dagger}(\mathbf{r})\widehat{\Psi}_{NC}(\mathbf{r})\widehat{\Psi}_{C}(\mathbf{r})\right\} )\label{eq:H3}\end{eqnarray}
\begin{equation}
\widehat{H}_{4}=\int d\mathbf{r\,}\frac{g_{N}}{N}\left\{ \widehat{\Psi}_{NC}^{\dagger}(\mathbf{r})\widehat{\Psi}_{NC}^{\dagger}(\mathbf{r})\widehat{\Psi}_{NC}(\mathbf{r})\widehat{\Psi}_{C}(\mathbf{r}))+\widehat{\Psi}_{C}^{\dagger}(\mathbf{r})\widehat{\Psi}_{NC}^{\dagger}(\mathbf{r})\widehat{\Psi}_{NC}(\mathbf{r})\widehat{\Psi}_{NC}(\mathbf{r})\right\} \label{Eq.H4}\end{equation}
\begin{equation}
\widehat{H}_{5}=\int d\mathbf{r\,}\frac{g_{N}}{2N}\left\{ \widehat{\Psi}_{NC}^{\dagger}(\mathbf{r})\widehat{\Psi}_{NC}^{\dagger}(\mathbf{r})\widehat{\Psi}_{NC}(\mathbf{r})\widehat{\Psi}_{NC}(\mathbf{r})\right\} \label{Eq.H5}\end{equation}
The term $\widehat{H}_{1}$ is of order $N=(\sqrt{N})^{2}$ and is
the Hamiltonian for the \emph{condensate}. The term $\widehat{H}_{2}$
is of order $\sqrt{N}=(\sqrt{N})^{1}$ and describes part of the interaction
between the \emph{condensate} and the \emph{non-condensate} that is
linear in the non-condensate field. To obtain this term spatial integration
by parts was used to have $\nabla$ only operate on $\widehat{\Psi}_{C}(\mathbf{r})$
or $\widehat{\Psi}_{C}(\mathbf{r})^{\dagger}$. This term needs further
development to avoid Fokker-Planck equations containing functional
derivatives with respect to spatial derivatives of field functions,
and this is accomplished in the next section. The term $\widehat{H}_{3}$
is of order $1=(\sqrt{N})^{0}$ and describes part of the interaction
between the condensate and the non-condensate that is quadratic in
the non-condensate field.plus the kinetic and trap potential terms
for the non-condensate. If the condensate fields are replaced by c-numbers,
this term describes Bogoliubov excitations \citep{Castin98a}, \citep{Gardiner97a}.
The term $\widehat{H}_{4}$ is of order $1/\sqrt{N}=(\sqrt{N})^{-1}$
and describes part of the interaction between the condensate and the
non-condensate that is cubic in the non-condensate field. The term
$\widehat{H}_{5}$ is of order $1/N=(\sqrt{N})^{-2}$ and describes
part of the interaction within the non-condensate, which is quartic
in the non-condensate field.

We now make an approximation and neglect the terms $\widehat{H}_{4}$
and $\widehat{H}_{5}$. This leads to the so-called \emph{Bogoliubov
Hamiltonian}, albeit still in a number conserving form. This Hamiltonian
would be adequate to describe Bogoliubov excitations, so we will use
in to treat BEC interferometry in the weak interaction regime. The
Bogoliubov Hamiltonian is \begin{equation}
\widehat{H}_{B}=\widehat{H}_{1}+\widehat{H}_{2}+\widehat{H}_{3}\label{Eq.BogolHamiltonian}\end{equation}
The neglected terms would be needed in a theory for BEC interferometry
in the strong interaction regime.

\subsection{Two-Mode Theory and Generalised Gross-Pitaevski Equations}

The development of a suitable form for the $\widehat{H}_{2}$ term
in the case where \emph{two condensate modes} are involved is based
on a general two mode theory for one component BECs similar to that
in \citep{Dalton07a}, though here we apply the Dirac-Frenkel principle
to the dynamic action and obtain Gross-Pitaevski equations for time-dependent
mode functions, rather than the time independent Gross-Pitaevskii
equations for adiabatic mode functions obtained in \citep{Dalton07a}
by applying a variational principle to the adiabatic action and involving
Lagrange multipliers. In two-mode theories we write the \emph{quantum
state} $\left\vert \,\Phi(t)\right\rangle $ of the $N$ boson system
as a superposition of the $N+1$ \emph{basis states} $\left\vert \,\frac{{\small N}}{{\small2}},k\right\rangle $,
where there are $\frac{{\small N}}{{\small2}}-k$ and $\frac{{\small N}}{{\small2}}+k$
bosons (respectively) occupying the two modes with (time dependent)
\emph{mode functions} $\phi_{1}(\mathbf{r},t)$ and $\phi_{2}(\mathbf{r},t)$.
The \emph{amplitude} for this basis state is $b_{k}(t)$. \begin{equation}
\left\vert \,\Phi(t)\right\rangle =\sum\limits _{k=-\frac{\mathbf{N}}{\mathbf{2}}}^{\frac{\mathbf{N}}{\mathbf{2}}}\, b_{k}(t)\,\left\vert \,\frac{{\small N}}{{\small2}},k\right\rangle .\label{Eq.TwoModeQState}\end{equation}
and the basis states are \emph{Fock states} given by \begin{equation}
\left\vert \,\frac{{\small N}}{{\small2}},k\right\rangle =\frac{\left(\widehat{a}_{1}(t)^{\dagger}\right)^{(\frac{{\LARGE N}}{{\LARGE2}}-k)}}{[(\frac{N}{2}-k)!]^{\frac{\mathbf{1}}{\mathbf{2}}}}\frac{\left(\widehat{a}_{2}(t)^{\dagger}\right)^{(\frac{{\LARGE N}}{{\LARGE2}}+k)}}{[(\frac{N}{2}+k)!]^{\frac{\mathbf{1}}{\mathbf{2}}}}\left\vert \,0\right\rangle \qquad(k=-N/2,-N/2+1,..,+N/2)\label{Eq.OrthonormalBasisStates}\end{equation}
These basis states are \emph{fragmented} or \emph{number squeezed}
states, allowing for both modes to have \emph{macroscopic occupancy
}when $|k|\ll N/2$.

The notation $\frac{{\small N}}{{\small2}},k$ for the basis states
reflects the feature that the two mode Bose condensate behaves like
a giant spin system. \emph{Spin angular momentum} operators can be
defined by \begin{eqnarray}
\widehat{S}_{x} & = & (\widehat{a}_{2}^{\dagger}\widehat{a}_{1}+\widehat{a}_{1}^{\dagger}\widehat{a}_{2})/2\nonumber \\
\widehat{S}_{y} & = & (\widehat{a}_{2}^{\dagger}\widehat{a}_{1}-\widehat{a}_{1}^{\dagger}\widehat{a}_{2})/2i\nonumber \\
\widehat{S}_{z} & = & (\widehat{a}_{2}^{\dagger}\widehat{a}_{2}-\widehat{a}_{1}^{\dagger}\widehat{a}_{1})/2\label{Eq.SpinOprs}\end{eqnarray}
which satisfy the standard angular momentum commutation rules. The
\emph{square} of the angular momentum $(\underrightarrow{\widehat{S}})^{2}$
is related to the total two mode boson number operator $\widehat{N}=(\widehat{a}_{2}^{\dagger}\widehat{a}_{2}+\widehat{a}_{1}^{\dagger}\widehat{a}_{1})$
\begin{equation}
(\underrightarrow{\widehat{S}})^{2}=\sum\limits _{a}(\widehat{S}_{a})^{2}=\frac{\widehat{N}}{2}(\frac{\widehat{N}}{2}+1)\label{Eq.AngMtmSquared}\end{equation}
and the Fock state $\left\vert \,\frac{{\small N}}{{\small2}},k\right\rangle $
is a simultaneous eigenstate of $(\underrightarrow{\widehat{S}})^{2},\widehat{S}_{z}$
\begin{eqnarray}
(\underrightarrow{\widehat{S}})^{2}\,\left\vert \frac{{\small N}}{{\small2}},k\right\rangle  & = & \frac{{\small N}}{{\small2}}(\frac{{\small N}}{{\small2}}+{\small1})\,\left\vert \frac{{\small N}}{{\small2}},k\right\rangle \nonumber \\
\widehat{S}_{z}\,\left\vert \frac{{\small N}}{{\small2}},k\right\rangle  & = & k\,\left\vert \frac{{\small N}}{{\small2}},k\right\rangle \label{Eq.SpinEigenstates}\end{eqnarray}
Hence the \emph{total angular momentum} quantum number $j=\frac{{\small N}}{{\small2}}$
is \emph{macroscopic}, and $k=-\frac{{\small N}}{{\small2}},-\frac{{\small N}}{{\small2}}{\small+1,..,}\frac{{\small N}}{{\small2}}{\small-1,+}\frac{{\small N}}{{\small2}}$
specifies the \emph{magnetic} quantum number as well as $2k$ determining
the difference in \emph{mode occupancy}.

In \citep{Dalton07a} equations for the amplitudes and adiabatic mode
functions have been determined by applying Principles of Least Action
, involving minimising the \emph{dynamic} and \emph{adiabatic actions}
respectively for the state vector given by (\ref{Eq.TwoModeQState}),
subject to the normalisation constraints for the amplitude and orthonormality
constraints for the mode functions \begin{eqnarray}
\sum\limits _{k=-\frac{\mathbf{N}}{\mathbf{2}}}^{\frac{\mathbf{N}}{\mathbf{2}}}\left\vert b_{k}(t)\right\vert ^{2} & = & 1\nonumber \\
\int d{\small\mathbf{r}}\mathbf{\,}\phi_{i}^{\ast}(\mathbf{r},t)\,\phi_{j}(\mathbf{r},t) & {\small=} & {\small\,}\delta_{ij}\qquad i,j=1,2\label{Eq.OrthoNorm}\end{eqnarray}
In the present treatment, the Dirac-Frenkel principle \citep{Dirac30a},
\citep{Frenkel34a} is applied to the dynamic action to obtain equations
for the amplitudes $b_{k}(t)$.and time-dependent mode functions $\phi_{i}(\mathbf{r},t)$
$(i=1,2)$. In applying the Dirac-Frenkel principle no Lagrange multipliers
associated with the equations of constraint (\ref{Eq.OrthoNorm})
are introduced, however mode orthonormality is used in the treatment
and the final amplitude and mode equations can be shown to be consistent
with both these constraints. Such variational principles are well-known
in quantum physics, the Dirac-Frenkel principle applied to the dynamic
action for an arbitrary unnormalised state vector gives the time-dependent
Schrodinger equation. \ The Hartree-Fock equations for electrons
in atoms and molecules and the Gross-Pitaevskii equations for a single
mode condensate are two examples of their application based on state
vectors with restricted forms. In the latter case the state vector
assumed is a special case of (\ref{Eq.TwoModeQState}) such as with
just the single term $\left\vert \,\frac{{\small N}}{{\small2}},-\frac{{\small N}}{{\small2}}\right\rangle $
or a special superposition (\emph{binomial state}) corresponding to
all bosons being in the same single particle state \citep{Dalton07a},
itself a linear combination of the two original modes.

In the present case of\emph{\ two condensate modes}, the mode functions
satisfy the \emph{coupled generalised} \emph{Gross-Pitaevskii }equations
\begin{eqnarray}
i\hbar\sum\limits _{j}X_{ij}\,\frac{\partial}{\partial t}\phi_{j} & = & \sum\limits _{j}X_{ij}(-\frac{\hbar^{2}}{2m}\nabla^{2}+V)\,\phi_{j}\nonumber \\
 &  & +\sum\limits _{j}(g\sum\limits _{mn}Y_{im\, jn}\,\phi_{m}^{\ast}\,\phi_{n})\,\phi_{j}\qquad\qquad(i=1,2).\label{Eq.GenGrossPitEqns}\end{eqnarray}
These mode functions allow for boson-boson interactions and are time-dependent.
They follow the changes in the time dependent potential $V(\mathbf{r},t)$.
The quantities $X_{ij}$ and $Y_{im\, jn}$ are \emph{one-body} and
\emph{two-body} \emph{correlation functions} \begin{eqnarray}
X_{ij} & = & \left\langle \Phi\right\vert \widehat{a}_{i}^{\dagger}\widehat{a}_{j}\left\vert \Phi\right\rangle \label{Eq.OneBodyCorrFn}\\
Y_{im\, jn} & = & \left\langle \Phi\right\vert \widehat{a}_{i}^{\dagger}\widehat{a}_{m}^{\dagger}\widehat{a}_{j}\widehat{a}_{n}\left\vert \Phi\right\rangle \label{Eq.TwoBodyCorrFn}\end{eqnarray}
Detailed expressions given in the \ref{App: Quantities-1} in Eqs.
(\ref{eq:OneBodyCorrnFn}) and (\ref{eq:TwoBodyCorrnFn}), showing
that $X_{ij}$ and $Y_{im\, jn}$ are quadratic forms of the amplitudes
$b_{k}$. These are of order $N$ and $N^{2}$ respectively. The one
body correlation functions can be expressed in terms of matrix elements
between the Fock states of the spin operators, and the two body correlation
functions as matrix elements products of two spin operators. The coupled
Gross-Pitaevskii equations are non-linear in the mode functions. The
non-linear terms $(g\sum\limits _{mn}Y_{im\, jn}\,\phi_{m}^{\ast}\,\phi_{n})$
that are present due to the boson-boson interactions scale like the
boson particle density and behave as generalised mean fields. Hence
the approach that produces generalised Gross-Pitaevskii equations
is a form of \emph{mean field theory}, though not of course as simple
as in the case of a single mode theory. The kinetic energy and trap
potential terms and the mean field terms may also be written as \begin{eqnarray}
 &  & \sum\limits _{j}X_{ij}(-\frac{\hbar^{2}}{2m}\nabla^{2}+V)\,\phi_{j}(\mathbf{r},t)\nonumber \\
 & = & \sum\limits _{j}\left\langle \Phi\right\vert \widehat{a}_{i}^{\dagger}\left\{ -\frac{\hbar^{2}}{2m}\nabla_{\mathbf{r}}^{2}+V(\mathbf{r},t)\right\} \widehat{a}_{j}\left\vert \Phi\right\rangle \,\phi_{j}(\mathbf{r},t)\label{Eq.KineticTrapTerms}\\
 &  & \sum\limits _{j}(g\sum\limits _{mn}Y_{im\, jn}\,\phi_{m}^{\ast}\,\phi_{n})\,\phi_{j}(\mathbf{r},t)\nonumber \\
 & = & \sum\limits _{j}\left\langle \Phi\right\vert \widehat{a}_{i}^{\dagger}\left\{ g\widehat{\Psi}_{C}(\mathbf{r})^{\dagger}\widehat{\Psi}_{C}(\mathbf{r})\right\} \widehat{a}_{j}\left\vert \Phi\right\rangle \,\phi_{j}(\mathbf{r},t)\label{Eq.MeanFieldTerms}\end{eqnarray}
showing the formal relationship of the terms to the state vector $\left\vert \Phi\right\rangle $.

For the present two mode condensate case the amplitudes satisfy \emph{matrix
mechanics} equations, as in \citep{Dalton07a} \begin{equation}
i\hbar\frac{\partial b_{k}}{\partial t}=\sum\limits _{l}(H_{kl}-\hbar U_{kl})b_{l}\qquad(k=-N/2,..,N/2).\label{Eq.AmpEqns}\end{equation}
These $N+1$ equations (\ref{Eq.AmpEqns}) describe the system dynamics
as it evolves amongst the possible fragmented states. The equations
are similar to the standard amplitude equations obtained from matrix
mechanics. In these equations the matrix elements $H_{kl}$, $U_{kl}\ $depend
on the mode functions $\phi_{i}(r,t)$. Detailed expressions for $H_{kl}$,
$U_{kl}$ are given in \ref{App: Quantities-1} in Eqs.(\ref{Eq.HamMatrixExpn-1})
and (\ref{Eq.RotMatExpn-1}). The matrix elements $H_{kl}$ are in
fact the matrix elements of the Hamiltonian $\widehat{H}$ in equation
(\ref{Eq.H1}) between the fragmented states $\left\vert \,\frac{{\small N}}{{\small2}},k\right\rangle $,
$\left\vert \,\frac{{\small N}}{{\small2}},l\right\rangle $. The
matrix elements $U_{kl}$ are elements of the so-called \emph{rotation
matrix}, and allow for the time dependence of the mode functions.
\begin{eqnarray}
H_{kl} & = & \left\langle \frac{{\small N}}{{\small2}},k\right\vert \widehat{H}\left\vert \frac{{\small N}}{{\small2}},l\right\rangle \label{Eq.HamMatrix}\\
U_{kl} & = & \frac{1}{2i}\left(\left\{ \partial_{t}\left\langle \frac{{\small N}}{{\small2}},k\right\vert \right\} \left\vert \frac{{\small N}}{{\small2}},l\right\rangle -\left\langle \frac{{\small N}}{{\small2}},k\right\vert \left\{ \partial_{t}\left\vert \frac{{\small N}}{{\small2}},l\right\rangle \right\} \right)\label{Eq.RotationMatrix}\end{eqnarray}
The specific forms of the $X_{ij}$, $Y_{im\, jn}$, $H_{kl}$, $U_{kl}$
are not important in what follows, all that is required is that they
have been determined. Equations for the mode functions and amplitudes
similar to (\ref{Eq.GenGrossPitEqns}) and (\ref{Eq.AmpEqns}) have
been obtained by Alon et al \citep{Alon08a} for single component
BECs.

From the amplitude and mode equations it can be shown that\begin{eqnarray}
\frac{\partial}{\partial t}\sum\limits _{k=-\frac{\mathbf{N}}{\mathbf{2}}}^{\frac{\mathbf{N}}{\mathbf{2}}}\left\vert b_{k}(t)\right\vert ^{2} & = & 0\label{Eq.AmplitudeNorm}\\
i\hbar\sum\limits _{ij}X_{ij}\,\frac{\partial}{\partial t}\int d{\small\mathbf{r}}\mathbf{\,}\phi_{i}^{\ast}(\mathbf{r},t)\,\phi_{j}(\mathbf{r},t) & = & 0\label{Eq.ModeOrthonorm}\end{eqnarray}
The first result shows that the amplitudes would remain normalised
to unity and the second result is consistent with the modes remaining
orthogonal and normalised, assuming they were so chosen at $t=0$.
The second result involves the trace of the product of a positive
definite invertible matrix $X$ with a matrix which is the time derivative
of the mode orthogonality matrix.

Adiabatic solutions to the time dependent coupled Gross-Pitaevskii
equations can be obtained for slowly varying trap potentials via the
transformation to new \emph{adiabatic modes} $\xi_{k}(\mathbf{r},t)$
$(k=1,2)$ in the form\begin{equation}
\phi_{i}(\mathbf{r},t)=\sum_{k}\alpha_{ki}\,\exp(-i\mu_{k}t)\,\xi_{k}(\mathbf{r},t)\label{Eq.TransfAdiabaticModes}\end{equation}
where it is assumed that the coefficients $\alpha_{ki}$ and the new
modes $\xi_{k}(\mathbf{r},t)$ are so slowly varying with time that
their time derivatives can be ignored. All the time dependence is
assumed to be carried in the oscillating exponential factors. The
new modes are required to be orthonormal, and the frequency factors
$\mu_{k}$ are required to be real, so that the transformation does
not diverge for large $|t|$. The orthonormality condition shows that
the $\alpha_{ki}$ form a unitary matrix.\begin{equation}
\sum_{k}\alpha_{ki}^{\ast}\,\alpha_{kj}=\delta_{ij}\qquad\sum_{i}\alpha_{ki}^{\ast}\,\alpha_{li}=\delta_{kl}\label{Eq.UnitaryCondns}\end{equation}
The condensate field operator can also be expressed in terms of the
adiabatic mode functions and their associated annihilation, creation
operators as\begin{equation}
\widehat{\Psi}_{C}(\mathbf{r})=\widehat{b}_{1}\xi_{1}(\mathbf{r})+\widehat{b}_{2}\xi_{2}(\mathbf{r})\qquad\widehat{\Psi}_{C}^{\dagger}(\mathbf{r})=\xi_{1}^{\ast}(\mathbf{r})\widehat{b}_{1}^{\dagger}+\xi_{2}^{\ast}(\mathbf{r})\widehat{b}_{2}^{\dagger}\label{Eq.CondFieldOprsAdiabModes}\end{equation}
where\begin{equation}
\widehat{b}_{k}=\sum_{i}\alpha_{ki}\,\exp(-i\mu_{k}t)\,\widehat{a}_{i}\qquad\widehat{b}_{k}^{\dagger}=\sum_{i}\alpha_{ki}^{\ast}\,\exp(+i\mu_{k}t)\,\widehat{a}_{i}^{\dagger}\label{Eq.AdiabaticModeOprs}\end{equation}
and the standard commutation rules apply $[\widehat{b}_{k},\widehat{b}_{l}^{\dagger}]=\delta_{kl}$.

Substituting for $\phi_{j}(\mathbf{r},t)$ in the coupled Gross-Pitaevskii
equations (\ref{Eq.GenGrossPitEqns}), multiplying by $\alpha_{li}^{\ast}\,\exp(+i\mu_{l}t)$
and summing over $i$ gives a pair of \emph{time independent coupled
Gross-Pitaevskii equations} for the adiabatic mode functions \begin{eqnarray}
\sum\limits _{k}P_{lk}\,\hbar\mu_{k}\,\xi_{k} & = & \sum\limits _{k}P_{lk}(-\frac{\hbar^{2}}{2m}\nabla^{2}+V)\,\xi_{k}\nonumber \\
 &  & +\sum\limits _{k}(g\sum\limits _{mn}Q_{lr\, ks}\,\xi_{r}^{\ast}\,\xi_{s})\,\xi_{k}\qquad\qquad(l=1,2).\label{Eq.AdiabaticGenGrossPitEqns}\end{eqnarray}
where now the one and two body correlation functions are\begin{eqnarray}
P_{lk} & = & \left\langle \Phi\right\vert \widehat{b}_{l}^{\dagger}\widehat{b}_{k}\left\vert \Phi\right\rangle \label{Eq.OneBodyCorrFnAdia}\\
Q_{lr\, ks} & = & \left\langle \Phi\right\vert \widehat{b}_{l}^{\dagger}\widehat{b}_{r}^{\dagger}\widehat{b}_{k}\widehat{b}_{s}\left\vert \Phi\right\rangle \label{Eq.TwoBodyCorrFnAdia}\end{eqnarray}
Equations (\ref{Eq.KineticTrapTerms}) and (\ref{Eq.MeanFieldTerms})
were also used in the derivation. These equations are only meaningful
if the trap potential and the adiabatic mode functions are in fact
slowly varying with time. The frequencies $\mu_{1}$, $\mu_{2}$ play
the role of generalised chemical potentials. Noting that the $N$
dependence in the mode equations is carried in the one and two body
correlation functions - these being of $O(N)$ and $O(N^{2})$ respectively,
it is then possible to show that the chemical potential is given by
\begin{equation}
\mu=\frac{\partial}{\partial N}\left\langle \Phi\right\vert \widehat{H}\left\vert \Phi\right\rangle =\sum_{k}\hbar\mu_{k}\,\frac{\left\langle \Phi\right\vert \widehat{b}_{k}^{\dagger}\widehat{b}_{k}\left\vert \Phi\right\rangle }{N}.\label{Eq.ChemicalPtl}\end{equation}
As the quantity $\left\langle \Phi\right\vert \widehat{b}_{k}^{\dagger}\widehat{b}_{k}\left\vert \Phi\right\rangle /N$
is the fractional number of bosons occupying the adiabatic mode $\xi_{k}$
it follows that $\hbar\mu_{k}$ is the \emph{chemical potential} associated
with that mode.

\subsection{Single-Mode Theory and Standard Gross-Pitaevski Equation}

For the case where there is just a \emph{single condensate mode} the
state vector becomes\begin{equation}
\left\vert \,\Phi(t)\right\rangle =\frac{\left(\widehat{a}_{1}(t)^{\dagger}\right)^{N}}{[(N)!]^{\frac{\mathbf{1}}{\mathbf{2}}}}\left\vert \,0\right\rangle \label{Eq.OneModeQuantumState}\end{equation}
The general Gross-Pitaevskii equations (\ref{Eq.GenGrossPitEqns})
then reduce to the single \emph{Gross-Pitaevskii equation}. \begin{equation}
\left(-\frac{\hbar^{2}}{2m}\nabla^{2}+V+g({\small N-1})\left\vert \phi_{1}\right\vert ^{2}\right)\phi_{1}=i\hbar\frac{\partial}{\partial t}\,\phi_{1}\label{Eq.Gross-Pitaevskii}\end{equation}
Note that in the regime of interest with $N$ becoming very large,
the factor $gY_{im\, jn}/N=g_{N}Y_{im\, jn}/N^{2}$ becomes approximately
equal to $g_{N}$ tiimes a factor of order unity. In deriving the
single Gross-Pitaevskii equation from (\ref{Eq.GenGrossPitEqns}),
the matrices with elements $X_{ij}$ and $Y_{im\, jn}$ reduce to
$1x1$ matrices with non-zero elements\begin{equation}
X_{11}=N\qquad Y_{1111}=N(N-1)\label{Eq.SingleModeMatrices}\end{equation}
since in this case $b_{k}=\delta_{k,-N/2}$. As there is now only
one mode, there is now a single Fock state so amplitude equations,
spin operators no longer apply.

For the single mode case an adiabatic solution can be obtained via
the transformation \begin{equation}
\phi_{1}(\mathbf{r},t)=\exp(-i\mu_{1}t)\,\xi_{1}(\mathbf{r},t)\label{Eq.TransfAdiabaticMode2}\end{equation}
applied to (\ref{Eq.Gross-Pitaevskii}), where it is assumed that
the new mode $\xi_{1}(\mathbf{r},t)$ is so slowly varying with time
that its time derivative can be ignored. The time independent Gross-Pitaevskii
equation for the adiabatic mode function becomes\begin{equation}
\hbar\mu_{1}\,\xi_{1}=(-\frac{\hbar^{2}}{2m}\nabla^{2}+V)\,\xi_{1}+g({\small N-1})\left\vert \xi_{1}\right\vert ^{2}\,\xi_{1}\label{Eq.AdiabaticGenGrossPitEqn2}\end{equation}
and $\mu_{1}$ is the chemical potential $\mu_{1}=\frac{{\LARGE\partial}}{{\LARGE\partial N}}\left\langle \Phi\right\vert \widehat{H}\left\vert \Phi\right\rangle $.

\subsection{Modified Form for $\widehat{H}_{2}$ Term}

The previous form (\ref{Eq.H2}) of the $\widehat{H}_{2}$ term contains
spatial derivatives and these would produce Fokker-Planck equations
with functional derivatives with respect to spatial derivatives of
field functions, which cannot be treated in the standard approach.
However, the $\widehat{H}_{2}$ term can be put in a form in which
spatial derivatives are absent.

\subsubsection{Two-Mode Case}

It is straightforward to show that the eigenvalues of the $2\times2$
matrix of the $X_{ij}$ are both real, positive and their sum equals
$N$. Apart from special cases where one of the eigenvalues is zero
we see that the matrix of $X_{ij}$ is invertible and hence we can
write\begin{equation}
(-\frac{\hbar^{2}}{2m}\sum\limits _{\mu=x,y,z}\partial_{\mu}^{2}+V)\,\phi_{l}=i\hbar\frac{\partial}{\partial t}\phi_{l}-\sum_{ij}X_{li}^{-1}Z_{ij}\phi_{j}\end{equation}
where the \emph{generalised mean field} that occurs in the mode equations
is defined by\begin{equation}
Z_{ij}=g\sum_{mn}Y_{im\, jn}\,\phi_{m}^{\ast}\,\phi_{n}\label{Eq.MeanFieldGenGPE}\end{equation}
and is quadratic in the mode functions. Thus we find that the condensate
field annihilation operator satisfies the equation\begin{eqnarray}
\left(-\frac{\hbar^{2}}{2m}\nabla^{2}+V\right)\widehat{\Psi}_{C}(\mathbf{r}) & = & \, i\hbar\sum_{l}\widehat{a}_{l}\frac{\partial}{\partial t}\phi_{l}-\sum_{ijl}X_{li}^{-1}Z_{ij}\phi_{j}\widehat{a}_{l}\nonumber \\
 & = & -i\hbar\sum_{l}\phi_{l}\frac{\partial}{\partial t}\widehat{a}_{l}\nonumber \\
 &  & -g\sum_{ijmnl}X_{li}^{-1}Y_{im\, jn}\,\phi_{m}^{\ast}\phi_{n}\phi_{j}\,\widehat{a}_{l}\end{eqnarray}
using $\frac{{\LARGE\partial}}{{\LARGE\partial t}}\widehat{\Psi}_{C}(\mathbf{r})=0$.

However from (\ref{Eq.CondOpr})\begin{equation}
\widehat{a}_{l}=\int d\mathbf{s\,}\phi_{l}^{\ast}(\mathbf{s\,)}\widehat{\Psi}_{C}(\mathbf{s})\qquad(l=1,2)\label{Eq.CondModeAnnihOpr}\end{equation}
so that the condensate field operator satisfies the integro-differential
equation\begin{eqnarray}
\left(-\frac{\hbar^{2}}{2m}\nabla^{2}+V\right)\widehat{\Psi}_{C}(\mathbf{r}) & = & -i\hbar\sum_{l}\phi_{l}(\mathbf{r})\int d\mathbf{s\,}\frac{\partial}{\partial t}\phi_{l}^{\ast}(\mathbf{s\,)\,}\widehat{\Psi}_{C}(\mathbf{s})\nonumber \\
 &  & -g\sum_{ijmnl}X_{li}^{-1}Y_{im\, jn}\,\phi_{m}^{\ast}(\mathbf{r})\phi_{n}(\mathbf{r})\phi_{j}(\mathbf{r})\,\nonumber \\
 &  & \times\int d\mathbf{s\,}\phi_{l}^{\ast}(\mathbf{s\,)\,}\widehat{\Psi}_{C}(\mathbf{s})\end{eqnarray}
Hence operating from the left with $\widehat{\Psi}_{NC}(\mathbf{r})^{\dagger}\,$and
integrating over $\mathbf{r}$ we find that \begin{eqnarray}
 &  & \int d\mathbf{r\,}\widehat{\Psi}_{NC}(\mathbf{r})^{\dagger}\,\left(-\frac{\hbar^{2}}{2m}\nabla^{2}+V\right)\widehat{\Psi}_{C}(\mathbf{r})\nonumber \\
 & = & -i\hbar\sum_{l}\int d\mathbf{r\,}\widehat{\Psi}_{NC}(\mathbf{r})^{\dagger}\,\phi_{l}(\mathbf{r})\int d\mathbf{s\,}\left(\frac{\partial}{\partial t}\phi_{l}^{\ast}(\mathbf{s\,)}\right)\mathbf{\,}\widehat{\Psi}_{C}(\mathbf{s})\nonumber \\
 &  & -g\sum_{ijmnl}X_{li}^{-1}Y_{im\, jn}\,\int d\mathbf{r\,}\widehat{\Psi}_{NC}(\mathbf{r})^{\dagger}\,\phi_{m}^{\ast}(\mathbf{r})\phi_{n}(\mathbf{r})\phi_{j}(\mathbf{r})\,\int d\mathbf{s\,}\phi_{l}^{\ast}(\mathbf{s\,)}\widehat{\Psi}_{C}(\mathbf{s})\nonumber \\
 & = & -g\sum_{ijmnl}X_{li}^{-1}Y_{im\, jn}\,\int d\mathbf{r\,}\widehat{\Psi}_{NC}(\mathbf{r})^{\dagger}\,\phi_{m}^{\ast}(\mathbf{r})\phi_{n}(\mathbf{r})\phi_{j}(\mathbf{r})\nonumber \\
 &  & \,\times\int d\mathbf{s\,}\phi_{l}^{\ast}(\mathbf{s\,)}\widehat{\Psi}_{C}(\mathbf{s})\end{eqnarray}
since the first term on the right hand side is zero because the condensate
mode functions $\phi_{j}(\mathbf{r})$ are orthogonal to the non-condensate
mode functions $\phi_{k}^{\ast}(\mathbf{r\,)}$ that are present in
the expansion of the non-condensate field operator $\widehat{\Psi}_{NC}(\mathbf{r})^{\dagger}$.
Thus we can write\begin{equation}
\int d\mathbf{r\,}\widehat{\Psi}_{NC}(\mathbf{r})^{\dagger}\,\left\{ \left(-\frac{\hbar^{2}}{2m}\nabla^{2}+V\right)\widehat{\Psi}_{C}(\mathbf{r})\right\} =-\frac{g_{N}}{N}\int\int d\mathbf{r\,}d\mathbf{s\,}F(\mathbf{r},\mathbf{s})\widehat{\Psi}_{NC}(\mathbf{r})^{\dagger}\,\widehat{\Psi}_{C}(\mathbf{s})\label{Eq.H2KinPtlEnergyResult1}\end{equation}
where the kernel $F(\mathbf{r},\mathbf{s})$ is an ordinary spatial
function of two positions and is defined by \begin{equation}
F(\mathbf{r},\mathbf{s})=\sum_{ijmnl}X_{li}^{-1}Y_{im\, jn}\phi_{m}^{\ast}(\mathbf{r})\phi_{n}(\mathbf{r})\phi_{j}(\mathbf{r})\,\phi_{l}^{\ast}(\mathbf{s\,)}\label{Eq.Kernel}\end{equation}
Note that this kernel is \emph{not} symmetric in $\mathbf{r},\mathbf{s}$.

Taking the adjoint of the last equation gives\begin{equation}
\int d\mathbf{r\,}\left\{ \left(-\frac{\hbar^{2}}{2m}\nabla^{2}+V\right)\widehat{\Psi}_{C}^{\dagger}(\mathbf{r})\right\} \widehat{\Psi}_{NC}(\mathbf{r})=-\frac{g_{N}}{N}\int\int d\mathbf{r\,}d\mathbf{s\,}F^{\ast}(\mathbf{s},\mathbf{r})\widehat{\Psi}_{C}(\mathbf{r})^{\dagger}\widehat{\Psi}_{NC}(\mathbf{s})\,\label{Eq.H2KinPtlEnergyResult2}\end{equation}
so the term $\widehat{H}_{2}$ can now be written as \begin{eqnarray}
\widehat{H}_{2} & = & -\int\int d\mathbf{r\,}d\mathbf{s\,}\frac{g_{N}}{N}F(\mathbf{r},\mathbf{s})\widehat{\Psi}_{NC}(\mathbf{r})^{\dagger}\,\widehat{\Psi}_{C}(\mathbf{s})\nonumber \\
 &  & +\int d\mathbf{r\,}\widehat{\Psi}_{NC}(\mathbf{r})^{\dagger}\,\left\{ +\frac{g_{N}}{N}\widehat{\Psi}_{C}(\mathbf{r})^{\dagger}\widehat{\Psi}_{C}(\mathbf{r})\widehat{\Psi}_{C}(\mathbf{r})\right\} \nonumber \\
 &  & +\int d\mathbf{r\,}\left\{ +\frac{g_{N}}{N}\widehat{\Psi}_{C}(\mathbf{r})^{\dagger}\widehat{\Psi}_{C}(\mathbf{r})^{\dagger}\widehat{\Psi}_{C}(\mathbf{r})\right\} \widehat{\Psi}_{NC}(\mathbf{r}))\nonumber \\
 &  & -\int\int d\mathbf{r\,}d\mathbf{s\,}\frac{g_{N}}{N}F^{\ast}(\mathbf{s},\mathbf{r})\widehat{\Psi}_{C}(\mathbf{r})^{\dagger}\widehat{\Psi}_{NC}(\mathbf{s})\,\label{eq:H2TermResult}\end{eqnarray}
This eliminates the awkward terms involving integrals of $\widehat{\Psi}_{NC}$
with the spatial derivative of $\widehat{\Psi}_{C}$ (and the adjoint
expressions). These would lead to Fokker-Planck equations with functional
derivatives with respect to spatial derivatives of field functions,
which cannot be treated in the standard approach. However, the term
$\widehat{H}_{2}$ now involves \emph{double spatial integrals} of
field operators, and these require special treatment.

We can write $\widehat{H}_{2}$ as the sum of two terms, one involving
field operators to the second order, the other involving field operators
to the fourth order. Thus\begin{eqnarray}
\widehat{H}_{2} & = & \widehat{H}_{2U4}+\widehat{H}_{2U2}\label{Eq.NewH2}\\
\widehat{H}_{2U4} & = & \frac{g_{N}}{N}\int d\mathbf{r\,(}\widehat{\Psi}_{NC}^{\dagger}(\mathbf{r})\widehat{\Psi}_{C}^{\dagger}(\mathbf{r})\,\widehat{\Psi}_{C}(\mathbf{r})\widehat{\Psi}_{C}(\mathbf{r}))\nonumber \\
 &  & +\frac{g_{N}}{N}\int d\mathbf{r\,(}\widehat{\Psi}_{C}^{\dagger}(\mathbf{r})\widehat{\Psi}_{C}^{\dagger}(\mathbf{r})\,\widehat{\Psi}_{C}(\mathbf{r})\widehat{\Psi}_{NC}(\mathbf{r}))\label{eq:NewH24}\\
\widehat{H}_{2U2} & = & -\frac{g_{N}}{N}\int\int d\mathbf{r\,}d\mathbf{s\,}F(\mathbf{r},\mathbf{s})\widehat{\Psi}_{NC}(\mathbf{r})^{\dagger}\,\widehat{\Psi}_{C}(\mathbf{s})\nonumber \\
 &  & -\frac{g_{N}}{N}\int\int d\mathbf{r\,}d\mathbf{s\,}F^{\ast}(\mathbf{s},\mathbf{r})\widehat{\Psi}_{C}(\mathbf{r})^{\dagger}\widehat{\Psi}_{NC}(\mathbf{s})\label{eq:NewH22}\end{eqnarray}
Note that both terms are proportional to the factor $g_{N}/N$. Thus
we see that $\widehat{H}_{2}$ is now associated only with terms analogous
to those for \emph{boson-boson} interactions, both $\widehat{H}_{2U4}$
and $\widehat{H}_{2U2}$ being proportional to $g_{N}/N$.

\subsubsection{Single Mode Case}

If only a \emph{single condensate mode} was involved the development
of $\widehat{H}_{2}$ is simpler. From (\ref{Eq.Gross-Pitaevskii})
and (\ref{Eq.CondFieldOprsSingleMode}) similar procedure to the two
mode case gives \begin{equation}
\left(-\frac{\hbar^{2}}{2m}\nabla^{2}+V\right)\widehat{\Psi}_{C}(\mathbf{r})=-i\hbar\phi_{1}(\mathbf{r})\int d\mathbf{s\,}\left(\frac{\partial}{\partial t}\phi_{1}^{\ast}(\mathbf{s\,)}\right)\,\widehat{\Psi}_{C}(\mathbf{s})-g({\small N-1})\left\vert \phi_{1}(\mathbf{r})\right\vert ^{2}\widehat{\Psi}_{C}(\mathbf{r})\end{equation}
so that using the orthogonality of the condensate mode to all the
non-condensate modes \begin{eqnarray}
 &  & \int d\mathbf{r\,}\widehat{\Psi}_{NC}^{\dagger}(\mathbf{r})\left(-\frac{\hbar^{2}}{2m}\nabla^{2}+V\right)\widehat{\Psi}_{C}(\mathbf{r})\nonumber \\
 & = & -i\hbar\int d\mathbf{r\,}\widehat{\Psi}_{NC}^{\dagger}(\mathbf{r})\,\phi_{1}(\mathbf{r})\int d\mathbf{s\,}\left(\frac{\partial}{\partial t}\phi_{1}^{\ast}(\mathbf{s\,)}\right)\,\widehat{\Psi}_{C}(\mathbf{s})\nonumber \\
 &  & -\int d\mathbf{r\,}\widehat{\Psi}_{NC}^{\dagger}(\mathbf{r})g({\small N-1})\left\vert \phi_{1}(\mathbf{r})\right\vert ^{2}\widehat{\Psi}_{C}(\mathbf{r})\nonumber \\
 & = & -\int d\mathbf{r\,}\widehat{\Psi}_{NC}^{\dagger}(\mathbf{r})g({\small N-1})\left\vert \phi_{1}(\mathbf{r})\right\vert ^{2}\widehat{\Psi}_{C}(\mathbf{r})\label{Eq.H2Simplification}\end{eqnarray}
This result may also be recognised as a special case of (\ref{Eq.H2KinPtlEnergyResult2}).
Using the special forms in (\ref{Eq.SingleModeMatrices}) for the
$X_{li}^{-1}Y_{im\, jn}$ we have \begin{eqnarray}
F(\mathbf{r},\mathbf{s}) & = & \frac{1}{N}N(N-1)\phi_{1}^{\ast}(\mathbf{r})\phi_{1}(\mathbf{r})\phi_{1}(\mathbf{r})\,\phi_{1}^{\ast}(\mathbf{s\,)}\nonumber \\
 & = & (N-1)\phi_{1}^{\ast}(\mathbf{r})\phi_{1}(\mathbf{r})\delta_{C}(\mathbf{r},\mathbf{s\,)}\label{Eq.SpecialSingleCondF}\\
-\frac{g_{N}}{N}\int\int d\mathbf{r\,}d\mathbf{s\,}F(\mathbf{r},\mathbf{s})\widehat{\Psi}_{NC}(\mathbf{r})^{\dagger}\,\widehat{\Psi}_{C}(\mathbf{s}) & = & -\int d\mathbf{r\,}\widehat{\Psi}_{NC}(\mathbf{r})^{\dagger}g(N-1)\left\vert \phi_{1}(\mathbf{r})\right\vert ^{2}\,\widehat{\Psi}_{C}(\mathbf{r})\nonumber \end{eqnarray}
as before, where a result from (\ref{Eq.RestrictedDeltaFns}) has
been used to evaluate the $\mathbf{s}$ integral..

We can use (\ref{Eq.H2Simplification}) and the related adjoint equation
involving $\widehat{\Psi}_{C}^{\dagger}(\mathbf{r})g({\small N-1})\left\vert \phi_{1}(\mathbf{r})\right\vert ^{2}\widehat{\Psi}_{NC}(\mathbf{r})$
to simplify $\widehat{H}_{2}$ into a form

\begin{eqnarray}
\widehat{H}_{2} & = & \int d\mathbf{r\,(}\widehat{\Psi}_{NC}^{\dagger}(\mathbf{r})\frac{g_{N}}{N}\{\widehat{\Psi}_{C}^{\dagger}(\mathbf{r})\,\widehat{\Psi}_{C}(\mathbf{r})-\left\langle \widehat{\Psi}_{C}^{\dagger}(\mathbf{r})\,\widehat{\Psi}_{C}(\mathbf{r})\right\rangle \}\widehat{\Psi}_{C}(\mathbf{r}))\nonumber \\
 &  & +\int d\mathbf{r\,(}\widehat{\Psi}_{C}^{\dagger}(\mathbf{r})\frac{g_{N}}{N}\{\widehat{\Psi}_{C}^{\dagger}(\mathbf{r})\,\widehat{\Psi}_{C}(\mathbf{r})-\left\langle \widehat{\Psi}_{C}^{\dagger}(\mathbf{r})\,\widehat{\Psi}_{C}(\mathbf{r})\right\rangle \}\widehat{\Psi}_{NC}(\mathbf{r}))\label{Eq.H2New}\end{eqnarray}
where we use the notation $\left\langle \widehat{\Psi}_{C}(\mathbf{r})^{\dagger}\widehat{\Psi}_{C}(\mathbf{r})\right\rangle =({\small N-1})\left\vert \phi_{1}(\mathbf{r})\right\vert ^{2}$.
We see that for the single mode condensate case $\widehat{H}_{2}$
is also the sum of a term $\widehat{H}_{2U4}$ which is fourth order
in the field operators and a term $\widehat{H}_{2U2}$ which is second
order. \begin{eqnarray}
\widehat{H}_{2} & = & \widehat{H}_{2U4}+\widehat{H}_{2U2}\label{Eq.H2NewParts}\\
\widehat{H}_{2U4} & = & \frac{g_{N}}{N}\int d\mathbf{r\,(}\widehat{\Psi}_{NC}^{\dagger}(\mathbf{r})\widehat{\Psi}_{C}^{\dagger}(\mathbf{r})\,\widehat{\Psi}_{C}(\mathbf{r})\widehat{\Psi}_{C}(\mathbf{r}))\nonumber \\
 &  & +\frac{g_{N}}{N}\int d\mathbf{r\,(}\widehat{\Psi}_{C}^{\dagger}(\mathbf{r})\widehat{\Psi}_{C}^{\dagger}(\mathbf{r})\,\widehat{\Psi}_{C}(\mathbf{r})\widehat{\Psi}_{NC}(\mathbf{r}))\label{eq:H24}\\
\widehat{H}_{2U2} & = & -\frac{g_{N}}{N}\int d\mathbf{r\,(}\widehat{\Psi}_{NC}^{\dagger}(\mathbf{r})\{\left\langle \widehat{\Psi}_{C}^{\dagger}(\mathbf{r})\,\widehat{\Psi}_{C}(\mathbf{r})\right\rangle \}\widehat{\Psi}_{C}(\mathbf{r}))\nonumber \\
 &  & -\frac{g_{N}}{N}\int d\mathbf{r\,(}\widehat{\Psi}_{C}^{\dagger}(\mathbf{r})\{\left\langle \widehat{\Psi}_{C}^{\dagger}(\mathbf{r})\,\widehat{\Psi}_{C}(\mathbf{r})\right\rangle \}\widehat{\Psi}_{NC}(\mathbf{r}))\label{eq:H22}\end{eqnarray}
Thus we see that $\widehat{H}_{2}$ is now associated only with \emph{boson-boson}
interaction terms. However, unlike the two mode condensate case there
is no double spatial integral involved. The general form of the development
for $\widehat{H}_{2}$ is a simpler version of the result (\ref{Eq.NewH2})
for the two mode case.\pagebreak{}

\section{Phase space distribution functional}

\label{Sect Phase Space Dist Fnal}

In this section the phase space distribution functional is introduced
starting with the characteristic functional. The distribution functional
is of a mixed type, with the condensate fields involving a generalised
Wigner form, whilst the non-condensate fields involving a positive
P form. This is to reflect the feature that many bosons occupy the
condensate modes, so a Wigner distribution is better suited since
it descibes fields whose behaviour is close to a classical mean field
situation. On the other hand, there will be few bosons occupying the
non-condensate modes, hence a positive P distribution is better, since
the non-condensate fields may display quantum behaviour. In this section
we emphasise how the phase space distribution functionals determine
the quantum correlation functions which are used to describe the probabilities
for bosonic position measurements. The theory in this section is set
out for the two-mode situation, but can be easily modified for the
single mode condensate by just restricting the sums over condensate
modes to a single term.

\subsection{Characteristic Functional}

From the density operator $\widehat{\rho}$ and by introducing four
distinct functions $\xi_{C}^{+}(\mathbf{r}),\xi_{C}(\mathbf{r}),\xi_{NC}^{+}(\mathbf{r})$
and $\xi_{NC}(\mathbf{r})$ associated with the field operators\\
 $\widehat{\Psi}_{C}(\mathbf{r}),\widehat{\Psi}_{C}^{\dagger}(\mathbf{r}),\widehat{\Psi}_{NC}(\mathbf{r})$
and $\widehat{\Psi}_{NC}^{\dagger}(\mathbf{r})$ respectively, we
define the \emph{characteristic functional} $\chi\lbrack\xi_{C}(\mathbf{r}),\xi_{C}^{+}(\mathbf{r}),\xi_{NC}(\mathbf{r}),\xi_{NC}^{+}(\mathbf{r})]$
as\begin{equation}
\chi\lbrack\xi_{C},\xi_{C}^{+},\xi_{NC},\xi_{NC}^{+}]=Tr(\widehat{\rho}\,\widehat{\Omega}[\xi_{C},\xi_{C}^{+},\xi_{NC},\xi_{NC}^{+}])\label{Eq.CharFnal}\end{equation}
with\begin{eqnarray}
\widehat{\Omega} & = & \,\widehat{\Omega}_{C}\,\widehat{\Omega}_{NC}\label{Eq.Omega}\\
\widehat{\Omega}_{C} & = & \exp\int d\mathbf{r}\, i\{\xi_{C}(\mathbf{r})\widehat{\Psi}_{C}^{\dagger}(\mathbf{r})+\widehat{\Psi}_{C}(\mathbf{r})\xi_{C}^{+}(\mathbf{r})\}\,\label{Eq.OmegaC}\\
\widehat{\Omega}_{NC} & = & \exp\int d\mathbf{r}\, i\{\xi_{NC}(\mathbf{r})\widehat{\Psi}_{NC}^{\dagger}(\mathbf{r})\}\,\exp\int d\mathbf{r}\, i\{\widehat{\Psi}_{NC}(\mathbf{r})\xi_{NC}^{+}(\mathbf{r})\}\label{Eq.OmegaNC}\end{eqnarray}
Thus this mixed characteristic functional is of the \emph{Wigner}
type for the condensate modes and of the \emph{Positive P (P}$^{+})$
type for the non-condensate modes. The basic idea of a \emph{functional}
is explained in Appendix B (\citep{Dalton10b}) but essentially a
functional $F[\psi(x)]$ of a field function $\psi(x)$ just defines
a process that results in a c-number which depends on \emph{all} the
values of the field function, that is over the entire range of positions
$x$.

If the mode expansions are used with\begin{eqnarray}
\xi_{C}(\mathbf{r}) & = & \xi_{1}\phi_{1}(\mathbf{r})+\xi_{2}\phi_{2}(\mathbf{r})\label{Eq.XiMode1}\\
\xi_{C}^{+}(\mathbf{r}) & = & \phi_{1}^{\ast}(\mathbf{r})\xi_{1}^{+}+\phi_{2}^{\ast}(\mathbf{r})\xi_{2}^{+}\label{Eq.XiMode2}\\
\xi_{NC}(\mathbf{r}) & = & \sum\limits _{k\neq1,2}^{K}\xi_{k}\,\phi_{k}(\mathbf{r})\label{Eq.XiMode3}\\
\xi_{NC}^{+}(\mathbf{r}) & = & \sum\limits _{k\neq1,2}^{K}\phi_{k}^{\ast}(\mathbf{r})\,\xi_{k}^{+}\label{Eq.XiMode4}\end{eqnarray}
then we have\begin{eqnarray}
\widehat{\Omega}_{C} & = & \exp i\{\xi_{1}\widehat{a}_{1}^{\dagger}+\widehat{a}_{1}\xi_{1}^{+}+\xi_{2}\widehat{a}_{2}^{\dagger}+\widehat{a}_{2}\xi_{2}^{+}\}\label{Eq.OmegaMode1}\\
\widehat{\Omega}_{NC} & = & \exp i\sum\limits _{k\neq1,2}\xi_{k}\widehat{a}_{k}^{\dagger}\,\exp i\sum\limits _{k\neq1,2}\widehat{a}_{k}\xi_{k}^{+}\label{Eq.OmegaMode2}\end{eqnarray}
This shows that the characteristic \emph{functional} $\chi\lbrack\xi_{C},\xi_{C}^{+},\xi_{NC},\xi_{NC}^{+}]$
is also a characteristic \emph{function} $\chi_{b}(\xi_{1},\xi_{1}^{+},\xi_{2},\xi_{2}^{+},..,\xi_{k},\xi_{k}^{+},..)$
of the \emph{c-number} expansion coefficients, a result that is important
in deriving expressions based on functionals.

\subsection{Quasi-Distribution Functional}

For double phase space distributions as in the present case the \emph{quasi-distribution
functional} $P[\psi_{C}(\mathbf{r}),\psi_{C}^{+}(\mathbf{r}),\psi_{NC}(\mathbf{r}),\psi_{NC}^{+}(\mathbf{r}),\psi_{C}^{\ast}(\mathbf{r}),\psi_{C}^{+\ast}(\mathbf{r}),\psi_{NC}^{\ast}(\mathbf{r}),\psi_{NC}^{+\ast}(\mathbf{r})]$
involves four field functions $\psi_{C}(\mathbf{r}),\psi_{C}^{+}(\mathbf{r}),\psi_{NC}(\mathbf{r}),\psi_{NC}^{+}(\mathbf{r})$
corresponding to the field operators $\widehat{\Psi}_{C}(\mathbf{r}),\widehat{\Psi}_{C}^{\dagger}(\mathbf{r}),\widehat{\Psi}_{NC}(\mathbf{r})$
and $\widehat{\Psi}_{NC}^{\dagger}(\mathbf{r})$ respectively, plus
their complex conjugate fields $\psi_{C}^{\ast}(\mathbf{r}),\psi_{C}^{+\ast}(\mathbf{r}),\psi_{NC}^{\ast}(\mathbf{r}),\psi_{NC}^{+\ast}(\mathbf{r})$.
It is chosen to give the characteristic functional $\chi\lbrack\xi_{C}(\mathbf{r}),\xi_{C}^{+}(\mathbf{r}),\xi_{NC}(\mathbf{r}),\xi_{NC}^{+}(\mathbf{r})]$
via a \emph{functional integration} process over the four complex
field functions, the integration also incorporating an exponential
factor, which may be written as\\
 $\exp i\int d\mathbf{r\,\{}\xi_{C}(\mathbf{r})\psi_{C}^{+}(\mathbf{r})+\psi_{C}(\mathbf{r})\xi_{C}^{+}(\mathbf{r})\}\,\,\exp i\int d\mathbf{r\,\{}\xi_{NC}(\mathbf{r})\psi_{NC}^{+}(\mathbf{r})\}\exp i\int d\mathbf{r\,\{}\psi_{NC}(\mathbf{r})\xi_{NC}^{+}(\mathbf{r})\}$.
\\
Thus\begin{eqnarray}
 &  & \chi\lbrack\xi_{C}(\mathbf{r}),\xi_{C}^{+}(\mathbf{r}),\xi_{NC}(\mathbf{r}),\xi_{NC}^{+}(\mathbf{r})]\nonumber \\
 & = & \iiiint D^{2}\psi_{C}\, D^{2}\psi_{C}^{+}\, D^{2}\psi_{NC}\, D^{2}\psi_{NC}^{+}\,\nonumber \\
 &  & \times P[\psi_{C}(\mathbf{r}),\psi_{C}^{+}(\mathbf{r}),\psi_{NC}(\mathbf{r}),\psi_{NC}^{+}(\mathbf{r}),\psi_{C}^{\ast}(\mathbf{r}),\psi_{C}^{+\ast}(\mathbf{r}),\psi_{NC}^{\ast}(\mathbf{r}),\psi_{NC}^{+\ast}(\mathbf{r})]\nonumber \\
 &  & \times\exp i\int d\mathbf{r\,\{}\xi_{C}(\mathbf{r})\psi_{C}^{+}(\mathbf{r})+\psi_{C}(\mathbf{r})\xi_{C}^{+}(\mathbf{r})\}\,\,\nonumber \\
 &  & \times\exp i\int d\mathbf{r\,\{}\xi_{NC}(\mathbf{r})\psi_{NC}^{+}(\mathbf{r})\}\exp i\int d\mathbf{r\,\{}\psi_{NC}(\mathbf{r})\xi_{NC}^{+}(\mathbf{r})\}.\label{Eq.DistribnFnal}\end{eqnarray}
The justification of this important result is set out below. Note
that the quasi-distribution functional is not necessarily unique,
it is only required that the above functional integral gives the characteristic
functional, (which is unique). In the present case we will use a weight
function of the form based on Eq.(B.60), but adapted to there being
eight real fields, rather than four as in Appendix B.8.\begin{equation}
w(\psi_{1},\psi_{1}^{+},..,\psi_{i},\psi_{i}^{+},..,\psi_{n},\psi_{n}^{+},\psi_{1}^{\ast},\psi_{1}^{+\ast},..,\psi_{i}^{\ast},\psi_{i}^{+\ast},..,\psi_{n}^{\ast},\psi_{n}^{+\ast})=\prod\limits _{i}(\Delta\mathbf{r}_{i})^{4}\label{Eq.WeightFn3D}\end{equation}
The power law $(\Delta\mathbf{r}_{i})^{4}$ arises because each field
contributes $(\Delta\mathbf{r}_{i})^{1/2}$.

Functional integration is fully explained in Appendix B (\citep{Dalton10b}),
but a brief summary is as follows. If there are $n$ modes then the
range for each function $\psi(x)$ is divided up into $n$ small intervals
$\Delta x_{i}=x_{i+1}-x_{i}$ (the $i$th interval, where $\epsilon>|\Delta x_{i}|$),
then we may specify the \emph{value} $\psi_{i}$ of the function $\psi(x)$
in the $i$th interval via the \emph{average}\begin{equation}
\psi_{i}=\frac{1}{\Delta x_{i}}\int\limits _{\Delta x_{i}}dx\,\psi(x)\label{Eq.AverageField}\end{equation}
and then any functional $F[\psi(x)]$ may be regarded as a \emph{function}
$F(\psi_{1},\psi_{2},..,\psi_{i},..,\psi_{n})$ of all the $\psi_{i}$,
and ordinary integration over the $\psi_{i}$ is used to define the
functional integral. If each function $\psi(x)=\psi_{x}(x)+i\psi_{y}(x)$.is
written in terms of its real and imaginary parts, then the functional
integration becomes an ordinary integration over the values $\psi_{ix}$,
$\psi_{iy}$ of these components in each interval $i$ of the function
$F(\psi_{1},\psi_{2},..,\psi_{i},..,\psi_{n})$ multiplied by a suitably
chosen weight function $w(\psi_{1},\psi_{2},..,\psi_{i},..,\psi_{n})$.
Thus \begin{eqnarray}
\int D^{2}\psi\, F[\psi(x)] & = & \lim_{n\rightarrow\infty}\lim_{\epsilon\rightarrow0}\int\ldots\int d^{2}\psi_{1}d^{2}\psi_{2}..d^{2}\psi_{i}..d^{2}\psi_{n}\, w(\psi_{1},\psi_{2},..,\psi_{i},..,\psi_{n})\,\nonumber \\
 &  & \times F(\psi_{1},\psi_{2},..,\psi_{i},..,\psi_{n})\label{Eq.FnalIntegral}\end{eqnarray}
where the number of modes is increased to infinity along with the
space interval decreasing to zero. The symbol $D^{2}\psi$ stands
for $d^{2}\psi_{1}d^{2}\psi_{2}..d^{2}\psi_{i}..d^{2}\psi_{n}\, w(\psi_{1},..,\psi_{i},..,\psi_{n})$,
where the quantity $d^{2}\psi_{i}$ means $d\psi_{ix}d\psi_{iy}$.
The present case involves a generalisation to treat four complex fields
$\psi_{C}(\mathbf{r}),\psi_{C}^{+}(\mathbf{r}),\psi_{NC}(\mathbf{r}),\psi_{NC}^{+}(\mathbf{r})$.

To justify the characteristic functional result (\ref{Eq.DistribnFnal})
mode expansions for the field functions are used with \emph{c-number}
expansion coefficients $\alpha_{k},\alpha_{k}^{+}$

\begin{eqnarray}
\psi_{C}(\mathbf{r}) & = & \alpha_{1}\,\phi_{1}(\mathbf{r})+\alpha_{2}\,\phi_{2}(\mathbf{r})\label{Eq.FieldFn1}\\
\psi_{C}^{+}(\mathbf{r}) & = & \phi_{1}^{\ast}(\mathbf{r})\,\alpha_{1}^{+}+\phi_{2}^{\ast}(\mathbf{r})\,\alpha_{2}^{+}\label{Eq.FieldFn2}\\
\psi_{NC}(\mathbf{r}) & = & \sum\limits _{k\neq1,2}^{K}\alpha_{k}\,\phi_{k}(\mathbf{r})\label{Eq.FieldFn3}\\
\psi_{NC}^{+}(\mathbf{r}) & = & \sum\limits _{k\neq1,2}^{K}\phi_{k}^{\ast}(\mathbf{r})\alpha_{k}^{+}\label{Eq.FieldFn4}\end{eqnarray}
The P$^{+}$ quasi-distribution \emph{functional}\\
 $P[\psi_{C}(\mathbf{r}),\psi_{C}^{+}(\mathbf{r}),\psi_{NC}(\mathbf{r}),\psi_{NC}^{+}(\mathbf{r}),\psi_{C}^{\ast}(\mathbf{r}),\psi_{C}^{+\ast}(\mathbf{r}),\psi_{NC}^{\ast}(\mathbf{r}),\psi_{NC}^{+\ast}(\mathbf{r})]$
would then be equivalent to a distribution \emph{function} $P_{b}(\alpha_{1},\alpha_{1}^{+},..,\alpha_{k},\alpha_{k}^{+},..,\alpha_{1}^{\ast},\alpha_{1}^{+\ast},..,\alpha_{k}^{\ast},\alpha_{k}^{+\ast},..)$
of the c-number expansion coefficients and their complex conjugates.
For double phase space representations of bosonic systems the connection
between the characteristic function $\chi_{b}(\xi_{1},\xi_{1}^{+},\xi_{2},\xi_{2}^{+}..,\xi_{k},\xi_{k}^{+},..)$
and the distribution function via a phase space integral has been
established by Drummond and Gardiner \citep{Drummond80a}, \citep{Gardiner91a}.
The characteristic function is given by

\begin{eqnarray}
 &  & \chi_{b}(\xi_{1},\xi_{1}^{+},\xi_{2},\xi_{2}^{+}..,\xi_{k},\xi_{k}^{+},..)\nonumber \\
 & = & \int\ldots\int d^{2}\alpha_{1}d^{2}\alpha_{1}^{+}d^{2}\alpha_{2}d^{2}\alpha_{2}^{+}..d^{2}\alpha_{k}d^{2}\alpha_{k}^{+}..d^{2}\alpha_{n}d^{2}\alpha_{n}^{+}\,\nonumber \\
 &  & \times P_{b}(\alpha_{1},\alpha_{1}^{+},..,\alpha_{k},\alpha_{k}^{+},..,\alpha_{1}^{\ast},\alpha_{1}^{+\ast},..,\alpha_{k}^{\ast},\alpha_{k}^{+\ast},..)\nonumber \\
 &  & \times\exp i\sum\limits _{k=1}^{n}\{\xi_{k}\,\alpha_{k}^{+}\}\,\exp i\sum\limits _{k=1}^{n}\{\alpha_{k}\,\xi_{k}^{+}\}\,\label{Eq.CharFnDistnFn}\end{eqnarray}
where $\alpha_{k}=\alpha_{kx}+i\alpha_{ky}$, $\alpha_{k}^{+}=\alpha_{kx}^{+}+i\alpha_{ky}^{+}$
and $d^{2}\alpha_{k}=d\alpha_{kx}\, d\alpha_{ky}$, $d^{2}\alpha_{k}^{+}=d\alpha_{kx}^{+}\, d\alpha_{ky}^{+}.$
If the phase space integration is replaced by functional integration
we can show that Eq.(\ref{Eq.CharFnDistnFn}) leads to the result
(\ref{Eq.DistribnFnal}), which thus demonstrates that the distribution
functional exists. The change from phase space integration to functional
integration is outlined in Appendix B.8 (see Appendix B, \citep{Dalton10b}).
In deriving the functional integration result for the characteristic
function the expressions \begin{eqnarray*}
\exp i\sum\limits _{k=1}^{n}\{\xi_{k}\,\alpha_{k}^{+}\} & = & \exp i\sum\limits _{j=1,2}\{\xi_{j}\,\alpha_{j}^{+}\}\,\exp i\sum\limits _{k\neq1,2}^{n}\{\xi_{k}\,\alpha_{k}^{+}\}\\
 & = & \exp i\int d\mathbf{r\,\{}\xi_{C}(\mathbf{r})\psi_{C}^{+}(\mathbf{r})\}\,\exp i\int d\mathbf{r\,\{}\xi_{NC}(\mathbf{r})\psi_{NC}^{+}(\mathbf{r})\}\\
\exp i\sum\limits _{k=1}^{n}\{\alpha_{k}\,\xi_{k}^{+}\} & = & \exp i\sum\limits _{j=1,2}\{\alpha_{j}\,\xi_{j}^{+}\}\,\exp i\sum\limits _{k\neq1,2}^{n}\{\alpha_{k}\,\xi_{k}^{+}\}\,\\
 & = & \exp i\int d\mathbf{r\,\{}\psi_{C}(\mathbf{r})\xi_{C}^{+}(\mathbf{r})\}\,\exp i\int d\mathbf{r\,\{}\psi_{NC}(\mathbf{r})\xi_{NC}^{+}(\mathbf{r})\}\end{eqnarray*}
are used.

Note that as each field can be expressed in terms of its real and
imaginary components, the distribution functional involving the four
fields and their conjugates may also be considered as a functional
of the eight real, imaginary components. \begin{eqnarray*}
 &  & P[\psi_{C}(\mathbf{r}),\psi_{C}^{+}(\mathbf{r}),\psi_{NC}(\mathbf{r}),\psi_{NC}^{+}(\mathbf{r}),\psi_{C}^{\ast}(\mathbf{r}),\psi_{C}^{+\ast}(\mathbf{r}),\psi_{NC}^{\ast}(\mathbf{r}),\psi_{NC}^{+\ast}(\mathbf{r})]\\
 & \equiv & F[\psi_{CX}(\mathbf{r}),\psi_{CX}^{+}(\mathbf{r}),\psi_{NCX}(\mathbf{r}),\psi_{NCX}^{+}(\mathbf{r}),\psi_{CY}(\mathbf{r}),\psi_{CY}^{+}(\mathbf{r}),\psi_{NCY}(\mathbf{r}),\psi_{NCY}^{+}(\mathbf{r})\end{eqnarray*}
This form of the distribution functional is analogous the corresponding
form for the distribution function, which has been used as the basis
for deriving Ito stochastic equations for the real, imaginary parts
of the phase variables \citep{Drummond80a}, \citep{Gardiner91a}.\begin{eqnarray}
 &  & P_{b}(\alpha_{1},\alpha_{1}^{+},..,\alpha_{k},\alpha_{k}^{+},..,\alpha_{1}^{\ast},\alpha_{1}^{+\ast},..,\alpha_{k}^{\ast},\alpha_{k}^{+\ast},..)\nonumber \\
 & \equiv & F_{b}(\alpha_{1X},\alpha_{1X}^{+},..,\alpha_{kX},\alpha_{kX}^{+},..,\alpha_{1Y},\alpha_{1Y}^{+},..,\alpha_{kY},\alpha_{kY}^{+},..)\label{Eq.DistnFnRealImagCompts}\end{eqnarray}

\subsection{Interferometric Measurements}

Coherence effects in BECs are described via \emph{Quantum Correlation
Functions}\begin{eqnarray}
 &  & G^{N}(\mathbf{r}_{1}\mathbf{,r}_{2}\mathbf{,..,r}_{N};\mathbf{s}_{N}\mathbf{,..,s}_{2}\mathbf{,s}_{1})=\left\langle \,\widehat{\Psi}\,(\mathbf{r}_{1})^{\dagger}\,..\widehat{\Psi}\,(\mathbf{r}_{N})^{\dagger}\,\widehat{\Psi}\,(\mathbf{s}_{N})\,..\,\widehat{\Psi}\,(\mathbf{s}_{1})\right\rangle \nonumber \\
 & = & Tr(\widehat{\rho}(t)\,\widehat{\Psi}\,(\mathbf{r}_{1})^{\dagger}\,..\widehat{\Psi}\,(\mathbf{r}_{N})^{\dagger}\,\widehat{\Psi}\,(\mathbf{s}_{N})\,..\,\widehat{\Psi}\,(\mathbf{s}_{1}))\label{Eq.QuantCorrFns}\end{eqnarray}
Various BEC \emph{spatial interference effects} can be described via
quantum correlation functions, which thereby specify the spatial coherence
effects.

If we interchange coordinates of a pair of \emph{bosons}, say $(\mathbf{r}_{i}\mathbf{,s}_{i})\leftrightarrow(\mathbf{r}_{j}\mathbf{,s}_{j})$
we see that because the commutation properties of the bosonic field
operators, the quantum correlation function is unchanged. The \emph{symmetrization
principle} for bosonic systems is consistent with measured quantities
remaining unchanged due to interchange of identical particles.

The quantum correlation function with $\mathbf{r}_{i}=\mathbf{s}_{i}\,(i=1,...,N)$
measures the simultaneous probability of detecting one boson at $\mathbf{r}_{1}$,
a second at $\mathbf{r}_{2}$, .., the $N$th at $\mathbf{r}_{N}$,
(\citep{Bach04a}. Actual measurements of quantum correlation functions
may be made via scattering a weak probe beam (atoms, light) off the
system, (\citep{CohenTannoudji01a}. If the field operators are written
as the sum of condensate and non-condensate terms, then the quantum
correlation functions will contain purely condensate terms, purely
non-condensate terms and mixed terms involving both condensate and
non-condensate operators.

The quantum averages of symmetrically ordered products of the condensate
field operators $\{\widehat{\Psi}_{C}^{\dagger}(\mathbf{r}_{1})\widehat{\Psi}_{C}^{\dagger}(\mathbf{r}_{2})....\widehat{\Psi}_{C}^{\dagger}(\mathbf{r}_{p})\widehat{\Psi}_{C}(\mathbf{s}_{q})..\widehat{\Psi}_{C}(\mathbf{s}_{1})\}$
and normally ordered products of the non-condensate field operators\\
 $\widehat{\Psi}_{NC}^{\dagger}(\mathbf{u}_{1})\widehat{\Psi}_{NC}^{\dagger}(\mathbf{u}_{2})....\widehat{\Psi}_{NC}^{\dagger}(\mathbf{u}_{r})\widehat{\Psi}_{NC}(\mathbf{v}_{s})..\widehat{\Psi}_{NC}(\mathbf{v}_{1})$
may then be expressed as functional integrals of the quasi-distribution
functional\\
 $P[\psi_{C}(\mathbf{r}),\psi_{C}^{+}(\mathbf{r}),\psi_{NC}(\mathbf{r}),\psi_{NC}^{+}(\mathbf{r}),\psi_{C}^{\ast}(\mathbf{r}),\psi_{C}^{+\ast}(\mathbf{r}),\psi_{NC}^{\ast}(\mathbf{r}),\psi_{NC}^{+\ast}(\mathbf{r})]$
with products of the field functions. Thus, with $\left\langle \widehat{\Xi}\right\rangle \equiv Tr(\widehat{\rho}\widehat{\Xi})$

\begin{eqnarray}
 &  & \left\langle \{\widehat{\Psi}_{C}^{\dagger}(\mathbf{r}_{1})....\widehat{\Psi}_{C}^{\dagger}(\mathbf{r}_{p})\widehat{\Psi}_{C}(\mathbf{s}_{q})..\widehat{\Psi}_{C}(\mathbf{s}_{1})\}\,\widehat{\Psi}_{NC}^{\dagger}(\mathbf{u}_{1})....\widehat{\Psi}_{NC}^{\dagger}(\mathbf{u}_{r})\widehat{\Psi}_{NC}(\mathbf{v}_{s})..\widehat{\Psi}_{NC}(\mathbf{v}_{1})\right\rangle \nonumber \\
 & = & Tr\left(\widehat{\rho}\,\{\widehat{\Psi}_{C}^{\dagger}(\mathbf{r}_{1})....\widehat{\Psi}_{C}^{\dagger}(\mathbf{r}_{p})\widehat{\Psi}_{C}(\mathbf{s}_{q})..\widehat{\Psi}_{C}(\mathbf{s}_{1})\}\,\widehat{\Psi}_{NC}^{\dagger}(\mathbf{u}_{1})....\widehat{\Psi}_{NC}^{\dagger}(\mathbf{u}_{r})\widehat{\Psi}_{NC}(\mathbf{v}_{s})..\widehat{\Psi}_{NC}(\mathbf{v}_{1})\right)\nonumber \\
 & = & \iiiint D^{2}\psi_{C}\, D^{2}\psi_{C}^{+}\, D^{2}\psi_{NC}\, D^{2}\psi_{NC}^{+}\,\nonumber \\
 &  & \times P[\psi_{C}(\mathbf{r}),\psi_{C}^{+}(\mathbf{r}),\psi_{NC}(\mathbf{r}),\psi_{NC}^{+}(\mathbf{r}),\psi_{C}^{\ast}(\mathbf{r}),\psi_{C}^{+\ast}(\mathbf{r}),\psi_{NC}^{\ast}(\mathbf{r}),\psi_{NC}^{+\ast}(\mathbf{r})]\nonumber \\
 &  & \times\psi_{C}^{+}(\mathbf{r}_{1})\,\psi_{C}^{+}(\mathbf{r}_{2})\,..\psi_{C}^{+}(\mathbf{r}_{p})\,\psi_{C}(\mathbf{s}_{q})\,..\psi_{C}(\mathbf{s}_{2}).\psi_{C}(\mathbf{s}_{1})\nonumber \\
 &  & \times\psi_{NC}^{+}(\mathbf{u}_{1})\,\psi_{NC}^{+}(\mathbf{u}_{2})\,..\psi_{NC}^{+}(\mathbf{u}_{r})\,\psi_{NC}(\mathbf{v}_{s})\,..\psi_{NC}(\mathbf{v}_{2})\,\psi_{NC}(\mathbf{v}_{1})\label{Eq.QuantumAverages}\end{eqnarray}
and where \begin{eqnarray}
 &  & \{\widehat{\Psi}^{\dagger}(\mathbf{r}_{1})\widehat{\Psi}^{\dagger}(\mathbf{r}_{2})....\widehat{\Psi}^{\dagger}(\mathbf{r}_{p})\widehat{\Psi}(\mathbf{s}_{q})..\widehat{\Psi}(\mathbf{s}_{1})\}\nonumber \\
 & = & \frac{1}{(p+q)!}\sum\limits _{R}\Re(\widehat{\Psi}^{\dagger}(\mathbf{r}_{1})\widehat{\Psi}^{\dagger}(\mathbf{r}_{2})....\widehat{\Psi}^{\dagger}(\mathbf{r}_{p})\widehat{\Psi}(\mathbf{s}_{q})..\widehat{\Psi}(\mathbf{s}_{1})).\label{Eq.SymmOrder}\end{eqnarray}
In Eq.(\ref{Eq.SymmOrder}) the sum over $R$ is over all $(p+q)!$
orderings $\Re$ of the factors $\widehat{\Psi}^{\dagger}(\mathbf{r}_{1})\widehat{\Psi}^{\dagger}(\mathbf{r}_{2})....\widehat{\Psi}^{\dagger}(\mathbf{r}_{p})\widehat{\Psi}(\mathbf{s}_{q})..\widehat{\Psi}(\mathbf{s}_{1})$.
T hus, the condensate field operator $\widehat{\Psi}_{C}^{\dagger}(\mathbf{r}_{i})$
is replaced by $\psi_{C}^{+}(\mathbf{r}_{i})$ and $\widehat{\Psi}_{C}(\mathbf{s}_{j})$
is replaced by $\psi(\mathbf{s}_{j})$, with analogous replacements
for the non-condensate field operators. The proof is given in Appendix
C (\citep{Dalton10b}) and involves functional differentiation, which
is explained in Appendix B (\citep{Dalton10b}).

These results together with the equal time commutation rules give
the quantum correlation functions. For example, the \emph{first order}
quantum correlation function (which is used to exhibit macroscopic
spatial coherence in a BEC) is given by\begin{eqnarray}
 &  & G^{1}(\mathbf{r}_{1};\mathbf{s}_{1})\nonumber \\
 & = & \left\langle \,\widehat{\Psi}(\mathbf{r}_{1})^{\dagger}\,\widehat{\Psi}(\mathbf{s}_{1})\right\rangle \nonumber \\
 & = & \left\langle \,\widehat{\Psi}_{C}(\mathbf{r}_{1})^{\dagger}\,\widehat{\Psi}_{C}(\mathbf{s}_{1})\right\rangle +\left\langle \,\widehat{\Psi}_{C}(\mathbf{r}_{1})^{\dagger}\,\widehat{\Psi}_{NC}(\mathbf{s}_{1})\right\rangle \nonumber \\
 &  & +\left\langle \,\widehat{\Psi}_{NC}(\mathbf{r}_{1})^{\dagger}\,\widehat{\Psi}_{C}(\mathbf{s}_{1})\right\rangle +\left\langle \,\widehat{\Psi}_{NC}(\mathbf{r}_{1})^{\dagger}\,\widehat{\Psi}_{NC}(\mathbf{s}_{1})\right\rangle \nonumber \\
 & = & \left\langle \left(\{\,\widehat{\Psi}_{C}(\mathbf{r}_{1})^{\dagger}\,\widehat{\Psi}_{C}(\mathbf{s}_{1})\}-\frac{1}{2}\delta(\mathbf{r}_{1}-\mathbf{s}_{1})\right)\right\rangle +\left\langle \,\{\widehat{\Psi}_{C}(\mathbf{r}_{1})^{\dagger}\}\,\widehat{\Psi}_{NC}(\mathbf{s}_{1})\right\rangle \nonumber \\
 &  & +\left\langle \,\widehat{\Psi}_{NC}(\mathbf{r}_{1})^{\dagger}\,\{\widehat{\Psi}_{C}(\mathbf{s}_{1})\}\right\rangle +\left\langle \,\widehat{\Psi}_{NC}(\mathbf{r}_{1})^{\dagger}\,\widehat{\Psi}_{NC}(\mathbf{s}_{1})\right\rangle \nonumber \\
 & = & -\frac{1}{2}\delta(\mathbf{r}_{1}-\mathbf{s}_{1})\nonumber \\
 &  & +\iiiint D^{2}\psi_{C}\, D^{2}\psi_{C}^{+}\, D^{2}\psi_{NC}\, D^{2}\psi_{NC}^{+}\,\nonumber \\
 &  & \times P[\psi_{C}(\mathbf{r}),\psi_{C}^{+}(\mathbf{r}),\psi_{NC}(\mathbf{r}),\psi_{NC}^{+}(\mathbf{r}),\psi_{C}^{\ast}(\mathbf{r}),\psi_{C}^{+\ast}(\mathbf{r}),\psi_{NC}^{\ast}(\mathbf{r}),\psi_{NC}^{+\ast}(\mathbf{r})]\nonumber \\
 &  & \times(\psi_{C}^{+}(\mathbf{r}_{1})+\,\psi_{NC}^{+}(\mathbf{r}_{1}))(\psi_{C}(\mathbf{s}_{1})+\,\psi_{NC}(\mathbf{s}_{1}))\label{Eq.1stOrderQCF}\end{eqnarray}
and includes pure condensate terms, pure non-condensate terms and
mixed terms. Note the delta function term which arises because of
the difference between normal and symmetric ordering that applies
for the condensate terms.

It is worth noting that some authors \citep{Castin98a}, \citep{Sorensen02a},
\citep{Gardiner07a} determine the mode functions as the eigenfunctions
of the first order quantum correlation function $\left\langle \,\widehat{\Psi}(\mathbf{r}_{1})^{\dagger}\,\widehat{\Psi}(\mathbf{s}_{1})\right\rangle $.
Thus the mode functions $\varphi_{i}(\mathbf{r})$ satisfy the eigenvalue
equations\begin{equation}
\int d\mathbf{s}_{1}\left\langle \,\widehat{\Psi}(\mathbf{r}_{1})^{\dagger}\,\widehat{\Psi}(\mathbf{s}_{1})\right\rangle \varphi_{i}(\mathbf{s}_{1})=\lambda_{i}\varphi_{i}(\mathbf{r}_{1})\label{Eq.CondensateModes}\end{equation}
The mode functions can be shown to be orthonormal and the eigenvalues
real and positive. The eigenvalues give the occupancy of the modes.
For two mode condensates in a general fragmented state, two such eigenvalues
will have macroscopic values $\symbol{126}N$ and the other modes
will have small eigenvalues. This approach to determining the mode
functions has certain formal advantages, such as leading to the $\widehat{H}_{2}$
term in the Hamiltonian being zero. However, the method would require
knowing the first order correlation function, and it is not clear
how this could be done prior to knowing the mode functions. In the
present approach the formalism is designed as a way to determine all
the quantum correlation functions.\pagebreak{}

\section{Functional Fokker-Planck equation}

\label{Sect Func Fokker Planck}

In this section we show how the Liouville-von Neumann equation for
the quantum density operator describing the state of the bosonic system
is equivalent to a functional Fokker-Planck equation for the phase
space distribution functional. This is accomplished via the use of
correspondence rules, wherein the product of the quantum density operator
with the various condensate and non-condensate field operators (for
both product orders) is equivalent to the operation of functional
derivatives or field functions on the distribution function. The actual
results for the functional Fokker-Planck equation in the case of the
present two mode BEC condensate system are set out at the end of the
section. For completeness the corresponding simpler results for a
single mode condensate are also obtained.

\subsection{Dynamics}

The state of the bosonic system is described by the \emph{density
operator} $\widehat{\rho}$ which satisfies the \emph{Liouville-von
Neumann equation}\begin{equation}
i\hbar\frac{\partial}{\partial t}\widehat{\rho}=[\widehat{H},\widehat{\rho}]\label{Eq.LiouvilleVN}\end{equation}
where the Bogoliubov Hamiltonian \ref{Eq.BogolHamiltonian} will be
used.

The approach used will be to turn the Liouville-von Neumann equation
for the density operator $\widehat{\rho}$ into a \emph{functional
Fokker-Planck equation} for a quasi distribution functional $P[\psi_{C}(\mathbf{r}),\psi_{C}^{+}(\mathbf{r}),\psi_{NC}(\mathbf{r}),\psi_{NC}^{+}(\mathbf{r}),\psi_{C}^{\ast}(\mathbf{r}),\psi_{C}^{+\ast}(\mathbf{r}),\psi_{NC}^{\ast}(\mathbf{r}),\psi_{NC}^{+\ast}(\mathbf{r})]$
and then replace this by \emph{stochastic equations} for stochastic
field functions\\
 $\widetilde{\psi}_{C}(\mathbf{r,}t),\widetilde{\psi}_{C}^{+}(\mathbf{r,}t),\widetilde{\psi}_{NC}(\mathbf{r,}t),\widetilde{\psi}_{NC}^{+}(\mathbf{r\mathbf{,}}t\mathbf{)}$.
The latter are c-number Langevin equations of the Ito type, and in
general will contain \emph{random noise} terms as well as \emph{deterministic}
terms coupling the field functions.

\subsection{Correspondence Rules}

We now wish to replace the Liouville-von Neumann equation for the
density operator by a \emph{Functional Fokker-Planck Equation} for
the quasi distribution functional. To do this we make use of so-called
\emph{correspondence rules}, in which the effect of a field operator
on the density operator corresponds to the effects of \emph{functional
differentiation} and/or function \emph{multiplication} on the distribution
functional.\emph{\ }

Functional differentiation is fully explained in Appendix B (\citep{Dalton10b}),
but a summary is as follows. For a functional $F[\psi(x)]$ of a field
$\psi(x)$ the functional derivative $\frac{{\LARGE\delta F[\psi(x)]}}{{\LARGE\delta\psi(x)}}$
is defined by \begin{equation}
F[\psi(x)+\delta\psi(x)]\doteqdot F[\psi(x)]+\int dx\,\delta\psi(x)\,\left(\frac{\delta F[\psi(x)]}{\delta\psi(x)}\right)_{x}\label{Eq.FnalDeriv}\end{equation}
where $\delta\psi(x)$ is small. In this equation the left side is
a functional of $\psi(x)+\delta\psi(x)$ and the first term on the
right side is a functional of $\psi(x)$. The second term on the right
side is a functional of $\delta\psi(x)$ and thus the functional derivative
must be a function of $x$, hence the subscript $x$. In most situations
this subscript will be left understood. If we write $\delta\psi(x)=\epsilon\delta(x-y)$
for small $\epsilon$ then an equivalent result for the functional
derivative at $x=y$ is\begin{equation}
\left(\frac{\delta F[\psi(x)]}{\delta\psi(x)}\right)_{x=y}=\lim_{\epsilon\rightarrow0}\left(\frac{F[\psi(x)+\epsilon\delta(x-y)]-F[\psi(x)]}{\epsilon}\right).\label{Eq.FnalDeriv2}\end{equation}
Note that for functionals involving both $\psi(x)$, $\psi^{\ast}(x)$
we treat these complex fields as independent, and functional derivatives
with respect to both $\psi(x)$, $\psi^{\ast}(x)$ exist. Thus\begin{eqnarray}
 &  & F[\psi(x)+\delta\psi(x),\psi^{\ast}(x)+\delta\psi^{\ast}(x)]\nonumber \\
 & \doteqdot & F[\psi(x),\psi^{\ast}(x)]+\int dx\,\delta\psi(x)\,\left(\frac{\delta F[\psi(x),\psi^{\ast}(x)]}{\delta\psi(x)}\right)_{x}\nonumber \\
 &  & +\int dx\,\delta\psi^{\ast}(x)\,\left(\frac{\delta F[\psi(x),\psi^{\ast}(x)]}{\delta\psi^{\ast}(x)}\right)_{x}\label{eq:FnalDeriv3}\end{eqnarray}
For the equivalent functional $G[\psi_{X}(x),\psi_{Y}(x)]\equiv F[\psi(x),\psi^{\ast}(x)]$
involving the real, imaginary components $\psi_{X}(x)$, $\psi_{Y}(x)$
the functional derivatives are defined by\begin{eqnarray}
 &  & G[\psi_{X}(x)+\delta\psi_{X}(x),\psi_{Y}(x)+\delta\psi_{Y}(x)]\nonumber \\
 & \doteqdot & G[\psi_{X}(x),\psi_{Y}(x)]+\int dx\,\delta\psi_{X}(x)\,\left(\frac{\delta G[\psi_{X}(x),\psi_{Y}(x)]}{\delta\psi_{X}(x)}\right)_{x}\nonumber \\
 &  & +\int dx\,\delta\psi_{Y}(x)\,\left(\frac{\delta G[\psi_{X}(x),\psi_{Y}(x)]}{\delta\psi_{Y}(x)}\right)_{x}\label{eq:FnalDer4}\end{eqnarray}
The present case involves a generalisation to treat four complex fields\\
 $\psi_{C}(\mathbf{r}),\psi_{C}^{+}(\mathbf{r}),\psi_{NC}(\mathbf{r}),\psi_{NC}^{+}(\mathbf{r})$.

\subsubsection{Notation}

As the notation is now getting rather cumbersome we will designate
\begin{eqnarray}
\underrightarrow{\psi}(\mathbf{r}) & \equiv & \{\psi_{C}(\mathbf{r}),\psi_{C}^{+}(\mathbf{r}),\psi_{NC}(\mathbf{r}),\psi_{NC}^{+}(\mathbf{r})\}\label{Eq.NotationFields}\\
\underrightarrow{\psi^{\ast}}(\mathbf{r}) & \equiv & \{\psi_{C}^{\ast}(\mathbf{r}),\psi_{C}^{+\ast}(\mathbf{r}),\psi_{NC}^{\ast}(\mathbf{r}),\psi_{NC}^{+\ast}(\mathbf{r})\}\label{Eq.NotationConjFields}\\
P[\underrightarrow{\psi}(\mathbf{r}),\underrightarrow{\psi^{\ast}}(\mathbf{r})] & \equiv & P[\psi_{C}(\mathbf{r}),\psi_{C}^{+}(\mathbf{r}),\psi_{NC}(\mathbf{r}),\psi_{NC}^{+}(\mathbf{r}),\psi_{C}^{\ast}(\mathbf{r}),\psi_{C}^{+\ast}(\mathbf{r}),\psi_{NC}^{\ast}(\mathbf{r}),\psi_{NC}^{+\ast}(\mathbf{r})]\nonumber \\
\label{eq:NotationDistnFnal}\\\underrightarrow{\xi}(\mathbf{r}) & \equiv & \{\xi_{C}(\mathbf{r}),\xi_{C}^{+}(\mathbf{r}),\xi_{NC}(\mathbf{r}),\xi_{NC}^{+}(\mathbf{r})\}\label{Eq.NotationXiFields}\\
\chi\lbrack\underrightarrow{\xi}(\mathbf{r})] & \equiv & \chi\lbrack\xi_{C},\xi_{C}^{+},\xi_{NC},\xi_{NC}^{+}]\label{Eq.NotationCharFnal}\end{eqnarray}
for the fields and the distribution, characteristic functionals. For
the expansion coefficients and the distribution function we introduce
the notation \begin{eqnarray}
\underrightarrow{\alpha} & \equiv & \{\alpha_{k},\alpha_{k}^{+}\}\label{Eq.NotationAmplitudes}\\
\underrightarrow{\alpha}^{\ast} & \equiv & \{\alpha_{k}^{\ast},\alpha_{k}^{+\ast}\}\label{Eq.NotationConjAmplitudes}\\
P_{b}(\underrightarrow{\alpha},\underrightarrow{\alpha}^{\ast}) & \equiv & P_{b}(\alpha_{k},\alpha_{k}^{+},\alpha_{k}^{\ast},\alpha_{k}^{+\ast})\label{Eq.NotationDistnFunction}\\
P[\underrightarrow{\psi}(\mathbf{r}),\underrightarrow{\psi^{\ast}}(\mathbf{r})] & \equiv & P_{b}(\underrightarrow{\alpha},\underrightarrow{\alpha}^{\ast})\label{Eq.EquivalenceDistnFnalFn}\end{eqnarray}
where the original functional\\
 $P[\psi_{C}(\mathbf{r}),\psi_{C}^{+}(\mathbf{r}),\psi_{NC}(\mathbf{r}),\psi_{NC}^{+}(\mathbf{r}),\psi_{C}^{\ast}(\mathbf{r}),\psi_{C}^{+\ast}(\mathbf{r}),\psi_{NC}^{\ast}(\mathbf{r}),\psi_{NC}^{+\ast}(\mathbf{r})]$
of the fields\\
 $\psi_{C}(\mathbf{r}),\psi_{C}^{+}(\mathbf{r}),\psi_{NC}(\mathbf{r}),\psi_{NC}^{+}(\mathbf{r})$
and their complex conjugates $\psi_{C}^{\ast}(\mathbf{r}),\psi_{C}^{+\ast}(\mathbf{r}),\psi_{NC}^{\ast}(\mathbf{r}),\psi_{NC}^{+\ast}(\mathbf{r})$
is equivalent to the function $P_{b}(\alpha_{k},\alpha_{k}^{+},\alpha_{k}^{\ast},\alpha_{k}^{+\ast})$
of the expansion amplitudes $\alpha_{k},\alpha_{k}^{+}$ and their
complex conjugates $\alpha_{k}^{\ast},\alpha_{k}^{+\ast}$.

\subsubsection{Correspondence Rules for Condensate and Non-Condensate Fields}

For the condensate operators we have\begin{eqnarray}
\widehat{\Psi}_{C}(\mathbf{s})\widehat{\rho} & \leftrightarrow & \left(\psi_{C}(\mathbf{s})+\frac{1}{2}\frac{\delta}{\delta\psi_{C}^{+}(\mathbf{s})}\right)P[\underrightarrow{\psi}(\mathbf{r}),\underrightarrow{\psi^{\ast}}(\mathbf{r})]\nonumber \\
\widehat{\rho}\widehat{\Psi}_{C}(\mathbf{s}) & \leftrightarrow & \left(\psi_{C}(\mathbf{s})-\frac{1}{2}\frac{\delta}{\delta\psi_{C}^{+}(\mathbf{s})}\right)P[\underrightarrow{\psi}(\mathbf{r}),\underrightarrow{\psi^{\ast}}(\mathbf{r})]\nonumber \\
\widehat{\Psi}_{C}^{\dagger}(\mathbf{s})\widehat{\rho} & \leftrightarrow & \left(\psi_{C}^{+}(\mathbf{s})-\frac{1}{2}\frac{\delta}{\delta\psi_{C}(\mathbf{s})}\right)P[\underrightarrow{\psi}(\mathbf{r}),\underrightarrow{\psi^{\ast}}(\mathbf{r})]\nonumber \\
\widehat{\rho}\widehat{\Psi}_{C}^{\dagger}(\mathbf{s}) & \leftrightarrow & \left(\psi_{C}^{+}(\mathbf{s})+\frac{1}{2}\frac{\delta}{\delta\psi_{C}(\mathbf{s})}\right)P[\underrightarrow{\psi}(\mathbf{r}),\underrightarrow{\psi^{\ast}}(\mathbf{r})]\label{Eq.CorrespRulesCond}\end{eqnarray}
and for the non-condensate operators \begin{eqnarray}
\widehat{\Psi}_{NC}(\mathbf{s})\widehat{\rho} & \leftrightarrow & \left(\psi_{NC}(\mathbf{s})\right)P[\underrightarrow{\psi}(\mathbf{r}),\underrightarrow{\psi^{\ast}}(\mathbf{r})]\nonumber \\
\widehat{\rho}\widehat{\Psi}_{NC}(\mathbf{s}) & \leftrightarrow & \left(\psi_{NC}(\mathbf{s})-\frac{\delta}{\delta\psi_{NC}^{+}(\mathbf{s})}\right)P[\underrightarrow{\psi}(\mathbf{r}),\underrightarrow{\psi^{\ast}}(\mathbf{r})]\nonumber \\
\widehat{\Psi}_{NC}^{\dagger}(\mathbf{s})\widehat{\rho} & \leftrightarrow & \left(\psi_{NC}^{+}(\mathbf{s})-\frac{\delta}{\delta\psi_{NC}(\mathbf{s})}\right)P[\underrightarrow{\psi}(\mathbf{r}),\underrightarrow{\psi^{\ast}}(\mathbf{r})]\nonumber \\
\widehat{\rho}\widehat{\Psi}_{NC}^{\dagger}(\mathbf{s}) & \leftrightarrow & \left(\psi_{NC}^{+}(\mathbf{s})\right)P[\underrightarrow{\psi}(\mathbf{r}),\underrightarrow{\psi^{\ast}}(\mathbf{r})]\label{Eq.CorrespRulesNonCond}\end{eqnarray}
whilst for the density operator\begin{equation}
\frac{\partial\widehat{\rho}}{\partial t}\rightarrow\frac{\partial P[\underrightarrow{\psi}(\mathbf{r}),\underrightarrow{\psi^{\ast}}(\mathbf{r})]}{\partial t}\label{Eq.CorrespRuleDensityOpr}\end{equation}

\subsubsection{Deriving the Correspondence Rules}

The proof of these correspondence rules is dealt with in Appendix
D (\citep{Dalton10b}). Key steps in the derivation include first
establishing the following changes to the characteristic functional.
For the condensate modes we have\begin{eqnarray}
\widehat{\Psi}_{C}(\mathbf{s})\widehat{\rho} & \leftrightarrow & \frac{1}{i}\,\left(\frac{\delta}{\delta\xi_{C}^{+}(\mathbf{s})}+\frac{1}{2}\xi_{C}(\mathbf{s})\right)\,\chi\lbrack\underrightarrow{\xi}(\mathbf{r})]\nonumber \\
\widehat{\rho}\widehat{\Psi}_{C}(\mathbf{s}) & \leftrightarrow & \frac{1}{i}\,\left(\frac{\delta}{\delta\xi_{C}^{+}(\mathbf{s})}-\frac{1}{2}\xi_{C}(\mathbf{s})\right)\,\chi\lbrack\underrightarrow{\xi}(\mathbf{r})]\nonumber \\
\widehat{\Psi}_{C}^{\dagger}(\mathbf{s})\widehat{\rho} & \leftrightarrow & \frac{1}{i}\,\left(\frac{\delta}{\delta\xi_{C}(\mathbf{s})}-\frac{1}{2}\xi_{C}^{+}(\mathbf{s})\right)\,\chi\lbrack\underrightarrow{\xi}(\mathbf{r})]\nonumber \\
\widehat{\rho}\widehat{\Psi}_{C}^{\dagger}(\mathbf{s}) & \leftrightarrow & \frac{1}{i}\,\left(\frac{\delta}{\delta\xi_{C}(\mathbf{s})}+\frac{1}{2}\xi_{C}^{+}(\mathbf{s})\right)\,\chi\lbrack\underrightarrow{\xi}(\mathbf{r})]\label{Eq.CondCharFnalCorres}\end{eqnarray}
and for the non-condensate modes

\begin{eqnarray}
\widehat{\Psi}_{NC}(\mathbf{s})\widehat{\rho} & \leftrightarrow & \frac{1}{i}\,\left(\frac{\delta}{\delta\xi_{NC}^{+}(\mathbf{s})}\right)\,\chi\lbrack\underrightarrow{\xi}(\mathbf{r})]\nonumber \\
\widehat{\rho}\widehat{\Psi}_{NC}(\mathbf{s}) & \leftrightarrow & \frac{1}{i}\,\left(\frac{\delta}{\delta\xi_{NC}^{+}(\mathbf{s})}-\xi_{NC}(\mathbf{s})\right)\,\chi\lbrack\underrightarrow{\xi}(\mathbf{r})]\nonumber \\
\widehat{\Psi}_{NC}^{\dagger}(\mathbf{s})\widehat{\rho} & \leftrightarrow & \frac{1}{i}\,\left(\frac{\delta}{\delta\xi_{NC}(\mathbf{s})}-\xi_{NC}^{+}(\mathbf{s})\right)\,\chi\lbrack\underrightarrow{\xi}(\mathbf{r})]\nonumber \\
\widehat{\rho}\widehat{\Psi}_{NC}^{\dagger}(\mathbf{s}) & \leftrightarrow & \frac{1}{i}\,\left(\frac{\delta}{\delta\xi_{NC}(\mathbf{s})}\right)\,\chi\lbrack\underrightarrow{\xi}(\mathbf{r})]\label{Eq.NonCondCharFnalCorres}\end{eqnarray}

As can be seen from eqs. (\ref{Eq.FieldFn1}, \ref{Eq.FieldFn2},
\ref{Eq.FieldFn3}, \ref{Eq.FieldFn4}) the distribution function
$P[\psi_{C}(\mathbf{r}),\psi_{C}^{+}(\mathbf{r}),\psi_{NC}(\mathbf{r}),\psi_{NC}^{+}(\mathbf{r}),\psi_{C}^{\ast}(\mathbf{r}),\psi_{C}^{+\ast}(\mathbf{r}),\psi_{NC}^{\ast}(\mathbf{r}),\psi_{NC}^{+\ast}(\mathbf{r})]$
is a functional of restricted functions (see Appendix B, \citep{Dalton10b}).
It can also be considered as a function $P_{b}(\alpha_{k},\alpha_{k}^{+},\alpha_{k}^{\ast},\alpha_{k}^{+\ast})$
of all the expansion coefficients $\alpha_{k},\alpha_{k}^{+}$ in
eqs. (\ref{Eq.FieldFn1}, \ref{Eq.FieldFn2}, \ref{Eq.FieldFn3},
\ref{Eq.FieldFn4}) and their complex conjugates $\alpha_{k}^{\ast},\alpha_{k}^{+\ast}$.
Hence in applying the correspondence rules the following operator
identities for the various functional derivatives can be used \begin{eqnarray}
\left(\frac{\delta}{\delta\psi_{C}(\mathbf{s})}\right)_{\mathbf{s}} & \equiv & \sum\limits _{k=1,2}\phi_{k}^{\ast}(\mathbf{s})\,\frac{\partial}{\partial\alpha_{k}}\qquad\left(\frac{\delta}{\delta\psi_{NC}(\mathbf{s})}\right)_{\mathbf{s}}\equiv\sum\limits _{k\neq1,2}^{K}\phi_{k}^{\ast}(\mathbf{s})\,\frac{\partial}{\partial\alpha_{k}}\nonumber \\
\left(\frac{\delta}{\delta\psi_{C}^{+}(\mathbf{s})}\right)_{\mathbf{s}} & \equiv & \sum\limits _{k=1,2}\phi_{k}(\mathbf{s})\,\frac{\partial}{\partial\alpha_{k}^{+}}\qquad\left(\frac{\delta}{\delta\psi_{NC}^{+}(\mathbf{s})}\right)_{\mathbf{s}}\equiv\sum\limits _{k\neq1,2}^{K}\phi_{k}(\mathbf{s})\,\frac{\partial}{\partial\alpha_{k}^{+}}\nonumber \\
 &  & \,\label{eq:RestrictFnalDerivIdent1}\end{eqnarray}
where it is understood that the left side operates on the distribution
functional $P[\psi_{C}(\mathbf{r}),\psi_{C}^{+}(\mathbf{r}),\psi_{NC}(\mathbf{r}),\psi_{NC}^{+}(\mathbf{r}),\psi_{C}^{\ast}(\mathbf{r}),\psi_{C}^{+\ast}(\mathbf{r}),\psi_{NC}^{\ast}(\mathbf{r}),\psi_{NC}^{+\ast}(\mathbf{r})]$
of the restricted functions $\psi_{C}(\mathbf{r}),\psi_{C}^{+}(\mathbf{r}),\psi_{NC}(\mathbf{r}),\psi_{NC}^{+}(\mathbf{r}),\psi_{C}^{\ast}(\mathbf{r}),\psi_{C}^{+\ast}(\mathbf{r}),\psi_{NC}^{\ast}(\mathbf{r}),\psi_{NC}^{+\ast}(\mathbf{r})$
and the right side operates on the equivalent function $P_{b}(\alpha_{k},\alpha_{k}^{+},\alpha_{k}^{\ast},\alpha_{k}^{+\ast})$.
The related identities for the functional differentiation with respect
to the complex conjugate fields also exist, but are not needed because
the correspondence rules only involve functional derivation with respect
to $\psi_{C}(\mathbf{r}),\psi_{C}^{+}(\mathbf{r}),\psi_{NC}(\mathbf{r}),\psi_{NC}^{+}(\mathbf{r})$,
and any functions arising from the multiplications are only functions
of these fields and not their complex conjugates.

In deriving the correspondence rules that result in functional differentiation
a key step involves a functional integration by parts of the form

\begin{eqnarray}
 &  & \iiiint D^{2}\psi_{C}\, D^{2}\psi_{C}^{+}\, D^{2}\psi_{NC}\, D^{2}\psi_{NC}^{+}\,\left(\frac{\delta G[\underrightarrow{\psi}(\mathbf{r})]}{\delta\psi(\mathbf{r})}\right)P[\underrightarrow{\psi}(\mathbf{r}),\underrightarrow{\psi^{\ast}}(\mathbf{r})]\nonumber \\
 & = & -\iiiint D^{2}\psi_{C}\, D^{2}\psi_{C}^{+}\, D^{2}\psi_{NC}\, D^{2}\psi_{NC}^{+}\, G[\underrightarrow{\psi}(\mathbf{r})]\left(\frac{\delta P[\underrightarrow{\psi}(\mathbf{r}),\underrightarrow{\psi^{\ast}}(\mathbf{r})]}{\delta\psi(\mathbf{r})}\right)\nonumber \\
 &  & \,\label{eq:KeyIntegPartsStepReal}\end{eqnarray}
where \begin{eqnarray}
G[\underrightarrow{\psi}(\mathbf{r})] & = & \exp i\int d\mathbf{r\,\{}\xi_{C}(\mathbf{r})\psi_{C}^{+}(\mathbf{r})+\psi_{C}(\mathbf{r})\xi_{C}^{+}(\mathbf{r})\}\,\nonumber \\
 &  & \times\exp i\int d\mathbf{r\,\{}\xi_{C}(\mathbf{r})\psi_{C}^{+}(\mathbf{r})\}\,\exp i\int d\mathbf{r\,\{}\xi_{NC}(\mathbf{r})\psi_{NC}^{+}(\mathbf{r})\}\label{Eq.FnalGiving CharFnal}\end{eqnarray}
is a functional of the four fields $\psi_{C}(\mathbf{r}),\psi_{C}^{+}(\mathbf{r}),\psi_{NC}(\mathbf{r}),\psi_{NC}^{+}(\mathbf{r})$
and $\psi(\mathbf{r})$ refers to any one of these. This step relies
on the distribution function $P_{b}(\alpha_{k},\alpha_{k}^{+},\alpha_{k}^{\ast},\alpha_{k}^{+\ast})$
going to zero on the boundaries of phase space, an assumption common
to all correspondence rule derivations. Note that the functional differentiation
of $P[\psi_{C}(\mathbf{r}),\psi_{C}^{+}(\mathbf{r}),\psi_{NC}(\mathbf{r}),\psi_{NC}^{+}(\mathbf{r}),\psi_{C}^{\ast}(\mathbf{r}),\psi_{C}^{+\ast}(\mathbf{r}),\psi_{NC}^{\ast}(\mathbf{r}),\psi_{NC}^{+\ast}(\mathbf{r})]$
is well-defined, since $P[\underrightarrow{\psi}(\mathbf{r}),\underrightarrow{\psi^{\ast}}(\mathbf{r})]$
is a functional of both the fields and their complex conjugates.

\subsubsection{Real and Imaginary Field Components}

Note that because $G[\underrightarrow{\psi}(\mathbf{r})]$ does not
depend on the conjugate fields, its functional derivative with respect
to any $\psi^{\ast}(\mathbf{r})$ is zero. Thus we also have \begin{eqnarray}
 &  & 0=\iiiint D^{2}\psi_{C}\, D^{2}\psi_{C}^{+}\, D^{2}\psi_{NC}\, D^{2}\psi_{NC}^{+}\,\left(\frac{\delta G[\underrightarrow{\psi}(\mathbf{r})]}{\delta\psi^{\ast}(\mathbf{r})}\right)P[\underrightarrow{\psi}(\mathbf{r}),\underrightarrow{\psi^{\ast}}(\mathbf{r})]\nonumber \\
 & = & -\iiiint D^{2}\psi_{C}\, D^{2}\psi_{C}^{+}\, D^{2}\psi_{NC}\, D^{2}\psi_{NC}^{+}\, G[\underrightarrow{\psi}(\mathbf{r})]\left(\frac{\delta P[\underrightarrow{\psi}(\mathbf{r}),\underrightarrow{\psi^{\ast}}(\mathbf{r})]}{\delta\psi^{\ast}(\mathbf{r})}\right)\label{Eq.KeyIntegPartsStepComplex}\end{eqnarray}
Adding an arbitrary multiple $\lambda$ of the last equation to each
side of (\ref{eq:KeyIntegPartsStepReal}) gives\begin{eqnarray}
 &  & \iiiint D^{2}\psi_{C}\, D^{2}\psi_{C}^{+}\, D^{2}\psi_{NC}\, D^{2}\psi_{NC}^{+}\,\left(\frac{\delta G[\underrightarrow{\psi}(\mathbf{r})]}{\delta\psi(\mathbf{r})}\right)P[\underrightarrow{\psi}(\mathbf{r}),\underrightarrow{\psi^{\ast}}(\mathbf{r})]\nonumber \\
 & = & -\iiiint D^{2}\psi_{C}\, D^{2}\psi_{C}^{+}\, D^{2}\psi_{NC}\, D^{2}\psi_{NC}^{+}\, G[\underrightarrow{\psi}(\mathbf{r})]\nonumber \\
 &  & \times\left(\frac{\delta P[\underrightarrow{\psi}(\mathbf{r}),\underrightarrow{\psi^{\ast}}(\mathbf{r})]}{\delta\psi(\mathbf{r})}+\lambda\frac{\delta P[\underrightarrow{\psi}(\mathbf{r}),\underrightarrow{\psi^{\ast}}(\mathbf{r})]}{\delta\psi^{\ast}(\mathbf{r})}\right)\end{eqnarray}
Noting that we can write the field in terms of its real, imaginary
components and replace the distribution functional $P[\underrightarrow{\psi}(\mathbf{r}),\underrightarrow{\psi^{\ast}}(\mathbf{r})]$
with an equivalent functional $F[\underrightarrow{\psi_{X}}(\mathbf{r}),\underrightarrow{\psi_{Y}}(\mathbf{r})]$
of the components \begin{eqnarray}
\psi(\mathbf{r}) & = & \psi_{X}(\mathbf{r})+i\psi_{Y}(\mathbf{r})\qquad\psi^{\ast}(\mathbf{r})=\psi_{X}(\mathbf{r})-i\psi_{Y}(\mathbf{r})\nonumber \\
\psi_{X}(\mathbf{r}) & = & (\psi(\mathbf{r})+\psi^{\ast}(\mathbf{r}))/2\qquad\psi_{Y}(\mathbf{r})=(\psi(\mathbf{r})-\psi^{\ast}(\mathbf{r}))/2i\nonumber \\
\underrightarrow{\psi_{X}}(\mathbf{r}) & \equiv & \{\psi_{CX}(\mathbf{r}),\psi_{CX}^{+}(\mathbf{r}),\psi_{NCX}(\mathbf{r}),\psi_{NCX}^{+}(\mathbf{r})\}\nonumber \\
\underrightarrow{\psi_{Y}}(\mathbf{r}) & \equiv & \{\psi_{CY}(\mathbf{r}),\psi_{CY}^{+}(\mathbf{r}),\psi_{NCY}(\mathbf{r}),\psi_{NCY}^{+}(\mathbf{r})\}\label{Eq.NotationRealImagCompts}\\
P[\underrightarrow{\psi}(\mathbf{r}),\underrightarrow{\psi^{\ast}}(\mathbf{r})] & \equiv & F[\underrightarrow{\psi_{X}}(\mathbf{r}),\underrightarrow{\psi_{Y}}(\mathbf{r})]\label{Eq.NotationDistnFnalReImagFlds}\end{eqnarray}
then a straightford application of functional differentiation rules
shows that by choosing $\lambda=1$ or $\lambda=-1$ we have \begin{eqnarray}
 &  & \iiiint D^{2}\psi_{C}\, D^{2}\psi_{C}^{+}\, D^{2}\psi_{NC}\, D^{2}\psi_{NC}^{+}\,\left(\frac{\delta G[\underrightarrow{\psi}(\mathbf{r})]}{\delta\psi(\mathbf{r})}\right)P[\underrightarrow{\psi}(\mathbf{r}),\underrightarrow{\psi^{\ast}}(\mathbf{r})]\nonumber \\
 & = & -\iiiint D^{2}\psi_{C}\, D^{2}\psi_{C}^{+}\, D^{2}\psi_{NC}\, D^{2}\psi_{NC}^{+}\, G[\underrightarrow{\psi}(\mathbf{r})]\left(\frac{\delta F[\underrightarrow{\psi_{X}}(\mathbf{r}),\underrightarrow{\psi_{Y}}(\mathbf{r})]}{\delta\psi_{X}(\mathbf{r})}\right)\nonumber \\
\label{eq:KeyIntegPartsStepRealCompt}\\ & = & -\iiiint D^{2}\psi_{C}\, D^{2}\psi_{C}^{+}\, D^{2}\psi_{NC}\, D^{2}\psi_{NC}^{+}\, G[\underrightarrow{\psi}(\mathbf{r})]\left(\frac{\delta F[\underrightarrow{\psi_{X}}(\mathbf{r}),\underrightarrow{\psi_{Y}}(\mathbf{r})]}{\delta(i\psi_{Y}(\mathbf{r}))}\right)\nonumber \\
 &  & \,\label{eq:KeyIntegPartsStepImagCompt}\end{eqnarray}
This shows that functional differentiation of the distribution functional
with respect to $\psi(\mathbf{r})$ is equivalent to functional differentiation
of the related distribution functional $F[\underrightarrow{\psi_{X}}(\mathbf{r}),\underrightarrow{\psi_{Y}}(\mathbf{r})]$
with respect to either $\psi_{X}(\mathbf{r})$ or $i\psi_{Y}(\mathbf{r})$.
This feature is useful if we wish to replace the fields by their real,
imaginary components.

\subsubsection{Applying the Correspondence Rules}

In dealing with terms in the Liouville-von Neumann equation the density
operator is often operated on by more than one field operator. To
determine the overall effect on the quasi distribution functional
it is necessary to carry out the above replacements in succession.
A couple of examples illustrate the proceedure.\begin{eqnarray*}
 &  & \widehat{\Psi}_{NC}^{\dagger}(\mathbf{s}_{1})\widehat{\rho}\widehat{\Psi}_{C}(\mathbf{s}_{2})\\
 & \rightarrow & \left(\psi_{NC}^{+}(\mathbf{s}_{1})-\frac{\delta}{\delta\psi_{NC}(\mathbf{s}_{1})}\right)\left(\psi_{C}(\mathbf{s}_{2})-\frac{1}{2}\frac{\delta}{\delta\psi_{C}^{+}(\mathbf{s}_{2})}\right)P[\underrightarrow{\psi}(\mathbf{r}),\underrightarrow{\psi^{\ast}}(\mathbf{r})]\\
 &  & \widehat{\Psi}_{C}^{\dagger}(\mathbf{s}_{1})\widehat{\rho}\widehat{\Psi}_{C}(\mathbf{s}_{2})\\
 & \rightarrow & \left(\psi_{C}^{+}(\mathbf{s}_{1})-\frac{1}{2}\frac{\delta}{\delta\psi_{C}(\mathbf{s}_{1})}\right)\left(\psi_{C}(\mathbf{s}_{2})-\frac{1}{2}\frac{\delta}{\delta\psi_{C}^{+}(\mathbf{s}_{2})}\right)P[\underrightarrow{\psi}(\mathbf{r}),\underrightarrow{\psi^{\ast}}(\mathbf{r})].\end{eqnarray*}
Using the rules for functional differentiation we see that the differentiations
can be carried out in either order. \ 

In applying these rules to the BEC problem, the following functional
derivative results can be obtained (see Appendix B, \citep{Dalton10b})
The general functions $\psi(\mathbf{r})$ and $\psi^{+}(\mathbf{r})$
each were used to cover the results for condensate and non-condensate
modes. For the case where $\psi(\mathbf{r})\equiv\psi_{C}(\mathbf{r})$
the restricted set $K$ refers to the modes $\phi_{1}(\mathbf{r})$
and $\phi_{2}(\mathbf{r})$, and for the non-condensate case where
$\psi(\mathbf{r})\equiv\psi_{NC}(\mathbf{r})$ the restricted set
refers to the remaining modes $\phi_{k}(\mathbf{r})$. For the case
where $\psi^{+}(\mathbf{r})\equiv\psi_{C}^{+}(\mathbf{r})$ the restricted
set $K$ refers to the conjugate modes $\phi_{1}^{\ast}(\mathbf{r})$
and $\phi_{2}^{\ast}(\mathbf{r})$, and for the non-condensate case
where $\psi^{+}(\mathbf{r})\equiv\psi_{NC}^{+}(\mathbf{r})$ the restricted
set $K^{\ast}$refers to the remaining conjugate modes $\phi_{k}^{\ast}(\mathbf{r})$.
Because the coefficients are unrelated we are dealing with functionals
such as the distribution functional $P[\underrightarrow{\psi}(\mathbf{r}),\underrightarrow{\psi^{\ast}}(\mathbf{r})]$
in which the functions $\psi_{C}(\mathbf{r}),\psi_{C}^{+}(\mathbf{r}),\psi_{NC}(\mathbf{r}),\psi_{NC}^{+}(\mathbf{r})$
are mutually independent. \begin{eqnarray}
\frac{\delta}{\delta\psi_{C}(\mathbf{s})}\psi_{C}(\mathbf{r}) & = & \delta_{C}(\mathbf{r,s})\nonumber \\
\frac{\delta}{\delta\psi_{C}^{+}(\mathbf{s})}\psi_{C}^{+}(\mathbf{r}) & = & \delta_{C+}(\mathbf{r,s})=\delta_{C}(\mathbf{s,r})\nonumber \\
\frac{\delta}{\delta\psi_{C}(\mathbf{s})}\psi_{C}^{+}(\mathbf{r}) & = & 0\qquad\frac{\delta}{\delta\psi_{C}^{+}(\mathbf{s})}\psi_{C}(\mathbf{r})=0\label{Eq.FuncDerivativeRule1}\end{eqnarray}
with four other results obtained by replacing $C$ by $NC$. Note
the reverse order of $\mathbf{r,s}$ in the second result. Similarly
the functional derivatives of condensate fields with respect to non-condensate
fields are zero, and vice-versa. Thus\begin{eqnarray}
\frac{\delta}{\delta\psi_{C}(\mathbf{s})}\psi_{NC}(\mathbf{r}) & = & 0\qquad\frac{\delta}{\delta\psi_{C}^{+}(\mathbf{s})}\psi_{NC}^{+}(\mathbf{r})=0\nonumber \\
\frac{\delta}{\delta\psi_{C}(\mathbf{s})}\psi_{NC}^{+}(\mathbf{r}) & = & 0\qquad\frac{\delta}{\delta\psi_{C}^{+}(\mathbf{s})}\psi_{NC}(\mathbf{r})=0\label{Eq.FuncDerivativeRule2}\end{eqnarray}
with four other results obtained by interchanging $C$ and $NC$.

The product rule for functional derivatives\begin{eqnarray}
 &  & \frac{\delta}{\delta\psi(\mathbf{s})}(F[\psi(\mathbf{r}),\psi^{+}(\mathbf{r})]G[\psi(\mathbf{r}),\psi^{+}(\mathbf{r})])\nonumber \\
 & = & (\frac{\delta}{\delta\psi(\mathbf{s})}F[\psi(\mathbf{r}),\psi^{+}(\mathbf{r})])G[\psi(\mathbf{r}),\psi^{+}(\mathbf{r})]+F[\psi(\mathbf{r}),\psi^{+}(\mathbf{r})](\frac{\delta}{\delta\psi(\mathbf{s})}G[\psi(\mathbf{r}),\psi^{+}(\mathbf{r})])\nonumber \\
 &  & \frac{\delta}{\delta\psi^{+}(\mathbf{s})}(F[\psi(\mathbf{r}),\psi^{+}(\mathbf{r})]G[\psi(\mathbf{r}),\psi^{+}(\mathbf{r})])\nonumber \\
 & = & (\frac{\delta}{\delta\psi^{+}(\mathbf{s})}F[\psi(\mathbf{r}),\psi^{+}(\mathbf{r})])G[\psi(\mathbf{r}),\psi^{+}(\mathbf{r})]+F[\psi(\mathbf{r}),\psi^{+}(\mathbf{r})](\frac{\delta}{\delta\psi^{+}(\mathbf{s})}G[\psi(\mathbf{r}),\psi^{+}(\mathbf{r})])\nonumber \\
 &  & \,\label{eq:ProdRuleFnalDeriv}\end{eqnarray}
is also needed. Here $\psi(\mathbf{r})$ refers to either $\psi_{C}(\mathbf{r})$
or $\psi_{NC}(\mathbf{r})$ and $\psi^{+}(\mathbf{r})$ refers to
either $\psi_{C}^{+}(\mathbf{r})$ or $\psi_{NC}^{+}(\mathbf{r})$.

In addition the standard approach to space integration gives the result\begin{equation}
\int d\mathbf{s\,\{}\partial_{\mu}C(\mathbf{s})\,\}=0\label{Eq.SpatialIntnResultB}\end{equation}
for functions $C(\mathbf{s})$ that become zero on the boundary. This
then leads to the useful result involving product functions $C(\mathbf{s})=A(\mathbf{s})B(\mathbf{s})$
enabling the spatial derivative to be applied to either $A(\mathbf{s})$
or $B(\mathbf{s})$\begin{equation}
\int d\mathbf{s\,\{}\partial_{\mu}A(\mathbf{s})\,\}B(\mathbf{s})=-\int d\mathbf{s\,}A(\mathbf{s})\,\mathbf{\{}\partial_{\mu}B(\mathbf{s})\,\}\label{Eq.SpatialIntnResultC}\end{equation}
We can assume that the $\psi(\mathbf{s})$ and $\psi^{+}(\mathbf{s})$
become zero on the boundary, since they both involve condensate mode
functions or their conjugates that are localised due to the trap potential.
Also the functional derivatives produce linear combinations of either
the condensate mode functions or their conjugates (see (\ref{eq:RestrictFnalDerivIdent1}))
so the various $C(\mathbf{s})$ that will be involved should become
zero on the boundary.

The results in this section also apply to the single mode case with
obvious modifications, the sums over condensate modes now restricted
to $k=1$.

\subsection{Condensate Functional Fokker-Planck Equation}

The \emph{functional Fokker-Planck equation} may be written in the
form

\begin{eqnarray}
 &  & \left(\frac{\partial}{\partial t}P[\underrightarrow{\psi}(\mathbf{r}),\underrightarrow{\psi^{\ast}}(\mathbf{r})]\right)\nonumber \\
 & = & \left(\frac{\partial}{\partial t}P[\underrightarrow{\psi}(\mathbf{r}),\underrightarrow{\psi^{\ast}}(\mathbf{r})]\right)_{H1}+\left(\frac{\partial}{\partial t}P[\underrightarrow{\psi}(\mathbf{r}),\underrightarrow{\psi^{\ast}}(\mathbf{r})]\right)_{H2}\nonumber \\
 &  & +\left(\frac{\partial}{\partial t}P[\underrightarrow{\psi}(\mathbf{r}),\underrightarrow{\psi^{\ast}}(\mathbf{r})]\right)_{H3}\label{Eq.FokkerPlanck}\end{eqnarray}
This is the sum from the terms in the Bogoliubov Hamiltonian of order
$N$, $\sqrt{N}$ and $1/\sqrt{N}$ respectively. The derivation of
the results for the Fokker-Planck equation is carried out in Appendix
E (\citep{Dalton10b}).

\subsubsection{The $\widehat{H}_{1}$ Terms}

The contributions to the \emph{functional Fokker-Planck equation}
from the $\widehat{H}_{1}$ term, which is equal to the \emph{condensate
Hamiltonian}, may be written in the form

\begin{eqnarray}
 &  & \left(\frac{\partial}{\partial t}P[\underrightarrow{\psi}(\mathbf{r}),\underrightarrow{\psi^{\ast}}(\mathbf{r})]\right)_{H1}\nonumber \\
 & = & \left(\frac{\partial}{\partial t}P[\underrightarrow{\psi}(\mathbf{r}),\underrightarrow{\psi^{\ast}}(\mathbf{r})]\right)_{H1K}+\left(\frac{\partial}{\partial t}P[\underrightarrow{\psi}(\mathbf{r}),\underrightarrow{\psi^{\ast}}(\mathbf{r})]\right)_{H1V}\nonumber \\
 &  & +\left(\frac{\partial}{\partial t}P[\underrightarrow{\psi}(\mathbf{r}),\underrightarrow{\psi^{\ast}}(\mathbf{r})]\right)_{H1U}\label{Eq.FnalFPTermH1}\end{eqnarray}
of the sum of terms from the kinetic energy, the trap potential and
the boson-boson interaction. Derivations of the form for each term
are given in Appendix E (\citep{Dalton10b}). Here and elsewhere $\partial_{\mu}$
is short for $\frac{{\LARGE\partial}}{{\LARGE\partial s}_{\mu}}$.

\paragraph{H1K Terms - Single and Two-Mode Condensates}

The contribution to the functional Fokker-Planck equation from the
\emph{condensate kinetic energy} is given by\begin{eqnarray}
 &  & \left(\frac{\partial}{\partial t}P[\underrightarrow{\psi}(\mathbf{r}),\underrightarrow{\psi^{\ast}}(\mathbf{r})]\right)_{H1K}\nonumber \\
 & = & \frac{-i}{\hbar}\left\{ -\int d\mathbf{s}\left\{ \frac{\delta}{\delta\psi_{C}^{+}(\mathbf{s})}\left(\sum\limits _{\mu}\frac{\hbar^{2}}{2m}\partial_{\mu}^{2}\psi_{C}^{+}(\mathbf{s})\right)P[\underrightarrow{\psi}(\mathbf{r}),\underrightarrow{\psi^{\ast}}(\mathbf{r})]\right\} \right\} \nonumber \\
 &  & +\frac{-i}{\hbar}\left\{ +\int d\mathbf{s}\left\{ \frac{\delta}{\delta\psi_{C}(\mathbf{s})}\left(\sum\limits _{\mu}\frac{\hbar^{2}}{2m}\partial_{\mu}^{2}\psi_{C}(\mathbf{s})\right)P[\underrightarrow{\psi}(\mathbf{r}),\underrightarrow{\psi^{\ast}}(\mathbf{r})]\right\} \right\} \label{eq:FnalFPH1K}\end{eqnarray}

\paragraph{H1V Terms - Single and Two-Mode Condensates}

The contribution to the functional Fokker-Planck equation from the
\emph{condensate trap potential} is given by\begin{eqnarray}
 &  & \left(\frac{\partial}{\partial t}P[\underrightarrow{\psi}(\mathbf{r}),\underrightarrow{\psi^{\ast}}(\mathbf{r})]\right)_{H1V}\nonumber \\
 & = & \frac{-i}{\hbar}\left\{ -\int d\mathbf{s}\left\{ \frac{\delta}{\delta\psi_{C}(\mathbf{s})}\{V(\mathbf{s})\psi_{C}(\mathbf{s})\}\right\} P[\underrightarrow{\psi}(\mathbf{r}),\underrightarrow{\psi^{\ast}}(\mathbf{r})]\right\} \nonumber \\
 &  & +\frac{-i}{\hbar}\left\{ +\int d\mathbf{s}\left\{ \frac{\delta}{\delta\psi_{C}^{+}(\mathbf{s})}\{V(\mathbf{s})\psi_{C}^{+}(\mathbf{s})\}\right\} P[\underrightarrow{\psi}(\mathbf{r}),\underrightarrow{\psi^{\ast}}(\mathbf{r})]\right\} \label{eq:FnalFPH1V}\end{eqnarray}

\paragraph{H1U Terms - Single and Two-Mode Condensates }

The contribution to the functional Fokker-Planck equation from the
\emph{condensate boson-boson interaction} is given by\begin{eqnarray}
 &  & \left(\frac{\partial}{\partial t}P[\underrightarrow{\psi}(\mathbf{r}),\underrightarrow{\psi^{\ast}}(\mathbf{r})]\right)_{H1U}\nonumber \\
 & = & \frac{-i}{\hbar}\left\{ -\frac{g_{N}}{N}\int d\mathbf{s}\frac{\delta}{\delta\psi_{C}(\mathbf{s})}\left\{ (\psi_{C}^{+}(\mathbf{s})\psi_{C}(\mathbf{s})-\delta_{C}(\mathbf{s},\mathbf{s}))\psi_{C}(\mathbf{s})\right\} P[\underrightarrow{\psi}(\mathbf{r}),\underrightarrow{\psi^{\ast}}(\mathbf{r})]\right\} \nonumber \\
 &  & +\frac{-i}{\hbar}\left\{ +\frac{g_{N}}{N}\int d\mathbf{s}\frac{\delta}{\delta\psi_{C}^{+}(\mathbf{s})}\left\{ (\psi_{C}^{+}(\mathbf{s})\psi_{C}(\mathbf{s})-\delta_{C}(\mathbf{s},\mathbf{s}))\psi_{C}^{+}(\mathbf{s})\right\} P[\underrightarrow{\psi}(\mathbf{r}),\underrightarrow{\psi^{\ast}}(\mathbf{r})]\right\} \nonumber \\
 &  & +\frac{-i}{\hbar}\left\{ \frac{g_{N}}{N}\int d\mathbf{s}\frac{\delta}{\delta\psi_{C}(\mathbf{s})}\frac{\delta}{\delta\psi_{C}(\mathbf{s})}\frac{\delta}{\delta\psi_{C}^{+}(\mathbf{s})}\{\frac{1}{4}\psi_{C}(\mathbf{s})\}P[\underrightarrow{\psi}(\mathbf{r}),\underrightarrow{\psi^{\ast}}(\mathbf{r})]\right\} \nonumber \\
 &  & +\frac{-i}{\hbar}\left\{ -\frac{g_{N}}{N}\int d\mathbf{s}\frac{\delta}{\delta\psi_{C}^{+}(\mathbf{s})}\frac{\delta}{\delta\psi_{C}^{+}(\mathbf{s})}\frac{\delta}{\delta\psi_{C}(\mathbf{s})}\{\frac{1}{4}\psi_{C}^{+}(\mathbf{s})\}P[\underrightarrow{\psi}(\mathbf{r}),\underrightarrow{\psi^{\ast}}(\mathbf{r})]\right\} \label{eq:FnalFPTermH1U}\end{eqnarray}
which involves first order and third order functional derivatives.
The quantity $\delta_{C}(\mathbf{s},\mathbf{s})$ is a diagonal element
of the restricted delta function for condensate modes.

For the one mode case we note that\begin{eqnarray}
\int d\mathbf{s\,}\delta_{C}(\mathbf{s},\mathbf{s}) & = & 1\label{Eq.NormCondDeltaFnOneMode}\\
\delta_{C}(\mathbf{s},\mathbf{s}) & = & \left\vert \phi_{1}(\mathbf{s})\right\vert ^{2}\label{Eq.CondDeltaResultOneMode}\end{eqnarray}
corresponding to there being a single occupied condensate mode.

For the two mode case we have instead \begin{eqnarray}
\int d\mathbf{s\,}\delta_{C}(\mathbf{s},\mathbf{s}) & = & 2\label{Eq.NormCondDeltaFnTwoMode}\\
\delta_{C}(\mathbf{s},\mathbf{s}) & = & \left\vert \phi_{1}(\mathbf{s})\right\vert ^{2}+\left\vert \phi_{2}(\mathbf{s})\right\vert ^{2}\label{Eq.CondDeltaResultTwoMode}\end{eqnarray}
corresponding to there being two occupied condensate modes.

The total condensate number given by\begin{equation}
\iiiint D^{2}\psi_{C}\, D^{2}\psi_{C}^{+}\, D^{2}\psi_{NC}\, D^{2}\psi_{NC}^{+}\int d\mathbf{s}(\psi_{C}^{+}(\mathbf{s})\psi_{C}(\mathbf{s}))P[\underrightarrow{\psi}(\mathbf{r}),\underrightarrow{\psi^{\ast}}(\mathbf{r})]\label{eq:TotalCondNumber}\end{equation}
which is of order $N$. The result of order $N$ for the last expression
indicates that the important contributions to the functional integral
are where the condensate fields are of order $\sqrt{N}$. Similar
considerations for the much smaller total non-condensate number indicate
that the most important contributions are where the non-condensate
fields are much smaller than $\sqrt{N}$.

Similar expressions for the functional Fokker-Planck equation in the
case of a pure Wigner representation (but not involving a doubled
phase space) are given in the paper by Steel et al \citep{Steel98b}
(see Eq. (23)). Comparisons can be made after substituting $\psi_{C}^{+}(\mathbf{s})$
with $\psi_{C}^{\ast}(\mathbf{s})$. In their result however, the
restricted delta function $\delta_{C}(\mathbf{s},\mathbf{s})$ term
in the condensate interaction contribution is replaced by $1$. For
the single condensate mode case unity is of course the integral of
the restricted delta function, but it is not equal to it.

\subsubsection{The $\widehat{H}_{2}$ Term}

The contributions to the \emph{functional Fokker-Planck equation}
from the $\widehat{H}_{2}$ term, which is equal to terms in the interaction
between the \emph{condensate }and \emph{non-condensate Hamiltonian}
that are \emph{linear} in the non-condensate fields, may be written
in the form

\begin{eqnarray}
 &  & \left(\frac{\partial}{\partial t}P[\underrightarrow{\psi}(\mathbf{r}),\underrightarrow{\psi^{\ast}}(\mathbf{r})]\right)_{H2}\nonumber \\
 & = & \left(\frac{\partial}{\partial t}P[\underrightarrow{\psi}(\mathbf{r}),\underrightarrow{\psi^{\ast}}(\mathbf{r})]\right)_{H2U4}+\left(\frac{\partial}{\partial t}P[\underrightarrow{\psi}(\mathbf{r}),\underrightarrow{\psi^{\ast}}(\mathbf{r})]\right)_{H2U2}\label{Eq.FnalFPTermH2}\end{eqnarray}
These two contributions may be written as the sum of terms which are
linear, quadratic, cubic and quartic in the number of functional derivatives.
Derivations of the form for each term are given in Appendix E (\citep{Dalton10b}).

\paragraph{H2U4 Terms - Single and Two-Mode Condensates}

The contribution to the functional Fokker-Planck equation from the
$\widehat{H}_{2U4}$ term is given by\begin{eqnarray}
 &  & \left(\frac{\partial}{\partial t}P[\underrightarrow{\psi}(\mathbf{r}),\underrightarrow{\psi^{\ast}}(\mathbf{r})]\right)_{H2U4}\nonumber \\
 & = & \left(\frac{\partial}{\partial t}P[\underrightarrow{\psi}(\mathbf{r}),\underrightarrow{\psi^{\ast}}(\mathbf{r})]\right)_{H2U4}^{1}+\left(\frac{\partial}{\partial t}P[\underrightarrow{\psi}(\mathbf{r}),\underrightarrow{\psi^{\ast}}(\mathbf{r})]\right)_{H2U4}^{2}\nonumber \\
 &  & +\left(\frac{\partial}{\partial t}P[\underrightarrow{\psi}(\mathbf{r}),\underrightarrow{\psi^{\ast}}(\mathbf{r})]\right)_{H2U4}^{3}+\left(\frac{\partial}{\partial t}P[\underrightarrow{\psi}(\mathbf{r}),\underrightarrow{\psi^{\ast}}(\mathbf{r})]\right)_{H2U4}^{4}\label{Eq.FnalFPTermH24}\end{eqnarray}

where\begin{eqnarray}
 &  & \left(\frac{\partial}{\partial t}P[\underrightarrow{\psi}(\mathbf{r}),\underrightarrow{\psi^{\ast}}(\mathbf{r})]\right)_{H2U4}^{1}\nonumber \\
 & = & \frac{-i}{\hbar}\left\{ +\frac{g_{N}}{N}\int d\mathbf{s\,}\left\{ \left(\frac{\delta}{\delta\psi_{C}^{+}(\mathbf{s})}\right)\{[2\psi_{C}^{+}(\mathbf{s})\psi_{C}(\mathbf{s})-\delta_{C}(\mathbf{s},\mathbf{s})]\psi_{NC}^{+}(\mathbf{s})\}\right\} P[\underrightarrow{\psi}(\mathbf{r}),\underrightarrow{\psi^{\ast}}(\mathbf{r})]\right\} \nonumber \\
 &  & +\frac{-i}{\hbar}\left\{ +\frac{g_{N}}{N}\int d\mathbf{s\,}\left\{ \mathbf{\left(\frac{\delta}{\delta\psi_{C}^{+}(\mathbf{s})}\right)\{[}\psi_{C}^{+}\mathbf{(\mathbf{s})}\psi_{C}^{+}\mathbf{(\mathbf{s})]}\psi_{NC}\mathbf{(\mathbf{s})\}}\right\} P[\underrightarrow{\psi}(\mathbf{r}),\underrightarrow{\psi^{\ast}}(\mathbf{r})]\right\} \nonumber \\
 &  & +\frac{-i}{\hbar}\left\{ -\frac{g_{N}}{N}\int d\mathbf{s\,}\left\{ \mathbf{\left(\frac{\delta}{\delta\psi_{C}(\mathbf{s})}\right)\{[}2\psi\mathbf{_{C}(\mathbf{s})}\psi_{C}^{+}\mathbf{(\mathbf{s})-\delta_{C}(\mathbf{s},\mathbf{s})]\psi}_{NC}\mathbf{(\mathbf{s})\}}\right\} P[\underrightarrow{\psi}(\mathbf{r}),\underrightarrow{\psi^{\ast}}(\mathbf{r})]\right\} \nonumber \\
 &  & +\frac{-i}{\hbar}\left\{ -\frac{g_{N}}{N}\int d\mathbf{s\,}\left\{ \left(\frac{\delta}{\delta\psi_{C}(\mathbf{s})}\right)\{[\psi_{C}(\mathbf{s})\psi_{C}(\mathbf{s})]\psi_{NC}^{+}(\mathbf{s})\}\right\} P[\underrightarrow{\psi}(\mathbf{r}),\underrightarrow{\psi^{\ast}}(\mathbf{r})]\right\} \nonumber \\
 &  & +\frac{-i}{\hbar}\left\{ -\frac{g_{N}}{N}\int d\mathbf{s\,}\left\{ \left(\frac{\delta}{\delta\psi_{NC}(\mathbf{s})}\right)\{[\psi_{C}^{+}(\mathbf{s})\psi_{C}(\mathbf{s})-\delta_{C}(\mathbf{s},\mathbf{s)}]\psi_{C}(\mathbf{s})\}\right\} P[\underrightarrow{\psi}(\mathbf{r}),\underrightarrow{\psi^{\ast}}(\mathbf{r})]\right\} \nonumber \\
 &  & +\frac{-i}{\hbar}\left\{ +\frac{g_{N}}{N}\int d\mathbf{s\,}\left\{ \left(\frac{\delta}{\delta\psi_{NC}^{+}(\mathbf{s})}\right)\{[\psi_{C}(\mathbf{s})\psi_{C}^{+}(\mathbf{s})-\delta_{C}(\mathbf{s},\mathbf{s})]\psi_{C}^{+}(\mathbf{s})\}\right\} P[\underrightarrow{\psi}(\mathbf{r}),\underrightarrow{\psi^{\ast}}(\mathbf{r})]\right\} \nonumber \\
 &  & \,\label{eq:FnalFPTermH2U4Linear}\end{eqnarray}
\begin{eqnarray}
 &  & \left(\frac{\partial}{\partial t}P[\underrightarrow{\psi}(\mathbf{r}),\underrightarrow{\psi^{\ast}}(\mathbf{r})]\right)_{H2U4}^{2}\nonumber \\
 & = & \frac{-i}{\hbar}\left\{ -\frac{g_{N}}{N}\int d\mathbf{s\,}\left\{ \left(\frac{\delta}{\delta\psi_{C}^{+}(\mathbf{s})}\right)\left(\frac{\delta}{\delta\psi_{NC}(\mathbf{s})}\right)\{\psi_{C}^{+}(\mathbf{s})\psi_{C}(\mathbf{s})\}\right\} P[\underrightarrow{\psi}(\mathbf{r}),\underrightarrow{\psi^{\ast}}(\mathbf{r})]\right\} \nonumber \\
 &  & +\frac{-i}{\hbar}\left\{ +\frac{g_{N}}{N}\int d\mathbf{s\,}\left\{ \left(\frac{\delta}{\delta\psi_{C}(\mathbf{s})}\right)\left(\frac{\delta}{\delta\psi_{NC}^{+}(\mathbf{s})}\right)\{\psi_{C}(\mathbf{s})\psi_{C}^{+}(\mathbf{s})\}\right\} P[\underrightarrow{\psi}(\mathbf{r}),\underrightarrow{\psi^{\ast}}(\mathbf{r})]\right\} \nonumber \\
 &  & +\frac{-i}{\hbar}\left\{ +\frac{g_{N}}{N}\int d\mathbf{s\,}\left\{ \left(\frac{\delta}{\delta\psi_{C}^{+}(\mathbf{s})}\right)\left(\frac{\delta}{\delta\psi_{NC}(\mathbf{s})}\right)\left\{ \frac{1}{2}\delta_{C}(\mathbf{s},\mathbf{s})\right\} \right\} P[\underrightarrow{\psi}(\mathbf{r}),\underrightarrow{\psi^{\ast}}(\mathbf{r})]\right\} \nonumber \\
 &  & +\frac{-i}{\hbar}\left\{ -\frac{g_{N}}{N}\int d\mathbf{s\,}\left\{ \left(\frac{\delta}{\delta\psi_{C}(\mathbf{s})}\right)\left(\frac{\delta}{\delta\psi_{NC}^{+}(\mathbf{s})}\right)\left\{ \frac{1}{2}\delta_{C}(\mathbf{s},\mathbf{s})\right\} \right\} P[\underrightarrow{\psi}(\mathbf{r}),\underrightarrow{\psi^{\ast}}(\mathbf{r})]\right\} \nonumber \\
 &  & +\frac{-i}{\hbar}\left\{ +\frac{g_{N}}{N}\int d\mathbf{s\,}\left\{ \left(\frac{\delta}{\delta\psi_{C}(\mathbf{s})}\right)\left(\frac{\delta}{\delta\psi_{NC}(\mathbf{s})}\right)\{\frac{1}{2}\psi_{C}(\mathbf{s})\psi_{C}(\mathbf{s})\}\right\} P[\underrightarrow{\psi}(\mathbf{r}),\underrightarrow{\psi^{\ast}}(\mathbf{r})]\right\} \nonumber \\
 &  & +\frac{-i}{\hbar}\left\{ -\frac{g_{N}}{N}\int d\mathbf{s\,}\left\{ \left(\frac{\delta}{\delta\psi_{C}^{+}(\mathbf{s})}\right)\left(\frac{\delta}{\delta\psi_{NC}^{+}(\mathbf{s})}\right)\{\frac{1}{2}\psi_{C}^{+}(\mathbf{s})\psi_{C}^{+}(\mathbf{s})\}\right\} P[\underrightarrow{\psi}(\mathbf{r}),\underrightarrow{\psi^{\ast}}(\mathbf{r})]\right\} \nonumber \\
 &  & \,\label{eq:FnalFPTermH2U4Quadratic}\end{eqnarray}

\begin{eqnarray}
 &  & \left(\frac{\partial}{\partial t}P[\underrightarrow{\psi}(\mathbf{r}),\underrightarrow{\psi^{\ast}}(\mathbf{r})]\right)_{H2U4}^{3}\nonumber \\
 & = & \frac{-i}{\hbar}\left\{ -\frac{g_{N}}{N}\int d\mathbf{s\,}\left\{ \left(\frac{\delta}{\delta\psi_{C}^{+}(\mathbf{s})}\right)\left(\frac{\delta}{\delta\psi_{C}^{+}(\mathbf{s})}\right)\left(\frac{\delta}{\delta\psi_{C}(\mathbf{s})}\right)\{\frac{1}{4}\psi_{NC}^{+}(\mathbf{s})\}\right\} P[\underrightarrow{\psi}(\mathbf{r}),\underrightarrow{\psi^{\ast}}(\mathbf{r})]\right\} \nonumber \\
 &  & +\frac{-i}{\hbar}\left\{ +\frac{g_{N}}{N}\int d\mathbf{s\,}\left\{ \left(\frac{\delta}{\delta\psi_{C}(\mathbf{s})}\right)\left(\frac{\delta}{\delta\psi_{C}(\mathbf{s})}\right)\left(\frac{\delta}{\delta\psi_{C}^{+}(\mathbf{s})}\right)\{\frac{1}{4}\psi_{NC}(\mathbf{s})\}\right\} P[\underrightarrow{\psi}(\mathbf{r}),\underrightarrow{\psi^{\ast}}(\mathbf{r})]\right\} \nonumber \\
 &  & +\frac{-i}{\hbar}\left\{ -\frac{g_{N}}{N}\int d\mathbf{s\,}\left\{ \left(\frac{\delta}{\delta\psi_{C}^{+}(\mathbf{s})}\right)\left(\frac{\delta}{\delta\psi_{C}^{+}(\mathbf{s})}\right)\left(\frac{\delta}{\delta\psi_{NC}(\mathbf{s})}\right)\{\frac{1}{4}\psi_{C}^{+}(\mathbf{s})\}\right\} P[\underrightarrow{\psi}(\mathbf{r}),\underrightarrow{\psi^{\ast}}(\mathbf{r})]\right\} \nonumber \\
 &  & +\frac{-i}{\hbar}\left\{ +\frac{g_{N}}{N}\int d\mathbf{s\,}\left\{ \left(\frac{\delta}{\delta\psi_{C}(\mathbf{s})}\right)\left(\frac{\delta}{\delta\psi_{C}(\mathbf{s})}\right)\left(\frac{\delta}{\delta\psi_{NC}^{+}(\mathbf{s})}\right)\{\frac{1}{4}\psi_{C}(\mathbf{s})\}\right\} P[\underrightarrow{\psi}(\mathbf{r}),\underrightarrow{\psi^{\ast}}(\mathbf{r})]\right\} \nonumber \\
 &  & +\frac{-i}{\hbar}\left\{ +\frac{g_{N}}{N}\int d\mathbf{s\,}\left\{ \left(\frac{\delta}{\delta\psi_{C}^{+}(\mathbf{s})}\right)\left(\frac{\delta}{\delta\psi_{C}(\mathbf{s})}\right)\left(\frac{\delta}{\delta\psi_{NC}(\mathbf{s})}\right)\{\frac{1}{2}\psi_{C}(\mathbf{s})\}\right\} P[\underrightarrow{\psi}(\mathbf{r}),\underrightarrow{\psi^{\ast}}(\mathbf{r})]\right\} \nonumber \\
 &  & +\frac{-i}{\hbar}\left\{ -\frac{g_{N}}{N}\int d\mathbf{s\,}\left\{ \left(\frac{\delta}{\delta\psi_{C}(\mathbf{s})}\right)\left(\frac{\delta}{\delta\psi_{C}^{+}(\mathbf{s})}\right)\left(\frac{\delta}{\delta\psi_{NC}^{+}(\mathbf{s})}\right)\{\frac{1}{2}\psi_{C}^{+}(\mathbf{s})\}\right\} P[\underrightarrow{\psi}(\mathbf{r}),\underrightarrow{\psi^{\ast}}(\mathbf{r})]\right\} \nonumber \\
 &  & \,\label{eq:FnalFPTermH2U4Cubic}\end{eqnarray}
\begin{eqnarray}
 &  & \left(\frac{\partial}{\partial t}P[\underrightarrow{\psi}(\mathbf{r}),\underrightarrow{\psi^{\ast}}(\mathbf{r})]\right)_{H2U4}^{4}\nonumber \\
 & = & \frac{-i}{\hbar}\left\{ \frac{g_{N}}{N}\int d\mathbf{s\,}\left\{ \left(\frac{\delta}{\delta\psi_{C}^{+}(\mathbf{s})}\right)\left(\frac{\delta}{\delta\psi_{C}^{+}(\mathbf{s})}\right)\left(\frac{\delta}{\delta\psi_{C}(\mathbf{s})}\right)\left(\frac{\delta}{\delta\psi_{NC}(\mathbf{s})}\right)\{\frac{1}{8}\}\right\} P[\underrightarrow{\psi}(\mathbf{r}),\underrightarrow{\psi^{\ast}}(\mathbf{r})]\right\} \nonumber \\
 &  & +\frac{-i}{\hbar}\left\{ -\frac{g_{N}}{N}\int d\mathbf{s}\left\{ \left(\frac{\delta}{\delta\psi_{C}(\mathbf{s})}\right)\left(\frac{\delta}{\delta\psi_{C}(\mathbf{s})}\right)\left(\frac{\delta}{\delta\psi_{C}^{+}(\mathbf{s})}\right)\left(\frac{\delta}{\delta\psi_{NC}^{+}(\mathbf{s})}\right)\{\frac{1}{8}\}\right\} P[\underrightarrow{\psi}(\mathbf{r}),\underrightarrow{\psi^{\ast}}(\mathbf{r})]\right\} \nonumber \\
 &  & \,\label{eq:FnalFPTermH2U4Quartic}\end{eqnarray}

The contribution to the functional Fokker-Planck equation from the
$\widehat{H}_{2U2}$ term is

\begin{eqnarray}
 &  & \left(\frac{\partial}{\partial t}P[\underrightarrow{\psi}(\mathbf{r}),\underrightarrow{\psi^{\ast}}(\mathbf{r})]\right)_{H2U2}\nonumber \\
 & = & \left(\frac{\partial}{\partial t}P[\underrightarrow{\psi}(\mathbf{r}),\underrightarrow{\psi^{\ast}}(\mathbf{r})]\right)_{H22}^{1}+\left(\frac{\partial}{\partial t}P[\underrightarrow{\psi}(\mathbf{r}),\underrightarrow{\psi^{\ast}}(\mathbf{r})]\right)_{H22}^{2}\label{Eq.FnalFPTermH22}\end{eqnarray}

\paragraph{H2U2 Terms - Two-Mode Condensate}

For the two mode condensate case\begin{eqnarray*}
 &  & \left(\frac{\partial}{\partial t}P[\underrightarrow{\psi}(\mathbf{r}),\underrightarrow{\psi^{\ast}}(\mathbf{r})]\right)_{H2U2}^{1}\\
 & = & \frac{-i}{\hbar}\left\{ -\frac{g_{N}}{N}\int\int d\mathbf{s\,}d\mathbf{u}\left\{ \left(\frac{\delta}{\delta\psi_{C}^{+}(\mathbf{u})}\right)\{F(\mathbf{s},\mathbf{u})\,\psi_{NC}^{+}(\mathbf{s})\}\right\} \, P[\underrightarrow{\psi}(\mathbf{r}),\underrightarrow{\psi^{\ast}}(\mathbf{r})]\right\} \\
 &  & +\frac{-i}{\hbar}\left\{ +\frac{g_{N}}{N}\int\int d\mathbf{s\,}d\mathbf{u}\left\{ \left(\frac{\delta}{\delta\psi_{C}(\mathbf{s})}\right)\{F(\mathbf{u},\mathbf{s})^{\ast}\,\psi_{NC}(\mathbf{u})\}\right\} \, P[\underrightarrow{\psi}(\mathbf{r}),\underrightarrow{\psi^{\ast}}(\mathbf{r})]\right\} \\
 &  & +\frac{-i}{\hbar}\left\{ -\frac{g_{N}}{N}\int\int d\mathbf{s\, d\mathbf{u\,}}\left\{ \left(\frac{\delta}{\delta\psi_{NC}^{+}(\mathbf{u})}\right)\{F\mathbf{(\mathbf{u},\mathbf{s})^{\ast}}\,\psi_{C}^{+}(\mathbf{s})\}\right\} \, P[\underrightarrow{\psi}(\mathbf{r}),\underrightarrow{\psi^{\ast}}(\mathbf{r})]\right\} \\
 &  & +\frac{-i}{\hbar}\left\{ +\frac{g_{N}}{N}\int\int d\mathbf{s\,}d\mathbf{u}\left\{ \left(\frac{\delta}{\delta\psi_{NC}(\mathbf{s})}\right)\{F(\mathbf{s},\mathbf{u})\,\psi_{C}(\mathbf{u})\}\right\} \, P[\underrightarrow{\psi}(\mathbf{r}),\underrightarrow{\psi^{\ast}}(\mathbf{r})]\right\} \end{eqnarray*}
and\begin{eqnarray*}
 &  & \left(\frac{\partial}{\partial t}P[\underrightarrow{\psi}(\mathbf{r}),\underrightarrow{\psi^{\ast}}(\mathbf{r})]\right)_{H2U2}^{2}\\
 & = & \frac{-i}{\hbar}\left\{ +\frac{g_{N}}{N}\int\int d\mathbf{s\,}d\mathbf{u}\left\{ \left(\frac{\delta}{\delta\psi_{C}^{+}(\mathbf{u})}\right)\left(\frac{\delta}{\delta\psi_{NC}(\mathbf{s})}\right)\{\frac{1}{2}F(\mathbf{s},\mathbf{u})\}\right\} \, P[\underrightarrow{\psi}(\mathbf{r}),\underrightarrow{\psi^{\ast}}(\mathbf{r})]\right\} \\
 &  & +\frac{-i}{\hbar}\left\{ -\frac{g_{N}}{N}\int\int d\mathbf{s\, d\mathbf{u\,}}\left\{ \left(\frac{\delta}{\delta\psi_{C}(\mathbf{s})}\right)\left(\frac{\delta}{\delta\psi_{NC}^{+}(\mathbf{u})}\right)\{\frac{1}{2}\mathbf{F(\mathbf{u},\mathbf{s})^{\ast}}\}\right\} \, P[\underrightarrow{\psi}(\mathbf{r}),\underrightarrow{\psi^{\ast}}(\mathbf{r})]\right\} \end{eqnarray*}
These terms now involve double spatial integrals, and in the case
of the quadratic term there are second order functional derivatives
with respect to field functions at different spatial positions. This
is different to the standard functional Fokker-Planck equation and
requires special considerations for conversion to Ito stochastic equations
for the field functions. The linear term is not so difficult to treat,
though it still leads to an integro-differential equation. By changing
the spatial variables we see that the linear term is\begin{eqnarray}
 &  & \left(\frac{\partial}{\partial t}P[\underrightarrow{\psi}(\mathbf{r}),\underrightarrow{\psi^{\ast}}(\mathbf{r})]\right)_{H2U2}^{1}\nonumber \\
 & = & \frac{-i}{\hbar}\left\{ -\frac{g_{N}}{N}\int d\mathbf{s\,}\left\{ \left(\frac{\delta}{\delta\psi_{C}^{+}(\mathbf{s})}\right)\{\int d\mathbf{u\,}F(\mathbf{u},\mathbf{s})\,\psi_{NC}^{+}(\mathbf{u})\}\right\} \, P[\underrightarrow{\psi}(\mathbf{r}),\underrightarrow{\psi^{\ast}}(\mathbf{r})]\right\} \nonumber \\
 &  & +\frac{-i}{\hbar}\left\{ +\frac{g_{N}}{N}\int d\mathbf{s\,}\left\{ \left(\frac{\delta}{\delta\psi_{C}(\mathbf{s})}\right)\{\int d\mathbf{u\,}F(\mathbf{u},\mathbf{s})^{\ast}\,\psi_{NC}(\mathbf{u})\}\right\} \, P[\underrightarrow{\psi}(\mathbf{r}),\underrightarrow{\psi^{\ast}}(\mathbf{r})]\right\} \nonumber \\
 &  & +\frac{-i}{\hbar}\left\{ -\frac{g_{N}}{N}\int d\mathbf{s\,}\left\{ \left(\frac{\delta}{\delta\psi_{NC}^{+}(\mathbf{s})}\right)\{\int d\mathbf{u\,}F\mathbf{(s,u)^{\ast}}\,\psi_{C}^{+}(\mathbf{u})\}\right\} \, P[\underrightarrow{\psi}(\mathbf{r}),\underrightarrow{\psi^{\ast}}(\mathbf{r})]\right\} \nonumber \\
 &  & +\frac{-i}{\hbar}\left\{ +\frac{g_{N}}{N}\int d\mathbf{s\,}\left\{ \left(\frac{\delta}{\delta\psi_{NC}(\mathbf{s})}\right)\{\int d\mathbf{u\,}F(\mathbf{s},\mathbf{u})\,\psi_{C}(\mathbf{u})\}\right\} \, P[\underrightarrow{\psi}(\mathbf{r}),\underrightarrow{\psi^{\ast}}(\mathbf{r})]\right\} \nonumber \\
 &  & \,\label{eq:FnalFPTermH2U2Linear}\end{eqnarray}
so the quantity inside the inner brackets is just another functional.
The quadratic term is left unchanged except for interchanging positions
to make the expression more symmetrical\begin{eqnarray}
 &  & \left(\frac{\partial}{\partial t}P[\underrightarrow{\psi}(\mathbf{r}),\underrightarrow{\psi^{\ast}}(\mathbf{r})]\right)_{H2U2}^{2}\nonumber \\
 & = & \frac{-i}{\hbar}\left\{ +\frac{g_{N}}{N}\int\int d\mathbf{s\,}d\mathbf{u}\left\{ \left(\frac{\delta}{\delta\psi_{C}^{+}(\mathbf{s})}\right)\left(\frac{\delta}{\delta\psi_{NC}(\mathbf{u})}\right)\{\frac{1}{2}F(\mathbf{u},\mathbf{s})\}\right\} \, P[\underrightarrow{\psi}(\mathbf{r}),\underrightarrow{\psi^{\ast}}(\mathbf{r})]\right\} \nonumber \\
 &  & +\frac{-i}{\hbar}\left\{ -\frac{g_{N}}{N}\int\int d\mathbf{s\, d\mathbf{u\,}}\left\{ \left(\frac{\delta}{\delta\psi_{C}(\mathbf{s})}\right)\left(\frac{\delta}{\delta\psi_{NC}^{+}(\mathbf{u})}\right)\{\frac{1}{2}\mathbf{F(\mathbf{u},\mathbf{s})^{\ast}}\}\right\} \, P[\underrightarrow{\psi}(\mathbf{r}),\underrightarrow{\psi^{\ast}}(\mathbf{r})]\right\} \nonumber \\
 &  & \,\label{eq:FnalFPTermH2U2Quadratic}\end{eqnarray}

\paragraph{H2U2 Term - Single Mode Condensate}

For the case of a single mode condensate the result is simpler\begin{eqnarray}
 &  & \left(\frac{\partial}{\partial t}P[\underrightarrow{\psi}(\mathbf{r}),\underrightarrow{\psi^{\ast}}(\mathbf{r})]\right)_{H2U2}^{1}\nonumber \\
 & = & \frac{-i}{\hbar}\left\{ -\frac{g_{N}}{N}\int d\mathbf{s\,}\left\{ \left(\frac{\delta}{\delta\psi_{C}^{+}(\mathbf{s})}\right)\{\left\langle \widehat{\Psi}_{C}(\mathbf{s})^{\dagger}\widehat{\Psi}_{C}(\mathbf{s})\right\rangle \,\psi_{NC}^{+}(\mathbf{s})\}\right\} \, P[\underrightarrow{\psi}(\mathbf{r}),\underrightarrow{\psi^{\ast}}(\mathbf{r})]\right\} \nonumber \\
 &  & +\frac{-i}{\hbar}\left\{ +\frac{g_{N}}{N}\int d\mathbf{s\,}\left\{ \left(\frac{\delta}{\delta\psi_{C}(\mathbf{s})}\right)\{\left\langle \widehat{\Psi}_{C}(\mathbf{s})^{\dagger}\widehat{\Psi}_{C}(\mathbf{s})\right\rangle \,\psi_{NC}(\mathbf{s})\}\right\} \, P[\underrightarrow{\psi}(\mathbf{r}),\underrightarrow{\psi^{\ast}}(\mathbf{r})]\right\} \nonumber \\
 &  & +\frac{-i}{\hbar}\left\{ -\frac{g_{N}}{N}\int d\mathbf{s\,}\left\{ \left(\frac{\delta}{\delta\psi_{NC}^{+}(\mathbf{s})}\right)\{\left\langle \widehat{\Psi}_{C}(\mathbf{s})^{\dagger}\widehat{\Psi}_{C}(\mathbf{s})\right\rangle \,\psi_{C}^{+}(\mathbf{s})\}\right\} \, P[\underrightarrow{\psi}(\mathbf{r}),\underrightarrow{\psi^{\ast}}(\mathbf{r})]\right\} \nonumber \\
 &  & +\frac{-i}{\hbar}\left\{ +\frac{g_{N}}{N}\int d\mathbf{s\,}\left\{ \left(\frac{\delta}{\delta\psi_{NC}(\mathbf{s})}\right)\{\left\langle \widehat{\Psi}_{C}(\mathbf{s})^{\dagger}\widehat{\Psi}_{C}(\mathbf{s})\right\rangle \,\psi_{C}(\mathbf{s})\}\right\} \, P[\underrightarrow{\psi}(\mathbf{r}),\underrightarrow{\psi^{\ast}}(\mathbf{r})]\right\} \nonumber \\
 &  & \,\label{eq:FnalFPTermH2U2LinearSingleMode}\end{eqnarray}
\begin{eqnarray}
 &  & \left(\frac{\partial}{\partial t}P[\underrightarrow{\psi}(\mathbf{r}),\underrightarrow{\psi^{\ast}}(\mathbf{r})]\right)_{H2U2}^{2}\nonumber \\
 & = & \frac{-i}{\hbar}\left\{ +\frac{g_{N}}{N}\int d\mathbf{s\,}\left\{ \left(\frac{\delta}{\delta\psi_{C}^{+}(\mathbf{s})}\right)\left(\frac{\delta}{\delta\psi_{NC}(\mathbf{s})}\right)\{\frac{1}{2}\left\langle \widehat{\Psi}_{C}(\mathbf{s})^{\dagger}\widehat{\Psi}_{C}(\mathbf{s})\right\rangle \}\right\} \, P[\underrightarrow{\psi}(\mathbf{r}),\underrightarrow{\psi^{\ast}}(\mathbf{r})]\right\} \nonumber \\
 &  & +\frac{-i}{\hbar}\left\{ -\frac{g_{N}}{N}\int d\mathbf{s\,}\left\{ \left(\frac{\delta}{\delta\psi_{C}(\mathbf{s})}\right)\left(\frac{\delta}{\delta\psi_{NC}^{+}(\mathbf{s})}\right)\{\frac{1}{2}\left\langle \widehat{\Psi}_{C}(\mathbf{s})^{\dagger}\widehat{\Psi}_{C}(\mathbf{s})\right\rangle \}\right\} \, P[\underrightarrow{\psi}(\mathbf{r}),\underrightarrow{\psi^{\ast}}(\mathbf{r})]\right\} \nonumber \\
 &  & \,\label{eq:FnalFPTermH2U2QuadraticSingleMode}\end{eqnarray}
Derivations of the form for each term are given in Appendix E (\citep{Dalton10b}).
We can show using the particular form of $F(\mathbf{r},\mathbf{s})$
for a single mode condensate, that the results for the single mode
condensate can be obtained from those for the two mode condensate
(see Appendix E, \citep{Dalton10b}).

\subsubsection{The $\widehat{H}_{3}$ Term}

The contributions to the \emph{functional Fokker-Planck equation}
from the $\widehat{H}_{3}$ term, which is equal to the sum of the
kinetic energy and trap potential terms in the \emph{non-condensate}
Hamiltonian plus the terms in the interaction between the \emph{condensate
}and \emph{non-condensate} that are \emph{quadratic} in the non-condensate
fields, may be written in the form\begin{eqnarray}
 &  & \left(\frac{\partial}{\partial t}P[\underrightarrow{\psi}(\mathbf{r}),\underrightarrow{\psi^{\ast}}(\mathbf{r})]\right)_{H3}\nonumber \\
 & = & \left(\frac{\partial}{\partial t}P[\underrightarrow{\psi}(\mathbf{r}),\underrightarrow{\psi^{\ast}}(\mathbf{r})]\right)_{H3K}+\left(\frac{\partial}{\partial t}P[\underrightarrow{\psi}(\mathbf{r}),\underrightarrow{\psi^{\ast}}(\mathbf{r})]\right)_{H3V}\nonumber \\
 &  & +\left(\frac{\partial}{\partial t}P[\underrightarrow{\psi}(\mathbf{r}),\underrightarrow{\psi^{\ast}}(\mathbf{r})]\right)_{H3U}\label{Eq.FnalFPTermH3}\end{eqnarray}
Derivations of the form for each term are given in Appendix E (\citep{Dalton10b}).

\paragraph{H3K Terms - Single and Two-Mode Condensates}

The contribution to the functional Fokker-Planck equation from the
\emph{non-condensate kinetic energy} term is\begin{eqnarray}
 &  & \left(\frac{\partial}{\partial t}P[\underrightarrow{\psi}(\mathbf{r}),\underrightarrow{\psi^{\ast}}(\mathbf{r})]\right)_{H3K}\nonumber \\
 & = & \frac{-i}{\hbar}\left\{ -\int d\mathbf{s}\left\{ \frac{\delta}{\delta\psi_{NC}^{+}(\mathbf{s})}\left(\sum\limits _{\mu}\frac{\hbar^{2}}{2m}\partial_{\mu}^{2}\psi_{NC}^{+}(\mathbf{s})\right)P[\underrightarrow{\psi}(\mathbf{r}),\underrightarrow{\psi^{\ast}}(\mathbf{r})]\right\} \right\} \nonumber \\
 &  & +\frac{-i}{\hbar}\left\{ +\int d\mathbf{s}\left\{ \frac{\delta}{\delta\psi_{NC}(\mathbf{s})}\left(\sum\limits _{\mu}\frac{\hbar^{2}}{2m}\partial_{\mu}^{2}\psi_{NC}(\mathbf{s})\right)P[\underrightarrow{\psi}(\mathbf{r}),\underrightarrow{\psi^{\ast}}(\mathbf{r})])\right\} \right\} \nonumber \\
 &  & \,\label{eq:FnalFPTermH3K}\end{eqnarray}

\paragraph{H3V Terms - Single and Two-Mode Condensates}

The contribution to the functional Fokker-Planck equation from the
\emph{non-condensate trap potential }term is\begin{eqnarray}
 &  & \left(\frac{\partial}{\partial t}P[\underrightarrow{\psi}(\mathbf{r}),\underrightarrow{\psi^{\ast}}(\mathbf{r})]\right)_{H3V}\nonumber \\
 & = & \frac{-i}{\hbar}\left\{ -\int d\mathbf{s}\left\{ \frac{\delta}{\delta\psi_{NC}(\mathbf{s})}\{V(\mathbf{s})\psi_{NC}(\mathbf{s})\}\right\} P[\underrightarrow{\psi}(\mathbf{r}),\underrightarrow{\psi^{\ast}}(\mathbf{r})]\right\} \nonumber \\
 &  & +\frac{-i}{\hbar}\left\{ +\int d\mathbf{s}\left\{ \frac{\delta}{\delta\psi_{NC}^{+}(\mathbf{s})}V(\mathbf{s})\psi_{NC}^{+}(\mathbf{s})\right\} P[\underrightarrow{\psi}(\mathbf{r}),\underrightarrow{\psi^{\ast}}(\mathbf{r})]\right\} \label{eq:FnalFPTermH3V}\end{eqnarray}

\paragraph{H3U Terms - Single and Two-Mode Condensates}

The contribution to the functional Fokker-Planck equation from the
$\widehat{H}_{3U}$ term is \begin{eqnarray}
 &  & \left(\frac{\partial}{\partial t}P[\underrightarrow{\psi}(\mathbf{r}),\underrightarrow{\psi^{\ast}}(\mathbf{r})]\right)_{H3U}\nonumber \\
 & = & \left(\frac{\partial}{\partial t}P[\underrightarrow{\psi}(\mathbf{r}),\underrightarrow{\psi^{\ast}}(\mathbf{r})]\right)_{H3U}^{1}+\left(\frac{\partial}{\partial t}P[\underrightarrow{\psi}(\mathbf{r}),\underrightarrow{\psi^{\ast}}(\mathbf{r})]\right)_{H3U}^{2}\nonumber \\
 &  & +\left(\frac{\partial}{\partial t}P[\underrightarrow{\psi}(\mathbf{r}),\underrightarrow{\psi^{\ast}}(\mathbf{r})]\right)_{H3U}^{3}+\left(\frac{\partial}{\partial t}P[\underrightarrow{\psi}(\mathbf{r}),\underrightarrow{\psi^{\ast}}(\mathbf{r})]\right)_{H3U}^{4}\label{Eq.FnalFPTermH3U}\end{eqnarray}
where\begin{eqnarray}
 &  & \left(\frac{\partial}{\partial t}P[\underrightarrow{\psi}(\mathbf{r}),\underrightarrow{\psi^{\ast}}(\mathbf{r})]\right)_{H3U}^{1}\nonumber \\
 & = & \frac{-i}{\hbar}\left\{ +\frac{g_{N}}{N}\int d\mathbf{s}\left\{ \left(\frac{\delta}{\delta\psi_{C}^{+}(\mathbf{s})}\right)\{[\psi_{NC}^{+}(\mathbf{s})\psi_{C}(\mathbf{s})+2\psi_{C}^{+}(\mathbf{s})\psi_{NC}(\mathbf{s})]\psi_{NC}^{+}(\mathbf{s})\}\right\} P[\underrightarrow{\psi}(\mathbf{r}),\underrightarrow{\psi^{\ast}}(\mathbf{r})]\right\} \nonumber \\
 &  & +\frac{-i}{\hbar}\left\{ -\frac{g_{N}}{N}\int d\mathbf{s}\left\{ \left(\frac{\delta}{\delta\psi_{C}(\mathbf{s})}\right)\{[\psi_{NC}(\mathbf{s})\psi_{C}^{+}(\mathbf{s})+2\psi_{C}(\mathbf{s})\psi_{NC}^{+}(\mathbf{s})]\psi_{NC}(\mathbf{s})\}\right\} P[\underrightarrow{\psi}(\mathbf{r}),\underrightarrow{\psi^{\ast}}(\mathbf{r})]\right\} \nonumber \\
 &  & +\frac{-i}{\hbar}\left\{ +\frac{g_{N}}{N}\int d\mathbf{s}\left\{ \left(\frac{\delta}{\delta\psi_{NC}^{+}(\mathbf{s})}\right)\{[\psi_{C}^{+}(\mathbf{s})\psi_{C}^{+}(\mathbf{s})]\psi_{NC}(\mathbf{s})\}\right\} P[\underrightarrow{\psi}(\mathbf{r}),\underrightarrow{\psi^{\ast}}(\mathbf{r})]\right\} \nonumber \\
 &  & +\frac{-i}{\hbar}\left\{ -\frac{g_{N}}{N}\int d\mathbf{s}\left\{ \left(\frac{\delta}{\delta\psi_{NC}(\mathbf{s})}\right)\{[\psi_{C}(\mathbf{s})\psi_{C}(\mathbf{s})]\psi_{NC}^{+}(\mathbf{s})\}\right\} P[\underrightarrow{\psi}(\mathbf{r}),\underrightarrow{\psi^{\ast}}(\mathbf{r})]\right\} \nonumber \\
 &  & +\frac{-i}{\hbar}\left\{ +\frac{g_{N}}{N}\int d\mathbf{s}\left\{ \left(\frac{\delta}{\delta\psi_{NC}^{+}(\mathbf{s})}\right)\{[2\psi_{C}(\mathbf{s})\psi_{C}^{+}(\mathbf{s})-\delta_{C}(\mathbf{s},\mathbf{s})]\psi_{NC}^{+}(\mathbf{s})\}\right\} P[\underrightarrow{\psi}(\mathbf{r}),\underrightarrow{\psi^{\ast}}(\mathbf{r})]\right\} \nonumber \\
 &  & +\frac{-i}{\hbar}\left\{ -\frac{g_{N}}{N}\int d\mathbf{s}\left\{ \left(\frac{\delta}{\delta\psi_{NC}(\mathbf{s})}\right)\{[2\psi_{C}^{+}(\mathbf{s})\psi_{C}(\mathbf{s})-\delta_{C}(\mathbf{s,s})]\psi_{NC}(\mathbf{s})\}\right\} P[\underrightarrow{\psi}(\mathbf{r}),\underrightarrow{\psi^{\ast}}(\mathbf{r})]\right\} \nonumber \\
 &  & \,\label{eq:FnalFPTermH3ULinear}\end{eqnarray}
\begin{eqnarray}
 &  & \left(\frac{\partial}{\partial t}P[\underrightarrow{\psi}(\mathbf{r}),\underrightarrow{\psi^{\ast}}(\mathbf{r})]\right)_{H3U}^{2}\nonumber \\
 & = & \frac{-i}{\hbar}\left\{ -\frac{g_{N}}{N}\int d\mathbf{s}\left\{ \left(\frac{\delta}{\delta\psi_{C}^{+}(\mathbf{s})}\right)\left(\frac{\delta}{\delta\psi_{NC}(\mathbf{s})}\right)\{\psi_{NC}^{+}(\mathbf{s})\psi_{C}(\mathbf{s})+\psi_{C}^{+}(\mathbf{s})\psi_{NC}(\mathbf{s})\}\right\} P[\underrightarrow{\psi}(\mathbf{r}),\underrightarrow{\psi^{\ast}}(\mathbf{r})]\right\} \nonumber \\
 &  & +\frac{-i}{\hbar}\left\{ +\frac{g_{N}}{N}\int d\mathbf{s}\left\{ \left(\frac{\delta}{\delta\psi_{C}(\mathbf{s})}\right)\left(\frac{\delta}{\delta\psi_{NC}^{+}(\mathbf{s})}\right)\{\psi_{NC}(\mathbf{s})\psi_{C}^{+}(\mathbf{s})+\psi_{C}(\mathbf{s})\psi_{NC}^{+}(\mathbf{s})\}\right\} P[\underrightarrow{\psi}(\mathbf{r}),\underrightarrow{\psi^{\ast}}(\mathbf{r})]\right\} \nonumber \\
 &  & +\frac{-i}{\hbar}\left\{ +\frac{g_{N}}{N}\int d\mathbf{s}\left\{ \left(\frac{\delta}{\delta\psi_{C}(\mathbf{s})}\right)\left(\frac{\delta}{\delta\psi_{NC}(\mathbf{s})}\right)\{\psi_{NC}(\mathbf{s})\psi_{C}(\mathbf{s})\}\right\} P[\underrightarrow{\psi}(\mathbf{r}),\underrightarrow{\psi^{\ast}}(\mathbf{r})]\right\} \nonumber \\
 &  & +\frac{-i}{\hbar}\left\{ -\frac{g_{N}}{N}\int d\mathbf{s}\left\{ \left(\frac{\delta}{\delta\psi_{C}^{+}(\mathbf{s})}\right)\left(\frac{\delta}{\delta\psi_{NC}^{+}(\mathbf{s})}\right)\{\psi_{C}^{+}(\mathbf{s})\psi_{NC}^{+}(\mathbf{s})\}\right\} P[\underrightarrow{\psi}(\mathbf{r}),\underrightarrow{\psi^{\ast}}(\mathbf{r})]\right\} \nonumber \\
 &  & +\frac{-i}{\hbar}\left\{ +\frac{g_{N}}{N}\int d\mathbf{s}\left\{ \left(\frac{\delta}{\delta\psi_{NC}(\mathbf{s})}\right)\left(\frac{\delta}{\delta\psi_{NC}(\mathbf{s})}\right)\{\frac{1}{2}\psi_{C}(\mathbf{s})\psi_{C}(\mathbf{s})\}\right\} P[\underrightarrow{\psi}(\mathbf{r}),\underrightarrow{\psi^{\ast}}(\mathbf{r})]\right\} \nonumber \\
 &  & +\frac{-i}{\hbar}\left\{ -\frac{g_{N}}{N}\int d\mathbf{s}\left\{ \left(\frac{\delta}{\delta\psi_{NC}^{+}(\mathbf{s})}\right)\left(\frac{\delta}{\delta\psi_{NC}^{+}(\mathbf{s})}\right)\{\frac{1}{2}\psi_{C}^{+}(\mathbf{s})\psi_{C}^{+}(\mathbf{s})\}\right\} P[\underrightarrow{\psi}(\mathbf{r}),\underrightarrow{\psi^{\ast}}(\mathbf{r})]\right\} \nonumber \\
 &  & \,\label{eq:FnalFPTermH3UQuadratic}\end{eqnarray}
\begin{eqnarray}
 &  & \left(\frac{\partial}{\partial t}P[\underrightarrow{\psi}(\mathbf{r}),\underrightarrow{\psi^{\ast}}(\mathbf{r})]\right)_{H3U}^{3}\nonumber \\
 & = & \frac{-i}{\hbar}\left\{ -\frac{g_{N}}{N}\int d\mathbf{s}\left\{ \left(\frac{\delta}{\delta\psi_{C}^{+}(\mathbf{s})}\right)\left(\frac{\delta}{\delta\psi_{C}^{+}(\mathbf{s})}\right)\left(\frac{\delta}{\delta\psi_{NC}(\mathbf{s})}\right)\{\frac{1}{4}\psi_{NC}^{+}(\mathbf{s})\}\right\} P[\underrightarrow{\psi}(\mathbf{r}),\underrightarrow{\psi^{\ast}}(\mathbf{r})]\right\} \nonumber \\
 &  & +\frac{-i}{\hbar}\left\{ +\frac{g_{N}}{N}\int d\mathbf{s}\left\{ \left(\frac{\delta}{\delta\psi_{C}(\mathbf{s})}\right)\left(\frac{\delta}{\delta\psi_{C}(\mathbf{s})}\right)\left(\frac{\delta}{\delta\psi_{NC}^{+}(\mathbf{s})}\right)\{\frac{1}{4}\psi_{NC}(\mathbf{s})\}\right\} P[\underrightarrow{\psi}(\mathbf{r}),\underrightarrow{\psi^{\ast}}(\mathbf{r})]\right\} \nonumber \\
 &  & +\frac{-i}{\hbar}\left\{ +\frac{g_{N}}{N}\int d\mathbf{s}\left\{ \left(\frac{\delta}{\delta\psi_{C}^{+}(\mathbf{s})}\right)\left(\frac{\delta}{\delta\psi_{NC}(\mathbf{s})}\right)\left(\frac{\delta}{\delta\psi_{NC}(\mathbf{s})}\right)\{\frac{1}{2}\psi_{C}(\mathbf{s})\}\right\} P[\underrightarrow{\psi}(\mathbf{r}),\underrightarrow{\psi^{\ast}}(\mathbf{r})]\right\} \nonumber \\
 &  & +\frac{-i}{\hbar}\left\{ -\frac{g_{N}}{N}\int d\mathbf{s}\left\{ \left(\frac{\delta}{\delta\psi_{C}(\mathbf{s})}\right)\left(\frac{\delta}{\delta\psi_{NC}^{+}(\mathbf{s})}\right)\left(\frac{\delta}{\delta\psi_{NC}^{+}(\mathbf{s})}\right)\{\frac{1}{2}\psi_{C}^{+}(\mathbf{s})\}\right\} P[\underrightarrow{\psi}(\mathbf{r}),\underrightarrow{\psi^{\ast}}(\mathbf{r})]\right\} \nonumber \\
 &  & +\frac{-i}{\hbar}\left\{ +\frac{g_{N}}{N}\int d\mathbf{s}\left\{ \left(\frac{\delta}{\delta\psi_{C}(\mathbf{s})}\right)\left(\frac{\delta}{\delta\psi_{C}^{+}(\mathbf{s})}\right)\left(\frac{\delta}{\delta\psi_{NC}(\mathbf{s})}\right)\{\frac{1}{2}\psi_{NC}(\mathbf{s})\}\right\} P[\underrightarrow{\psi}(\mathbf{r}),\underrightarrow{\psi^{\ast}}(\mathbf{r})]\right\} \nonumber \\
 &  & +\frac{-i}{\hbar}\left\{ -\frac{g_{N}}{N}\int d\mathbf{s}\left\{ \left(\frac{\delta}{\delta\psi_{C}^{+}(\mathbf{s})}\right)\left(\frac{\delta}{\delta\psi_{C}(\mathbf{s})}\right)\left(\frac{\delta}{\delta\psi_{NC}^{+}(\mathbf{s})}\right)\{\frac{1}{2}\psi_{NC}^{+}(\mathbf{s})\}\right\} P[\underrightarrow{\psi}(\mathbf{r}),\underrightarrow{\psi^{\ast}}(\mathbf{r})]\right\} \nonumber \\
 &  & \,\label{eq:FnalFPTermH3UCubic}\end{eqnarray}
\begin{eqnarray}
 &  & \left(\frac{\partial}{\partial t}P[\underrightarrow{\psi}(\mathbf{r}),\underrightarrow{\psi^{\ast}}(\mathbf{r})]\right)_{H3U}^{4}\nonumber \\
 & = & \frac{-i}{\hbar}\left\{ \frac{g_{N}}{N}\int d\mathbf{s}\left\{ \left(\frac{\delta}{\delta\psi_{C}^{+}(\mathbf{s})}\right)\left(\frac{\delta}{\delta\psi_{C}^{+}(\mathbf{s})}\right)\left(\frac{\delta}{\delta\psi_{NC}(\mathbf{s})}\right)\left(\frac{\delta}{\delta\psi_{NC}(\mathbf{s})}\right)\{\frac{1}{8}\}\right\} P[\underrightarrow{\psi}(\mathbf{r}),\underrightarrow{\psi^{\ast}}(\mathbf{r})]\right\} \nonumber \\
 &  & +\frac{-i}{\hbar}\left\{ -\frac{g_{N}}{N}\int d\mathbf{s}\left\{ \left(\frac{\delta}{\delta\psi_{C}(\mathbf{s})}\right)\left(\frac{\delta}{\delta\psi_{C}(\mathbf{s})}\right)\left(\frac{\delta}{\delta\psi_{NC}^{+}(\mathbf{s})}\right)\left(\frac{\delta}{\delta\psi_{NC}^{+}(\mathbf{s})}\right)\{\frac{1}{8}\}\right\} P[\underrightarrow{\psi}(\mathbf{r}),\underrightarrow{\psi^{\ast}}(\mathbf{r})]\right\} \nonumber \\
 &  & \,\label{eq:FnalFPTermH3UQuartic}\end{eqnarray}
Derivations of the form for each term are given in Appendix E (\citep{Dalton10b}).\pagebreak{}

\section{Ito stochastic equations}

\label{Sect Ito Stochastic Eqns}

In this section we show how the functional Fokker-Planck equations
for the phase space distribution functional are equivalent to Ito
stochastic equations for stochastic fields. This first involves truncating
the Fokker-Planck equations to only include terms with at most second
order functional derivatives. The stochastic fields are defined via
the expansion of the phase space field functions in terms of mode
functions and then treating the expansion coefficients as stochastic
variables. The derivation of the Ito equations for the stochastic
fields is based on well-known Ito equations for stochastic expansion
coefficients. The Ito stochastic field equations are the sum of a
deterministic term associated with the first order functional derivatives
in the FFPE (the drift terms) and a quantum noise term associated
with the second order functional derivatives in the FFPE (the diffusion
terms). The two mode condendsate case results in non-local drift and
diffusion terms, so a special treatment is required to derive the
Ito equations. Results for the Ito equations for the stochastic condensate
and non-condensate fields are obtained for the two mode condensate
case. Also, the corresponding simpler Ito equations for the single
mode condensate case are presented. In this section we emphasise how
the phase space distribution functionals which determine the quantum
correlation functions can then be replaced by stochastic averages
involving products of the stochastic condensate and non-condensate
fields.\

\subsection{General Results}

The derivation of Ito stochastic equations the the condensate and
non-condensate fields is based on approximating the functional Fokker-Planck
equation by neglecting all terms involving \emph{third} and \emph{fourth
order} functional derivatives. The justification for this is as follows.
The condensate fields are of order $\sqrt{N}$ in the regions of phase
space important to the determination of the correlation functions
via the functional integrals (\ref{Eq.QuantumAverages}), whereas
the non-condensate fields are much smaller. Hence terms like the third
order functional derivatives in (\ref{eq:FnalFPTermH1U}) scale like
$1/N^{2}$ whereas the second order functional derivatives in (\ref{eq:FnalFPTermH2U4Quadratic})
scale like $1/\sqrt{N}$. This enables all such third and fourth order
terms from the functional Fokker-Planck equation based on the Bogoliubov
Hamiltonian to be discarded. The resulting functional Fokker-Planck
equation is then in a standard form involving just first and second
order functional derivatives, from which Ito stochastic equations
can be obtained.

The remaining \emph{first} and \emph{second order} functional derivative
terms that are left are referred to as the drift and diffusion terms
respectively, and the Ito stochastic equations for the stochastic
fields can expressed in terms of the drift and diffusion terms. The
stochastic fields will be indicated with a tilde, $\widetilde{\psi}_{C}(\mathbf{s,}t)$,
..,$\widetilde{\psi}_{NC}^{+}(\mathbf{s,}t)$. The Ito stochastic
field equations are the sum of two terms. The first is obtained from
the drift term in the functional Fokker-Planck equation and is the
so-called deterministic term, the second is obtained from the diffusion
term and is the stochastic noise term. The stochastic fields are expanded
in terms of a convenient set of real,orthonormal mode functions, with
the expansion coefficients regarded as stochastic quantities. The
original stochastic noise terms in the Ito stochastic field equations
depends on two types of stochastic quantities. One type are stochastic
space dependent fields that involve the mode functions and quantities
depending on the stochastic expansion coefficients that are obtained
from the diffusion terms. The other type are time dependent stochastic
Gaussian-Markov noise terms that would be the noise terms in Ito equations
for the expansion coefficients. The derivation of the Ito equations
for the stochastic fields is based on well-known Ito equations for
stochastic expansion coefficients. Details of the derivation of the
Ito stochastic equations are given in Appendix F (\citep{Dalton10b}).
Here we will summarise the key features and results.

\subsubsection{Symmetric Form of Functional Fokker-Planck Equation}

The derivation begins with the functional Fokker-Planck equation set
out in Section \ref{Sect Func Fokker Planck}, but now with all terms
having functional derivatives of third and fourth order ignored. For
convenience we now introduce a simpler notation for listing the fields,
namely we list $\psi_{C}\equiv\psi_{C-},\psi_{C}^{+}\equiv\psi_{C+},\psi_{NC}\equiv\psi_{NC-},\psi_{NC}^{+}\equiv\psi_{NC+}$
as $\psi_{1},\psi_{2},\psi_{3},\psi_{4}$ respectively. Now with $\underrightarrow{\psi}(\mathbf{r})\equiv\{\psi_{1}(\mathbf{r}),\psi_{2}(\mathbf{r}),\psi_{3}(\mathbf{r}),\psi_{4}(\mathbf{r})\}\equiv\{\psi_{K}(\mathbf{r})\}$
and $\underrightarrow{\psi^{\ast}}(\mathbf{r})\equiv\{\psi_{1}^{\ast}(\mathbf{r}),\psi_{2}^{\ast}(\mathbf{r}),\psi_{3}^{\ast}(\mathbf{r}),\psi_{4}^{\ast}(\mathbf{r})\}\equiv\{\psi_{K}^{\ast}(\mathbf{r})\}$
the functional Fokker-Planck equations from Section \ref{Sect Func Fokker Planck}
are as follows.

For the \emph{two mode} condensate case we have. \begin{eqnarray}
\frac{\partial P}{\partial t} & = & \sum_{A}\int dx\,\frac{\delta}{\delta\psi_{A}(x)}A_{A}(\underrightarrow{\psi}(x),x)\, P\nonumber \\
 &  & +\sum_{A\leq\, B}\int\int dx\, dy\frac{\delta}{\delta\psi_{A}(x)}\frac{\delta}{\delta\psi_{B}(y)}H_{AB}(\underrightarrow{\psi}(x),x,\underrightarrow{\psi}(y),y)\, P\label{Eq.FFPETwoStart}\end{eqnarray}
and for the \emph{single mode} condensate case\begin{eqnarray}
\frac{\partial P}{\partial t} & = & \sum_{A}\int dx\,\frac{\delta}{\delta\psi_{A}(x)}A_{A}(\underrightarrow{\psi}(x),x)\, P\nonumber \\
 &  & +\sum_{A\leq\, B}\int dx\,\frac{\delta}{\delta\psi_{A}(x)}\frac{\delta}{\delta\psi_{B}(x)}H_{AB}(\underrightarrow{\psi}(x),x)\, P\label{EQ.FFPESingleStart}\end{eqnarray}
Here we use $x$, $y$ to denote the spatial variables and in accord
with the expressions in Section \ref{Sect Func Fokker Planck} the
restriction to $A\leq\, B$ in the double sum is to avoid repetition
of double functional derivatives. Since there are four fields involved
$\int A,B=1,2,3,4$. In both cases the distribution functional is
$P[\underrightarrow{\psi},\underrightarrow{\psi}^{\ast}]$ and $A_{A}(\underrightarrow{\psi}(x),x)$
is the $A$ element of a \emph{drift column vector}. For the single
mode condensate case $H_{AB}(\underrightarrow{\psi}(x),x)$ is the
$A,B$ element of a \emph{local diffusion matrix}, and for the two
mode condensate case $H_{AB}(\underrightarrow{\psi}(x),x,\underrightarrow{\psi}(y),y)$
is the $A;B$ element of a \emph{non-local diffusion matrix}. In the
latter case a double spatial integral is involved. Also, $A_{A}$
and $H_{AB}$ may depend on spatial derivatives $\partial_{x}\psi_{K}(x).$etc$.$
but in order to avoid too many symbols we have not shown this. For
simplicity the Fokker-Planck equation has been written with just one-dimensional
spatial variables $x,y$, but the generalisation to three dimensional
variables $\mathbf{r},\mathbf{s}$ is straight-forward.

To proceed further the functional Fokker-Planck equations need to
be recast with a symmetrical diffusion term. The details are covered
in Appendix F (\citep{Dalton10b}). If we define a new diffusion matrix
such that\begin{eqnarray}
D_{AB}(\underrightarrow{\psi}(x),x,\underrightarrow{\psi}(y),y) & = & H_{AB}(\underrightarrow{\psi}(x),x,\underrightarrow{\psi}(y),y)\qquad A<\, B\nonumber \\
D_{AB}(\underrightarrow{\psi}(x),x,\underrightarrow{\psi}(y),y) & = & H_{BA}(\underrightarrow{\psi}(y),y,\underrightarrow{\psi}(x),x)\qquad A>\, B\nonumber \\
D_{AA}(\underrightarrow{\psi}(x),x,\underrightarrow{\psi}(y),y) & = & H_{AA}(\underrightarrow{\psi}(x),x,\underrightarrow{\psi}(y),y)+H_{AA}(\underrightarrow{\psi}(y),y,\underrightarrow{\psi}(x),x)\qquad A=B\nonumber \\
 &  & \,\label{eq:DiffusionMatrixTwoMode}\end{eqnarray}
we see that the functional Fokker-Planck equation for the two mode
case becomes\begin{eqnarray}
\frac{\partial P}{\partial t} & = & \sum_{A}\int dx\,\frac{\delta}{\delta\psi_{A}(x)}A_{A}(\underrightarrow{\psi}(x),x)\, P\nonumber \\
 &  & +\frac{1}{2}\sum_{A,B}\int\int dx\, dy\frac{\delta}{\delta\psi_{A}(x)}\frac{\delta}{\delta\psi_{B}(y)}D_{AB}(\underrightarrow{\psi}(x),x,\underrightarrow{\psi}(y),y)\, P\label{Eq.FFPEDoubleIntegralForm}\end{eqnarray}
The expressions have been defined so that $D_{AB}$ is symmetric.
For the two mode condensate case\begin{equation}
D_{AB}(\underrightarrow{\psi}(x),x,\underrightarrow{\psi}(y),y)=D_{BA}(\underrightarrow{\psi}(y),y,\underrightarrow{\psi}(x),x)\label{Eq.SymmCondnDoubleIntDiffusion}\end{equation}

For the \emph{single mode} condensate case we may also write the functional
Fokker-Planck equation in the symmetric form\begin{eqnarray}
\frac{\partial P}{\partial t} & = & \sum_{A}\int dx\,\frac{\delta}{\delta\psi_{A}(x)}A_{A}(\underrightarrow{\psi}(x),x)\, P\nonumber \\
 &  & +\frac{1}{2}\sum_{A,\, B}\int dx\,\frac{\delta}{\delta\psi_{A}(x)}\frac{\delta}{\delta\psi_{B}(x)}D_{AB}(\underrightarrow{\psi}(x),x)\, P\label{Eq.FFPESingleIntegralForm}\end{eqnarray}
The proof is similar but now \begin{eqnarray}
D_{AB}(\underrightarrow{\psi}(x),x) & = & H_{AB}(\underrightarrow{\psi}(x),x)\qquad A<\, B\nonumber \\
D_{AB}(\underrightarrow{\psi}(x),x) & = & H_{BA}(\underrightarrow{\psi}(x),x)\qquad A>\, B\nonumber \\
D_{AA}(\underrightarrow{\psi}(x),x) & = & 2H_{AA}(\underrightarrow{\psi}(x),x)\qquad A=B\label{Eq.DiffusionMatrixOneMode}\end{eqnarray}
and again $D_{AB}$ is symmetric. \begin{equation}
D_{AB}(\underrightarrow{\psi}(x),x)=D_{BA}(\underrightarrow{\psi}(x),x)\label{Eq.SymmCondnSingleIntDiffusion}\end{equation}
Results (\ref{Eq.DiffusionMatrixOneMode}) and (\ref{eq:DiffusionMatrixTwoMode})
enable us to identify the diffusion coefficients in the general forms
(\ref{Eq.FFPESingleIntegralForm}) and (\ref{Eq.FFPEDoubleIntegralForm})
from those in the original functional Fokker-Planck equation forms
(\ref{EQ.FFPESingleStart}) and (\ref{Eq.FFPETwoStart}).

\subsubsection{Fokker-Planck Equation for Distribution Function}

The field functions $\psi_{A}(x)$ may be expanded \begin{equation}
\psi_{A}(x)=\sum_{i}\alpha_{i}^{A}\,\xi_{i}^{A}(x)\label{Eq.FieldFnGeneral}\end{equation}
where the $\xi_{i}^{A}(x)$ are a convenient set of orthonormal mode
functions for the $A$ field satisfying \begin{eqnarray}
\int dx\xi_{i}^{A}(x)^{\ast}\xi_{j}^{A}(x) & = & \delta_{ij}\label{Eq.OrhogGeneral}\\
\sum_{i}\xi_{i}^{A}(x)\xi_{i}^{A}(y)^{\ast} & = & \delta(x-y)\label{Eq.CompletenessGeneral}\end{eqnarray}
For the various $\psi_{A}(x)$ these orthonormal mode functions may
be interrelated. Thus if for $\psi_{1}(x)\equiv\psi_{C}(x)$ the mode
functions are $\xi_{i}(x)$ $(i=1,2)$, then those for $\psi_{2}(x)\equiv\psi_{C}^{+}(x)$
are $\xi_{i}(x)^{\ast}$ $(i=1,2)$. Mode functions for different
fields also may be orthogonal, thus for $\psi_{3}(x)\equiv\psi_{NC}(x)$
if the mode functions are $\xi_{i}(x)$ $(i\neq1,2)$, and those for
$\psi_{4}(x)\equiv\psi_{NC}^{+}(x)$ are $\xi_{i}(x)^{\ast}$ $(i\neq1,2)$,
then the $\xi_{i}^{1}(x)$ and $\xi_{i}^{3}(x)$ are mutually orthogonal,
as are $\xi_{i}^{2}(x)$ and $\xi_{i}^{4}(x)$. However, these features
are not required, the main requirement is that the mode functions
for each specific field are orthonormal. The mode functions may be
time dependent, but this will not be made explicit.

The derivation of the Ito stochastic field equation is based on first
converting the functional Fokker-Planck equation to an ordinary \emph{Fokker-Planck
equation} via expanding the field functions and replacing the functional
derivatives with ordinary derivatives \begin{eqnarray}
A_{A}(\underrightarrow{\psi}(x),x) & \rightarrow & \mathcal{A}^{A}(\underrightarrow{\alpha})\nonumber \\
D_{AB}(\underrightarrow{\psi}(x),x,\underrightarrow{\psi}(y),y)\; or\; D_{AB}(\underrightarrow{\psi}(x),x)\, & \rightarrow & \mathcal{D}^{AB}(\underrightarrow{\alpha})\nonumber \\
P[\underrightarrow{\psi},\underrightarrow{\psi}^{\ast}] & \rightarrow & P_{b}(\underrightarrow{\alpha},\underrightarrow{\alpha}^{\ast})\nonumber \\
\frac{\delta}{\delta\psi_{A}(x)} & \rightarrow & \sum_{i}\xi_{i}^{A}(x)^{\ast}\frac{\partial}{\partial\alpha_{i}^{A}}\label{Eq.EquivalencesGeneral}\end{eqnarray}
where $\mathcal{A}^{A}$ is the \emph{drift vector}, $\mathcal{D}^{AB}$
is the symmetric \emph{diffusion matrix} and $P_{b}(\underrightarrow{\alpha},\underrightarrow{\alpha}^{\ast})$
is the phase space \emph{distribution function}. The drift and diffusion
elements depend on the expansion coefficients $\underrightarrow{\alpha}\equiv\{\alpha_{k},\alpha_{k}^{+}\}$
and the distribution function depends on $\underrightarrow{\alpha}^{\ast}\equiv\{\alpha_{k}^{\ast},\alpha_{k}^{+\ast}\}$
also. The explicit expressions are\begin{eqnarray}
\mathcal{A}_{i}^{A}(\underrightarrow{\alpha}) & = & \int dx\,\xi_{i}^{A}(x)^{\ast}A_{A}(\underrightarrow{\psi}(x),x)\label{Eq.DriftVectorCompts}\\
\mathcal{D}_{ij}^{AB}(\underrightarrow{\alpha}) & = & \int\int dx\, dy\,\xi_{i}^{A}(x)^{\ast}D_{AB}(\underrightarrow{\psi}(x),x,\underrightarrow{\psi}(y),y)\,\xi_{j}^{B}(y)^{\ast}\qquad Two\; Mode\nonumber \\
\mathcal{D}_{ij}^{AB}(\underrightarrow{\alpha}) & = & \int\int dx\,\xi_{i}^{A}(x)^{\ast}D_{AB}(\underrightarrow{\psi}(x),x)\,\xi_{j}^{B}(x)^{\ast}\qquad One\; Mode\label{Eq.DiffusionMatrixElements}\end{eqnarray}
These relationships can be inverted using the completeness relationships
to give\begin{eqnarray}
A_{A}(\underrightarrow{\psi}(x),x) & = & \sum_{i}\xi_{i}^{A}(x)\mathcal{A}_{i}^{A}(\underrightarrow{\alpha})\label{Eq.DriftVectInverse}\\
D_{AB}(\underrightarrow{\psi}(x),x,\underrightarrow{\psi}(y),y)\, & = & \sum_{ij}\xi_{i}^{A}(x)\mathcal{D}_{ij}^{AB}(\underrightarrow{\alpha})\xi_{j}^{B}(y)\qquad Two\; Mode\nonumber \\
D_{AB}(\underrightarrow{\psi}(x),x)\delta(x-y) & = & \sum_{ij}\xi_{i}^{A}(x)\mathcal{D}_{ij}^{AB}(\underrightarrow{\alpha})\xi_{j}^{B}(y)\qquad One\; Mode\label{Eq.DiffusionMatrixInverse}\end{eqnarray}
The diffusion matrix is symmetric \begin{equation}
\mathcal{D}_{ji}^{BA}(\underrightarrow{\alpha})=\mathcal{D}_{ij}^{AB}(\underrightarrow{\alpha})\label{Eq.SymmCondnDiffusionMatrix}\end{equation}
this result being easily obtained from (\ref{Eq.SymmCondnSingleIntDiffusion})
or (\ref{Eq.SymmCondnDoubleIntDiffusion}). As a result we can always
write the diffusion matrix $\mathcal{D}$ in the form \begin{equation}
\mathcal{D=BB}^{T}\label{eq:DiffusionMatrixFactorn}\end{equation}
where $\mathcal{B}$ has the same dimension as $\mathcal{D}$. This
result is known as the Takagi factorisation \citep{Takagi25a}. A
proof may be found in the textbook by Horn et al. \citep{Horn85a}.
A non-square matrix $\mathcal{B}$ can also be found, this is shown
in Appendix F (\citep{Dalton10b}).

The ordinary \emph{Fokker-Planck equation} that is obtained is given
by \begin{eqnarray}
\frac{\partial P_{b}(\underrightarrow{\alpha},\underrightarrow{\alpha}^{\ast})}{\partial t} & = & \sum_{Ai}\frac{\partial}{\partial\alpha_{i}^{A}}\mathcal{A}_{i}^{A}(\underrightarrow{\alpha})P_{b}(\underrightarrow{\alpha},\underrightarrow{\alpha}^{\ast})\nonumber \\
 &  & +\frac{1}{2}\sum_{Ai\, Bj}\frac{\partial}{\partial\alpha_{i}^{A}}\frac{\partial}{\partial\alpha_{j}^{B}}\mathcal{D}_{i\,;j}^{A;B}(\underrightarrow{\alpha})P_{b}(\underrightarrow{\alpha},\underrightarrow{\alpha}^{\ast})\label{Eq.FokkerPlanckGeneral}\end{eqnarray}
This Fokker-Planck equation is equivalent to Ito stochastic equations,
as is described in standard textbooks (see \citep{Drummond80a}, \citep{Gardiner91a}).
The procedure involves replacing the time independent phase space
variables $\alpha_{i}^{A}$ by time dependent stochastic variables
$\widetilde{\alpha}_{i}^{A}(t).$ The Ito stochastic equations for
the $\widetilde{\alpha}_{i}^{A}(t)$ are such that phase space averages
of functions of the $\alpha_{i}^{A}$ give the same result as stochastic
averages of the same functions of the $\widetilde{\alpha}_{i}^{A}(t)$.
The derivation of the Ito stochastic equations requires that the complex
diffusion matrix $\mathcal{D}$ is symmetric, a result we have now
obtained.

\subsubsection{Ito Equations for Stochastic Expansion Coefficients}

The \emph{Ito equations} for the stochastic expansion coefficients
$\widetilde{\alpha}_{i}^{C}$ can be written in several forms\begin{eqnarray}
\delta\widetilde{\alpha}_{i}^{A}(t) & = & \widetilde{\alpha}_{i}^{A}(t+\delta t)-\widetilde{\alpha}_{i}^{A}(t)\nonumber \\
 & = & -\mathcal{A}_{i}^{A}(\underrightarrow{\widetilde{\alpha}}(t))\delta t+\sum_{Dk}\mathcal{B}_{i;k}^{A;D}(\underrightarrow{\widetilde{\alpha}}(t))\,\int_{t}^{t+\delta t}dt_{1}\Gamma_{k}^{D}(t_{1})\label{Eq.ItoStochAlphaGeneral}\\
\frac{d}{dt}\widetilde{\alpha}_{i}^{A}(t) & = & -\mathcal{A}_{i}^{A}(\underrightarrow{\widetilde{\alpha}}(t))+\sum_{Dk}\mathcal{B}_{i;k}^{A;D}(\underrightarrow{\widetilde{\alpha}}(t))\frac{d}{dt}w_{k}^{D}(t)\label{Eq.ItoStochAlphaGeneral2}\\
 & = & -\mathcal{A}_{i}^{A}(\underrightarrow{\widetilde{\alpha}}(t))+\sum_{Dk}\mathcal{B}_{i;k}^{A;D}(\underrightarrow{\widetilde{\alpha}}(t))\Gamma_{k}^{D}(t_{+})\label{Eq.ItoStochAlphaGeneral3}\end{eqnarray}
where$\underrightarrow{\widetilde{\alpha}}(t)\equiv\{\widetilde{\alpha}_{i}^{A}(t)\}\equiv\{\widetilde{\alpha}_{k}(t),\widetilde{\alpha}_{k}^{+}(t)\}$
and the matrix $\mathcal{B}$ is related to the diffusion matrix $\mathcal{D}$
as in (\ref{eq:DiffusionMatrixFactorn}). \begin{equation}
\mathcal{D}_{i\, j}^{A;B}(\underrightarrow{\widetilde{\alpha}}(t))=\sum\limits _{Dk}\mathcal{B}_{i;k}^{A;D}(\underrightarrow{\widetilde{\alpha}}(t))\mathcal{B}_{j;k}^{B;D}(\underrightarrow{\widetilde{\alpha}}(t))\label{Eq.DiffusionResultGeneral}\end{equation}
The matrix elements $\mathcal{B}_{i;k}^{A;D}(\underrightarrow{\widetilde{\alpha}}(t))$
are functions of the $\widetilde{\alpha}_{i}^{A}(t).$ The quantity
$t_{+}$ is to indicate that if the Ito stochastic equation is integrated
from $t$ to $t+\delta t$, the Gaussian-Markoff noise term is integrated
over this interval whilst the $\mathcal{A}_{i}^{A}(\widetilde{\alpha}_{j}^{C}(t))$
and $\mathcal{B}_{i;k}^{A;D}(\widetilde{\alpha}_{j}^{C}(t))$ are
left at time $t$..

The quantities $w_{k}^{D}(t)$ and $\Gamma_{k}^{D}(t)$ are Wiener
and Gaussian-Markoff stochastic variables. The \emph{Gaussian-Markoff
}quantities $\Gamma_{k}^{D}$ satisfy the stochastic averaging results\begin{eqnarray}
\overline{\Gamma_{k}^{D}(t_{1})} & = & 0\nonumber \\
\overline{\{\Gamma_{k}^{D}(t_{1})\Gamma_{l}^{E}(t_{2})\}} & = & \delta_{DE}\delta_{kl}\delta(t_{1}-t_{2})\nonumber \\
\overline{\{\Gamma_{k}^{D}(t_{1})\Gamma_{l}^{E}(t_{2})\Gamma_{m}^{F}(t_{3})\}} & = & 0\nonumber \\
\overline{\{\Gamma_{k}^{D}(t_{1})\Gamma_{l}^{E}(t_{2})\Gamma_{m}^{F}(t_{3})\Gamma_{n}^{G}(t_{4})\}} & = & \overline{\{\Gamma_{k}^{D}(t_{1})\Gamma_{l}^{E}(t_{2})\}}\;\overline{\{\Gamma_{m}^{F}(t_{3})\Gamma_{n}^{G}(t_{4})\}}\nonumber \\
 &  & +\overline{\{\Gamma_{k}^{E}(t_{1})\Gamma_{m}^{F}(t_{3})\}}\;\overline{\{\Gamma_{l}^{E}(t_{2})\Gamma_{n}^{G}(t_{4})\}}\nonumber \\
 &  & +\overline{\{\Gamma_{k}^{D}(t_{1})\Gamma_{n}^{G}(t_{4})\}}\;\overline{\{\Gamma_{l}^{E}(t_{2})\Gamma_{m}^{F}(t_{3})\}}\nonumber \\
 &  & ...\label{Eq.GaussMarkGeneral}\end{eqnarray}
with stochastic averages being denote with a bar. The stochastic average
of an odd number of noise terms is always zero, whilst that for an
even number is the sum of all products of stochastic averages of two
noise terms. The Gaussian-Markoff noise terms $\Gamma_{k}^{D}$ are
related to the \emph{Wiener} stochastic variables $w_{k}^{D}$ via
\begin{eqnarray}
w_{k}^{D}(t) & = & \int_{0}^{t}dt_{1}\,\Gamma_{k}^{D}(t_{1})\label{Eq.WienerStochasticVariableGeneral}\\
\delta w_{k}^{D}(t) & = & w_{k}^{D}(t+\delta t)-w_{k}^{D}(t)=\int_{t}^{t+\delta t}dt_{1}\,\Gamma_{k}^{D}(t_{1})\label{Eq.WienerDifferentialGeneral}\\
\frac{d}{dt}w_{k}^{D}(t) & = & \lim_{{\LARGE\delta t\rightarrow0}}\left(\frac{\delta w_{k}^{D}(t)}{\delta t}\right)=\Gamma_{k}^{D}(t_{+})\label{Eq.WienerDerivativeGeneral}\end{eqnarray}

One of the rules in stochastic averaging is\begin{equation}
\overline{\sum_{a}F_{a}(\underrightarrow{\widetilde{\alpha}}(t))}=\sum_{a}\overline{F_{a}(\underrightarrow{\widetilde{\alpha}}(t))}\label{Eq.StochAverageSum}\end{equation}
so the stochastic average of the sum is the sum of the stochastic
averages. Also,in Ito stochastic calculus the noise terms $\Gamma_{k}^{D}(t_{1})$
within the interval $t,t+\delta t$ are uncorrelated with any function
of the $\widetilde{\alpha}_{i}^{C}(t)$ at the earlier time $t$,
so that the stochastic average of the product of such a function with
a product of the noise terms factorises\begin{eqnarray}
 &  & \overline{F(\underrightarrow{\widetilde{\alpha}}(t_{1}))\{\Gamma_{k}^{D}(t_{2})\Gamma_{l}^{E}(t_{3})\Gamma_{m}^{F}(t_{4})...\Gamma_{a}^{X}(t_{l})\}}\nonumber \\
 & = & \overline{F(\underrightarrow{\widetilde{\alpha}}(t_{1}))}\,\overline{\{\Gamma_{k}^{D}(t_{2})\Gamma_{l}^{E}(t_{3})\Gamma_{m}^{F}(t_{4})...\Gamma_{a}^{X}(t_{l})\}}\qquad t_{1}<t_{2},t_{3},..,t_{l}\label{Eq.StochAverProduct}\end{eqnarray}
These key features of Ito stochastic calculus are important in deriving
the properties of the noise fields in the stochastic field equations.

\subsubsection{Derivation of Ito Stochastic Field Equations}

The \emph{stochastic fields} $\widetilde{\psi}_{A}(x,t)$ are defined
via the same expansion as for the time independent field functions
$\psi_{A}(x)$ by replacing the time independent phase space variables
$\alpha_{i}^{A}$ by time dependent stochastic variables $\widetilde{\alpha}_{i}^{A}(t)$\begin{equation}
\widetilde{\psi}_{A}(x,t)=\sum_{i}\widetilde{\alpha}_{i}^{A}(t)\xi_{i}^{A}(x)\label{Eq.StochasticFieldFnGeneral}\end{equation}
The expansion coefficents in (\ref{Eq.StochasticFieldFnGeneral})
are restricted to those required in expanding the particular field
function $\psi_{A}(x)$. Also, stochastic variations in $\widetilde{\psi}_{A}(x,t)$
are chosen as to only being due to stochastic fluctuations in the
$\widetilde{\alpha}_{i}^{A}(t)$. Although the mode functions may
be time dependent, their time variations are not stochastic in origin,
so the stochastic field equations for the $\widetilde{\psi}_{A}(x,t)$
do not allow for time variations in the mode functions.

The \emph{Ito stochastic equation} for the stochastic fields $\widetilde{\psi}_{A}(x,t)$
can then be derived from the Ito stochastic equations for the expansion
coefficients. Using (\ref{Eq.DriftVectInverse}) the drift term in
the stochastic equation gives\begin{equation}
-\sum_{i}\mathcal{A}_{i}^{A}(\underrightarrow{\widetilde{\alpha}}(t))\,\xi_{i}^{A}(x)\,\delta t=-A_{A}(\underrightarrow{\widetilde{\psi}}(x,t))\delta t\label{Eq.DriftTerm}\end{equation}
which involves the drift vector $A_{A}$ evaluated at the stochastic
fields $\underrightarrow{\widetilde{\psi}}(x,t)$. The diffusion term
in the stochastic equation gives\begin{equation}
\sum_{i}\sum_{Dk}\mathcal{B}_{i;k}^{A;D}(\underrightarrow{\widetilde{\alpha}}(t))\,\xi_{i}^{A}(x)\int_{t}^{t+\delta t}dt_{1}\Gamma_{k}^{D}(t_{1})=\sum_{Dk}\eta_{k}^{A;D}(\underrightarrow{\widetilde{\psi}}(x,t))\int_{t}^{t+\delta t}dt_{1}\Gamma_{k}^{D}(t_{1})\label{Eq.DiffnTerm}\end{equation}
where\begin{equation}
\eta_{k}^{A;D}(\underrightarrow{\widetilde{\psi}}(x,t))=\sum_{i}\mathcal{B}_{i;k}^{A;D}(\underrightarrow{\widetilde{\alpha}}(t))\,\xi_{i}^{A}(x)\label{Eq.EtaStochasticTerm}\end{equation}
is related via $\mathcal{B}_{i;k}^{A;D}(\underrightarrow{\widetilde{\alpha}}(t))$
to the diffusion matrix $D_{AB}$ evaluated at the stochastic fields
$\underrightarrow{\widetilde{\psi}}(x,t)$ or $\underrightarrow{\widetilde{\psi}}(x,t),\underrightarrow{\widetilde{\psi}}(y,t)$.

The stochastic field equations can then be written in several ways\begin{eqnarray}
\delta\widetilde{\psi}_{A}(x,t) & = & \widetilde{\psi}_{A}(x,t+\delta t)-\widetilde{\psi}_{A}(x,t)\nonumber \\
 & = & -A_{A}(\underrightarrow{\widetilde{\psi}}(x,t))\delta t+\sum_{Dk}\eta_{k}^{A;D}(\underrightarrow{\widetilde{\psi}}(x,t))\int_{t}^{t+\delta t}dt_{1}\Gamma_{k}^{D}(t_{1})\nonumber \\
 & = & -A_{A}(\underrightarrow{\widetilde{\psi}}(x,t))\delta t+\delta\widetilde{G}_{A}(\underrightarrow{\widetilde{\psi}}(x,t),\underrightarrow{\Gamma}(t_{+}))\label{Eq.ItoStoFdEqnGeneralA}\\
\frac{\partial}{\partial t}\widetilde{\psi}_{A}(x,t) & = & -A_{A}(\underrightarrow{\widetilde{\psi}}(x,t))+\sum_{Dk}\eta_{k}^{A;D}(\underrightarrow{\widetilde{\psi}}(x,t))\,\frac{d}{dt}w_{k}^{D}(t)\nonumber \\
 & = & -A_{A}(\underrightarrow{\widetilde{\psi}}(x,t))+\sum_{Dk}\eta_{k}^{A;D}(\underrightarrow{\widetilde{\psi}}(x,t))\,\Gamma_{k}^{D}(t_{+})\nonumber \\
 & = & -A_{A}(\underrightarrow{\widetilde{\psi}}(x,t))+\frac{\partial}{\partial t}\widetilde{G}_{A}(\underrightarrow{\widetilde{\psi}}(x,t),\underrightarrow{\Gamma}(t_{+}))\label{Eq.ItoStoFdEqnGeneral}\end{eqnarray}
Here we denote $\underrightarrow{\widetilde{\psi}}(x,t)\equiv\{\widetilde{\psi}_{A}(x,t)\}\equiv\{\widetilde{\psi}_{1}(x,t),\widetilde{\psi}_{2}(x,t),\widetilde{\psi}_{3}(x,t),\widetilde{\psi}_{4}(x,t)\}$
and $\underrightarrow{\Gamma}(t_{+})\equiv\{\Gamma_{k}^{1}(t_{+}),\Gamma_{k}^{2}(t_{+}),\Gamma_{k}^{3}(t_{+}),\Gamma_{k}^{4}(t_{+})\}$.
The first form gives the change in the stochastic field over a small
time integral $t.t+\delta t$, the second is in the form of a partial
differential equation. The first term in the Ito equation for the
stochastic fields (\ref{Eq.ItoStoFdEqnGeneral}) $-A_{A}(\underrightarrow{\widetilde{\psi}}(x,t))$
is the \emph{deterministic term} and is obtained from the drift vector
in the functional Fokker-Planck equation and the second term $\frac{{\LARGE\partial}}{{\LARGE\partial t}}\widetilde{G}_{A}(\underrightarrow{\widetilde{\psi}}(x,t),\underrightarrow{\Gamma}(t_{+}))$
is the \emph{quantum noise field} whose statistical properties are
obtained from the diffusion matrix, and which depends both on the
stochastic fields $\underrightarrow{\widetilde{\psi}}(x,t)$ and on
the Gaussian-Markoff stochastic variables $\underrightarrow{\Gamma}(t_{+})$.

The \emph{noise field} term is \begin{equation}
\frac{\partial}{\partial t}\widetilde{G}_{A}(\underrightarrow{\widetilde{\psi}}(x,t),\underrightarrow{\Gamma}(t_{+}))=\sum_{Dk}\eta_{k}^{A;D}(\underrightarrow{\widetilde{\psi}}(x,t))\,\frac{d}{dt}w_{k}^{D}(t)=\sum_{Dk}\eta_{k}^{A;D}(\underrightarrow{\widetilde{\psi}}(x,t))\,\Gamma_{k}^{D}(t_{+})\label{Eq.NoiseFieldGeneral}\end{equation}
where the stochastic field $\eta_{k}^{A;D}(\underrightarrow{\widetilde{\psi}}(x,t))$
is related to the diffusion matrix expressed in terms of the stochastic
fields $\underrightarrow{\widetilde{\psi}}(x,t)$ or $\underrightarrow{\widetilde{\psi}}(x,t),\underrightarrow{\widetilde{\psi}}(y,t)$.

\subsubsection{Properties of Noise Fields}

\label{sub:Properties of Noise Fields}

To determine the properties of the noise field we first establish
the connection between the $\eta_{k}^{A;D}$ and the $D_{AB}$. \begin{eqnarray}
 &  & \sum_{Dk}\eta_{k}^{A;D}(\underrightarrow{\widetilde{\psi}}(x_{1},t))\eta_{k}^{B;D}(\underrightarrow{\widetilde{\psi}}(x_{2},t))\nonumber \\
 & = & \left[\eta(\underrightarrow{\widetilde{\psi}}(x_{1},t))\eta(\underrightarrow{\widetilde{\psi}}(x_{2},t))^{T}\right]_{AB}\nonumber \\
 & = & \sum_{Dkij}\xi_{i}^{A}(x_{1})\mathcal{B}_{ik}^{A;D}(\underrightarrow{\widetilde{\alpha}}(t))\,\xi_{j}^{B}(x_{2})\mathcal{B}_{jk}^{B;D}(\underrightarrow{\widetilde{\alpha}}(t))\nonumber \\
 & = & \sum_{ij}\xi_{i}^{A}(x_{1})\mathcal{D}_{ij}^{A;B}(\underrightarrow{\widetilde{\alpha}}(t))\,\xi_{j}^{B}(x_{2})\nonumber \\
 & = & D_{AB}(\underrightarrow{\widetilde{\psi}}(x_{1},t),x_{1},\underrightarrow{\widetilde{\psi}}(x_{2},t),x_{2})\qquad Two\; Mode\nonumber \\
\label{eq:SumEtasTwoMode}\\ & = & D_{AB}(\underrightarrow{\widetilde{\psi}}(x_{1,2},t),x_{1,2})\,\delta(x_{1}-x_{2})\qquad One\; Mode\nonumber \\
 &  & \,\label{eq:SumEtasOneMode}\end{eqnarray}
using (\ref{Eq.DiffusionResultGeneral}) and (\ref{Eq.DiffusionMatrixInverse}).
Thus for the single mode condensate $\left[\eta(\underrightarrow{\widetilde{\psi}}(x_{1},t))\eta(\underrightarrow{\widetilde{\psi}}(x_{2},t),t)^{T}\right]_{AB}$
is delta function correlated in space and equal to the local diffusion
matrix element, whereas in the two-mode condensate case this quantity
is equal to the non-local diffusion matrix element.

The \emph{stochastic averages} of the noise field terms can now be
obtained. These results follow from (\ref{eq:SumEtasOneMode}), (\ref{eq:SumEtasTwoMode})
and the properties (\ref{Eq.GaussMarkGeneral}), (\ref{Eq.StochAverageSum}),
(\ref{Eq.StochAverProduct}) and are derived in Appendix F (\citep{Dalton10b}).
For the stochastic average of each noise term\begin{equation}
\overline{\left(\frac{\partial}{\partial t}\widetilde{G}_{A}(\underrightarrow{\widetilde{\psi}}(x,t),\underrightarrow{\Gamma}(t_{+}))\right)}=0\label{Eq.StochAverNoise}\end{equation}
showing that the stochastic average of of each noise field is zero.
For the stochastic average of the product of two noise terms we have
\begin{eqnarray}
 &  & \overline{\left(\frac{\partial}{\partial t}\widetilde{G}_{A}(\underrightarrow{\widetilde{\psi}}(x_{1},t_{1}),\underrightarrow{\Gamma}(t_{1+}))\right)\left(\frac{\partial}{\partial t}\widetilde{G}_{B}(\underrightarrow{\widetilde{\psi}}(x_{2},t_{2}),\underrightarrow{\Gamma}(t_{2+}))\right)}\nonumber \\
 & = & \overline{D_{AB}(\underrightarrow{\widetilde{\psi}}(x_{1},t_{1,2}),x_{1},\underrightarrow{\widetilde{\psi}}(x_{2},t_{1,2}),x_{2})}\nonumber \\
 &  & \times\delta(t_{1}-t_{2})\,\,\,\,\,\,\,\,\,\,\,\,\,\,\,\,\,\,\,\,\,\,\,\,\,\,\,\,\,\,\,\,\,\, Two\; Mode\label{eq:StochAverNoiseTwoMode-1}\\
 & = & \overline{D_{AB}(\underrightarrow{\widetilde{\psi}}(x_{1,2},t_{1,2}),x_{1,2})}\nonumber \\
 &  & \times\delta(x_{1}-x_{2})\delta(t_{1}-t_{2})\,\,\,\,\,\,\,\,\,\,\,\, One\; Mode\label{eq:StochAverNoiseSingleMode-1}\end{eqnarray}
The stochastic average of the product of two noise terms is always
\emph{delta function correlated in time}. In the single mode condensate
case this average is also delta function correlated in space, and
the spatial correlation is given by the stochastic average of the
\emph{local diffusion term} $D_{AB}(\underrightarrow{\widetilde{\psi}}(x_{1,2},t),x_{1,2})$
in the original functional Fokker-Planck equation (\ref{Eq.FFPESingleIntegralForm}).
However for the two mode condensate it is not delta function correlated
in space. Instead the spatial correlation is given by the stochastic
average of the \emph{non-local diffusion term} $D_{AB}(\underrightarrow{\widetilde{\psi}}(x_{1},t),x_{1},\underrightarrow{\widetilde{\psi}}(x_{2},t),x_{2})$.in
the original functional Fokker-Planck equation (\ref{Eq.FFPEDoubleIntegralForm}).

However, although the noise fields have some of the features in (\ref{Eq.GaussMarkGeneral}),
they are not themselves Gaussian-Markov processes. The stochastic
averages of products of odd numbers of noise fields are indeed zero,
but although averages of products of even numbers of noise fields
can be written as sums of products of stochastic averages of pairs
of stochastic quantities with the same delta function time correlations
as in (\ref{Eq.GaussMarkGeneral}), the pairs involved are the diffusion
matrix elements $D_{AB}(\underrightarrow{\widetilde{\psi}}(x_{1},t),x_{1},\underrightarrow{\widetilde{\psi}}(x_{2},t),x_{2})$
rather than products of noise fields such as $\left(\frac{{\LARGE\partial}}{{\LARGE\partial t}}\widetilde{G}_{A}(\underrightarrow{\widetilde{\psi}}(x_{1},t_{1}),\underrightarrow{\Gamma}(t_{1+}))\right)\left(\frac{{\LARGE\partial}}{{\LARGE\partial t}}\widetilde{G}_{B}(\underrightarrow{\widetilde{\psi}}(x_{2},t_{2}),\underrightarrow{\Gamma}(t_{2+}))\right)$.
Nevertheless, the stochastic averages of the noise field terms are
either zero or are determined from stochastic averages only involving
the diffusion matrix elements $D_{AB}(\underrightarrow{\widetilde{\psi}}(x_{1},t),x_{1},\underrightarrow{\widetilde{\psi}}(x_{2},t),x_{2})$.
There is thus never any need to actually determine the matrices $\eta(\underrightarrow{\widetilde{\psi}}(x,t))$
such that $\eta(\underrightarrow{\widetilde{\psi}}(x_{1},t))\eta(\underrightarrow{\widetilde{\psi}}(x_{2},t))^{T}=D(\underrightarrow{\widetilde{\psi}}(x_{1},t),x_{1},\underrightarrow{\widetilde{\psi}}(x_{2},t),x_{2})$
or $D(\underrightarrow{\widetilde{\psi}}(x_{1,2},t),x_{1,2})\delta(x_{1}-x_{2})$,
so all the required expressions for treating the stochastic properties
of the noise fields are provided in the functional Fokker-Planck equation.
Detailed expressions for stochastic averages of more than two noise
fields are derived in Appendix F (\citep{Dalton10b}) as Eqns. (\ref{Eq.StochAverThreeNoise2Mode}),
(\ref{eq:StochAverFourNoise2Mode}), (\ref{eq:StochAverThreeNoise1Mode})
and (\ref{eq:StochAverFourNoise1Mode}).

For the two mode condensate case the results are\begin{eqnarray}
 &  & \overline{\begin{array}{c}
\{\left(\frac{{\LARGE\partial}}{{\LARGE\partial t}}\widetilde{G}_{A}(\underrightarrow{\widetilde{\psi}}(x_{1},t_{1}),\underrightarrow{\Gamma}(t_{1+}))\right)\left(\frac{{\LARGE\partial}}{{\LARGE\partial t}}\widetilde{G}_{B}(\underrightarrow{\widetilde{\psi}}(x_{2},t_{2}),\underrightarrow{\Gamma}(t_{2+}))\right)\\
\times\left(\frac{{\LARGE\partial}}{{\LARGE\partial t}}\widetilde{G}_{C}(\underrightarrow{\widetilde{\psi}}(x_{3},t_{3}),\underrightarrow{\Gamma}(t_{3+}))\right)\}\end{array}}\nonumber \\
 & = & 0\label{Eq.StochAverThreeNoiseTwoMode}\end{eqnarray}
for three noise fields and\begin{eqnarray}
 &  & \overline{\begin{array}{c}
\{\left(\frac{{\LARGE\partial}}{{\LARGE\partial t}}\widetilde{G}_{A}(\underrightarrow{\widetilde{\psi}}(x_{1},t_{1}),\underrightarrow{\Gamma}(t_{1+}))\right)\left(\frac{{\LARGE\partial}}{{\LARGE\partial t}}\widetilde{G}_{B}(\underrightarrow{\widetilde{\psi}}(x_{2},t_{2}),\underrightarrow{\Gamma}(t_{2+}))\right)\\
\times\left(\frac{{\LARGE\partial}}{{\LARGE\partial t}}\widetilde{G}_{C}(\underrightarrow{\widetilde{\psi}}(x_{3},t_{3}),\underrightarrow{\Gamma}(t_{3+}))\right)\left(\frac{{\LARGE\partial}}{{\LARGE\partial t}}\widetilde{G}_{D}(\underrightarrow{\widetilde{\psi}}(x_{4},t_{4}),\underrightarrow{\Gamma}(t_{4+}))\right)\}\end{array}}\nonumber \\
 & = & \overline{\left[D_{AB}(\underrightarrow{\widetilde{\psi}}(x_{1},t_{1,2}),x_{1},\underrightarrow{\widetilde{\psi}}(x_{2},t_{1,2}),x_{2})\right]\left[D_{CD}(\underrightarrow{\widetilde{\psi}}(x_{3},t_{3,4}),x_{3},\underrightarrow{\widetilde{\psi}}(x_{4},t_{3,4}),x_{4})\right]}\;\nonumber \\
 &  & \times\delta(t_{1}-t_{2})\delta(t_{3}-t_{4})\nonumber \\
 &  & +\overline{\left[D_{AC}(\underrightarrow{\widetilde{\psi}}(x_{1},t_{1,3}),x_{1},\underrightarrow{\widetilde{\psi}}(x_{3},t_{1,3}),x_{3})\right]\left[D_{BD}(\underrightarrow{\widetilde{\psi}}(x_{2},t_{2,4}),x_{2},\underrightarrow{\widetilde{\psi}}(x_{4},t_{2,4}),x_{4})\right]}\;\nonumber \\
 &  & \times\delta(t_{1}-t_{3})\delta(t_{2}-t_{4})\nonumber \\
 &  & +\overline{\left[D_{AD}(\underrightarrow{\widetilde{\psi}}(x_{1},t_{1,4}),x_{1},\underrightarrow{\widetilde{\psi}}(x_{4},t_{1,4}),x_{2})\right]\left[D_{BC}(\underrightarrow{\widetilde{\psi}}(x_{2},t_{2,3}),x_{2},\underrightarrow{\widetilde{\psi}}(x_{3},t_{2,3}),x_{3})\right]}\;\nonumber \\
 &  & \times\delta(t_{1}-t_{4})\delta(t_{2}-t_{3})\label{eq:StochAverFourNoiseTwoMode}\end{eqnarray}
for four noise fields. The result for the stochastic average of four
noise field terms is not quite the same as \begin{eqnarray}
 &  & \overline{\{\left(\frac{\partial}{\partial t}\widetilde{G}_{A}(\underrightarrow{\widetilde{\psi}}(x_{1},t_{1}),\underrightarrow{\Gamma}(t_{1+}))\right)\left(\frac{\partial}{\partial t}\widetilde{G}_{B}(\underrightarrow{\widetilde{\psi}}(x_{2},t_{2}),\underrightarrow{\Gamma}(t_{2+}))\right)\}}\nonumber \\
 &  & \times\overline{\{\left(\frac{\partial}{\partial t}\widetilde{G}_{C}(\underrightarrow{\widetilde{\psi}}(x_{3},t_{3}),\underrightarrow{\Gamma}(t_{3+}))\right)\left(\frac{\partial}{\partial t}\widetilde{G}_{D}(\underrightarrow{\widetilde{\psi}}(x_{4},t_{4}),\underrightarrow{\Gamma}(t_{4+}))\right)\}}\nonumber \\
 &  & +\overline{\{\left(\frac{\partial}{\partial t}\widetilde{G}_{A}(\underrightarrow{\widetilde{\psi}}(x_{1},t_{1}),\underrightarrow{\Gamma}(t_{1+}))\right)\left(\frac{\partial}{\partial t}\widetilde{G}_{C}(\underrightarrow{\widetilde{\psi}}(x_{3},t_{3}),\underrightarrow{\Gamma}(t_{3+}))\right)\}}\nonumber \\
 &  & \times\overline{\{\left(\frac{\partial}{\partial t}\widetilde{G}_{B}(\underrightarrow{\widetilde{\psi}}(x_{2},t_{2}),\underrightarrow{\Gamma}(t_{2+}))\right)\left(\frac{\partial}{\partial t}\widetilde{G}_{D}(\underrightarrow{\widetilde{\psi}}(x_{4},t_{4}),\underrightarrow{\Gamma}(t_{4+}))\right)\}}\nonumber \\
 &  & +\overline{\{\left(\frac{\partial}{\partial t}\widetilde{G}_{A}(\underrightarrow{\widetilde{\psi}}(x_{1},t_{1}),\underrightarrow{\Gamma}(t_{1+}))\right)\left(\frac{\partial}{\partial t}\widetilde{G}_{D}(\underrightarrow{\widetilde{\psi}}(x_{4},t_{4}),\underrightarrow{\Gamma}(t_{4+}))\right)\}}\nonumber \\
 &  & \times\overline{\{\left(\frac{\partial}{\partial t}\widetilde{G}_{B}(\underrightarrow{\widetilde{\psi}}(x_{2},t_{2}),\underrightarrow{\Gamma}(t_{2+}))\right)\left(\frac{\partial}{\partial t}\widetilde{G}_{C}(\underrightarrow{\widetilde{\psi}}(x_{3},t_{3}),\underrightarrow{\Gamma}(t_{3+}))\right)\}}\end{eqnarray}
because in general the stochastic average of a product of two diffusion
matrix elements is not the same as the product of the stochastic averages
of each element. Results analogous to (\ref{Eq.StochAverThreeNoiseTwoMode})
and (\ref{eq:StochAverFourNoiseTwoMode}) apply also for the single
mode condensate case. For four noise fields factors such as $D_{AB}(\underrightarrow{\widetilde{\psi}}(x_{1},t_{1,2}),x_{1},\underrightarrow{\widetilde{\psi}}(x_{2},t_{1,2}),x_{2})$
are just replaced by $D_{AB}(\underrightarrow{\widetilde{\psi}}(x_{1,2},t_{1,2}),x_{1})\delta(x_{1}-x_{2})$
etc., (see Appendix F (\citep{Dalton10b}), Eqs.(\ref{eq:StochAverThreeNoise1Mode})
and (\ref{eq:StochAverFourNoise1Mode}).

\subsubsection{Classical Field Equations}

\emph{Classical field} equations can be obtained from the Ito equations
by ignoring the quantum noise term. The classical field equations
are\begin{equation}
\frac{\partial\psi_{A}^{class}(x,t)}{\partial t}=-A_{A}(\underrightarrow{\psi}^{class}(x,t),x)\label{Eq.ClassicalFieldEqnGen}\end{equation}
for both the single and two mode condensate cases. Such equations
are not of course really classical as they involve Planck's constant.
As will be seen in specific cases (see Eq. (\ref{Eq.ClassicalFieldEqnSingleCondC1}))
their leading terms are often similar to Gross-Pitaevskii equations,
so they could be referred to as generalised mean field equations.

\subsubsection{Noise Fields for Single Mode Condensate }

Having now established the general results for the stochastic averages
of products of one, two, .. noise fields we can show for single mode
condensates that the noise field terms can be written in a different
form in which the noise fields are just functions of the stochastic
fields $\underrightarrow{\widetilde{\psi}}(x,t)$ and new fundamental
Gaussian-Markoff stochastic fields $\underrightarrow{\Theta}(x,\, t_{+})\equiv\{\Theta_{k}(x,t_{+})\}$,
pairs of which are delta function correlated in both space and time
\citep{Steel98b}. These now replace the $\underrightarrow{\Gamma}(t_{+})$.
Similarly to the $\underrightarrow{\Gamma}(t_{+})$ the $\underrightarrow{\Theta}(x,\, t_{+})$
are defined by their stochastic averages

\begin{eqnarray}
\overline{\Theta_{k}(x_{1},t_{1})} & = & 0\nonumber \\
\overline{\{\Theta_{k}(x_{1},t_{1})\Theta_{l}(x_{2},t_{2})\}} & = & \delta_{kl}\delta(t_{1}-t_{2})\nonumber \\
\overline{\{\Theta_{k}(x_{1},t_{1})\Theta_{l}(x_{2},t_{2})\Theta_{m}(x_{3},t_{3})\}} & = & 0\nonumber \\
\overline{\{\Theta_{k}(x_{1},t_{1})\Theta_{l}(x_{2},t_{2})\Theta_{m}(x_{3},t_{3})\Theta_{n}(x_{4},t_{4})\}} & = & \overline{\{\Theta_{k}(x_{1},t_{1})\Theta_{l}(x_{2},t_{2})\}}\;\overline{\{\Theta_{m}(x_{3},t_{3})\Theta_{n}(x_{4},t_{4})\}}\nonumber \\
 &  & +\overline{\{\Theta_{k}(x_{1},t_{1})\Theta_{m}(x_{3},t_{3})\}}\;\overline{\{\Theta_{l}(x_{2},t_{2})\Theta_{n}(x_{4},t_{4})\}}\nonumber \\
 &  & +\overline{\{\Theta_{k}(x_{1},t_{1})\Theta_{n}(x_{4},t_{24})\}}\;\overline{\{\Theta_{l}(x_{2},t_{2})\Theta_{m}(x_{3},t_{3})\}}\nonumber \\
 &  & ...\label{eq:GaussMarkStochasticFieldsGeneral}\end{eqnarray}
with stochastic averages being denoted with a bar. The stochastic
average of an odd number of noise field terms is always zero, whilst
that for an even number is the sum of all products of stochastic averages
of two noise field terms. Also,in Ito stochastic calculus the noise
terms $\Theta_{k}(x,t)$ within the interval $t,\, t+\lyxmathsym{\textgreek{d}}t$
are uncorrelated with any function of the $\underrightarrow{\widetilde{\psi}}(x,t)$
at the earlier time $t$, so that the stochastic average of the product
of such a function with a product of the noise field terms factorises\begin{eqnarray}
 &  & \overline{F(\underrightarrow{\widetilde{\psi}}(x_{1},t_{1}))\{\Theta_{k}(x_{2},t_{2})\Theta_{l}(x_{3},t_{3})\Theta_{m}(x_{4},t_{4})...\Theta_{a}(x_{l},t_{l})\}}\nonumber \\
 & = & \overline{F(\underrightarrow{\widetilde{\psi}}(x_{1},t_{1}))}\,\overline{\{\Theta_{k}(x_{2},t_{2})\Theta_{l}(x_{3},t_{3})\Theta_{m}(x_{4},t_{4})...\Theta_{a}(x_{l},t_{l})\}}\qquad t_{1}<t_{2},t_{3},...,t_{_{l}}\nonumber \\
 &  & \,\label{eq:StochAverProd2}\end{eqnarray}
As previously, the stochastic average of a sum is the sum of stochastic
averages. These key features of Ito stochastic calculus are important
in deriving the properties of the noise fields in the stochastic field
equations. In the case of the single mode condensate the diffusion
matrix is symmetric (\ref{Eq.SymmCondnSingleIntDiffusion}). Hence
we can write the diffusion matrix $D$ in the form $D(\underrightarrow{\psi}(x,t),x)=B(\underrightarrow{\psi}(x,t),x)B(\underrightarrow{\psi}(x,t),x)^{T}$
so that

\begin{equation}
D_{AB}(\underrightarrow{\psi}(x,t),x))=\sum_{k}B_{k}^{A}(\underrightarrow{\psi}(x,t),x))B_{k}^{B}(\underrightarrow{\psi}(x,t),x))\label{eq:DiffusionMatrixFactnSingleMode}\end{equation}
Note that in this case only a single space variable is involved. Now
consider the new stochastic noise field terms defined by\begin{equation}
\frac{\partial}{\partial t}\widetilde{H}_{A}(\underrightarrow{\widetilde{\psi}}(x,t),\underrightarrow{\Theta}(x,\, t_{+}))=\sum_{k}B_{k}^{A}(\underrightarrow{\widetilde{\psi}}(x,t),x)\,\Theta_{k}(x,t_{+})\label{eq:NewStochasticNoiseField}\end{equation}
This is a function of the stochastic fields $\underrightarrow{\lyxmathsym{\textgreek{y}}}(x,t)$
and the Gaussian-Markoff stochastic fields $\underrightarrow{\lyxmathsym{\textgreek{J}}}(x,t)$.
It is straightforward to determine results for the new stochastic
noise field terms. For the stochastic average of each noise term

\begin{equation}
\overline{\left(\frac{\partial}{\partial t}\widetilde{H}_{A}(\underrightarrow{\widetilde{\psi}}(x,t),\underrightarrow{\Theta}(x,t_{+})\right)}=0\label{eq:StochAverOneNoiseSingleMode}\end{equation}
showing that the stochastic average of of each new noise field is
zero as before. For the stochastic average of the product of two new
noise field terms we have

\begin{eqnarray}
 &  & \overline{\left(\frac{\partial}{\partial t}\widetilde{H}_{A}(\underrightarrow{\widetilde{\psi}}(x_{1},t_{1}),\underrightarrow{\Theta}(x_{1},t_{1+}))\right)\left(\frac{\partial}{\partial t}\widetilde{H}_{B}(\underrightarrow{\widetilde{\psi}}(x_{2},t_{2}),\underrightarrow{\Theta}(x_{2},t_{2+}))\right)}\nonumber \\
 & = & \overline{D_{AB}(\underrightarrow{\widetilde{\psi}}(x_{1,2},t_{1,2}),x_{1,2})}\times\delta(x_{1}-x_{2})\delta(t_{1}-t_{2})\label{eq:StochAverTwoNoiseSingleMode}\end{eqnarray}
giving the same result as before. For products of three, four, ..
new noise field terms the results are again as before, so we can now
write the original noise field term as\begin{align}
\frac{\partial}{\partial t}\widetilde{G}_{A}(\underrightarrow{\widetilde{\psi}}(x,t),\underrightarrow{\Gamma}(t_{+})) & =\frac{\partial}{\partial t}\widetilde{H}_{A}(\underrightarrow{\widetilde{\psi}}(x,t),\underrightarrow{\Theta}(x,\, t_{+}))\nonumber \\
 & =\sum_{k}B_{k}^{A}(\underrightarrow{\widetilde{\psi}}(x,t),x)\,\Theta_{k}(x,t_{+})\label{eq:NoiseFieldSingleModeCond}\end{align}
This form of the noise field is useful when the diffusion matrix $D(\underrightarrow{\psi}(x,t),x)$is
easily factorised, as in Section \ref{sub:ApproxSolnSingleMode}.

\subsection{Ito Equations for Two-Mode Condensate}

The theory involved in writing down Ito stochastic equation for the
condensate and non-condensate fields is non-standard. From above,
the terms can be written down from the general form (\ref{Eq.ItoStoFdEqnGeneral})
by identifying the relevant terms in the functional Fokker-Planck
equations set out in Section \ref{Sect Func Fokker Planck}. All stochastic
fields depend on $t$, but this is left implicit.

For the \emph{condensate stochastic field} the Ito equation is\begin{eqnarray}
 &  & \frac{\partial}{\partial t}\widetilde{\psi}_{C}(\mathbf{s,}t)\nonumber \\
 & = & -\frac{i}{\hbar}[-\frac{\hbar^{2}}{2m}\nabla^{2}\widetilde{\psi}_{C}(\mathbf{s})+V(\mathbf{s})\widetilde{\psi}_{C}(\mathbf{s})+\frac{g_{N}}{N}\{\widetilde{\psi}_{C}^{+}(\mathbf{s})\widetilde{\psi}_{C}(\mathbf{s})-\left\vert \phi_{1}(\mathbf{s})\right\vert ^{2}-\left\vert \phi_{2}(\mathbf{s})\right\vert ^{2}\}\widetilde{\psi}_{C}(\mathbf{s})\nonumber \\
 &  & +\frac{g_{N}}{N}\mathbf{\{}2\widetilde{\psi}_{C}^{+}\mathbf{(\mathbf{s})}\widetilde{\psi}_{C}\mathbf{(\mathbf{s})-}\left\vert \phi_{1}(\mathbf{s})\right\vert ^{2}\mathbf{-}\left\vert \phi_{2}(\mathbf{s})\right\vert ^{2}\mathbf{\}}\widetilde{\psi}_{NC}\mathbf{(\mathbf{s})-}\frac{g_{N}}{N}\int d\mathbf{u\,}F(\mathbf{u},\mathbf{s})^{\ast}\widetilde{\psi}_{NC}(\mathbf{u})\nonumber \\
 &  & +\frac{g_{N}}{N}\{\widetilde{\psi}_{C}(\mathbf{s})\widetilde{\psi}_{C}(\mathbf{s})\}\widetilde{\psi}_{NC}^{+}(\mathbf{s})\nonumber \\
 &  & +\frac{g_{N}}{N}\{2\widetilde{\psi}_{NC}^{+}(\mathbf{s})\widetilde{\psi}_{NC}(\mathbf{s})\}\widetilde{\psi}_{C}(\mathbf{s})+\frac{g_{N}}{N}\{\widetilde{\psi}_{NC}(\mathbf{s})\widetilde{\psi}_{NC}(\mathbf{s})\}\widetilde{\psi}_{C}^{+}(\mathbf{s})]\nonumber \\
 &  & +\frac{\partial}{\partial t}\widetilde{G}_{C}(\underrightarrow{\widetilde{\psi}}(\mathbf{s},t),\underrightarrow{\Gamma}(t_{+}))\label{Eq.ItoBogoliubovCondensateFieldTwoMode}\end{eqnarray}
where $\frac{{\LARGE\partial}}{{\LARGE\partial t}}\widetilde{G}_{C}(\underrightarrow{\widetilde{\psi}}(\mathbf{s},t),\underrightarrow{\Gamma}(t_{+}))$
is the \emph{noise field}.

For the \emph{non-condensate stochastic field }the Ito equation is\begin{eqnarray}
 &  & \frac{\partial}{\partial t}\widetilde{\psi}_{NC}(\mathbf{s,}t)\nonumber \\
 & = & -\frac{i}{\hbar}[+\frac{g_{N}}{N}\{\widetilde{\psi}_{C}^{+}(\mathbf{s})\widetilde{\psi}_{C}(\mathbf{s})-\left\vert \phi_{1}(\mathbf{s})\right\vert ^{2}-\left\vert \phi_{2}(\mathbf{s})\right\vert ^{2}\}\widetilde{\psi}_{C}(\mathbf{s})\mathbf{-}\frac{g_{N}}{N}\int d\mathbf{u\,}F(\mathbf{s},\mathbf{u})\widetilde{\psi}_{C}(\mathbf{u})\nonumber \\
 &  & -\frac{\hbar^{2}}{2m}\nabla^{2}\widetilde{\psi}_{NC}(\mathbf{s})+V(\mathbf{s})\widetilde{\psi}_{NC}(\mathbf{s})+\frac{g_{N}}{N}\{2\widetilde{\psi}_{C}^{+}(\mathbf{s})\widetilde{\psi}_{C}(\mathbf{s})-\left\vert \phi_{1}(\mathbf{s})\right\vert ^{2}-\left\vert \phi_{2}(\mathbf{s})\right\vert ^{2}\}\widetilde{\psi}_{NC}(\mathbf{s})\nonumber \\
 &  & +\frac{g_{N}}{N}\{\widetilde{\psi}_{C}(\mathbf{s})\widetilde{\psi}_{C}(\mathbf{s})\}\widetilde{\psi}_{NC}^{+}(\mathbf{s})]\nonumber \\
 &  & +\frac{\partial}{\partial t}\widetilde{G}_{NC}(\underrightarrow{\widetilde{\psi}}(\mathbf{s},t),\underrightarrow{\Gamma}(t_{+}))\label{Eq.ItoBogoliubovNonCondensateFieldTwoMode}\end{eqnarray}
where $\frac{{\LARGE\partial}}{{\LARGE\partial t}}\widetilde{G}_{NC}(\underrightarrow{\widetilde{\psi}}(\mathbf{s},t),\underrightarrow{\Gamma}(t_{+}))$
is the \emph{noise field}. Similar equations apply for $\widetilde{\psi}_{C}^{+}\mathbf{(\mathbf{s})}$
and $\widetilde{\psi}_{NC}^{+}\mathbf{(\mathbf{s})}$. The stochastic
condensate and non-condensate fields are \emph{coupled} together and
each is affected by \emph{stochastic noise fields}. For the condensate
field, the first line in the equation reads like a \emph{time-dependent
Gross-Pitaevskii equation} if $\widetilde{\psi}_{C}(\mathbf{s,}t)$
is regarded as the \emph{order function}. The three terms are the
kinetic energy, the trap potential energy and the non-linear mean
field energy contributions. Note that for the condensate equation
the condensate density $\widetilde{\psi}_{C}^{+}(\mathbf{s})\widetilde{\psi}_{C}(\mathbf{s})$
is depleted by two bosons due to the $\left\vert \phi_{1}(\mathbf{s})\right\vert ^{2}$
and $\left\vert \phi_{2}(\mathbf{s})\right\vert ^{2}$ terms. Both
Ito stochastic equations are integro-differential equations due to
the terms involving $\int d\mathbf{u\,}F(\mathbf{s},\mathbf{u})$
or $\int d\mathbf{u\,}F(\mathbf{u},\mathbf{s})^{\ast}$ - thus on
the right side there are terms depending on stochastic fields at different
spatial points. The first line in the condensate equation comes from
the $\widehat{H}_{1}$ term, the second and third from the $\widehat{H}_{2}$
term and the fourth from the $\widehat{H}_{3}$ term. The first line
in the non-condensate equation is a term coupling in the condensate
field and comes from the $\widehat{H}_{2}$ term, the second and third
from the $\widehat{H}_{3}$ term. The latter two lines differ somewhat
from the form of a time-dependent Gross-Pitaevskii equation, which
is not surprising since these refer to the relatively unoccupied non-condensate
modes.

The \emph{stochastic averages} of the noise fields are given in (\ref{eq:StochAverNoiseTwoMode-1}),
where the non-zero \emph{diffusion matrix elements} are\begin{eqnarray}
 &  & D_{C+;NC-}(\underrightarrow{\widetilde{\psi}}(\mathbf{s}_{1},t),\mathbf{s}_{1},\underrightarrow{\widetilde{\psi}}(\mathbf{s}_{2},t),\mathbf{s}_{2})\nonumber \\
 & = & +\frac{i}{\hbar}\frac{g_{N}}{N}\{\widetilde{\psi}_{C}^{+}(\mathbf{s}_{1,2})\widetilde{\psi}_{C}(\mathbf{s}_{1,2})-\frac{{\small1}}{{\small2}}(\left\vert \phi_{1}(\mathbf{s}_{1,2})\right\vert ^{2}+\left\vert \phi_{2}(\mathbf{s}_{1,2})\right\vert ^{2})\}\delta(\mathbf{s}_{1}-\mathbf{s}_{2})\nonumber \\
 &  & +\frac{i}{\hbar}\frac{g_{N}}{N}\{\widetilde{\psi}_{NC}^{+}(\mathbf{s}_{1,2})\widetilde{\psi}_{C}(\mathbf{s}_{1,2})+\widetilde{\psi}_{C}^{+}(\mathbf{s}_{1,2})\widetilde{\psi}_{NC}(\mathbf{s}_{1,2})\}\delta(\mathbf{s}_{1}-\mathbf{s}_{2})\nonumber \\
 &  & -\frac{i}{\hbar}\frac{g_{N}}{N}\{\frac{{\small1}}{{\small2}}F(\mathbf{s}_{2},\mathbf{s}_{1})\}\nonumber \\
 & = & D_{NC-;C+}(\underrightarrow{\widetilde{\psi}}(\mathbf{s}_{2},t),\mathbf{s}_{2},\underrightarrow{\widetilde{\psi}}(\mathbf{s}_{1},t),\mathbf{s}_{1})\\
 &  & D_{C-;NC+}(\underrightarrow{\widetilde{\psi}}(\mathbf{s}_{1},t),\mathbf{s}_{1},\underrightarrow{\widetilde{\psi}}(\mathbf{s}_{2},t),\mathbf{s}_{2})\nonumber \\
 & = & -\frac{i}{\hbar}\frac{g_{N}}{N}\{\widetilde{\psi}_{C}(\mathbf{s}_{1,2})\widetilde{\psi}_{C}^{+}(\mathbf{s}_{1,2})-\frac{{\small1}}{{\small2}}(\left\vert \phi_{1}(\mathbf{s}_{1,2})\right\vert ^{2}+\left\vert \phi_{2}(\mathbf{s}_{1,2})\right\vert ^{2})\}\delta(\mathbf{s}_{1}-\mathbf{s}_{2})\nonumber \\
 &  & -\frac{i}{\hbar}\frac{g_{N}}{N}\{\widetilde{\psi}_{NC}(\mathbf{s}_{1,2})\widetilde{\psi}_{C}^{+}(\mathbf{s}_{1,2})+\widetilde{\psi}_{C}(\mathbf{s}_{1,2})\widetilde{\psi}_{NC}^{+}(\mathbf{s}_{1,2})\}\delta(\mathbf{s}_{1}-\mathbf{s}_{2})\nonumber \\
 &  & +\frac{i}{\hbar}\frac{g_{N}}{N}\{\frac{{\small1}}{{\small2}}F(\mathbf{s}_{2},\mathbf{s}_{1})^{\ast}\}\nonumber \\
 & = & D_{NC+;C-}(\underrightarrow{\widetilde{\psi}}(\mathbf{s}_{2},t),\mathbf{s}_{2},\underrightarrow{\widetilde{\psi}}(\mathbf{s}_{1},t),\mathbf{s}_{1})\\
 &  & D_{C-;NC-}(\underrightarrow{\widetilde{\psi}}(\mathbf{s}_{1},t),\mathbf{s}_{1},\underrightarrow{\widetilde{\psi}}(\mathbf{s}_{2},t),\mathbf{s}_{2})\nonumber \\
 & = & -\frac{i}{\hbar}\frac{g_{N}}{N}\{\frac{{\small1}}{{\small2}}\widetilde{\psi}_{C}(\mathbf{s}_{1,2})\widetilde{\psi}_{C}(\mathbf{s}_{1,2})+\widetilde{\psi}_{NC}(\mathbf{s}_{1,2})\widetilde{\psi}_{C}(\mathbf{s}_{1,2})\}\delta(\mathbf{s}_{1}-\mathbf{s}_{2})\nonumber \\
 & = & D_{NC-;C-}(\underrightarrow{\widetilde{\psi}}(\mathbf{s}_{2},t),\mathbf{s}_{2},\underrightarrow{\widetilde{\psi}}(\mathbf{s}_{1},t),\mathbf{s}_{1})\\
 &  & D_{C+;NC+}(\underrightarrow{\widetilde{\psi}}(\mathbf{s}_{1},t),\mathbf{s}_{1},\underrightarrow{\widetilde{\psi}}(\mathbf{s}_{2},t),\mathbf{s}_{2})\nonumber \\
 & = & +\frac{i}{\hbar}\frac{g_{N}}{N}\{\frac{{\small1}}{{\small2}}\widetilde{\psi}_{C}^{+}(\mathbf{s}_{1,2})\widetilde{\psi}_{C}^{+}(\mathbf{s}_{1,2})+\widetilde{\psi}_{NC}^{+}(\mathbf{s}_{1,2})\widetilde{\psi}_{C}^{+}(\mathbf{s}_{1,2})\}\delta(\mathbf{s}_{1}-\mathbf{s}_{2})\nonumber \\
 & = & D_{NC+;C+}(\underrightarrow{\widetilde{\psi}}(\mathbf{s}_{2},t),\mathbf{s}_{2},\underrightarrow{\widetilde{\psi}}(\mathbf{s}_{1},t),\mathbf{s}_{1})\\
 &  & D_{NC-;NC-}(\underrightarrow{\widetilde{\psi}}(\mathbf{s}_{1},t),\mathbf{s}_{1},\underrightarrow{\widetilde{\psi}}(\mathbf{s}_{2},t),\mathbf{s}_{2})\nonumber \\
 & = & -\frac{i}{\hbar}\frac{g_{N}}{N}\{\widetilde{\psi}_{C}(\mathbf{s}_{1,2})\widetilde{\psi}_{C}(\mathbf{s}_{1,2}\mathbf{)\}}\delta(\mathbf{s}_{1}-\mathbf{s}_{2})\\
 &  & D_{NC+;NC+}(\underrightarrow{\widetilde{\psi}}(\mathbf{s}_{1},t),\mathbf{s}_{1},\underrightarrow{\widetilde{\psi}}(\mathbf{s}_{2},t),\mathbf{s}_{2})\nonumber \\
 & = & +\frac{i}{\hbar}\frac{g_{N}}{N}\{\widetilde{\psi}_{C}^{+}(\mathbf{s}_{1,2})\widetilde{\psi}_{C}^{+}(\mathbf{s}_{1,2}\mathbf{)\}}\delta(\mathbf{s}_{1}-\mathbf{s}_{2})\label{Eq.DiffusionResultTwoModeCondGeneral}\end{eqnarray}
with the notation $D_{AB}(\underrightarrow{\widetilde{\psi}}(\mathbf{s}_{1},t),\mathbf{s}_{1},\underrightarrow{\widetilde{\psi}}(\mathbf{s}_{2},t),\mathbf{s}_{2})$
for\\
 $D_{AB}(\widetilde{\psi}_{1}(\mathbf{s}_{1},t),\widetilde{\psi}_{2}(\mathbf{s}_{1},t),\widetilde{\psi}_{3}(\mathbf{s}_{1},t),\widetilde{\psi}_{4}(\mathbf{s}_{1},t),\mathbf{s}_{1}\mathbf{,}\widetilde{\psi}_{1}(\mathbf{s}_{2},t),\widetilde{\psi}_{2}(\mathbf{s}_{2},t),\widetilde{\psi}_{3}(\mathbf{s}_{2},t),\widetilde{\psi}_{4}(\mathbf{s}_{2},t),\mathbf{s}_{2})$
and replacements for $AB$ as follows: $1\equiv C-,2\equiv C+,3\equiv NC-,4\equiv NC+$.
Also we write $\mathbf{s}_{1,2}=\mathbf{s}_{1}=\mathbf{s}_{2}$ for
the delta function \ terms. The presence of the terms $F(\mathbf{s}_{2},\mathbf{s}_{1}),F(\mathbf{s}_{2},\mathbf{s}_{1})^{\ast}$
reflects the non-local nature of the diffusion matrix and also give
an explicit $\mathbf{s}_{1},\mathbf{s}_{2}$ dependence. We see that
the average of the product of any pair of noise fields is delta function
correlated in time but not in space, and is then given by the diffusion
matrix element that appears in the functional Fokker-Planck equation.
The stochastic averages of products of odd numbers of noise fields
is zero and the stochastic averages of products of even numbers of
noise fields can be written as sums of products of stochastic averages
of pairs of diffusion matrix elements in accordance with (\ref{Eq.StochAverThreeNoiseTwoMode})
and (\ref{eq:StochAverFourNoiseTwoMode}).

The \emph{classical field equations} for the condensate field are\begin{eqnarray}
 &  & \frac{\partial}{\partial t}\psi_{C}^{class}(\mathbf{s,}t)\nonumber \\
 & = & -\frac{i}{\hbar}[-\frac{\hbar^{2}}{2m}\nabla^{2}\psi_{C}(\mathbf{s})+V(\mathbf{s})\psi_{C}(\mathbf{s})+\frac{g_{N}}{N}\{\psi_{C}^{+}(\mathbf{s})\psi_{C}(\mathbf{s})-\left\vert \phi_{1}(\mathbf{s})\right\vert ^{2}-\left\vert \phi_{2}(\mathbf{s})\right\vert ^{2}\}\psi_{C}(\mathbf{s})\nonumber \\
 &  & +\frac{g_{N}}{N}\mathbf{\{}2\psi_{C}^{+}\mathbf{(\mathbf{s})}\psi_{C}\mathbf{(\mathbf{s})\mathbf{-}}\left\vert \phi_{1}(\mathbf{s})\right\vert ^{2}-\left\vert \phi_{2}(\mathbf{s})\right\vert ^{2}\mathbf{\}}\psi_{NC}\mathbf{(\mathbf{s})-}\frac{g_{N}}{N}\int d\mathbf{u\,}F(\mathbf{u},\mathbf{s})^{\ast}\psi_{NC}(\mathbf{u})\nonumber \\
 &  & +\frac{g_{N}}{N}\{\psi_{C}(\mathbf{s})\psi_{C}(\mathbf{s})\}\psi_{NC}^{+}(\mathbf{s})\nonumber \\
 &  & +\frac{g_{N}}{N}\{2\psi_{NC}^{+}(\mathbf{s})\psi_{NC}(\mathbf{s})\}\psi_{C}(\mathbf{s})+\frac{g_{N}}{N}\{\psi_{NC}(\mathbf{s})\psi_{NC}(\mathbf{s})\}\psi_{C}^{+}(\mathbf{s})]\nonumber \\
 &  & \,\label{eq:ClassicalFieldEqnCondTwoMode}\end{eqnarray}
and for the non-condensate stochastic field\begin{eqnarray}
 &  & \frac{\partial}{\partial t}\psi_{NC}^{class}(\mathbf{s,}t)\nonumber \\
 & = & -\frac{i}{\hbar}[+\frac{g_{N}}{N}\{\psi_{C}^{+}(\mathbf{s})\psi_{C}(\mathbf{s})-\left\vert \phi_{1}(\mathbf{s})\right\vert ^{2}-\left\vert \phi_{2}(\mathbf{s})\right\vert ^{2}\}\psi_{C}(\mathbf{s})\mathbf{-}\frac{g_{N}}{N}\int d\mathbf{u\,}F(\mathbf{s},\mathbf{u})\psi_{C}(\mathbf{u})\nonumber \\
 &  & -\frac{\hbar^{2}}{2m}\nabla^{2}\psi_{NC}(\mathbf{s})+V(\mathbf{s})\psi_{NC}(\mathbf{s})+\frac{g_{N}}{N}\{2\psi_{C}^{+}(\mathbf{s})\psi_{C}(\mathbf{s})-\left\vert \phi_{1}(\mathbf{s})\right\vert ^{2}-\left\vert \phi_{2}(\mathbf{s})\right\vert ^{2}\}\psi_{NC}(\mathbf{s})\nonumber \\
 &  & +\frac{g_{N}}{N}\{\psi_{C}(\mathbf{s})\psi_{C}(\mathbf{s})\}\psi_{NC}^{+}(\mathbf{s})]\nonumber \\
 &  & \,\label{eq:ClassicalFieldEqnNonCondTwoMode}\end{eqnarray}
with corresponding equations for $\psi_{C}^{+\; class}$ and $\psi_{NC}^{+\; class}$.
These also are integro-differential equations.

\subsection{Ito Equations for Single Mode Condensate}

We will next consider the simpler case where the BEC only involves
a single mode. Here the Ito stochastic equations are relatively standard.
From above, the terms can be written down from the general form (\ref{Eq.ItoStoFdEqnGeneral})
by identifying the relevant terms in the functional Fokker-Planck
equations set out in Section \ref{Sect Func Fokker Planck}. All stochastic
fields depend on $t$, but this is left implicit.

For the \emph{condensate stochastic field} the Ito stochastic equation
is\begin{eqnarray}
 &  & \frac{\partial}{\partial t}\widetilde{\psi}_{C}(\mathbf{s,}t)\nonumber \\
 & = & -\frac{i}{\hbar}[-\frac{\hbar^{2}}{2m}\nabla^{2}\widetilde{\psi}_{C}(\mathbf{s})+V(\mathbf{s})\widetilde{\psi}_{C}(\mathbf{s})+\frac{g_{N}}{N}\{\widetilde{\psi}_{C}^{+}(\mathbf{s})\widetilde{\psi}_{C}(\mathbf{s})-\left\vert \phi_{1}(\mathbf{s})\right\vert ^{2}\}\widetilde{\psi}_{C}(\mathbf{s})\nonumber \\
 &  & +\frac{g_{N}}{N}\mathbf{\{}2\widetilde{\psi}_{C}^{+}\mathbf{(\mathbf{s})}\widetilde{\psi}_{C}\mathbf{(\mathbf{s})-}N{\small\,}\left\vert \phi_{1}(\mathbf{s})\right\vert ^{2}\mathbf{\}}\widetilde{\psi}_{NC}\mathbf{(\mathbf{s})}+\frac{g_{N}}{N}\{\widetilde{\psi}_{C}(\mathbf{s})\widetilde{\psi}_{C}(\mathbf{s})\}\widetilde{\psi}_{NC}^{+}(\mathbf{s})\nonumber \\
 &  & +\frac{g_{N}}{N}\{2\widetilde{\psi}_{NC}^{+}(\mathbf{s})\widetilde{\psi}_{NC}(\mathbf{s})\}\widetilde{\psi}_{C}(\mathbf{s})+\frac{g_{N}}{N}\{\widetilde{\psi}_{NC}(\mathbf{s})\widetilde{\psi}_{NC}(\mathbf{s})\}\widetilde{\psi}_{C}^{+}(\mathbf{s})]\nonumber \\
 &  & +\frac{\partial}{\partial t}\widetilde{G}_{C}(\underrightarrow{\widetilde{\psi}}(\mathbf{s},t),\underrightarrow{\Gamma}(t_{+}))\label{Eq.ItoBogoliubovCondensateFieldOneMode}\end{eqnarray}
where $\frac{{\LARGE\partial}}{{\LARGE\partial t}}\widetilde{G}_{C}(\underrightarrow{\widetilde{\psi}}(\mathbf{s},t),\underrightarrow{\Gamma}(t_{+}))$
is the \emph{noise field}.

For the \emph{non-condensate stochastic field }the Ito stochastic
equation is\begin{eqnarray}
 &  & \frac{\partial}{\partial t}\widetilde{\psi}_{NC}(\mathbf{s,}t)\nonumber \\
 & = & -\frac{i}{\hbar}[+\frac{g_{N}}{N}\{\widetilde{\psi}_{C}^{+}(\mathbf{s})\widetilde{\psi}_{C}(\mathbf{s})-{\small N}\,\left\vert \phi_{1}(\mathbf{s})\right\vert ^{2}\}\widetilde{\psi}_{C}(\mathbf{s})\nonumber \\
 &  & -\frac{\hbar^{2}}{2m}\nabla^{2}\widetilde{\psi}_{NC}(\mathbf{s})+V(\mathbf{s})\widetilde{\psi}_{NC}(\mathbf{s})+\frac{g_{N}}{N}\{2\widetilde{\psi}_{C}^{+}(\mathbf{s})\widetilde{\psi}_{C}(\mathbf{s})-\left\vert \phi_{1}(\mathbf{s})\right\vert ^{2}\}\widetilde{\psi}_{NC}(\mathbf{s})\nonumber \\
 &  & +\frac{g_{N}}{N}\{\widetilde{\psi}_{C}(\mathbf{s})\widetilde{\psi}_{C}(\mathbf{s})\}\widetilde{\psi}_{NC}^{+}(\mathbf{s})]\nonumber \\
 &  & +\frac{\partial}{\partial t}\widetilde{G}_{NC}(\underrightarrow{\widetilde{\psi}}(\mathbf{s},t),\underrightarrow{\Gamma}(t_{+}))\label{Eq.ItoBogoliubovNonCondensateFieldOneMode}\end{eqnarray}
where $\frac{{\LARGE\partial}}{{\LARGE\partial t}}\widetilde{G}_{NC}(\underrightarrow{\widetilde{\psi}}(\mathbf{s},t),\underrightarrow{\Gamma}(t_{+}))$
is the \emph{noise field}. Similar equations apply for $\widetilde{\psi}_{C}^{+}\mathbf{(\mathbf{s})}$
and $\widetilde{\psi}_{NC}^{+}\mathbf{(\mathbf{s})}$. The stochastic
condensate and non-condensate fields are \emph{coupled} together and
each is affected by \emph{stochastic noise fields}. For the condensate
field, the first line in the equation reads like a \emph{time-dependent
Gross-Pitaevskii equation} if $\widetilde{\psi}_{C}(\mathbf{s,}t)$
is regarded as the \emph{order function}. The three terms are the
kinetic energy, the trap potential energy and the non-linear mean
field energy contributions. Note that for the condensate equation
the condensate density $\widetilde{\psi}_{C}^{+}(\mathbf{s})\widetilde{\psi}_{C}(\mathbf{s})$
is depleted by one boson due to the $\left\vert \phi_{1}(\mathbf{s})\right\vert ^{2}$
term. The first line in the condensate equation comes from the $\widehat{H}_{1}$
term, the second from the $\widehat{H}_{2}$ term and the third from
the $\widehat{H}_{3}$ term. The first line in the non-condensate
equation is a term coupling in the condensate field and comes from
the $\widehat{H}_{2}$ term, the second and third from the $\widehat{H}_{3}$
term. The latter two lines differ somewhat from the form of a time-dependent
Gross-Pitaevskii equation, which is not surprising since these refer
to the relatively unoccupied non-condensate modes.

The \emph{stochastic averages} of the noise fields are given in (\ref{eq:StochAverNoiseSingleMode-1}),
where the non-zero \emph{diffusion matrix elements} are\begin{eqnarray}
D_{C+;NC-}(\underrightarrow{\widetilde{\psi}}(\mathbf{s},t),\mathbf{s}) & = & \frac{i}{\hbar}\frac{g_{N}}{N}\{\widetilde{\psi}_{C}^{+}(\mathbf{s})\widetilde{\psi}_{C}(\mathbf{s})-\frac{{\small1}}{{\small2}}N\,\left\vert \phi_{1}(\mathbf{s})\right\vert ^{2}\}\nonumber \\
 &  & +\frac{i}{\hbar}\frac{g_{N}}{N}\{\widetilde{\psi}_{NC}^{+}(\mathbf{s})\widetilde{\psi}_{C}(\mathbf{s})+\widetilde{\psi}_{C}^{+}(\mathbf{s})\widetilde{\psi}_{NC}(\mathbf{s})\}\nonumber \\
 & = & D_{NC-;C+}(\underrightarrow{\widetilde{\psi}}(\mathbf{s},t),\mathbf{s})\\
D_{C-;NC+}(\underrightarrow{\widetilde{\psi}}(\mathbf{s},t),\mathbf{s}) & = & -\frac{i}{\hbar}\frac{g_{N}}{N}\{\widetilde{\psi}_{C}(\mathbf{s})\widetilde{\psi}_{C}^{+}(\mathbf{s})-\frac{{\small1}}{{\small2}}N\,\left\vert \phi_{1}(\mathbf{s})\right\vert ^{2}\}\nonumber \\
 &  & -\frac{i}{\hbar}\frac{g_{N}}{N}\{\widetilde{\psi}_{NC}(\mathbf{s})\widetilde{\psi}_{C}^{+}(\mathbf{s})+\widetilde{\psi}_{C}(\mathbf{s})\widetilde{\psi}_{NC}^{+}(\mathbf{s})\}\nonumber \\
 & = & D_{NC+;C-}(\underrightarrow{\widetilde{\psi}}(\mathbf{s},t),\mathbf{s})\\
D_{C-;NC-}(\underrightarrow{\widetilde{\psi}}(\mathbf{s},t),\mathbf{s}) & = & -\frac{i}{\hbar}\frac{g_{N}}{N}\{\frac{{\small1}}{{\small2}}\widetilde{\psi}_{C}(\mathbf{s})\widetilde{\psi}_{C}(\mathbf{s})+\widetilde{\psi}_{NC}(\mathbf{s})\widetilde{\psi}_{C}(\mathbf{s})\}\nonumber \\
 & = & D_{NC-;C-}(\underrightarrow{\widetilde{\psi}}(\mathbf{s},t),\mathbf{s})\\
D_{C+;NC+}(\underrightarrow{\widetilde{\psi}}(\mathbf{s},t),\mathbf{s}) & = & +\frac{i}{\hbar}\frac{g_{N}}{N}\{\frac{{\small1}}{{\small2}}\widetilde{\psi}_{C}^{+}(\mathbf{s})\widetilde{\psi}_{C}^{+}(\mathbf{s})+\widetilde{\psi}_{NC}^{+}(\mathbf{s})\widetilde{\psi}_{C}^{+}(\mathbf{s})\}\nonumber \\
 & = & D_{NC+;C+}(\underrightarrow{\widetilde{\psi}}(\mathbf{s},t),\mathbf{s})\\
D_{NC-;NC-}(\underrightarrow{\widetilde{\psi}}(\mathbf{s},t),\mathbf{s}) & = & -\frac{i}{\hbar}\frac{g_{N}}{N}\{\widetilde{\psi}_{C}(\mathbf{s})\widetilde{\psi}_{C}(\mathbf{s)\}}\\
D_{NC+;NC+}(\underrightarrow{\widetilde{\psi}}(\mathbf{s},t),\mathbf{s}) & = & +\frac{i}{\hbar}\frac{g_{N}}{N}\{\widetilde{\psi}_{C}^{+}(\mathbf{s})\widetilde{\psi}_{C}^{+}(\mathbf{s)\}}\label{Eq.DiffusionResultSingleCondGen}\end{eqnarray}
with the notation $D_{AB}(\underrightarrow{\widetilde{\psi}}(\mathbf{s},t),\mathbf{s})$
for\\
 $D_{AB}(\widetilde{\psi}_{1}(\mathbf{s},t),\widetilde{\psi}_{2}(\mathbf{s},t),\widetilde{\psi}_{3}(\mathbf{s},t),\widetilde{\psi}_{4}(\mathbf{s},t),\mathbf{s})$
and replacements for $AB$ as follows: $1\equiv C-,2\equiv C+,3\equiv NC-,4\equiv NC+$.
Note that the $\left\vert \phi_{1}(\mathbf{s})\right\vert ^{2}$ terms
give an explicit $\mathbf{s}$ dependence as well as that in the stochastic
fields. We see that the average of the product of any pair of noise
fields is delta function correlated in both space and time, and is
then given by the diffusion matrix element that appears in the functional
Fokker-Planck equation. The stochastic averages of products of odd
numbers of noise fields is zero and the stochastic averages of products
of even numbers of noise fields can be written as sums of products
of stochastic averages of pairs of diffusion matrix elements analogous
to (\ref{Eq.StochAverThreeNoiseTwoMode}) and (\ref{eq:StochAverFourNoiseTwoMode})
(see \ref{sub:Properties of Noise Fields}). 

The \emph{classical field equations} for the condensate field are\begin{eqnarray}
 &  & \frac{\partial}{\partial t}\psi_{C}^{class}(\mathbf{s,}t)\nonumber \\
 & = & -\frac{i}{\hbar}[-\frac{\hbar^{2}}{2m}\nabla^{2}\psi_{C}(\mathbf{s})+V(\mathbf{s})\psi_{C}(\mathbf{s})+\frac{g_{N}}{N}\{\psi_{C}^{+}(\mathbf{s})\psi_{C}(\mathbf{s})-\left\vert \phi_{1}(\mathbf{s})\right\vert ^{2}\}\psi_{C}(\mathbf{s})\nonumber \\
 &  & +\frac{g_{N}}{N}\mathbf{\{}2\psi_{C}^{+}(\mathbf{s})\psi_{C}(\mathbf{s})\mathbf{-}N{\small\,}\left\vert \phi_{1}(\mathbf{s})\right\vert ^{2}\mathbf{\}}\psi_{NC}(\mathbf{s})+\frac{g_{N}}{N}\{\psi_{C}(\mathbf{s})\psi_{C}(\mathbf{s})\}\psi_{NC}^{+}(\mathbf{s})\nonumber \\
 &  & +\frac{g_{N}}{N}\{2\psi_{NC}^{+}(\mathbf{s})\psi_{NC}(\mathbf{s})\}\psi_{C}(\mathbf{s})+\frac{g_{N}}{N}\{\psi_{NC}(\mathbf{s})\psi_{NC}(\mathbf{s})\}\psi_{C}^{+}(\mathbf{s})]\label{Eq.ClassicalFieldEqnSingleCondC1}\end{eqnarray}
and for the non-condensate stochastic field\begin{eqnarray}
 &  & \frac{\partial}{\partial t}\psi_{NC}^{class}(\mathbf{s,}t)\nonumber \\
 & = & -\frac{i}{\hbar}[+\frac{g_{N}}{N}\{\psi_{C}^{+}(\mathbf{s})\psi_{C}(\mathbf{s})-{\small N}\,\left\vert \phi_{1}(\mathbf{s})\right\vert ^{2}\}\psi_{C}(\mathbf{s})\nonumber \\
 &  & -\frac{\hbar^{2}}{2m}\nabla^{2}\psi_{NC}(\mathbf{s})+V(\mathbf{s})\psi_{NC}(\mathbf{s})+\frac{g_{N}}{N}\{2\psi_{C}^{+}(\mathbf{s})\psi_{C}(\mathbf{s})-\left\vert \phi_{1}(\mathbf{s})\right\vert ^{2}\}\psi_{NC}(\mathbf{s})\nonumber \\
 &  & +\frac{g_{N}}{N}\{\psi_{C}(\mathbf{s})\psi_{C}(\mathbf{s})\}\psi_{NC}^{+}(\mathbf{s})]\label{Eq.ClassicalFieldEqnSingleCondNC1}\end{eqnarray}
with corresponding equations for $\psi_{C}^{+\; class}$ and $\psi_{NC}^{+\; class}$.
If the coupling terms to the non-condensate modes are ignored then
the equation for $\psi_{C}^{class}(\mathbf{s,}t)$ has a solution
$\psi_{C}^{class}(\mathbf{s,}t)=\sqrt{N}\,\phi_{1}(\mathbf{s}),\psi_{C}^{+\; class}(\mathbf{s,}t)=\sqrt{N}\,\phi_{1}^{\ast}(\mathbf{s})$
for large $N$ , where $\phi_{1}(\mathbf{s})$ satisfies the standard
single mode Gross-Pitaevskii equation (\ref{Eq.Gross-Pitaevskii}).
Assuming the effects of coupling with the non-condensate field are
small, this result shows that $\psi_{C}^{class}(\mathbf{s,}t)$ is
similar to the usual mean field solution.

\subsection*{5.4 Approximate Solutions - Single Mode Condensate}

\label{sub:ApproxSolnSingleMode}

In general the coupled stochastic field equations are difficult to
solve, even numerically. Approximate solutions can however be obtained
which enable some features of the physics to be explored. As an illustration
of how such approximate solutions can be obtained we consider the
single mode condensate case for large $N$. By applying certain approximations
to (\ref{Eq.ItoBogoliubovCondensateFieldOneMode}) - (\ref{Eq.ClassicalFieldEqnSingleCondNC1})
the equations obtained by Krachmalnicoff et al. \citep{Krachmalnicoff10a}
can be obtained. Their approach is also based on a hybrid Wigner P+
distribution functional.

Firstly, we ignore all but the first line in of the Ito equation for
the stochastic condensate field (\ref{Eq.ItoBogoliubovCondensateFieldOneMode}).
Thus the noise field term is ignored as are the coupling terms involving
non-condensate stochastic fields. The latter are higher order in $(\sqrt{N})^{-1}$,
so this a reasonable first approximation. Consistency in neglecting
the noise field term then requires that the only non-zero diffusion
matrix elements in (\ref{Eq.ItoBogoliubovCondensateFieldOneMode})
that are retained are those just involving the non-condensate stochastic
fields, $D_{NC-;NC-}$ and $D_{NC+;NC+}$ . Consistency with the classical
condensate field equation (\ref{Eq.ClassicalFieldEqnSingleCondC1})
also requires neglecting the coupling terms involving the non-condensate
fields. The condensate stochastic field then satisfies \begin{eqnarray}
 &  & i\hbar\frac{\partial}{\partial t}\widetilde{\psi}_{C}(\mathbf{s,}t)\nonumber \\
 & = & -\frac{\hbar^{2}}{2m}\nabla^{2}\widetilde{\psi}_{C}(\mathbf{s})+V(\mathbf{s})\widetilde{\psi}_{C}(\mathbf{s})+\frac{g_{N}}{N}\{\widetilde{\psi}_{C}^{+}(\mathbf{s})\widetilde{\psi}_{C}(\mathbf{s})-\left\vert \phi_{1}(\mathbf{s})\right\vert ^{2}\}\widetilde{\psi}_{C}(\mathbf{s})\nonumber \\
 &  & \,\label{eq:ApproxCondFieldOneMode}\end{eqnarray}
We see from the Gross-Pitaevskii equation (\ref{Eq.Gross-Pitaevskii})
that a solution is given by $\widetilde{\psi}_{C}(\mathbf{s,}t)=\psi_{C}^{class}(\mathbf{s,}t)=\sqrt{N}\,\phi_{1}(\mathbf{s}),\widetilde{\psi}_{C}^{+}(\mathbf{s,}t)=\psi_{C}^{+\; class}(\mathbf{s,}t)=\sqrt{N}\,\phi_{1}^{\ast}(\mathbf{s})$.
Hence the condensate field now becomes non-stochastic. 

Secondly, the first line in the Ito equation (\ref{Eq.ItoBogoliubovNonCondensateFieldOneMode})
for the stochastic non-condensate field then becomes zero leaving
just the second and third line together with the noise field term.
As the diffusion matrix is now diagonal then using (\ref{eq:NoiseFieldSingleModeCond})
we can write the noise field as \begin{align}
\frac{\partial}{\partial t}\widetilde{G}_{NC}(\underrightarrow{\widetilde{\psi}}(x,t),\underrightarrow{\Gamma}(t_{+})) & =\sqrt{\frac{-i}{\hbar}\frac{g_{N}}{N}(\widetilde{\psi}_{C}})^{2}\,\Theta_{NC}^{-}(x,t)\,\\
\frac{\partial}{\partial t}\widetilde{G}_{NC}^{+}(\underrightarrow{\widetilde{\psi}}(x,t),\underrightarrow{\Gamma}(t_{+})) & =\sqrt{\frac{+i}{\hbar}\frac{g_{N}}{N}(\widetilde{\psi}_{C}^{+}})^{2}\,\Theta_{NC}^{+}(x,t)\,\label{eq:ApproxNonCondNoiseField}\end{align}
where with $a,b=+,-$ we introduce two Gaussian-Markoff stochastic
fields $\Theta_{NC}^{\pm}(x,t)$. The stochastic average for two stochastic
fields is \begin{equation}
\overline{\Theta_{NC}^{a}(x_{1},t_{1})\,\Theta_{NC}^{b}(x_{2},t_{2})\,}=\delta(x_{1}-x_{2})\,\delta(t_{1}-t_{2})\,\delta_{ab}\,\,\,\,\,(a,b=+,-)\label{eq:GaussMarkStochFields}\end{equation}
and the results for products of other numbers of fields satisfy the
standard Gaussian-Markoff rules. It is then straightforward to show
that the two noise fields $\frac{\partial}{\partial t}\widetilde{G}_{NC}$$ $
and $\frac{\partial}{\partial t}\widetilde{G}_{NC}^{+}$ satisfy the
correct results in (\ref{eq:StochAverNoiseSingleMode-1}) etc. for
stochastic averages.

For large $N$ the $-\left\vert \phi_{1}(\mathbf{s})\right\vert ^{2}$
term can be neglected, so the Ito equation (\ref{Eq.ItoBogoliubovNonCondensateFieldOneMode})
for the stochastic non-condensate field is then\begin{align}
i\hbar\frac{\partial}{\partial t}\widetilde{\psi}_{NC}(\mathbf{s,}t) & =-\frac{\hbar^{2}}{2m}\nabla^{2}\widetilde{\psi}_{NC}(\mathbf{s})+V(\mathbf{s})\widetilde{\psi}_{NC}(\mathbf{s})+2\frac{g_{N}}{N}\{\widetilde{\psi}_{C}^{+}(\mathbf{s})\widetilde{\psi}_{C}(\mathbf{s})\}\widetilde{\psi}_{NC}(\mathbf{s})\nonumber \\
 & +\frac{g_{N}}{N}\{\widetilde{\psi}_{C}(\mathbf{s})\widetilde{\psi}_{C}(\mathbf{s})\}\widetilde{\psi}_{NC}^{+}(\mathbf{s})+\sqrt{+i\hbar\,\frac{g_{N}}{N}(\widetilde{\psi}_{C}})^{2}\,\Theta_{NC}^{-}(x,t)\label{eq:ApproxNonCondFieldOneMode}\end{align}
This equation is equivalent to Eq.(5) in the paper by Krachmalnicoff
et al. \citep{Krachmalnicoff10a}. Note however that the derivation
involves making approximations to the actual stochastic field equations
for single mode condensates, in particular the neglect of noise terms
in the equation for the stochastic condensate field.

\subsection{Stochastic Averages for Quantum Correlation Functions}

The quantum averages of symmetrically ordered products of the condensate
field operators $\{\widehat{\Psi}_{C}^{\dagger}(\mathbf{r}_{1})\widehat{\Psi}_{C}^{\dagger}(\mathbf{r}_{2})....\widehat{\Psi}_{C}^{\dagger}(\mathbf{r}_{p})\widehat{\Psi}_{C}(\mathbf{s}_{q})..\widehat{\Psi}_{C}(\mathbf{s}_{1})\}$
and normally ordered products of the non-condensate field operators\\
 $\widehat{\Psi}_{NC}^{\dagger}(\mathbf{u}_{1})\widehat{\Psi}_{NC}^{\dagger}(\mathbf{u}_{2})....\widehat{\Psi}_{NC}^{\dagger}(\mathbf{u}_{r})\widehat{\Psi}_{NC}(\mathbf{v}_{s})..\widehat{\Psi}_{NC}(\mathbf{v}_{1})$
are now given by \emph{stochastic averages. }These replace the \emph{functional
integrals} involving \emph{quasi distribution functional} given above
in (\ref{Eq.QuantumAverages}). We have

\begin{eqnarray}
 &  & Tr[\widehat{\rho}\,\{\widehat{\Psi}_{C}^{\dagger}(\mathbf{r}_{1})..\widehat{\Psi}_{C}^{\dagger}(\mathbf{r}_{p})\widehat{\Psi}_{C}(\mathbf{s}_{q})..\widehat{\Psi}_{C}(\mathbf{s}_{1})\}\,\nonumber \\
 &  & \times\widehat{\Psi}_{NC}^{\dagger}(\mathbf{u}_{1})..\widehat{\Psi}_{NC}^{\dagger}(\mathbf{u}_{r})\widehat{\Psi}_{NC}(\mathbf{v}_{s})..\widehat{\Psi}_{NC}(\mathbf{v}_{1})]\nonumber \\
\nonumber \\ & = & \overline{\begin{array}{c}
\psi_{C}^{+}(\mathbf{r}_{1})\,..\psi_{C}^{+}(\mathbf{r}_{p})\,\psi_{C}(\mathbf{s}_{q})\,..\psi_{C}(\mathbf{s}_{1})\times\\
\times\,\psi_{NC}^{+}(\mathbf{u}_{1})\,..\psi_{NC}^{+}(\mathbf{u}_{r})\,\psi_{NC}(\mathbf{v}_{s})..\psi_{NC}(\mathbf{v}_{1})\end{array}}\label{Eq.QuantumAveragesStochastic}\end{eqnarray}
where the bar denotes a stochastic average.\pagebreak{}

\section{Summary}

\label{Sect Summary}

The present paper sets up a general approach for treating both dephasing
and decoherence effects due to collisions in interferometry experiments
using single component Bose-Einstein condensates in double well situations,
where two condensate modes may be involved. The treatment starts from
a description of dephasing and fragmentation effects in two mode condensates
in which the two modes satisfy generalised coupled Gross-Pitaevskii
equations, and the amplitudes describing the fragmentation of the
condensate into the two modes satisfy matrix equations. The two sets
of equations, which are coupled and self-consistent, are derived from
the Dirac-Frenkel variational principle. The treatment of decoherence
effects requires the consideration of non-condensate modes and a full
phase space method involving a distribution functional is used, where
the highly occupied condensate modes are described via a truncated
Wigner representation (since the bosons in condensate modes behave
like a classical mean field), whilst the basically unoccupied non-condensate
modes are described via a positive P representation (these bosons
should exhibit quantum effects). The functional Fokker-Planck equation
is derived using the correspondence rules and then Ito equations for
the stochastic fields associated with the condensate and non-condensate
field annihilation and creation field operators are determined. The
Ito stochastic field equations contain a deterministic term which
is obtained from the drift term in the functional Fokker-Planck equation,
and a noise field term whose stochastic properties are obtained from
the diffusion term in the functional Fokker-Planck equation. The link
with interferometry experiments is via the quantum correlation functions,
which are shown to be equal to phase space functional integrals of
products of field functions with the distribution functional. These
phase space functional integrals are then shown to be determined by
stochastic averages of products of the stochastic fields, and in the
present approach the quantum correlation functions would be evaluated
numerically via such stochastic averages. Clearly, the general approach
presented here is rather complex, so in order that the reader can
understand what is involved this paper contains a full coverage of
all the important steps in the derivations of the key expressions
obtained for the quantum correlation functions, correspondence rules,
functional Fokker-Planck equations and Ito stochastic field equations.
These are not covered in any of the standard textbooks and previous
papers only provide a brief outline of how such results are obtained.

For the condensate field, the first line in the Ito stochastic field
equation reads like a time-dependent Gross-Pitaevskii equation if
the condensate field is regarded as the order function. The first
line in the non-condensate equation is a term coupling in the condensate
field. The results for the two mode condensate have unusual features
such as the Ito stochastic field equations being integro-differential
equations and the diffusion matrix being non-local. These features
are not found in the situation where there is only one condensate
mode, where the Ito equations are differential equations and the diffusion
matrix is local. The stochastic properties of the noise field terms
are determined and are similar to those for Gaussian-Markov processes
in that the stochastic averages of odd numbers of noise fields are
zero and those for even numbers of noise field terms are the sums
of products of stochastic averages associated with pairs of noise
fields. However each pair is represented by an element of the diffusion
matrix rather than products of the noise fields themselves, as in
the case of Gaussian-Markov processes. Hence it is only stochastic
averages involving diffusion matrix elements that determine all the
stochastic properties. Results for both two mode condensates and the
simpler single mode condensate case are presented here.

The Ito stochastic field equations for single mode condensate have
been compared to similar equations in the recent paper by Krachmalnicoff
et al. \citep{Krachmalnicoff10a}. We see that their equations are
an approximate version for large $N$ of those presented here, the
approximation involving the neglect of noise terms and higher order
terms in the condensate stochastic field equations - which requires
ignoring off-diagonal terms in the diffusion matrix. In this approximation
the condensate fields are non-stochastic and given by the $\sqrt{N}$
times the normalised solution to the single mode Gross-Pitaevskii
equation, or its complex conjugate. The non-condensate fields are
stochastic and involves two Gaussian-Markoff delta correlated stochastic
fields. 

Numerical applications to a range of actual and potential interferometry
experiments with Bose-Einstein condensates are planned. These include
the Heisenberg-limited interferometry experiment proposed by Dunningham
and Burnett \citep{Dunningham04a}, where the existing theory is based
on the Josephson Hamiltonian in which the two mode functions are unchanged
during each stage of the process. A more comprehensive analysis of
this potentially important experiment by a theory that allows for
changes to the two mode functions and decoherence effects would be
of interest. Future theoretical work will involve the extension of
the present theory to two component condensates in single wells, where
there are also two spatial mode functions involved, and where interferometry
experiments of the Ramsey type have already been performed \citep{Anderson09a}.
However, the current theoretical treatment \citep{Anderson09a} ignores
decoherence and is based on a single mode theory. A theory along the
lines of that presented here for single component condensates would
enable both decoherence effects and the possibility of fragmentation
effects to be studied.

\section{Acknowlegements}

This work was supported by the Australian Research Council Centre
of Excellence for Quantum-Atom Optics (ACQAO). The author wishes to
thank R. Ballagh, S. Barnett, T. Busch, J. Corney, P. Drummond, B.
Garraway, S. Gardiner, A. Griffin, P. Hannaford, K. Molmer, M. Olsen,
L. Plimak, K. Rzazewski, A. Sorensen, T. Vaughan and R. Walser for
helpful discussions.\pagebreak{}

\appendix

\section{- Amplitude and Mode Equations for Two-Mode Theory}

\label{App: Quantities-1}

\subsection{Angular Momentum Quantities}

In the two-mode approximation the $N$ boson system behaves like a
giant spin system with spin quantum number $j=N/2$ and which can
be described via angular momentum eigenstates $\left\vert \,\frac{{\small N}}{{\small2}},k\right\rangle $,
where $k=-N/2,..,+N/2$ is a magnetic quantum number which describes
fragmented states of the bosonic system with $(\frac{N}{2}-k)$\
bosons in mode $\phi_{{\small1}}(r,t)$\ and $(\frac{N}{2}+k)$\ bosons
in mode $\phi_{{\small2}}(r,t)$. Details of the spin operator treatment
for two mode theory are given in {[}17{]}. It is therefore not surprising
that the basic equations will involve expressions arising from angular
momentum theory. These are the quantities $X_{kl}^{ij}$ and $Y_{kl}^{im\, jn}$
which are defined as

\begin{eqnarray}
X_{kl}^{{\small11}} & = & {\small(}\frac{{\small N}}{{\small2}}{\small-k)\delta}_{kl}\qquad X_{kl}^{{\small12}}={\small\{(}\frac{{\small N}}{{\small2}}{\small-k)(\frac{N}{2}+}l{\small)\}}^{\frac{1}{2}}{\small\delta}_{k,l-1}\nonumber \\
X_{kl}^{{\small21}} & = & {\small\{(}\frac{{\small N}}{{\small2}}{\small-}l{\small)(\frac{N}{2}+k)\}}^{\frac{1}{2}}{\small\delta}_{l,k-1}\qquad X_{kl}^{{\small22}}={\small(}\frac{{\small N}}{{\small2}}{\small+k)\delta}_{kl}\label{Eq.AngMtmXCoeftDefn-1}\\
Y_{kl}^{{\small11}\,{\small11}} & = & {\small(}\frac{{\small N}}{{\small2}}{\small-k)(}\frac{{\small N}}{{\small2}}{\small-k-1)\delta}_{kl}\nonumber \\
Y_{kl}^{{\small22}\,{\small22}} & = & {\small(}\frac{{\small N}}{{\small2}}{\small+k)(}\frac{{\small N}}{{\small2}}{\small+k-1)\delta}_{kl}\nonumber \\
Y_{kl}^{{\small12}\,{\small12}} & = & Y_{kl}^{{\small12}\,{\small21}}=Y_{kl}^{{\small21}\,{\small12}}=Y_{kl}^{{\small21}\,{\small21}}={\small(}\frac{{\small N}}{{\small2}}{\small-k)(}\frac{{\small N}}{{\small2}}{\small+k)\delta}_{kl}\nonumber \\
Y_{kl}^{{\small11}\,{\small12}} & = & Y_{kl}^{{\small11}\,{\small21}}={\small(}\frac{{\small N}}{{\small2}}{\small-}l{\small)\{(}\frac{{\small N}}{{\small2}}{\small-k)(\frac{N}{2}+}l{\small)\}}^{\frac{1}{2}}{\small\delta}_{k,l-1}\nonumber \\
Y_{kl}^{{\small12}\,{\small22}} & = & Y_{kl}^{{\small21}\,{\small22}}={\small(}\frac{{\small N}}{{\small2}}{\small+k)\{(}\frac{{\small N}}{{\small2}}{\small-k)(\frac{N}{2}+}l{\small)\}}^{\frac{1}{2}}{\small\delta}_{k,l-1}\nonumber \\
Y_{kl}^{{\small12}\,{\small11}} & = & Y_{kl}^{{\small21}\,{\small11}}={\small(}\frac{{\small N}}{{\small2}}{\small-k)\{(}\frac{{\small N}}{{\small2}}{\small-}l{\small)(\frac{N}{2}+k)\}}^{\frac{1}{2}}{\small\delta}_{l,k-1}\nonumber \\
Y_{kl}^{{\small22}\,{\small12}} & = & Y_{kl}^{{\small22}\,{\small21}}={\small(}\frac{{\small N}}{{\small2}}{\small+}l{\small)\{(}\frac{{\small N}}{{\small2}}{\small-}l{\small)(\frac{N}{2}+k)\}}^{\frac{1}{2}}{\small\delta}_{l,k-1}\nonumber \\
Y_{kl}^{{\small11}\,{\small22}} & = & {\small\{(\frac{N}{2}-}l{\small+1)(}\frac{{\small N}}{{\small2}}{\small-k)(\frac{N}{2}+}l{\small)(\frac{N}{2}+k+1)\}}^{\frac{1}{2}}{\small\delta}_{k,l-2}\nonumber \\
Y_{kl}^{{\small22}\,{\small11}} & = & {\small\{(\frac{N}{2}-k+1)(}\frac{{\small N}}{{\small2}}{\small-}l{\small)(\frac{N}{2}+k)(\frac{N}{2}+}l{\small+1)\}}^{\frac{1}{2}}{\small\delta}_{l,k-2}.\label{Eq.AngMtmYCoeftDefn-1}\end{eqnarray}
These results would apply for the general two-mode theory before the
localisation assumption is made.

\subsection{Hamiltonian and Rotation Matrices}

The Hamiltonian and rotation matrix elements $H_{kl}$ and $U_{kl}$
that occur in the amplitude equations (\ref{Eq.AmpEqns-1}) involve
spatial integrals involving the mode functions $\phi_{1}$ and $\phi_{2}$.
They are therefore functionals of the mode functions. The expressions
depend also on the spatial and time derivatives of the mode functions
through the quantities $\widetilde{W}_{ij}({\small\mathbf{r}},t)$,
$\widetilde{V}_{im\, jn}({\small\mathbf{r}},t)$ and $\widetilde{T}_{ij}({\small\mathbf{r}},t)$,
where $(i,j,m,n=1,2)$, and which are defined by \begin{eqnarray}
\widetilde{W}_{ij}({\small\mathbf{r}},t) & = & \frac{\hbar^{2}}{2m}\sum\limits _{\mu=x,y,z}\partial_{\mu}\phi_{i}^{\ast}\,\partial_{\mu}\phi_{j}+\phi_{i}^{\ast}V\phi_{j}\label{Eq.WtildeDefn-1}\\
\widetilde{V}_{im\, jn}({\small\mathbf{r}},t) & = & \frac{g}{2}\phi_{i}^{\ast}\,\phi_{m}^{\ast}\,\phi_{j}\,\phi_{n}\label{Eq.VtildeDefn-1}\\
\widetilde{T}_{ij}({\small\mathbf{r}},t) & = & \frac{1}{2i}(\partial_{t}\phi_{i}^{\ast}\,\phi_{j}-\phi_{i}^{\ast}\,\partial_{t}\phi_{j})\label{Eq.TtildeDefn-1}\end{eqnarray}

The rotation matrix elements $U_{kl}$ $(-\frac{N}{2}\leq k,l\leq+\frac{N}{2})$
are given by\begin{eqnarray}
U_{kl} & = & \frac{1}{2i}[(\partial_{t}\left\langle \frac{{\small N}}{{\small2}},k\,\right\vert \mathbf{)\,}\left\vert \frac{{\small N}}{{\small2}},l\right\rangle -\left\langle \frac{{\small N}}{{\small2}},k\right\vert \mathbf{\,(}\partial_{t}\left\vert \,\frac{{\small N}}{{\small2}},l\right\rangle )]=U_{lk}^{\ast}\label{Eq.RotMatDefn-1}\\
 & = & \int d{\small\mathbf{r\,}}\widetilde{U}_{kl}(\phi_{i}{\small,}\phi_{i}^{\ast}{\small,}\partial_{t}\phi_{i}{\small,}\partial_{t}\phi_{i}^{\ast}).\label{Eq.RotMatExpn-1}\end{eqnarray}
In the expression (\ref{Eq.RotMatExpn-1}) for the rotation matrix
the quantity $\widetilde{{\small U}}_{kl}$ is\begin{equation}
\widetilde{{\small U}}_{kl}=\sum_{ij}\, X_{kl}^{ij}\,\widetilde{T}_{ij}.\label{Eq.RotMatrixResult1-1}\end{equation}
The result involves the angular momentum theory quantities $X_{kl}^{ij}$.
Thus for the rotation matrix, space integrals of the mode functions
and their time derivatives are involved.

The Hamiltonian matrix elements $H_{kl}$ $(-\frac{N}{2}\leq k,l\leq+\frac{N}{2})$
are given by\begin{eqnarray}
H_{kl} & = & \left\langle \frac{{\small N}}{{\small2}},k|\mathbf{\,}\widehat{H}\,|\frac{{\small N}}{{\small2}},l\right\rangle =H_{lk}^{\ast}\label{Eq.HamMatrixDefn-1}\\
 & = & \int d{\small\mathbf{r\,}}\widetilde{H}_{kl}(\phi_{i}{\small,}\phi_{i}^{\ast}{\small,}\partial_{\mu}\phi_{i}{\small,}\partial_{\mu}\phi_{i}^{\ast}).\label{Eq.HamMatrixExpn-1}\end{eqnarray}
In the expression (\ref{Eq.HamMatrixExpn-1}) for the Hamiltonian
matrix the quantity $\widetilde{{\small H}}_{kl}$ is a Hamiltonian
density and is given by\begin{equation}
\widetilde{{\small H}}_{kl}=\sum_{ij}\, X_{kl}^{ij}\,\widetilde{W}_{ij}+\sum_{ijmn}\, Y_{kl}^{im\, jn}\,\widetilde{V}_{im\, jn}.\label{Eq.HamMatrixResult1-1}\end{equation}
This result involves the angular momentum theory quantities $X_{kl}^{ij}$
and $Y_{kl}^{im\, jn}$. Thus for the Hamiltonian matrix, space integrals
of the mode functions and their spatial derivatives are involved.

The coefficients $X_{ij}$ and $Y_{im\, jn}$ $(i,j,m,n=1,2)$ that
occur in the generalized Gross-Pitaevskii equations (\ref{Eq.GenGrossPitEqns-1})
for the mode functions are quadratic functions of the amplitudes\ $b_{k}$
$(-\frac{N}{2}\leq k,l\leq+\frac{N}{2})${\small \ }\begin{eqnarray}
X_{ij} & = & \sum\limits _{k,l}b_{k}^{\ast}\, X_{kl}^{ij}\, b_{l}=X_{ji}^{\ast}\sim N\label{eq:OneBodyCorrnFn}\\
Y_{im\, jn} & = & \sum\limits _{k,l}b_{k}^{\ast}\, Y_{kl}^{im\, jn}\, b_{l}=Y_{jn\, im}^{\ast}\sim N^{2}\label{eq:TwoBodyCorrnFn}\end{eqnarray}
Note the Hermitian properties of these quantities and the $N$ dependence
of their order of magnitude.

\subsection{Supplementary Equations}

~

Amplitude Equations\begin{equation}
i\hbar\frac{\partial b_{k}}{\partial t}=\sum\limits _{l}(H_{kl}-\hbar U_{kl})b_{l}\qquad(k=-N/2,..,N/2).\label{Eq.AmpEqns-1}\end{equation}

Mode Equations\begin{eqnarray}
i\hbar\sum\limits _{j}X_{ij}\,\frac{\partial}{\partial t}\phi_{j} & = & \sum\limits _{j}X_{ij}(-\frac{\hbar^{2}}{2m}\nabla^{2}+V)\,\phi_{j}\nonumber \\
 &  & +\sum\limits _{j}(g\sum\limits _{mn}Y_{im\, jn}\,\phi_{m}^{\ast}\,\phi_{n})\,\phi_{j}\qquad\qquad(i=1,2).\label{Eq.GenGrossPitEqns-1}\end{eqnarray}
\pagebreak{}

\section{- Functional Calculus}

\label{Appendix Func Calculus}

The basic ideas of functional calculus are outlined here for the case
of c-number quantities. The two main processes of interest are \emph{functional
differentiation} and \emph{functional integration}, but we begin by
explaining what is meant by a functional.

\subsection{Definition of Functional}

A \emph{functional} $F[\psi(x)]$ maps a c-number function $\psi(x)$
onto a c-number that depends on \emph{all} the values of $\psi(x)$
over its entire range. The independent variable $x$ could in some
cases refer to a position coordinate, in other cases it may refer
to time. If $x$ does refer to position then $\psi(x)$ is refered
to as a \emph{field function}. Note that the functional is written
with square brackets to distinguish it from a \emph{function}, written
with round brackets.

We will assume that c-number functions $\psi(x)$ can be expanded
in terms of a suitable orthonormal set of mode functions with \emph{c-number}
expansion coefficients $\alpha_{k}$

\begin{equation}
\psi(x)=\sum\limits _{k}\alpha_{k}\,\phi_{k}(x)\label{Eq.FieldFn6}\end{equation}
where the orthonormality conditions are\begin{equation}
\int dx\,\phi_{k}^{\ast}(x)\phi_{l}(x)=\delta_{kl}\label{Eq.Orhog}\end{equation}

This gives the well-known result for the expansion coefficients\begin{equation}
\alpha_{k}\,=\int dx\,\phi_{k}^{\ast}(x)\psi(x)\label{Eq.ExpnCoefts}\end{equation}
and the completeness relationship is\begin{equation}
\sum\limits _{k}\phi_{k}(x)\phi_{k}^{\ast}(y)=\delta(x-y).\label{Eq.Completeness}\end{equation}

As the value of the function at any point in the range for $x$ is
determined uniquely by the expansion coefficients $\{\alpha_{k}\}$,
then the functional $F[\psi(x)]$ must therefore also just depend
on the expansion coefficients, and hence may also be viewed as a \emph{function}
$f(\alpha_{1},\alpha_{2},..,\alpha_{k},..\alpha_{n})$ of the \emph{expansion
coefficients}, a useful equivalence when functional differentiation
and integration are considered.\begin{equation}
F[\psi(x)]\equiv f(\alpha_{1},\alpha_{2},..,\alpha_{k},..\alpha_{n})\label{Eq.EquivFunction}\end{equation}

It is sometimes convenient to expand a field function in terms of
the \emph{complex conjugate} modes $\phi_{k}^{\ast}(x)$. Thus $\psi^{+}(x)$
given by\begin{equation}
\psi^{+}(x)=\sum\limits _{k}\,\phi_{k}^{\ast}(x)\alpha_{k}^{+}\label{Eq.FieldFn5}\end{equation}
is also a field function, and if $\alpha_{k}^{+}=\alpha_{k}^{\ast}$
then $\psi^{+}(x)=\psi^{\ast}(x)$, the complex conjugate field.

The idea of a functional can be extended to cases of the form $F[\psi(x_{1},x_{2},..,x_{n})]$
where $\psi(x_{1},x_{2},..,x_{n})$ is a function of several variables
$x_{1},x_{2},..,x_{n}$. For 3D fields the situation $x_{1}=x,x_{2}=y,x_{3}=z$
is such an application. In addition, cases $F[\widehat{\psi}(x)]$
where $\widehat{\psi}(x)$ is an operator function rather than a c-number
function occur. For example, $\widehat{\psi}(x)$ may be a bosonic
field operator. In this case $F[\widehat{\psi}(x)]$ maps the operator
function onto an operator. Also functionals $F[\psi_{1}(x),\psi_{2}(x),..,\psi_{i}(x),..\psi_{n}(x)]$
involving several functions $\psi_{1}(x),\psi_{2}(x),..,\psi_{i}(x),..\psi_{n}(x)$
occur. For example, a bosonic field operator $\widehat{\psi}(x)$
may be associated with a field function $\psi_{1}(x)=\psi(x)$ and
the field operator $\widehat{\psi}(x)^{\dagger}$ may be associated
with a different field function $\psi_{2}(x)=\psi^{+}(x)$, so functionals
of the form $F[\psi(x),\psi^{+}(x)]$ are involved. Of particular
relevance are cases where the functional involves fields and their
complex conjugates, such as $F[\psi(x),\psi^{+}(x),\psi^{\ast}(x),\psi^{+\ast}(x)]$.
Functional derivatives and functional integrals can be defined for
all of these cases.

\subsection{Examples of Functionals}

A typical example of a functional involves an integration process:\begin{equation}
F[\psi(x)]=\int\limits _{a}^{b}dx\,\phi(\psi(x))\label{Eq.FnalExample1}\end{equation}
where $\phi(\psi(x))$ is some function of $\psi(x)$.

The \emph{scalar product} of $\psi(x)$ with a fixed function $\chi(x)$
is a typical example of a functional (written $\chi\lbrack\psi(x)]$)
since

\begin{equation}
\chi\lbrack\psi(x)]=\int dx\,\chi^{\ast}(x)\,\psi(x).\label{Eq.FnalExample2}\end{equation}

A functional $F[\psi(x)]$ may take the form of an integral of a function
$\mathcal{F}(\psi(x),\partial_{_{x}}\psi(x))$ involving the \emph{spatial
derivative} $\partial_{x}\psi(x)$ as well as $\psi(x)$ \begin{equation}
F[\psi(x)]=\int dx\,\mathcal{F}(\psi(x),\partial_{x}\psi(x))\label{Eq.FnalExample4}\end{equation}

A \emph{function} $\psi(y)$ may also be expressed as a functional
$F_{y}[\psi(x)]$\begin{eqnarray}
F_{y}[\psi(x)] & = & \int dx\,\delta(x-y)\,\psi(x).\nonumber \\
 & = & \psi(y)\label{Eq.FnalExample5}\end{eqnarray}

Another example involves the \emph{spatial derivative} $\triangledown_{y}\psi(y)$
which may also be expressed as a functional $F_{\nabla y}[\psi(x)]$\begin{eqnarray}
F_{\nabla y}[\psi(x)] & = & \int dx\,\delta(x-y)\,\smallskip\triangledown_{x}\psi(x)\nonumber \\
 & = & \triangledown_{y}\psi(y)\label{Eq.FnalExample6}\end{eqnarray}

A functional is said to be \emph{linear} if \begin{equation}
F[c_{1}\psi_{1}(x)+c_{2}\psi_{2}(x)]=c_{1}F[\psi_{1}(x)]+c_{2}F[\psi_{2}(x)]\label{Eq.LinearFnal}\end{equation}
where $c_{1},c_{2}$ are constants. The scalar product is a linear
functional.

\subsection{Functional Differentiation}

The \emph{functional derivative} $\frac{{\LARGE\delta F[\psi(x)]}}{{\LARGE\delta\psi(x)}}$
is defined by \begin{equation}
F[\psi(x)+\delta\psi(x)]\doteqdot F[\psi(x)]+\int dx\,\delta\psi(x)\,\left(\frac{\delta F[\psi(x)]}{\delta\psi(x)}\right)_{x}\label{Eq.FuncDeriv1}\end{equation}
where $\delta\psi(x)$ is small. In this equation the left side is
a functional of $\psi(x)+\delta\psi(x)$ and the first term on the
right side is a functional of $\psi(x)$. The second term on the right
side is a functional of $\delta\psi(x)$ and thus the functional derivative
must be a function of $x$, hence the subscript $x$. In most situations
this subscript will be left understood. If we write $\delta\psi(x)=\epsilon\delta(x-y)$
for small $\epsilon$ then an equivalent result for the functional
derivative at $x=y$ is\begin{equation}
\left(\frac{\delta F[\psi(x)]}{\delta\psi(x)}\right)_{x=y}=\lim_{\epsilon\rightarrow0}\left(\frac{F[\psi(x)+\epsilon\delta(x-y)]-F[\psi(x)]}{\epsilon}\right).\label{Eq.FuncDeriv2}\end{equation}

This definition of a functional derivative can be extended to cases
where $\psi(x_{1},x_{2},..,x_{n})$ is a function of several variables
or where $\widehat{\psi}(x)$ is an operator function rather than
a c-number function. Also functionals $F[\psi_{1}(x),\psi_{2}(x),..,\psi_{i}(x),..\psi_{n}(x)]$
involving several functions $\psi_{1}(x),\psi_{2}(x),..,\psi_{i}(x),..\psi_{n}(x)$
occur, and functional derivatives with respect to any of these functions
can be defined. For example, the functional $F[\psi(x),\psi^{+}(x),\psi^{\ast}(x),\psi^{+\ast}(x)]$
leads to functional derivatives with respect to all four fields defined
via an obvious generalisation of (\ref{Eq.FuncDeriv1}), the conjugate
fields $\psi(x),\psi^{\ast}(x)$ and $\psi^{+}(x),\psi^{+\ast}(x)$
being regarded as independent of each other.

Finally, \emph{higher order functional derivatives} can be defined
by applying the basic definitions to lower order functional derivatives.

\subsection{Examples of Functional Derivatives}

For the case of the functional $F_{y}[\psi(x)]$ in Eq.(\ref{Eq.FnalExample5})
that gives the function $\psi(y)$ \begin{eqnarray}
\left(\frac{\delta F_{y}[\psi(x)]}{\delta\psi(x)}\right)_{x} & = & \left(\frac{\delta\psi(y)}{\delta\psi(x)}\right)_{x}\nonumber \\
 & = & \lim_{\epsilon\rightarrow0}\left(\frac{\int dz\,\delta(z-y)\,\{\psi(z)+\epsilon\delta(z-x)\}-\int dz\,\delta(z-y)\,\psi(z)}{\epsilon}\right)\nonumber \\
 & = & \delta(x-y)\label{Eq.FuncDerivExample2}\end{eqnarray}
so here the functional derivative is a delta function.

A similar situation applies to the case where the functional $F_{\nabla y}[\psi(x)]$
in Eq.(\ref{Eq.FnalExample6}) that gives the spatial derivative function
$\triangledown_{y}\psi(y)$. Using integration by parts\begin{eqnarray*}
F_{\nabla y}[\psi(x)+\delta\psi(x)] & = & \int dx\,\delta(x-y)\,\smallskip\triangledown_{x}(\psi(x)+\delta\psi(x))\\
 & = & F_{\nabla y}[\psi(x)]+\int dx\,\delta(x-y)\,\smallskip\triangledown_{x}\delta\psi(x)\\
 & = & F_{\nabla y}[\psi(x)]-\int dx\,\triangledown_{x}\delta(x-y)\,\smallskip\delta\psi(x)\\
 & = & F_{\nabla y}[\psi(x)]+\int dx\,\triangledown_{y}\delta(x-y)\,\smallskip\delta\psi(x)\end{eqnarray*}
Hence\begin{eqnarray}
\left(\frac{\delta F_{\nabla y}[\psi(x)]}{\delta\psi(x)}\right)_{x} & = & \left(\frac{\delta\triangledown_{y}\psi(y)}{\delta\psi(x)}\right)_{x}\nonumber \\
 & = & \triangledown_{y}\delta(x-y)=-\triangledown_{x}\delta(x-y)\nonumber \\
 &  & \,\label{eq:FuncDerivExample3}\end{eqnarray}
so here the functional derivative is the derivative of a delta function.

\subsection{Functional Derivative and Mode Functions}

If a mode expansion for $\psi(x)$ as in Eq.(\ref{Eq.FieldFn1-1})
etc. is used, then we can obtain an expression for the functional
derivative in terms of mode functions. By writing \[
\delta\psi(x)=\sum\limits _{k}\delta\alpha_{k}\,\phi_{k}(x)\]
we see that \begin{eqnarray*}
F[\psi(x)+\delta\psi(x)]-F[\psi(x)] & \doteqdot & \int dx\,\delta\psi(x)\,\left(\frac{\delta F[\psi(x)]}{\delta\psi(x)}\right)_{x}\\
 & \doteqdot & \sum\limits _{k}\delta\alpha_{k}\int dx\,\phi_{k}(x)\,\left(\frac{\delta F[\psi(x)]}{\delta\psi(x)}\right)_{x}\end{eqnarray*}
But the left side is the same as\[
f(\alpha_{1}+\delta\alpha_{1},..,\alpha_{k}+\delta\alpha_{k},..)-f(\alpha_{1},..,\alpha_{k},..)\doteqdot\sum\limits _{k}\delta\alpha_{k}\,\frac{\partial f(\alpha_{1},..,\alpha_{k},..)}{\partial\alpha_{k}}\]
Equating the coefficients of the independent $\delta\alpha_{k}$ and
then using the completeness relationship in Eq.(\ref{Eq.Completeness})
gives the key result\begin{eqnarray}
\left(\frac{\delta F[\psi(x)]}{\delta\psi(x)}\right)_{x} & = & \sum\limits _{k}\phi_{k}^{\ast}(x)\,\frac{\partial f(\alpha_{1},..,\alpha_{k},..)}{\partial\alpha_{k}}\label{Eq.FnalDerivResult}\\
\frac{\partial f(\alpha_{1},..,\alpha_{k},..)}{\partial\alpha_{k}} & = & \int dx\,\phi_{k}(x)\,\left(\frac{\delta F[\psi(x)]}{\delta\psi(x)}\right)_{x}\label{Eq.ModeDerivResult}\end{eqnarray}
These relate the functional derivative to the mode functions and to
the ordinary partial derivatives of the function $f(\alpha_{1},\alpha_{2},..,\alpha_{k},..\alpha_{n})$
that was equivalent to the original functional $F[\psi(x)]$. Again,
we see that the result for the functional derivative is a function
of $x$.

For the case of the functional $F[\psi(x),\psi^{+}(x),\psi^{\ast}(x),\psi^{+\ast}(x)]$
whose equivalent function based on the expansions (\ref{Eq.FieldFn6})
and (\ref{Eq.FieldFn5}) is $f(\alpha_{k},\alpha_{k}^{+},\alpha_{k}^{\ast},\alpha_{k}^{+\ast})$,
the generalisation of (\ref{Eq.FnalDerivResult}) is\begin{eqnarray}
\left(\frac{\delta F[\psi(x),\psi^{+}(x),\psi^{\ast}(x),\psi^{+\ast}(x)]}{\delta\psi(x)}\right)_{x} & = & \sum\limits _{k}\phi_{k}^{\ast}(x)\,\frac{\partial f(\alpha_{k},\alpha_{k}^{+},\alpha_{k}^{\ast},\alpha_{k}^{+\ast})}{\partial\alpha_{k}}\nonumber \\
\left(\frac{\delta F[\psi(x),\psi^{+}(x),\psi^{\ast}(x),\psi^{+\ast}(x)]}{\delta\psi^{+}(x)}\right)_{x} & = & \sum\limits _{k}\phi_{k}(x)\,\frac{\partial f(\alpha_{k},\alpha_{k}^{+},\alpha_{k}^{\ast},\alpha_{k}^{+\ast})}{\partial\alpha_{k}^{+}}\nonumber \\
\left(\frac{\delta F[\psi(x),\psi^{+}(x),\psi^{\ast}(x),\psi^{+\ast}(x)]}{\delta\psi^{\ast}(x)}\right)_{x} & = & \sum\limits _{k}\phi_{k}(x)\,\frac{\partial f(\alpha_{k},\alpha_{k}^{+},\alpha_{k}^{\ast},\alpha_{k}^{+\ast})}{\partial\alpha_{k}^{\ast}}\nonumber \\
\left(\frac{\delta F[\psi(x),\psi^{+}(x),\psi^{\ast}(x),\psi^{+\ast}(x)]}{\delta\psi^{+\ast}(x)}\right)_{x} & = & \sum\limits _{k}\phi_{k}^{\ast}(x)\,\frac{\partial f(\alpha_{k},\alpha_{k}^{+},\alpha_{k}^{\ast},\alpha_{k}^{+\ast})}{\partial\alpha_{k}^{+\ast}}\nonumber \\
 &  & \,\label{eq:FnalDerivResultGeneral}\end{eqnarray}
and \begin{eqnarray}
\frac{\partial f(\alpha_{k},\alpha_{k}^{+},\alpha_{k}^{\ast},\alpha_{k}^{+\ast})}{\partial\alpha_{k}} & = & \int dx\,\phi_{k}(x)\,\left(\frac{\delta F[\psi(x),\psi^{+}(x),\psi^{\ast}(x),\psi^{+\ast}(x)]}{\delta\psi(x)}\right)_{x}\nonumber \\
\frac{\partial f(\alpha_{k},\alpha_{k}^{+},\alpha_{k}^{\ast},\alpha_{k}^{+\ast})}{\partial\alpha_{k}^{+}} & = & \int dx\,\phi_{k}^{\ast}(x)\,\left(\frac{\delta F[\psi(x),\psi^{+}(x),\psi^{\ast}(x),\psi^{+\ast}(x)]}{\delta\psi^{+}(x)}\right)_{x}\nonumber \\
\frac{\partial f(\alpha_{k},\alpha_{k}^{+},\alpha_{k}^{\ast},\alpha_{k}^{+\ast})}{\partial\alpha_{k}^{\ast}} & = & \int dx\,\phi_{k}^{\ast}(x)\,\left(\frac{\delta F[\psi(x),\psi^{+}(x),\psi^{\ast}(x),\psi^{+\ast}(x)]}{\delta\psi^{\ast}(x)}\right)_{x}\nonumber \\
\frac{\partial f(\alpha_{k},\alpha_{k}^{+},\alpha_{k}^{\ast},\alpha_{k}^{+\ast})}{\partial\alpha_{k}^{+\ast}} & = & \int dx\,\phi_{k}(x)\,\left(\frac{\delta F[\psi(x),\psi^{+}(x),\psi^{\ast}(x),\psi^{+\ast}(x)]}{\delta\psi^{+\ast}(x)}\right)_{x}\nonumber \\
 &  & \,\label{eq:ModeDerivResultGeneral}\end{eqnarray}
which relate the functional derivatives and the derivatives with respect
to the mode amplitudes.

\subsection{Rules for Functional Derivatives}

Rules can be established for the functional derivative of the \emph{sum}
of two functionals. It is easily shown that\begin{equation}
\left(\frac{\delta\{F[\psi(x)]+G[\psi(x)]\}}{\delta\psi(x)}\right)_{x=y}=\left(\frac{\delta\{F[\psi(x)]\}}{\delta\psi(x)}\right)_{x=y}+\left(\frac{\delta\{G[\psi(x)]\}}{\delta\psi(x)}\right)_{x=y}\label{Eq.SumRule}\end{equation}

Also, rules can be established for functional derivative of the \emph{product}
of two functionals. We will keep these in order to cover the case
where the functionals are operators\begin{equation}
\left(\frac{\delta\{F[\psi(x)]G[\psi(x)]\}}{\delta\psi(x)}\right)_{x=y}=\left(\frac{\delta F[\psi(x)]}{\delta\psi(x)}\right)_{x=y}G[\psi(x)]+F[\psi(x)]\left(\frac{\delta G[\psi(x)]}{\delta\psi(x)}\right)_{x=y}\label{Eq.ProdRule}\end{equation}

A \emph{chain} rule for functional differentiation can also be derived
for the case where a functional $G[\psi_{y}(x)]$ involves not just
one function $\psi(x)$, but a set of functions each labelled by a
variable $y$. Since $G[\psi_{y}(x)]$ maps $\psi_{y}(x)$ onto a
c-number which depends on $y$, we can regard the functional $G[\psi_{y}(x)]$
also as a function $G(y)$ of the variable $y.$Now consider a second
functional $F[G(y)]$ of this function $G(y)$, and we could determine
the functional derivative $\left(\frac{\delta F[G(y)]}{\delta G(y)}\right)_{y}$.
But $F[G(y)]$ is also a functional of the $\psi_{y}(x)$ via \[
F[G[\psi_{y}(x)]]\equiv F[G(y)]\]
We obtain the chain rule\begin{equation}
\left(\frac{\delta F[G[\psi_{y}(x)]]}{\delta\psi_{y}(x)}\right)_{x}=\int dy\,\left(\frac{\delta F[G(y)]}{\delta G(y)}\right)_{y}\,\left(\frac{\delta F[G(y)]}{\delta G(y)}\right)_{x}\label{Eq.ChainRule}\end{equation}
where we have left the order of the factors as they appeared in order
to allow for operator cases.

We may also define the \emph{spatial derivative} of the functional
derivative. Thus\begin{eqnarray}
\partial_{y}\left(\frac{\delta F[\psi(x)]}{\delta\psi(x)}\right)_{x=y} & = & \lim_{\Delta y\rightarrow0}\left(\frac{\left(\frac{\delta F[\psi(x)]}{\delta\psi(x)}\right)_{x=y+\Delta y}-\left(\frac{\delta F[\psi(x)]}{\delta\psi(x)}\right)_{x=y}}{\Delta y}\right)\nonumber \\
 & = & -\int dx\left(\frac{\partial}{\partial x}\delta(x-y)\right)_{x=y}\left(\frac{\delta F[\psi(x)]}{\delta\psi(x)}\right)_{x}\label{Eq.SpatialDeriv2}\end{eqnarray}
This expresses the spatial derivative as an integral involving the
functional derivative and the spatial derivative of the delta function.
The result will be a function of $s$.

A number of \emph{other rules} may also be established.

(1) Power rule\begin{eqnarray}
F[\psi(x)] & = & \int dx\,\psi(x)^{n}\nonumber \\
\frac{\delta F[\psi(x)]}{\delta\psi(x)} & = & n\psi(x)^{n-1}\label{Eq.PowerRule}\end{eqnarray}

(2) Function rule\begin{eqnarray}
F[\psi(x)] & = & \int dx\,\phi(\psi(x))\nonumber \\
\frac{\delta F[\psi(x)]}{\delta\psi(x)} & = & \phi^{\prime}(\psi(x))\label{Eq.FuncRule}\end{eqnarray}

(3) Power derivative rule

\begin{eqnarray}
F[\psi(x)] & = & \int dx\,(\frac{d\psi(x)}{dx})^{n}\nonumber \\
\frac{\delta F[\psi(x)]}{\delta\psi(x)} & = & -n\frac{d}{dx}((\frac{d\psi(x)}{dx})^{n-1})\label{Eq.PowerDerivRule}\end{eqnarray}

(4) Function derivative rule\begin{eqnarray}
F[\psi(x)] & = & \int dx\,\phi(\frac{d\psi(x)}{dx})\nonumber \\
\frac{\delta F[\psi(x)]}{\delta\psi(x)} & = & -\frac{d}{dx}((\frac{d\phi}{d(\frac{d\psi}{dx})}))\label{Eq.FuncDerivRule}\end{eqnarray}

(5) Convolution rule\begin{eqnarray}
F_{y}[\psi(x)] & = & \int dx\, K(y,x)\,\psi(x)\nonumber \\
\left(\frac{\delta F_{y}[\psi(x)]}{\delta\psi(x)}\right)_{x} & = & K(y,x)\,\label{Eq.ConvolRule}\end{eqnarray}

(6) Trivial rule\begin{eqnarray}
F_{y}[\psi(x)] & = & \psi(y)\nonumber \\
\left(\frac{\delta F_{y}[\psi(x)]}{\delta\psi(x)}\right)_{x} & = & \left(\frac{\delta\psi(y)}{\delta\psi(x)}\right)_{x}\nonumber \\
 & = & \delta(x-y)\label{Eq.TrivialRule}\end{eqnarray}
This was proved above.

(7) Gradient rule\begin{eqnarray}
F_{\nabla y}[\psi(x)] & = & \triangledown_{y}\psi(y)\nonumber \\
\left(\frac{\delta F_{\nabla y}[\psi(x)]}{\delta\psi(x)}\right)_{x} & = & \triangledown_{y}\delta(x-y)=-\triangledown_{x}\delta(x-y)\label{Eq.GradRule}\end{eqnarray}
This was proved above.

(8) Exponential rule\begin{eqnarray}
F[\psi(x)] & = & \exp G[\psi(x)]\nonumber \\
\frac{\delta F[\psi(x)]}{\delta\psi(x)} & = & \exp G[\psi(x)]\,\frac{\delta G[\psi(x)]}{\delta\psi(x)}\label{Eq.ExpRule}\end{eqnarray}
The exponential rule only applies in this form if $F[\psi(x)]$ and
$G[\psi(x)]$ are c-numbers.

All these rules have obvious generalisations for functionals involving
several fields, such as $F[\psi(x),\psi^{+}(x),\psi^{\ast}(x),\psi^{+\ast}(x)]$.

\subsection{Functional Integration}

If the range for the function $\psi(x)$ is divided up into $n$ small
intervals $\Delta x_{i}=x_{i+1}-x_{i}$ (the $i$th interval), then
we may specify the \emph{value} $\psi_{i}$ of the function $\psi(x)$
in the $i$th interval via the \emph{average}\begin{equation}
\psi_{i}=\frac{1}{\Delta x_{i}}\int\limits _{\Delta x_{i}}dx\,\psi(x)\label{Eq.AverageValue}\end{equation}
and then the functional $F[\psi(x)]$ may be regarded as a \emph{function}
$F(\psi_{1},\psi_{2},..,\psi_{i},..,\psi_{n})$ of all the $\psi_{i}$.

Introducing a suitable \emph{weight function} $w(\psi_{1},\psi_{2},..,\psi_{i},..,\psi_{n})$
we may then define the \emph{functional integral} for the case of
\emph{real} functions as \begin{eqnarray}
\int D\psi\, F[\psi(x)] & = & \lim_{n\rightarrow\infty}\lim_{\epsilon\rightarrow0}\int\ldots\int d\psi_{1}d\psi_{2}..d\psi_{i}..d\psi_{n}\, w(\psi_{1},\psi_{2},..,\psi_{i},..,\psi_{n})\,\nonumber \\
 &  & \times F(\psi_{1},\psi_{2},..,\psi_{i},..,\psi_{n})\label{Eq.FuncIntegral}\end{eqnarray}
where $\epsilon>\Delta x_{i}$. Thus the symbol $D\psi$ stands for
$d\psi_{1}d\psi_{2}..d\psi_{i}..d\psi_{n}\, w(\psi_{1},\psi_{2},..,\psi_{i},..,\psi_{n})$.

If the functions are \emph{complex} then the functional integral is
\begin{eqnarray}
\int D^{2}\psi\, F[\psi(x)] & = & \lim_{n\rightarrow\infty}\lim_{\epsilon\rightarrow0}\int\ldots\int d^{2}\psi_{1}d^{2}\psi_{2}..d^{2}\psi_{i}..d^{2}\psi_{n}\, w(\psi_{1},\psi_{2},..,\psi_{i},..,\psi_{n})\,\nonumber \\
 &  & \times F(\psi_{1},\psi_{2},..,\psi_{i},..,\psi_{n})\label{Eq.FuncIntegral2}\end{eqnarray}
The symbol $D^{2}\psi$ stands for $d^{2}\psi_{1}d^{2}\psi_{2}..d^{2}\psi_{i}..d^{2}\psi_{n}\, w(\psi_{1},\psi_{2},..,\psi_{i},..,\psi_{n})$,
where with $\psi_{i}=\psi_{ix}+i\psi_{iy}$ the quantity $d^{2}\psi_{i}$
means $d\psi_{ix}d\psi_{iy}$, involving integration over the real,
imaginary parts of the complex function.

For cases involving several complex functions such as $F[\psi(x),\psi^{+}(x),\psi^{\ast}(x),\psi^{+\ast}(x)]$
the functional integrals are of the form \begin{eqnarray}
 &  & \int\int D^{2}\psi\, D^{2}\psi^{+}\, F[\psi(x),\psi^{+}(x),\psi^{\ast}(x),\psi^{+\ast}(x)]\nonumber \\
 & = & \lim_{n\rightarrow\infty}\lim_{\epsilon\rightarrow0}\int\ldots\int d^{2}\psi_{1}d^{2}\psi_{2}..d^{2}\psi_{i}..d^{2}\psi_{n}\,\lim_{n\rightarrow\infty}\lim_{\epsilon\rightarrow0}\int\ldots\int d^{2}\psi_{1}^{+}d^{2}\psi_{2}^{+}..d^{2}\psi_{i}^{+}..d^{2}\psi_{n}^{+}\,\nonumber \\
 &  & \times w(\psi_{1},..,\psi_{i},..,\psi_{n},\psi_{1}^{+},..,\psi_{i}^{+},..,\psi_{n}^{+},\psi_{1}^{\ast},..,\psi_{i}^{\ast},..,\psi_{n}^{\ast},\psi_{1}^{+\ast},..,\psi_{i}^{+\ast},..,\psi_{n}^{+\ast})\nonumber \\
 &  & \times F(\psi_{1},..,\psi_{i},..,\psi_{n},\psi_{1}^{+},..,\psi_{i}^{+},..,\psi_{n}^{+},\psi_{1}^{\ast},..,\psi_{i}^{\ast},..,\psi_{n}^{\ast},\psi_{1}^{+\ast},..,\psi_{i}^{+\ast},..,\psi_{n}^{+\ast})\label{Eq.FuncIntegral3}\end{eqnarray}
where $D^{2}\psi$ $D^{2}\psi^{+}\,$stands for \begin{align*}
d^{2}\psi_{1}..d^{2}\psi_{i}..d^{2}\psi_{n}\, d^{2}\psi_{1}^{+}..d^{2}\psi_{i}^{+}..d^{2}\psi_{n}^{+}\, w(\psi_{1},..,\psi_{i},..,\psi_{n},\psi_{1}^{+},..,\psi_{i}^{+},..,\psi_{n}^{+},\psi_{1}^{\ast},..,\psi_{i}^{\ast},..,\psi_{n}^{\ast},\psi_{1}^{+\ast},..,\psi_{i}^{+\ast},..,\psi_{n}^{+\ast})\end{align*}
 and where with $\psi_{i}^{+}=\psi_{ix}^{+}+i\psi_{iy}^{+}$, the
quantity $d^{2}\psi_{i}^{+}$ means $d\psi_{ix}^{+}d\psi_{iy}^{+}$.

A functional integral of a functional of a c-number function gives
a c-number. Unlike ordinary calculus, functional integration and differentiation
are not related as inverse processes.

\subsection{Functional Integrals and Phase Space Integrals}

\label{SectA1.8 Functional and Phase Space Integ}

We first consider the case of a functional $F[\psi(x)]$ of a real
function $\psi(x)$, which we expand in terms of real, orthogonal
mode functions. The expansion coefficients in this case will be real
also. If a mode expansion such as in Eq.(\ref{Eq.FieldFn1-1}) etc.
is used then the value $\phi_{ki}$ of the mode function in the $i$th
interval is also defined via the \emph{average}\begin{equation}
\phi_{ki}=\frac{1}{\Delta x_{i}}\int\limits _{\Delta x_{i}}dx\,\phi_{k}(x)\label{Eq.AverModeFn}\end{equation}
and hence\begin{equation}
\psi_{i}=\sum\limits _{k}\alpha_{k}\,\phi_{ki}.\label{Eq.AverValInter1}\end{equation}
This shows that the values in the $i$th interval of the function
$\psi_{i}$ and the mode function $\phi_{ki}$ are related via the
expansion coefficients $\alpha_{k}$. For simplicity we will choose
the \emph{same number} $n$ of intervals as mode functions. Using
the expression Eq.(\ref{Eq.ExpnCoefts}) for the expansion coefficients
we then obtain the inverse formula to Eq.(\ref{Eq.AverValInter1})
\begin{equation}
\alpha_{k}=\sum\limits _{i}\Delta x_{i}\,\phi_{ki}.\psi_{i}.\label{Eq.AverValInter2}\end{equation}
Note that this involves a sum over intervals $i$ and the interval
size $\Delta x_{i}$ is also involved.

The relationship in Eq.(\ref{Eq.AverValInter1}) shows that the functions
$F(\psi_{1},\psi_{2},..,\psi_{i},..,\psi_{n})$ and $w(\psi_{1},\psi_{2},..,\psi_{i},..,\psi_{n})$
of all the interval values $\psi_{i}$ can also be regarded as functions
of the expansion coefficients $\alpha_{k}$ which we may write as\begin{eqnarray}
\, f(\alpha_{1},..,\alpha_{k},..\alpha_{n}) & \equiv & F(\psi_{1}(\alpha_{1},..,\alpha_{k},..\alpha_{n}),...,\psi_{i}(\alpha_{1},..,\alpha_{k},..\alpha_{n}),..,\psi_{n})\nonumber \\
\label{eq:PhaseSpaceFn}\\v(\alpha_{1},..,\alpha_{k},..\alpha_{n}) & \equiv & w(\psi_{1}(\alpha_{1},..,\alpha_{k},..\alpha_{n}),...,\psi_{i}(\alpha_{1},..,\alpha_{k},..\alpha_{n}),..,\psi_{n})\nonumber \\
 &  & \,\label{eq:WeightFn}\end{eqnarray}
Thus the various \emph{values} $\psi_{1},\psi_{2},..,\psi_{1},\psi_{2},..,\psi_{i},..,\psi_{n},..,\psi_{n}$
of that the function $\psi(x)$ takes on in the $n$ intervals - and
which are integrated over in the functional integration process -
are \emph{all} determined by the choice of the expansion coefficients
$\alpha_{1},\alpha_{2},..,\alpha_{k},..\alpha_{n}$. Hence integration
over all the $\psi_{i}$ will be equivalent to integration over all
the $\alpha_{k}$.

This enables us to express the functional integral in Eq.(\ref{Eq.FuncIntegral})
as a \emph{phase space integral} over the expansion coefficients $\alpha_{1},\alpha_{2},..,\alpha_{k},..\alpha_{n}$.
We have \begin{eqnarray}
\int D\psi\, F[\psi(x)] & = & \lim_{n\rightarrow\infty}\lim_{\epsilon\rightarrow0}\int\ldots\int d\alpha_{1}d\alpha_{2}..d\alpha_{k}..d\alpha_{n}\,||J(\alpha_{1},\alpha_{2},..,\alpha_{k},..\alpha_{n})||\nonumber \\
 &  & \times v(\alpha_{1},\alpha_{2},..,\alpha_{k},..\alpha_{n})\, f(\alpha_{1},\alpha_{2},..,\alpha_{k},..\alpha_{n})\label{Eq.PhaseSpaceIntegral1}\end{eqnarray}
where the Jacobian is given by\begin{equation}
||J(\alpha_{1},\alpha_{2},..,\alpha_{k},..\alpha_{n})||=\left\Vert \begin{tabular}{cccc}
 \ensuremath{\frac{{\LARGE\partial\psi}_{1}}{{\LARGE\partial\alpha}_{1}}} &  \ensuremath{\frac{{\LARGE\partial\psi}_{1}}{{\LARGE\partial\alpha}_{2}}} &  \ensuremath{...} &  \ensuremath{\frac{{\LARGE\partial\psi}_{1}}{{\LARGE\partial\alpha}_{n}}}\\
\ensuremath{\frac{{\LARGE\partial\psi}_{2}}{{\LARGE\partial\alpha}_{1}}} &  \ensuremath{\frac{{\LARGE\partial\psi}_{2}}{{\LARGE\partial\alpha}_{2}}} &  \ensuremath{...} &  \ensuremath{\frac{{\LARGE\partial\psi}_{2}}{{\LARGE\partial\alpha}_{n}}}\\
\ensuremath{...} &  \ensuremath{...} &  \ensuremath{...} &  \ensuremath{...}\\
\ensuremath{\frac{{\LARGE\partial\psi}_{n}}{{\LARGE\partial\alpha}_{1}}} &  \ensuremath{\frac{{\LARGE\partial\psi}_{n}}{{\LARGE\partial\alpha}_{2}}} &   &  \ensuremath{\frac{{\LARGE\partial\psi}_{n}}{{\LARGE\partial\alpha}_{n}}}\end{tabular}\right\Vert \label{Eq.Jacobian}\end{equation}
Now using Eq.(\ref{Eq.AverValInter1}) \begin{equation}
\frac{\partial\psi_{i}}{\partial\alpha_{k}}=\phi_{ki}\label{Eq.JacobianMatrix}\end{equation}
and evaluating the Jacobian using after showing that $(JJ^{T})_{ik}=\delta_{ik}/\Delta x_{i}$
using the completeness relationship in Eq.(\ref{Eq.Completeness})
we find that \begin{equation}
||J(\alpha_{1},\alpha_{2},..,\alpha_{k},..\alpha_{n})||=\prod\limits _{i}\frac{1}{(\Delta x_{i})^{1/2}}\label{Eq.Jacobian2}\end{equation}
and thus \begin{eqnarray}
\int D\psi\, F[\psi(x)] & = & \lim_{n\rightarrow\infty}\lim_{\epsilon\rightarrow0}\int\ldots\int d\alpha_{1}d\alpha_{2}..d\alpha_{k}..d\alpha_{n}\,\prod\limits _{i}\frac{1}{(\Delta x_{i})^{1/2}}\nonumber \\
 &  & \times v(\alpha_{1},\alpha_{2},..,\alpha_{k},..\alpha_{n})\, f(\alpha_{1},\alpha_{2},..,\alpha_{k},..\alpha_{n})\label{Eq.PhaseSpaceIntegral2}\end{eqnarray}
This key result expresses the original functional integral as a phase
space integral over the expansion coefficients $\alpha_{k}$ of the
function $\psi(x)$ in terms of the mode functions $\phi_{k}(x)$.

The general result can be simplified with a \emph{special choice}
of the weight function\begin{equation}
w(\psi_{1},\psi_{2},..,\psi_{i},..,\psi_{n})=\prod\limits _{i}(\Delta x_{i})^{1/2}\label{Eq.SimpleWeightFn}\end{equation}
and we then get a simple expression for the functional integral \begin{equation}
\int D\psi\, F[\psi(x)]=\lim_{n\rightarrow\infty}\lim_{\epsilon\rightarrow0}\int\ldots\int d\alpha_{1}d\alpha_{2}..d\alpha_{k}..d\alpha_{n}\,\, f(\alpha_{1},\alpha_{2},..,\alpha_{k},..\alpha_{n})\label{Eq.PhaseSpaceIntegral3}\end{equation}
In this form of the functional integral the original functional $F[\psi(x)]$
has been replaced by the equivalent function $f(\alpha_{1},\alpha_{2},..,\alpha_{k},..\alpha_{n})$
of the expansion coefficients $\alpha_{k}$, and the functional integration
is now replaced by a phase space integration over the expansion coefficients.

The relationship between the functional integral and the phase space
integral can be generalised to cases involving several complex functions.
For the case of the functional $F[\psi(x),\psi^{+}(x),\psi^{\ast}(x),\psi^{+\ast}(x)]$,
where $\psi(x),\psi^{+}(x)$ are expanded in terms of complex mode
functions as in (\ref{Eq.FieldFn6}), (\ref{Eq.FieldFn5}) and $\psi_{i},\psi_{i}^{+}$
defined as in (\ref{Eq.AverageValue}) we have \begin{eqnarray}
\psi_{i} & = & \sum\limits _{k}\alpha_{k}\,\phi_{ki}.\qquad\alpha_{k}=\sum\limits _{i}\Delta x_{i}\,\phi_{ki}^{\ast}.\psi_{i}.\nonumber \\
\psi_{i}^{+} & = & \sum\limits _{k}\alpha_{k}^{+}\,\phi_{ki}^{\ast}.\qquad\alpha_{k}^{+}=\sum\limits _{i}\Delta x_{i}\,\phi_{ki}.\psi_{i}^{+}.\nonumber \\
 &  & \,\label{eq:CellAverFieldsModes}\end{eqnarray}
For variety we will turn the phase space integral into a functional
integral. We first have the transformation involving real quantities
\begin{eqnarray}
\alpha_{kX} & = & \sum\limits _{i}\Delta x_{i}\,(\phi_{kiX}.\psi_{iX}+\phi_{kiY}.\psi_{iY})\nonumber \\
\alpha_{kY} & = & \sum\limits _{i}\Delta x_{i}\,(\phi_{kiX}.\psi_{iY}-\phi_{kiY}.\psi_{iX})\nonumber \\
\alpha_{kX}^{+} & = & \sum\limits _{i}\Delta x_{i}\,(\phi_{kiX}.\psi_{iX}^{+}-\phi_{kiY}.\psi_{iY}^{+})\nonumber \\
\alpha_{kY}^{+} & = & \sum\limits _{i}\Delta x_{i}\,(\phi_{kiX}.\psi_{iY}^{+}+\phi_{kiY}.\psi_{iX}^{+})\label{Eq.ModeAmpFunctionValueTransfn}\end{eqnarray}
In the standard notation with $\alpha_{k}=\alpha_{kX}+i\alpha_{kY}$,
$\alpha_{k}^{+}=\alpha_{kX}^{+}+i\alpha_{kY}^{+}$ and $d^{2}\alpha_{k}=d\alpha_{kX}d\alpha_{kY}$,
$d^{2}\alpha_{k}^{+}=d\alpha_{kX}^{+}d\alpha_{kY}^{+}$ the phase
space integral is of the form \begin{eqnarray}
 &  & \int\int d^{2}\alpha\; d^{2}\alpha^{+}\; f(\alpha,\alpha^{+},\alpha^{\ast},\alpha^{+\ast})\nonumber \\
 & = & \int\ldots\int d^{2}\alpha_{1}d^{2}\alpha_{2}..d^{2}\alpha_{k}..d^{2}\alpha_{n}\int\ldots\int d^{2}\alpha_{1}^{+}d^{2}\alpha_{2}^{+}..d^{2}\alpha_{k}^{+}..d^{2}\alpha_{n}^{+}\; f(\alpha_{k},\alpha_{k}^{+},\alpha_{k}^{\ast},\alpha_{k}^{+\ast})\nonumber \\
 &  & \,\end{eqnarray}
and after transforming to the new variables $\psi_{iX},\psi_{iY},\psi_{iX}^{+},\psi_{iY}^{+}$
we get \begin{eqnarray}
 &  & \int\ldots\int d^{2}\alpha_{1}d^{2}\alpha_{2}..d^{2}\alpha_{k}..d^{2}\alpha_{n}\int\ldots\int d^{2}\alpha_{1}^{+}d^{2}\alpha_{2}^{+}..d^{2}\alpha_{k}^{+}..d^{2}\alpha_{n}^{+}\; f(\alpha_{k},\alpha_{k}^{+},\alpha_{k}^{\ast},\alpha_{k}^{+\ast})\nonumber \\
 & = & \int\ldots\int d^{2}\psi_{1}d^{2}\psi_{2}..d^{2}\psi_{i}..d^{2}\psi_{n}\int\ldots\int d^{2}\psi_{1}^{+}d^{2}\psi_{2}^{+}..d^{2}\psi_{i}^{+}..d^{2}\psi_{n}^{+}\;||J(\alpha_{k},\alpha_{k}^{+},\alpha_{k}^{\ast},\alpha_{k}^{+\ast})||\;\nonumber \\
 &  & \times F(\psi_{1},..,\psi_{i},..,\psi_{n},\psi_{1}^{+},..,\psi_{i}^{+},..,\psi_{n}^{+},\psi_{1}^{\ast},..,\psi_{i}^{\ast},..,\psi_{n}^{\ast},\psi_{1}^{+\ast},..,\psi_{i}^{+\ast},..,\psi_{n}^{+\ast})\end{eqnarray}
where the Jacobian can be written in terms of the notation $\alpha_{kX}\rightarrow\alpha_{k1},\alpha_{kY}\rightarrow\alpha_{k2},\alpha_{kX}^{+}\rightarrow\alpha_{k3},\alpha_{kY}^{+}\rightarrow\alpha_{k4}$
and $\psi_{iX}\rightarrow\psi_{i1,}\psi_{iY}\rightarrow\psi_{i2,}\psi_{iX}^{+}\rightarrow\psi_{i3,}\psi_{iY}^{+}\rightarrow\psi_{i4}$
in which the Jacobian is the determinent of the matrix $J$ where
\begin{eqnarray}
J_{k\mu\; i\nu} & = & \frac{\partial\alpha_{k\mu}}{\partial\psi_{i\nu}}\qquad(k=1,..,n;i=1,..,n;\mu=1,..,4;\nu=1,..,4)\nonumber \\
||J(\alpha_{k},\alpha_{k}^{+},\alpha_{k}^{\ast},\alpha_{k}^{+\ast})|| & = & \left\Vert J_{k\mu\; i\nu}\right\Vert \label{Eq.JacobianMatrix2}\end{eqnarray}
The elements in the $4x4$ submatrix $J_{k\; i}$ are obtained from
(\ref{Eq.ModeAmpFunctionValueTransfn}) and are\begin{equation}
\left[J_{k\; i}\right]=\left[\begin{tabular}{cccc}
 \ensuremath{\Delta x_{i}\,\phi_{kiX}} &  \ensuremath{\Delta x_{i}\,\phi_{kiY}} &  \ensuremath{0} &  \ensuremath{0}\\
\ensuremath{-\Delta x_{i}\,\phi_{kiY}} &  \ensuremath{\Delta x_{i}\,\phi_{kiX}} &  \ensuremath{0} &  \ensuremath{0}\\
\ensuremath{0} &  \ensuremath{0} &  \ensuremath{\Delta x_{i}\,\phi_{kiX}} &  \ensuremath{-\Delta x_{i}\,\phi_{kiY}}\\
\ensuremath{0} &  \ensuremath{0} &  \ensuremath{\Delta x_{i}\,\phi_{kiY}} &  \ensuremath{\Delta x_{i}\,\phi_{kiX}}\end{tabular}\right]\label{Eq.JacobianSubMatrix2}\end{equation}
The completeness relationship (\ref{Eq.Completeness}) can then be
used to show that \begin{eqnarray}
\Delta x_{i}\Delta x_{j}\sum_{k}(\phi_{kiX}\,\phi_{kjX}+\phi_{kiY}\,\phi_{kjY}) & = & \Delta x_{i}\,\delta_{i\; j}\nonumber \\
\Delta x_{i}\Delta x_{j}\sum_{k}(-\phi_{kiX}\,\phi_{kjY}+\phi_{kiY}\,\phi_{kjX}) & = & 0\label{Eq.Completeness3}\end{eqnarray}
which is the same as\begin{eqnarray}
\sum_{k\mu}\, J_{k\mu\; i\nu}J_{k\mu\; j\xi} & = & \Delta x_{i}\,\delta_{i\; j}\delta_{\nu\;\xi}\nonumber \\
\left[J^{T}\, J\right]_{i\nu\; j\xi} & = & \Delta x_{i}\,\delta_{i\; j}\delta_{\nu\;\xi}\label{Eq.JacobianMatrixResult2}\end{eqnarray}
Hence \begin{equation}
\left\Vert J_{k\mu\; i\nu}\right\Vert =\prod\limits _{i=1}^{n}(\Delta x_{i})^{2}\label{Eq.JacobianResult}\end{equation}
so that we have finally after letting $n\rightarrow\infty$ and $\Delta x_{i}\rightarrow0$
and with $d^{2}\alpha=\prod\limits _{k}d^{2}\alpha_{k}$, $d^{2}\alpha^{+}=\prod\limits _{k}d^{2}\alpha_{k}^{+}$
\begin{eqnarray}
 &  & \int\int d^{2}\alpha\; d^{2}\alpha^{+}\; f(\alpha,\alpha^{+},\alpha^{\ast},\alpha^{+\ast})\label{Eq.PhaseIntegral3}\\
 & = & \lim_{n\rightarrow\infty}\lim_{\epsilon\rightarrow0}\int\ldots\int d^{2}\alpha_{1}d^{2}\alpha_{2}..d^{2}\alpha_{k}..d^{2}\alpha_{n}\nonumber \\
 &  & \times\int\ldots\int d^{2}\alpha_{1}^{+}d^{2}\alpha_{2}^{+}..d^{2}\alpha_{k}^{+}..d^{2}\alpha_{n}^{+}\; f(\alpha_{k},\alpha_{k}^{+},\alpha_{k}^{\ast},\alpha_{k}^{+\ast})\nonumber \\
 & = & \lim_{n\rightarrow\infty}\lim_{\epsilon\rightarrow0}\int\ldots\int d^{2}\psi_{1}d^{2}\psi_{2}..d^{2}\psi_{i}..d^{2}\psi_{n}\int\ldots\int d^{2}\psi_{1}^{+}d^{2}\psi_{2}^{+}..d^{2}\psi_{i}^{+}..d^{2}\psi_{n}^{+}\;\nonumber \\
 &  & \times w(\psi_{1},..,\psi_{i},..,\psi_{n},\psi_{1}^{+},..,\psi_{i}^{+},..,\psi_{n}^{+},\psi_{1}^{\ast},..,\psi_{i}^{\ast},..,\psi_{n}^{\ast},\psi_{1}^{+\ast},..,\psi_{i}^{+\ast},..,\psi_{n}^{+\ast})\nonumber \\
 &  & \times F(\psi_{1},..,\psi_{i},..,\psi_{n},\psi_{1}^{+},..,\psi_{i}^{+},..,\psi_{n}^{+},\psi_{1}^{\ast},..,\psi_{i}^{\ast},..,\psi_{n}^{\ast},\psi_{1}^{+\ast},..,\psi_{i}^{+\ast},..,\psi_{n}^{+\ast})\nonumber \\
 & = & \int\int D^{2}\psi\; D^{2}\psi^{+}\; F[\psi(x),\psi^{+}(x),\psi^{\ast}(x),\psi^{+\ast}(x)]\label{Eq.FnalIntegral3}\end{eqnarray}
where $D^{2}\psi D^{2}\psi^{+}=\prod\limits _{i}d^{2}\psi_{i}\prod\limits _{i}d^{2}\psi_{i}\; w(\psi_{1},..,\psi_{n},\psi_{1}^{+},..,\psi_{n}^{+},\psi_{1}^{\ast},..,\psi_{n}^{\ast},\psi_{1}^{+\ast},..,\psi_{n}^{+\ast})$
and the weight function is \begin{equation}
w(\psi_{1},..,\psi_{n},\psi_{1}^{+},..,\psi_{n}^{+},\psi_{1}^{\ast},..,\psi_{n}^{\ast},\psi_{1}^{+\ast},..,\psi_{n}^{+\ast})=\prod\limits _{i=1}^{n}(\Delta x_{i})^{2}\label{Eq.WeightFn3}\end{equation}
and is independent of the functions. The power law $(\Delta x_{i})^{2}$
is consistent with there being four real functions involved instead
of the single function as previously.

\subsection{Functional Integration Rules}

A useful \emph{integration by parts }rule can often be established
from Eq.(\ref{Eq.ProdRule}). Consider the functional $H[\psi(x)]=F[\psi(x)]G[\psi(x)]$.
Then\[
F[\psi(x)]\left(\frac{\delta G[\psi(x)]}{\delta\psi(x)}\right)=\left(\frac{\delta\{F[\psi(x)]G[\psi(x)]\}}{\delta\psi(x)}\right)-\left(\frac{\delta F[\psi(x)]}{\delta\psi(x)}\right)G[\psi(x)]\]
Then\[
\int D\psi\, F[\psi(x)]\left(\frac{\delta G[\psi(x)]}{\delta\psi(x)}\right)=\int D\psi\,\left(\frac{\delta H[\psi(x)]}{\delta\psi(x)}\right)-\int D\psi\,\left(\frac{\delta F[\psi(x)]}{\delta\psi(x)}\right)G[\psi(x)]\]
If we now introduce mode expansions and use Eq.(\ref{Eq.FnalDerivResult})
for the functional derivative of $H[\psi(x)]$ and Eq.(\ref{Eq.PhaseSpaceIntegral3})
for the first of the two functional integrals on the right hand side
of the last equation then\begin{eqnarray*}
\int D\psi\,\left(\frac{\delta H[\psi(x)]}{\delta\psi(x)}\right) & = & \lim_{n\rightarrow\infty}\lim_{\epsilon\rightarrow0}\int\ldots\int d\alpha_{1}d\alpha_{2}..d\alpha_{k}..d\alpha_{n}\,\sum\limits _{k}\phi_{k}^{\ast}(x)\,\frac{\partial h(\alpha_{1},..,\alpha_{k},..)}{\partial\alpha_{k}}\\
 & = & \lim_{n\rightarrow\infty}\lim_{\epsilon\rightarrow0}\sum\limits _{k}\phi_{k}^{\ast}(x)\,\int\ldots\int d\alpha_{1}d\alpha_{2}..\\
 &  & \times\{h(\alpha_{1},..,\alpha_{k},..)_{\alpha_{k}\rightarrow+\infty}-h(\alpha_{1},..,\alpha_{k},..)_{\alpha_{k}\rightarrow-\infty}\}..d\alpha_{n}\end{eqnarray*}
so that the functional integral of this term reduces to contributions
on the boundaries of phase space. Hence if $h(\alpha_{1},..,\alpha_{k},..)\rightarrow0$
as all $\alpha_{k}\rightarrow\pm\infty$ then the functional integral
involving the functional derivative of $H[\psi(x)]$ vanishes and
we have the integration by parts result\begin{equation}
\int D\psi\, F[\psi(x)]\left(\frac{\delta G[\psi(x)]}{\delta\psi(x)}\right)=-\int D\psi\,\left(\frac{\delta F[\psi(x)]}{\delta\psi(x)}\right)G[\psi(x)]\label{Eq.FnlIntegParts}\end{equation}

All these rules have obvious generalisations for functionals such
as\\
 $F[\psi(x),\psi^{+}(x),\psi^{\ast}(x),\psi^{+\ast}(x)]$ involving
several fields.

\subsection{Restricted Functions}

It is necessary to also consider functionals involving c-number field
functions $\psi^{K}(x)$ which are still based on an expansion in
terms of orthonormal mode functions $\phi_{k}(x)$, but where there
is some restriction on the modes that are included. Such functions
will be referred to as \emph{restricted functions}. Examples include
the fields $\psi_{C}(\mathbf{r}),\psi_{C}^{+}(\mathbf{r}),\psi_{NC}(\mathbf{r}),\psi_{NC}^{+}(\mathbf{r})$
used for condensate and non-condensate modes in the theory of Bose
condensates, where even the combined condensate and non-condensate
modes are subject to a restriction, in that modes associated with
a momentum greater than a cut-off value are excluded.

Thus we have

\begin{equation}
\psi^{K}(x)=\sum\limits _{k}^{K}\beta_{k}\,\phi_{k}(x)\label{Eq.RestrictedFieldFn2}\end{equation}
where the specific restricted mode expansion for the \emph{restricted
set }$K$ is signified by the symbol $K$. Other restricted sets involving
different modes will be designated $L$, $M$ etc., with expansion
coefficients $\gamma_{k}$, $\delta_{k}$ etc.

Orthonormality conditions still apply to all modes \begin{equation}
\int dx\,\phi_{k}^{\ast}(x)\phi_{l}(x)=\delta_{kl}\label{Eq.Orthog2}\end{equation}
and this gives the well-known result for the expansion coefficients\begin{equation}
\beta_{k}\,=\int dx\,\phi_{k}^{\ast}(x)\psi^{K}(x)\label{Eq.RestrictedExpnCoefts2}\end{equation}

However the completeness relationship is now\begin{equation}
\sum\limits _{k}^{K}\phi_{k}(y)\phi_{k}^{\ast}(x)=\delta_{K}(y,x).\label{Eq.RestrictedCompleteness}\end{equation}
which defines the \emph{restricted delta function} $\delta_{K}(y,x)$
for the $K$ set. This is a function of two variables $x$ and $y$,
and does not depend on $y-x$.

The restricted delta functions have some interesting properties\begin{eqnarray}
\iint dx\, dy\,\phi_{l}^{\ast}(y)\,\delta_{K}(y,x)\,\phi_{m}(x) & = & \delta_{l,m}\qquad(l,m\in\, K)\nonumber \\
 & = & 0\qquad(l\notin K,\, m\notin K)\label{Eq.RestrictedDelta0}\end{eqnarray}
\begin{eqnarray}
\int dx\,\delta_{K}(y,x).\delta_{L}(x,z) & = & \int dx\,\sum\limits _{k}^{K}\phi_{k}(y)\phi_{k}^{\ast}(x)\,\sum\limits _{l}^{L}\phi_{l}(x)\phi_{l}^{\ast}(z)\nonumber \\
 & = & \,\sum\limits _{k}^{K}\phi_{k}(y)\,\delta_{k,l}\,\sum\limits _{l}^{L}\phi_{l}^{\ast}(z)\nonumber \\
 & = & \delta_{K,L}\,\delta_{K}(y,z)\label{Eq.RestrictedDelta1}\end{eqnarray}
and\begin{equation}
\int dx\,\delta_{K}(x,x)=N_{K}\label{Eq.RestrictedDelta2}\end{equation}
where $N_{K}$\ is the number of mode functions in the set $K$.
Unlike the normal delta function the restricted delta functions are
non-singular and can be treated as standard c-number functions within
expressions.

\subsection{Functionals of Restricted Functions}

As for general functions, a \emph{functional} $F[\psi^{K}(x)]$ of
restricted functions $\psi^{K}(x)$ maps the \emph{c-number function}
$\psi^{K}(x)$ onto a \emph{c-number} that depends on \emph{all} the
values of $\psi^{K}(x)$ over its entire range.

The \emph{restricted function} $\psi^{K}(y)$ can be expressed as
a functional $F_{y}[\psi^{K}(x)]$ of the restricted function $\psi^{K}(x)$.
In terms of the restricted delta function we have \begin{eqnarray}
\psi^{K}(y) & = & \int dx\,\delta_{K}(y,x).\psi^{K}(x)\nonumber \\
 & = & F_{y}[\psi^{K}(x)]\label{Eq.RestrictedFnalExample5}\end{eqnarray}
showing how $\psi^{K}(y)$ can still be written as a functional $F_{y}[\psi^{K}(x)]$
of $\psi^{K}(x)$, but now Eq. (\ref{Eq.RestrictedFnalExample5})
applies which involves the restricted delta function $\delta_{K}(y,x)$
as a kernal, rather than Eq. (\ref{Eq.FnalExample5}) which involved
the normal delta function and applied to functions $\psi(x)$ with
unrestricted mode expansions.

The \emph{spatial derivative }$\nabla_{y}\psi^{K}(y)$ of the restricted
function $\psi^{K}(y)$ can also be expressed as a functional $F_{\nabla y}[\psi^{K}(x)]$
of $\psi^{K}(x)$. Using (\ref{Eq.RestrictedFnalExample5}) we have
\begin{eqnarray}
\nabla_{y}\psi^{K}(y) & = & \int dx\,\nabla_{y}\delta_{K}(y,x).\psi^{K}(x)\nonumber \\
 & = & F_{\nabla y}[\psi^{K}(x)]\label{Eq.RestrictedFnalExample6}\end{eqnarray}
which now involves $\nabla_{y}\delta_{K}(y,x)$ as a kernal. We can
confirm the validity of (\ref{Eq.RestrictedFnalExample6}) by substituting
for $\psi^{K}(x)$ from (\ref{Eq.RestrictedFieldFn2}) which gives\begin{eqnarray*}
\int dx\,\nabla_{y}\delta_{K}(y,x).\psi^{K}(x) & = & \sum\limits _{k}^{K}\beta_{k}\,\int dx\,\nabla_{y}\delta_{K}(y,x).\phi_{k}(x)\\
 & = & \sum\limits _{k}^{K}\beta_{k}\,\sum\limits _{l}^{K}\int dx\,\nabla_{y}\phi_{l}(y)\phi_{l}^{\ast}(x).\phi_{k}(x)\\
 & = & \sum\limits _{k}^{K}\beta_{k}\,\nabla_{y}\phi_{k}(y)\\
 & = & \nabla_{y}\psi^{K}(y)\end{eqnarray*}
as required.

As the value of the function at any point in the range for $x$ is
determined uniquely by the expansion coefficients $\{\beta_{k}\}$,
then the functional $F[\psi^{K}(x)]$ must therefore also just depend
on the c-number expansion coefficients, and hence may also be viewed
as a \emph{function} $g(\beta_{1},\beta_{2},..,\beta_{k},..\beta_{n})$
of the \emph{expansion coefficients}, a useful equivalence when functional
differentiation and integration are considered.\begin{equation}
F[\psi^{K}(x)]\equiv g(\beta_{1},\beta_{2},..,\beta_{k},..\beta_{n})\label{Eq.EquivFunction2}\end{equation}

\subsection{Related Restricted Function Sets}

We may also consider restricted functions based on the \emph{conjugate
modes}. This set will be referred to as $K^{\ast}$ or $K^{+}$. Thus
the previous equations become\begin{eqnarray}
\psi^{K+}(x) & = & \sum\limits _{k}^{K}\phi_{k}^{\ast}(x)\beta_{k}^{+}\,\label{Eq.RestrictedFieldFn2Conj}\\
\beta_{k}^{+}\, & = & \int dx\,\phi_{k}(x)\psi^{K+}(x)\label{Eq.ExpnCoefts2Conj}\\
\delta_{K+}(y,x) & = & \sum\limits _{k}^{K}\phi_{k}^{\ast}(y)\phi_{k}(x)\label{Eq.Completeness2Conj}\end{eqnarray}
where the last equation defines the restricted delta function for
the $K^{+}$ case. We note that the restricted delta function $\delta_{K+}(y,x)$
is related to the previous one via\begin{equation}
\delta_{K+}(y,x)=\delta_{K}(x,y).\label{Eq.RestrictedDelta3}\end{equation}

We can again write the \emph{restricted function} $\psi^{K+}(y)$
as a functional $F_{y}[\psi^{K+}(x)]$ via \begin{eqnarray}
\psi^{K+}(y) & = & \int dx\,\delta_{K+}(y,x).\psi^{K+}(x)\nonumber \\
 & = & \int dx\,\delta_{K}(x,y).\psi^{K+}(y)\nonumber \\
 & = & F_{y}[\psi^{K+}(x)]\label{Eq.RestrictedFnalExample5Conj}\end{eqnarray}

Similarly the \emph{spatial derivative} $\nabla_{y}\psi^{K+}(y)$
of the restricted function is also a functional $F_{\nabla_{y}}[\psi^{K+}(x)]$
given by\begin{eqnarray}
\nabla_{y}\psi^{K+}(y) & = & \int dx\,\nabla_{y}\delta_{K+}(y,x).\psi^{K+}(x)\nonumber \\
 & = & \int dx\,\nabla_{y}\delta_{K}(x,y).\psi^{K+}(x)\nonumber \\
 & = & F_{\nabla_{y}}[\psi^{K+}(x)]\label{Eq.RestrictedFnalExample6Conj}\end{eqnarray}

Note that considered as a function of $y$, the restricted delta function
$\delta_{K}(y,x)$ is a member of the $K$ set of restricted functions
$\psi^{K}(y)$ (the expansion coefficients are $\phi_{k}^{\ast}(x)$).
On the other hand, considered as a function of $x$ the restricted
delta function $\delta_{K}(y,x)$ is a member of the \emph{conjugate
set} $K^{+}$ of mode functions $\phi_{k}^{\ast}(x)$ (the expansion
coefficients are $\phi_{k}(y)$).

As the value of the function at any point in the range for $x$ is
determined uniquely by the expansion coefficients $\{\beta_{k}^{+}\}$,
then the functional $F[\psi^{K+}(x)]$ must therefore also just depend
on the c-number expansion coefficients, and hence may also be viewed
as a \emph{function} $g^{+}(\beta_{1}^{+},\beta_{2}^{+},..,\beta_{k}^{+},..\beta_{n}^{+})$
of the \emph{expansion coefficients}, a useful equivalence when functional
differentiation and integration are considered.

\begin{equation}
F[\psi^{K+}(x)]\equiv g^{+}(\beta_{1}^{+},\beta_{2}^{+},..,\beta_{k}^{+},..\beta_{n}^{+})\label{Eq.EquivFunction3}\end{equation}

A second related restricted set is the \emph{complementary set }$\overline{K}$
which includes all the other orthonormal mode functions not included
in the $K$ set.

Clearly, \emph{any} function can be expanded in terms of modes in
the $K$ and $\overline{K}$ restricted sets. Thus we now have\begin{eqnarray}
\psi(x) & = & \sum\limits _{k}^{K}\gamma_{k}\,\phi_{k}(x)+\sum\limits _{k}^{\overline{K}}\gamma_{k}\,\phi_{k}(x)\label{Eq.RestrictedFieldFn3Compl}\\
\gamma_{k}\, & = & \int dx\,\phi_{k}^{\ast}(x)\psi(x)\qquad k\in K,\overline{K}\label{Eq.ExpnCoefts3}\\
\delta_{\overline{K}}(y,x) & = & \sum\limits _{k}^{\overline{K}}\phi_{k}(y)\phi_{k}^{\ast}(x)=\sum\limits _{L\neq K}\delta_{L}(y,x)\label{Eq.RestrictedDelta4}\end{eqnarray}
and now the \emph{full Dirac delta function} is \begin{equation}
\delta(y,x)=\sum\limits _{k}^{K}\phi_{k}(y)\phi_{k}^{\ast}(x)+\sum\limits _{k}^{\overline{K}}\phi_{k}(y)\phi_{k}^{\ast}(x)\label{Eq.FullDeltaFn}\end{equation}
The general function $\psi(y)$ may be written as a functional $F_{y}[\psi(x)]$
of $\psi(x)$ involving the full delta function \begin{eqnarray}
\psi(y) & = & \int dx\,\delta(y,x).\psi(x)\nonumber \\
 & = & F_{y}[\psi(x)]\label{Eq.RestrictedFnalExample6Compl}\end{eqnarray}

Applying (\ref{Eq.RestrictedDelta1}) we obtain the interesting result
\begin{equation}
\int dx\,\delta_{K}(y,x).\delta_{\overline{K}}(x,z)=0\label{Eq.RestrictedDelta5}\end{equation}

Note that the full delta function is still written as a function of
$x$ and $y$. Because the total set of functions is still restricted
it will have a narrow though finite width and can be treated like
a normal function.

\subsection{Functional Derivatives for Restricted Functions}

The \emph{functional derivative} $\frac{{\LARGE\delta F[\psi}^{K}{\LARGE(x)]}}{{\LARGE\delta\psi}^{K}{\LARGE(x)}}$
is defined by \begin{equation}
F[\psi^{K}(x)+\delta\psi^{K}(x)]\doteqdot F[\psi^{K}(x)]+\int dx\,\delta\psi^{K}(x)\,\left(\frac{\delta F[\psi^{K}(x)]}{\delta\psi^{K}(x)}\right)_{x}\label{Eq.RestrictedFuncDeriv1}\end{equation}
where $\delta\psi^{K}(x)$ is a small change in $\psi^{K}(x)$. Since
as in (\ref{Eq.EquivFunction2}) the functional is equivalent to a
function of the expansion coefficients $\beta_{k}$, the only meaningful
change to $\psi^{K}(x)$ would be associated with changes $\delta\beta_{k}$
in these expansion coefficients and thus $\delta\psi^{K}(x)$ will
be within the $K$ restricted function space. In this equation the
left side is a functional of $\psi^{K}(x)+\delta\psi^{K}(x)$ and
the first term on the right side is a functional of $\psi^{K}(x)$.
The second term on the right side is a functional of $\delta\psi^{K}(x)$
and thus the functional derivative must be a function of $x$, hence
the subscript $x$. In most situations this subscript will be left
understood.

Thus the functional derivative will be defined in terms of changes
to the restricted function of the form\begin{equation}
\delta\psi^{K}(x)=\sum\limits _{k}^{K}\delta\beta_{k}\,\phi_{k}(x)\label{Eq.RestrictedFuncFluctn}\end{equation}

As the functional derivative is just a function of $x$ we can expand
it in terms of \emph{all} the \emph{conjugate} modes (these also form
a full basis set of orthogonal functions)\[
\left(\frac{\delta F[\psi^{K}(x)]}{\delta\psi^{K}(x)}\right)_{x}=\sum\limits _{l}\eta_{l}\,\phi_{l}^{\ast}(x)\]
then we have\begin{eqnarray*}
\int dx\,\delta\psi^{K}(x)\,\left(\frac{\delta F[\psi^{K}(x)]}{\delta\psi^{K}(x)}\right)_{x} & = & \sum\limits _{l}\eta_{l}\int dx\,\delta\psi^{K}(x)\phi_{l}^{\ast}(x)\,\\
 & = & \sum\limits _{k}^{K}\eta_{k}\delta\beta_{k}\end{eqnarray*}
since the contributions from modes $l$ not in the $K^{+}$ set will
be zero using orthogonality. This shows that any contribution to the
functional derivative from modes $\phi_{l}^{\ast}(x)$ outside the
$K^{+}$ set cannot contribute to $F[\psi^{K}(x)+\delta\psi^{K}(x)]-F[\psi^{K}(x)]$,
and hence can be arbitarily set to zero in determining the functional
derivative with respect to restricted functions $\psi^{K}(x)$ in
the $K$ set. Thus we have\begin{equation}
\left(\frac{\delta F[\psi^{K}(x)]}{\delta\psi^{K}(x)}\right)_{x}=\sum\limits _{k}^{K}\eta_{k}\,\phi_{k}^{\ast}(x)\label{Eq.RestrictedFuncDeriv2}\end{equation}
showing that the functional derivative is a function in the $K^{+}\equiv K^{\ast}$
set.

Noting that the function $\delta_{K}(x,y)$ is within the restricted
function space, we may obtain a useful expression for the functional
derivative by applying (\ref{Eq.RestrictedFnalExample5}) for a function
in the $K^{\ast}$ set. Since $\delta_{K^{+}}(y,x)=\delta_{K}(x,y)$
this shows that the functional derivative may be obtained by choosing
$\delta\psi^{K}(x)=\epsilon\delta_{K}(x,y)$ for small $\epsilon$
in the definition (\ref{Eq.RestrictedFuncDeriv1})

\begin{eqnarray*}
\left(\frac{\delta F[\psi^{K}(x)]}{\delta\psi^{K}(x)}\right)_{y} & = & \int dx\,\delta_{K^{+}}(y,x)\left(\frac{\delta F[\psi^{K}(x)]}{\delta\psi^{K}(x)}\right)_{x}\\
 & = & \lim_{\epsilon\rightarrow0}\left(\frac{F[\psi^{K}(x)+\epsilon\delta_{K}(x,y)]-F[\psi^{K}(x)]}{\epsilon}\right).\end{eqnarray*}
To confirm that the right side of the last equation does in fact give
$\left(\frac{{\LARGE\delta F[\psi}^{K}{\LARGE(x)]}}{{\LARGE\delta\psi}^{K}{\LARGE(x)}}\right)_{y}$
we substitute from (\ref{Eq.RestrictedFuncDeriv2})\begin{eqnarray*}
\int dx\,\delta_{K}(x,y)\left(\frac{\delta F[\psi^{K}(x)]}{\delta\psi^{K}(x)}\right)_{x} & = & \sum\limits _{k}^{K}\eta_{k}\,\int dx\,\delta_{K}(x,y)\phi_{k}^{\ast}(x)\\
 & = & \sum\limits _{k}^{K}\eta_{k}\,\sum\limits _{l}^{K}\phi_{l}^{\ast}(y)\int dx\,\phi_{l}(x)\,\phi_{k}^{\ast}(x)\\
 & = & \sum\limits _{k}^{K}\eta_{k}\,\phi_{k}^{\ast}(y)\\
 & = & \left(\frac{\delta F[\psi^{K}(x)]}{\delta\psi^{K}(x)}\right)_{y}\end{eqnarray*}
as required. Thus we have the useful expression for the functional
derivative of restricted functions\begin{equation}
\left(\frac{\delta F[\psi^{K}(x)]}{\delta\psi^{K}(x)}\right)_{y}=\lim_{\epsilon\rightarrow0}\left(\frac{F[\psi^{K}(x)+\epsilon\delta_{K}(x,y)]-F[\psi^{K}(x)]}{\epsilon}\right)\label{Eq.RestrictedFuncDeriv3}\end{equation}

We may also have functionals $F[\psi^{K}(x),\psi^{L}(x)]$ that involve
two functions $\psi^{K}(x)$, $\psi^{L}(x)$ in two different restricted
sets $K$, $L$. The straight-forward generalisation of useful result
(\ref{Eq.RestrictedFuncDeriv3}) is\begin{eqnarray}
\left(\frac{\delta F[\psi^{K}(x),\psi^{L}(x)]}{\delta\psi^{K}(x)}\right)_{y} & = & \lim_{\epsilon\rightarrow0}\left(\frac{F[\psi^{K}(x)+\epsilon\delta_{K}(x,y),\psi^{L}(x)]-F[\psi^{K}(x),\psi^{L}(x)]}{\epsilon}\right)\nonumber \\
 &  & \,\label{eq:RestrictedFuncDeriv5}\\
\left(\frac{\delta F[\psi^{K}(x),\psi^{L}(x)]}{\delta\psi^{L}(x)}\right)_{y} & = & \lim_{\epsilon\rightarrow0}\left(\frac{F[\psi^{K}(x),\psi^{L}(x)+\epsilon\delta_{L}(x,y)]-F[\psi^{K}(x),\psi^{L}(x)]}{\epsilon}\right)\nonumber \\
 &  & \,\label{eq:RestrictedFuncDeriv6}\end{eqnarray}

Similar results apply for the functional derivative $\frac{{\LARGE\delta F[\psi^{K+}(x)]}}{{\LARGE\delta\psi^{K+}(x)}}$
with respect to the restricted function $\psi^{K+}(x)$ in the $K^{+}\equiv K^{\ast}$
set, which is defined by \begin{equation}
F[\psi^{K+}(x)+\delta\psi^{K+}(x)]\doteqdot F[\psi^{K+}(x)]+\int dx\,\delta\psi^{K+}(x)\,\left(\frac{\delta F[\psi^{K+}(x)]}{\delta\psi^{K+}(x)}\right)_{x}\label{Eq.RestrictedFuncDeriv1Conj}\end{equation}
where $\delta\psi^{K+}(x)$ is a small change in $\psi^{K+}(x)$.
The function $\delta\psi^{K+}(x)$ be associated with changes $\delta\beta_{k}^{+}$
in these expansion coefficients and thus $\delta\psi^{K+}(x)$ will
be within the $K^{+}$ restricted function space. We then have\begin{equation}
\left(\frac{\delta F[\psi^{K+}(x)]}{\delta\psi^{K+}(x)}\right)_{x}=\sum\limits _{k}^{K}\eta_{k}^{+}\,\phi_{k}(x)\label{Eq.RestrictedFuncDeriv2Conj}\end{equation}
showing that the functional derivative is a function in the $K$ set.

Also the function $\delta_{K+}(x,y)$ is within the restricted function
space, we may obtain a useful expression for the functional derivative
as\begin{eqnarray}
\left(\frac{\delta F[\psi^{K+}(x)]}{\delta\psi^{K+}(x)}\right)_{y} & = & \lim_{\epsilon\rightarrow0}\left(\frac{F[\psi^{K+}(x)+\epsilon\delta_{K+}(x,y)]-F[\psi^{K+}(x)]}{\epsilon}\right)\nonumber \\
 & = & \lim_{\epsilon\rightarrow0}\left(\frac{F[\psi^{K+}(x)+\epsilon\delta_{K}(y,x)]-F[\psi^{K+}(x)]}{\epsilon}\right)\label{Eq.RestrictedFuncDeriv3bConj}\end{eqnarray}

\subsection{Examples of Restricted Functional Derivatives}

To obtain the functional derivative of the \emph{function} $\psi^{K}(y)$
with respect to $\psi^{K}(x)$, we note that this derivative exists
as a function of $x$ in the $K^{+}\equiv K^{\ast}$ restricted set
since $\psi^{K}(y)$ is also a functional $\psi^{K}(y)=F_{y}[\psi^{K}(x)]$.
We can thus use the expression (\ref{Eq.RestrictedFuncDeriv3})\begin{eqnarray}
\left(\frac{\delta}{\delta\psi^{K}(x)}\psi^{K}(y)\right)_{x} & = & \lim_{\epsilon\rightarrow0}\left(\frac{F_{y}[\psi^{K}(u)+\epsilon\delta_{K}(u,x\mathbf{)}]-F_{y}[\psi^{K}(u)]}{\epsilon}\right)\nonumber \\
 & = & \lim_{\epsilon\rightarrow0}\left(\frac{\int du\,\mathbf{\delta}_{K}(y,u)\{\psi^{K}(u)+\epsilon\delta_{K}(u,x\mathbf{)}\}-\int du\mathbf{\,\delta}_{K}(y,u)\psi^{K}(u)}{\epsilon}\right)\nonumber \\
 & = & \lim_{\epsilon\rightarrow0}\left(\int du\,\mathbf{\delta}_{K}(y,u)\,\delta_{K}(u,x\mathbf{)}\right)\nonumber \\
 & = & \mathbf{\delta}_{K}(y,x)\label{Eq.RestrictedFuncDerivExample2}\end{eqnarray}
where (\ref{Eq.RestrictedDelta1}) has been used. As noted before
considered as a function of $x$, the derivative of $\psi^{K}(y)$
with respect to $\psi^{K}(x)$ is in the $K^{\ast}$ restricted set,
but is in the $K$ set considered as a function of $y$. This result
is the modification of (\ref{Eq.FuncDerivExample2}) for restricted
functions.

A further result can be derived for when the functional derivative
is with respect to $\psi^{L}(x)$ is in a different $L$ restricted
set. Applying (\ref{eq:RestrictedFuncDeriv6}) we get \begin{eqnarray}
\left(\frac{\delta}{\delta\psi^{L}(x)}\psi^{K}(y)\right)_{x} & = & \lim_{\epsilon\rightarrow0}\left(\frac{F_{y}[\psi^{K}(u),\psi^{L}(u)+\epsilon\delta_{L}(u,x\mathbf{)}]-F_{y}[\psi^{K}(u),\psi^{L}(u)]}{\epsilon}\right)\nonumber \\
 & = & \lim_{\epsilon\rightarrow0}\left(\frac{\int du\,\mathbf{\delta}_{K}(y,u)\psi^{K}(u)-\int du\mathbf{\,\delta}_{K}(y,u)\psi^{K}(u)}{\epsilon}\right)\nonumber \\
 & = & 0\label{Eq.RestrictedFuncDerivExample3}\end{eqnarray}
since the functional $\psi^{K}(y)=F_{y}[\psi^{K}(x),\psi^{L}(x)]$
does not involve $\psi^{L}(x)$ at all.

For the functional derivative of the \emph{spatial derivative} $\nabla_{y}\psi^{K}(y)$
of the \emph{function} $\psi^{K}(y)$ with respect to $\psi^{K}(x)$,
we note that this derivative exists as a function of $x$ in the $K^{+}\equiv K^{\ast}$
restricted set since $\nabla_{y}\psi^{K}(y)$ is also a functional
$\nabla_{y}\psi^{K}(y)=F_{\nabla_{y}}[\psi^{K}(x)]$. We can thus
use the expression (\ref{Eq.RestrictedFuncDeriv3})\begin{eqnarray}
\left(\frac{\delta}{\delta\psi^{K}(x)}\nabla_{y}\psi^{K}(y)\right)_{x} & = & \lim_{\epsilon\rightarrow0}\left(\frac{F_{y}[\psi^{K}(u)+\epsilon\delta_{K}(u,x\mathbf{)}]-F_{y}[\psi^{K}(u)]}{\epsilon}\right)\nonumber \\
 & = & \lim_{\epsilon\rightarrow0}\left(\frac{\int du\,\nabla_{y}\mathbf{\delta}_{K}(y,u)\{\psi^{K}(u)+\epsilon\delta_{K}(u,x\mathbf{)}\}-\int du\nabla_{y}\mathbf{\,\delta}_{K}(y,u)\psi^{K}(u)}{\epsilon}\right)\nonumber \\
 & = & \lim_{\epsilon\rightarrow0}\left(\int du\,\nabla_{y}\mathbf{\delta}_{K}(y,u)\,\delta_{K}(u,x\mathbf{)}\right)\nonumber \\
 & = & \nabla_{y}\left(\int du\,\mathbf{\delta}_{K}(y,u)\,\delta_{K}(u,x\mathbf{)}\right)\nonumber \\
 & = & \nabla_{y}\mathbf{\delta}_{K}(y,x)\label{Eq.RestrictedFuncDerivExample5}\end{eqnarray}
showing that the functional derivative involves a spatial derivative
of the restricted delta function with respect to $y$. As pointed
out previously, considered as a function of $x$ the functional derivative
is in the $K^{+}$ set. Note also that we see that the functional
derivative and the spatial derivative processes can be carried out
in either order\begin{eqnarray}
\left(\frac{\delta}{\delta\psi^{K}(x)}\nabla_{y}\psi^{K}(y)\right)_{x} & = & \nabla_{y}\left(\frac{\delta}{\delta\psi^{K}(x)}\psi^{K}(y)\right)_{x}\nonumber \\
 & = & \nabla_{y}\mathbf{\delta}_{K}(y,x)\label{Eq.RestrictedFuncDerivExample6}\end{eqnarray}

We also can obtain similar results for the \emph{function} $\psi^{K+}(y)$
which is in the $K^{+}\equiv K^{\ast}$ restricted set, and can be
written as a functional $F_{y}[\psi^{K+}(x)]\equiv\psi^{K+}(y)$.
Thus\begin{eqnarray}
\left(\frac{\delta}{\delta\psi^{K+}(x)}\psi^{K+}(y)\right)_{x} & = & \lim_{\epsilon\rightarrow0}\left(\frac{F_{y}[\psi^{K+}(u)+\epsilon\delta_{K+}(u,x\mathbf{)}]-F_{y}[\psi^{K+}(u)]}{\epsilon}\right)\nonumber \\
 & = & \lim_{\epsilon\rightarrow0}\left(\frac{\int du\,\mathbf{\delta}_{K^{+}}(y,u)\{\psi^{K+}(u)+\epsilon\delta_{K+}(u,x\mathbf{)}\}-\int du\mathbf{\,\delta}_{K+}(y,u)\psi^{K+}(u)}{\epsilon}\right)\nonumber \\
 & = & \lim_{\epsilon\rightarrow0}\left(\int du\,\mathbf{\delta}_{K+}(y,u)\,\delta_{K+}(u,x\mathbf{)}\right)\nonumber \\
 & = & \mathbf{\delta}_{K+}(y,x)\nonumber \\
 & = & \mathbf{\delta}_{K}(x,y)\label{Eq.RestrictedFnalDerivResult5B}\end{eqnarray}

For the \emph{spatial derivative} $\nabla_{y}\psi^{K+}(y)$ of the
\emph{function} $\psi^{K+}(y)$ we have\begin{eqnarray}
\left(\frac{\delta}{\delta\psi^{K+}(x)}\nabla_{y}\psi^{K+}(y)\right)_{x} & = & \lim_{\epsilon\rightarrow0}\left(\frac{F_{\nabla_{y}}[\psi^{K+}(u)+\epsilon\delta_{K+}(u,x\mathbf{)}]-F_{\nabla_{y}}[\psi^{K+}(u)]}{\epsilon}\right)\nonumber \\
 & = & \lim_{\epsilon\rightarrow0}\left(\frac{\int du\,\nabla_{y}\mathbf{\delta}_{K+}(y,u)\{\psi^{K+}(u)+\epsilon\delta_{K+}(u,x\mathbf{)}\}-\int du\mathbf{\,\nabla}_{y}\mathbf{\delta}_{K+}(y,u)\psi^{K+}(u)}{\epsilon}\right)\nonumber \\
 & = & \lim_{\epsilon\rightarrow0}\left(\int du\nabla_{y}\mathbf{\delta}_{K+}(y,u)\,\delta_{K+}(u,x\mathbf{)}\right)\nonumber \\
 & = & \left(\int du\nabla_{y}\mathbf{\delta}_{K}(u,y)\,\delta_{K}(x,u\mathbf{)}\right)\nonumber \\
 & = & \nabla_{y}\mathbf{\delta}_{K}(x,y)\nonumber \\
 & = & \nabla_{y}\mathbf{\delta}_{K+}(y,x)\label{Eq.RestrictedFuncDerivExample7}\end{eqnarray}
as expected. Note also \begin{eqnarray}
\left(\frac{\delta}{\delta\psi^{K+}(x)}\nabla_{y}\psi^{K+}(y)\right)_{x} & = & \nabla_{y}\left(\frac{\delta}{\delta\psi^{K+}(x)}\psi^{K+}(y)\right)_{x}\nonumber \\
 & = & \nabla_{y}\mathbf{\delta}_{K+}(y,x)\label{Eq.RestrictedFnalDerivResult8}\end{eqnarray}

\subsection{Restricted Functional Derivatives and Mode Functions}

We can obtain an expression for the functional derivative $\left(\frac{{\LARGE\delta F[\psi}^{K}{\LARGE(x)]}}{{\LARGE\delta\psi}^{K}{\LARGE(x)}}\right)_{x}$with
respect to restricted function $\psi^{K}(x)$ in terms of the ordinary
derivatives of the function (\ref{Eq.EquivFunction2}) that is equivalent
to the functional.

Substituting from (\ref{Eq.RestrictedFuncFluctn}) we see that \begin{eqnarray*}
F[\psi^{K}(x)+\delta\psi^{K}(x)]-F[\psi^{K}(x)] & \doteqdot & \int dx\,\delta\psi^{K}(x)\,\left(\frac{\delta F[\psi^{K}(x)]}{\delta\psi^{K}(x)}\right)_{x}\\
 & \doteqdot & \sum\limits _{k}^{K}\delta\beta_{k}\int dx\,\phi_{k}(x)\,\left(\frac{\delta F[\psi(x)]}{\delta\psi(x)}\right)_{x}\end{eqnarray*}
But the left side is the same as\[
g(\beta_{1}+\delta\beta_{1},..,\beta_{k}+\delta\beta_{k},..)-g(\beta_{1},..,\beta_{k},..)\doteqdot\sum\limits _{k}^{K}\delta\beta_{k}\,\frac{\partial g(\beta_{1},..,\beta_{k},..)}{\partial\beta_{k}}\]
Equating the coefficients of the independent $\delta\alpha_{k}$ and
then using the completeness relationship in Eq.(\ref{Eq.Completeness2Conj})
gives the key result\begin{equation}
\left(\frac{\delta F[\psi^{K}(x)]}{\delta\psi^{K}(x)}\right)_{x}=\sum\limits _{k}^{K}\phi_{k}^{\ast}(x)\,\frac{\partial g(\beta_{1},..,\beta_{k},..)}{\partial\beta_{k}}\label{Eq.RestrictedFnalDerivResult}\end{equation}
This relates the functional derivative to the mode functions and to
the ordinary partial derivatives of the function $g(\beta_{1},..,\beta_{k},..\beta_{n})$
that was equivalent to the original functional $F[\psi^{K}(x)]$.
Again, we see that the result is a function of $x$. Note that the
functional derivative involves an expansion in terms of the conjugate
mode functions $\phi_{k}^{\ast}(x)$ rather than the original modes
$\phi_{k}(x)$.

The last result can be put in the form of a useful \emph{operator
identity}\begin{equation}
\left(\frac{\delta}{\delta\psi^{K}(x)}\right)_{x}=\sum\limits _{k}^{K}\phi_{k}^{\ast}(x)\,\frac{\partial}{\partial\beta_{k}}\label{Eq.RestrictedFnalDerivResultB}\end{equation}
where it is understood that the left side operates on an arbitary
functional $F[\psi^{K}(x)]$ of the restricted function $\psi^{K}(x)$
and the right side operates on the equivalent function $g(\beta_{1},..,\beta_{k},..)$.

We can obtain a similar expression for the functional derivative $\left(\frac{{\LARGE\delta F[\psi^{K+}(x)]}}{{\LARGE\delta\psi^{K+}(x)}}\right)_{x}$with
respect to restricted function $\psi^{K+}(x)$ in the $K^{+}$ set
in terms of the ordinary derivatives of the function (\ref{Eq.EquivFunction3})
that is equivalent to the functional.\begin{equation}
\left(\frac{\delta F[\psi^{K+}(x)]}{\delta\psi^{K+}(x)}\right)_{x}=\sum\limits _{k}^{K}\phi_{k}(x)\,\frac{\partial g^{+}(\beta_{1}^{+},..,\beta_{k}^{+},..)}{\partial\beta_{k}^{+}}\label{Eq.RestrictedFnalDerivResult2}\end{equation}

The last result can be put in the form of a useful operator identity\begin{equation}
\left(\frac{\delta}{\delta\psi^{K+}(x)}\right)_{x}=\sum\limits _{k}^{K}\phi_{k}(x)\,\frac{\partial}{\partial\beta_{k}^{+}}\label{Eq.RestrictedFnalDerivResult2B}\end{equation}
where it is understood that the left side operates on an arbitary
functional $F[\psi^{K+}(x)]$ of the restricted function $\psi^{K+}(x)$
and the right side operates on the equivalent function $g^{+}(\beta_{1}^{+},..,\beta_{k}^{+},..)$.

The \emph{spatial derivative} of a \emph{functional derivative} can
be found from\begin{eqnarray}
\nabla_{x}\left(\frac{\delta F[\psi^{K}(x)]}{\delta\psi^{K}(x)}\right)_{x} & = & \sum\limits _{k}^{K}\{\nabla_{x}\phi_{k}^{\ast}(x)\}\,\frac{\partial g(\beta_{1},..,\beta_{k},..)}{\partial\beta_{k}}\label{Eq.RestrictedFnalDerivResult3}\\
\nabla_{x}\left(\frac{\delta F[\psi^{K+}(x)]}{\delta\psi^{K+}(x)}\right)_{x} & = & \sum\limits _{k}^{K}\{\nabla_{x}\phi_{k}(x)\}\,\frac{\partial g^{+}(\beta_{1}^{+},..,\beta_{k}^{+},..)}{\partial\beta_{k}^{+}}\label{Eq.RestrictedFnalDerivResult3B}\end{eqnarray}
in the two cases of functionals of $\psi^{K}(x)$ or $\psi^{K+}(x)$.
Clearly the spatial derivative acts \emph{only} on either the $\phi_{k}^{\ast}(x)$
or the $\phi_{k}(x)$.

The last results can be put in the form of \emph{operator identities}\begin{eqnarray}
\nabla_{x}\left(\frac{\delta}{\delta\psi^{K}(x)}\right)_{x} & = & \sum\limits _{k}^{K}\{\nabla_{x}\phi_{k}^{\ast}(x)\}\,\frac{\partial}{\partial\beta_{k}}\label{Eq.RestrictedFnalDerivResult4}\\
\nabla_{x}\left(\frac{\delta}{\delta\psi^{K+}(x)}\right)_{x} & = & \sum\limits _{k}^{K}\{\nabla_{x}\phi_{k}(x)\}\,\frac{\partial}{\partial\beta_{k}^{+}}\label{Eq.RestrictedFnalDerivResult4B}\end{eqnarray}
where it is understood that the left side operates on an arbitary
functional $F[\psi^{K}(x)]$ or $F[\psi^{K+}(x)]$ of the restricted
function $\psi^{K}(x)$ or $\psi^{K+}(x)$ respectively, and the right
side operates on the equivalent function $g(\beta_{1},..,\beta_{k},..)$
or $g^{+}(\beta_{1}^{+},..,\beta_{k}^{+},..)$. These operator forms
are useful in deriving results for applying functional derivatives
in succession.

As an example of applying these operator identities consider the case
of the functionals $F_{y}[\psi^{K}(x)]\equiv\psi^{K}(y)=\sum\limits _{k}\beta_{k}\phi_{k}(y)\,$
and $F_{y}[\psi^{K+}(x)]\equiv\psi^{K+}(y)=\sum\limits _{k}\phi_{k}^{\ast}(y)\beta_{k}^{+}\,$.
Since in these cases \[
\frac{\partial g(\beta_{1},..,\beta_{k},..)}{\partial\beta_{k}}=\phi_{k}(y)\qquad\frac{\partial g^{+}(\beta_{1}^{+},..,\beta_{k}^{+},..)}{\partial\beta_{k}^{+}}=\phi_{k}^{\ast}(y)\]
we have\begin{eqnarray*}
\left(\frac{\delta\psi^{K}(y)}{\delta\psi^{K}(x)}\right)_{x} & = & \sum\limits _{k}^{K}\phi_{k}^{\ast}(x)\,\phi_{k}(y)=\delta_{K}(y,x)\\
\left(\frac{\delta\psi^{K+}(y)}{\delta\psi^{K+}(x)}\right)_{x} & = & \sum\limits _{k}^{K}\phi_{k}(x)\,\phi_{k}^{\ast}(y)=\delta_{K+}(y,x)\end{eqnarray*}
for the functional derivatives as before, and \begin{eqnarray}
\nabla_{x}\left(\frac{\delta\psi^{K}(y)}{\delta\psi^{K}(x)}\right)_{x} & = & \sum\limits _{k}^{K}\{\nabla_{x}\phi_{k}^{\ast}(x)\}\phi_{k}(y)\,=\nabla_{x}\delta_{K}(y,x)\label{Eq.RestrictedFnalDerivResult5}\\
\nabla_{x}\left(\frac{\delta\psi^{K+}(y)}{\delta\psi^{K+}(x)}\right)_{x} & = & \sum\limits _{k}^{K}\{\nabla_{x}\phi_{k}(x)\}\,\phi_{k}^{\ast}(y)=\nabla_{x}\delta_{K+}(y,x)\label{Eq.RestrictedFnalDerivResult5b}\end{eqnarray}
for the spatial derivatives of the functional derivatives.

Similarly for the spatial derivative functionals $F_{\nabla y}[\psi^{K}(x)]\equiv\nabla_{y}\psi^{K}(y)=\sum\limits _{k}\beta_{k}\nabla_{y}\phi_{k}(y)\,$
and $F_{\nabla y}[\psi^{K+}(x)]\equiv\nabla_{y}\psi^{K+}(y)=\sum\limits _{k}\nabla_{y}\phi_{k}^{\ast}(y)\beta_{k}^{+}\,$.
Since in these cases \[
\frac{\partial g(\beta_{1},..,\beta_{k},..)}{\partial\beta_{k}}=\nabla_{y}\phi_{k}(y)\qquad\frac{\partial g^{+}(\beta_{1}^{+},..,\beta_{k}^{+},..)}{\partial\beta_{k}^{+}}=\nabla_{y}\phi_{k}^{\ast}(y)\]
we have\begin{eqnarray*}
\left(\frac{\delta\,\nabla_{y}\psi^{K}(y)}{\delta\psi^{K}(x)}\right)_{x} & = & \sum\limits _{k}^{K}\phi_{k}^{\ast}(x)\,\nabla_{y}\phi_{k}(y)=\nabla_{y}\delta_{K}(y,x)\\
\left(\frac{\delta\,\nabla_{y}\psi^{K+}(y)}{\delta\psi^{K+}(x)}\right)_{x} & = & \sum\limits _{k}^{K}\phi_{k}(x)\,\nabla_{y}\phi_{k}^{\ast}(y)=\nabla_{y}\delta_{K+}(y,x)\end{eqnarray*}
which are the same results as before. Note the distinction between
$\left(\frac{{\LARGE\delta\,\nabla}_{y}{\LARGE\psi}^{K}{\LARGE(y)}}{{\LARGE\delta\psi}^{K}{\LARGE(x)}}\right)_{x}$
and $\nabla_{x}\left(\frac{{\LARGE\delta\psi}^{K}{\LARGE(y)}}{{\LARGE\delta\psi}^{K}{\LARGE(x)}}\right)_{x}$-
the first being the functional derivative of the $y$ spatial derivative
$\nabla_{y}\psi^{K}(y)$ with respect to $\psi^{K}(x)$, the second
being the $x$ spatial derivative of the functional derivative of
$\psi^{K}(y)$ with respect to $\psi^{K}(x)$. \[
\]
\[
\]

\subsection{Functional Derivatives in Theory of Bose-Einstein Condensates}

The theory of Bose-Einstein condensates (BEC) often requires separate
consideration of certain highly occupied modes - the condensate modes,
and other sparsely occupied modes - the non-condensate modes. In phase
space distribution functional methods these two types of modes can
be used in defining condensate fields and non-condensate fields as
restricted functions, and the treatment presented in this section
can then be used in evaluating the various functional derivatives.

In applying these rules to the BEC problem, the following functional
derivative results can be obtained as straightforward generalisations
of (\ref{Eq.RestrictedFuncDerivExample2}) and (\ref{Eq.RestrictedFuncDerivExample3}).
The general functions $\psi(\mathbf{r})$ and $\psi^{+}(\mathbf{r})$
each will be used to cover the results for condensate and non-condensate
modes. For the case where $\psi(\mathbf{r})\equiv\psi_{C}(\mathbf{r})$
the restricted set $K$ refers to two modes $\phi_{1}(\mathbf{r})$,
$\phi_{2}(\mathbf{r})$, and for the non-condensate case where $\psi(\mathbf{r})\equiv\psi_{NC}(\mathbf{r})$
the restricted set $\overline{K}$ refers to the remaining modes $\phi_{k}(\mathbf{r})$.
For the case where $\psi^{+}(\mathbf{r})\equiv\psi_{C}^{+}(\mathbf{r})$
the restricted set $K^{+}$ $\equiv K^{\ast}$ refers to two conjugate
modes $\phi_{1}^{\ast}(\mathbf{r})$, $\phi_{2}^{\ast}(\mathbf{r})$,
and for the non-condensate case where $\psi^{+}(\mathbf{r})\equiv\psi_{NC}^{+}(\mathbf{r})$
the restricted set $\overline{K}^{+}$ refers to the remaining conjugate
modes $\phi_{k}^{\ast}(\mathbf{r})$. Because the coefficients are
unrelated we are dealing with functionals such as the distribution
functional\\
 $P[\psi_{C}(\mathbf{r}),\psi_{C}^{+}(\mathbf{r}),\psi_{NC}(\mathbf{r}),\psi_{NC}^{+}(\mathbf{r}),\psi_{C}^{\ast}(\mathbf{r}),\psi_{C}^{+\ast}(\mathbf{r}),\psi_{NC}^{\ast}(\mathbf{r}),\psi_{NC}^{+\ast}(\mathbf{r})]$
which involve eight independent functions, namely $\psi_{C}(\mathbf{r}),\psi_{C}^{+}(\mathbf{r}),\psi_{NC}(\mathbf{r}),\psi_{NC}^{+}(\mathbf{r})$
plus the complex conjugates $\psi_{C}^{\ast}(\mathbf{r}),\psi_{C}^{+\ast}(\mathbf{r}),\psi_{NC}^{\ast}(\mathbf{r}),\psi_{NC}^{+\ast}(\mathbf{r})$.
\begin{eqnarray}
\frac{\delta}{\delta\psi(\mathbf{s})}\psi(\mathbf{r}) & = & \delta_{K}(\mathbf{r,s})\qquad\frac{\delta}{\delta\psi^{+}(\mathbf{s})}\psi^{+}(\mathbf{r})=\delta_{K+}(\mathbf{r,s})=\delta_{K}(\mathbf{s,r})\nonumber \\
\frac{\delta}{\delta\psi(\mathbf{s})}\psi^{+}(\mathbf{r}) & = & 0\qquad\frac{\delta}{\delta\psi^{+}(\mathbf{s})}\psi(\mathbf{r})=0\end{eqnarray}
Note the reverse order of $\mathbf{r,s}$ in the second result, due
to (\ref{Eq.RestrictedFuncDerivExample3}). The functional $\psi(\mathbf{r})$
is not a functional of $\psi^{+}(\mathbf{s})$ and vice-versa, the
other two functional derivatives are zero. Similarly the functional
derivatives of condensate fiels with respect to non-condensate fields
are zero, and vice-versa. Thus\begin{eqnarray}
\frac{\delta}{\delta\psi_{C}(\mathbf{s})}\psi_{NC}(\mathbf{r}) & = & 0\qquad\frac{\delta}{\delta\psi_{C}^{+}(\mathbf{s})}\psi_{NC}^{+}(\mathbf{r})=0\nonumber \\
\frac{\delta}{\delta\psi_{C}(\mathbf{s})}\psi_{NC}^{+}(\mathbf{r}) & = & 0\qquad\frac{\delta}{\delta\psi_{C}^{+}(\mathbf{s})}\psi_{NC}(\mathbf{r})=0\end{eqnarray}
with four other results obtained by interchanging $C$ and $NC$.

\subsection{Supplementary Equations}

~

Field Expansions\begin{eqnarray}
\psi_{C}(\mathbf{r}) & = & \alpha_{1}\,\phi_{1}(\mathbf{r})+\alpha_{2}\,\phi_{2}(\mathbf{r})\label{Eq.FieldFn1-1}\\
\psi_{C}^{+}(\mathbf{r}) & = & \phi_{1}^{\ast}(\mathbf{r})\,\alpha_{1}^{+}+\phi_{2}^{\ast}(\mathbf{r})\,\alpha_{2}^{+}\label{Eq.FieldFn2-1}\\
\psi_{NC}(\mathbf{r}) & = & \sum\limits _{k\neq1,2}^{K}\alpha_{k}\,\phi_{k}(\mathbf{r})\label{Eq.FieldFn3-1}\\
\psi_{NC}^{+}(\mathbf{r}) & = & \sum\limits _{k\neq1,2}^{K}\phi_{k}^{\ast}(\mathbf{r})\alpha_{k}^{+}\label{Eq.FieldFn4-1}\end{eqnarray}
\pagebreak{}

\section{- Quantum Averages}

\label{Appendix Quantum Averages}

To prove Eq.(\ref{Eq.QuantumAverages-1}) it will be convenient to
treat the condensate operators first ignoring the non-condensate operators
and with the quasidistribution functional being purely of the Wigner
type. Following that we then reverse the process by treating the non-condensate
operators with the quasidistribution functional being of the positive
P type.

\subsection{The Condensate Averages}

The functional derivative of the symmetrically ordered characteristic
functional with respect to say, $\xi(\mathbf{r})$ is defined by

\[
\left(\frac{\delta\chi^{W}[\xi(\mathbf{r}),\xi^{+}(\mathbf{r})]}{\delta\xi(\mathbf{r})}\right)_{\mathbf{r=r}_{1}}=\lim_{\epsilon\rightarrow0}\left(\frac{\chi^{W}[\xi(\mathbf{r})+\epsilon\delta(\mathbf{r-r}_{1}),\xi^{+}(\mathbf{r})]-\chi^{W}[\xi(\mathbf{r}),\xi^{+}(\mathbf{r})]}{\epsilon}\right).\]

It is not difficult to see that\begin{eqnarray*}
 &  & \chi^{W}[\xi(\mathbf{r})+\epsilon\delta(\mathbf{r-r}_{1}),\xi^{+}(\mathbf{r})]\\
 & = & \iiiint D^{2}\psi\, D^{2}\psi^{+}\, W[\psi(\mathbf{r}),\psi^{+}(\mathbf{r})]\\
 &  & \times\exp i\int d\mathbf{r\,\{}(\xi(\mathbf{r})+\epsilon\delta(\mathbf{r-r}_{1}))\psi^{+}(\mathbf{r})+\psi(\mathbf{r})\xi^{+}(\mathbf{r})\}\,\\
 & = & \chi^{W}[\xi(\mathbf{r}),\xi^{+}(\mathbf{r})]\\
 &  & +i\epsilon\iiiint D^{2}\psi\, D^{2}\psi^{+}\, W[\psi(\mathbf{r}),\psi^{+}(\mathbf{r})]\,\psi^{+}(\mathbf{r}_{1})\exp i\int d\mathbf{r\,\{}\xi(\mathbf{r})\psi^{+}(\mathbf{r})+\psi(\mathbf{r})\xi^{+}(\mathbf{r})\}\end{eqnarray*}
Thus the functional derivative is \begin{eqnarray*}
\left(\frac{\delta\chi^{W}[\xi(\mathbf{r}),\xi^{+}(\mathbf{r})]}{\delta\xi(\mathbf{r})}\right)_{\mathbf{r=r}_{1}} & = & \iiiint D^{2}\psi\, D^{2}\psi^{+}\, W[\psi(\mathbf{r}),\psi^{+}(\mathbf{r})]\\
 &  & \times i\psi^{+}(\mathbf{r}_{1})\exp i\int d\mathbf{r\,\{}\xi(\mathbf{r})\psi^{+}(\mathbf{r})+\psi(\mathbf{r})\xi^{+}(\mathbf{r})\}.\end{eqnarray*}
Note that the field function at position $\mathbf{r}_{1}$ is still
subject to the functional integration.

Similarly\begin{eqnarray*}
\left(\frac{\delta\chi^{W}[\xi(\mathbf{r}),\xi^{+}(\mathbf{r})]}{\delta\xi^{+}(\mathbf{r})}\right)_{\mathbf{r=r}_{1}} & = & \iiiint D^{2}\psi\, D^{2}\psi^{+}\, W[\psi(\mathbf{r}),\psi^{+}(\mathbf{r})]\\
 &  & \times i\psi(\mathbf{r}_{1})\exp i\int d\mathbf{r\,\{}\xi(\mathbf{r})\psi^{+}(\mathbf{r})+\psi(\mathbf{r})\xi^{+}(\mathbf{r})\}.\end{eqnarray*}
Thus we see that these functional derivatives are in the form of expressions
for characteristic functionals in which $W[\psi(\mathbf{r}),\psi^{+}(\mathbf{r})]$
is replaced by $i\psi^{+}(\mathbf{r}_{1})W[\psi(\mathbf{r}),\psi^{+}(\mathbf{r})]$
or $i\psi(\mathbf{r}_{1})W[\psi(\mathbf{r}),\psi^{+}(\mathbf{r})]$.

Continuing in this way we may establish a result for higher order
functional derviatives

\begin{eqnarray*}
 &  & \left(\frac{\delta^{p+q}\chi^{W}[\xi(\mathbf{r}),\xi^{+}(\mathbf{r})]}{\,\delta^{p}\xi(\mathbf{r})\,\delta^{q}\xi^{+}(\mathbf{r})}\right)_{\mathbf{r}_{1},\mathbf{r}_{2},..,\mathbf{r}_{p};\mathbf{s}_{q},..,\mathbf{s}_{2},\mathbf{s}_{1};}\\
 & = & \iiiint D^{2}\psi\, D^{2}\psi^{+}\, W[\psi(\mathbf{r}),\psi^{+}(\mathbf{r})]\\
 &  & \times i^{p+q}\,\psi^{+}(\mathbf{r}_{1})\,\psi^{+}(\mathbf{r}_{2})\,..\psi^{+}(\mathbf{r}_{p})\,\psi(\mathbf{s}_{q})\,..\psi(\mathbf{s}_{2})\,\psi(\mathbf{s}_{1})\\
 &  & \times\exp i\int d\mathbf{r\,\{}\xi(\mathbf{r})\psi^{+}(\mathbf{r})+\psi(\mathbf{r})\xi^{+}(\mathbf{r})\}.\end{eqnarray*}
where for bosonic systems the functional differentiation can be carried
out in any order but with the differentiation with respect to $\xi(\mathbf{r})$
involving positions $\mathbf{r}_{1},\mathbf{r}_{2},..,\mathbf{r}_{p}$
and the $\xi^{+}(\mathbf{r})$ differentiation involving positions
$\mathbf{s}_{q},..,\mathbf{s}_{2},\mathbf{s}_{1}$.

Evaluating the functional derivatives and then letting $\xi(\mathbf{r}),\xi^{+}(\mathbf{r})$
all approach zero (symbolically $\mathbf{\xi\longrightarrow0}$),
we have for bosonic systems

\begin{eqnarray*}
 &  & \left(\frac{\delta^{p+q}\chi^{W}[\xi(\mathbf{r}),\xi^{+}(\mathbf{r})]}{\,\delta^{p}\xi(\mathbf{r})\,\delta^{q}\xi^{+}(\mathbf{r})}\right)_{\mathbf{r}_{1},\mathbf{r}_{2},..,\mathbf{r}_{p};\mathbf{s}_{q},..,\mathbf{s}_{2},\mathbf{s}_{1};}^{\mathbf{\xi\rightarrow0}}\\
 & = & \iiiint D^{2}\psi\, D^{2}\psi^{+}\, W[\psi(\mathbf{r}),\psi^{+}(\mathbf{r})]\\
 &  & \times i^{p+q}\,\psi^{+}(\mathbf{r}_{1})\,\psi^{+}(\mathbf{r}_{2})\,..\psi^{+}(\mathbf{r}_{p})\,\psi(\mathbf{s}_{q})\,..\psi(\mathbf{s}_{2}).\psi(\mathbf{s}_{1})\end{eqnarray*}

We then apply the same process to the definition of the characteristic
functional

\begin{eqnarray*}
\chi^{W}[\xi(\mathbf{r}),\xi^{+}(\mathbf{r})] & = & Tr(\widehat{\rho}\,\exp\int d\mathbf{r}\, i\{\xi(\mathbf{r})\widehat{\Psi}^{\dagger}(\mathbf{r})+\widehat{\Psi}(\mathbf{r)}\xi^{+}(\mathbf{r})\}\,\\
 & = & \sum\limits _{n}\frac{1}{n!}Tr(\widehat{\rho}\,\left(\int d\mathbf{r}\, i\xi(\mathbf{r})\,\widehat{\Psi}^{\dagger}(\mathbf{r})+\int d\mathbf{r}\,\widehat{\Psi}(\mathbf{r})\, i\xi^{+}(\mathbf{r})\right)^{n}.\end{eqnarray*}
Now with $\widehat{A}[\xi(\mathbf{r})]=\int d\mathbf{r}\, i\xi(\mathbf{r})\,\widehat{\Psi}^{\dagger}(\mathbf{r})$
and $\widehat{B}[\xi^{+}(\mathbf{r})]=\int d\mathbf{r}\,\widehat{\Psi}(\mathbf{r})\, i\xi^{+}(\mathbf{r})$
there are $N(p,q)=(p+q)!/p!q!$ ways that the operator $\widehat{A}$
appears $p$ times and the operator $\widehat{B}$ appears $q$ times
when we expand $(\widehat{A}+\widehat{B})^{n}$ (where $n=p+q$) and
each order of these operators appears once. We can introduce the symbol
$\{(\widehat{A})^{p}\,(\widehat{B})^{q}\}$ to denote the average
of these $N(p,q)$ ordered products \[
\{(\widehat{A})^{p}\,(\widehat{B})^{q}\}=\frac{1}{N(p,q)}\left((\widehat{A})^{p}\,(\widehat{B})^{q}+(\widehat{A})^{p-1}\,(\widehat{B})^{q}\,(\widehat{A})+...+(\widehat{B})^{q}\,(\widehat{A})^{p}\right)\]
and write

\[
\chi^{W}[\xi(\mathbf{r}),\xi^{+}(\mathbf{r})]=\sum\limits _{p,q}\frac{1}{p!q!}Tr(\widehat{\rho}\,\{(\widehat{A})^{p}\,(\widehat{B})^{q}\}).\]

In this form it is convenient to calculate the functional derivatives,
since $\widehat{A}[\xi(\mathbf{r})]$ and $\widehat{B}[\xi^{+}(\mathbf{r})]$
are functionals only of $\xi(\mathbf{r})$ and $\xi^{+}(\mathbf{r})$.respectively,
so their functional derivatives with respect to the other function
will be zero. Then

\begin{eqnarray*}
\left(\frac{\delta\widehat{A}[\xi(\mathbf{r})]}{\delta\xi(\mathbf{r})}\right)_{\mathbf{r=r}_{1}} & = & \lim_{\epsilon\rightarrow0}\left(\frac{\widehat{A}[\xi(\mathbf{r})+\epsilon\delta(\mathbf{r-r}_{1})]-\widehat{A}[\xi(\mathbf{r})]}{\epsilon}\right)\\
 & = & \lim_{\epsilon\rightarrow0}\left(\frac{i\epsilon\widehat{\Psi}^{\dagger}(\mathbf{r}_{1})}{\epsilon}\right)\\
 & = & i\widehat{\Psi}^{\dagger}(\mathbf{r}_{1})\end{eqnarray*}
Similarly\[
\left(\frac{\delta\widehat{B}[\xi^{+}(\mathbf{r})]}{\delta\xi^{+}(\mathbf{r})}\right)_{\mathbf{r=s}_{1}}=i\widehat{\Psi}(\mathbf{s}_{1})\]
We see that each time that either $\widehat{A}[\xi(\mathbf{r})]$
is differentiated with respect to $\xi(\mathbf{r})$ or $\widehat{B}[\xi^{+}(\mathbf{r})]$
is differentiated with respect to $\xi^{+}(\mathbf{r})$ an operator
results, and therefore no further functional differentiation can occur.
We also note that as $\mathbf{\xi\longrightarrow0}$ both $\widehat{A}[\xi(\mathbf{r})]$
and $\widehat{B}[\xi^{+}(\mathbf{r})]$ become zero. To proceed further
we need to calculate functional derivatives of products of $\widehat{A}[\xi(\mathbf{r})]$
and $\widehat{B}[\xi^{+}(\mathbf{r})]$. This can be carried out by
applying the general rule for functional derivatives of products of
functionals. Consider the term $\{(\widehat{A})^{p}\,(\widehat{B})^{q}\}$
(which is the average of the $N(p,q)$ ordered products where $\widehat{A}[\xi(\mathbf{r})]$
appears $p$ times and $\widehat{B}[\xi^{+}(\mathbf{r})]$ appears
$q$ times).If each of the terms in $\{(\widehat{A})^{p}\,(\widehat{B})^{q}\}$
is differentiated \emph{less} than $p$ times with respect to $\xi(\mathbf{r})$
then there will be at least one factor $\widehat{A}[\xi(\mathbf{r})]$
still remaining and thus as $\mathbf{\xi\longrightarrow0}$ the result
of the differentiation will be zero. Similar conclusions apply if
the term in $\{(\widehat{A})^{p}\,(\widehat{B})^{q}\}$ is differentiated
less than $q$ times with respect to $\xi^{+}(\mathbf{r})$. On the
other hand if ach of the terms in $\{(\widehat{A})^{p}\,(\widehat{B})^{q}\}$
is differentiated \emph{more} than $p$ times with respect to $\xi(\mathbf{r})$
then the result of the differentiation must be zero because after
the $p$th differentiation all of the $\widehat{A}[\xi(\mathbf{r})]$
will have been replaced by a factor $i\widehat{\Psi}^{\dagger}(\mathbf{r}_{i})$
and therefore further functional differentiation with respect to $\xi(\mathbf{r})$
will give zero. Similar conclusions apply if if the term in $\{(\widehat{A})^{p}\,(\widehat{B})^{q}\}$
is differentiated more than $q$ times with respect to $\xi^{+}(\mathbf{r})$.
\ Hence only the $p,q$ term in the last expression for $\chi\lbrack\xi(\mathbf{r}),\xi^{+}(\mathbf{r})]$
contributes in the required result for

\begin{align*}
 & \left(\frac{\delta^{p+q}\chi^{W}[\xi(\mathbf{r}),\xi^{+}(\mathbf{r})]}{\,\delta^{p}\xi(\mathbf{r})\,\delta^{q}\xi^{+}(\mathbf{r})}\right)_{\mathbf{r}_{1},\mathbf{r}_{2},..,\mathbf{r}_{p};\mathbf{s}_{q},..,\mathbf{s}_{2},\mathbf{s}_{1};}^{\mathbf{\xi\rightarrow0}}\\
= & \frac{1}{p!q!}Tr\left(\widehat{\rho}\,\left(\frac{\delta^{p+q}\{(\widehat{A})^{p}\,(\widehat{B})^{q}\}}{\,\delta^{p}\xi(\mathbf{r})\,\delta^{q}\xi^{+}(\mathbf{r})}\right)_{\mathbf{r}_{1},\mathbf{r}_{2},..,\mathbf{r}_{p};\mathbf{s}_{q},..,\mathbf{s}_{2},\mathbf{s}_{1};}^{\mathbf{\xi\rightarrow0}}\right)\end{align*}

Now consider the $p$th functional derivative of any term in $\{(\widehat{A})^{p}\,(\widehat{B})^{q}\}$
with respect to the $\xi(\mathbf{r})$, where the $q$ factors $\widehat{B}[\xi^{+}(\mathbf{r})]$
are just represented by dots. The result will be the sum of products
of factors $i\widehat{\Psi}^{\dagger}(\mathbf{r}_{1}),i\widehat{\Psi}^{\dagger}(\mathbf{r}_{2}),..,i\widehat{\Psi}^{\dagger}(\mathbf{r}_{p})$
in all $p!$ orders \begin{align*}
 & \left(\frac{\delta^{p}(\widehat{A}[\xi(\mathbf{r})]..\widehat{A}[\xi(\mathbf{r})]..\widehat{A}[\xi(\mathbf{r})]..\widehat{A}[\xi(\mathbf{r})])^{p,q}}{\delta^{p}\xi(\mathbf{r})}\right)_{\mathbf{r}_{1},\mathbf{r}_{2},..,\mathbf{r}_{p}}^{\mathbf{\xi\rightarrow0}}\\
= & \, i^{p}\sum\limits _{P}(\widehat{\Psi}^{\dagger}(\mathbf{r}_{\mu_{1}})..\widehat{\Psi}^{\dagger}(\mathbf{r}_{\mu_{2}})..\widehat{\Psi}^{\dagger}(\mathbf{r}_{\mu_{i}})..\widehat{\Psi}^{\dagger}(\mathbf{r}_{\mu_{p}}))\end{align*}
where the sum is over all permutations $P=\uparrow\left(\frac{\mu_{1}}{1}\frac{\mu_{2}}{2}..\frac{\mu_{i}}{i}..\frac{\mu_{p}}{p}\right)$
of $1,2,..,i,..p$. Considering also the $q$th functional derivative
of the same term in $\{(\widehat{A})^{p}\,(\widehat{B})^{q}\}$ with
respect to the $\xi^{+}(\mathbf{r})$ we get overall

\begin{eqnarray*}
 &  & \left(\frac{\delta^{p+q}(\widehat{A}[\xi(\mathbf{r})]..\widehat{B}[\xi^{+}(\mathbf{r})]..\widehat{A}[\xi(\mathbf{r})]..\widehat{B}[\xi^{+}(\mathbf{r})]..\widehat{A}[\xi(\mathbf{r})..\widehat{B}[\xi^{+}(\mathbf{r})]])^{p,q}}{\delta^{p}\xi(\mathbf{r})\,\delta^{q}\xi^{+}(\mathbf{r})}\right)_{\mathbf{r}_{1},\mathbf{r}_{2},..,\mathbf{r}_{p};\mathbf{s}_{q},..,\mathbf{s}_{2},\mathbf{s}_{1}}^{\mathbf{\xi\rightarrow0}}\\
 & = & i^{p+q}\sum\limits _{P,Q}(\widehat{\Psi}^{\dagger}(\mathbf{r}_{\mu_{1}})..\widehat{\Psi}(\mathbf{s}_{\lambda_{q}})..\widehat{\Psi}^{\dagger}(\mathbf{r}_{\mu_{2}})..\widehat{\Psi}(\mathbf{s}_{\lambda_{2}})..\widehat{\Psi}^{\dagger}(\mathbf{r}_{\mu_{p}})..\widehat{\Psi}(\mathbf{s}_{\lambda_{1}}))\end{eqnarray*}
where the second sum is over all permutations $Q=\uparrow\left(\frac{\lambda_{q}}{q}..\frac{\lambda j}{j}..\frac{\lambda_{2}}{2}\frac{\lambda_{1}}{1}\right)$
of $q,..,i,..,2,1$. Now within each of the $N(p,q)=(p+q)!/p!q!$orderings
of products of $\widehat{A}$ and $\widehat{B}$ where $\widehat{A}$
appears $p$ times and $\widehat{B}$ appears $q$ times, there are
$p!$ orderings of the $\widehat{\Psi}^{\dagger}$ operators and $q!$orderings
of the $\widehat{\Psi}$ operators, giving a total of $M(p,q)=N(p,q)\, p!q!=(p+q)!$
different orderings of the $p$ operators $\widehat{\Psi}^{\dagger}$
and the $q$ operators $\widehat{\Psi}$, and all possible orderings
are present in view of the sum over the permutations $P,Q$. Taking
into account the factor $1/p!q!$ we see that the when the differentiation
is applied to the quantity $\{(\widehat{A})^{p}\,(\widehat{B})^{q}\}$
itself we see that we just get $i^{p+q}\{\widehat{\Psi}^{\dagger}(\mathbf{r}_{1})\widehat{\Psi}^{\dagger}(\mathbf{r}_{2})....\widehat{\Psi}^{\dagger}(\mathbf{r}_{p})\widehat{\Psi}(\mathbf{s}_{1})..\widehat{\Psi}(\mathbf{s}_{q})\}$.
Thus

\begin{align*}
 & \left(\frac{\delta^{p+q}\chi^{W}[\xi(\mathbf{r}),\xi^{+}(\mathbf{r})]}{\,\delta^{p}\xi(\mathbf{r})\,\delta^{q}\xi^{+}(\mathbf{r})}\right)_{\mathbf{r}_{1},\mathbf{r}_{2},..,\mathbf{r}_{p};\mathbf{s}_{q},..,\mathbf{s}_{2},\mathbf{s}_{1};}^{\mathbf{\xi\rightarrow0}}\\
= & i^{p+q}Tr\left(\widehat{\rho}\,\{\widehat{\Psi}^{\dagger}(\mathbf{r}_{1})\widehat{\Psi}^{\dagger}(\mathbf{r}_{2})....\widehat{\Psi}^{\dagger}(\mathbf{r}_{p})\widehat{\Psi}(\mathbf{s}_{q})..\widehat{\Psi}(\mathbf{s}_{1})\}\right)\end{align*}
where the symmetric ordering symbol is given by \begin{align*}
 & \{\widehat{\Psi}^{\dagger}(\mathbf{r}_{1})\widehat{\Psi}^{\dagger}(\mathbf{r}_{2})....\widehat{\Psi}^{\dagger}(\mathbf{r}_{p})\widehat{\Psi}(\mathbf{s}_{q})..\widehat{\Psi}(\mathbf{s}_{1})\}\\
= & \frac{1}{(p+q)!}\sum\limits _{R}\Re(\widehat{\Psi}^{\dagger}(\mathbf{r}_{1})\widehat{\Psi}^{\dagger}(\mathbf{r}_{2})....\widehat{\Psi}^{\dagger}(\mathbf{r}_{p})\widehat{\Psi}(\mathbf{s}_{q})..\widehat{\Psi}(\mathbf{s}_{1}))\end{align*}
In the the sum over $R$ is over all $(p+q)!$ orderings $\Re$ of
the factors\\
 $\widehat{\Psi}^{\dagger}(\mathbf{r}_{1})\widehat{\Psi}^{\dagger}(\mathbf{r}_{2})....\widehat{\Psi}^{\dagger}(\mathbf{r}_{p})\widehat{\Psi}(\mathbf{s}_{q})..\widehat{\Psi}(\mathbf{s}_{1})$.

Hence we obtain the following key result for the quantum average of
the symmetrically ordered product $\{\widehat{\Psi}^{\dagger}(\mathbf{r}_{1})\widehat{\Psi}^{\dagger}(\mathbf{r}_{2})....\widehat{\Psi}^{\dagger}(\mathbf{r}_{p})\widehat{\Psi}(\mathbf{s}_{q})..\widehat{\Psi}(\mathbf{s}_{1})\}$
of the field operators\begin{eqnarray*}
 &  & \left\langle \{\widehat{\Psi}^{\dagger}(\mathbf{r}_{1})\widehat{\Psi}^{\dagger}(\mathbf{r}_{2})....\widehat{\Psi}^{\dagger}(\mathbf{r}_{p})\widehat{\Psi}(\mathbf{s}_{q})..\widehat{\Psi}(\mathbf{s}_{1})\}\right\rangle \\
 & = & Tr\left(\widehat{\rho}\,\{\widehat{\Psi}^{\dagger}(\mathbf{r}_{1})\widehat{\Psi}^{\dagger}(\mathbf{r}_{2})....\widehat{\Psi}^{\dagger}(\mathbf{r}_{p})\widehat{\Psi}(\mathbf{s}_{q})..\widehat{\Psi}(\mathbf{s}_{1})\}\right)\\
 & = & \iiiint D^{2}\psi\, D^{2}\psi^{+}\, W[\psi(\mathbf{r}),\psi^{+}(\mathbf{r})]\\
 &  & \times\psi^{+}(\mathbf{r}_{1})\,\psi^{+}(\mathbf{r}_{2})\,..\psi^{+}(\mathbf{r}_{p})\,\psi(\mathbf{s}_{q})\,..\psi(\mathbf{s}_{2})\,\psi(\mathbf{s}_{1})\end{eqnarray*}
This result gives the required synmmetrically ordered average as a
functional integral involving the quasi-distribution functional $W[\psi(\mathbf{r}),\psi^{+}(\mathbf{r})]$
times the product of the field functions, with the field operator
$\widehat{\Psi}^{\dagger}(\mathbf{r}_{i})$ being replaced by $\psi^{+}(\mathbf{r}_{i})$
and $\widehat{\Psi}(\mathbf{s}_{j})$ being replaced by $\psi(\mathbf{s}_{j})$.

\subsection{The Non-Condensate Averages}

The functional derivative of a normally ordered characteristic functional
with respect to say, $\xi(\mathbf{r})$ is defined by

\[
\left(\frac{\delta\chi_{N}[\xi(\mathbf{r}),\xi^{+}(\mathbf{r})]}{\delta\xi(\mathbf{r})}\right)_{\mathbf{r=r}_{1}}=\lim_{\epsilon\rightarrow0}\left(\frac{\chi_{N}[\xi(\mathbf{r})+\epsilon\delta(\mathbf{r-r}_{1}),\xi^{+}(\mathbf{r})]-\chi_{N}[\xi(\mathbf{r}),\xi^{+}(\mathbf{r})]}{\epsilon}\right).\]

It is not difficult to see that\begin{eqnarray*}
 &  & \chi_{N}[\xi(\mathbf{r})+\epsilon\delta(\mathbf{r-r}_{1}),\xi^{+}(\mathbf{r})]\\
 & = & \iiiint D^{2}\psi\, D^{2}\psi^{+}\, P^{+}[\psi(\mathbf{r}),\psi^{+}(\mathbf{r})]\\
 &  & \times\exp i\int d\mathbf{r\,}\{(\xi(\mathbf{r})+\epsilon\delta(\mathbf{r-r}_{1}))\psi^{+}(\mathbf{r})\}\exp i\int d\mathbf{r\,\{}\psi(\mathbf{r})\xi^{+}(\mathbf{r})\}\,\\
 & = & \chi_{N}[\xi(\mathbf{r}),\xi^{+}(\mathbf{r})]\\
 &  & +i\epsilon\iiiint D^{2}\psi\, D^{2}\psi^{+}\, P^{+}[\psi(\mathbf{r}),\psi^{+}(\mathbf{r})]\,\psi^{+}(\mathbf{r}_{1})\exp i\int d\mathbf{r\,}\{\xi(\mathbf{r})\psi^{+}(\mathbf{r})\}\exp i\int d\mathbf{r\,\{}\psi(\mathbf{r})\xi^{+}(\mathbf{r})\}\,\end{eqnarray*}
Thus the functional derivative is \begin{eqnarray*}
\left(\frac{\delta\chi_{N}[\xi(\mathbf{r}),\xi^{+}(\mathbf{r})]}{\delta\xi(\mathbf{r})}\right)_{\mathbf{r=r}_{1}} & = & \iiiint D^{2}\psi\, D^{2}\psi^{+}\, P^{+}[\psi(\mathbf{r}),\psi^{+}(\mathbf{r})]\\
 &  & \times i\psi^{+}(\mathbf{r}_{1})\exp i\int d\mathbf{r\,}\{\xi(\mathbf{r})\psi^{+}(\mathbf{r})\}\exp i\int d\mathbf{r\,\{}\psi(\mathbf{r})\xi^{+}(\mathbf{r})\}\,.\end{eqnarray*}
Note that the field function at position $\mathbf{r}_{1}$ is still
subject to the functional integration.

Similarly\begin{eqnarray*}
\left(\frac{\delta\chi_{N}[\xi(\mathbf{r}),\xi^{+}(\mathbf{r})]}{\delta\xi^{+}(\mathbf{r})}\right)_{\mathbf{r=r}_{1}} & = & \iiiint D^{2}\psi\, D^{2}\psi^{+}\, P^{+}[\psi(\mathbf{r}),\psi^{+}(\mathbf{r})]\\
 &  & \times i\psi(\mathbf{r}_{1})\exp i\int d\mathbf{r\,}\{\xi(\mathbf{r})\psi^{+}(\mathbf{r})\exp i\int d\mathbf{r\,\{}\psi(\mathbf{r})\xi^{+}(\mathbf{r})\}\,.\end{eqnarray*}
Thus we see that these functional derivatives are in the form of expressions
for characteristic functionals in which $P^{+}[\psi(\mathbf{r}),\psi^{+}(\mathbf{r})]$
is replaced by $i\psi^{+}(\mathbf{r}_{1})P^{+}[\psi(\mathbf{r}),\psi^{+}(\mathbf{r})]$
or $i\psi(\mathbf{r}_{1})P^{+}[\psi(\mathbf{r}),\psi^{+}(\mathbf{r})]$.

Continuing in this way we may establish a result for higher order
functional derviatives

\begin{eqnarray*}
 &  & \left(\frac{\delta^{p+q}\chi_{N}[\xi(\mathbf{r}),\xi^{+}(\mathbf{r})]}{\,\delta^{p}\xi(\mathbf{r})\,\delta^{q}\xi^{+}(\mathbf{r})}\right)_{\mathbf{r}_{1},\mathbf{r}_{2},..,\mathbf{r}_{p};\mathbf{s}_{q},..,\mathbf{s}_{2},\mathbf{s}_{1};}\\
 & = & \iiiint D^{2}\psi\, D^{2}\psi^{+}\, P^{+}[\psi(\mathbf{r}),\psi^{+}(\mathbf{r})]\\
 &  & \times i^{p+q}\,\psi^{+}(\mathbf{r}_{1})\,\psi^{+}(\mathbf{r}_{2})\,..\psi^{+}(\mathbf{r}_{p})\,\psi(\mathbf{s}_{q})\,..\psi(\mathbf{s}_{2})\,\psi(\mathbf{s}_{1})\\
 &  & \times\exp i\int d\mathbf{r\,}\{\xi(\mathbf{r})\psi^{+}(\mathbf{r})\,\exp i\int d\mathbf{r\,\{}\psi(\mathbf{r})\xi^{+}(\mathbf{r})\}\,.\end{eqnarray*}
where for bosonic systems the functional differentiation can be carried
out in any order but with the differentiation with respect to $\xi(\mathbf{r})$
involving positions $\mathbf{r}_{1},\mathbf{r}_{2},..,\mathbf{r}_{p}$
and the $\xi^{+}(\mathbf{r})$ differentiation involving positions
$\mathbf{s}_{q},..,\mathbf{s}_{2},\mathbf{s}_{1}$.

Evaluating the functional derivatives and then letting $\xi(\mathbf{r}),\xi^{+}(\mathbf{r})$
all approach zero (symbolically $\mathbf{\xi\longrightarrow0}$),
we have for bosonic systems

\begin{eqnarray*}
 &  & \left(\frac{\delta^{p+q}\chi_{N}[\xi(\mathbf{r}),\xi^{+}(\mathbf{r})]}{\,\delta^{p}\xi(\mathbf{r})\,\delta^{q}\xi^{+}(\mathbf{r})}\right)_{\mathbf{r}_{1},\mathbf{r}_{2},..,\mathbf{r}_{p};\mathbf{s}_{q},..,\mathbf{s}_{2},\mathbf{s}_{1};}^{\mathbf{\xi\rightarrow0}}\\
 & = & \iiiint D^{2}\psi\, D^{2}\psi^{+}\, P^{+}[\psi(\mathbf{r}),\psi^{+}(\mathbf{r})]\\
 &  & \times i^{p+q}\,\psi^{+}(\mathbf{r}_{1})\,\psi^{+}(\mathbf{r}_{2})\,..\psi^{+}(\mathbf{r}_{p})\,\psi(\mathbf{s}_{q})\,..\psi(\mathbf{s}_{2})\,\psi(\mathbf{s}_{1})\end{eqnarray*}

We then apply the same process to the definition of the characteristic
functional

\begin{eqnarray*}
 &  & \chi_{N}[\xi(\mathbf{r}),\xi^{+}(\mathbf{r})]\\
 & = & Tr(\widehat{\rho}\,\exp\int d\mathbf{r}\, i\{\xi(\mathbf{r})\widehat{\Psi}^{\dagger}(\mathbf{r})\}\,\exp\int d\mathbf{r}\, i\{\widehat{\Psi}(\mathbf{r)}\xi^{+}(\mathbf{r})\}\,\\
 & = & \sum\limits _{p,q}\frac{1}{p!q!}Tr(\widehat{\rho}\,\left(\int d\mathbf{r}\, i\xi(\mathbf{r})\,\widehat{\Psi}^{\dagger}(\mathbf{r})\right)^{p}\left(\int d\mathbf{r}\,\widehat{\Psi}(\mathbf{r})\, i\xi^{+}(\mathbf{r})\right)^{q}).\end{eqnarray*}
Now with $\widehat{A}[\xi(\mathbf{r})]=\int d\mathbf{r}\, i\xi(\mathbf{r})\,\widehat{\Psi}^{\dagger}(\mathbf{r})$
and $\widehat{B}[\xi^{+}(\mathbf{r})]=\int d\mathbf{r}\,\widehat{\Psi}(\mathbf{r})\, i\xi^{+}(\mathbf{r})$
we see that

\[
\chi_{N}[\xi(\mathbf{r}),\xi^{+}(\mathbf{r})]=\sum\limits _{p,q}\frac{1}{p!q!}Tr(\widehat{\rho}\,(\widehat{A})^{p}\,(\widehat{B})^{q}).\]
keeping strictly to the operator order.

In this form it is convenient to calculate the functional derivatives,
since $\widehat{A}[\xi(\mathbf{r})]$ and $\widehat{B}[\xi^{+}(\mathbf{r})]$
are functionals only of $\xi(\mathbf{r})$ and $\xi^{+}(\mathbf{r})$.respectively,
so their functional derivatives with respect to the other function
will be zero. Then

\begin{eqnarray*}
\left(\frac{\delta\widehat{A}[\xi(\mathbf{r})]}{\delta\xi(\mathbf{r})}\right)_{\mathbf{r=r}_{1}} & = & \lim_{\epsilon\rightarrow0}\left(\frac{\widehat{A}[\xi(\mathbf{r})+\epsilon\delta(\mathbf{r-r}_{1})]-\widehat{A}[\xi(\mathbf{r})]}{\epsilon}\right)\\
 & = & \lim_{\epsilon\rightarrow0}\left(\frac{i\epsilon\widehat{\Psi}^{\dagger}(\mathbf{r}_{1})}{\epsilon}\right)\\
 & = & i\widehat{\Psi}^{\dagger}(\mathbf{r}_{1})\end{eqnarray*}
Similarly\[
\left(\frac{\delta\widehat{B}[\xi^{+}(\mathbf{r})]}{\delta\xi^{+}(\mathbf{r})}\right)_{\mathbf{r=s}_{1}}=i\widehat{\Psi}(\mathbf{s}_{1})\]
We see that each time that either $\widehat{A}[\xi(\mathbf{r})]$
is differentiated with respect to $\xi(\mathbf{r})$ or $\widehat{B}[\xi^{+}(\mathbf{r})]$
is differentiated with respect to $\xi^{+}(\mathbf{r})$ an operator
results, and therefore no further functional differentiation can occur.
We also note that as $\mathbf{\xi\longrightarrow0}$ both $\widehat{A}[\xi(\mathbf{r})]$
and $\widehat{B}[\xi^{+}(\mathbf{r})]$ become zero. To proceed further
we need to calculate functional derivatives of powers of $\widehat{A}[\xi(\mathbf{r})]$
and $\widehat{B}[\xi^{+}(\mathbf{r})]$. This can be carried out by
applying the general rule for functional derivatives of products of
functionals. Consider the term $(\widehat{A})^{p}\,(\widehat{B})^{q}$
where $\widehat{A}[\xi(\mathbf{r})]$ appears $p$ times and $\widehat{B}[\xi^{+}(\mathbf{r})]$
appears $q$ times.If each of the terms in $(\widehat{A})^{p}\,(\widehat{B})^{q}$
is differentiated \emph{less} than $p$ times with respect to $\xi(\mathbf{r})$
then there will be at least one factor $\widehat{A}[\xi(\mathbf{r})]$
still remaining and thus as $\mathbf{\xi\longrightarrow0}$ the result
of the differentiation will be zero. Similar conclusions apply if
the term $(\widehat{A})^{p}\,(\widehat{B})^{q}$ is differentiated
less than $q$ times with respect to $\xi^{+}(\mathbf{r})$. On the
other hand if each of the terms in $(\widehat{A})^{p}\,(\widehat{B})^{q}$
is differentiated \emph{more} than $p$ times with respect to $\xi(\mathbf{r})$
then the result of the differentiation must be zero because after
the $p$th differentiation all of the $\widehat{A}[\xi(\mathbf{r})]$
will have been replaced by a factor $i\widehat{\Psi}^{\dagger}(\mathbf{r}_{i})$
and therefore further functional differentiation with respect to $\xi(\mathbf{r})$
will give zero. Similar conclusions apply if $(\widehat{A})^{p}\,(\widehat{B})^{q}$
is differentiated more than $q$ times with respect to $\xi^{+}(\mathbf{r})$.
\ Hence only the $p,q$ term in the last expression for $\chi_{N}[\xi(\mathbf{r}),\xi^{+}(\mathbf{r})]$
contributes in the required result for

\begin{align*}
 & \left(\frac{\delta^{p+q}\chi_{N}[\xi(\mathbf{r}),\xi^{+}(\mathbf{r})]}{\,\delta^{p}\xi(\mathbf{r})\,\delta^{q}\xi^{+}(\mathbf{r})}\right)_{\mathbf{r}_{1},\mathbf{r}_{2},..,\mathbf{r}_{p};\mathbf{s}_{q},..,\mathbf{s}_{2},\mathbf{s}_{1};}^{\mathbf{\xi\rightarrow0}}\\
= & \frac{1}{p!q!}Tr\left(\widehat{\rho}\,\left(\frac{\delta^{p+q}((\widehat{A})^{p}\,(\widehat{B})^{q})}{\,\delta^{p}\xi(\mathbf{r})\,\delta^{q}\xi^{+}(\mathbf{r})}\right)_{\mathbf{r}_{1},\mathbf{r}_{2},..,\mathbf{r}_{p};\mathbf{s}_{q},..,\mathbf{s}_{2},\mathbf{s}_{1};}^{\mathbf{\xi\rightarrow0}}\right)\end{align*}

Now consider the $p$th functional derivative of $(\widehat{A})^{p}\,(\widehat{B})^{q}$
with respect to the $\xi(\mathbf{r})$, where the $q$ factors $\widehat{B}[\xi^{+}(\mathbf{r})]$
are always to the right of the $\widehat{A}[\xi(\mathbf{r})]$. The
result will be the sum of products of factors $i\widehat{\Psi}^{\dagger}(\mathbf{r}_{1}),i\widehat{\Psi}^{\dagger}(\mathbf{r}_{2}),..,i\widehat{\Psi}^{\dagger}(\mathbf{r}_{p})$
in all $p!$ orders \begin{eqnarray*}
 &  & \left(\frac{\delta^{p}(\widehat{A}[\xi(\mathbf{r})]\,\widehat{A}[\xi(\mathbf{r})]...\widehat{A}][\xi(\mathbf{r})]\,)^{p}\widehat{B}[\xi^{+}(\mathbf{r})]^{q}}{\delta^{p}\xi(\mathbf{r})}\right)_{\mathbf{r}_{1},\mathbf{r}_{2},..,\mathbf{r}_{p}}^{\mathbf{\xi\rightarrow0}}\\
 & = & i^{p}\sum\limits _{P}(\widehat{\Psi}^{\dagger}(\mathbf{r}_{\mu_{1}})..\widehat{\Psi}^{\dagger}(\mathbf{r}_{\mu_{2}})..\widehat{\Psi}^{\dagger}(\mathbf{r}_{\mu_{i}})..\widehat{\Psi}^{\dagger}(\mathbf{r}_{\mu_{p}})\widehat{B}[\xi^{+}(\mathbf{r})]^{q})\end{eqnarray*}
where the sum is over all permutations $P=\uparrow\left(\frac{\mu_{1}}{1}\frac{\mu_{2}}{2}..\frac{\mu_{i}}{i}..\frac{\mu_{p}}{p}\right)$
of $1,2,..,i,..p$. Considering also the $q$th functional derivative
of $(\widehat{A})^{p}\,(\widehat{B})^{q}$ with respect to the $\xi^{+}(\mathbf{r})$
we get overall

\begin{eqnarray*}
 &  & \left(\frac{\delta^{p+q}(\widehat{A}[\xi(\mathbf{r})]..\widehat{A}[\xi(\mathbf{r})]...\widehat{A}[\xi(\mathbf{r})].\widehat{B}[\xi^{+}(\mathbf{r})]..\widehat{B}[\xi^{+}(\mathbf{r})]..\widehat{B}[\xi^{+}(\mathbf{r})]])^{p,q}}{\delta^{p}\xi(\mathbf{r})\,\delta^{q}\xi^{+}(\mathbf{r})}\right)_{\mathbf{r}_{1},\mathbf{r}_{2},..,\mathbf{r}_{p};\mathbf{s}_{q},..,\mathbf{s}_{2},\mathbf{s}_{1}}^{\mathbf{\xi\rightarrow0}}\\
 & = & i^{p+q}\sum\limits _{P,Q}(\widehat{\Psi}^{\dagger}(\mathbf{r}_{\mu_{1}})..\widehat{\Psi}^{\dagger}(\mathbf{r}_{\mu_{2}})...\widehat{\Psi}^{\dagger}(\mathbf{r}_{\mu_{p}}).\widehat{\Psi}(\mathbf{s}_{\lambda_{q}})...\widehat{\Psi}(\mathbf{s}_{\lambda_{2}})\,\widehat{\Psi}(\mathbf{s}_{\lambda_{1}}))\end{eqnarray*}
where the second sum is over all permutations $Q=\uparrow\left(\frac{\lambda_{q}}{q}..\frac{\lambda j}{j}..\frac{\lambda_{2}}{2}\frac{\lambda_{1}}{1}\right)$
of $q,..,i,..,2,1$. Now all of the $p!$ products of the $\widehat{\Psi}^{\dagger}$
operators commute with each other and can therefore be set out in
the order $\widehat{\Psi}^{\dagger}(\mathbf{r}_{1}).\widehat{\Psi}^{\dagger}(\mathbf{r}_{2})...\widehat{\Psi}^{\dagger}(\mathbf{r}_{p})$.
Similarly, all of the $q!$ products of the $\widehat{\Psi}$ operators
commute with each other and can therefore be set out in the order
$\widehat{\Psi}(\mathbf{s}_{q})...\widehat{\Psi}(\mathbf{s}_{2})\,\widehat{\Psi}(\mathbf{s}_{1})$.
Thus the sum over the permutations $P,Q$ just cancells out the $1/p!q!$
factor and we just get $i^{p+q}\{\widehat{\Psi}^{\dagger}(\mathbf{r}_{1})\widehat{\Psi}^{\dagger}(\mathbf{r}_{2})....\widehat{\Psi}^{\dagger}(\mathbf{r}_{p})\widehat{\Psi}(\mathbf{s}_{q})..\widehat{\Psi}(\mathbf{s}_{1})\}$.
Thus

\begin{align*}
 & \left(\frac{\delta^{p+q}\chi_{N}[\xi(\mathbf{r}),\xi^{+}(\mathbf{r})]}{\,\delta^{p}\xi(\mathbf{r})\,\delta^{q}\xi^{+}(\mathbf{r})}\right)_{\mathbf{r}_{1},\mathbf{r}_{2},..,\mathbf{r}_{p};\mathbf{s}_{q},..,\mathbf{s}_{2},\mathbf{s}_{1};}^{\mathbf{\xi\rightarrow0}}\\
= & i^{p+q}Tr\left(\widehat{\rho}\,\widehat{\Psi}^{\dagger}(\mathbf{r}_{1})\widehat{\Psi}^{\dagger}(\mathbf{r}_{2})....\widehat{\Psi}^{\dagger}(\mathbf{r}_{p})\widehat{\Psi}(\mathbf{s}_{q})..\widehat{\Psi}(\mathbf{s}_{1})\right)\end{align*}

Hence we obtain the following key result for the quantum average of
the normally ordered product $\widehat{\Psi}^{\dagger}(\mathbf{r}_{1})\widehat{\Psi}^{\dagger}(\mathbf{r}_{2})....\widehat{\Psi}^{\dagger}(\mathbf{r}_{p})\widehat{\Psi}(\mathbf{s}_{q})..\widehat{\Psi}(\mathbf{s}_{1})$
of the field operators\begin{eqnarray*}
 &  & \left\langle \widehat{\Psi}^{\dagger}(\mathbf{r}_{1})\widehat{\Psi}^{\dagger}(\mathbf{r}_{2})....\widehat{\Psi}^{\dagger}(\mathbf{r}_{p})\widehat{\Psi}(\mathbf{s}_{q})..\widehat{\Psi}(\mathbf{s}_{1})\right\rangle \\
 & = & Tr\left(\widehat{\rho}\,\widehat{\Psi}^{\dagger}(\mathbf{r}_{1})\widehat{\Psi}^{\dagger}(\mathbf{r}_{2})....\widehat{\Psi}^{\dagger}(\mathbf{r}_{p})\widehat{\Psi}(\mathbf{s}_{q})..\widehat{\Psi}(\mathbf{s}_{1})\right)\\
 & = & \iiiint D^{2}\psi\, D^{2}\psi^{+}\, P^{+}[\psi(\mathbf{r}),\psi^{+}(\mathbf{r})]\\
 &  & \times\psi^{+}(\mathbf{r}_{1})\,\psi^{+}(\mathbf{r}_{2})\,..\psi^{+}(\mathbf{r}_{p})\,\psi(\mathbf{s}_{q})\,..\psi(\mathbf{s}_{2})\,\psi(\mathbf{s}_{1})\end{eqnarray*}
This result gives the required synmmetrically ordered average as a
functional integral involving the quasi-distribution functional $P^{+}[\psi(\mathbf{r}),\psi^{+}(\mathbf{r})]$
times the product of the field functions, with the field operator
$\widehat{\Psi}^{\dagger}(\mathbf{r}_{i})$ being replaced by $\psi^{+}(\mathbf{r}_{i})$
and $\widehat{\Psi}(\mathbf{s}_{j})$ being replaced by $\psi(\mathbf{s}_{j})$.

\subsection{Supplementary Equations}

~

Quantum Correlation Function\begin{eqnarray}
 &  & \left\langle \{\widehat{\Psi}_{C}^{\dagger}(\mathbf{r}_{1})....\widehat{\Psi}_{C}^{\dagger}(\mathbf{r}_{p})\widehat{\Psi}_{C}(\mathbf{s}_{q})..\widehat{\Psi}_{C}(\mathbf{s}_{1})\}\,\widehat{\Psi}_{NC}^{\dagger}(\mathbf{u}_{1})....\widehat{\Psi}_{NC}^{\dagger}(\mathbf{u}_{r})\widehat{\Psi}_{NC}(\mathbf{v}_{s})..\widehat{\Psi}_{NC}(\mathbf{v}_{1})\right\rangle \nonumber \\
 & = & Tr\left(\widehat{\rho}\,\{\widehat{\Psi}_{C}^{\dagger}(\mathbf{r}_{1})....\widehat{\Psi}_{C}^{\dagger}(\mathbf{r}_{p})\widehat{\Psi}_{C}(\mathbf{s}_{q})..\widehat{\Psi}_{C}(\mathbf{s}_{1})\}\,\widehat{\Psi}_{NC}^{\dagger}(\mathbf{u}_{1})....\widehat{\Psi}_{NC}^{\dagger}(\mathbf{u}_{r})\widehat{\Psi}_{NC}(\mathbf{v}_{s})..\widehat{\Psi}_{NC}(\mathbf{v}_{1})\right)\nonumber \\
 & = & \iiiint D^{2}\psi_{C}\, D^{2}\psi_{C}^{+}\, D^{2}\psi_{NC}\, D^{2}\psi_{NC}^{+}\,\nonumber \\
 &  & \times P[\psi_{C}(\mathbf{r}),\psi_{C}^{+}(\mathbf{r}),\psi_{NC}(\mathbf{r}),\psi_{NC}^{+}(\mathbf{r}),\psi_{C}^{\ast}(\mathbf{r}),\psi_{C}^{+\ast}(\mathbf{r}),\psi_{NC}^{\ast}(\mathbf{r}),\psi_{NC}^{+\ast}(\mathbf{r})]\nonumber \\
 &  & \times\psi_{C}^{+}(\mathbf{r}_{1})\,\psi_{C}^{+}(\mathbf{r}_{2})\,..\psi_{C}^{+}(\mathbf{r}_{p})\,\psi_{C}(\mathbf{s}_{q})\,..\psi_{C}(\mathbf{s}_{2}).\psi_{C}(\mathbf{s}_{1})\nonumber \\
 &  & \times\psi_{NC}^{+}(\mathbf{u}_{1})\,\psi_{NC}^{+}(\mathbf{u}_{2})\,..\psi_{NC}^{+}(\mathbf{u}_{r})\,\psi_{NC}(\mathbf{v}_{s})\,..\psi_{NC}(\mathbf{v}_{2})\,\psi_{NC}(\mathbf{v}_{1})\label{Eq.QuantumAverages-1}\end{eqnarray}
\pagebreak{}

\section{- Correspondence Rules}

\label{App 4}

As the expressions can get cumbersome we find it convenient at times
to use the following notation:\begin{eqnarray}
\underrightarrow{\xi}(\mathbf{r}) & \equiv & \{\xi_{C}(\mathbf{r}),\xi_{C}^{+}(\mathbf{r}),\xi_{NC}(\mathbf{r}),\xi_{NC}^{+}(\mathbf{r})\}\\
\underrightarrow{\xi^{C}}(\mathbf{r}) & \equiv & \{\xi_{C}(\mathbf{r}),\xi_{C}^{+}(\mathbf{r})\}\qquad\underrightarrow{\xi}^{NC}(\mathbf{r})\equiv\{\xi_{NC}(\mathbf{r}),\xi_{NC}^{+}(\mathbf{r})\}\\
\chi\lbrack\underrightarrow{\xi}(\mathbf{r})] & \equiv & \chi\lbrack\xi_{C},\xi_{C}^{+},\xi_{NC},\xi_{NC}^{+}]\\
\underrightarrow{\psi}(\mathbf{r}) & \equiv & \{\psi_{C}(\mathbf{r}),\psi_{C}^{+}(\mathbf{r}),\psi_{NC}(\mathbf{r}),\psi_{NC}^{+}(\mathbf{r})\}\\
\underrightarrow{\psi^{\ast}}(\mathbf{r}) & \equiv & \{\psi_{C}^{\ast}(\mathbf{r}),\psi_{C}^{+\ast}(\mathbf{r}),\psi_{NC}^{\ast}(\mathbf{r}),\psi_{NC}^{+\ast}(\mathbf{r})\}\\
P[\underrightarrow{\psi}(\mathbf{r}),\underrightarrow{\psi^{\ast}}(\mathbf{r})] & \equiv & P[\psi_{C}(\mathbf{r}),\psi_{C}^{+}(\mathbf{r}),\psi_{NC}(\mathbf{r}),\psi_{NC}^{+}(\mathbf{r}),\psi_{C}^{\ast}(\mathbf{r}),\psi_{C}^{+\ast}(\mathbf{r}),\psi_{NC}^{\ast}(\mathbf{r}),\psi_{NC}^{+\ast}(\mathbf{r})]\nonumber \\
 &  & \,\end{eqnarray}

\subsection{Functional Derivative Rules - Condensate Operators}

To proceed further we need to establish some rules for functional
derivatives of operator expressions.Consider\begin{eqnarray*}
\widehat{\Omega}_{C}[\xi_{C},\xi_{C}^{+}] & = & \exp\widehat{G}[\xi_{C},\xi_{C}^{+}]\\
\widehat{G}[\xi_{C},\xi_{C}^{+}] & = & \int d\mathbf{r}\, i\{\xi_{C}(\mathbf{r})\widehat{\Psi}_{C}^{\dagger}(\mathbf{r})+\widehat{\Psi}_{C}(\mathbf{r})\xi_{C}^{+}(\mathbf{r})\}\end{eqnarray*}

(1) We first establish a result for $\widehat{\Omega}_{C}[\xi_{C},\xi_{C}^{+}]\,\widehat{\Psi}_{C}^{\dagger}(\mathbf{s})$.
Now \begin{eqnarray*}
\left(\frac{\delta\widehat{\Omega}_{C}[\xi_{C},\xi_{C}^{+}]}{\delta\xi_{C}}\right)_{\mathbf{r=s}} & = & \lim_{\epsilon\rightarrow0}\left(\frac{\exp\widehat{G}[\xi_{C}(\mathbf{r)+}\epsilon\delta(\mathbf{r-s)},\xi_{C}^{+}(\mathbf{r})]-\exp\widehat{G}[\xi_{C}(\mathbf{r)},\xi_{C}^{+}(\mathbf{r})]}{\epsilon}\right)\\
 & = & \lim_{\epsilon\rightarrow0}\left(\frac{\exp\{\widehat{G}[\xi_{C}(\mathbf{r)},\xi_{C}^{+}(\mathbf{r})]+\epsilon i\widehat{\Psi}_{C}^{\dagger}(\mathbf{s})\}-\exp\widehat{G}[\xi_{C}(\mathbf{r)},\xi_{C}^{+}(\mathbf{r})]}{\epsilon}\right)\end{eqnarray*}
Now we can use the Baker-Haussdorf theorem which is that $\exp(\widehat{A}+\widehat{B})=\exp(\widehat{A})\,\exp(\widehat{B})\,\exp\{-\frac{1}{2}[\widehat{A},\widehat{B}]\}$,
if the commutator commutes with $\widehat{A}$ and $\widehat{B}$,
so with $\widehat{A}=\widehat{G}[\xi_{C}(\mathbf{r)},\xi_{C}^{+}(\mathbf{r})]$
and $\widehat{B}=\epsilon i\widehat{\Psi}_{C}^{\dagger}(\mathbf{s})$
we have \begin{eqnarray*}
\exp\{\widehat{G}[\xi_{C}(\mathbf{r)},\xi_{C}^{+}(\mathbf{r})]+\epsilon i\widehat{\Psi}_{C}^{\dagger}(\mathbf{s})\} & = & \exp\widehat{G}[\xi_{C}(\mathbf{r)},\xi_{C}^{+}(\mathbf{r})]\,\exp\epsilon i\widehat{\Psi}_{C}^{\dagger}(\mathbf{s})\,\exp\frac{1}{2}\epsilon\xi_{C}^{+}(\mathbf{s})\\
 & \doteqdot & \exp\widehat{G}[\xi_{C}(\mathbf{r)},\xi_{C}^{+}(\mathbf{r})]\,\{1+\epsilon(i\widehat{\Psi}_{C}^{\dagger}(\mathbf{s})+\frac{1}{2}\xi_{C}^{+}(\mathbf{s}))\}\end{eqnarray*}
since using Eqs.(19, 22) \begin{eqnarray*}
\lbrack\widehat{G}[\xi_{C}(\mathbf{r)},\xi_{C}^{+}(\mathbf{r})],\epsilon i\widehat{\Psi}_{C}^{\dagger}(\mathbf{s})\,] & = & \epsilon i\int d\mathbf{r}\, i[\{\xi_{C}(\mathbf{r})\widehat{\Psi}_{C}^{\dagger}(\mathbf{r})+\widehat{\Psi}_{C}(\mathbf{r})\xi_{C}^{+}(\mathbf{r})\},\widehat{\Psi}_{C}^{\dagger}(\mathbf{s})]\\
 & = & \epsilon i^{2}\int d\mathbf{r}\,\xi_{C}^{+}(\mathbf{r})[\widehat{\Psi}_{C}(\mathbf{r}),\widehat{\Psi}_{C}^{\dagger}(\mathbf{s})]\\
 & = & \epsilon i^{2}\int d\mathbf{r}\,\xi_{C}^{+}(\mathbf{r})\delta_{C}(\mathbf{r},\mathbf{s})\\
 & = & -\epsilon\xi_{C}^{+}(\mathbf{s})\end{eqnarray*}
noting that the $\xi_{C}^{+}(\mathbf{r})$ only involve complex conjugates
of condensate modes. Hence\begin{eqnarray}
\left(\frac{\delta\widehat{\Omega}_{C}[\xi_{C},\xi_{C}^{+}]}{\delta\xi_{C}}\right)_{\mathbf{r=s}} & = & \widehat{\Omega}_{C}[\xi_{C},\xi_{C}^{+}]\,(i\widehat{\Psi}_{C}^{\dagger}(\mathbf{s})+\frac{1}{2}\xi_{C}^{+}(\mathbf{s}))\nonumber \\
\widehat{\Omega}_{C}[\xi_{C},\xi_{C}^{+}]\,\widehat{\Psi}_{C}^{\dagger}(\mathbf{s}) & = & \frac{1}{i}\left(\frac{\delta\widehat{\Omega}_{C}[\xi_{C},\xi_{C}^{+}]}{\delta\xi_{C}}\right)_{\mathbf{r=s}}-\frac{1}{i}\widehat{\Omega}_{C}[\xi_{C},\xi_{C}^{+}]\,\frac{1}{2}\xi_{C}^{+}(\mathbf{s})\nonumber \\
 &  & \,\label{eq:OmegaCPsiCDagRule}\end{eqnarray}

(2) We next establish a result for $\widehat{\Omega}_{C}[\xi_{C},\xi_{C}^{+}]\,\widehat{\Psi}_{C}(\mathbf{s})$.
Similarly

\begin{eqnarray*}
\left(\frac{\delta\widehat{\Omega}_{C}[\xi_{C},\xi_{C}^{+}]}{\delta\xi_{C}^{+}}\right)_{\mathbf{r=s}} & = & \lim_{\epsilon\rightarrow0}\left(\frac{\exp\widehat{G}[\xi_{C}(\mathbf{r)},\xi_{C}^{+}(\mathbf{r})\mathbf{+}\epsilon\delta(\mathbf{r-s)}]-\exp\widehat{G}[\xi_{C}(\mathbf{r)},\xi_{C}^{+}(\mathbf{r})]}{\epsilon}\right)\\
 & = & \lim_{\epsilon\rightarrow0}\left(\frac{\exp\{\widehat{G}[\xi_{C}(\mathbf{r)},\xi_{C}^{+}(\mathbf{r})]+\epsilon i\widehat{\Psi}_{C}(\mathbf{s})\}-\exp\widehat{G}[\xi_{C}(\mathbf{r)},\xi_{C}^{+}(\mathbf{r})]}{\epsilon}\right)\end{eqnarray*}
Using the Baker-Haussdorf theorem again but now with $\widehat{A}=\widehat{G}[\xi_{C}(\mathbf{r)},\xi_{C}^{+}(\mathbf{r})]$
and $\widehat{B}=\epsilon i\widehat{\Psi}_{C}(\mathbf{s})$ we have
\begin{eqnarray*}
\exp\{\widehat{G}[\xi_{C}(\mathbf{r)},\xi_{C}^{+}(\mathbf{r})]+\epsilon i\widehat{\Psi}_{C}(\mathbf{s})\} & = & \exp\widehat{G}[\xi_{C}(\mathbf{r)},\xi_{C}^{+}(\mathbf{r})]\,\exp\epsilon i\widehat{\Psi}_{C}(\mathbf{s})\,\exp-\frac{1}{2}\epsilon\xi_{C}(\mathbf{s})\\
 & \doteqdot & \exp\widehat{G}[\xi_{C}(\mathbf{r)},\xi_{C}^{+}(\mathbf{r})]\,\{1+\epsilon(i\widehat{\Psi}_{C}(\mathbf{s})-\frac{1}{2}\xi_{C}(\mathbf{s}))\}\end{eqnarray*}
since using Eqs.(19, 22) \begin{eqnarray*}
\lbrack\widehat{G}[\xi_{C}(\mathbf{r)},\xi_{C}^{+}(\mathbf{r})],\epsilon i\widehat{\Psi}_{C}(\mathbf{s})\,] & = & \epsilon i\int d\mathbf{r}\, i[\{\xi_{C}(\mathbf{r})\widehat{\Psi}_{C}^{\dagger}(\mathbf{r})+\widehat{\Psi}_{C}(\mathbf{r})\xi_{C}^{+}(\mathbf{r})\},\widehat{\Psi}_{C}(\mathbf{s})]\\
 & = & \epsilon i^{2}\int d\mathbf{r}\,\xi_{C}(\mathbf{r})[\widehat{\Psi}_{C}^{\dagger}(\mathbf{r}),\widehat{\Psi}_{C}(\mathbf{s})]\\
 & = & \epsilon\int d\mathbf{r}\,\xi_{C}(\mathbf{r})\delta_{C}(\mathbf{s},\mathbf{r})\\
 & = & \epsilon\xi_{C}(\mathbf{s})\end{eqnarray*}
noting that the $\xi_{C}(\mathbf{r})$ only involve condensate modes.
Hence\begin{eqnarray}
\left(\frac{\delta\widehat{\Omega}_{C}[\xi_{C},\xi_{C}^{+}]}{\delta\xi_{C}^{+}}\right)_{\mathbf{r=s}} & = & \widehat{\Omega}_{C}[\xi_{C},\xi_{C}^{+}]\,(i\widehat{\Psi}_{C}(\mathbf{s})-\frac{1}{2}\xi_{C}(\mathbf{s}))\nonumber \\
\widehat{\Omega}_{C}[\xi_{C},\xi_{C}^{+}]\,\widehat{\Psi}_{C}(\mathbf{s}) & = & \frac{1}{i}\left(\frac{\delta\widehat{\Omega}_{C}[\xi_{C},\xi_{C}^{+}]}{\delta\xi_{C}^{+}}\right)_{\mathbf{r=s}}+\frac{1}{i}\widehat{\Omega}_{C}[\xi_{C},\xi_{C}^{+}]\,\frac{1}{2}\xi_{C}(\mathbf{s})\nonumber \\
 &  & \,\label{eq:OmegaCPsiCRule}\end{eqnarray}

(3) We next establish a result for $\widehat{\Psi}_{C}^{\dagger}(\mathbf{s})$
$\widehat{\Omega}_{C}[\xi_{C},\xi_{C}^{+}]\,$. From above\[
\left(\frac{\delta\widehat{\Omega}_{C}[\xi_{C},\xi_{C}^{+}]}{\delta\xi_{C}}\right)_{\mathbf{r=s}}=\lim_{\epsilon\rightarrow0}\left(\frac{\exp\{\widehat{G}[\xi_{C}(\mathbf{r)},\xi_{C}^{+}(\mathbf{r})]+\epsilon i\widehat{\Psi}_{C}^{\dagger}(\mathbf{s})\}-\exp\widehat{G}[\xi_{C}(\mathbf{r)},\xi_{C}^{+}(\mathbf{r})]}{\epsilon}\right)\]
Now we use the Baker-Haussdorf theorem with $\widehat{A}=\epsilon i\widehat{\Psi}_{C}^{\dagger}(\mathbf{s})$
and $\widehat{B}=$ $\widehat{G}[\xi_{C}(\mathbf{r)},\xi_{C}^{+}(\mathbf{r})]$
we have \begin{eqnarray*}
\exp\{\widehat{G}[\xi_{C}(\mathbf{r)},\xi_{C}^{+}(\mathbf{r})]+\epsilon i\widehat{\Psi}_{C}^{\dagger}(\mathbf{s})\} & = & \exp\epsilon i\widehat{\Psi}_{C}^{\dagger}(\mathbf{s})\,\exp\widehat{G}[\xi_{C}(\mathbf{r)},\xi_{C}^{+}(\mathbf{r})]\,\exp-\frac{1}{2}\epsilon\xi_{C}^{+}(\mathbf{s})\\
 & \doteqdot & \{1+\epsilon(i\widehat{\Psi}_{C}^{\dagger}(\mathbf{s})-\frac{1}{2}\xi_{C}^{+}(\mathbf{s}))\}\exp\widehat{G}[\xi_{C}(\mathbf{r)},\xi_{C}^{+}(\mathbf{r})]\,\end{eqnarray*}
using the commutation result derived earlier

Hence\begin{eqnarray}
\left(\frac{\delta\widehat{\Omega}_{C}[\xi_{C},\xi_{C}^{+}]}{\delta\xi_{C}}\right)_{\mathbf{r=s}} & = & (i\widehat{\Psi}_{C}^{\dagger}(\mathbf{s})-\frac{1}{2}\xi_{C}^{+}(\mathbf{s}))\,\widehat{\Omega}_{C}[\xi_{C},\xi_{C}^{+}]\,\nonumber \\
\widehat{\Psi}_{C}^{\dagger}(\mathbf{s})\,\widehat{\Omega}_{C}[\xi_{C},\xi_{C}^{+}]\, & = & \frac{1}{i}\left(\frac{\delta\widehat{\Omega}_{C}[\xi_{C},\xi_{C}^{+}]}{\delta\xi_{C}}\right)_{\mathbf{r=s}}+\frac{1}{i}\frac{1}{2}\xi_{C}^{+}(\mathbf{s})\,\widehat{\Omega}_{C}[\xi_{C},\xi_{C}^{+}]\,\nonumber \\
 &  & \,\label{eq:CDagOmegaRule}\end{eqnarray}

(4) We next establish a result for $\widehat{\Psi}_{C}(\mathbf{s})$
$\widehat{\Omega}_{C}[\xi_{C},\xi_{C}^{+}]\,$. From above

\[
\left(\frac{\delta\widehat{\Omega}_{C}[\xi_{C},\xi_{C}^{+}]}{\delta\xi_{C}^{+}}\right)_{\mathbf{r=s}}=\lim_{\epsilon\rightarrow0}\left(\frac{\exp\{\widehat{G}[\xi_{C}(\mathbf{r)},\xi_{C}^{+}(\mathbf{r})]+\epsilon i\widehat{\Psi}_{C}(\mathbf{s})\}-\exp\widehat{G}[\xi_{C}(\mathbf{r)},\xi_{C}^{+}(\mathbf{r})]}{\epsilon}\right)\]
Using the Baker-Haussdorf theorem again but now with $\widehat{A}=\epsilon i\widehat{\Psi}_{C}(\mathbf{s})$
and $\widehat{B}=$ $\widehat{G}[\xi_{C}(\mathbf{r)},\xi_{C}^{+}(\mathbf{r})]$
we have \begin{eqnarray*}
\exp\{\widehat{G}[\xi_{C}(\mathbf{r)},\xi_{C}^{+}(\mathbf{r})]+\epsilon i\widehat{\Psi}_{C}(\mathbf{s})\} & = & \exp\epsilon i\widehat{\Psi}_{C}(\mathbf{s})\,\exp\widehat{G}[\xi_{C}(\mathbf{r)},\xi_{C}^{+}(\mathbf{r})]\,\exp+\frac{1}{2}\epsilon\xi_{C}(\mathbf{s})\\
 & \doteqdot & \{1+\epsilon(i\widehat{\Psi}_{C}(\mathbf{s})+\frac{1}{2}\xi_{C}(\mathbf{s}))\}\exp\widehat{G}[\xi_{C}(\mathbf{r)},\xi_{C}^{+}(\mathbf{r})]\,\end{eqnarray*}
using the commutation rule derived earlier.

Hence\begin{eqnarray}
\left(\frac{\delta\widehat{\Omega}_{C}[\xi_{C},\xi_{C}^{+}]}{\delta\xi_{C}^{+}}\right)_{\mathbf{r=s}} & = & (i\widehat{\Psi}_{C}(\mathbf{s})+\frac{1}{2}\xi_{C}(\mathbf{s}))\,\widehat{\Omega}_{C}[\xi_{C},\xi_{C}^{+}]\,\nonumber \\
\widehat{\Psi}_{C}(\mathbf{s})\,\widehat{\Omega}_{C}[\xi_{C},\xi_{C}^{+}]\, & = & \frac{1}{i}\left(\frac{\delta\widehat{\Omega}_{C}[\xi_{C},\xi_{C}^{+}]}{\delta\xi_{C}^{+}}\right)_{\mathbf{r=s}}-\frac{1}{i}\frac{1}{2}\xi_{C}(\mathbf{s})\,\widehat{\Omega}_{C}[\xi_{C},\xi_{C}^{+}]\,\nonumber \\
 &  & \,\label{eq:PsiCOmegaCRule}\end{eqnarray}

(5) To establish a result for $\widehat{\Omega}_{C}[\xi_{C},\xi_{C}^{+}]\,\partial_{\mu}\widehat{\Psi}_{C}^{\dagger}(\mathbf{s})$
we start with\[
\partial_{\mu}\widehat{\Psi}_{C}^{\dagger}(\mathbf{s})=\lim_{\Delta s_{\mu}\rightarrow0}\left(\frac{\widehat{\Psi}_{C}^{\dagger}(\mathbf{s+\Delta s}_{\mu})-\widehat{\Psi}_{C}^{\dagger}(\mathbf{s})}{\mathbf{\Delta s}_{\mu}}\right)\]
so we can use previous results in Eq.(\ref{eq:OmegaCPsiCDagRule})
for $\widehat{\Omega}_{C}[\xi_{C},\xi_{C}^{+}]\,\widehat{\Psi}_{C}^{\dagger}(\mathbf{s})$.
Using the previous results we have\begin{eqnarray*}
\widehat{\Omega}_{C}[\xi_{C},\xi_{C}^{+}]\,\partial_{\mu}\widehat{\Psi}_{C}^{\dagger}(\mathbf{s}) & = & \lim_{\Delta s_{\mu}\rightarrow0}\frac{1}{i}\frac{1}{\mathbf{\Delta s}_{\mu}}\left(\left(\frac{\delta\widehat{\Omega}_{C}[\xi_{C},\xi_{C}^{+}]}{\delta\xi_{C}}\right)_{\mathbf{r=s+\Delta s}_{\mu}}-\left(\frac{\delta\widehat{\Omega}_{C}[\xi_{C},\xi_{C}^{+}]}{\delta\xi_{C}}\right)_{\mathbf{r=s}}\right)\\
 &  & -\lim_{\Delta s_{\mu}\rightarrow0}\frac{1}{i}\widehat{\Omega}_{C}[\xi_{C},\xi_{C}^{+}]\,\frac{1}{2}\left(\frac{1}{\mathbf{\Delta s}_{\mu}}\left(\xi_{C}^{+}(\mathbf{s+\Delta s}_{\mu})-\xi_{C}^{+}(\mathbf{s})\right)\right)\\
 & = & \frac{1}{i}\left(\partial_{\mu}\left(\frac{\delta\widehat{\Omega}_{C}[\xi_{C},\xi_{C}^{+}]}{\delta\xi_{C}}\right)\right)_{_{\mathbf{r=s}}}-\frac{1}{i}\widehat{\Omega}_{C}[\xi_{C},\xi_{C}^{+}]\,\frac{1}{2}\left(\partial_{\mu}\xi_{C}^{+}\right)_{\mathbf{r=s}}\end{eqnarray*}
so that from the definition of the spatial derivative we obtain the
result\begin{equation}
\widehat{\Omega}_{C}[\xi_{C},\xi_{C}^{+}]\,(\partial_{\mu}\widehat{\Psi}_{C}^{\dagger}(\mathbf{s}))=\frac{1}{i}\left(\partial_{\mu}\frac{\delta\widehat{\Omega}_{C}[\xi_{C},\xi_{C}^{+}]}{\delta\xi_{C}}\right)_{\mathbf{r=s}}-\frac{1}{i}\,\widehat{\Omega}_{C}[\xi_{C},\xi_{C}^{+}]\,\frac{1}{2}\partial_{\mu}\xi_{C}^{+}(\mathbf{s})\label{Eq.OmegaCDerivPsiCDag}\end{equation}

(6) To establish a result for $\widehat{\Omega}_{C}[\xi_{C},\xi_{C}^{+}]\,\partial_{\mu}\widehat{\Psi}_{C}(\mathbf{s})$
we start with\[
\partial_{\mu}\widehat{\Psi}_{C}(\mathbf{s})=\lim_{\Delta s_{\mu}\rightarrow0}\left(\frac{\widehat{\Psi}_{C}(\mathbf{s+\Delta s}_{\mu})-\widehat{\Psi}_{C}(\mathbf{s})}{\mathbf{\Delta s}_{\mu}}\right)\]
so we can use previous results in Eq.(\ref{eq:OmegaCPsiCRule}) for
$\widehat{\Omega}_{C}[\xi_{C},\xi_{C}^{+}]\,\widehat{\Psi}_{C}(\mathbf{s})$.
Using the previous results we have\begin{eqnarray*}
\widehat{\Omega}_{C}[\xi_{C},\xi_{C}^{+}]\,\partial_{\mu}\widehat{\Psi}_{C}(\mathbf{s}) & = & \lim_{\Delta s_{\mu}\rightarrow0}\frac{1}{i}\frac{1}{\mathbf{\Delta s}_{\mu}}\left(\left(\frac{\delta\widehat{\Omega}_{C}[\xi_{C},\xi_{C}^{+}]}{\delta\xi_{C}^{+}}\right)_{\mathbf{r=s+\Delta s}_{\mu}}-\left(\frac{\delta\widehat{\Omega}_{C}[\xi_{C},\xi_{C}^{+}]}{\delta\xi_{C}^{+}}\right)_{\mathbf{r=s}}\right)\\
 &  & +\lim_{\Delta s_{\mu}\rightarrow0}\frac{1}{i}\widehat{\Omega}_{C}[\xi_{C},\xi_{C}^{+}]\,\frac{1}{2}\left(\frac{1}{\mathbf{\Delta s}_{\mu}}\left(\xi_{C}(\mathbf{s+\Delta s}_{\mu})-\xi_{C}(\mathbf{s})\right)\right)\\
 & = & \frac{1}{i}\left(\partial_{\mu}\left(\frac{\delta\widehat{\Omega}_{C}[\xi_{C},\xi_{C}^{+}]}{\delta\xi_{C}^{+}}\right)\right)_{_{\mathbf{r=s}}}+\frac{1}{i}\widehat{\Omega}_{C}[\xi_{C},\xi_{C}^{+}]\,\frac{1}{2}\left(\partial_{\mu}\xi_{C}\right)_{\mathbf{r=s}}\end{eqnarray*}
so that from the definition of the spatial derivative we obtain the
result\begin{equation}
\widehat{\Omega}_{C}[\xi_{C},\xi_{C}^{+}]\,(\partial_{\mu}\widehat{\Psi}_{C}(\mathbf{s}))=\frac{1}{i}\left(\partial_{\mu}\frac{\delta\widehat{\Omega}_{C}[\xi_{C},\xi_{C}^{+}]}{\delta\xi_{C}^{+}}\right)_{\mathbf{r=s}}+\frac{1}{i}\,\widehat{\Omega}_{C}[\xi_{C},\xi_{C}^{+}]\,\frac{1}{2}\partial_{\mu}\xi_{C}(\mathbf{s})\label{Eq.OmegaCDerivPsiC}\end{equation}

(7) To establish a result for $\partial_{\mu}\widehat{\Psi}_{C}^{\dagger}(\mathbf{s})$
$\widehat{\Omega}_{C}[\xi_{C},\xi_{C}^{+}]\,$\ we can use Eq.(\ref{eq:CDagOmegaRule})
and follow the previous procedure to obtain the result\begin{align}
\partial_{\mu}\widehat{\Psi}_{C}^{\dagger}(\mathbf{s})\,\widehat{\Omega}_{C}[\xi_{C},\xi_{C}^{+}]\, & =\frac{1}{i}\left(\partial_{\mu}\frac{\delta\widehat{\Omega}_{C}[\xi_{C},\xi_{C}^{+}]}{\delta\xi_{C}}\right)_{\mathbf{r=s}}+\frac{1}{i}\frac{1}{2}\partial_{\mu}\xi_{C}^{+}(\mathbf{s})\,\widehat{\Omega}_{C}[\xi_{C},\xi_{C}^{+}]\,\nonumber \\
 & \,\label{eq:DerivPsiCDagOmegaC}\end{align}

(8) To establish a result for $\partial_{\mu}\widehat{\Psi}_{C}(\mathbf{s})$
$\widehat{\Omega}_{C}[\xi_{C},\xi_{C}^{+}]\,$\ we can use Eq.(\ref{eq:PsiCOmegaCRule})
and follow the previous procedure to obtain the result\begin{align}
\partial_{\mu}\widehat{\Psi}_{C}(\mathbf{s})\,\widehat{\Omega}_{C}[\xi_{C},\xi_{C}^{+}]\, & =\frac{1}{i}\left(\partial_{\mu}\frac{\delta\widehat{\Omega}_{C}[\xi_{C},\xi_{C}^{+}]}{\delta\xi_{C}^{+}}\right)_{\mathbf{r=s}}-\frac{1}{i}\frac{1}{2}\partial_{\mu}\xi_{C}(\mathbf{s})\,\widehat{\Omega}_{C}[\xi_{C},\xi_{C}^{+}]\,\nonumber \\
 & \,\label{eq:DerivPsiCOmegaC}\end{align}

\subsection{Condensate Operators}

(1) If $\widehat{\rho}$ is replaced by $\widehat{\Psi}_{C}(\mathbf{s})\widehat{\rho}$
then the characteristic function becomes\begin{eqnarray*}
\chi\lbrack\xi_{C},\xi_{C}^{+},\xi_{NC},\xi_{NC}^{+}] & \rightarrow & Tr(\widehat{\Psi}_{C}(\mathbf{s})\widehat{\rho}\,\widehat{\Omega}[\xi_{C},\xi_{C}^{+},\xi_{NC},\xi_{NC}^{+}])\\
 & = & Tr(\widehat{\rho}\,\widehat{\Omega}[\xi_{C},\xi_{C}^{+},\xi_{NC},\xi_{NC}^{+}]\widehat{\Psi}_{C}(\mathbf{s}))\\
 & = & Tr(\widehat{\rho}\,\,\widehat{\Omega}_{C}[\xi_{C},\xi_{C}^{+}]\,\widehat{\Psi}_{C}(\mathbf{s})\,\widehat{\Omega}_{NC})\end{eqnarray*}
using the cyclic property of the trace and the feature that condensate
operators commute with non-condensate operators.

Hence from Eq.(\ref{eq:OmegaCPsiCRule}) \[
\chi\lbrack\xi_{C},\xi_{C}^{+},\xi_{NC},\xi_{NC}^{+}]\rightarrow Tr(\widehat{\rho}\,\frac{1}{i}\,\left(\frac{\delta\widehat{\Omega}_{C}[\xi_{C},\xi_{C}^{+}]}{\delta\xi_{C}^{+}(\mathbf{s})}+\widehat{\Omega}_{C}[\xi_{C},\xi_{C}^{+}]\,\frac{1}{2}\xi_{C}(\mathbf{s})\right)\,\widehat{\Omega}_{NC})\]
Hence\[
\chi\lbrack\xi_{C},\xi_{C}^{+},\xi_{NC},\xi_{NC}^{+}]\rightarrow\frac{1}{i}\,\left(\frac{\delta}{\delta\xi_{C}^{+}(\mathbf{s})}+\frac{1}{2}\xi_{C}(\mathbf{s})\right)\,\chi\lbrack\xi_{C},\xi_{C}^{+},\xi_{NC},\xi_{NC}^{+}]\]
Then using the relationship to the distribution functional we see
that \begin{eqnarray*}
\chi\lbrack\xi_{C},\xi_{C}^{+},\xi_{NC},\xi_{NC}^{+}] & \rightarrow & \frac{1}{i}\,\left(\frac{\delta}{\delta\xi_{C}^{+}(\mathbf{s})}+\frac{1}{2}\xi_{C}(\mathbf{s})\right)\iiiint D^{2}\psi_{C}\, D^{2}\psi_{C}^{+}\, D^{2}\psi_{NC}\, D^{2}\psi_{NC}^{+}\, P[\underrightarrow{\psi}(\mathbf{r}),\underrightarrow{\psi^{\ast}}(\mathbf{r})]\\
 &  & \times\exp i\int d\mathbf{r\,\{}\xi_{C}(\mathbf{r})\psi_{C}^{+}(\mathbf{r})+\psi_{C}(\mathbf{r})\xi_{C}^{+}(\mathbf{r})\}\,\\
 &  & \times\exp i\int d\mathbf{r\,\{}\xi_{NC}(\mathbf{r})\psi_{NC}^{+}(\mathbf{r})\}\exp i\int d\mathbf{r\,\{}\psi_{NC}(\mathbf{r})\xi_{NC}^{+}(\mathbf{r})\}\,\\
 & = & \iiiint D^{2}\psi_{C}\, D^{2}\psi_{C}^{+}\, D^{2}\psi_{NC}\, D^{2}\psi_{NC}^{+}\, P[\underrightarrow{\psi}(\mathbf{r}),\underrightarrow{\psi^{\ast}}(\mathbf{r})]\\
 &  & \times\lbrack\left(\psi_{C}(\mathbf{s})-\frac{1}{2}\frac{\delta}{\delta\psi_{C}^{+}(\mathbf{s})}\right)\exp i\int d\mathbf{r\,\{}\xi_{C}(\mathbf{r})\psi_{C}^{+}(\mathbf{r})+\psi_{C}(\mathbf{r})\xi_{C}^{+}(\mathbf{r})\}]\\
 &  & \times\,\exp i\int d\mathbf{r\,\{}\xi_{NC}(\mathbf{r})\psi_{NC}^{+}(\mathbf{r})\}\exp i\int d\mathbf{r\,\{}\psi_{NC}(\mathbf{r})\xi_{NC}^{+}(\mathbf{r})\}\,\end{eqnarray*}
since from the functional differentiation rules with $G[\psi_{C},\xi_{C}^{+},\xi_{C},\psi_{C}^{+}]=i\int d\mathbf{r\,\{}\xi_{C}(\mathbf{r})\psi_{C}^{+}(\mathbf{r})+\psi_{C}(\mathbf{r})\xi_{C}^{+}(\mathbf{r})\}$\begin{eqnarray}
 &  & \frac{1}{i}\frac{\delta}{\delta\xi_{C}^{+}(\mathbf{s})}\exp i\int d\mathbf{r\,\{}\xi_{C}(\mathbf{r})\psi_{C}^{+}(\mathbf{r})+\psi_{C}(\mathbf{r})\xi_{C}^{+}(\mathbf{r})\}\nonumber \\
 & = & \frac{1}{i}\exp G[\psi_{C},\xi_{C}^{+},\xi_{C},\psi_{C}^{+}]\,\frac{\delta G[\psi_{C},\xi_{C}^{+},\xi_{C},\psi_{C}^{+}]}{\delta\xi_{C}^{+}(\mathbf{s})}\nonumber \\
 & = & \frac{1}{i}\exp G[\psi_{C},\xi_{C}^{+},\xi_{C},\psi_{C}^{+}]\, i\psi_{C}(\mathbf{s})\nonumber \\
 & = & \psi_{C}(\mathbf{s})\,\exp G[\psi_{C},\xi_{C}^{+},\xi_{C},\psi_{C}^{+}]\\
 &  & -\frac{1}{2}\frac{\delta}{\delta\psi_{C}^{+}(\mathbf{s})}\exp i\int d\mathbf{r\,\mathbf{\{}\xi_{C}(\mathbf{r})\psi_{C}^{+}(\mathbf{r})+\psi_{C}(\mathbf{r})\xi_{C}^{+}(\mathbf{r})\}}\nonumber \\
 & = & -\frac{1}{2}\exp G[\psi_{C},\xi_{C}^{+},\xi_{C},\psi_{C}^{+}]\,\frac{\delta G[\psi_{C},\xi_{C}^{+},\xi_{C},\psi_{C}^{+}]}{\delta\psi_{C}^{+}(\mathbf{s})}\nonumber \\
 & = & -\frac{1}{2}\,\exp G[\psi_{C},\xi_{C}^{+},\xi_{C},\psi_{C}^{+}]\, i\xi_{C}(\mathbf{s})\\
 &  & \frac{1}{i}\frac{1}{2}\xi_{C}(\mathbf{s)\,}\exp G[\psi_{C},\xi_{C}^{+},\xi_{C},\psi_{C}^{+}]\nonumber \\
 & = & -\frac{1}{2}\frac{\delta}{\delta\psi_{C}^{+}(\mathbf{s})}\exp i\int d\mathbf{r\,\mathbf{\{}\xi_{C}(\mathbf{r})\psi_{C}^{+}(\mathbf{r})+\psi_{C}(\mathbf{r})\xi_{C}^{+}(\mathbf{r})\}}\end{eqnarray}

To proceed further we need to replace the functional differentiation
of the exponential functional with a functional differentiation of
the quasi distribution functional itself. This can be accomplished
using a functional integration by parts result, which requires the
condition that the mode expansion form of the product functional $P[\underrightarrow{\psi}(\mathbf{r}),\underrightarrow{\psi^{\ast}}(\mathbf{r})]\exp i\int d\mathbf{r\,\mathbf{\{}\xi_{C}(\mathbf{r})\psi_{C}^{+}(\mathbf{r})+\psi_{C}(\mathbf{r})\xi_{C}^{+}(\mathbf{r})\}}$
goes to zero as the expansion coefficients become large (note that
there is no normalisation condition on the $\psi_{C}(\mathbf{r}),\psi_{C}^{+}(\mathbf{r})$
that bounds the expansion coefficients). Using this integration by
parts result we then find that\begin{eqnarray*}
\chi\lbrack\xi_{C},\xi_{C}^{+},\xi_{NC},\xi_{NC}^{+}] & \rightarrow & \iiiint D^{2}\psi_{C}\, D^{2}\psi_{C}^{+}\, D^{2}\psi_{NC}\, D^{2}\psi_{NC}^{+}\\
 &  & \times\{\left(\psi_{C}(\mathbf{s})+\frac{1}{2}\frac{\delta}{\delta\psi_{C}^{+}(\mathbf{s})}\right)P[\underrightarrow{\psi}(\mathbf{r}),\underrightarrow{\psi^{\ast}}(\mathbf{r})]\}\\
 &  & \times\exp i\int d\mathbf{r\,\mathbf{\{}\xi_{C}(\mathbf{r})\psi_{C}^{+}(\mathbf{r})+\psi_{C}(\mathbf{r})\xi_{C}^{+}(\mathbf{r})\}}]\,\\
 &  & \times\exp i\int d\mathbf{r\,\{}\xi_{NC}(\mathbf{r})\psi_{NC}^{+}(\mathbf{r})\}\exp i\int d\mathbf{r\,\{}\psi_{NC}(\mathbf{r})\xi_{NC}^{+}(\mathbf{r})\}\,\end{eqnarray*}

Hence the change to the characteristic functional if $\widehat{\rho}$
is replaced by $\widehat{\Psi}_{C}(\mathbf{s})\widehat{\rho}$ is
equivalent to then the quasi distribution functional is replaced as
follows \begin{equation}
P[\underrightarrow{\psi}(\mathbf{r}),\underrightarrow{\psi^{\ast}}(\mathbf{r})]\rightarrow\left(\psi_{C}(\mathbf{s})+\frac{1}{2}\frac{\delta}{\delta\psi_{C}^{+}(\mathbf{s})}\right)P[\underrightarrow{\psi}(\mathbf{r}),\underrightarrow{\psi^{\ast}}(\mathbf{r})]\label{Eq.PsiCRhoPRule}\end{equation}
Thus $P[\underrightarrow{\psi}(\mathbf{r}),\underrightarrow{\psi^{\ast}}(\mathbf{r})]$
is both multiplied by $\psi_{C}(\mathbf{s})$, the field function
that the operator $\widehat{\Psi}_{C}(\mathbf{s})$ is mapped onto
and functionally differentiated with respect to $\psi_{C}^{+}(\mathbf{s})$,
the field function that the operator $\widehat{\Psi}_{C}^{\dagger}(\mathbf{s})$
is mapped onto.

(2) If $\widehat{\rho}$ is replaced by $\widehat{\Psi}_{C}^{\dagger}(\mathbf{s})\widehat{\rho}$
then the characteristic function becomes\begin{eqnarray*}
\chi\lbrack\xi_{C},\xi_{C}^{+},\xi_{NC},\xi_{NC}^{+}] & \rightarrow & Tr(\widehat{\Psi}_{C}^{\dagger}(\mathbf{s})\widehat{\rho}\,\widehat{\Omega}[\xi_{C},\xi_{C}^{+},\xi_{NC},\xi_{NC}^{+}])\\
 & = & Tr(\widehat{\rho}\,\widehat{\Omega}[\xi_{C},\xi_{C}^{+},\xi_{NC},\xi_{NC}^{+}]\widehat{\Psi}_{C}^{\dagger}(\mathbf{s}))\\
 & = & Tr(\widehat{\rho}\,\,\widehat{\Omega}_{C}[\xi_{C},\xi_{C}^{+}]\,\widehat{\Psi}_{C}^{\dagger}(\mathbf{s})\,\widehat{\Omega}_{NC})\end{eqnarray*}

using the cyclic property of the trace and the feature that condensate
operators commute with non-condensate operators.

Hence from Eq.(\ref{eq:OmegaCPsiCDagRule}) \[
\chi\lbrack\xi_{C},\xi_{C}^{+},\xi_{NC},\xi_{NC}^{+}]\rightarrow Tr(\widehat{\rho}\,\frac{1}{i}\,\left(\frac{\delta\widehat{\Omega}_{C}[\xi_{C},\xi_{C}^{+}]}{\delta\xi_{C}(\mathbf{s})}-\widehat{\Omega}_{C}[\xi_{C},\xi_{C}^{+}]\,\frac{1}{2}\xi_{C}^{+}(\mathbf{s})\right)\,\widehat{\Omega}_{NC})\]
Hence\[
\chi\lbrack\xi_{C},\xi_{C}^{+},\xi_{NC},\xi_{NC}^{+}]\rightarrow\frac{1}{i}\,\left(\frac{\delta}{\delta\xi_{C}(\mathbf{s})}-\frac{1}{2}\xi_{C}^{+}(\mathbf{s})\right)\,\chi\lbrack\xi_{C},\xi_{C}^{+},\xi_{NC},\xi_{NC}^{+}]\]
Then using the relationship to the distribution functional we see
that \begin{eqnarray*}
\chi\lbrack\xi_{C},\xi_{C}^{+},\xi_{NC},\xi_{NC}^{+}] & \rightarrow & \frac{1}{i}\,\left(\frac{\delta}{\delta\xi_{C}(\mathbf{s})}-\frac{1}{2}\xi_{C}^{+}(\mathbf{s})\right)\iiiint D^{2}\psi_{C}\, D^{2}\psi_{C}^{+}\, D^{2}\psi_{NC}\, D^{2}\psi_{NC}^{+}\, P[\underrightarrow{\psi}(\mathbf{r}),\underrightarrow{\psi^{\ast}}(\mathbf{r})]\\
 &  & \times\exp i\int d\mathbf{r\,\{}\xi_{C}(\mathbf{r})\psi_{C}^{+}(\mathbf{r})+\psi_{C}(\mathbf{r})\xi_{C}^{+}(\mathbf{r})\}\,\\
 &  & \times\exp i\int d\mathbf{r\,\{}\xi_{NC}(\mathbf{r})\psi_{NC}^{+}(\mathbf{r})\}\exp i\int d\mathbf{r\,\{}\psi_{NC}(\mathbf{r})\xi_{NC}^{+}(\mathbf{r})\}\,\\
 & = & \iiiint D^{2}\psi_{C}\, D^{2}\psi_{C}^{+}\, D^{2}\psi_{NC}\, D^{2}\psi_{NC}^{+}\, P[\underrightarrow{\psi}(\mathbf{r}),\underrightarrow{\psi^{\ast}}(\mathbf{r})]\\
 &  & \times\lbrack\left(\psi_{C}^{+}(\mathbf{s})+\frac{1}{2}\frac{\delta}{\delta\psi_{C}(\mathbf{s})}\right)\exp i\int d\mathbf{r\,\{}\xi_{C}(\mathbf{r})\psi_{C}^{+}(\mathbf{r})+\psi_{C}(\mathbf{r})\xi_{C}^{+}(\mathbf{r})\}]\\
 &  & \times\,\exp i\int d\mathbf{r\,\{}\xi_{NC}(\mathbf{r})\psi_{NC}^{+}(\mathbf{r})\}\exp i\int d\mathbf{r\,\{}\psi_{NC}(\mathbf{r})\xi_{NC}^{+}(\mathbf{r})\}\end{eqnarray*}
where the proof of the second step is similar to that in (1).

To proceed further we use integration by parts result we then find
that\begin{eqnarray*}
\chi\lbrack\xi_{C},\xi_{C}^{+},\xi_{NC},\xi_{NC}^{+}] & \rightarrow & \iiiint D^{2}\psi_{C}\, D^{2}\psi_{C}^{+}\, D^{2}\psi_{NC}\, D^{2}\psi_{NC}^{+}\\
 &  & \times\{\left(\psi_{C}^{+}(\mathbf{s})-\frac{1}{2}\frac{\delta}{\delta\psi_{C}(\mathbf{s})}\right)P[\underrightarrow{\psi}(\mathbf{r}),\underrightarrow{\psi^{\ast}}(\mathbf{r})]\}\\
 &  & \times\exp i\int d\mathbf{r\,\{}\xi_{C}(\mathbf{r})\psi_{C}^{+}(\mathbf{r})+\psi_{C}(\mathbf{r})\xi_{C}^{+}(\mathbf{r})\}\,\\
 &  & \times\exp i\int d\mathbf{r\,\{}\xi_{NC}(\mathbf{r})\psi_{NC}^{+}(\mathbf{r})\}\exp i\int d\mathbf{r\,\{}\psi_{NC}(\mathbf{r})\xi_{NC}^{+}(\mathbf{r})\}\end{eqnarray*}

Hence the change to the characteristic functional if $\widehat{\rho}$
is replaced by $\widehat{\Psi}_{C}^{\dagger}(\mathbf{s})\widehat{\rho}$
is equivalent to then the quasi distribution functional is replaced
as follows \begin{equation}
P[\underrightarrow{\psi}(\mathbf{r}),\underrightarrow{\psi^{\ast}}(\mathbf{r})]\rightarrow\left(\psi_{C}^{+}(\mathbf{s})-\frac{1}{2}\frac{\delta}{\delta\psi_{C}(\mathbf{s})}\right)P[\underrightarrow{\psi}(\mathbf{r}),\underrightarrow{\psi^{\ast}}(\mathbf{r})]\label{Eq.PsiCDagRhoPRule}\end{equation}
Thus $P[\underrightarrow{\psi}(\mathbf{r}),\underrightarrow{\psi^{\ast}}(\mathbf{r})]$
is both multiplied by $\psi_{C}^{+}(\mathbf{r})$, the field function
that the operator $\widehat{\Psi}_{C}^{\dagger}(\mathbf{r})$ is mapped
onto and functionally differentiated with respect to $\psi_{C}(\mathbf{r})$,
the field function that the operator $\widehat{\Psi}_{C}(\mathbf{r})$
is mapped onto.

(3) If $\widehat{\rho}$ is replaced by $\widehat{\rho}$ $\widehat{\Psi}_{C}(\mathbf{s})$
then the characteristic function becomes\begin{eqnarray*}
\chi\lbrack\xi_{C},\xi_{C}^{+},\xi_{NC},\xi_{NC}^{+}] & \rightarrow & Tr(\widehat{\rho}\,\widehat{\Psi}_{C}(\mathbf{s})\,\widehat{\Omega}[\xi_{C},\xi_{C}^{+},\xi_{NC},\xi_{NC}^{+}])\\
 & = & Tr(\widehat{\rho}\,\widehat{\Psi}_{C}(\mathbf{s})\,\widehat{\Omega}_{C}[\xi_{C},\xi_{C}^{+}]\,\,\widehat{\Omega}_{NC})\end{eqnarray*}
using the feature that condensate operators commute with non-condensate
operators.

Hence from Eq.(\ref{eq:PsiCOmegaCRule})\[
\chi\lbrack\xi_{C},\xi_{C}^{+},\xi_{NC},\xi_{NC}^{+}]\rightarrow Tr(\widehat{\rho}\,\frac{1}{i}\,\left(\frac{\delta\widehat{\Omega}_{C}[\xi_{C},\xi_{C}^{+}]}{\delta\xi_{C}^{+}(\mathbf{s})}-\frac{1}{2}\xi_{C}(\mathbf{s})\,\widehat{\Omega}_{C}[\xi_{C},\xi_{C}^{+}]\right)\,\widehat{\Omega}_{NC})\]
Hence\[
\chi\lbrack\xi_{C},\xi_{C}^{+},\xi_{NC},\xi_{NC}^{+}]\rightarrow\frac{1}{i}\,\left(\frac{\delta}{\delta\xi_{C}^{+}(\mathbf{s})}-\frac{1}{2}\xi_{C}(\mathbf{s})\right)\,\chi\lbrack\xi_{C},\xi_{C}^{+},\xi_{NC},\xi_{NC}^{+}]\]
Then using the relationship to the distribution functional we see
that \begin{eqnarray*}
\chi\lbrack\xi_{C},\xi_{C}^{+},\xi_{NC},\xi_{NC}^{+}] & \rightarrow & \frac{1}{i}\,\left(\frac{\delta}{\delta\xi_{C}^{+}(\mathbf{s})}-\frac{1}{2}\xi_{C}(\mathbf{s})\right)\iiiint D^{2}\psi_{C}\, D^{2}\psi_{C}^{+}\, D^{2}\psi_{NC}\, D^{2}\psi_{NC}^{+}\, P[\underrightarrow{\psi}(\mathbf{r}),\underrightarrow{\psi^{\ast}}(\mathbf{r})]\\
 &  & \times\exp i\int d\mathbf{r\,\{}\xi_{C}(\mathbf{r})\psi_{C}^{+}(\mathbf{r})+\psi_{C}(\mathbf{r})\xi_{C}^{+}(\mathbf{r})\}\,\\
 &  & \times\exp i\int d\mathbf{r\,\{}\xi_{NC}(\mathbf{r})\psi_{NC}^{+}(\mathbf{r})\}\exp i\int d\mathbf{r\,\{}\psi_{NC}(\mathbf{r})\xi_{NC}^{+}(\mathbf{r})\}\,\\
 & = & \iiiint D^{2}\psi_{C}\, D^{2}\psi_{C}^{+}\, D^{2}\psi_{NC}\, D^{2}\psi_{NC}^{+}\, P[\underrightarrow{\psi}(\mathbf{r}),\underrightarrow{\psi^{\ast}}(\mathbf{r})]\\
 &  & \times\lbrack\left(\psi_{C}(\mathbf{s})+\frac{1}{2}\frac{\delta}{\delta\psi_{C}^{+}(\mathbf{s})}\right)\exp i\int d\mathbf{r\,\{}\xi_{C}(\mathbf{r})\psi_{C}^{+}(\mathbf{r})+\psi_{C}(\mathbf{r})\xi_{C}^{+}(\mathbf{r})\}]\,\\
 &  & \times\exp i\int d\mathbf{r\,\{}\xi_{NC}(\mathbf{r})\psi_{NC}^{+}(\mathbf{r})\}\exp i\int d\mathbf{r\,\{}\psi_{NC}(\mathbf{r})\xi_{NC}^{+}(\mathbf{r})\}\,\end{eqnarray*}
To proceed further we use the integration by parts result and then
find that\begin{eqnarray*}
\chi\lbrack\xi_{C},\xi_{C}^{+},\xi_{NC},\xi_{NC}^{+}] & \rightarrow & \iiiint D^{2}\psi_{C}\, D^{2}\psi_{C}^{+}\, D^{2}\psi_{NC}\, D^{2}\psi_{NC}^{+}\\
 &  & \times\{\left(\psi_{C}(\mathbf{s})-\frac{1}{2}\frac{\delta}{\delta\psi_{C}^{+}(\mathbf{s})}\right)P[\underrightarrow{\psi}(\mathbf{r}),\underrightarrow{\psi^{\ast}}(\mathbf{r})]P[\psi_{C}(\mathbf{r}),\psi_{C}^{+}(\mathbf{r}),\psi_{NC}(\mathbf{r}),\psi_{NC}^{+}(\mathbf{r})]\}\\
 &  & \times\exp i\int d\mathbf{r\,\{}\xi_{C}(\mathbf{r})\psi_{C}^{+}(\mathbf{r})+\psi_{C}(\mathbf{r})\xi_{C}^{+}(\mathbf{r})\}\,\\
 &  & \times\exp i\int d\mathbf{r\,\{}\xi_{NC}(\mathbf{r})\psi_{NC}^{+}(\mathbf{r})\}\exp i\int d\mathbf{r\,\{}\psi_{NC}(\mathbf{r})\xi_{NC}^{+}(\mathbf{r})\}\,\end{eqnarray*}

Hence the change to the characteristic functional if $\widehat{\rho}$
is replaced by $\widehat{\rho}\,\widehat{\Psi}_{C}(\mathbf{s})$ is
equivalent to then the quasi distribution functional is replaced as
follows \begin{equation}
P[\underrightarrow{\psi}(\mathbf{r}),\underrightarrow{\psi^{\ast}}(\mathbf{r})]\rightarrow\left(\psi_{C}(\mathbf{s})-\frac{1}{2}\frac{\delta}{\delta\psi_{C}^{+}(\mathbf{s})}\right)P[\underrightarrow{\psi}(\mathbf{r}),\underrightarrow{\psi^{\ast}}(\mathbf{r})]\label{Eq.RhoPsiCPRule}\end{equation}
Thus $P[\underrightarrow{\psi}(\mathbf{r}),\underrightarrow{\psi^{\ast}}(\mathbf{r})]$
is both multiplied by $\psi_{C}(\mathbf{s})$, the field function
that the operator $\widehat{\Psi}_{C}(\mathbf{s})$ is mapped onto
and functionally differentiated with respect to $\psi_{C}^{+}(\mathbf{s})$,
the field function that the operator $\widehat{\Psi}_{C}^{\dagger}(\mathbf{s})$
is mapped onto.

(4) If $\widehat{\rho}$ is replaced by $\widehat{\rho}$ $\widehat{\Psi}_{C}^{\dagger}(\mathbf{s})$
then the characteristic function becomes\begin{eqnarray*}
\chi\lbrack\xi_{C},\xi_{C}^{+},\xi_{NC},\xi_{NC}^{+}] & \rightarrow & Tr(\widehat{\rho}\,\widehat{\Psi}_{C}^{\dagger}(\mathbf{s})\,\widehat{\Omega}[\xi_{C},\xi_{C}^{+},\xi_{NC},\xi_{NC}^{+}])\\
 & = & Tr(\widehat{\rho}\,\widehat{\Psi}_{C}^{\dagger}(\mathbf{s})\,\widehat{\Omega}_{C}[\xi_{C},\xi_{C}^{+}]\,\,\widehat{\Omega}_{NC})\end{eqnarray*}
using the feature that condensate operators commute with non-condensate
operators.

Hence from Eq.(\ref{eq:CDagOmegaRule})\[
\chi\lbrack\xi_{C},\xi_{C}^{+},\xi_{NC},\xi_{NC}^{+}]\rightarrow Tr(\widehat{\rho}\,\frac{1}{i}\,\left(\frac{\delta\widehat{\Omega}_{C}[\xi_{C},\xi_{C}^{+}]}{\delta\xi_{C}(\mathbf{s})}+\frac{1}{2}\xi_{C}^{+}(\mathbf{s})\,\widehat{\Omega}_{C}[\xi_{C},\xi_{C}^{+}]\right)\,\widehat{\Omega}_{NC})\]
Hence\[
\chi\lbrack\xi_{C},\xi_{C}^{+},\xi_{NC},\xi_{NC}^{+}]\rightarrow\frac{1}{i}\,\left(\frac{\delta}{\delta\xi_{C}(\mathbf{s})}+\frac{1}{2}\xi_{C}^{+}(\mathbf{s})\right)\,\chi\lbrack\xi_{C},\xi_{C}^{+},\xi_{NC},\xi_{NC}^{+}]\]

Then using the relationship to the distribution functional we see
that \begin{eqnarray*}
\chi\lbrack\xi_{C},\xi_{C}^{+},\xi_{NC},\xi_{NC}^{+}] & \rightarrow & \frac{1}{i}\,\left(\frac{\delta}{\delta\xi_{C}(\mathbf{s})}+\frac{1}{2}\xi_{C}^{+}(\mathbf{s})\right)\iiiint D^{2}\psi_{C}\, D^{2}\psi_{C}^{+}\, D^{2}\psi_{NC}\, D^{2}\psi_{NC}^{+}\, P[\underrightarrow{\psi}(\mathbf{r}),\underrightarrow{\psi^{\ast}}(\mathbf{r})]\\
 &  & \times\exp i\int d\mathbf{r\,\{}\xi_{C}(\mathbf{r})\psi_{C}^{+}(\mathbf{r})+\psi_{C}(\mathbf{r})\xi_{C}^{+}(\mathbf{r})\}\,\\
 &  & \times\exp i\int d\mathbf{r\,\{}\xi_{NC}(\mathbf{r})\psi_{NC}^{+}(\mathbf{r})\}\exp i\int d\mathbf{r\,\{}\psi_{NC}(\mathbf{r})\xi_{NC}^{+}(\mathbf{r})\}\,\\
 & = & \iiiint D^{2}\psi_{C}\, D^{2}\psi_{C}^{+}\, D^{2}\psi_{NC}\, D^{2}\psi_{NC}^{+}\, P[\underrightarrow{\psi}(\mathbf{r}),\underrightarrow{\psi^{\ast}}(\mathbf{r})]\\
 &  & \times\lbrack\left(\psi_{C}^{+}(\mathbf{s})-\frac{1}{2}\frac{\delta}{\delta\psi_{C}(\mathbf{s})}\right)\exp i\int d\mathbf{r\,\{}\xi_{C}(\mathbf{r})\psi_{C}^{+}(\mathbf{r})+\psi_{C}(\mathbf{r})\xi_{C}^{+}(\mathbf{r})\}]\,\\
 &  & \times\exp i\int d\mathbf{r\,\{}\xi_{NC}(\mathbf{r})\psi_{NC}^{+}(\mathbf{r})\}\exp i\int d\mathbf{r\,\{}\psi_{NC}(\mathbf{r})\xi_{NC}^{+}(\mathbf{r})\}\end{eqnarray*}

To proceed further we use the integration by parts result and then
find that\begin{eqnarray*}
\chi\lbrack\xi_{C},\xi_{C}^{+},\xi_{NC},\xi_{NC}^{+}] & \rightarrow & \iiiint D^{2}\psi_{C}\, D^{2}\psi_{C}^{+}\, D^{2}\psi_{NC}\, D^{2}\psi_{NC}^{+}\\
 &  & \times\{\left(\psi_{C}^{+}(\mathbf{s})+\frac{1}{2}\frac{\delta}{\delta\psi_{C}(\mathbf{s})}\right)P[\underrightarrow{\psi}(\mathbf{r}),\underrightarrow{\psi^{\ast}}(\mathbf{r})]\}\\
 &  & \times\exp i\int d\mathbf{r\,\{}\xi_{C}(\mathbf{r})\psi_{C}^{+}(\mathbf{r})+\psi_{C}(\mathbf{r})\xi_{C}^{+}(\mathbf{r})\}\,\\
 &  & \times\exp i\int d\mathbf{r\,\{}\xi_{NC}(\mathbf{r})\psi_{NC}^{+}(\mathbf{r})\}\exp i\int d\mathbf{r\,\{}\psi_{NC}(\mathbf{r})\xi_{NC}^{+}(\mathbf{r})\}\end{eqnarray*}

Hence the change to the characteristic functional if $\widehat{\rho}$
is replaced by $\widehat{\rho}$ $\widehat{\Psi}_{C}^{\dagger}(\mathbf{s})$
is equivalent to then the quasi distribution functional is replaced
as follows \begin{equation}
P[\underrightarrow{\psi}(\mathbf{r}),\underrightarrow{\psi^{\ast}}(\mathbf{r})]\rightarrow\left(\psi_{C}^{+}(\mathbf{s})+\frac{1}{2}\frac{\delta}{\delta\psi_{C}(\mathbf{s})}\right)P[\underrightarrow{\psi}(\mathbf{r}),\underrightarrow{\psi^{\ast}}(\mathbf{r})]\label{Eq.RhoPsiCDagPRule}\end{equation}
Thus $P[\underrightarrow{\psi}(\mathbf{r}),\underrightarrow{\psi^{\ast}}(\mathbf{r})]$
is both multiplied by $\psi_{C}^{+}(\mathbf{s})$, the field function
that the operator $\widehat{\Psi}_{C}^{\dagger}(\mathbf{s})$ is mapped
onto and functionally differentiated with respect to $\psi_{C}(\mathbf{s})$,
the field function that the operator $\widehat{\Psi}_{C}(\mathbf{s})$
is mapped onto.

(5) A summary of these key results is as follows:\begin{eqnarray*}
\widehat{\rho} & \rightarrow & \widehat{\Psi}_{C}(\mathbf{s})\widehat{\rho}\qquad P[\underrightarrow{\psi}(\mathbf{r}),\underrightarrow{\psi^{\ast}}(\mathbf{r})]\rightarrow\left(\psi_{C}(\mathbf{s})+\frac{1}{2}\frac{\delta}{\delta\psi_{C}^{+}(\mathbf{s})}\right)P[\underrightarrow{\psi}(\mathbf{r}),\underrightarrow{\psi^{\ast}}(\mathbf{r})]\\
\widehat{\rho} & \rightarrow & \widehat{\Psi}_{C}^{\dagger}(\mathbf{s})\widehat{\rho}\qquad P[\underrightarrow{\psi}(\mathbf{r}),\underrightarrow{\psi^{\ast}}(\mathbf{r})]\rightarrow\left(\psi_{C}^{+}(\mathbf{s})-\frac{1}{2}\frac{\delta}{\delta\psi_{C}(\mathbf{s})}\right)P[\underrightarrow{\psi}(\mathbf{r}),\underrightarrow{\psi^{\ast}}(\mathbf{r})]\\
\widehat{\rho} & \rightarrow & \widehat{\rho}\,\widehat{\Psi}_{C}(\mathbf{s})\qquad P[\underrightarrow{\psi}(\mathbf{r}),\underrightarrow{\psi^{\ast}}(\mathbf{r})]\rightarrow\left(\psi_{C}(\mathbf{s})-\frac{1}{2}\frac{\delta}{\delta\psi_{C}^{+}(\mathbf{s})}\right)P[\underrightarrow{\psi}(\mathbf{r}),\underrightarrow{\psi^{\ast}}(\mathbf{r})]\\
\widehat{\rho} & \rightarrow & \widehat{\rho}\,\widehat{\Psi}_{C}^{\dagger}(\mathbf{s})\qquad P[\underrightarrow{\psi}(\mathbf{r}),\underrightarrow{\psi^{\ast}}(\mathbf{r})]\rightarrow\left(\psi_{C}^{+}(\mathbf{s})+\frac{1}{2}\frac{\delta}{\delta\psi_{C}(\mathbf{s})}\right)P[\underrightarrow{\psi}(\mathbf{r}),\underrightarrow{\psi^{\ast}}(\mathbf{r})]\end{eqnarray*}

(6) If $\widehat{\rho}$ is replaced by $\partial_{\mu}\widehat{\Psi}_{C}(\mathbf{s})\widehat{\rho}$
then the characteristic function becomes\begin{eqnarray*}
\chi\lbrack\xi_{C},\xi_{C}^{+},\xi_{NC},\xi_{NC}^{+}] & \rightarrow & Tr(\partial_{\mu}\widehat{\Psi}_{C}(\mathbf{s})\widehat{\rho}\,\widehat{\Omega}[\xi_{C},\xi_{C}^{+},\xi_{NC},\xi_{NC}^{+}])\\
 & = & Tr(\widehat{\rho}\,\,\widehat{\Omega}_{C}[\xi_{C},\xi_{C}^{+}]\,\partial_{\mu}\widehat{\Psi}_{C}(\mathbf{s})\,\widehat{\Omega}_{NC})\end{eqnarray*}
using the cyclic property of the trace and the feature that condensate
operators commute with non-condensate operators.

Hence from Eq.(\ref{Eq.OmegaCDerivPsiC}) \[
\chi\lbrack\xi_{C},\xi_{C}^{+},\xi_{NC},\xi_{NC}^{+}]\rightarrow Tr(\widehat{\rho}\,\frac{1}{i}\,\left(\left(\partial_{\mu}\frac{\delta\widehat{\Omega}_{C}[\xi_{C},\xi_{C}^{+}]}{\delta\xi_{C}^{+}(\mathbf{r})}\right)_{\mathbf{r=s}}+\widehat{\Omega}_{C}[\xi_{C},\xi_{C}^{+}]\,\frac{1}{2}\partial_{\mu}\xi_{C}(\mathbf{s})\right)\,\widehat{\Omega}_{NC})\]
Hence\[
\chi\lbrack\xi_{C},\xi_{C}^{+},\xi_{NC},\xi_{NC}^{+}]\rightarrow\frac{1}{i}\,\left(\left(\partial_{\mu}\frac{\delta\widehat{\Omega}_{C}[\xi_{C},\xi_{C}^{+}]}{\delta\xi_{C}^{+}(\mathbf{r})}\right)_{\mathbf{r=s}}+\widehat{\Omega}_{C}[\xi_{C},\xi_{C}^{+}]\,\frac{1}{2}\partial_{\mu}\xi_{C}(\mathbf{s})\right)\,\chi\lbrack\xi_{C},\xi_{C}^{+},\xi_{NC},\xi_{NC}^{+}]\]

\subsection{Functional Derivative Rules - Non-Condensate Operators}

To proceed further we need to establish some rules for functional
derivatives of operator expressions.Consider\begin{eqnarray*}
\widehat{\Omega}_{NC}[\xi_{NC},\xi_{NC}^{+}] & = & \exp\widehat{F}[\xi_{NC}]\,\exp\widehat{H}[\xi_{NC}^{+}]\\
\widehat{F}[\xi_{NC}] & = & \int d\mathbf{r}\, i\{\xi_{NC}(\mathbf{r})\widehat{\Psi}_{NC}^{\dagger}(\mathbf{r})\}\qquad\widehat{H}[\xi_{NC}^{+}]=\int d\mathbf{r}\, i\{\widehat{\Psi}_{NC}(\mathbf{r})\xi_{NC}^{+}(\mathbf{r})\}\end{eqnarray*}

(1) We first establish a result for $\widehat{\Omega}_{NC}[\xi_{NC},\xi_{NC}^{+}]\,\widehat{\Psi}_{NC}^{\dagger}(\mathbf{s})$.
Now using the product rule and noting that $\widehat{H}[\xi_{NC}^{+}]$
is not a functional of $\xi_{NC}$ \begin{eqnarray*}
\left(\frac{\delta\widehat{\Omega}_{NC}[\xi_{NC},\xi_{NC}^{+}]}{\delta\xi_{NC}}\right)_{\mathbf{r=s}} & = & \lim_{\epsilon\rightarrow0}\left(\frac{\exp\widehat{F}[\xi_{NC}(\mathbf{r)+}\epsilon\delta(\mathbf{r-s)}]-\exp\widehat{F}[\xi_{NC}(\mathbf{r)}]}{\epsilon}\right)\exp\widehat{H}[\xi_{NC}^{+}]\\
 & = & \lim_{\epsilon\rightarrow0}\left(\frac{\exp\{\widehat{F}[\xi_{NC}(\mathbf{r)}]+\epsilon i\widehat{\Psi}_{NC}^{\dagger}(\mathbf{s})\}-\exp\widehat{F}[\xi_{NC}(\mathbf{r)}]}{\epsilon}\right)\exp\widehat{H}[\xi_{NC}^{+}]\end{eqnarray*}
Now we can use the Baker-Haussdorf theorem which is that $\exp(\widehat{A}+\widehat{B})=\exp(\widehat{A})\,\exp(\widehat{B})\,\exp\{-\frac{1}{2}[\widehat{A},\widehat{B}]\}$,
if the commutator commutes with $\widehat{A}$ and $\widehat{B}$,
so with $\widehat{A}=\widehat{F}[\xi_{NC}(\mathbf{r)}]$ and $\widehat{B}=\epsilon i\widehat{\Psi}_{NC}^{\dagger}(\mathbf{s})$
we have \begin{eqnarray*}
\exp\{\widehat{F}[\xi_{NC}(\mathbf{r)}]+\epsilon i\widehat{\Psi}_{NC}^{\dagger}(\mathbf{s})\} & = & \exp\widehat{F}[\xi_{NC}(\mathbf{r)}]\,\exp\epsilon i\widehat{\Psi}_{NC}^{\dagger}(\mathbf{s})\\
 & \doteqdot & \exp\widehat{F}[\xi_{NC}(\mathbf{r)}]\,\{1+\epsilon(i\widehat{\Psi}_{NC}^{\dagger}(\mathbf{s}))\}\end{eqnarray*}
since using Eqs.(\ref{Eq.BoseCommutationRules2-1}, \ref{Eq.RestrictedDeltaFns-1})
\begin{eqnarray*}
\lbrack\widehat{F}[\xi_{NC}(\mathbf{r)}],\epsilon i\widehat{\Psi}_{NC}^{\dagger}(\mathbf{s})\,] & = & \epsilon i\int d\mathbf{r}\, i[\{\xi_{NC}(\mathbf{r})\widehat{\Psi}_{NC}^{\dagger}(\mathbf{r})\},\widehat{\Psi}_{NC}^{\dagger}(\mathbf{s})]\\
 & = & 0\end{eqnarray*}
Hence\[
\left(\frac{\delta\widehat{\Omega}_{NC}[\xi_{NC},\xi_{C}^{+}]}{\delta\xi_{NC}}\right)_{\mathbf{r=s}}=\exp\widehat{F}[\xi_{C}]\,(i\widehat{\Psi}_{NC}^{\dagger}(\mathbf{s}))\,\exp\widehat{H}[\xi_{NC}^{+}]\]
But although $i\widehat{\Psi}_{NC}^{\dagger}(\mathbf{s})$ does not
commute with $\exp\widehat{H}[\xi_{NC}^{+}]$ we can use the identity
$\widehat{\Xi}\,\exp\widehat{S}=\exp\widehat{S}\,\{\widehat{\Xi}-[\widehat{S},\widehat{\Xi}]+\frac{1}{2!}[\widehat{S},[\widehat{S},\widehat{\Xi}]]-\frac{1}{3!}[\widehat{S},[\widehat{S},[\widehat{S},\widehat{\Xi}]]]+..\}$
to place the exponential on the left. Here we have $\widehat{S}=\widehat{H}[\xi_{NC}^{+}]$
and $\widehat{\Xi}=i\widehat{\Psi}_{NC}^{\dagger}(\mathbf{s})$. Using
Eqs.(\ref{Eq.BoseCommutationRules2-1}, \ref{Eq.RestrictedDeltaFns-1})
we have on noting that $\,\xi_{NC}^{+}(\mathbf{r})$ only involves
the complex conjugates of non-condensate modes \begin{eqnarray*}
\lbrack\widehat{H}[\xi_{NC}^{+}],i\widehat{\Psi}_{NC}^{\dagger}(\mathbf{s})] & = & i^{2}\int d\mathbf{r}\,[\widehat{\Psi}_{NC}(\mathbf{r}),\widehat{\Psi}_{NC}^{\dagger}(\mathbf{s})]\,\xi_{NC}^{+}(\mathbf{r})\\
 & = & -\int d\mathbf{r}\,\delta_{NC}(\mathbf{r},\mathbf{s})\,\xi_{NC}^{+}(\mathbf{r})\\
 & = & -\xi_{NC}^{+}(\mathbf{s})\end{eqnarray*}
Thus we see that the series terminates after the second term giving\begin{eqnarray}
\left(\frac{\delta\widehat{\Omega}_{NC}[\xi_{NC},\xi_{NC}^{+}]}{\delta\xi_{NC}}\right)_{\mathbf{r=s}} & = & \exp\widehat{F}[\xi_{NC}]\,\exp\widehat{H}[\xi_{NC}^{+}]\,\{i\widehat{\Psi}_{NC}^{\dagger}(\mathbf{s}))+\xi_{NC}^{+}(\mathbf{s})\}\nonumber \\
 & = & \widehat{\Omega}_{NC}[\xi_{NC},\xi_{C}^{+}]\,\{i\widehat{\Psi}_{NC}^{\dagger}(\mathbf{s}))+\xi_{NC}^{+}(\mathbf{s})\}\nonumber \\
\widehat{\Omega}_{NC}[\xi_{NC},\xi_{NC}^{+}]\,\widehat{\Psi}_{NC}^{\dagger}(\mathbf{s})) & = & \frac{1}{i}\left(\frac{\delta\widehat{\Omega}_{NC}[\xi_{NC},\xi_{NC}^{+}]}{\delta\xi_{NC}}\right)_{\mathbf{r=s}}-\frac{1}{i}\widehat{\Omega}_{NC}[\xi_{NC},\xi_{NC}^{+}]\,\xi_{NC}^{+}(\mathbf{s})\nonumber \\
 & = & \frac{1}{i}\left(\frac{\delta\widehat{\Omega}_{NC}[\xi_{NC},\xi_{NC}^{+}]}{\delta\xi_{NC}}\right)_{\mathbf{r=s}}-\frac{1}{i}\xi_{NC}^{+}(\mathbf{s})\,\widehat{\Omega}_{NC}[\xi_{NC},\xi_{NC}^{+}]\nonumber \\
 &  & \,\label{eq:NCPsiNCDagRule}\end{eqnarray}

(2) We next establish a result for $\widehat{\Omega}_{NC}[\xi_{NC},\xi_{NC}^{+}]\,\widehat{\Psi}_{NC}(\mathbf{s})$.
Similarly using the product rule and noting that $\widehat{F}[\xi_{NC}]$
is not a functional of $\xi_{NC}^{+}$

\begin{eqnarray*}
\left(\frac{\delta\widehat{\Omega}_{NC}[\xi_{NC},\xi_{NC}^{+}]}{\delta\xi_{NC}^{+}}\right)_{\mathbf{r=s}} & = & \exp\widehat{F}[\xi_{NC}]\,\lim_{\epsilon\rightarrow0}\left(\frac{\exp\widehat{H}[\xi_{NC}^{+}(\mathbf{r})\mathbf{+}\epsilon\delta(\mathbf{r-s)}]-\exp\widehat{H}[\xi_{NC}^{+}(\mathbf{r})]}{\epsilon}\right)\\
 & = & \exp\widehat{F}[\xi_{NC}]\,\lim_{\epsilon\rightarrow0}\left(\frac{\exp\{\widehat{H}[\xi_{NC}^{+}(\mathbf{r})]+\epsilon i\widehat{\Psi}_{NC}(\mathbf{s})\}-\exp\widehat{H}[\xi_{NC}^{+}(\mathbf{r})]}{\epsilon}\right)\end{eqnarray*}
Using the Baker-Haussdorf theorem again but now with $\widehat{A}=\widehat{H}[\xi_{NC}^{+}(\mathbf{r})]$
and $\widehat{B}=\epsilon i\widehat{\Psi}_{NC}(\mathbf{s})$ we have
\begin{eqnarray*}
\exp\{\widehat{H}[\xi_{NC}^{+}(\mathbf{r})]+\epsilon i\widehat{\Psi}_{NC}(\mathbf{s})\} & = & \exp\widehat{H}[\xi_{NC}^{+}(\mathbf{r})]\,\exp\epsilon i\widehat{\Psi}_{NC}(\mathbf{s})\\
 & = & \exp\widehat{H}[\xi_{NC}^{+}(\mathbf{r})]\,\{1+\epsilon i\widehat{\Psi}_{NC}(\mathbf{s})\}\end{eqnarray*}
since from Eqs.(\ref{Eq.BoseCommutationRules2-1}, \ref{Eq.RestrictedDeltaFns-1})\begin{eqnarray*}
\lbrack\widehat{H}[\xi_{NC}^{+}(\mathbf{r)}],\epsilon i\widehat{\Psi}_{NC}(\mathbf{s})\,] & = & \epsilon i\int d\mathbf{r}\, i[\{\widehat{\Psi}_{NC}(\mathbf{r})\xi_{NC}^{+}(\mathbf{r})\},\widehat{\Psi}_{NC}(\mathbf{s})]\\
 & = & 0\end{eqnarray*}
Hence\begin{eqnarray}
\left(\frac{\delta\widehat{\Omega}_{NC}[\xi_{NC},\xi_{NC}^{+}]}{\delta\xi_{NC}^{+}}\right)_{\mathbf{r=s}} & = & \exp\widehat{F}[\xi_{NC}]\,\exp\widehat{H}[\xi_{NC}^{+}(\mathbf{r})]\,(i\widehat{\Psi}_{NC}(\mathbf{s}))\nonumber \\
 & = & \widehat{\Omega}_{NC}[\xi_{NC},\xi_{NC}^{+}]\,(i\widehat{\Psi}_{NC}(\mathbf{s}))\nonumber \\
\widehat{\Omega}_{NC}[\xi_{NC},\xi_{NC}^{+}]\,\widehat{\Psi}_{NC}(\mathbf{s}) & = & \frac{1}{i}\left(\frac{\delta\widehat{\Omega}_{NC}[\xi_{NC},\xi_{NC}^{+}]}{\delta\xi_{NC}^{+}}\right)_{\mathbf{r=s}}\label{Eq.OmegaNCPsiNCRule}\end{eqnarray}

(3) We next establish a result for $\widehat{\Psi}_{NC}^{\dagger}(\mathbf{s})$
$\widehat{\Omega}_{NC}[\xi_{NC},\xi_{NC}^{+}]\,$. From above and
now using the result that $i\widehat{\Psi}_{NC}^{\dagger}(\mathbf{s})$
commutes with $\widehat{F}[\xi_{C}]$ \begin{eqnarray}
\left(\frac{\delta\widehat{\Omega}_{NC}[\xi_{NC},\xi_{C}^{+}]}{\delta\xi_{NC}}\right)_{\mathbf{r=s}} & = & \exp\widehat{F}[\xi_{C}]\,(i\widehat{\Psi}_{NC}^{\dagger}(\mathbf{s}))\,\exp\widehat{H}[\xi_{NC}^{+}]\nonumber \\
 & = & (i\widehat{\Psi}_{NC}^{\dagger}(\mathbf{s}))\,\exp\widehat{F}[\xi_{C}]\,\exp\widehat{H}[\xi_{NC}^{+}]\nonumber \\
 & = & (i\widehat{\Psi}_{NC}^{\dagger}(\mathbf{s}))\,\widehat{\Omega}_{NC}[\xi_{NC},\xi_{C}^{+}]\nonumber \\
\widehat{\Psi}_{NC}^{\dagger}(\mathbf{s})\,\widehat{\Omega}_{NC}[\xi_{NC},\xi_{C}^{+}] & = & \frac{1}{i}\left(\frac{\delta\widehat{\Omega}_{NC}[\xi_{NC},\xi_{C}^{+}]}{\delta\xi_{NC}}\right)_{\mathbf{r=s}}\label{Eq.PsiNCDagOmegaNCRule}\end{eqnarray}

(4) We next establish a result for $\widehat{\Psi}_{NC}(\mathbf{s})$
$\widehat{\Omega}_{NC}[\xi_{NC},\xi_{NC}^{+}]\,$. From above and
now using the result that $i\widehat{\Psi}_{NC}(\mathbf{s})$ commutes
with $\widehat{H}[\xi_{C}^{+}]$

\begin{eqnarray*}
\left(\frac{\delta\widehat{\Omega}_{NC}[\xi_{NC},\xi_{NC}^{+}]}{\delta\xi_{NC}^{+}}\right)_{\mathbf{r=s}} & = & \exp\widehat{F}[\xi_{NC}]\,\exp\widehat{H}[\xi_{NC}^{+}(\mathbf{r})]\,(i\widehat{\Psi}_{NC}(\mathbf{s}))\\
 & = & \exp\widehat{F}[\xi_{NC}]\,(i\widehat{\Psi}_{NC}(\mathbf{s}))\,\exp\widehat{H}[\xi_{NC}^{+}(\mathbf{r})]\end{eqnarray*}
But although $i\widehat{\Psi}_{NC}(\mathbf{s})$ does not commute
with $\exp\widehat{F}[\xi_{NC}]$ we can use the identity $\,\exp\widehat{S}\,\widehat{\Xi}=\{\widehat{\Xi}+[\widehat{S},\widehat{\Xi}]+\frac{1}{2!}[\widehat{S},[\widehat{S},\widehat{\Xi}]]+\frac{1}{3!}[\widehat{S},[\widehat{S},[\widehat{S},\widehat{\Xi}]]]+..\}\exp\widehat{S}\,$
to place the exponential on the right. Here we have $\widehat{S}=\widehat{F}[\xi_{NC}]$
and $\widehat{\Xi}=i\widehat{\Psi}_{NC}(\mathbf{s}).$ Using Eqs.(\ref{Eq.BoseCommutationRules2-1},
\ref{Eq.RestrictedDeltaFns-1}) we have on noting that $\xi_{NC}(\mathbf{r})$
only involves non-condensate modes \begin{eqnarray*}
\lbrack\widehat{F}[\xi_{NC}],i\widehat{\Psi}_{NC}(\mathbf{s})] & = & i^{2}\int d\mathbf{r}\,[\xi_{NC}(\mathbf{r})\widehat{\Psi}_{NC}^{\dagger}(\mathbf{r}),\widehat{\Psi}_{NC}(\mathbf{s})]\\
 & = & \int d\mathbf{r}\,\xi_{NC}(\mathbf{r})\delta_{NC}(\mathbf{s,r)}\\
 & = & +\xi_{NC}(\mathbf{s})\end{eqnarray*}
Thus we see that the series terminates after the second term giving

Hence\begin{eqnarray}
\left(\frac{\delta\widehat{\Omega}_{NC}[\xi_{NC},\xi_{NC}^{+}]}{\delta\xi_{NC}^{+}}\right)_{\mathbf{r=s}} & = & (i\widehat{\Psi}_{NC}(\mathbf{s})+\xi_{NC}(\mathbf{s}))\,\exp\widehat{F}[\xi_{NC}]\,\exp\widehat{H}[\xi_{NC}^{+}(\mathbf{r})]\nonumber \\
 & = & (i\widehat{\Psi}_{NC}(\mathbf{s})+\xi_{NC}(\mathbf{s}))\,\widehat{\Omega}_{NC}[\xi_{NC},\xi_{NC}^{+}]\,\nonumber \\
\widehat{\Psi}_{NC}(\mathbf{s})\,\widehat{\Omega}_{NC}[\xi_{NC},\xi_{NC}^{+}]\, & = & \frac{1}{i}\left(\frac{\delta\widehat{\Omega}_{NC}[\xi_{NC},\xi_{NC}^{+}]}{\delta\xi_{NC}^{+}}\right)_{\mathbf{r=s}}-\frac{1}{i}\xi_{NC}(\mathbf{s})\,\widehat{\Omega}_{NC}[\xi_{NC},\xi_{NC}^{+}]\nonumber \\
 & \, & \,\label{eq:PsiNCOmegaNCRule}\end{eqnarray}

\subsection{Non-Condensate Operators}

(1) If $\widehat{\rho}$ is replaced by $\widehat{\Psi}_{NC}(\mathbf{s})\widehat{\rho}$
then the characteristic function becomes\begin{eqnarray*}
\chi\lbrack\xi_{C},\xi_{C}^{+},\xi_{NC},\xi_{NC}^{+}] & \rightarrow & Tr(\widehat{\Psi}_{NC}(\mathbf{s})\widehat{\rho}\,\widehat{\Omega}[\xi_{C},\xi_{C}^{+},\xi_{NC},\xi_{NC}^{+}])\\
 & = & Tr(\widehat{\rho}\,\widehat{\Omega}[\xi_{C},\xi_{C}^{+},\xi_{NC},\xi_{NC}^{+}]\widehat{\Psi}_{NC}(\mathbf{s}))\\
 & = & Tr(\widehat{\rho}\,\widehat{\Omega}_{C}\,\widehat{\Omega}_{NC}[\xi_{NC},\xi_{NC}^{+}]\,\widehat{\Psi}_{NC}(\mathbf{s}))\end{eqnarray*}
using the cyclic property of the trace and the feature that condensate
operators commute with non-condensate operators.

Hence from Eq.(\ref{Eq.OmegaNCPsiNCRule}) \[
\chi\lbrack\xi_{C},\xi_{C}^{+},\xi_{NC},\xi_{NC}^{+}]\rightarrow Tr(\widehat{\rho}\,\widehat{\Omega}_{C}\,\frac{1}{i}\left(\frac{\delta\widehat{\Omega}_{NC}[\xi_{NC},\xi_{NC}^{+}]}{\delta\xi_{NC}^{+}}\right)_{\mathbf{r=s}})\]
Hence\[
\chi\lbrack\xi_{C},\xi_{C}^{+},\xi_{NC},\xi_{NC}^{+}]\rightarrow\frac{1}{i}\,\left(\frac{\delta}{\delta\xi_{NC}^{+}(\mathbf{s})}\right)\,\chi\lbrack\xi_{C},\xi_{C}^{+},\xi_{NC},\xi_{NC}^{+}]\]
Then using the relationship to the distribution functional we see
that \begin{eqnarray*}
\chi\lbrack\xi_{C},\xi_{C}^{+},\xi_{NC},\xi_{NC}^{+}] & \rightarrow & \frac{1}{i}\,\left(\frac{\delta}{\delta\xi_{NC}^{+}(\mathbf{s})}\right)\iiiint D^{2}\psi_{C}\, D^{2}\psi_{C}^{+}\, D^{2}\psi_{NC}\, D^{2}\psi_{NC}^{+}\, P[\underrightarrow{\psi}(\mathbf{r}),\underrightarrow{\psi^{\ast}}(\mathbf{r})]\\
 &  & \times\exp i\int d\mathbf{r\,\{}\xi_{C}(\mathbf{r})\psi_{C}^{+}(\mathbf{r})+\psi_{C}(\mathbf{r})\xi_{C}^{+}(\mathbf{r})\}\,\\
 &  & \times\exp i\int d\mathbf{r\,\{}\xi_{NC}(\mathbf{r})\psi_{NC}^{+}(\mathbf{r})\}\exp i\int d\mathbf{r\,\{}\psi_{NC}(\mathbf{r})\xi_{NC}^{+}(\mathbf{r})\}\,\\
 & = & \iiiint D^{2}\psi_{C}\, D^{2}\psi_{C}^{+}\, D^{2}\psi_{NC}\, D^{2}\psi_{NC}^{+}\, P[\underrightarrow{\psi}(\mathbf{r}),\underrightarrow{\psi^{\ast}}(\mathbf{r})]\\
 &  & \times\lbrack\left(\psi_{NC}(\mathbf{s})\right)\exp i\int d\mathbf{r\,\{}\xi_{C}(\mathbf{r})\psi_{C}^{+}(\mathbf{r})+\psi_{C}(\mathbf{r})\xi_{C}^{+}(\mathbf{r})\}]\,\\
 &  & \times\exp i\int d\mathbf{r\,\{}\xi_{NC}(\mathbf{r})\psi_{NC}^{+}(\mathbf{r})\}\exp i\int d\mathbf{r\,\{}\psi_{NC}(\mathbf{r})\xi_{NC}^{+}(\mathbf{r})\}\,\end{eqnarray*}

To proceed further we only need to place the multiplicative term $\psi_{NC}(\mathbf{s})$
next to the quasi distribution functional itself. We then find that\begin{eqnarray*}
\chi\lbrack\xi_{C},\xi_{C}^{+},\xi_{NC},\xi_{NC}^{+}] & \rightarrow & \iiiint D^{2}\psi_{C}\, D^{2}\psi_{C}^{+}\, D^{2}\psi_{NC}\, D^{2}\psi_{NC}^{+}\\
 &  & \times\{\left(\psi_{NC}(\mathbf{s})\right)P[\underrightarrow{\psi}(\mathbf{r}),\underrightarrow{\psi^{\ast}}(\mathbf{r})]\}\\
 &  & \times\exp i\int d\mathbf{r\,\{}\xi_{C}(\mathbf{r})\psi_{C}^{+}(\mathbf{r})+\psi_{C}(\mathbf{r})\xi_{C}^{+}(\mathbf{r})\}\,\\
 &  & \times\exp i\int d\mathbf{r\,\{}\xi_{NC}(\mathbf{r})\psi_{NC}^{+}(\mathbf{r})\}\exp i\int d\mathbf{r\,\{}\psi_{NC}(\mathbf{r})\xi_{NC}^{+}(\mathbf{r})\}\,\end{eqnarray*}

Hence the change to the characteristic functional if $\widehat{\rho}$
is replaced by $\widehat{\Psi}_{NC}(\mathbf{s})\widehat{\rho}$ is
equivalent to then the quasi distribution functional is replaced as
follows \begin{equation}
P[\underrightarrow{\psi}(\mathbf{r}),\underrightarrow{\psi^{\ast}}(\mathbf{r})]\rightarrow\left(\psi_{NC}(\mathbf{s})\right)P[\underrightarrow{\psi}(\mathbf{r}),\underrightarrow{\psi^{\ast}}(\mathbf{r})]\label{Eq.PsiNCRhoPRule}\end{equation}
Thus $P[\underrightarrow{\psi}(\mathbf{r}),\underrightarrow{\psi^{\ast}}(\mathbf{r})]$
is multiplied by $\psi_{NC}(\mathbf{s})$, the field function that
the operator $\widehat{\Psi}_{NC}(\mathbf{s})$ is mapped onto.

(2) If $\widehat{\rho}$ is replaced by $\widehat{\Psi}_{NC}^{\dagger}(\mathbf{s})\widehat{\rho}$
then the characteristic function becomes\begin{eqnarray*}
\chi\lbrack\xi_{C},\xi_{C}^{+},\xi_{NC},\xi_{NC}^{+}] & \rightarrow & Tr(\widehat{\Psi}_{NC}^{\dagger}(\mathbf{s})\widehat{\rho}\,\widehat{\Omega}[\xi_{C},\xi_{C}^{+},\xi_{NC},\xi_{NC}^{+}])\\
 & = & Tr(\widehat{\rho}\,\widehat{\Omega}[\xi_{C},\xi_{C}^{+},\xi_{NC},\xi_{NC}^{+}]\widehat{\Psi}_{NC}^{\dagger}(\mathbf{s}))\\
 & = & Tr(\widehat{\rho}\,\widehat{\Omega}_{C}\,\widehat{\Omega}_{NC}[\xi_{NC},\xi_{NC}^{+}]\,\widehat{\Psi}_{NC}^{\dagger}(\mathbf{s})\,)\end{eqnarray*}
using the cyclic property of the trace and the feature that condensate
operators commute with non-condensate operators.

Hence from Eq.(\ref{eq:NCPsiNCDagRule}) \[
\chi\lbrack\xi_{C},\xi_{C}^{+},\xi_{NC},\xi_{NC}^{+}]\rightarrow Tr(\widehat{\rho}\,\widehat{\Omega}_{C}\,\frac{1}{i}\left(\frac{\delta\widehat{\Omega}_{NC}[\xi_{NC},\xi_{NC}^{+}]}{\delta\xi_{NC}}\right)_{\mathbf{r=s}}-\frac{1}{i}\xi_{NC}^{+}(\mathbf{s})\,\widehat{\Omega}_{NC}[\xi_{NC},\xi_{NC}^{+}])\]
Hence\[
\chi\lbrack\xi_{C},\xi_{C}^{+},\xi_{NC},\xi_{NC}^{+}]\rightarrow\frac{1}{i}\,\left(\frac{\delta}{\delta\xi_{NC}(\mathbf{s})}-\xi_{NC}^{+}(\mathbf{s})\right)\,\chi\lbrack\xi_{C},\xi_{C}^{+},\xi_{NC},\xi_{NC}^{+}]\]
Then using the relationship to the distribution functional we see
that \begin{eqnarray*}
\chi\lbrack\xi_{C},\xi_{C}^{+},\xi_{NC},\xi_{NC}^{+}] & \rightarrow & \frac{1}{i}\,\left(\frac{\delta}{\delta\xi_{NC}(\mathbf{s})}-\xi_{NC}^{+}(\mathbf{s})\right)\iiiint D^{2}\psi_{C}\, D^{2}\psi_{C}^{+}\, D^{2}\psi_{NC}\, D^{2}\psi_{NC}^{+}\, P[\underrightarrow{\psi}(\mathbf{r}),\underrightarrow{\psi^{\ast}}(\mathbf{r})]\\
 &  & \times\exp i\int d\mathbf{r\,\{}\xi_{C}(\mathbf{r})\psi_{C}^{+}(\mathbf{r})+\psi_{C}(\mathbf{r})\xi_{C}^{+}(\mathbf{r})\}\,\\
 &  & \times\exp i\int d\mathbf{r\,\{}\xi_{NC}(\mathbf{r})\psi_{NC}^{+}(\mathbf{r})\}\exp i\int d\mathbf{r\,\{}\psi_{NC}(\mathbf{r})\xi_{NC}^{+}(\mathbf{r})\}\,\\
 & = & \iiiint D^{2}\psi_{C}\, D^{2}\psi_{C}^{+}\, D^{2}\psi_{NC}\, D^{2}\psi_{NC}^{+}\, P[\underrightarrow{\psi}(\mathbf{r}),\underrightarrow{\psi^{\ast}}(\mathbf{r})]\\
 &  & \times\lbrack\left(\psi_{NC}^{+}(\mathbf{s})+\frac{\delta}{\delta\psi_{NC}(\mathbf{s})}\right)\exp i\int d\mathbf{r\,\{}\xi_{NC}(\mathbf{r})\psi_{NC}^{+}(\mathbf{r})\}\exp i\int d\mathbf{r\,\{}\psi_{NC}(\mathbf{r})\xi_{NC}^{+}(\mathbf{r})\}]\\
 &  & \times\,\exp i\int d\mathbf{r\,\{}\xi_{C}(\mathbf{r})\psi_{C}^{+}(\mathbf{r})+\psi_{C}(\mathbf{r})\xi_{C}^{+}(\mathbf{r})\}\,\end{eqnarray*}

To proceed further we use the integration by parts result we then
find that\begin{eqnarray*}
\chi\lbrack\xi_{C},\xi_{C}^{+},\xi_{NC},\xi_{NC}^{+}] & \rightarrow & \iiiint D^{2}\psi_{C}\, D^{2}\psi_{C}^{+}\, D^{2}\psi_{NC}\, D^{2}\psi_{NC}^{+}\\
 &  & \times\{\left(\psi_{NC}^{+}(\mathbf{s})-\frac{\delta}{\delta\psi_{NC}(\mathbf{s})}\right)P[\underrightarrow{\psi}(\mathbf{r}),\underrightarrow{\psi^{\ast}}(\mathbf{r})]\}\\
 &  & \times\exp i\int d\mathbf{r\,\{}\xi_{C}(\mathbf{r})\psi_{C}^{+}(\mathbf{r})+\psi_{C}(\mathbf{r})\xi_{C}^{+}(\mathbf{r})\}\,\\
 &  & \times\exp i\int d\mathbf{r\,\{}\xi_{NC}(\mathbf{r})\psi_{NC}^{+}(\mathbf{r})\}\exp i\int d\mathbf{r\,\{}\psi_{NC}(\mathbf{r})\xi_{NC}^{+}(\mathbf{r})\}\,\end{eqnarray*}

Hence the change to the characteristic functional if $\widehat{\rho}$
is replaced by $\widehat{\Psi}_{NC}^{\dagger}(\mathbf{s})\widehat{\rho}$
is equivalent to then the quasi distribution functional is replaced
as follows \begin{equation}
P[\underrightarrow{\psi}(\mathbf{r}),\underrightarrow{\psi^{\ast}}(\mathbf{r})]\rightarrow\left(\psi_{NC}^{+}(\mathbf{s})-\frac{\delta}{\delta\psi_{NC}(\mathbf{s})}\right)P[\underrightarrow{\psi}(\mathbf{r}),\underrightarrow{\psi^{\ast}}(\mathbf{r})]\label{Eq.PsiNCDagRhoPRule}\end{equation}
Thus $P[\underrightarrow{\psi}(\mathbf{r}),\underrightarrow{\psi^{\ast}}(\mathbf{r})]$
is both multiplied by $\psi_{NC}^{+}(\mathbf{s})$, the field function
that the operator $\widehat{\Psi}_{NC}^{\dagger}(\mathbf{s})$ is
mapped onto and functionally differentiated with respect to $\psi_{NC}(\mathbf{s})$,
the field function that the operator $\widehat{\Psi}_{NC}(\mathbf{s})$
is mapped onto.

(3) If $\widehat{\rho}$ is replaced by $\widehat{\rho}$ $\widehat{\Psi}_{NC}(\mathbf{s})$
then the characteristic function becomes\begin{eqnarray*}
\chi\lbrack\xi_{C},\xi_{C}^{+},\xi_{NC},\xi_{NC}^{+}] & \rightarrow & Tr(\widehat{\rho}\,\widehat{\Psi}_{NC}(\mathbf{s})\,\widehat{\Omega}[\xi_{C},\xi_{C}^{+},\xi_{NC},\xi_{NC}^{+}])\\
 & = & Tr(\widehat{\rho}\,\,\widehat{\Omega}_{C}\,\widehat{\Psi}_{NC}(\mathbf{s})\,\widehat{\Omega}_{NC}[\xi_{NC},\xi_{NC}^{+}])\end{eqnarray*}
using the feature that condensate operators commute with non-condensate
operators.

Hence from Eq.(\ref{eq:PsiNCOmegaNCRule})\[
\chi\lbrack\xi_{C},\xi_{C}^{+},\xi_{NC},\xi_{NC}^{+}]\rightarrow Tr(\widehat{\rho}\,\widehat{\Omega}_{C}\,\frac{1}{i}\left(\frac{\delta\widehat{\Omega}_{NC}[\xi_{NC},\xi_{NC}^{+}]}{\delta\xi_{NC}^{+}}\right)_{\mathbf{r=s}}-\frac{1}{i}\xi_{NC}(\mathbf{s})\,\widehat{\Omega}_{NC}[\xi_{NC},\xi_{NC}^{+}])\]
Hence\[
\chi\lbrack\xi_{C},\xi_{C}^{+},\xi_{NC},\xi_{NC}^{+}]\rightarrow\frac{1}{i}\,\left(\frac{\delta}{\delta\xi_{NC}^{+}(\mathbf{s})}-\xi_{NC}(\mathbf{s})\right)\,\chi\lbrack\xi_{C},\xi_{C}^{+},\xi_{NC},\xi_{NC}^{+}]\]
Then using the relationship to the distribution functional we see
that \begin{eqnarray*}
\chi\lbrack\xi_{C},\xi_{C}^{+},\xi_{NC},\xi_{NC}^{+}] & \rightarrow & \frac{1}{i}\,\left(\frac{\delta}{\delta\xi_{NC}^{+}(\mathbf{s})}-\xi_{NC}(\mathbf{s})\right)\iiiint D^{2}\psi_{C}\, D^{2}\psi_{C}^{+}\, D^{2}\psi_{NC}\, D^{2}\psi_{NC}^{+}\, P[\underrightarrow{\psi}(\mathbf{r}),\underrightarrow{\psi^{\ast}}(\mathbf{r})]\\
 &  & \times\exp i\int d\mathbf{r\,\{}\xi_{C}(\mathbf{r})\psi_{C}^{+}(\mathbf{r})+\psi_{C}(\mathbf{r})\xi_{C}^{+}(\mathbf{r})\}\,\\
 &  & \times\exp i\int d\mathbf{r\,\{}\xi_{NC}(\mathbf{r})\psi_{NC}^{+}(\mathbf{r})\}\exp i\int d\mathbf{r\,\{}\psi_{NC}(\mathbf{r})\xi_{NC}^{+}(\mathbf{r})\}\,\\
 & = & \iiiint D^{2}\psi_{C}\, D^{2}\psi_{C}^{+}\, D^{2}\psi_{NC}\, D^{2}\psi_{NC}^{+}\, P[\underrightarrow{\psi}(\mathbf{r}),\underrightarrow{\psi^{\ast}}(\mathbf{r})]\\
 &  & \times\lbrack\left(\psi_{NC}(\mathbf{s})+\frac{\delta}{\delta\psi_{NC}^{+}(\mathbf{s})}\right)\,\exp i\int d\mathbf{r\,\{}\xi_{NC}(\mathbf{r})\psi_{NC}^{+}(\mathbf{r})\}\exp i\int d\mathbf{r\,\{}\psi_{NC}(\mathbf{r})\xi_{NC}^{+}(\mathbf{r})\}]\,\\
 &  & \times\exp i\int d\mathbf{r\,\{}\xi_{C}(\mathbf{r})\psi_{C}^{+}(\mathbf{r})+\psi_{C}(\mathbf{r})\xi_{C}^{+}(\mathbf{r})\}\end{eqnarray*}
To proceed further we use the integration by parts result and then
find that\begin{eqnarray*}
\chi\lbrack\xi_{C},\xi_{C}^{+},\xi_{NC},\xi_{NC}^{+}] & \rightarrow & \iiiint D^{2}\psi_{C}\, D^{2}\psi_{C}^{+}\, D^{2}\psi_{NC}\, D^{2}\psi_{NC}^{+}\\
 &  & \times\{\left(\psi_{NC}(\mathbf{s})-\frac{\delta}{\delta\psi_{NC}^{+}(\mathbf{s})}\right)P[\underrightarrow{\psi}(\mathbf{r}),\underrightarrow{\psi^{\ast}}(\mathbf{r})]\}\\
 &  & \times\exp i\int d\mathbf{r\,\{}\xi_{C}(\mathbf{r})\psi_{C}^{+}(\mathbf{r})+\psi_{C}(\mathbf{r})\xi_{C}^{+}(\mathbf{r})\}\,\\
 &  & \times\exp i\int d\mathbf{r\,\{}\xi_{NC}(\mathbf{r})\psi_{NC}^{+}(\mathbf{r})\}\exp i\int d\mathbf{r\,\{}\psi_{NC}(\mathbf{r})\xi_{NC}^{+}(\mathbf{r})\}\,\end{eqnarray*}

Hence the change to the characteristic functional if $\widehat{\rho}$
is replaced by $\widehat{\rho}\,\widehat{\Psi}_{NC}(\mathbf{s})$
is equivalent to then the quasi distribution functional is replaced
as follows \begin{equation}
P[\underrightarrow{\psi}(\mathbf{r}),\underrightarrow{\psi^{\ast}}(\mathbf{r})]\rightarrow\left(\psi_{NC}(\mathbf{s})-\frac{\delta}{\delta\psi_{NC}^{+}(\mathbf{s})}\right)P[\underrightarrow{\psi}(\mathbf{r}),\underrightarrow{\psi^{\ast}}(\mathbf{r})]\label{Eq.RhoPsiNCPRule}\end{equation}
Thus $P[\underrightarrow{\psi}(\mathbf{r}),\underrightarrow{\psi^{\ast}}(\mathbf{r})]$
is both multiplied by $\psi_{NC}(\mathbf{s})$, the field function
that the operator $\widehat{\Psi}_{NC}(\mathbf{s})$ is mapped onto
and functionally differentiated with respect to $\psi_{NC}^{+}(\mathbf{s})$,
the field function that the operator $\widehat{\Psi}_{NC}^{\dagger}(\mathbf{s})$
is mapped onto.

(4) If $\widehat{\rho}$ is replaced by $\widehat{\rho}$ $\widehat{\Psi}_{NC}^{\dagger}(\mathbf{s})$
then the characteristic function becomes\begin{eqnarray*}
\chi\lbrack\xi_{C},\xi_{C}^{+},\xi_{NC},\xi_{NC}^{+}] & \rightarrow & Tr(\widehat{\rho}\,\widehat{\Psi}_{NC}^{\dagger}(\mathbf{s})\,\widehat{\Omega}[\xi_{C},\xi_{C}^{+},\xi_{NC},\xi_{NC}^{+}])\\
 & = & Tr(\widehat{\rho}\,\widehat{\Omega}_{C}\,\widehat{\Psi}_{NC}^{\dagger}(\mathbf{s})\,\widehat{\Omega}_{NC}[\xi_{NC},\xi_{NC}^{+}])\end{eqnarray*}
using the feature that condensate operators commute with non-condensate
operators.

Hence from Eq.(\ref{Eq.PsiNCDagOmegaNCRule})\[
\chi\lbrack\xi_{C},\xi_{C}^{+},\xi_{NC},\xi_{NC}^{+}]\rightarrow Tr(\widehat{\rho}\,\widehat{\Omega}_{C}\,\frac{1}{i}\left(\frac{\delta\widehat{\Omega}_{NC}[\xi_{NC},\xi_{C}^{+}]}{\delta\xi_{NC}}\right)_{\mathbf{r=s}})\]
Hence\[
\chi\lbrack\xi_{C},\xi_{C}^{+},\xi_{NC},\xi_{NC}^{+}]\rightarrow\frac{1}{i}\,\left(\frac{\delta}{\delta\xi_{NC}(\mathbf{s})}\right)\,\chi\lbrack\xi_{C},\xi_{C}^{+},\xi_{NC},\xi_{NC}^{+}]\]

Then using the relationship to the distribution functional we see
that \begin{eqnarray*}
\chi\lbrack\xi_{C},\xi_{C}^{+},\xi_{NC},\xi_{NC}^{+}] & \rightarrow & \frac{1}{i}\,\left(\frac{\delta}{\delta\xi_{NC}(\mathbf{s})}\right)\iiiint D^{2}\psi_{C}\, D^{2}\psi_{C}^{+}\, D^{2}\psi_{NC}\, D^{2}\psi_{NC}^{+}\, P[\underrightarrow{\psi}(\mathbf{r}),\underrightarrow{\psi^{\ast}}(\mathbf{r})]\\
 &  & \times\exp i\int d\mathbf{r\,\{}\xi_{C}(\mathbf{r})\psi_{C}^{+}(\mathbf{r})+\psi_{C}(\mathbf{r})\xi_{C}^{+}(\mathbf{r})\}\,\\
 &  & \times\exp i\int d\mathbf{r\,\{}\xi_{NC}(\mathbf{r})\psi_{NC}^{+}(\mathbf{r})\}\exp i\int d\mathbf{r\,\{}\psi_{NC}(\mathbf{r})\xi_{NC}^{+}(\mathbf{r})\}\,\\
 & = & \iiiint D^{2}\psi_{C}\, D^{2}\psi_{C}^{+}\, D^{2}\psi_{NC}\, D^{2}\psi_{NC}^{+}\, P[\underrightarrow{\psi}(\mathbf{r}),\underrightarrow{\psi^{\ast}}(\mathbf{r})]\\
 &  & \times\lbrack\left(\psi_{NC}^{+}(\mathbf{s})\right)\,\exp i\int d\mathbf{r\,\{}\xi_{NC}(\mathbf{r})\psi_{NC}^{+}(\mathbf{r})\}\exp i\int d\mathbf{r\,\{}\psi_{NC}(\mathbf{r})\xi_{NC}^{+}(\mathbf{r})\}]\\
 &  & \times\exp i\int d\mathbf{r\,\{}\xi_{C}(\mathbf{r})\psi_{C}^{+}(\mathbf{r})+\psi_{C}(\mathbf{r})\xi_{C}^{+}(\mathbf{r})\}\end{eqnarray*}

To proceed further we only need to place the multiplicative term $\psi_{NC}^{+}(\mathbf{s})$
next to the quasi distribution functional itself. We then find that\begin{eqnarray*}
\chi\lbrack\xi_{C},\xi_{C}^{+},\xi_{NC},\xi_{NC}^{+}] & \rightarrow & \iiiint D^{2}\psi_{C}\, D^{2}\psi_{C}^{+}\, D^{2}\psi_{NC}\, D^{2}\psi_{NC}^{+}\\
 &  & \times\{\left(\psi_{NC}^{+}(\mathbf{s})\right)P[\underrightarrow{\psi}(\mathbf{r}),\underrightarrow{\psi^{\ast}}(\mathbf{r})]\}\\
 &  & \times\exp i\int d\mathbf{r\,\{}\xi_{C}(\mathbf{r})\psi_{C}^{+}(\mathbf{r})+\psi_{C}(\mathbf{r})\xi_{C}^{+}(\mathbf{r})\}\,\\
 &  & \times\exp i\int d\mathbf{r\,\{}\xi_{NC}(\mathbf{r})\psi_{NC}^{+}(\mathbf{r})\}\exp i\int d\mathbf{r\,\{}\psi_{NC}(\mathbf{r})\xi_{NC}^{+}(\mathbf{r})\}\end{eqnarray*}

Hence the change to the characteristic functional if $\widehat{\rho}$
is replaced by $\widehat{\rho}$ $\widehat{\Psi}_{NC}^{\dagger}(\mathbf{s})$
is equivalent to then the quasi distribution functional is replaced
as follows \begin{equation}
P[\underrightarrow{\psi}(\mathbf{r}),\underrightarrow{\psi^{\ast}}(\mathbf{r})]\rightarrow\left(\psi_{NC}^{+}(\mathbf{s})\right)P[\underrightarrow{\psi}(\mathbf{r}),\underrightarrow{\psi^{\ast}}(\mathbf{r})]\label{Eq.RhoPsiNCDagPRule}\end{equation}
Thus $P[\underrightarrow{\psi}(\mathbf{r}),\underrightarrow{\psi^{\ast}}(\mathbf{r})]$
is multiplied by $\psi_{NC}^{+}(\mathbf{s})$, the field function
that the operator $\widehat{\Psi}_{NC}^{\dagger}(\mathbf{s})$ is
mapped onto.

(5) A summary of these key results is as follows:\begin{eqnarray*}
\widehat{\rho} & \rightarrow & \widehat{\Psi}_{NC}(\mathbf{s})\widehat{\rho}\qquad P[\underrightarrow{\psi}(\mathbf{r}),\underrightarrow{\psi^{\ast}}(\mathbf{r})]\rightarrow\left(\psi_{NC}(\mathbf{s})\right)P[\underrightarrow{\psi}(\mathbf{r}),\underrightarrow{\psi^{\ast}}(\mathbf{r})]\\
\widehat{\rho} & \rightarrow & \widehat{\Psi}_{NC}^{\dagger}(\mathbf{s})\widehat{\rho}\qquad P[\underrightarrow{\psi}(\mathbf{r}),\underrightarrow{\psi^{\ast}}(\mathbf{r})]\rightarrow\left(\psi_{NC}^{+}(\mathbf{s})-\frac{\delta}{\delta\psi_{NC}(\mathbf{s})}\right)P[\underrightarrow{\psi}(\mathbf{r}),\underrightarrow{\psi^{\ast}}(\mathbf{r})]\\
\widehat{\rho} & \rightarrow & \widehat{\rho}\,\widehat{\Psi}_{NC}(\mathbf{s})\qquad P[\underrightarrow{\psi}(\mathbf{r}),\underrightarrow{\psi^{\ast}}(\mathbf{r})]\rightarrow\left(\psi_{NC}(\mathbf{s})-\frac{\delta}{\delta\psi_{NC}^{+}(\mathbf{s})}\right)P[\underrightarrow{\psi}(\mathbf{r}),\underrightarrow{\psi^{\ast}}(\mathbf{r})]\\
\widehat{\rho} & \rightarrow & \widehat{\rho}\,\widehat{\Psi}_{NC}^{\dagger}(\mathbf{s})\qquad P[\underrightarrow{\psi}(\mathbf{r}),\underrightarrow{\psi^{\ast}}(\mathbf{r})]\rightarrow\left(\psi_{NC}^{+}(\mathbf{s})\right)P[\underrightarrow{\psi}(\mathbf{r}),\underrightarrow{\psi^{\ast}}(\mathbf{r})]\end{eqnarray*}

\subsection{Time Derivative}

If $\widehat{\rho}$ is replaced by $\frac{\partial\widehat{\rho}}{\partial t}$
then \begin{eqnarray*}
\chi\lbrack\xi_{C},\xi_{C}^{+},\xi_{NC},\xi_{NC}^{+}] & \rightarrow & Tr(\frac{\partial}{\partial t}\widehat{\rho}\,\,\widehat{\Omega}[\xi_{C},\xi_{C}^{+},\xi_{NC},\xi_{NC}^{+}])\\
 & = & \frac{\partial}{\partial t}Tr(\widehat{\rho}\,\widehat{\Omega}[\xi_{C},\xi_{C}^{+},\xi_{NC},\xi_{NC}^{+}])\\
 & = & \frac{\partial}{\partial t}\chi\lbrack\xi_{C},\xi_{C}^{+},\xi_{NC},\xi_{NC}^{+}]\end{eqnarray*}

Hence\begin{eqnarray*}
\chi\lbrack\xi_{C},\xi_{C}^{+},\xi_{NC},\xi_{NC}^{+}] & \rightarrow & \frac{\partial}{\partial t}\iiiint D^{2}\psi_{C}\, D^{2}\psi_{C}^{+}\, D^{2}\psi_{NC}\, D^{2}\psi_{NC}^{+}\, P[\underrightarrow{\psi}(\mathbf{r}),\underrightarrow{\psi^{\ast}}(\mathbf{r})]\\
 &  & \times\exp i\int d\mathbf{r\,\{}\xi_{C}(\mathbf{r})\psi_{C}^{+}(\mathbf{r})+\psi_{C}(\mathbf{r})\xi_{C}^{+}(\mathbf{r})\}\,\\
 &  & \times\exp i\int d\mathbf{r\,\{}\xi_{NC}(\mathbf{r})\psi_{NC}^{+}(\mathbf{r})\}\exp i\int d\mathbf{r\,\{}\psi_{NC}(\mathbf{r})\xi_{NC}^{+}(\mathbf{r})\}\,\\
 & = & \iiiint D^{2}\psi_{C}\, D^{2}\psi_{C}^{+}\, D^{2}\psi_{NC}\, D^{2}\psi_{NC}^{+}\,\frac{\partial}{\partial t}P[\underrightarrow{\psi}(\mathbf{r}),\underrightarrow{\psi^{\ast}}(\mathbf{r})]\\
 &  & \times\exp i\int d\mathbf{r\,\{}\xi_{C}(\mathbf{r})\psi_{C}^{+}(\mathbf{r})+\psi_{C}(\mathbf{r})\xi_{C}^{+}(\mathbf{r})\}\,\\
 &  & \times\exp i\int d\mathbf{r\,\{}\xi_{NC}(\mathbf{r})\psi_{NC}^{+}(\mathbf{r})\}\exp i\int d\mathbf{r\,\{}\psi_{NC}(\mathbf{r})\xi_{NC}^{+}(\mathbf{r})\}\end{eqnarray*}
Thus the change to the characteristic functional if $\widehat{\rho}$
is replaced by $\frac{{\LARGE\partial}\widehat{{\LARGE\rho}}}{{\LARGE\partial t}}$
is equivalent to then the quasi distribution functional is replaced
as follows \begin{equation}
P[\underrightarrow{\psi}(\mathbf{r}),\underrightarrow{\psi^{\ast}}(\mathbf{r})]\rightarrow\frac{\partial}{\partial t}P[\underrightarrow{\psi}(\mathbf{r}),\underrightarrow{\psi^{\ast}}(\mathbf{r})]\label{Eq.DerivRhoRule}\end{equation}
Thus $P[\underrightarrow{\psi}(\mathbf{r}),\underrightarrow{\psi^{\ast}}(\mathbf{r})]$
is replaced by its time derivative.

\subsection{Supplementary Equations}

~

Commutation Rules and Delta Functions\begin{eqnarray}
\lbrack\widehat{\Psi}_{C}(\mathbf{r}),\widehat{\Psi}_{NC}^{\dagger}(\mathbf{r})] & = & 0\nonumber \\
\lbrack\widehat{\Psi}_{C}(\mathbf{r}),\widehat{\Psi}_{C}^{\dagger}(\mathbf{r}^{\prime})] & = & \phi_{1}(\mathbf{r})\phi_{1}^{\ast}(\mathbf{r}^{\prime})+\phi_{2}(\mathbf{r})\phi_{2}^{\ast}(\mathbf{r}^{\prime})\nonumber \\
 & = & \delta_{C}(\mathbf{r,r}^{\prime})\label{Eq.RestrictedCondDeltaTwoMode-1}\\
\lbrack\widehat{\Psi}_{NC}(\mathbf{r}),\widehat{\Psi}_{NC}^{\dagger}(\mathbf{r}^{\prime})] & = & \sum\limits _{k\neq1,2}\phi_{k}(\mathbf{r})\phi_{k}^{\ast}(\mathbf{r}^{\prime})\nonumber \\
 & = & \delta_{NC}(\mathbf{r,r}^{\prime})\label{Eq.BoseCommutationRules2-1}\end{eqnarray}

Field Expansions and Delta Functions\begin{eqnarray}
\psi_{C}(\mathbf{r}) & = & \alpha_{1}\phi_{1}(\mathbf{r})+\alpha_{2}\phi_{2}(\mathbf{r})\qquad\psi_{C}^{+}(\mathbf{r})=\phi_{1}^{\ast}(\mathbf{r})\alpha_{1}^{+}+\phi_{2}^{\ast}(\mathbf{r})\alpha_{2}^{+}\label{eq:CondFieldFn-1}\\
\psi_{NC}(\mathbf{r}) & = & \sum\limits _{k\neq1,2}\alpha_{k}\phi_{k}(\mathbf{r})\qquad\psi_{NC}^{+}(\mathbf{r})=\sum\limits _{k\neq1,2}\phi_{k}^{\ast}(\mathbf{r})\alpha_{k}^{+}\label{eq:NonCondFieldFn-1}\\
\psi_{C}(\mathbf{r}) & = & \int d\mathbf{r}^{\prime}\mathbf{\,}\delta_{C}(\mathbf{r,r}^{\prime})\psi_{C}(\mathbf{r}^{\prime})\qquad\psi_{C}^{+}(\mathbf{r})=\int d\mathbf{r}^{\prime}\mathbf{\,}\psi_{C}^{+}(\mathbf{r}^{\prime})\delta_{C}(\mathbf{r}^{\prime}\mathbf{,r})\nonumber \\
\psi_{NC}(\mathbf{r}) & = & \int d\mathbf{r}^{\prime}\mathbf{\,}\delta_{NC}(\mathbf{r,r}^{\prime})\psi_{NC}(\mathbf{r}^{\prime})\,\,\,\,\,\,\psi(\mathbf{r})=\int d\mathbf{r}^{\prime}\mathbf{\,}\psi_{C}^{+}(\mathbf{r}^{\prime})\delta_{C}(\mathbf{r}^{\prime}\mathbf{,r})\label{Eq.RestrictedDeltaFns-1}\end{eqnarray}
\pagebreak{}

\section{- Functional Fokker-Planck Equation}

\label{App 5}

In this Appendix we derive the Functional Fokker-Planck equation.
We will derive it based on the full Hamiltonian including the $\widehat{H}_{4}$
and $\widehat{H}_{5}$ terms. This gives the exact equation. We can
then write down the corresponding FFPE for the case of the Bogoliubov
Hamiltonian (\ref{Eq.BogolHamiltonian-1}) by discarding terms for
the exact FFPE - which would be needed for the stong interaction regime.
For this derivation it will be convenient to write the Hamiltonian
in the form \begin{equation}
\widehat{H}=\widehat{H}_{C}+\widehat{H}_{NC}+\widehat{V}\label{Eq.Ham}\end{equation}
where \begin{eqnarray}
\widehat{H}_{C} & = & \int d\mathbf{r(}\frac{\hbar^{2}}{2m}\nabla\widehat{\Psi}_{C}(\mathbf{r})^{\dagger}\cdot\nabla\widehat{\Psi}_{C}(\mathbf{r})+\widehat{\Psi}_{C}(\mathbf{r})^{\dagger}V\widehat{\Psi}_{C}(\mathbf{r})\nonumber \\
 &  & +\frac{g}{2}\widehat{\Psi}_{C}(\mathbf{r})^{\dagger}\widehat{\Psi}_{C}(\mathbf{r})^{\dagger}\widehat{\Psi}_{C}(\mathbf{r})\widehat{\Psi}_{C}(\mathbf{r}))\label{Eq.HCond}\\
\widehat{H}_{NC} & = & \int d\mathbf{r(}\frac{\hbar^{2}}{2m}\nabla\widehat{\Psi}_{NC}(\mathbf{r})^{\dagger}\cdot\nabla\widehat{\Psi}_{NC}(\mathbf{r})+\widehat{\Psi}_{NC}(\mathbf{r})^{\dagger}V\widehat{\Psi}_{NC}(\mathbf{r})\nonumber \\
 &  & +\frac{g}{2}\widehat{\Psi}_{NC}(\mathbf{r})^{\dagger}\widehat{\Psi}_{NC}(\mathbf{r})^{\dagger}\widehat{\Psi}_{NC}(\mathbf{r})\widehat{\Psi}_{NC}(\mathbf{r}))\label{Eq.HNonCond}\end{eqnarray}
are Hamiltonians for the \emph{condensate} and \emph{non-condensate}.
The \emph{interaction} between condensate and non-condensate is written
as the sum of three contributions which are linear, quadratic and
cubic in the non-condensate operators\begin{eqnarray}
\widehat{V} & = & \widehat{V}_{1}+\widehat{V}_{2}+\widehat{V}_{3}\label{Eq.V}\\
\widehat{V}_{1} & = & \int d\mathbf{r(}\frac{\hbar^{2}}{2m}\nabla\widehat{\Psi}_{NC}(\mathbf{r})^{\dagger}\cdot\nabla\widehat{\Psi}_{C}(\mathbf{r})+\frac{\hbar^{2}}{2m}\nabla\widehat{\Psi}_{C}(\mathbf{r})^{\dagger}\cdot\nabla\widehat{\Psi}_{NC}(\mathbf{r})\nonumber \\
 &  & +\widehat{\Psi}_{NC}(\mathbf{r})^{\dagger}V\widehat{\Psi}_{C}(\mathbf{r})+\widehat{\Psi}_{C}(\mathbf{r})^{\dagger}V\widehat{\Psi}_{NC}(\mathbf{r})\nonumber \\
 &  & +g\widehat{\Psi}_{NC}(\mathbf{r})^{\dagger}\widehat{\Psi}_{C}(\mathbf{r})^{\dagger}\widehat{\Psi}_{C}(\mathbf{r})\widehat{\Psi}_{C}(\mathbf{r})+g\widehat{\Psi}_{C}(\mathbf{r})^{\dagger}\widehat{\Psi}_{C}(\mathbf{r})^{\dagger}\widehat{\Psi}_{C}(\mathbf{r})\widehat{\Psi}_{NC}(\mathbf{r}))\label{Eq.V1}\\
\widehat{V}_{2} & = & \int d\mathbf{r(}\frac{g}{2}\widehat{\Psi}_{NC}(\mathbf{r})^{\dagger}\widehat{\Psi}_{NC}(\mathbf{r})^{\dagger}\widehat{\Psi}_{C}(\mathbf{r})\widehat{\Psi}_{C}(\mathbf{r})+\frac{g}{2}\widehat{\Psi}_{C}(\mathbf{r})^{\dagger}\widehat{\Psi}_{C}(\mathbf{r})^{\dagger}\widehat{\Psi}_{NC}(\mathbf{r})\widehat{\Psi}_{NC}(\mathbf{r})\nonumber \\
 &  & +2g\widehat{\Psi}_{NC}(\mathbf{r})^{\dagger}\widehat{\Psi}_{C}(\mathbf{r})^{\dagger}\widehat{\Psi}_{NC}(\mathbf{r})\widehat{\Psi}_{C}(\mathbf{r}))\label{Eq.V2}\\
\widehat{V}_{3} & = & \int d\mathbf{r(}g\widehat{\Psi}_{NC}(\mathbf{r})^{\dagger}\widehat{\Psi}_{NC}(\mathbf{r})^{\dagger}\widehat{\Psi}_{NC}(\mathbf{r})\widehat{\Psi}_{C}(\mathbf{r}))+g\widehat{\Psi}_{C}(\mathbf{r})^{\dagger}\widehat{\Psi}_{NC}(\mathbf{r})^{\dagger}\widehat{\Psi}_{NC}(\mathbf{r})\widehat{\Psi}_{NC}(\mathbf{r}))\nonumber \\
 &  & \,\label{Eq:V3}\end{eqnarray}

However, using the coupled generalised Gross-Pitaevskii equation we
can make the simplifications\begin{eqnarray}
\int d\mathbf{r\,}\widehat{\Psi}_{NC}(\mathbf{r})^{\dagger}\,\left\{ \left(-\frac{\hbar^{2}}{2m}\nabla^{2}+V\right)\widehat{\Psi}_{C}(\mathbf{r})\right\}  & = & -\frac{g_{N}}{N}\int\int d\mathbf{r\,}d\mathbf{s\,}F(\mathbf{r},\mathbf{s})\widehat{\Psi}_{NC}(\mathbf{r})^{\dagger}\,\widehat{\Psi}_{C}(\mathbf{s})\nonumber \\
\int d\mathbf{r\,}\left\{ \left(-\frac{\hbar^{2}}{2m}\nabla^{2}+V\right)\widehat{\Psi}_{C}^{\dagger}(\mathbf{r})\right\} \widehat{\Psi}_{NC}(\mathbf{r}) & = & -\frac{g_{N}}{N}\int\int d\mathbf{r\,}d\mathbf{s\,}F^{\ast}(\mathbf{s},\mathbf{r})\widehat{\Psi}_{C}(\mathbf{r})^{\dagger}\widehat{\Psi}_{NC}(\mathbf{s})\nonumber \\
 & \, & \,\label{eq:NewV1Simplication}\end{eqnarray}
to write $\widehat{V}_{1}$ in a form \begin{eqnarray}
\widehat{V}_{1} & = & \int d\mathbf{r\,}\widehat{\Psi}_{NC}^{\dagger}(\mathbf{r})g\{\widehat{\Psi}_{C}^{\dagger}(\mathbf{r})\,\widehat{\Psi}_{C}(\mathbf{r})\}\widehat{\Psi}_{C}(\mathbf{r})-\int\int d\mathbf{r\,}d\mathbf{s\,}gF(\mathbf{r},\mathbf{s})\widehat{\Psi}_{NC}(\mathbf{r})^{\dagger}\,\widehat{\Psi}_{C}(\mathbf{s})\nonumber \\
 &  & +\int d\mathbf{r\,}\widehat{\Psi}_{C}^{\dagger}(\mathbf{r})g\{\widehat{\Psi}_{C}^{\dagger}(\mathbf{r})\,\widehat{\Psi}_{C}(\mathbf{r})\}\widehat{\Psi}_{NC}(\mathbf{r})-\int\int d\mathbf{r\,}d\mathbf{s\,}gF^{\ast}(\mathbf{s},\mathbf{r})\widehat{\Psi}_{C}(\mathbf{r})^{\dagger}\widehat{\Psi}_{NC}(\mathbf{s})\,\nonumber \\
 &  & \,\label{eq:NewV1New}\end{eqnarray}

We see that $\widehat{V}_{1}$ is the sum of term $\widehat{V}_{14}$
which is fourth order in the field operators and a term $\widehat{V}_{12}$
which is second order. \begin{eqnarray}
\widehat{V}_{1} & = & \widehat{V}_{14}+\widehat{V}_{12}\label{Eq.V1NewParts}\\
\widehat{V}_{14} & = & g\int d\mathbf{r\,(}\widehat{\Psi}_{NC}^{\dagger}(\mathbf{r})\widehat{\Psi}_{C}^{\dagger}(\mathbf{r})\,\widehat{\Psi}_{C}(\mathbf{r})\widehat{\Psi}_{C}(\mathbf{r}))+g\int d\mathbf{r\,(}\widehat{\Psi}_{C}^{\dagger}(\mathbf{r})\widehat{\Psi}_{C}^{\dagger}(\mathbf{r})\,\widehat{\Psi}_{C}(\mathbf{r})\widehat{\Psi}_{NC}(\mathbf{r}))\nonumber \\
 &  & \,\label{eq:V14}\\
\widehat{V}_{12} & = & -g\int\int d\mathbf{r\,}d\mathbf{s\,}F(\mathbf{r},\mathbf{s})\widehat{\Psi}_{NC}(\mathbf{r})^{\dagger}\,\widehat{\Psi}_{C}(\mathbf{s})-g\int\int d\mathbf{r\,}d\mathbf{s\,}F^{\ast}(\mathbf{s},\mathbf{r})\widehat{\Psi}_{C}(\mathbf{r})^{\dagger}\widehat{\Psi}_{NC}(\mathbf{s})\,\nonumber \\
 &  & \,\label{eq:NewV12}\end{eqnarray}
Thus we see that $\widehat{V}$ is now associated only with \emph{boson-boson}
interaction terms.

From Eqs. (\ref{Eq.Ham}), (\ref{Eq.HCond}), (\ref{Eq.HNonCond})
and (\ref{Eq.V}) we see that there are a total of seventeen distinct
contributions to the Hamiltonian to be considered, ranging from the
kinetic energy contribution to the condensate Hamiltonian to an interaction
term between the condensate and non-condensate fields which is third
order in the non-condensate field. For the Bogoliubov Hamiltonian
for which we derive the functional Fokker-Planck equation the terms
$\widehat{V}_{3}$ and the boson-boson interaction in the non-condensate
Hamiltonian $\widehat{H}_{NC}$ are discarded.

In order to avoid using too many superscripts and subscripts, in considering
each term a simplified notation will be used, which is as follows.
For terms which only involve condensate fields or only non-condensate
fields we will use $\psi(\mathbf{s})$ and $\psi^{+}(\mathbf{s})$
for $\psi_{C}(\mathbf{s})$ and $\psi_{C}^{+}(\mathbf{s})$ or $\psi_{NC}(\mathbf{s})$
and $\psi_{NC}^{+}(\mathbf{s})$. We will write $W[\psi(\mathbf{r}),\psi^{+}(\mathbf{r})]$
or $P[\psi(\mathbf{r}),\psi^{+}(\mathbf{r})]$ (and sometimes just
$W$ or $P$ in large expressions) instead of the complete expression\\
 $P[\psi_{C}(\mathbf{r}),\psi_{C}^{+}(\mathbf{r}),\psi_{NC}(\mathbf{r}),\psi_{NC}^{+}(\mathbf{r}),\psi_{C}^{\ast}(\mathbf{r}),\psi_{C}^{+\ast}(\mathbf{r}),\psi_{NC}^{\ast}(\mathbf{r}),\psi_{NC}^{+\ast}(\mathbf{r})]\equiv P[\underrightarrow{\psi}(\mathbf{r}),\underrightarrow{\psi^{\ast}}(\mathbf{r})]$
\\
in the pure condensate and pure non-condensate cases. For the
interaction between condensate and non-condensate we will write $WP[\psi(\mathbf{r}),\psi^{+}(\mathbf{r}),\phi(\mathbf{r}),\phi^{+}(\mathbf{r})]$
(and sometimes just $WP$ in large expressions) instead of the complete
expression\\
 $P[\psi_{C}(\mathbf{r}),\psi_{C}^{+}(\mathbf{r}),\psi_{NC}(\mathbf{r}),\psi_{NC}^{+}(\mathbf{r}),\psi_{C}^{\ast}(\mathbf{r}),\psi_{C}^{+\ast}(\mathbf{r}),\psi_{NC}^{\ast}(\mathbf{r}),\psi_{NC}^{+\ast}(\mathbf{r})]\equiv P[\underrightarrow{\psi}(\mathbf{r}),\underrightarrow{\psi^{\ast}}(\mathbf{r})]$,
\\
using $\psi(\mathbf{s})$, $\psi^{+}(\mathbf{s})$ for $\psi_{C}(\mathbf{s})$,
$\psi_{C}^{+}(\mathbf{s})$, and $\phi(\mathbf{s})$, $\phi^{+}(\mathbf{s})$
for $\psi_{NC}(\mathbf{s})$, $\psi_{NC}^{+}(\mathbf{s})$ in the
expressions, since both condensate and non-condensate fields will
be present and must be distinguished. In this notation the fact that
the distribution functionals also depend on the complex conjugate
fields $\psi^{\ast}(\mathbf{s})$, $\psi^{+\ast}(\mathbf{s})$ and
$\phi^{\ast}(\mathbf{s})$, $\phi^{+\ast}(\mathbf{s})$ has been ignored.
This is because functional derivatives or functions involving these
complex conjugate fields are \emph{not} involved in the derivation
as a consequence of their absence from the correspondence rules. As
in \ref{App 4} the general notation that applies is\begin{eqnarray}
\underrightarrow{\psi}(\mathbf{r}) & \equiv & \{\psi_{C}(\mathbf{r}),\psi_{C}^{+}(\mathbf{r}),\psi_{NC}(\mathbf{r}),\psi_{NC}^{+}(\mathbf{r})\}\\
\underrightarrow{\psi^{\ast}}(\mathbf{r}) & \equiv & \{\psi_{C}^{\ast}(\mathbf{r}),\psi_{C}^{+\ast}(\mathbf{r}),\psi_{NC}^{\ast}(\mathbf{r}),\psi_{NC}^{+\ast}(\mathbf{r})\}\\
P[\underrightarrow{\psi}(\mathbf{r}),\underrightarrow{\psi^{\ast}}(\mathbf{r})] & \equiv & P[\psi_{C}(\mathbf{r}),\psi_{C}^{+}(\mathbf{r}),\psi_{NC}(\mathbf{r}),\psi_{NC}^{+}(\mathbf{r}),\psi_{C}^{\ast}(\mathbf{r}),\psi_{C}^{+\ast}(\mathbf{r}),\psi_{NC}^{\ast}(\mathbf{r}),\psi_{NC}^{+\ast}(\mathbf{r})]\nonumber \\
 &  & \,\\
\underrightarrow{\alpha} & \equiv & \{\alpha_{k},\alpha_{k}^{+}\}\\
\underrightarrow{\alpha}^{\ast} & \equiv & \{\alpha_{k}^{\ast},\alpha_{k}^{+\ast}\}\\
P_{b}(\underrightarrow{\alpha},\underrightarrow{\alpha}^{\ast}) & \equiv & P_{b}(\alpha_{k},\alpha_{k}^{+},\alpha_{k}^{\ast},\alpha_{k}^{+\ast})\equiv P[\underrightarrow{\psi}(\mathbf{r}),\underrightarrow{\psi^{\ast}}(\mathbf{r})]\end{eqnarray}
At the completion of the determination of the contribution to the
Fokker-Planck equation, the original notation\\
 $P[\underrightarrow{\psi}(\mathbf{r}),\underrightarrow{\psi^{\ast}}(\mathbf{r})]\equiv P[\psi_{C}(\mathbf{r}),\psi_{C}^{+}(\mathbf{r}),\psi_{NC}(\mathbf{r}),\psi_{NC}^{+}(\mathbf{r}),\psi_{C}^{\ast}(\mathbf{r}),\psi_{C}^{+\ast}(\mathbf{r}),\psi_{NC}^{\ast}(\mathbf{r}),\psi_{NC}^{+\ast}(\mathbf{r})]$
for the distribution functional will be reintroduced.

Also, to avoid too many nested brackets we will adopt the convention
that a functional derivative will operate on \textit{everything} to
the right of it unless otherwise indicated. Note that spatial derivatives
do not operate on functionals, only on functions. 

The terms in the Bogoliubov Hamiltonian that we need to consider are\begin{eqnarray}
\widehat{H}_{1} & = & \int d\mathbf{r(}\frac{\hbar^{2}}{2m}\nabla\widehat{\Psi}_{C}^{\dagger}(\mathbf{r})\cdot\nabla\widehat{\Psi}_{C}(\mathbf{r})+\widehat{\Psi}_{C}^{\dagger}(\mathbf{r})V\widehat{\Psi}_{C}(\mathbf{r})\nonumber \\
 &  & +\frac{g_{N}}{2N}\widehat{\Psi}_{C}^{\dagger}(\mathbf{r})\widehat{\Psi}_{C}^{\dagger}(\mathbf{r})\widehat{\Psi}_{C}(\mathbf{r})\widehat{\Psi}_{C}(\mathbf{r}))\end{eqnarray}
The term $\widehat{H}_{1}$ is the sum of the condensate kinetic energy,
condensate trap potential energy and condensate boson-boson interaction.\begin{eqnarray}
\widehat{H}_{2} & = & \int d\mathbf{r(}\widehat{\Psi}_{NC}(\mathbf{r})^{\dagger}\,\left\{ \mathbf{-}\frac{\hbar^{2}}{2m}\nabla^{2}\widehat{\Psi}_{C}(\mathbf{r})+V\widehat{\Psi}_{C}(\mathbf{r})+\frac{g_{N}}{N}\widehat{\Psi}_{C}^{\dagger}(\mathbf{r)}\widehat{\Psi}_{C}(\mathbf{r})\widehat{\Psi}_{C}(\mathbf{r})\right\} \nonumber \\
 &  & +\left\{ -\frac{\hbar^{2}}{2m}\nabla^{2}\widehat{\Psi}_{C}^{\dagger}(\mathbf{r)}+\widehat{\Psi}_{C}(\mathbf{r})^{\dagger}V+\frac{g_{N}}{N}\widehat{\Psi}_{C}^{\dagger}(\mathbf{r})\widehat{\Psi}_{C}^{\dagger}(\mathbf{r})\widehat{\Psi}_{C}(\mathbf{r})\right\} \widehat{\Psi}_{NC}(\mathbf{r}))\nonumber \\
 &  & \,\end{eqnarray}
The term $\widehat{H}_{2}$ is the coupling between the condensate
and non-condensate fields that is linear in the non-condensate field.
It can be put into different forms not involving the spatial derivatives.
Thus\begin{eqnarray}
\widehat{H}_{2} & = & \widehat{H}_{2U4}+\widehat{H}_{2U2}\\
\widehat{H}_{2U4} & = & \frac{g_{N}}{N}\int d\mathbf{r\,(}\widehat{\Psi}_{NC}^{\dagger}(\mathbf{r})\widehat{\Psi}_{C}^{\dagger}(\mathbf{r})\,\widehat{\Psi}_{C}(\mathbf{r})\widehat{\Psi}_{C}(\mathbf{r}))+\frac{g_{N}}{N}\int d\mathbf{r\,(}\widehat{\Psi}_{C}^{\dagger}(\mathbf{r})\widehat{\Psi}_{C}^{\dagger}(\mathbf{r})\,\widehat{\Psi}_{C}(\mathbf{r})\widehat{\Psi}_{NC}(\mathbf{r}))\nonumber \\
 &  & \,\\
\widehat{H}_{2U2} & = & -\frac{g_{N}}{N}\int\int d\mathbf{r\,}d\mathbf{s\,}F(\mathbf{r},\mathbf{s})\widehat{\Psi}_{NC}(\mathbf{r})^{\dagger}\,\widehat{\Psi}_{C}(\mathbf{s})-\frac{g_{N}}{N}\int\int d\mathbf{r\,}d\mathbf{s\,}F^{\ast}(\mathbf{s},\mathbf{r})\widehat{\Psi}_{C}(\mathbf{r})^{\dagger}\widehat{\Psi}_{NC}(\mathbf{s})\nonumber \\
 &  & \,\end{eqnarray}
 \begin{eqnarray}
\widehat{H}_{3} & = & \int d\mathbf{r}\left\{ \frac{\hbar^{2}}{2m}\nabla\widehat{\Psi}_{NC}^{\dagger}(\mathbf{r})\cdot\nabla\widehat{\Psi}_{NC}(\mathbf{r})+\widehat{\Psi}_{NC}^{\dagger}(\mathbf{r)}V\widehat{\Psi}_{NC}(\mathbf{r})\right\} \nonumber \\
 &  & +\frac{g_{N}}{2N}\int d\mathbf{r}\left\{ \widehat{\Psi}_{NC}^{\dagger}(\mathbf{r})\widehat{\Psi}_{NC}^{\dagger}(\mathbf{r})\widehat{\Psi}_{C}(\mathbf{r})\widehat{\Psi}_{C}(\mathbf{r})+\widehat{\Psi}_{C}^{\dagger}(\mathbf{r})\widehat{\Psi}_{C}^{\dagger}(\mathbf{r})\widehat{\Psi}_{NC}\widehat{\Psi}_{NC}\right\} \nonumber \\
 &  & +\frac{g_{N}}{2N}\int d\mathbf{r}\left\{ 4\widehat{\Psi}_{NC}^{\dagger}(\mathbf{r})\widehat{\Psi}_{C}^{\dagger}(\mathbf{r})\widehat{\Psi}_{NC}(\mathbf{r})\widehat{\Psi}_{C}(\mathbf{r})\right\} \end{eqnarray}
\

\subsection{Condensate Kinetic Energy Terms}

We write the kinetic energy as \begin{equation}
\widehat{T}=\frac{\hbar^{2}}{2m}\sum\limits _{\mu}\int d\mathbf{s\,}\partial_{\mu}\widehat{\Psi}(\mathbf{s})^{\dagger}\,\partial_{\mu}\widehat{\Psi}(\mathbf{s})\end{equation}
Now if \begin{equation}
\widehat{\rho}\rightarrow\widehat{T}\widehat{\rho}=\frac{\hbar^{2}}{2m}\sum\limits _{\mu}\int d\mathbf{s(}\partial_{\mu}\widehat{\Psi}(\mathbf{s})^{\dagger}\,\partial_{\mu}\widehat{\Psi}(\mathbf{s}))\widehat{\rho}\end{equation}
then\begin{eqnarray}
 &  & W[\psi(\mathbf{r}),\psi^{+}(\mathbf{r})]\nonumber \\
 & \rightarrow & \frac{\hbar^{2}}{2m}\sum\limits _{\mu}\int d\mathbf{s}\left\{ \left(\partial_{\mu}\psi^{+}(\mathbf{s})-\frac{1}{2}\partial_{\mu}\frac{\delta}{\delta\psi(\mathbf{s})}\right)\left(\partial_{\mu}\psi(\mathbf{s})+\frac{1}{2}\partial_{\mu}\frac{\delta}{\delta\psi^{+}(\mathbf{s})}\right)\right\} W[\psi,\psi^{+}]\nonumber \\
 &  & \,\end{eqnarray}

After expanding we find that \begin{eqnarray}
 &  & W[\psi(\mathbf{r}),\psi^{+}(\mathbf{r})]\nonumber \\
 & \rightarrow & \frac{\hbar^{2}}{2m}\sum\limits _{\mu}\int d\mathbf{s}\left\{ \left(\partial_{\mu}\psi^{+}(\mathbf{s})\right)\left(\partial_{\mu}\psi(\mathbf{s})\right)\right\} W[\psi,\psi^{+}]\nonumber \\
 &  & +\frac{\hbar^{2}}{2m}\sum\limits _{\mu}\int d\mathbf{s}\frac{1}{2}\left\{ \left(\partial_{\mu}\psi^{+}(\mathbf{s})\right)\left(\partial_{\mu}\frac{\delta}{\delta\psi^{+}(\mathbf{s})}\right)\right\} W[\psi,\psi^{+}]\nonumber \\
 &  & -\frac{\hbar^{2}}{2m}\sum\limits _{\mu}\int d\mathbf{s}\frac{1}{2}\left\{ \left(\partial_{\mu}\frac{\delta}{\delta\psi(\mathbf{s})}\right)\left(\partial_{\mu}\psi(\mathbf{s})\right)\right\} W[\psi,\psi^{+}]\nonumber \\
 &  & -\frac{\hbar^{2}}{2m}\sum\limits _{\mu}\int d\mathbf{s}\frac{1}{4}\left\{ \left(\partial_{\mu}\frac{\delta}{\delta\psi(\mathbf{s})}\right)\left(\partial_{\mu}\frac{\delta}{\delta\psi^{+}(\mathbf{s})}\right)\right\} W[\psi,\psi^{+}]\nonumber \\
 &  & \,\end{eqnarray}

Now the standard approach to space integration gives the result\begin{equation}
\int d\mathbf{s\,\{}\partial_{\mu}C(\mathbf{s})\,\}=0\end{equation}
for functions $C(\mathbf{s})$ that become zero on the boundary. This
then leads to the useful result involving product functions $C(\mathbf{s})=A(\mathbf{s})B(\mathbf{s})$
enabling the spatial derivative to be applied to either $A(\mathbf{s})$
or $B(\mathbf{s})$\begin{equation}
\int d\mathbf{s\,\{}\partial_{\mu}A(\mathbf{s})\,\}B(\mathbf{s})=-\int d\mathbf{s\,}A(\mathbf{s})\,\mathbf{\{}\partial_{\mu}B(\mathbf{s})\,\}\label{eq:SpatialIntnResultCC-1}\end{equation}
We can assume that the $\psi(\mathbf{s})$ and $\psi^{+}(\mathbf{s})$
become zero on the boundary, since they both involve condensate mode
functions or their conjugates that are localised due to the trap potential.
Also the functional derivatives produce linear combinations of either
the condensate mode functions or their conjugates (see (\ref{Eq.RestrictedFuncDeriv2}),
(\ref{Eq.RestrictedFuncDeriv2Conj})) so the various $C(\mathbf{s})$
that will be involved should become zero on the boundary.

For the first term, the product of the spatial functions can be written
in opposite order so that \begin{eqnarray}
 &  & \int d\mathbf{s}\left\{ \left(\partial_{\mu}\psi^{+}(\mathbf{s})\right)\left(\partial_{\mu}\psi(\mathbf{s})\right)\right\} W[\psi,\psi^{+}]\nonumber \\
 & = & \int d\mathbf{s}\left\{ \left(\partial_{\mu}\psi(\mathbf{s})\right)\left(\partial_{\mu}\psi^{+}(\mathbf{s})\right)\right\} W[\psi,\psi^{+}]\label{Eq.SimplifnResult1}\end{eqnarray}

We can then use (\ref{eq:SpatialIntnResultCC-1}) together with the
explicit forms (\ref{eq:RestrictFnalDerivIdent1-1}) for the functional
derivatives and their spatial derivatives to modify the terms in the
new $W[\psi,\psi^{+}]$, which is equivalent to the function $w(\alpha_{k},\alpha_{k}^{+})$
if $\psi(\mathbf{s)}$ and $\psi^{+}(\mathbf{s)}$ are expanded in
terms of modes $\phi_{k}(\mathbf{s)}$ or $\phi_{k}^{\ast}(\mathbf{s)}$,
as in (\ref{eq:CondFieldFn-1}) and (\ref{eq:NonCondFieldFn-1}) with
expansion coefficients $\alpha_{k}$ and $\alpha_{k}^{\ast}$.

In the second term, the spatial derivative of the functional derivative
can be removed and applied to the spatial function

\begin{eqnarray}
 &  & \int d\mathbf{s}\left\{ \left(\partial_{\mu}\psi^{+}(\mathbf{s})\right)\left(\partial_{\mu}\frac{\delta}{\delta\psi^{+}(\mathbf{s})}\right)\right\} W[\psi,\psi^{+}]\nonumber \\
 & = & \int d\mathbf{s}\sum\limits _{k=1,2}\{\partial_{\mu}\phi_{k}^{\ast}(\mathbf{s})\}\alpha_{k}^{+}\,\sum\limits _{l=1,2}\{\partial_{\mu}\phi_{l}(\mathbf{s})\}\frac{\partial}{\partial\alpha_{l}^{+}}w(\alpha_{k},\alpha_{k}^{+})\nonumber \\
 & = & -\int d\mathbf{s}\sum\limits _{k=1,2}\{\partial_{\mu}^{2}\phi_{k}^{\ast}(\mathbf{s})\}\alpha_{k}^{+}\,\sum\limits _{l=1,2}\{\phi_{l}(\mathbf{s})\}\frac{\partial}{\partial\alpha_{l}^{+}}w(\alpha_{k},\alpha_{k}^{+})\nonumber \\
 & = & -\int d\mathbf{s}\left\{ \left(\partial_{\mu}^{2}\psi^{+}(\mathbf{s})\right)\left(\frac{\delta}{\delta\psi^{+}(\mathbf{s})}\right)\right\} W[\psi,\psi^{+}]\nonumber \\
 &  & \,\end{eqnarray}
Applying the product rule (\ref{eq:ProdRuleFnalDeriv-1}) to the product
of $\left(\partial_{\mu}^{2}\psi^{+}(\mathbf{s})\right)$ with the
distribution functional gives \begin{eqnarray}
 &  & \left(\partial_{\mu}^{2}\psi^{+}(\mathbf{s})\right)\left(\frac{\delta}{\delta\psi^{+}(\mathbf{s})}W[\psi,\psi^{+}]\right)\nonumber \\
 & = & \left(\frac{\delta}{\delta\psi^{+}(\mathbf{s})}\left(\partial_{\mu}^{2}\psi^{+}(\mathbf{s})\right)W[\psi,\psi^{+}])\right)-\left(\frac{\delta}{\delta\psi^{+}(\mathbf{s})}\left(\partial_{\mu}^{2}\psi^{+}(\mathbf{s})\right)\right)W[\psi,\psi^{+}]\nonumber \\
 & = & \left(\frac{\delta}{\delta\psi^{+}(\mathbf{s})}\left(\partial_{\mu}^{2}\psi^{+}(\mathbf{s})\right)W[\psi,\psi^{+}])\right)-\left(\chi(\mathbf{s})\right)W[\psi,\psi^{+}]\nonumber \\
 &  & \,\end{eqnarray}
using\begin{align}
\left(\frac{\delta}{\delta\psi^{+}(\mathbf{s})}\left(\partial_{\mu}^{2}\psi^{+}(\mathbf{s})\right)\right) & =\sum\limits _{l=1,2}\{\phi_{l}(\mathbf{s})\}\frac{\partial}{\partial\alpha_{l}^{+}}\sum\limits _{k=1,2}\{\partial_{\mu}^{2}\phi_{k}^{\ast}(\mathbf{s})\}\alpha_{k}^{+}\nonumber \\
 & =\sum\limits _{k=1,2}\{\phi_{k}(\mathbf{s})\}\{\partial_{\mu}^{2}\phi_{k}^{\ast}(\mathbf{s})\}\nonumber \\
 & \equiv\omega_{C}(\mathbf{s})\label{eq:ChiFn}\end{align}
Note that the function $\omega_{C}(\mathbf{s})$ just defined only
depends on condensate mode functions. Thus the second term becomes\begin{eqnarray}
 &  & \int d\mathbf{s}\left\{ \left(\partial_{\mu}\psi^{+}(\mathbf{s})\right)\left(\partial_{\mu}\frac{\delta}{\delta\psi^{+}(\mathbf{s})}\right)\right\} W[\psi,\psi^{+}]\nonumber \\
 & = & -\int d\mathbf{s}\left\{ \frac{\delta}{\delta\psi^{+}(\mathbf{s})}\left(\partial_{\mu}^{2}\psi^{+}(\mathbf{s})\right)\right\} W[\psi,\psi^{+}]+\int d\mathbf{s}\left\{ \omega_{C}(\mathbf{s})\right\} W[\psi,\psi^{+}]\nonumber \\
 &  & \,\label{eq:SimplifnResult2}\end{eqnarray}

In the third term \begin{eqnarray}
 &  & \int d\mathbf{s}\left\{ \left(\partial_{\mu}\frac{\delta}{\delta\psi(\mathbf{s})}\right)\left(\partial_{\mu}\psi(\mathbf{s})\right)\right\} W[\psi,\psi^{+}]\nonumber \\
 & = & \int d\mathbf{s}\sum\limits _{k=1,2}\{\partial_{\mu}\phi_{k}^{\ast}(\mathbf{s})\}\,\frac{\partial}{\partial\alpha_{k}}\sum\limits _{l=1,2}\alpha_{l}\{\partial_{\mu}\phi_{l}(\mathbf{s})\}w(\alpha_{k},\alpha_{k}^{+})\nonumber \\
 & = & -\int d\mathbf{s}\sum\limits _{k=1,2}\{\phi_{k}^{\ast}(\mathbf{s})\}\,\frac{\partial}{\partial\alpha_{k}}\sum\limits _{l=1,2}\alpha_{l}\{\partial_{\mu}^{2}\phi_{l}(\mathbf{s})\}w(\alpha_{k},\alpha_{k}^{+})\nonumber \\
 & = & -\int d\mathbf{s}\left\{ \left(\frac{\delta}{\delta\psi(\mathbf{s})}\right)\left(\partial_{\mu}^{2}\psi(\mathbf{s})\right)\right\} W[\psi,\psi^{+}]\label{Eq.SimplifnResult3}\end{eqnarray}

For the fourth term, the double functional derivative term can be
written in the opposite order\begin{eqnarray}
 &  & \int d\mathbf{s}\left\{ \left(\partial_{\mu}\frac{\delta}{\delta\psi(\mathbf{s})}\right)\left(\partial_{\mu}\frac{\delta}{\delta\psi^{+}(\mathbf{s})}\right)\right\} W[\psi,\psi^{+}]\nonumber \\
 & = & \int d\mathbf{s}\sum\limits _{k=1,2}\{\partial_{\mu}\phi_{k}^{\ast}(\mathbf{s})\}\,\frac{\partial}{\partial\alpha_{k}}\sum\limits _{l=1,2}\{\partial_{\mu}\phi_{l}(\mathbf{s})\}\frac{\partial}{\partial\alpha_{l}^{+}}w(\alpha_{k},\alpha_{k}^{+})\nonumber \\
 & = & \int d\mathbf{s}\sum\limits _{l=1,2}\{\partial_{\mu}\phi_{l}(\mathbf{s})\}\frac{\partial}{\partial\alpha_{l}^{+}}\sum\limits _{k=1,2}\{\partial_{\mu}\phi_{k}^{\ast}(\mathbf{s})\}\,\frac{\partial}{\partial\alpha_{k}}w(\alpha_{k},\alpha_{k}^{+})\nonumber \\
 & = & \int d\mathbf{s}\left\{ \left(\partial_{\mu}\frac{\delta}{\delta\psi^{+}(\mathbf{s})}\right)\left(\partial_{\mu}\frac{\delta}{\delta\psi(\mathbf{s})}\right)\right\} W[\psi,\psi^{+}]\label{Eq.SimplifnResult4}\end{eqnarray}

Using results (\ref{Eq.SimplifnResult1}), (\ref{eq:SimplifnResult2}),
(\ref{Eq.SimplifnResult3}) and (\ref{Eq.SimplifnResult4}) we find
that\begin{eqnarray}
 &  & W[\psi(\mathbf{r}),\psi^{+}(\mathbf{r})]\nonumber \\
 & \rightarrow & \frac{\hbar^{2}}{2m}\sum\limits _{\mu}\int d\mathbf{s\,}\left\{ \left(\partial_{\mu}\psi(\mathbf{s})\right)\left(\partial_{\mu}\psi^{+}(\mathbf{s})\right)\right\} W[\psi,\psi^{+}]\nonumber \\
 &  & -\frac{\hbar^{2}}{2m}\sum\limits _{\mu}\int d\mathbf{s}\frac{1}{2}\left\{ \frac{\delta}{\delta\psi^{+}(\mathbf{s})}\left(\partial_{\mu}^{2}\psi^{+}(\mathbf{s})\right)\right\} W[\psi,\psi^{+}]\nonumber \\
 &  & +\frac{\hbar^{2}}{2m}\sum\limits _{\mu}\int d\mathbf{s}\frac{1}{2}\left\{ \omega_{C}(\mathbf{s})\right\} W[\psi,\psi^{+}]\nonumber \\
 &  & +\frac{\hbar^{2}}{2m}\sum\limits _{\mu}\int d\mathbf{s}\frac{1}{2}\left\{ \left(\frac{\delta}{\delta\psi(\mathbf{s})}\right)\left(\partial_{\mu}^{2}\psi(\mathbf{s})\right)\right\} W[\psi,\psi^{+}]\nonumber \\
 &  & -\frac{\hbar^{2}}{2m}\sum\limits _{\mu}\int d\mathbf{s}\frac{1}{4}\left\{ \left(\partial_{\mu}\frac{\delta}{\delta\psi^{+}(\mathbf{s})}\right)\left(\partial_{\mu}\frac{\delta}{\delta\psi(\mathbf{s})}\right)\right\} W[\psi,\psi^{+}]\label{Eq.TRhoResult}\end{eqnarray}

Now if \begin{equation}
\widehat{\rho}\rightarrow\widehat{\rho}\widehat{T}=\frac{\hbar^{2}}{2m}\sum\limits _{\mu}\int d\mathbf{s\,\widehat{\rho}(}\partial_{\mu}\widehat{\Psi}(\mathbf{s})^{\dagger}\,\partial_{\mu}\widehat{\Psi}(\mathbf{s}))\end{equation}
then\begin{eqnarray}
 &  & W[\psi(\mathbf{r}),\psi^{+}(\mathbf{r})]\nonumber \\
 & \rightarrow & \frac{\hbar^{2}}{2m}\sum\limits _{\mu}\int d\mathbf{s}\left\{ \left(\partial_{\mu}\psi(\mathbf{s})-\frac{1}{2}\partial_{\mu}\frac{\delta}{\delta\psi^{+}(\mathbf{s})}\right)\left(\partial_{\mu}\psi^{+}(\mathbf{s})+\frac{1}{2}\partial_{\mu}\frac{\delta}{\delta\psi(\mathbf{s})}\right)\right\} W[\psi,\psi^{+}]\nonumber \\
 &  & \,\end{eqnarray}

After expanding we find that \begin{eqnarray}
 &  & W[\psi(\mathbf{r}),\psi^{+}(\mathbf{r})]\nonumber \\
 & \rightarrow & \frac{\hbar^{2}}{2m}\sum\limits _{\mu}\int d\mathbf{s}\left\{ \left(\partial_{\mu}\psi(\mathbf{s})\right)\left(\partial_{\mu}\psi^{+}(\mathbf{s})\right)\right\} W[\psi,\psi^{+}]\nonumber \\
 &  & +\frac{\hbar^{2}}{2m}\sum\limits _{\mu}\int d\mathbf{s}\frac{1}{2}\left\{ \left(\partial_{\mu}\psi(\mathbf{s})\right)\left(\partial_{\mu}\frac{\delta}{\delta\psi(\mathbf{s})}\right)\right\} W[\psi,\psi^{+}]\nonumber \\
 &  & -\frac{\hbar^{2}}{2m}\sum\limits _{\mu}\int d\mathbf{s}\frac{1}{2}\left\{ \left(\partial_{\mu}\frac{\delta}{\delta\psi^{+}(\mathbf{s})}\right)\left(\partial_{\mu}\psi^{+}(\mathbf{s})\right)\right\} W[\psi,\psi^{+}]\nonumber \\
 &  & -\frac{\hbar^{2}}{2m}\sum\limits _{\mu}\int d\mathbf{s}\frac{1}{4}\left\{ \left(\partial_{\mu}\frac{\delta}{\delta\psi^{+}(\mathbf{s})}\right)\left(\partial_{\mu}\frac{\delta}{\delta\psi(\mathbf{s})}\right)\right\} W[\psi,\psi^{+}]\nonumber \\
 &  & \,\end{eqnarray}

Applying the same approach as above we find that the second term becomes\begin{eqnarray}
 &  & \int d\mathbf{s}\left\{ \left(\partial_{\mu}\psi(\mathbf{s})\right)\left(\partial_{\mu}\frac{\delta}{\delta\psi(\mathbf{s})}\right)\right\} W[\psi,\psi^{+}]\nonumber \\
 & = & -\int d\mathbf{s}\left\{ \frac{\delta}{\delta\psi(\mathbf{s})}\left(\partial_{\mu}^{2}\psi(\mathbf{s})\right)\right\} W[\psi,\psi^{+}]+\int d\mathbf{s}\left\{ \omega_{C}(\mathbf{s})^{*}\right\} W[\psi,\psi^{+}]\nonumber \\
 &  & \,\label{eq:SimplifnResult2B}\end{eqnarray}
and the third term is given by\begin{eqnarray}
 &  & \int d\mathbf{s}\left\{ \left(\partial_{\mu}\frac{\delta}{\delta\psi^{+}(\mathbf{s})}\right)\left(\partial_{\mu}\psi^{+}(\mathbf{s})\right)\right\} W[\psi,\psi^{+}]\nonumber \\
 & = & -\int d\mathbf{s}\left\{ \left(\frac{\delta}{\delta\psi^{+}(\mathbf{s})}\right)\left(\partial_{\mu}^{2}\psi^{+}(\mathbf{s})\right)\right\} W[\psi,\psi^{+}]\label{Eq.SimplifnResult3B}\end{eqnarray}

Using results (\ref{eq:SimplifnResult2B}) and (\ref{Eq.SimplifnResult3B})
and using the result obtained from integration by parts \begin{equation}
\int d\mathbf{s}\left\{ \omega_{C}(\mathbf{s})^{*}\right\} =\int d\mathbf{s}\left\{ \omega_{C}(\mathbf{s})\right\} \label{eq:ChiFnResult}\end{equation}
\[
\]
we find that\begin{eqnarray}
 &  & W[\psi(\mathbf{r}),\psi^{+}(\mathbf{r})]\nonumber \\
 & \rightarrow & \frac{\hbar^{2}}{2m}\sum\limits _{\mu}\int d\mathbf{s\,}\left\{ \left(\partial_{\mu}\psi(\mathbf{s})\right)\left(\partial_{\mu}\psi^{+}(\mathbf{s})\right)\right\} W[\psi,\psi^{+}]\nonumber \\
 &  & -\frac{\hbar^{2}}{2m}\sum\limits _{\mu}\int d\mathbf{s}\frac{1}{2}\left\{ \frac{\delta}{\delta\psi(\mathbf{s})}\left(\partial_{\mu}^{2}\psi(\mathbf{s})\right)\right\} W[\psi,\psi^{+}]\nonumber \\
 &  & +\frac{\hbar^{2}}{2m}\sum\limits _{\mu}\int d\mathbf{s}\frac{1}{2}\left\{ \omega_{C}(\mathbf{s})\right\} W[\psi,\psi^{+}]\nonumber \\
 &  & +\frac{\hbar^{2}}{2m}\sum\limits _{\mu}\int d\mathbf{s}\frac{1}{2}\left\{ \left(\frac{\delta}{\delta\psi^{+}(\mathbf{s})}\right)\left(\partial_{\mu}^{2}\psi^{+}(\mathbf{s})\right)\right\} W[\psi,\psi^{+}]\nonumber \\
 &  & -\frac{\hbar^{2}}{2m}\sum\limits _{\mu}\int d\mathbf{s}\frac{1}{4}\left\{ \left(\partial_{\mu}\frac{\delta}{\delta\psi^{+}(\mathbf{s})}\right)\left(\partial_{\mu}\frac{\delta}{\delta\psi(\mathbf{s})}\right)\right\} W[\psi,\psi^{+}]\nonumber \\
 &  & \,\label{eq:RhoTResult}\end{eqnarray}

We now combine the contributions so that when \begin{equation}
\widehat{\rho}\rightarrow\lbrack\widehat{T},\widehat{\rho}]=[\frac{\hbar^{2}}{2m}\sum\limits _{\mu}\int d\mathbf{s(}\partial_{\mu}\widehat{\Psi}(\mathbf{s})^{\dagger}\,\partial_{\mu}\widehat{\Psi}(\mathbf{s})),\widehat{\rho}]\end{equation}
then

\begin{equation}
W[\psi(\mathbf{r}),\psi^{+}(\mathbf{r})]\rightarrow W^{1}\end{equation}
where\begin{eqnarray}
W^{1} & = & -\int d\mathbf{s}\left\{ \frac{\delta}{\delta\psi^{+}(\mathbf{s})}\left(\sum\limits _{\mu}\frac{\hbar^{2}}{2m}\partial_{\mu}^{2}\psi^{+}(\mathbf{s})\right))W[\psi,\psi^{+}]\right\} \nonumber \\
 &  & +\int d\mathbf{s}\left\{ \frac{\delta}{\delta\psi(\mathbf{s})}\left(\sum\limits _{\mu}\frac{\hbar^{2}}{2m}\partial_{\mu}^{2}\psi(\mathbf{s})\right)W[\psi,\psi^{+}])\right\} \nonumber \\
 &  & \,\end{eqnarray}
where the $\left(\omega_{C}(\mathbf{s})\right)$, the $\left(\partial_{\mu}\psi(\mathbf{s})\right)\left(\partial_{\mu}\psi^{+}(\mathbf{s})\right)$
and the $\left(\partial_{\mu}\frac{{\LARGE\delta}}{{\LARGE\delta\psi}^{+}{\LARGE(s)}}\right)\left(\partial_{\mu}\frac{{\LARGE\delta}}{{\LARGE\delta\psi(s)}}\right)$
terms cancel and the first order functional derivative terms combine
to remove the $\frac{1}{2}$ factors. Thus only a first order functional
derivative term occurs.

Overall, the contribution to the functional Fokker-Planck equation
from the kinetic energy term is given by\begin{eqnarray}
 &  & \left(\frac{\partial}{\partial t}W[\psi,\psi^{+}]\right)_{K}\nonumber \\
 & = & \frac{-i}{\hbar}\left\{ -\int d\mathbf{s}\left\{ \frac{\delta}{\delta\psi^{+}(\mathbf{s})}\left(\sum\limits _{\mu}\frac{\hbar^{2}}{2m}\partial_{\mu}^{2}\psi^{+}(\mathbf{s})\right)W[\psi,\psi^{+}]\right\} \right\} \nonumber \\
 &  & +\frac{-i}{\hbar}\left\{ +\int d\mathbf{s}\left\{ \frac{\delta}{\delta\psi(\mathbf{s})}\left(\sum\limits _{\mu}\frac{\hbar^{2}}{2m}\partial_{\mu}^{2}\psi(\mathbf{s})\right)W[\psi,\psi^{+}])\right\} \right\} \end{eqnarray}

Reverting to the original notation, the contribution to the functional
Fokker-Planck equation from the kinetic energy term is given by\begin{eqnarray}
 &  & \left(\frac{\partial}{\partial t}P[\underrightarrow{\psi}(\mathbf{r}),\underrightarrow{\psi^{\ast}}(\mathbf{r})]\right)_{K}\nonumber \\
 & = & \frac{+i}{\hbar}\left\{ \int d\mathbf{s}\left\{ \frac{\delta}{\delta\psi_{C}^{+}(\mathbf{s})}\left(\sum\limits _{\mu}\frac{\hbar^{2}}{2m}\partial_{\mu}^{2}\psi_{C}^{+}(\mathbf{s})\right)P[\underrightarrow{\psi}(\mathbf{r}),\underrightarrow{\psi^{\ast}}(\mathbf{r})]\right\} \right\} \nonumber \\
 &  & -\frac{i}{\hbar}\left\{ \int d\mathbf{s}\left\{ \frac{\delta}{\delta\psi_{C}(\mathbf{s})}\left(\sum\limits _{\mu}\frac{\hbar^{2}}{2m}\partial_{\mu}^{2}\psi_{C}(\mathbf{s})\right)P[\underrightarrow{\psi}(\mathbf{r}),\underrightarrow{\psi^{\ast}}(\mathbf{r})])\right\} \right\} \nonumber \\
 &  & \,\label{eq:FFPECondKE}\end{eqnarray}

\subsection{Condensate Trap Potential Terms}

We write the trap potential as \begin{equation}
\widehat{V}=\int d\mathbf{s(}\widehat{\Psi}(\mathbf{s})^{\dagger}V\widehat{\Psi}(\mathbf{s}))\end{equation}
Now if \begin{equation}
\widehat{\rho}\rightarrow\widehat{V}\widehat{\rho}=\int d\mathbf{s(}\widehat{\Psi}(\mathbf{s})^{\dagger}V\widehat{\Psi}(\mathbf{s}))\widehat{\rho}\end{equation}
then\begin{eqnarray}
 &  & W[\psi(\mathbf{r}),\psi^{+}(\mathbf{r})]\nonumber \\
 & \rightarrow & \int d\mathbf{s}\left\{ \left(\psi^{+}(\mathbf{s})-\frac{1}{2}\frac{\delta}{\delta\psi(\mathbf{s})}\right)V(\mathbf{s})\left(\psi(\mathbf{s})+\frac{1}{2}\frac{\delta}{\delta\psi^{+}(\mathbf{s})}\right)\right\} W[\psi,\psi^{+}]\nonumber \\
 &  & \,\end{eqnarray}
After expanding we find that\begin{eqnarray}
 &  & W[\psi(\mathbf{r}),\psi^{+}(\mathbf{r})]\nonumber \\
 & \rightarrow & \int d\mathbf{s}\left\{ \psi^{+}(\mathbf{s})V(\mathbf{s})\psi(\mathbf{s})\right\} W[\psi,\psi^{+}]\nonumber \\
 &  & +\int d\mathbf{s}\frac{1}{2}\left\{ \psi^{+}(\mathbf{s})V(\mathbf{s})\frac{\delta}{\delta\psi^{+}(\mathbf{s})}\right\} W[\psi,\psi^{+}]\nonumber \\
 &  & -\int d\mathbf{s}\frac{1}{2}\left\{ \frac{\delta}{\delta\psi(\mathbf{s})}V(\mathbf{s})\psi(\mathbf{s})\right\} W[\psi,\psi^{+}]\nonumber \\
 &  & -\int d\mathbf{s}\frac{1}{4}\left\{ \frac{\delta}{\delta\psi(\mathbf{s})}V(\mathbf{s})\frac{\delta}{\delta\psi^{+}(\mathbf{s})}\right\} W[\psi,\psi^{+}]\nonumber \\
 &  & \,\end{eqnarray}

We can now use the product rule for functional derivatives (\ref{eq:ProdRuleFnalDeriv-1})
together with (\ref{Eq.FuncDerivativeRule1-1}) and (\ref{Eq.FuncDerivativeRule2-1})
to place all the derivatives on the left of the expression and obtain\begin{eqnarray}
 &  & W[\psi(\mathbf{r}),\psi^{+}(\mathbf{r})]\nonumber \\
 & \rightarrow & \int d\mathbf{s}\left\{ \psi^{+}(\mathbf{s})V(\mathbf{s})\psi(\mathbf{s})\right\} W[\psi,\psi^{+}]\qquad T1\nonumber \\
 &  & +\int d\mathbf{s}\frac{1}{2}\left\{ \frac{\delta}{\delta\psi^{+}(\mathbf{s})}\{\psi^{+}(\mathbf{s})V(\mathbf{s})\}-\delta_{C}(\mathbf{s},\mathbf{s})V(\mathbf{s})\right\} W[\psi,\psi^{+}]\qquad T22,T21\nonumber \\
 &  & -\int d\mathbf{s}\frac{1}{2}\left\{ \frac{\delta}{\delta\psi(\mathbf{s})}V(\mathbf{s})\psi(\mathbf{s})\right\} W[\psi,\psi^{+}]\qquad T3\nonumber \\
 &  & -\int d\mathbf{s}\frac{1}{4}\left\{ \frac{\delta}{\delta\psi(\mathbf{s})}\frac{\delta}{\delta\psi^{+}(\mathbf{s})}V(\mathbf{s})\right\} W[\psi,\psi^{+}]\qquad T4\nonumber \\
 &  & \,\end{eqnarray}

Details are

T2\begin{eqnarray*}
 &  & \frac{\delta}{\delta\psi^{+}(\mathbf{s})}\{\psi^{+}(\mathbf{s})V(\mathbf{s})W\}\\
 & = & \{\frac{\delta}{\delta\psi^{+}(\mathbf{s})}\psi^{+}(\mathbf{s})\}V(\mathbf{s})W+\psi^{+}(\mathbf{s})V(\mathbf{s})\{\frac{\delta}{\delta\psi^{+}(\mathbf{s})}W\}\\
 & = & \delta_{K}(0)V(\mathbf{s})W+\psi^{+}(\mathbf{s})V(\mathbf{s})\{\frac{\delta}{\delta\psi^{+}(\mathbf{s})}W\}\\
 &  & \psi^{+}(\mathbf{s})V(\mathbf{s})\{\frac{\delta}{\delta\psi^{+}(\mathbf{s})}W[\psi,\psi^{+}]\}\\
 & = & \frac{\delta}{\delta\psi^{+}(\mathbf{s})}\{\psi^{+}(\mathbf{s})V(\mathbf{s})W[\psi,\psi^{+}]\}-\delta_{C}(\mathbf{s},\mathbf{s})V(\mathbf{s})W[\psi,\psi^{+}]\end{eqnarray*}

Now if \begin{equation}
\widehat{\rho}\rightarrow\widehat{\rho}\widehat{V}=\int d\mathbf{s}\widehat{\mathbf{\rho}}\widehat{\Psi}(\mathbf{s})^{\dagger}V\widehat{\Psi}(\mathbf{s})\end{equation}
then\begin{eqnarray}
 &  & W[\psi(\mathbf{r}),\psi^{+}(\mathbf{r})]\nonumber \\
 & \rightarrow & \int d\mathbf{s}\left(\psi(\mathbf{s})-\frac{1}{2}\frac{\delta}{\delta\psi^{+}(\mathbf{s})}\right)V(\mathbf{s})\left(\psi^{+}(\mathbf{s})+\frac{1}{2}\frac{\delta}{\delta\psi(\mathbf{s})}\right)W[\psi,\psi^{+}]\nonumber \\
 &  & \,\end{eqnarray}
After expanding we have\begin{eqnarray}
 &  & W[\psi(\mathbf{r}),\psi^{+}(\mathbf{r})]\nonumber \\
 & \rightarrow & \int d\mathbf{s\{}\psi(\mathbf{s})V(\mathbf{s})\psi^{+}(\mathbf{s})\}W[\psi,\psi^{+}]\nonumber \\
 &  & +\int d\mathbf{s}\frac{1}{2}\left\{ \psi(\mathbf{s})V(\mathbf{s})\frac{\delta}{\delta\psi(\mathbf{s})}\right\} W[\psi,\psi^{+}]\nonumber \\
 &  & -\int d\mathbf{s}\frac{1}{2}\left\{ \frac{\delta}{\delta\psi^{+}(\mathbf{s})}V(\mathbf{s})\psi^{+}(\mathbf{s})\right\} W[\psi,\psi^{+}]\nonumber \\
 &  & -\int d\mathbf{s}\frac{1}{4}\left\{ \frac{\delta}{\delta\psi^{+}(\mathbf{s})}V(\mathbf{s})\frac{\delta}{\delta\psi(\mathbf{s})}\right\} W[\psi,\psi^{+}]\nonumber \\
 &  & \,\end{eqnarray}

We can now use the product rule for functional derivatives (\ref{eq:ProdRuleFnalDeriv-1})
together with (\ref{Eq.FuncDerivativeRule1-1}) and (\ref{Eq.FuncDerivativeRule2-1})
to place all the derivatives on the left of the expression. However
the results can more easily be obtained by noticing that the $\widehat{\rho}\widehat{V}$
is the same as the $\widehat{V}\widehat{\rho}$ if we interchange
$\psi(\mathbf{s})$ and $\psi^{+}(\mathbf{s})$ everywhere. Hence\begin{eqnarray}
 &  & W[\psi(\mathbf{r}),\psi^{+}(\mathbf{r})]\nonumber \\
 & \rightarrow & \int d\mathbf{s}\left\{ \psi(\mathbf{s})V(\mathbf{s})\psi^{+}(\mathbf{s})\right\} W[\psi,\psi^{+}]\qquad T1\nonumber \\
 &  & +\int d\mathbf{s}\frac{1}{2}\left\{ \frac{\delta}{\delta\psi(\mathbf{s})}\{\psi(\mathbf{s})V(\mathbf{s})\}-\delta_{C}(\mathbf{s},\mathbf{s})V(\mathbf{s})\right\} W[\psi,\psi^{+}]\qquad T22,T21\nonumber \\
 &  & -\int d\mathbf{s}\frac{1}{2}\left\{ \frac{\delta}{\delta\psi^{+}(\mathbf{s})}V(\mathbf{s})\psi^{+}(\mathbf{s})\right\} W[\psi,\psi^{+}]\qquad T3\nonumber \\
 &  & -\int d\mathbf{s}\frac{1}{4}\left\{ \frac{\delta}{\delta\psi^{+}(\mathbf{s})}\frac{\delta}{\delta\psi(\mathbf{s})}V(\mathbf{s})\right\} W[\psi,\psi^{+}]\qquad T4\nonumber \\
 &  & \,\end{eqnarray}

We now combine the contributions so that when \begin{equation}
\widehat{\rho}\rightarrow\lbrack\widehat{V},\widehat{\rho}]=[\int d\mathbf{s}\,\,\widehat{\Psi}(\mathbf{s})^{\dagger}V(\mathbf{s})\widehat{\Psi}(\mathbf{s}),\widehat{\rho}]\end{equation}
then

\begin{equation}
W[\psi(\mathbf{r}),\psi^{+}(\mathbf{r})]\rightarrow W^{0}+W^{1}+W^{2}\end{equation}
where the terms are listed via the order of derivatives that occur\begin{eqnarray}
 &  & W^{0}\nonumber \\
 & = & \int d\mathbf{s}\left\{ \psi^{+}(\mathbf{s})V(\mathbf{s})\psi(\mathbf{s})\right\} W[\psi,\psi^{+}]-\int d\mathbf{s}\left\{ \psi(\mathbf{s})V(\mathbf{s})\psi^{+}(\mathbf{s})\right\} W[\psi,\psi^{+}]\nonumber \\
 &  & +\int d\mathbf{s}\frac{1}{2}\left\{ -\delta_{C}(\mathbf{s},\mathbf{s})V(\mathbf{s})\right\} W[\psi,\psi^{+}]-\int d\mathbf{s}\frac{1}{2}\left\{ -\delta_{C}(\mathbf{s},\mathbf{s})V(\mathbf{s})\right\} W[\psi,\psi^{+}]\nonumber \\
 & = & 0\nonumber \\
 &  & \,\end{eqnarray}
\begin{eqnarray}
 &  & W^{1}\nonumber \\
 & = & \int d\mathbf{s}\frac{1}{2}\left\{ \frac{\delta}{\delta\psi^{+}(\mathbf{s})}\{\psi^{+}(\mathbf{s})V(\mathbf{s})\}\right\} W[\psi,\psi^{+}]-\int d\mathbf{s}\frac{1}{2}\left\{ \frac{\delta}{\delta\psi(\mathbf{s})}\{\psi(\mathbf{s})V(\mathbf{s})\}\right\} W[\psi,\psi^{+}]\nonumber \\
 &  & -\int d\mathbf{s}\frac{1}{2}\left\{ \frac{\delta}{\delta\psi(\mathbf{s})}V(\mathbf{s})\psi(\mathbf{s})\right\} W[\psi,\psi^{+}]+\int d\mathbf{s}\frac{1}{2}\left\{ \frac{\delta}{\delta\psi^{+}(\mathbf{s})}V(\mathbf{s})\psi^{+}(\mathbf{s})\right\} W[\psi,\psi^{+}]\nonumber \\
 & = & -\int d\mathbf{s}\left\{ \frac{\delta}{\delta\psi(\mathbf{s})}\{V(\mathbf{s})\psi(\mathbf{s})\}\right\} W[\psi,\psi^{+}]+\int d\mathbf{s}\left\{ \frac{\delta}{\delta\psi^{+}(\mathbf{s})}V(\mathbf{s})\psi^{+}(\mathbf{s})\right\} W[\psi,\psi^{+}]\nonumber \\
 &  & \,\end{eqnarray}
\begin{eqnarray}
 &  & W^{2}\nonumber \\
 & = & -\int d\mathbf{s}\frac{1}{4}\left\{ \frac{\delta}{\delta\psi(\mathbf{s})}\frac{\delta}{\delta\psi^{+}(\mathbf{s})}V(\mathbf{s})\right\} W[\psi,\psi^{+}]+\int d\mathbf{s}\frac{1}{4}\left\{ \frac{\delta}{\delta\psi^{+}(\mathbf{s})}\frac{\delta}{\delta\psi(\mathbf{s})}V(\mathbf{s})\right\} W[\psi,\psi^{+}]\nonumber \\
 & = & 0\nonumber \\
 &  & \,\end{eqnarray}

Overall, the contribution to the functional Fokker-Planck equation
from the trap potential term is given by\begin{eqnarray}
 &  & \left(\frac{\partial}{\partial t}W[\psi,\psi^{+}]\right)_{V}\nonumber \\
 & = & \frac{-i}{\hbar}\left\{ -\int d\mathbf{s}\left\{ \frac{\delta}{\delta\psi(\mathbf{s})}\{V(\mathbf{s})\psi(\mathbf{s})\}\right\} W[\psi,\psi^{+}]+\int d\mathbf{s}\left\{ \frac{\delta}{\delta\psi^{+}(\mathbf{s})}V(\mathbf{s})\psi^{+}(\mathbf{s})\right\} W[\psi,\psi^{+}]\right\} \nonumber \\
 &  & \,\end{eqnarray}
which only involves first order functional derivatives.

Reverting to the original notation, the contribution to the functional
Fokker-Planck equation from the trap potential term is given by\begin{eqnarray}
 &  & \left(\frac{\partial}{\partial t}P[\underrightarrow{\psi}(\mathbf{r}),\underrightarrow{\psi^{\ast}}(\mathbf{r})]\right)_{V}\nonumber \\
 & = & \frac{-i}{\hbar}\left\{ -\int d\mathbf{s}\left\{ \frac{\delta}{\delta\psi_{C}(\mathbf{s})}\{V(\mathbf{s})\psi_{C}(\mathbf{s})\}\right\} P[\underrightarrow{\psi}(\mathbf{r}),\underrightarrow{\psi^{\ast}}(\mathbf{r})]\right\} \nonumber \\
 &  & +\frac{-i}{\hbar}\left\{ +\int d\mathbf{s}\left\{ \frac{\delta}{\delta\psi_{C}^{+}(\mathbf{s})}V(\mathbf{s})\psi_{C}^{+}(\mathbf{s})\right\} P[\underrightarrow{\psi}(\mathbf{r}),\underrightarrow{\psi^{\ast}}(\mathbf{r})]\right\} \nonumber \\
 &  & \,\label{eq:FFPECondTrap}\end{eqnarray}

\subsection{Condensate Boson-Boson Interaction Terms}

We write the boson-boson interaction potential as \begin{equation}
\widehat{U}=\frac{g}{2}\int d\mathbf{s}\widehat{\Psi}(\mathbf{s})^{\dagger}\widehat{\Psi}(\mathbf{s})^{\dagger}\widehat{\Psi}(\mathbf{s})\widehat{\Psi}(\mathbf{s})\end{equation}

Now if \begin{equation}
\widehat{\rho}\rightarrow\widehat{U}\widehat{\rho}=\frac{g}{2}\int d\mathbf{s}\widehat{\Psi}(\mathbf{s})^{\dagger}\widehat{\Psi}(\mathbf{s})^{\dagger}\widehat{\Psi}(\mathbf{s})\widehat{\Psi}(\mathbf{s})\widehat{\rho}\end{equation}

\begin{eqnarray}
 &  & W[\psi(\mathbf{r}),\psi^{+}(\mathbf{r})]\nonumber \\
 & \rightarrow & \frac{g}{2}\int d\mathbf{s}\left(\psi^{+}(\mathbf{s})-\frac{1}{2}\frac{\delta}{\delta\psi(\mathbf{s})}\right)\left(\psi^{+}(\mathbf{s})-\frac{1}{2}\frac{\delta}{\delta\psi(\mathbf{s})}\right)\left(\psi(\mathbf{s})+\frac{1}{2}\frac{\delta}{\delta\psi^{+}(\mathbf{s})}\right)\left(\psi(\mathbf{s})+\frac{1}{2}\frac{\delta}{\delta\psi^{+}(\mathbf{s})}\right)W\nonumber \\
 &  & \,\end{eqnarray}
After expanding we get

\begin{eqnarray}
 &  & W[\psi(\mathbf{r}),\psi^{+}(\mathbf{r})]\nonumber \\
 & \rightarrow & \frac{g}{2}\int d\mathbf{s}\left\{ \psi^{+}(\mathbf{s})\psi^{+}(\mathbf{s})\psi(\mathbf{s})\psi(\mathbf{s})\right\} W+\frac{g}{2}\int d\mathbf{s}\frac{1}{2}\left\{ \psi^{+}(\mathbf{s})\psi^{+}(\mathbf{s})\psi(\mathbf{s})\frac{\delta}{\delta\psi^{+}(\mathbf{s})}\right\} W\nonumber \\
 &  & +\frac{g}{2}\int d\mathbf{s}\frac{1}{2}\left\{ \psi^{+}(\mathbf{s})\psi^{+}(\mathbf{s})\frac{\delta}{\delta\psi^{+}(\mathbf{s})}\psi(\mathbf{s})\right\} W+\frac{g}{2}\int d\mathbf{s}\frac{1}{4}\left\{ \psi^{+}(\mathbf{s})\psi^{+}(\mathbf{s})\frac{\delta}{\delta\psi^{+}(\mathbf{s})}\frac{\delta}{\delta\psi^{+}(\mathbf{s})}\right\} W\nonumber \\
 &  & -\frac{g}{2}\int d\mathbf{s}\frac{1}{2}\left\{ \psi^{+}(\mathbf{s})\frac{\delta}{\delta\psi(\mathbf{s})}\psi(\mathbf{s})\psi(\mathbf{s})\right\} W-\frac{g}{2}\int d\mathbf{s}\frac{1}{4}\left\{ \psi^{+}(\mathbf{s})\frac{\delta}{\delta\psi(\mathbf{s})}\psi(\mathbf{s})\frac{\delta}{\delta\psi^{+}(\mathbf{s})}\right\} W\nonumber \\
 &  & -\frac{g}{2}\int d\mathbf{s}\frac{1}{4}\left\{ \psi^{+}(\mathbf{s})\frac{\delta}{\delta\psi(\mathbf{s})}\frac{\delta}{\delta\psi^{+}(\mathbf{s})}\psi(\mathbf{s})\right\} W-\frac{g}{2}\int d\mathbf{s}\frac{1}{8}\left\{ \psi^{+}(\mathbf{s})\frac{\delta}{\delta\psi(\mathbf{s})}\frac{\delta}{\delta\psi^{+}(\mathbf{s})}\frac{\delta}{\delta\psi^{+}(\mathbf{s})}\right\} W\nonumber \\
 &  & -\frac{g}{2}\int d\mathbf{s}\frac{1}{2}\left\{ \frac{\delta}{\delta\psi(\mathbf{s})}\psi^{+}(\mathbf{s})\psi(\mathbf{s})\psi(\mathbf{s})\right\} W-\frac{g}{2}\int d\mathbf{s}\frac{1}{4}\left\{ \frac{\delta}{\delta\psi(\mathbf{s})}\psi^{+}(\mathbf{s})\psi(\mathbf{s})\frac{\delta}{\delta\psi^{+}(\mathbf{s})}\right\} W\nonumber \\
 &  & -\frac{g}{2}\int d\mathbf{s}\frac{1}{4}\left\{ \frac{\delta}{\delta\psi(\mathbf{s})}\psi^{+}(\mathbf{s})\frac{\delta}{\delta\psi^{+}(\mathbf{s})}\psi(\mathbf{s})\right\} W-\frac{g}{2}\int d\mathbf{s}\frac{1}{8}\left\{ \frac{\delta}{\delta\psi(\mathbf{s})}\psi^{+}(\mathbf{s})\frac{\delta}{\delta\psi^{+}(\mathbf{s})}\frac{\delta}{\delta\psi^{+}(\mathbf{s})}\right\} W\nonumber \\
 &  & +\frac{g}{2}\int d\mathbf{s}\frac{1}{4}\left\{ \frac{\delta}{\delta\psi(\mathbf{s})}\frac{\delta}{\delta\psi(\mathbf{s})}\psi(\mathbf{s})\psi(\mathbf{s})\right\} W+\frac{g}{2}\int d\mathbf{s}\frac{1}{8}\left\{ \frac{\delta}{\delta\psi(\mathbf{s})}\frac{\delta}{\delta\psi(\mathbf{s})}\psi(\mathbf{s})\frac{\delta}{\delta\psi^{+}(\mathbf{s})}\right\} W\nonumber \\
 &  & +\frac{g}{2}\int d\mathbf{s}\frac{1}{8}\left\{ \frac{\delta}{\delta\psi(\mathbf{s})}\frac{\delta}{\delta\psi(\mathbf{s})}\frac{\delta}{\delta\psi^{+}(\mathbf{s})}\psi(\mathbf{s})\right\} W+\frac{g}{2}\int d\mathbf{s}\frac{1}{16}\left\{ \frac{\delta}{\delta\psi(\mathbf{s})}\frac{\delta}{\delta\psi(\mathbf{s})}\frac{\delta}{\delta\psi^{+}(\mathbf{s})}\frac{\delta}{\delta\psi^{+}(\mathbf{s})}\right\} W\nonumber \\
 &  & \,\end{eqnarray}

We can now use the product rule for functional derivatives (\ref{eq:ProdRuleFnalDeriv-1})
together with (\ref{Eq.FuncDerivativeRule1-1}) and (\ref{Eq.FuncDerivativeRule2-1})
to place all the derivatives on the left of the expression and obtain\begin{eqnarray}
 &  & W[\psi(\mathbf{r}),\psi^{+}(\mathbf{r})]\nonumber \\
 & \rightarrow & \frac{g}{2}\int d\mathbf{s}\left\{ \psi^{+}(\mathbf{s})\psi^{+}(\mathbf{s})\psi(\mathbf{s})\psi(\mathbf{s})\right\} W\qquad T1\nonumber \\
 &  & +\frac{g}{2}\int d\mathbf{s}\frac{1}{2}\left\{ \frac{\delta}{\delta\psi^{+}(\mathbf{s})}\psi^{+}(\mathbf{s})\psi^{+}(\mathbf{s})\psi(\mathbf{s})-2\delta_{C}(\mathbf{s},\mathbf{s})\psi^{+}(\mathbf{s})\psi(\mathbf{s})\right\} W\qquad T22,T21\nonumber \\
 &  & +\frac{g}{2}\int d\mathbf{s}\frac{1}{2}\left\{ \frac{\delta}{\delta\psi^{+}(\mathbf{s})}\psi^{+}(\mathbf{s})\psi^{+}(\mathbf{s})\psi(\mathbf{s})-2\delta_{C}(\mathbf{s},\mathbf{s})\psi^{+}(\mathbf{s})\psi(\mathbf{s})\right\} W\qquad T32,T31\nonumber \\
 &  & +\frac{g}{2}\int d\mathbf{s}\frac{1}{4}\left\{ \frac{\delta}{\delta\psi^{+}(\mathbf{s})}\frac{\delta}{\delta\psi^{+}(\mathbf{s})}\psi^{+}(\mathbf{s})\psi^{+}(\mathbf{s})-\frac{\delta}{\delta\psi^{+}(\mathbf{s})}4\delta_{C}(\mathbf{s},\mathbf{s})\psi^{+}(\mathbf{s})+2\delta_{K}(0)^{2}\right\} W\qquad T43,T42,T41\nonumber \\
 &  & -\frac{g}{2}\int d\mathbf{s}\frac{1}{2}\left\{ \frac{\delta}{\delta\psi(\mathbf{s})}\psi^{+}(\mathbf{s})\psi(\mathbf{s})\psi(\mathbf{s})\right\} W\qquad T5\nonumber \\
 &  & -\frac{g}{2}\int d\mathbf{s}\frac{1}{4}\left\{ \frac{\delta}{\delta\psi(\mathbf{s})}\frac{\delta}{\delta\psi^{+}(\mathbf{s})}\psi^{+}(\mathbf{s})\psi(\mathbf{s})-\frac{\delta}{\delta\psi(\mathbf{s})}\delta_{C}(\mathbf{s},\mathbf{s})\psi(\mathbf{s})\right\} W\qquad T62,T61\nonumber \\
 &  & -\frac{g}{2}\int d\mathbf{s}\frac{1}{4}\left\{ \frac{\delta}{\delta\psi(\mathbf{s})}\frac{\delta}{\delta\psi^{+}(\mathbf{s})}\psi^{+}(\mathbf{s})\psi(\mathbf{s})-\frac{\delta}{\delta\psi(\mathbf{s})}\delta_{C}(\mathbf{s},\mathbf{s})\psi(\mathbf{s})\right\} W\qquad T72,T71\nonumber \\
 &  & -\frac{g}{2}\int d\mathbf{s}\frac{1}{8}\left\{ \frac{\delta}{\delta\psi(\mathbf{s})}\frac{\delta}{\delta\psi^{+}(\mathbf{s})}\frac{\delta}{\delta\psi^{+}(\mathbf{s})}\psi^{+}(\mathbf{s})-\frac{\delta}{\delta\psi(\mathbf{s})}\frac{\delta}{\delta\psi^{+}(\mathbf{s})}2\delta_{C}(\mathbf{s},\mathbf{s})\right\} W\qquad T82,T81\nonumber \\
 &  & -\frac{g}{2}\int d\mathbf{s}\frac{1}{2}\left\{ \frac{\delta}{\delta\psi(\mathbf{s})}\psi^{+}(\mathbf{s})\psi(\mathbf{s})\psi(\mathbf{s})\right\} W\qquad T9\nonumber \\
 &  & -\frac{g}{2}\int d\mathbf{s}\frac{1}{4}\left\{ \frac{\delta}{\delta\psi(\mathbf{s})}\frac{\delta}{\delta\psi^{+}(\mathbf{s})}\psi^{+}(\mathbf{s})\psi(\mathbf{s})-\frac{\delta}{\delta\psi(\mathbf{s})}\delta_{C}(\mathbf{s},\mathbf{s})\psi(\mathbf{s})\right\} W\qquad T10.2,T10.1\nonumber \\
 &  & -\frac{g}{2}\int d\mathbf{s}\frac{1}{4}\left\{ \frac{\delta}{\delta\psi(\mathbf{s})}\frac{\delta}{\delta\psi^{+}(\mathbf{s})}\psi^{+}(\mathbf{s})\psi(\mathbf{s})-\frac{\delta}{\delta\psi(\mathbf{s})}\delta_{C}(\mathbf{s},\mathbf{s})\psi(\mathbf{s})\right\} W\qquad T11.2,T11.1\nonumber \\
 &  & -\frac{g}{2}\int d\mathbf{s}\frac{1}{8}\left\{ \frac{\delta}{\delta\psi(\mathbf{s})}\frac{\delta}{\delta\psi^{+}(\mathbf{s})}\frac{\delta}{\delta\psi^{+}(\mathbf{s})}\psi^{+}(\mathbf{s})-\frac{\delta}{\delta\psi(\mathbf{s})}\frac{\delta}{\delta\psi^{+}(\mathbf{s})}2\delta_{C}(\mathbf{s},\mathbf{s})\right\} W\qquad T12.2,T12.1\nonumber \\
 &  & +\frac{g}{2}\int d\mathbf{s}\frac{1}{4}\left\{ \frac{\delta}{\delta\psi(\mathbf{s})}\frac{\delta}{\delta\psi(\mathbf{s})}\psi(\mathbf{s})\psi(\mathbf{s})\right\} W\qquad T13\nonumber \\
 &  & +\frac{g}{2}\int d\mathbf{s}\frac{1}{8}\left\{ \frac{\delta}{\delta\psi(\mathbf{s})}\frac{\delta}{\delta\psi(\mathbf{s})}\frac{\delta}{\delta\psi^{+}(\mathbf{s})}\psi(\mathbf{s})\right\} W\qquad T14\nonumber \\
 &  & +\frac{g}{2}\int d\mathbf{s}\frac{1}{8}\left\{ \frac{\delta}{\delta\psi(\mathbf{s})}\frac{\delta}{\delta\psi(\mathbf{s})}\frac{\delta}{\delta\psi^{+}(\mathbf{s})}\psi(\mathbf{s})\right\} W\qquad T15\nonumber \\
 &  & +\frac{g}{2}\int d\mathbf{s}\frac{1}{16}\left\{ \frac{\delta}{\delta\psi(\mathbf{s})}\frac{\delta}{\delta\psi(\mathbf{s})}\frac{\delta}{\delta\psi^{+}(\mathbf{s})}\frac{\delta}{\delta\psi^{+}(\mathbf{s})}\right\} W\qquad T16\nonumber \\
 &  & \,\end{eqnarray}

Details include

T2\begin{eqnarray*}
 &  & \frac{\delta}{\delta\psi^{+}(\mathbf{s})}\{\psi^{+}(\mathbf{s})\psi^{+}(\mathbf{s})\psi(\mathbf{s})W\}\\
 & = & \{\frac{\delta}{\delta\psi^{+}(\mathbf{s})}\psi^{+}(\mathbf{s})\}\psi^{+}(\mathbf{s})\psi(\mathbf{s})W+\psi^{+}(\mathbf{s})\{\frac{\delta}{\delta\psi^{+}(\mathbf{s})}\psi^{+}(\mathbf{s})\}\psi(\mathbf{s})W\\
 &  & +\psi^{+}(\mathbf{s})\psi^{+}(\mathbf{s})\{\frac{\delta}{\delta\psi^{+}(\mathbf{s})}\psi(\mathbf{s})\}W+\psi^{+}(\mathbf{s})\psi^{+}(\mathbf{s})\psi(\mathbf{s})\{\frac{\delta}{\delta\psi^{+}(\mathbf{s})}W\}\\
 & = & \{\delta_{C}(\mathbf{s},\mathbf{s})\}\psi^{+}(\mathbf{s})\psi(\mathbf{s})W+\psi^{+}(\mathbf{s})\{\delta_{C}(\mathbf{s},\mathbf{s})\}\psi(\mathbf{s})W\\
 &  & +\psi^{+}(\mathbf{s})\psi^{+}(\mathbf{s})\psi(\mathbf{s})\{\frac{\delta}{\delta\psi^{+}(\mathbf{s})}W\}\\
 & = & 2\delta_{C}(\mathbf{s},\mathbf{s})\psi^{+}(\mathbf{s})\psi(\mathbf{s})W+\psi^{+}(\mathbf{s})\psi^{+}(\mathbf{s})\psi(\mathbf{s})\{\frac{\delta}{\delta\psi^{+}(\mathbf{s})}W\}\\
 &  & \psi^{+}(\mathbf{s})\psi^{+}(\mathbf{s})\psi(\mathbf{s})\{\frac{\delta}{\delta\psi^{+}(\mathbf{s})}W[\psi,\psi^{+}]\}\\
 & = & \frac{\delta}{\delta\psi^{+}(\mathbf{s})}\{\psi^{+}(\mathbf{s})\psi^{+}(\mathbf{s})\psi(\mathbf{s})W[\psi,\psi^{+}]\}-2\delta_{C}(\mathbf{s},\mathbf{s})\psi^{+}(\mathbf{s})\psi(\mathbf{s})W[\psi,\psi^{+}]\end{eqnarray*}

T3\begin{eqnarray*}
 &  & \frac{\delta}{\delta\psi^{+}(\mathbf{s})}\{\psi^{+}(\mathbf{s})\psi^{+}(\mathbf{s})\psi(\mathbf{s})W\}\\
 & = & \{\frac{\delta}{\delta\psi^{+}(\mathbf{s})}\psi^{+}(\mathbf{s})\}\psi^{+}(\mathbf{s})\psi(\mathbf{s})W+\psi^{+}(\mathbf{s})\{\frac{\delta}{\delta\psi^{+}(\mathbf{s})}\psi^{+}(\mathbf{s})\}\psi(\mathbf{s})W\}\\
 &  & +\psi^{+}(\mathbf{s})\psi^{+}(\mathbf{s})\{\frac{\delta}{\delta\psi^{+}(\mathbf{s})}\psi(\mathbf{s})W\}\\
 & = & \delta_{C}(\mathbf{s},\mathbf{s})\psi^{+}(\mathbf{s})\psi(\mathbf{s})W+\psi^{+}(\mathbf{s})\delta_{C}(\mathbf{s},\mathbf{s})\psi(\mathbf{s})W\}+\psi^{+}(\mathbf{s})\psi^{+}(\mathbf{s})\{\frac{\delta}{\delta\psi^{+}(\mathbf{s})}\psi(\mathbf{s})W\}\\
 &  & \psi^{+}(\mathbf{s})\psi^{+}(\mathbf{s})\{\frac{\delta}{\delta\psi^{+}(\mathbf{s})}\psi(\mathbf{s})W\}\\
 & = & \frac{\delta}{\delta\psi^{+}(\mathbf{s})}\{\psi^{+}(\mathbf{s})\psi^{+}(\mathbf{s})\psi(\mathbf{s})W[\psi,\psi^{+}]\}-2\delta_{C}(\mathbf{s},\mathbf{s})\psi^{+}(\mathbf{s})\psi(\mathbf{s})W[\psi,\psi^{+}]\end{eqnarray*}

T4\begin{eqnarray*}
 &  & \frac{\delta}{\delta\psi^{+}(\mathbf{s})}\frac{\delta}{\delta\psi^{+}(\mathbf{s})}\{\psi^{+}(\mathbf{s})\psi^{+}(\mathbf{s})W\}\\
 & = & \frac{\delta}{\delta\psi^{+}(\mathbf{s})}\left(\{\frac{\delta}{\delta\psi^{+}(\mathbf{s})}\psi^{+}(\mathbf{s})\}\psi^{+}(\mathbf{s})W+\psi^{+}(\mathbf{s})\{\frac{\delta}{\delta\psi^{+}(\mathbf{s})}\psi^{+}(\mathbf{s})\}W\right)\\
 &  & +\frac{\delta}{\delta\psi^{+}(\mathbf{s})}\left(\psi^{+}(\mathbf{s})\psi^{+}(\mathbf{s})\{\frac{\delta}{\delta\psi^{+}(\mathbf{s})}W\}\right)\\
 & = & \frac{\delta}{\delta\psi^{+}(\mathbf{s})}\left(2\delta_{C}(\mathbf{s},\mathbf{s})\psi^{+}(\mathbf{s})W+\psi^{+}(\mathbf{s})\psi^{+}(\mathbf{s})\{\frac{\delta}{\delta\psi^{+}(\mathbf{s})}W\}\right)\\
 & = & \frac{\delta}{\delta\psi^{+}(\mathbf{s})}\{2\delta_{C}(\mathbf{s},\mathbf{s})\psi^{+}(\mathbf{s})W\}+\\
 &  & +\{\frac{\delta}{\delta\psi^{+}(\mathbf{s})}\psi^{+}(\mathbf{s})\}\psi^{+}(\mathbf{s})\{\frac{\delta}{\delta\psi^{+}(\mathbf{s})}W\}+\psi^{+}(\mathbf{s})\{\frac{\delta}{\delta\psi^{+}(\mathbf{s})}\psi^{+}(\mathbf{s})\}\{\frac{\delta}{\delta\psi^{+}(\mathbf{s})}W\}\\
 &  & +\psi^{+}(\mathbf{s})\psi^{+}(\mathbf{s})\{\frac{\delta}{\delta\psi^{+}(\mathbf{s})}\frac{\delta}{\delta\psi^{+}(\mathbf{s})}W\}\\
 & = & \frac{\delta}{\delta\psi^{+}(\mathbf{s})}\{2\delta_{C}(\mathbf{s},\mathbf{s})\psi^{+}(\mathbf{s})W\}+\\
 &  & +2\delta_{C}(\mathbf{s},\mathbf{s})\psi^{+}(\mathbf{s})\{\frac{\delta}{\delta\psi^{+}(\mathbf{s})}W\}+\psi^{+}(\mathbf{s})\psi^{+}(\mathbf{s})\{\frac{\delta}{\delta\psi^{+}(\mathbf{s})}\frac{\delta}{\delta\psi^{+}(\mathbf{s})}W\}\\
 & = & \frac{\delta}{\delta\psi^{+}(\mathbf{s})}\{2\delta_{C}(\mathbf{s},\mathbf{s})\psi^{+}(\mathbf{s})W\}+\psi^{+}(\mathbf{s})\psi^{+}(\mathbf{s})\{\frac{\delta}{\delta\psi^{+}(\mathbf{s})}\frac{\delta}{\delta\psi^{+}(\mathbf{s})}W\}\\
 &  & +\frac{\delta}{\delta\psi^{+}(\mathbf{s})}\{2\delta_{C}(\mathbf{s},\mathbf{s})\psi^{+}(\mathbf{s})W\}-2\delta_{C}(\mathbf{s},\mathbf{s})^{2}W\\
 &  & \psi^{+}(\mathbf{s})\psi^{+}(\mathbf{s})\{\frac{\delta}{\delta\psi^{+}(\mathbf{s})}\frac{\delta}{\delta\psi^{+}(\mathbf{s})}W[\psi,\psi^{+}]\}\\
 & = & \frac{\delta}{\delta\psi^{+}(\mathbf{s})}\frac{\delta}{\delta\psi^{+}(\mathbf{s})}\{\psi^{+}(\mathbf{s})\psi^{+}(\mathbf{s})W[\psi,\psi^{+}]\}-\frac{\delta}{\delta\psi^{+}(\mathbf{s})}\{2\delta_{C}(\mathbf{s},\mathbf{s})\psi^{+}(\mathbf{s})W[\psi,\psi^{+}]\}\\
 &  & -\frac{\delta}{\delta\psi^{+}(\mathbf{s})}\{2\delta_{C}(\mathbf{s},\mathbf{s})\psi^{+}(\mathbf{s})W[\psi,\psi^{+}]\}+2\delta_{C}(\mathbf{s},\mathbf{s})^{2}W[\psi,\psi^{+}]\end{eqnarray*}

T5

\begin{eqnarray*}
 &  & \frac{\delta}{\delta\psi(\mathbf{s})}\{\psi^{+}(\mathbf{s})\psi(\mathbf{s})\psi(\mathbf{s})W\}\\
 & = & \frac{\delta}{\delta\psi(\mathbf{s})}\{\psi^{+}(\mathbf{s})\}\psi(\mathbf{s})\psi(\mathbf{s})W+\psi^{+}(\mathbf{s})\{\frac{\delta}{\delta\psi(\mathbf{s})}\psi(\mathbf{s})\psi(\mathbf{s})W\}\\
 & = & \psi^{+}(\mathbf{s})\{\frac{\delta}{\delta\psi(\mathbf{s})}\psi(\mathbf{s})\psi(\mathbf{s})W\}\\
 &  & \psi^{+}(\mathbf{s})\{\frac{\delta}{\delta\psi(\mathbf{s})}\psi(\mathbf{s})\psi(\mathbf{s})W[\psi,\psi^{+}]\}\\
 & = & \frac{\delta}{\delta\psi(\mathbf{s})}\{\psi^{+}(\mathbf{s})\psi(\mathbf{s})\psi(\mathbf{s})W[\psi,\psi^{+}]\}\end{eqnarray*}

T6\begin{eqnarray*}
 &  & \frac{\delta}{\delta\psi(\mathbf{s})}\frac{\delta}{\delta\psi^{+}(\mathbf{s})}\{\psi^{+}(\mathbf{s})\psi(\mathbf{s})W\}\\
 & = & \frac{\delta}{\delta\psi(\mathbf{s})}\left\{ \{\frac{\delta}{\delta\psi^{+}(\mathbf{s})}\psi^{+}(\mathbf{s})\}\psi(\mathbf{s})W+\psi^{+}(\mathbf{s})\{\frac{\delta}{\delta\psi^{+}(\mathbf{s})}\psi(\mathbf{s})\}W\right\} \\
 &  & +\frac{\delta}{\delta\psi(\mathbf{s})}\left\{ \psi^{+}(\mathbf{s})\psi(\mathbf{s})\{\frac{\delta}{\delta\psi^{+}(\mathbf{s})}W\}\right\} \\
 & = & \frac{\delta}{\delta\psi(\mathbf{s})}\left\{ \delta_{C}(\mathbf{s},\mathbf{s})\psi(\mathbf{s})W+\psi^{+}(\mathbf{s})\psi(\mathbf{s})\{\frac{\delta}{\delta\psi^{+}(\mathbf{s})}W\}\right\} \\
 & = & \{\frac{\delta}{\delta\psi(\mathbf{s})}\delta_{C}(\mathbf{s},\mathbf{s})\psi(\mathbf{s})W\}+\{\frac{\delta}{\delta\psi(\mathbf{s})}\psi^{+}(\mathbf{s})\}\psi(\mathbf{s})\{\frac{\delta}{\delta\psi^{+}(\mathbf{s})}W\}\\
 &  & +\psi^{+}(\mathbf{s})\{\frac{\delta}{\delta\psi(\mathbf{s})}\psi(\mathbf{s})\{\frac{\delta}{\delta\psi^{+}(\mathbf{s})}W\}\}\\
 & = & \{\frac{\delta}{\delta\psi(\mathbf{s})}\delta_{C}(\mathbf{s},\mathbf{s})\psi(\mathbf{s})W\}+\psi^{+}(\mathbf{s})\{\frac{\delta}{\delta\psi(\mathbf{s})}\psi(\mathbf{s})\{\frac{\delta}{\delta\psi^{+}(\mathbf{s})}W\}\}\\
 &  & \psi^{+}(\mathbf{s})\{\frac{\delta}{\delta\psi(\mathbf{s})}\psi(\mathbf{s})\{\frac{\delta}{\delta\psi^{+}(\mathbf{s})}W[\psi,\psi^{+}]\}\}\\
 & = & \frac{\delta}{\delta\psi(\mathbf{s})}\frac{\delta}{\delta\psi^{+}(\mathbf{s})}\{\psi^{+}(\mathbf{s})\psi(\mathbf{s})W[\psi,\psi^{+}]\}-\{\frac{\delta}{\delta\psi(\mathbf{s})}\delta_{C}(\mathbf{s},\mathbf{s})\psi(\mathbf{s})W[\psi,\psi^{+}]\}\end{eqnarray*}

T7\begin{eqnarray*}
 &  & \frac{\delta}{\delta\psi(\mathbf{s})}\frac{\delta}{\delta\psi^{+}(\mathbf{s})}\{\psi^{+}(\mathbf{s})\psi(\mathbf{s})W\}\\
 & = & \frac{\delta}{\delta\psi(\mathbf{s})}\left\{ \{\frac{\delta}{\delta\psi^{+}(\mathbf{s})}\psi^{+}(\mathbf{s})\}\psi(\mathbf{s})W+\psi^{+}(\mathbf{s})\{\frac{\delta}{\delta\psi^{+}(\mathbf{s})}\psi(\mathbf{s})W\}\right\} \\
 & = & \frac{\delta}{\delta\psi(\mathbf{s})}\left\{ \delta_{C}(\mathbf{s},\mathbf{s})\psi(\mathbf{s})W+\psi^{+}(\mathbf{s})\{\frac{\delta}{\delta\psi^{+}(\mathbf{s})}\psi(\mathbf{s})W\}\right\} \\
 & = & \{\frac{\delta}{\delta\psi(\mathbf{s})}\delta_{C}(\mathbf{s},\mathbf{s})\psi(\mathbf{s})W\}+\{\frac{\delta}{\delta\psi(\mathbf{s})}\psi^{+}(\mathbf{s})\}\{\frac{\delta}{\delta\psi^{+}(\mathbf{s})}\psi(\mathbf{s})W\}\\
 &  & +\psi^{+}(\mathbf{s})\{\frac{\delta}{\delta\psi(\mathbf{s})}\{\frac{\delta}{\delta\psi^{+}(\mathbf{s})}\psi(\mathbf{s})W\}\}\\
 & = & \frac{\delta}{\delta\psi(\mathbf{s})}\{\delta_{C}(\mathbf{s},\mathbf{s})\psi(\mathbf{s})W\}+\psi^{+}(\mathbf{s})\{\frac{\delta}{\delta\psi(\mathbf{s})}\{\frac{\delta}{\delta\psi^{+}(\mathbf{s})}\psi(\mathbf{s})W\}\}\\
 &  & \psi^{+}(\mathbf{s})\{\frac{\delta}{\delta\psi(\mathbf{s})}\{\frac{\delta}{\delta\psi^{+}(\mathbf{s})}\psi(\mathbf{s})W[\psi,\psi^{+}]\}\}\\
 & = & \frac{\delta}{\delta\psi(\mathbf{s})}\frac{\delta}{\delta\psi^{+}(\mathbf{s})}\{\psi^{+}(\mathbf{s})\psi(\mathbf{s})W[\psi,\psi^{+}]\}-\frac{\delta}{\delta\psi(\mathbf{s})}\{\delta_{C}(\mathbf{s},\mathbf{s})\psi(\mathbf{s})W[\psi,\psi^{+}]\}\end{eqnarray*}

T8\begin{eqnarray*}
 &  & \frac{\delta}{\delta\psi(\mathbf{s})}\frac{\delta}{\delta\psi^{+}(\mathbf{s})}\frac{\delta}{\delta\psi^{+}(\mathbf{s})}\{\psi^{+}(\mathbf{s})W\}\\
 & = & \frac{\delta}{\delta\psi(\mathbf{s})}\frac{\delta}{\delta\psi^{+}(\mathbf{s})}\left\{ \{\frac{\delta}{\delta\psi^{+}(\mathbf{s})}\psi^{+}(\mathbf{s})\}W+\psi^{+}(\mathbf{s})\{\frac{\delta}{\delta\psi^{+}(\mathbf{s})}W\}\right\} \\
 & = & \frac{\delta}{\delta\psi(\mathbf{s})}\frac{\delta}{\delta\psi^{+}(\mathbf{s})}\left\{ \delta_{C}(\mathbf{s},\mathbf{s})W+\psi^{+}(\mathbf{s})\{\frac{\delta}{\delta\psi^{+}(\mathbf{s})}W\}\right\} \\
 & = & \frac{\delta}{\delta\psi(\mathbf{s})}\frac{\delta}{\delta\psi^{+}(\mathbf{s})}\{\delta_{C}(\mathbf{s},\mathbf{s})W\}+\frac{\delta}{\delta\psi(\mathbf{s})}\left\{ \frac{\delta}{\delta\psi^{+}(\mathbf{s})}\{\psi^{+}(\mathbf{s})\{\frac{\delta}{\delta\psi^{+}(\mathbf{s})}W\}\}\right\} \\
 & = & \frac{\delta}{\delta\psi(\mathbf{s})}\frac{\delta}{\delta\psi^{+}(\mathbf{s})}\{\delta_{C}(\mathbf{s},\mathbf{s})W\}\\
 &  & +\frac{\delta}{\delta\psi(\mathbf{s})}\left\{ \{\frac{\delta}{\delta\psi^{+}(\mathbf{s})}\psi^{+}(\mathbf{s})\}\{\frac{\delta}{\delta\psi^{+}(\mathbf{s})}W\}+\psi^{+}(\mathbf{s})\{\frac{\delta}{\delta\psi^{+}(\mathbf{s})}\{\frac{\delta}{\delta\psi^{+}(\mathbf{s})}W\}\}\right\} \\
 & = & \frac{\delta}{\delta\psi(\mathbf{s})}\frac{\delta}{\delta\psi^{+}(\mathbf{s})}\{\delta_{C}(\mathbf{s},\mathbf{s})W\}\\
 &  & +\frac{\delta}{\delta\psi(\mathbf{s})}\left\{ \delta_{C}(\mathbf{s},\mathbf{s})\{\frac{\delta}{\delta\psi^{+}(\mathbf{s})}W\}+\psi^{+}(\mathbf{s})\{\frac{\delta}{\delta\psi^{+}(\mathbf{s})}\{\frac{\delta}{\delta\psi^{+}(\mathbf{s})}W\}\}\right\} \\
 & = & \frac{\delta}{\delta\psi(\mathbf{s})}\frac{\delta}{\delta\psi^{+}(\mathbf{s})}\{\delta_{C}(\mathbf{s},\mathbf{s})W\}+\frac{\delta}{\delta\psi(\mathbf{s})}\delta_{C}(\mathbf{s},\mathbf{s})\{\frac{\delta}{\delta\psi^{+}(\mathbf{s})}W\}\\
 &  & +\{\frac{\delta}{\delta\psi(\mathbf{s})}\psi^{+}(\mathbf{s})\}\{\frac{\delta}{\delta\psi^{+}(\mathbf{s})}\{\frac{\delta}{\delta\psi^{+}(\mathbf{s})}W\}\}\\
 &  & +\psi^{+}(\mathbf{s})\{\frac{\delta}{\delta\psi(\mathbf{s})}\{\frac{\delta}{\delta\psi^{+}(\mathbf{s})}\{\frac{\delta}{\delta\psi^{+}(\mathbf{s})}W\}\}\}\\
 & = & \frac{\delta}{\delta\psi(\mathbf{s})}\frac{\delta}{\delta\psi^{+}(\mathbf{s})}\{\delta_{C}(\mathbf{s},\mathbf{s})W\}+\frac{\delta}{\delta\psi(\mathbf{s})}\{\frac{\delta}{\delta\psi^{+}(\mathbf{s})}\delta_{C}(\mathbf{s},\mathbf{s})W\}\\
 &  & +\psi^{+}(\mathbf{s})\{\frac{\delta}{\delta\psi(\mathbf{s})}\{\frac{\delta}{\delta\psi^{+}(\mathbf{s})}\{\frac{\delta}{\delta\psi^{+}(\mathbf{s})}W\}\}\}\\
 &  & \psi^{+}(\mathbf{s})\{\frac{\delta}{\delta\psi(\mathbf{s})}\{\frac{\delta}{\delta\psi^{+}(\mathbf{s})}\{\frac{\delta}{\delta\psi^{+}(\mathbf{s})}W[\psi,\psi^{+}]\}\}\}\\
 & = & \frac{\delta}{\delta\psi(\mathbf{s})}\frac{\delta}{\delta\psi^{+}(\mathbf{s})}\frac{\delta}{\delta\psi^{+}(\mathbf{s})}\{\psi^{+}(\mathbf{s})W[\psi,\psi^{+}]\}-\frac{\delta}{\delta\psi(\mathbf{s})}\frac{\delta}{\delta\psi^{+}(\mathbf{s})}\{\delta_{C}(\mathbf{s},\mathbf{s})W[\psi,\psi^{+}]\}\\
 &  & -\frac{\delta}{\delta\psi(\mathbf{s})}\{\frac{\delta}{\delta\psi^{+}(\mathbf{s})}\delta_{C}(\mathbf{s},\mathbf{s})W[\psi,\psi^{+}]\}\end{eqnarray*}

T10\begin{eqnarray*}
 &  & \frac{\delta}{\delta\psi(\mathbf{s})}\frac{\delta}{\delta\psi^{+}(\mathbf{s})}\{\psi^{+}(\mathbf{s})\psi(\mathbf{s})W\}\\
 & = & \frac{\delta}{\delta\psi(\mathbf{s})}\left\{ \{\frac{\delta}{\delta\psi^{+}(\mathbf{s})}\psi^{+}(\mathbf{s})\}\psi(\mathbf{s})W+\psi^{+}(\mathbf{s})\psi(\mathbf{s})\frac{\delta}{\delta\psi^{+}(\mathbf{s})}W\right\} \\
 & = & \frac{\delta}{\delta\psi(\mathbf{s})}\left\{ \delta_{C}(\mathbf{s},\mathbf{s})\psi(\mathbf{s})W+\psi^{+}(\mathbf{s})\psi(\mathbf{s})\frac{\delta}{\delta\psi^{+}(\mathbf{s})}W\right\} \\
 & = & \frac{\delta}{\delta\psi(\mathbf{s})}\{\delta_{C}(\mathbf{s},\mathbf{s})\psi(\mathbf{s})W\}+\frac{\delta}{\delta\psi(\mathbf{s})}\{\psi^{+}(\mathbf{s})\psi(\mathbf{s})\frac{\delta}{\delta\psi^{+}(\mathbf{s})}W\}\\
 &  & \frac{\delta}{\delta\psi(\mathbf{s})}\{\psi^{+}(\mathbf{s})\psi(\mathbf{s})\frac{\delta}{\delta\psi^{+}(\mathbf{s})}W[\psi,\psi^{+}]\}\\
 & = & \frac{\delta}{\delta\psi(\mathbf{s})}\frac{\delta}{\delta\psi^{+}(\mathbf{s})}\{\psi^{+}(\mathbf{s})\psi(\mathbf{s})W[\psi,\psi^{+}]\}-\frac{\delta}{\delta\psi(\mathbf{s})}\{\delta_{C}(\mathbf{s},\mathbf{s})\psi(\mathbf{s})W[\psi,\psi^{+}]\}\end{eqnarray*}

T11\begin{eqnarray*}
 &  & \frac{\delta}{\delta\psi(\mathbf{s})}\frac{\delta}{\delta\psi^{+}(\mathbf{s})}\{\psi^{+}(\mathbf{s})\psi(\mathbf{s})W\}\\
 & = & \frac{\delta}{\delta\psi(\mathbf{s})}\left\{ \{\frac{\delta}{\delta\psi^{+}(\mathbf{s})}\psi^{+}(\mathbf{s})\}\psi(\mathbf{s})W+\psi^{+}(\mathbf{s})\frac{\delta}{\delta\psi^{+}(\mathbf{s})}\{\psi(\mathbf{s})W\}\right\} \\
 & = & \frac{\delta}{\delta\psi(\mathbf{s})}\left\{ \delta_{C}(\mathbf{s},\mathbf{s})\psi(\mathbf{s})W+\psi^{+}(\mathbf{s})\frac{\delta}{\delta\psi^{+}(\mathbf{s})}\{\psi(\mathbf{s})W\}\right\} \\
 & = & \frac{\delta}{\delta\psi(\mathbf{s})}\{\delta_{C}(\mathbf{s},\mathbf{s})\psi(\mathbf{s})W\}+\frac{\delta}{\delta\psi(\mathbf{s})}\{\psi^{+}(\mathbf{s})\frac{\delta}{\delta\psi^{+}(\mathbf{s})}\{\psi(\mathbf{s})W\}\}\\
 &  & \frac{\delta}{\delta\psi(\mathbf{s})}\{\psi^{+}(\mathbf{s})\frac{\delta}{\delta\psi^{+}(\mathbf{s})}\{\psi(\mathbf{s})W[\psi,\psi^{+}]\}\}\\
 & = & \frac{\delta}{\delta\psi(\mathbf{s})}\frac{\delta}{\delta\psi^{+}(\mathbf{s})}\{\psi^{+}(\mathbf{s})\psi(\mathbf{s})W[\psi,\psi^{+}]\}-\frac{\delta}{\delta\psi(\mathbf{s})}\{\delta_{C}(\mathbf{s},\mathbf{s})\psi(\mathbf{s})W[\psi,\psi^{+}]\}\end{eqnarray*}

T12\begin{eqnarray*}
 &  & \frac{\delta}{\delta\psi(\mathbf{s})}\frac{\delta}{\delta\psi^{+}(\mathbf{s})}\frac{\delta}{\delta\psi^{+}(\mathbf{s})}\{\psi^{+}(\mathbf{s})W\}\\
 & = & \frac{\delta}{\delta\psi(\mathbf{s})}\frac{\delta}{\delta\psi^{+}(\mathbf{s})}\left\{ \{\frac{\delta}{\delta\psi^{+}(\mathbf{s})}\psi^{+}(\mathbf{s})\}W\}+\psi^{+}(\mathbf{s})\{\frac{\delta}{\delta\psi^{+}(\mathbf{s})}W\}\right\} \\
 & = & \frac{\delta}{\delta\psi(\mathbf{s})}\frac{\delta}{\delta\psi^{+}(\mathbf{s})}\left\{ \delta_{C}(\mathbf{s},\mathbf{s})W\}+\psi^{+}(\mathbf{s})\{\frac{\delta}{\delta\psi^{+}(\mathbf{s})}W\}\right\} \\
 & = & \frac{\delta}{\delta\psi(\mathbf{s})}\frac{\delta}{\delta\psi^{+}(\mathbf{s})}\delta_{C}(\mathbf{s},\mathbf{s})W\}\\
 &  & +\frac{\delta}{\delta\psi(\mathbf{s})}\left\{ \{\frac{\delta}{\delta\psi^{+}(\mathbf{s})}\psi^{+}(\mathbf{s})\}\{\frac{\delta}{\delta\psi^{+}(\mathbf{s})}W\}+\psi^{+}(\mathbf{s})\{\frac{\delta}{\delta\psi^{+}(\mathbf{s})}\{\frac{\delta}{\delta\psi^{+}(\mathbf{s})}W\}\}\right\} \\
 & = & \frac{\delta}{\delta\psi(\mathbf{s})}\frac{\delta}{\delta\psi^{+}(\mathbf{s})}\delta_{C}(\mathbf{s},\mathbf{s})W\}\\
 &  & +\frac{\delta}{\delta\psi(\mathbf{s})}\left\{ \delta_{C}(\mathbf{s},\mathbf{s})\{\frac{\delta}{\delta\psi^{+}(\mathbf{s})}W\}+\psi^{+}(\mathbf{s})\{\frac{\delta}{\delta\psi^{+}(\mathbf{s})}\{\frac{\delta}{\delta\psi^{+}(\mathbf{s})}W\}\}\right\} \\
 & = & \frac{\delta}{\delta\psi(\mathbf{s})}\frac{\delta}{\delta\psi^{+}(\mathbf{s})}2\delta_{C}(\mathbf{s},\mathbf{s})W\}+\frac{\delta}{\delta\psi(\mathbf{s})}\{\psi^{+}(\mathbf{s})\{\frac{\delta}{\delta\psi^{+}(\mathbf{s})}\{\frac{\delta}{\delta\psi^{+}(\mathbf{s})}W\}\}\}\\
 &  & \frac{\delta}{\delta\psi(\mathbf{s})}\{\psi^{+}(\mathbf{s})\{\frac{\delta}{\delta\psi^{+}(\mathbf{s})}\{\frac{\delta}{\delta\psi^{+}(\mathbf{s})}W[\psi,\psi^{+}]\}\}\}\\
 & = & \frac{\delta}{\delta\psi(\mathbf{s})}\frac{\delta}{\delta\psi^{+}(\mathbf{s})}\frac{\delta}{\delta\psi^{+}(\mathbf{s})}\{\psi^{+}(\mathbf{s})W[\psi,\psi^{+}]\}-\frac{\delta}{\delta\psi(\mathbf{s})}\frac{\delta}{\delta\psi^{+}(\mathbf{s})}2\delta_{C}(\mathbf{s},\mathbf{s})W[\psi,\psi^{+}]\}\end{eqnarray*}

T14\begin{eqnarray*}
 &  & \frac{\delta}{\delta\psi(\mathbf{s})}\frac{\delta}{\delta\psi(\mathbf{s})}\frac{\delta}{\delta\psi^{+}(\mathbf{s})}\{\psi(\mathbf{s})W\}\\
 & = & \frac{\delta}{\delta\psi(\mathbf{s})}\frac{\delta}{\delta\psi(\mathbf{s})}\left\{ \frac{\delta}{\delta\psi^{+}(\mathbf{s})}\{\psi(\mathbf{s})W\right\} \\
 & = & \frac{\delta}{\delta\psi(\mathbf{s})}\frac{\delta}{\delta\psi(\mathbf{s})}\left\{ \psi(\mathbf{s})\frac{\delta}{\delta\psi^{+}(\mathbf{s})}W\right\} \\
 &  & \frac{\delta}{\delta\psi(\mathbf{s})}\frac{\delta}{\delta\psi(\mathbf{s})}\{\psi(\mathbf{s})\frac{\delta}{\delta\psi^{+}(\mathbf{s})}W[\psi,\psi^{+}]\}\\
 & = & \frac{\delta}{\delta\psi(\mathbf{s})}\frac{\delta}{\delta\psi(\mathbf{s})}\frac{\delta}{\delta\psi^{+}(\mathbf{s})}\{\psi(\mathbf{s})W[\psi,\psi^{+}]\}\end{eqnarray*}

Now if \begin{equation}
\widehat{\rho}\rightarrow\widehat{\rho}\widehat{U}=\frac{g}{2}\int d\mathbf{s}\widehat{\rho}\widehat{\Psi}(\mathbf{s})^{\dagger}\widehat{\Psi}(\mathbf{s})^{\dagger}\widehat{\Psi}(\mathbf{s})\widehat{\Psi}(\mathbf{s})\end{equation}
then\begin{eqnarray}
 &  & W[\psi(\mathbf{r}),\psi^{+}(\mathbf{r})]\nonumber \\
 & \rightarrow & \frac{g}{2}\int d\mathbf{s}\left(\psi(\mathbf{s})-\frac{1}{2}\frac{\delta}{\delta\psi^{+}(\mathbf{s})}\right)\left(\psi(\mathbf{s})-\frac{1}{2}\frac{\delta}{\delta\psi^{+}(\mathbf{s})}\right)\left(\psi^{+}(\mathbf{s})+\frac{1}{2}\frac{\delta}{\delta\psi(\mathbf{s})}\right)\left(\psi^{+}(\mathbf{s})+\frac{1}{2}\frac{\delta}{\delta\psi(\mathbf{s})}\right)W\nonumber \\
 &  & \,\end{eqnarray}
After expanding we obtain

\begin{eqnarray}
 &  & W[\psi(\mathbf{r}),\psi^{+}(\mathbf{r})]\nonumber \\
 & \rightarrow & \frac{g}{2}\int d\mathbf{s}\left\{ \psi(\mathbf{s})\psi(\mathbf{s})\psi^{+}(\mathbf{s})\psi^{+}(\mathbf{s})\right\} W[\psi,\psi^{+}]+\frac{g}{2}\int d\mathbf{s}\frac{1}{2}\left\{ \psi(\mathbf{s})\psi(\mathbf{s})\psi^{+}(\mathbf{s})\frac{\delta}{\delta\psi(\mathbf{s})}\right\} W\nonumber \\
 &  & +\frac{g}{2}\int d\mathbf{s}\frac{1}{2}\left\{ \psi(\mathbf{s})\psi(\mathbf{s})\frac{\delta}{\delta\psi(\mathbf{s})}\psi^{+}(\mathbf{s})\right\} W+\frac{g}{2}\int d\mathbf{s}\frac{1}{4}\left\{ \psi(\mathbf{s})\psi(\mathbf{s})\frac{\delta}{\delta\psi(\mathbf{s})}\frac{\delta}{\delta\psi(\mathbf{s})}\right\} W\nonumber \\
 &  & -\frac{g}{2}\int d\mathbf{s}\frac{1}{2}\left\{ \psi(\mathbf{s})\frac{\delta}{\delta\psi^{+}(\mathbf{s})}\psi^{+}(\mathbf{s})\psi^{+}(\mathbf{s})\right\} W-\frac{g}{2}\int d\mathbf{s}\frac{1}{4}\left\{ \psi(\mathbf{s})\frac{\delta}{\delta\psi^{+}(\mathbf{s})}\psi^{+}(\mathbf{s})\frac{\delta}{\delta\psi(\mathbf{s})}\right\} W\nonumber \\
 &  & -\frac{g}{2}\int d\mathbf{s}\frac{1}{4}\left\{ \psi(\mathbf{s})\frac{\delta}{\delta\psi^{+}(\mathbf{s})}\frac{\delta}{\delta\psi(\mathbf{s})}\psi^{+}(\mathbf{s})\right\} W-\frac{g}{2}\int d\mathbf{s}\frac{1}{8}\left\{ \psi(\mathbf{s})\frac{\delta}{\delta\psi^{+}(\mathbf{s})}\frac{\delta}{\delta\psi(\mathbf{s})}\frac{\delta}{\delta\psi(\mathbf{s})}\right\} W\nonumber \\
 &  & -\frac{g}{2}\int d\mathbf{s}\frac{1}{2}\left\{ \frac{\delta}{\delta\psi^{+}(\mathbf{s})}\psi(\mathbf{s})\psi^{+}(\mathbf{s})\psi^{+}(\mathbf{s})\right\} W-\frac{g}{2}\int d\mathbf{s}\frac{1}{4}\left\{ \frac{\delta}{\delta\psi^{+}(\mathbf{s})}\psi(\mathbf{s})\psi^{+}(\mathbf{s})\frac{\delta}{\delta\psi(\mathbf{s})}\right\} W\nonumber \\
 &  & -\frac{g}{2}\int d\mathbf{s}\frac{1}{4}\left\{ \frac{\delta}{\delta\psi^{+}(\mathbf{s})}\psi(\mathbf{s})\frac{\delta}{\delta\psi(\mathbf{s})}\psi^{+}(\mathbf{s}_{1})\right\} W-\frac{g}{2}\int d\mathbf{s}\frac{1}{8}\left\{ \frac{\delta}{\delta\psi^{+}(\mathbf{s})}\psi(\mathbf{s})\frac{\delta}{\delta\psi(\mathbf{s})}\frac{\delta}{\delta\psi(\mathbf{s})}\right\} W\nonumber \\
 &  & +\frac{g}{2}\int d\mathbf{s}\frac{1}{4}\left\{ \frac{\delta}{\delta\psi^{+}(\mathbf{s})}\frac{\delta}{\delta\psi^{+}(\mathbf{s})}\psi^{+}(\mathbf{s})\psi^{+}(\mathbf{s})\right\} W+\frac{g}{2}\int d\mathbf{s}\frac{1}{8}\left\{ \frac{\delta}{\delta\psi^{+}(\mathbf{s})}\frac{\delta}{\delta\psi^{+}(\mathbf{s})}\psi^{+}(\mathbf{s})\frac{\delta}{\delta\psi(\mathbf{s})}\right\} W\nonumber \\
 &  & +\frac{g}{2}\int d\mathbf{s}\frac{1}{8}\left\{ \frac{\delta}{\delta\psi^{+}(\mathbf{s})}\frac{\delta}{\delta\psi^{+}(\mathbf{s})}\frac{\delta}{\delta\psi(\mathbf{s})}\psi^{+}(\mathbf{s})\right\} W+\frac{g}{2}\int d\mathbf{s}\frac{1}{16}\left\{ \frac{\delta}{\delta\psi^{+}(\mathbf{s})}\frac{\delta}{\delta\psi^{+}(\mathbf{s})}\frac{\delta}{\delta\psi(\mathbf{s})}\frac{\delta}{\delta\psi(\mathbf{s})}\right\} W\nonumber \\
 &  & \,\end{eqnarray}

We can now use the product rule for functional derivatives (\ref{eq:ProdRuleFnalDeriv-1})
together with (\ref{Eq.FuncDerivativeRule1-1}) and (\ref{Eq.FuncDerivativeRule2-1})
to place all the derivatives on the left of the expression. However
the results can more easily be obtained by noticing that the $\widehat{\rho}\widehat{U}$
is the same as the $\widehat{U}\widehat{\rho}$ if we interchange
$\psi(\mathbf{s})$ and $\psi^{+}(\mathbf{s})$ everywhere. Hence

\begin{eqnarray}
 &  & W[\psi(\mathbf{r}),\psi^{+}(\mathbf{r})]\nonumber \\
 & \rightarrow & \frac{g}{2}\int d\mathbf{s}\left\{ \psi(\mathbf{s})\psi(\mathbf{s})\psi^{+}(\mathbf{s})\psi^{+}(\mathbf{s})\right\} W\qquad T1\nonumber \\
 &  & +\frac{g}{2}\int d\mathbf{s}\frac{1}{2}\left\{ \frac{\delta}{\delta\psi(\mathbf{s})}\psi(\mathbf{s})\psi(\mathbf{s})\psi^{+}(\mathbf{s})-2\delta_{C}(\mathbf{s},\mathbf{s})\psi(\mathbf{s})\psi^{+}(\mathbf{s})\right\} W\qquad T22,T21\nonumber \\
 &  & +\frac{g}{2}\int d\mathbf{s}\frac{1}{2}\left\{ \frac{\delta}{\delta\psi(\mathbf{s})}\psi(\mathbf{s})\psi(\mathbf{s})\psi^{+}(\mathbf{s})-2\delta_{C}(\mathbf{s},\mathbf{s})\psi(\mathbf{s})\psi^{+}(\mathbf{s})\right\} W\qquad T32,T31\nonumber \\
 &  & +\frac{g}{2}\int d\mathbf{s}\frac{1}{4}\left\{ \frac{\delta}{\delta\psi(\mathbf{s})}\frac{\delta}{\delta\psi(\mathbf{s})}\psi(\mathbf{s})\psi(\mathbf{s})-\frac{\delta}{\delta\psi(\mathbf{s})}4\delta_{C}(\mathbf{s},\mathbf{s})\psi(\mathbf{s})+2\delta_{C}(\mathbf{s},\mathbf{s})^{2}\right\} W\qquad T43,T42,T41\nonumber \\
 &  & -\frac{g}{2}\int d\mathbf{s}\frac{1}{2}\left\{ \frac{\delta}{\delta\psi^{+}(\mathbf{s})}\psi(\mathbf{s})\psi^{+}(\mathbf{s})\psi^{+}(\mathbf{s})\right\} W\qquad T5\nonumber \\
 &  & -\frac{g}{2}\int d\mathbf{s}\frac{1}{4}\left\{ \frac{\delta}{\delta\psi^{+}(\mathbf{s})}\frac{\delta}{\delta\psi(\mathbf{s})}\psi(\mathbf{s})\psi^{+}(\mathbf{s})-\frac{\delta}{\delta\psi^{+}(\mathbf{s})}\delta_{C}(\mathbf{s},\mathbf{s})\psi^{+}(\mathbf{s})\right\} W\qquad T62,T61\nonumber \\
 &  & -\frac{g}{2}\int d\mathbf{s}\frac{1}{4}\left\{ \frac{\delta}{\delta\psi^{+}(\mathbf{s})}\frac{\delta}{\delta\psi(\mathbf{s})}\psi(\mathbf{s})\psi^{+}(\mathbf{s})-\frac{\delta}{\delta\psi^{+}(\mathbf{s})}\delta_{C}(\mathbf{s},\mathbf{s})\psi^{+}(\mathbf{s})\right\} W\qquad T72\nonumber \\
 &  & -\frac{g}{2}\int d\mathbf{s}\frac{1}{8}\left\{ \frac{\delta}{\delta\psi^{+}(\mathbf{s})}\frac{\delta}{\delta\psi(\mathbf{s})}\frac{\delta}{\delta\psi(\mathbf{s})}\psi(\mathbf{s})-\frac{\delta}{\delta\psi^{+}(\mathbf{s})}\frac{\delta}{\delta\psi(\mathbf{s})}2\delta_{C}(\mathbf{s},\mathbf{s})\right\} W\qquad T82,T81\nonumber \\
 &  & -\frac{g}{2}\int d\mathbf{s}\frac{1}{2}\left\{ \frac{\delta}{\delta\psi^{+}(\mathbf{s})}\psi(\mathbf{s})\psi^{+}(\mathbf{s})\psi^{+}(\mathbf{s})\right\} W\qquad T9\nonumber \\
 &  & -\frac{g}{2}\int d\mathbf{s}\frac{1}{4}\left\{ \frac{\delta}{\delta\psi^{+}(\mathbf{s})}\frac{\delta}{\delta\psi(\mathbf{s})}\psi(\mathbf{s})\psi^{+}(\mathbf{s})-\frac{\delta}{\delta\psi^{+}(\mathbf{s})}\delta_{C}(\mathbf{s},\mathbf{s})\psi^{+}(\mathbf{s})\right\} W\qquad T10.2,T10.1\nonumber \\
 &  & -\frac{g}{2}\int d\mathbf{s}\frac{1}{4}\left\{ \frac{\delta}{\delta\psi^{+}(\mathbf{s})}\frac{\delta}{\delta\psi(\mathbf{s})}\psi(\mathbf{s})\psi^{+}(\mathbf{s})-\frac{\delta}{\delta\psi^{+}(\mathbf{s})}\delta_{C}(\mathbf{s},\mathbf{s})\psi^{+}(\mathbf{s})\right\} W\qquad T11.2,T11.1\nonumber \\
 &  & -\frac{g}{2}\int d\mathbf{s}\frac{1}{8}\left\{ \frac{\delta}{\delta\psi^{+}(\mathbf{s})}\frac{\delta}{\delta\psi(\mathbf{s})}\frac{\delta}{\delta\psi(\mathbf{s})}\psi(\mathbf{s})-\frac{\delta}{\delta\psi^{+}(\mathbf{s})}\frac{\delta}{\delta\psi(\mathbf{s})}2\delta_{C}(\mathbf{s},\mathbf{s})\right\} W\qquad T12.2,T12.1\nonumber \\
 &  & +\frac{g}{2}\int d\mathbf{s}\frac{1}{4}\left\{ \frac{\delta}{\delta\psi^{+}(\mathbf{s})}\frac{\delta}{\delta\psi^{+}(\mathbf{s})}\psi^{+}(\mathbf{s})\psi^{+}(\mathbf{s})\right\} W\qquad T13\nonumber \\
 &  & +\frac{g}{2}\int d\mathbf{s}\frac{1}{8}\left\{ \frac{\delta}{\delta\psi^{+}(\mathbf{s})}\frac{\delta}{\delta\psi^{+}(\mathbf{s})}\frac{\delta}{\delta\psi(\mathbf{s})}\psi^{+}(\mathbf{s})\right\} W\qquad T14\nonumber \\
 &  & +\frac{g}{2}\int d\mathbf{s}\frac{1}{8}\left\{ \frac{\delta}{\delta\psi^{+}(\mathbf{s})}\frac{\delta}{\delta\psi^{+}(\mathbf{s})}\frac{\delta}{\delta\psi(\mathbf{s})}\psi^{+}(\mathbf{s})\right\} W\qquad T15\nonumber \\
 &  & +\frac{g}{2}\int d\mathbf{s}\frac{1}{16}\left\{ \frac{\delta}{\delta\psi^{+}(\mathbf{s})}\frac{\delta}{\delta\psi^{+}(\mathbf{s})}\frac{\delta}{\delta\psi(\mathbf{s})}\frac{\delta}{\delta\psi(\mathbf{s})}\right\} W\qquad T16\nonumber \\
 &  & \,\end{eqnarray}

We now combine the contributions so that when \begin{equation}
\widehat{\rho}\rightarrow\lbrack\widehat{U},\widehat{\rho}]=[\frac{g}{2}\int d\mathbf{s}\,\,\widehat{\Psi}(\mathbf{s})^{\dagger}\widehat{\Psi}(\mathbf{s})^{\dagger}\widehat{\Psi}(\mathbf{s})\widehat{\Psi}(\mathbf{s}),\widehat{\rho}]\end{equation}
then

\begin{align}
W[\psi(\mathbf{r}),\psi^{+}(\mathbf{r})] & \rightarrow W^{0}+W^{1}+W^{2}+W^{3}+W^{4}\nonumber \\
 & \,\end{align}
where the terms are listed via the order of derivatives that occur\begin{eqnarray}
 &  & W^{0}\nonumber \\
 & = & \frac{g}{2}\int d\mathbf{s}\left\{ \psi^{+}(\mathbf{s})\psi^{+}(\mathbf{s})\psi(\mathbf{s})\psi(\mathbf{s})\right\} W-\frac{g}{2}\int d\mathbf{s}\left\{ \psi(\mathbf{s})\psi(\mathbf{s})\psi^{+}(\mathbf{s})\psi^{+}(\mathbf{s})\right\} W\nonumber \\
 &  & +\frac{g}{2}\int d\mathbf{s}\frac{1}{2}\left\{ -2\delta_{C}(\mathbf{s},\mathbf{s})\psi^{+}(\mathbf{s})\psi(\mathbf{s})\right\} W-\frac{g}{2}\int d\mathbf{s}\frac{1}{2}\left\{ -2\delta_{C}(\mathbf{s},\mathbf{s})\psi(\mathbf{s})\psi^{+}(\mathbf{s})\right\} W\nonumber \\
 &  & +\frac{g}{2}\int d\mathbf{s}\frac{1}{2}\left\{ -2\delta_{C}(\mathbf{s},\mathbf{s})\psi^{+}(\mathbf{s})\psi(\mathbf{s})\right\} W-\frac{g}{2}\int d\mathbf{s}\frac{1}{2}\left\{ -2\delta_{C}(\mathbf{s},\mathbf{s})\psi(\mathbf{s})\psi^{+}(\mathbf{s})\right\} W\nonumber \\
 &  & +\frac{g}{2}\int d\mathbf{s}\frac{1}{4}\left\{ +2\delta_{C}(\mathbf{s},\mathbf{s})^{2}\right\} W-\frac{g}{2}\int d\mathbf{s}\frac{1}{4}\left\{ +2\delta_{C}(\mathbf{s},\mathbf{s})^{2}\right\} W\nonumber \\
 & = & 0\nonumber \\
 &  & \,\end{eqnarray}
\begin{eqnarray}
 &  & W^{1}\nonumber \\
 & = & +\frac{g}{2}\int d\mathbf{s}\frac{1}{2}\left\{ \frac{\delta}{\delta\psi^{+}(\mathbf{s})}\psi^{+}(\mathbf{s})\psi^{+}(\mathbf{s})\psi(\mathbf{s})\right\} W-\frac{g}{2}\int d\mathbf{s}\frac{1}{2}\left\{ \frac{\delta}{\delta\psi(\mathbf{s})}\psi(\mathbf{s})\psi(\mathbf{s})\psi^{+}(\mathbf{s})\right\} W\nonumber \\
 &  & +\frac{g}{2}\int d\mathbf{s}\frac{1}{2}\left\{ \frac{\delta}{\delta\psi^{+}(\mathbf{s})}\psi^{+}(\mathbf{s})\psi^{+}(\mathbf{s})\psi(\mathbf{s})\right\} W-\frac{g}{2}\int d\mathbf{s}\frac{1}{2}\left\{ \frac{\delta}{\delta\psi(\mathbf{s})}\psi(\mathbf{s})\psi(\mathbf{s})\psi^{+}(\mathbf{s})\right\} W\nonumber \\
 &  & +\frac{g}{2}\int d\mathbf{s}\frac{1}{4}\left\{ -\frac{\delta}{\delta\psi^{+}(\mathbf{s})}4\delta_{C}(\mathbf{s},\mathbf{s})\psi^{+}(\mathbf{s})\right\} W-\frac{g}{2}\int d\mathbf{s}\frac{1}{4}\left\{ -\frac{\delta}{\delta\psi(\mathbf{s})}4\delta_{C}(\mathbf{s},\mathbf{s})\psi(\mathbf{s})\right\} W\nonumber \\
 &  & -\frac{g}{2}\int d\mathbf{s}\frac{1}{2}\left\{ \frac{\delta}{\delta\psi(\mathbf{s})}\psi^{+}(\mathbf{s})\psi(\mathbf{s})\psi(\mathbf{s})\right\} W+\frac{g}{2}\int d\mathbf{s}\frac{1}{2}\left\{ \frac{\delta}{\delta\psi^{+}(\mathbf{s})}\psi(\mathbf{s})\psi^{+}(\mathbf{s})\psi^{+}(\mathbf{s})\right\} W\nonumber \\
 &  & -\frac{g}{2}\int d\mathbf{s}\frac{1}{4}\left\{ -\frac{\delta}{\delta\psi(\mathbf{s})}\delta_{C}(\mathbf{s},\mathbf{s})\psi(\mathbf{s})\right\} W+\frac{g}{2}\int d\mathbf{s}\frac{1}{4}\left\{ -\frac{\delta}{\delta\psi^{+}(\mathbf{s})}\delta_{C}(\mathbf{s},\mathbf{s})\psi^{+}(\mathbf{s})\right\} W\nonumber \\
 &  & -\frac{g}{2}\int d\mathbf{s}\frac{1}{4}\left\{ -\frac{\delta}{\delta\psi(\mathbf{s})}\delta_{C}(\mathbf{s},\mathbf{s})\psi(\mathbf{s})\right\} W+\frac{g}{2}\int d\mathbf{s}\frac{1}{4}\left\{ -\frac{\delta}{\delta\psi^{+}(\mathbf{s})}\delta_{C}(\mathbf{s},\mathbf{s})\psi^{+}(\mathbf{s})\right\} W\nonumber \\
 &  & -\frac{g}{2}\int d\mathbf{s}\frac{1}{2}\left\{ \frac{\delta}{\delta\psi(\mathbf{s})}\psi^{+}(\mathbf{s})\psi(\mathbf{s})\psi(\mathbf{s})\right\} W+\frac{g}{2}\int d\mathbf{s}\frac{1}{2}\left\{ \frac{\delta}{\delta\psi^{+}(\mathbf{s})}\psi(\mathbf{s})\psi^{+}(\mathbf{s})\psi^{+}(\mathbf{s})\right\} W\nonumber \\
 &  & -\frac{g}{2}\int d\mathbf{s}\frac{1}{4}\left\{ -\frac{\delta}{\delta\psi(\mathbf{s})}\delta_{C}(\mathbf{s},\mathbf{s})\psi(\mathbf{s})\right\} W+\frac{g}{2}\int d\mathbf{s}\frac{1}{4}\left\{ -\frac{\delta}{\delta\psi^{+}(\mathbf{s})}\delta_{C}(\mathbf{s},\mathbf{s})\psi^{+}(\mathbf{s})\right\} W\nonumber \\
 &  & -\frac{g}{2}\int d\mathbf{s}\frac{1}{4}\left\{ -\frac{\delta}{\delta\psi(\mathbf{s})}\delta_{C}(\mathbf{s},\mathbf{s})\psi(\mathbf{s})\right\} W+\frac{g}{2}\int d\mathbf{s}\frac{1}{4}\left\{ -\frac{\delta}{\delta\psi^{+}(\mathbf{s})}\delta_{C}(\mathbf{s},\mathbf{s})\psi^{+}(\mathbf{s})\right\} W\nonumber \\
 & = & -g\int d\mathbf{s}\frac{\delta}{\delta\psi(\mathbf{s})}\left\{ (\psi^{+}(\mathbf{s})\psi(\mathbf{s})-\delta_{C}(\mathbf{s},\mathbf{s}))\psi(\mathbf{s})\right\} W[\psi,\psi^{+}]\nonumber \\
 &  & +g\int d\mathbf{s}\frac{\delta}{\delta\psi^{+}(\mathbf{s})}\left\{ (\psi^{+}(\mathbf{s})\psi(\mathbf{s})-\delta_{C}(\mathbf{s},\mathbf{s}))\psi^{+}(\mathbf{s})\right\} W[\psi,\psi^{+}]\nonumber \\
 &  & \,\end{eqnarray}
\begin{eqnarray}
 &  & W^{2}\nonumber \\
 & = & \frac{g}{2}\int d\mathbf{s}\frac{1}{4}\left\{ \frac{\delta}{\delta\psi^{+}(\mathbf{s})}\frac{\delta}{\delta\psi^{+}(\mathbf{s})}\psi^{+}(\mathbf{s})\psi^{+}(\mathbf{s})\right\} W-\frac{g}{2}\int d\mathbf{s}\frac{1}{4}\left\{ \frac{\delta}{\delta\psi(\mathbf{s})}\frac{\delta}{\delta\psi(\mathbf{s})}\psi(\mathbf{s})\psi(\mathbf{s})\right\} W\nonumber \\
 &  & -\frac{g}{2}\int d\mathbf{s}\frac{1}{4}\left\{ \frac{\delta}{\delta\psi(\mathbf{s})}\frac{\delta}{\delta\psi^{+}(\mathbf{s})}\psi^{+}(\mathbf{s})\psi(\mathbf{s})\right\} W+\frac{g}{2}\int d\mathbf{s}\frac{1}{4}\left\{ \frac{\delta}{\delta\psi^{+}(\mathbf{s})}\frac{\delta}{\delta\psi(\mathbf{s})}\psi(\mathbf{s})\psi^{+}(\mathbf{s})\right\} W\nonumber \\
 &  & -\frac{g}{2}\int d\mathbf{s}\frac{1}{4}\left\{ \frac{\delta}{\delta\psi(\mathbf{s})}\frac{\delta}{\delta\psi^{+}(\mathbf{s})}\psi^{+}(\mathbf{s})\psi(\mathbf{s})\right\} W+\frac{g}{2}\int d\mathbf{s}\frac{1}{4}\left\{ \frac{\delta}{\delta\psi^{+}(\mathbf{s})}\frac{\delta}{\delta\psi(\mathbf{s})}\psi(\mathbf{s})\psi^{+}(\mathbf{s})\right\} W\nonumber \\
 &  & -\frac{g}{2}\int d\mathbf{s}\frac{1}{8}\left\{ -\frac{\delta}{\delta\psi(\mathbf{s})}\frac{\delta}{\delta\psi^{+}(\mathbf{s})}2\delta_{C}(\mathbf{s},\mathbf{s})\right\} W+\frac{g}{2}\int d\mathbf{s}\frac{1}{8}\left\{ -\frac{\delta}{\delta\psi^{+}(\mathbf{s})}\frac{\delta}{\delta\psi(\mathbf{s})}2\delta_{C}(\mathbf{s},\mathbf{s})\right\} W\nonumber \\
 &  & -\frac{g}{2}\int d\mathbf{s}\frac{1}{4}\left\{ \frac{\delta}{\delta\psi(\mathbf{s})}\frac{\delta}{\delta\psi^{+}(\mathbf{s})}\psi^{+}(\mathbf{s})\psi(\mathbf{s})\right\} W+\frac{g}{2}\int d\mathbf{s}\frac{1}{4}\left\{ \frac{\delta}{\delta\psi^{+}(\mathbf{s})}\frac{\delta}{\delta\psi(\mathbf{s})}\psi(\mathbf{s})\psi^{+}(\mathbf{s})\right\} W\nonumber \\
 &  & -\frac{g}{2}\int d\mathbf{s}\frac{1}{4}\left\{ \frac{\delta}{\delta\psi(\mathbf{s})}\frac{\delta}{\delta\psi^{+}(\mathbf{s})}\psi^{+}(\mathbf{s})\psi(\mathbf{s})\right\} W+\frac{g}{2}\int d\mathbf{s}\frac{1}{4}\left\{ \frac{\delta}{\delta\psi^{+}(\mathbf{s})}\frac{\delta}{\delta\psi(\mathbf{s})}\psi(\mathbf{s})\psi^{+}(\mathbf{s})\right\} W\nonumber \\
 &  & -\frac{g}{2}\int d\mathbf{s}\frac{1}{8}\left\{ -\frac{\delta}{\delta\psi(\mathbf{s})}\frac{\delta}{\delta\psi^{+}(\mathbf{s})}2\delta_{C}(\mathbf{s},\mathbf{s})\right\} W+\frac{g}{2}\int d\mathbf{s}\frac{1}{8}\left\{ -\frac{\delta}{\delta\psi^{+}(\mathbf{s})}\frac{\delta}{\delta\psi(\mathbf{s})}2\delta_{C}(\mathbf{s},\mathbf{s})\right\} W\nonumber \\
 &  & +\frac{g}{2}\int d\mathbf{s}\frac{1}{4}\left\{ \frac{\delta}{\delta\psi(\mathbf{s})}\frac{\delta}{\delta\psi(\mathbf{s})}\psi(\mathbf{s})\psi(\mathbf{s})\right\} W-\frac{g}{2}\int d\mathbf{s}\frac{1}{4}\left\{ \frac{\delta}{\delta\psi^{+}(\mathbf{s})}\frac{\delta}{\delta\psi^{+}(\mathbf{s})}\psi^{+}(\mathbf{s})\psi^{+}(\mathbf{s})\right\} W\nonumber \\
 & = & 0\nonumber \\
 &  & \,\end{eqnarray}
\begin{eqnarray}
 &  & W^{3}\nonumber \\
 & = & -\frac{g}{2}\int d\mathbf{s}\frac{1}{8}\left\{ \frac{\delta}{\delta\psi(\mathbf{s})}\frac{\delta}{\delta\psi^{+}(\mathbf{s})}\frac{\delta}{\delta\psi^{+}(\mathbf{s})}\psi^{+}(\mathbf{s})\right\} W+\frac{g}{2}\int d\mathbf{s}\frac{1}{8}\left\{ \frac{\delta}{\delta\psi^{+}(\mathbf{s})}\frac{\delta}{\delta\psi(\mathbf{s})}\frac{\delta}{\delta\psi(\mathbf{s})}\psi(\mathbf{s})\right\} W\nonumber \\
 &  & -\frac{g}{2}\int d\mathbf{s}\frac{1}{8}\left\{ \frac{\delta}{\delta\psi(\mathbf{s})}\frac{\delta}{\delta\psi^{+}(\mathbf{s})}\frac{\delta}{\delta\psi^{+}(\mathbf{s})}\psi^{+}(\mathbf{s})\right\} W+\frac{g}{2}\int d\mathbf{s}\frac{1}{8}\left\{ \frac{\delta}{\delta\psi^{+}(\mathbf{s})}\frac{\delta}{\delta\psi(\mathbf{s})}\frac{\delta}{\delta\psi(\mathbf{s})}\psi(\mathbf{s})\right\} W\nonumber \\
 &  & +\frac{g}{2}\int d\mathbf{s}\frac{1}{8}\left\{ \frac{\delta}{\delta\psi(\mathbf{s})}\frac{\delta}{\delta\psi(\mathbf{s})}\frac{\delta}{\delta\psi^{+}(\mathbf{s})}\psi(\mathbf{s})\right\} W-\frac{g}{2}\int d\mathbf{s}\frac{1}{8}\left\{ \frac{\delta}{\delta\psi^{+}(\mathbf{s})}\frac{\delta}{\delta\psi^{+}(\mathbf{s})}\frac{\delta}{\delta\psi(\mathbf{s})}\psi^{+}(\mathbf{s})\right\} W\nonumber \\
 &  & +\frac{g}{2}\int d\mathbf{s}\frac{1}{8}\left\{ \frac{\delta}{\delta\psi(\mathbf{s})}\frac{\delta}{\delta\psi(\mathbf{s})}\frac{\delta}{\delta\psi^{+}(\mathbf{s})}\psi(\mathbf{s})\right\} W-\frac{g}{2}\int d\mathbf{s}\frac{1}{8}\left\{ \frac{\delta}{\delta\psi^{+}(\mathbf{s})}\frac{\delta}{\delta\psi^{+}(\mathbf{s})}\frac{\delta}{\delta\psi(\mathbf{s})}\psi^{+}(\mathbf{s})\right\} W\nonumber \\
 & = & g\int d\mathbf{s}\frac{\delta}{\delta\psi(\mathbf{s})}\frac{\delta}{\delta\psi(\mathbf{s})}\frac{\delta}{\delta\psi^{+}(\mathbf{s})}\{\frac{1}{4}\psi(\mathbf{s})\}W[\psi,\psi^{+}]\nonumber \\
 &  & -g\int d\mathbf{s}\frac{\delta}{\delta\psi^{+}(\mathbf{s})}\frac{\delta}{\delta\psi^{+}(\mathbf{s})}\frac{\delta}{\delta\psi(\mathbf{s})}\{\frac{1}{4}\psi^{+}(\mathbf{s})\}W[\psi,\psi^{+}]\nonumber \\
 &  & \,\end{eqnarray}
\begin{eqnarray}
 &  & W^{4}\nonumber \\
 & = & +\frac{g}{2}\int d\mathbf{s}\frac{1}{16}\left\{ \frac{\delta}{\delta\psi(\mathbf{s})}\frac{\delta}{\delta\psi(\mathbf{s})}\frac{\delta}{\delta\psi^{+}(\mathbf{s})}\frac{\delta}{\delta\psi^{+}(\mathbf{s})}\right\} W\nonumber \\
 &  & -\frac{g}{2}\int d\mathbf{s}\frac{1}{16}\left\{ \frac{\delta}{\delta\psi^{+}(\mathbf{s})}\frac{\delta}{\delta\psi^{+}(\mathbf{s})}\frac{\delta}{\delta\psi(\mathbf{s})}\frac{\delta}{\delta\psi(\mathbf{s})}\right\} W\nonumber \\
 & = & 0\nonumber \\
 &  & \,\end{eqnarray}

Overall, the contribution to the functional Fokker-Planck equation
from the boson-boson interaction is given by\begin{eqnarray}
 &  & \left(\frac{\partial}{\partial t}W[\psi,\psi^{+}]\right)_{U}\nonumber \\
 & = & \frac{-i}{\hbar}\left\{ -g\int d\mathbf{s}\frac{\delta}{\delta\psi(\mathbf{s})}\left\{ (\psi^{+}(\mathbf{s})\psi(\mathbf{s})-\delta_{C}(\mathbf{s},\mathbf{s}))\psi(\mathbf{s})\right\} W[\psi,\psi^{+}]\right\} \nonumber \\
 &  & \frac{-i}{\hbar}\left\{ +g\int d\mathbf{s}\frac{\delta}{\delta\psi^{+}(\mathbf{s})}\left\{ (\psi^{+}(\mathbf{s})\psi(\mathbf{s})--\delta_{C}(\mathbf{s},\mathbf{s}))\psi^{+}(\mathbf{s})\right\} W[\psi,\psi^{+}]\right\} \nonumber \\
 &  & \frac{-i}{\hbar}\left\{ g\int d\mathbf{s}\frac{1}{4}\frac{\delta}{\delta\psi(\mathbf{s})}\frac{\delta}{\delta\psi(\mathbf{s})}\frac{\delta}{\delta\psi^{+}(\mathbf{s})}\{\psi(\mathbf{s})\}W[\psi,\psi^{+}]\right\} \nonumber \\
 &  & \frac{-i}{\hbar}\left\{ -g\int d\mathbf{s}\frac{1}{4}\frac{\delta}{\delta\psi^{+}(\mathbf{s})}\frac{\delta}{\delta\psi^{+}(\mathbf{s})}\frac{\delta}{\delta\psi(\mathbf{s})}\{\psi^{+}(\mathbf{s})\}W[\psi,\psi^{+}]\right\} \nonumber \\
 &  & \,\end{eqnarray}
which involves first order and third order functional derivatives.
We have replaced $\delta_{K}(0)$ by its full form $\delta_{K}(\mathbf{s},\mathbf{s})$.

Reverting to the original notation we have \begin{eqnarray}
 &  & \left(\frac{\partial}{\partial t}P[\underrightarrow{\psi}(\mathbf{r}),\underrightarrow{\psi^{\ast}}(\mathbf{r})]\right)_{U}\nonumber \\
 & = & \frac{-i}{\hbar}\left\{ -g\int d\mathbf{s}\frac{\delta}{\delta\psi_{C}(\mathbf{s})}\left\{ (\psi_{C}^{+}(\mathbf{s})\psi_{C}(\mathbf{s})-\delta_{C}(\mathbf{s},\mathbf{s}))\psi_{C}(\mathbf{s})\right\} P[\underrightarrow{\psi}(\mathbf{r}),\underrightarrow{\psi^{\ast}}(\mathbf{r})]\right\} \nonumber \\
 &  & +\frac{-i}{\hbar}\left\{ +g\int d\mathbf{s}\frac{\delta}{\delta\psi_{C}^{+}(\mathbf{s})}\left\{ (\psi_{C}^{+}(\mathbf{s})\psi_{C}(\mathbf{s})-\delta_{C}(\mathbf{s},\mathbf{s}))\psi_{C}^{+}(\mathbf{s})\right\} P[\underrightarrow{\psi}(\mathbf{r}),\underrightarrow{\psi^{\ast}}(\mathbf{r})]\right\} \nonumber \\
 &  & +\frac{-i}{\hbar}\left\{ g\int d\mathbf{s}\frac{\delta}{\delta\psi_{C}(\mathbf{s})}\frac{\delta}{\delta\psi_{C}(\mathbf{s})}\frac{\delta}{\delta\psi_{C}^{+}(\mathbf{s})}\{\frac{1}{4}\psi_{C}(\mathbf{s})\}P[\underrightarrow{\psi}(\mathbf{r}),\underrightarrow{\psi^{\ast}}(\mathbf{r})]\right\} \nonumber \\
 &  & +\frac{-i}{\hbar}\left\{ -g\int d\mathbf{s}\frac{\delta}{\delta\psi_{C}^{+}(\mathbf{s})}\frac{\delta}{\delta\psi_{C}^{+}(\mathbf{s})}\frac{\delta}{\delta\psi_{C}(\mathbf{s})}\{\frac{1}{4}\psi_{C}^{+}(\mathbf{s})\}P[\underrightarrow{\psi}(\mathbf{r}),\underrightarrow{\psi^{\ast}}(\mathbf{r})]\right\} \nonumber \\
 &  & \,\label{eq:FFPECondIntn}\end{eqnarray}

\subsection{Non-Condensate Kinetic Energy Terms}

We write the kinetic energy as \begin{equation}
\widehat{T}=\frac{\hbar^{2}}{2m}\sum\limits _{\mu}\int d\mathbf{s\,}\partial_{\mu}\widehat{\Psi}(\mathbf{s})^{\dagger}\,\partial_{\mu}\widehat{\Psi}(\mathbf{s})\end{equation}

Now if \begin{equation}
\widehat{\rho}\rightarrow\widehat{T}\widehat{\rho}=\frac{\hbar^{2}}{2m}\sum\limits _{\mu}\int d\mathbf{s(}\partial_{\mu}\widehat{\Psi}(\mathbf{s})^{\dagger}\,\partial_{\mu}\widehat{\Psi}(\mathbf{s}))\widehat{\rho}\end{equation}
then\begin{eqnarray}
 &  & P[\psi(\mathbf{r}),\psi^{+}(\mathbf{r})]\nonumber \\
 & \rightarrow & \frac{\hbar^{2}}{2m}\sum\limits _{\mu}\int d\mathbf{s}\left\{ \left(\partial_{\mu}\psi^{+}(\mathbf{s})-\partial_{\mu}\frac{\delta}{\delta\psi(\mathbf{s})}\right)\left(\partial_{\mu}\psi(\mathbf{s})\right)\right\} P[\psi,\psi^{+}]\nonumber \\
 &  & \,\end{eqnarray}

After expanding we find that \begin{eqnarray}
 &  & P[\psi(\mathbf{r}),\psi^{+}(\mathbf{r})]\nonumber \\
 & \rightarrow & \frac{\hbar^{2}}{2m}\sum\limits _{\mu}\int d\mathbf{s}\left\{ \left(\partial_{\mu}\psi^{+}(\mathbf{s})\right)\left(\partial_{\mu}\psi(\mathbf{s})\right)\right\} P[\psi,\psi^{+}]\nonumber \\
 &  & -\frac{\hbar^{2}}{2m}\sum\limits _{\mu}\int d\mathbf{s}\left\{ \left(\partial_{\mu}\frac{\delta}{\delta\psi(\mathbf{s})}\right)\left(\partial_{\mu}\psi(\mathbf{s})\right)\right\} P[\psi,\psi^{+}]\nonumber \\
 &  & \,\end{eqnarray}

For the first term, the product of the spatial functions can be written
in opposite order so that \begin{eqnarray}
 &  & \int d\mathbf{s}\left\{ \left(\partial_{\mu}\psi^{+}(\mathbf{s})\right)\left(\partial_{\mu}\psi(\mathbf{s})\right)\right\} P[\psi,\psi^{+}]\nonumber \\
 & = & \int d\mathbf{s}\left\{ \left(\partial_{\mu}\psi(\mathbf{s})\right)\left(\partial_{\mu}\psi^{+}(\mathbf{s})\right)\right\} P[\psi,\psi^{+}]\label{Eq.SimplifnResult1NC}\end{eqnarray}

We can use (\ref{eq:SpatialIntnResultCC-1}) together with the explicit
forms (\ref{eq:RestrictFnalDerivIdent1-1}) and for the functional
derivatives to modify the terms in the new $P[\psi,\psi^{+}]$, which
is equivalent to the function $p(\alpha_{k},\alpha_{k}^{+})$ if $\psi(\mathbf{s)}$
and $\psi^{+}(\mathbf{s)}$ are expanded in terms of modes $\phi_{k}(\mathbf{s)}$
or $\phi_{k}^{\ast}(\mathbf{s)}$, as in (\ref{eq:CondFieldFn-1})
and (\ref{eq:NonCondFieldFn-1}) with expansion coefficients $\alpha_{k}$
and $\alpha_{k}^{\ast}$.

In the second term we use (\ref{eq:SpatialIntnResultCC-1}) to apply
the spatial derivative to the $\psi(\mathbf{s})$ factor \begin{eqnarray}
 &  & \int d\mathbf{s}\left\{ \left(\partial_{\mu}\frac{\delta}{\delta\psi(\mathbf{s})}\right)\left(\partial_{\mu}\psi(\mathbf{s})\right)\right\} P[\psi,\psi^{+}]\nonumber \\
 & = & \int d\mathbf{s}\sum\limits _{k\neq1,2}^{K}\{\partial_{\mu}\phi_{k}^{\ast}(\mathbf{s})\}\,\frac{\partial}{\partial\alpha_{k}}\sum\limits _{l\neq1,2}^{K}\alpha_{l}\{\partial_{\mu}\phi_{l}(\mathbf{s})\}p(\alpha_{k},\alpha_{k}^{+})\nonumber \\
 & = & -\int d\mathbf{s}\sum\limits _{k\neq1,2}^{K}\{\phi_{k}^{\ast}(\mathbf{s})\}\,\frac{\partial}{\partial\alpha_{k}}\sum\limits _{l\neq1,2}^{K}\alpha_{l}\{\partial_{\mu}^{2}\phi_{l}(\mathbf{s})\}p(\alpha_{k},\alpha_{k}^{+})\nonumber \\
 & = & -\int d\mathbf{s}\left\{ \left(\frac{\delta}{\delta\psi(\mathbf{s})}\right)\left(\partial_{\mu}^{2}\psi(\mathbf{s})\right)\right\} P[\psi,\psi^{+}]\label{Eq.SimplifnResult3NC}\end{eqnarray}

Using results (\ref{Eq.SimplifnResult1NC}) and (\ref{Eq.SimplifnResult3NC})
we find that\begin{eqnarray}
 &  & P[\psi(\mathbf{r}),\psi^{+}(\mathbf{r})]\nonumber \\
 & \rightarrow & \frac{\hbar^{2}}{2m}\sum\limits _{\mu}\int d\mathbf{s\,}\left\{ \left(\partial_{\mu}\psi(\mathbf{s})\right)\left(\partial_{\mu}\psi^{+}(\mathbf{s})\right)\right\} P[\psi,\psi^{+}]\nonumber \\
 &  & +\frac{\hbar^{2}}{2m}\sum\limits _{\mu}\int d\mathbf{s}\left\{ \left(\frac{\delta}{\delta\psi(\mathbf{s})}\right)\left(\partial_{\mu}^{2}\psi(\mathbf{s})\right)\right\} P[\psi,\psi^{+}]\label{Eq.TRhoResultNC}\end{eqnarray}

Now if \begin{equation}
\widehat{\rho}\rightarrow\widehat{\rho}\widehat{T}=\frac{\hbar^{2}}{2m}\sum\limits _{\mu}\int d\mathbf{s\,\widehat{\rho}(}\partial_{\mu}\widehat{\Psi}(\mathbf{s})^{\dagger}\,\partial_{\mu}\widehat{\Psi}(\mathbf{s}))\end{equation}
then\begin{eqnarray}
 &  & P[\psi(\mathbf{r}),\psi^{+}(\mathbf{r})]\nonumber \\
 & \rightarrow & \frac{\hbar^{2}}{2m}\sum\limits _{\mu}\int d\mathbf{s}\left\{ \left(\partial_{\mu}\psi(\mathbf{s})-\partial_{\mu}\frac{\delta}{\delta\psi^{+}(\mathbf{s})}\right)\left(\partial_{\mu}\psi^{+}(\mathbf{s})\right)\right\} P[\psi,\psi^{+}]\nonumber \\
 &  & \,\end{eqnarray}

After expanding we find that \begin{eqnarray}
 &  & P[\psi(\mathbf{r}),\psi^{+}(\mathbf{r})]\nonumber \\
 & \rightarrow & \frac{\hbar^{2}}{2m}\sum\limits _{\mu}\int d\mathbf{s}\left\{ \left(\partial_{\mu}\psi(\mathbf{s})\right)\left(\partial_{\mu}\psi^{+}(\mathbf{s})\right)\right\} P[\psi,\psi^{+}]\nonumber \\
 &  & -\frac{\hbar^{2}}{2m}\sum\limits _{\mu}\int d\mathbf{s}\left\{ \left(\partial_{\mu}\frac{\delta}{\delta\psi^{+}(\mathbf{s})}\right)\left(\partial_{\mu}\psi^{+}(\mathbf{s})\right)\right\} P[\psi,\psi^{+}]\nonumber \\
 &  & \,\end{eqnarray}

Applying the same approach as before the second term is given by\begin{eqnarray}
 &  & \int d\mathbf{s}\left\{ \left(\partial_{\mu}\frac{\delta}{\delta\psi^{+}(\mathbf{s})}\right)\left(\partial_{\mu}\psi^{+}(\mathbf{s})\right)\right\} P[\psi,\psi^{+}]\nonumber \\
 & = & -\int d\mathbf{s}\left\{ \left(\frac{\delta}{\delta\psi^{+}(\mathbf{s})}\right)\left(\partial_{\mu}^{2}\psi^{+}(\mathbf{s})\right)\right\} P[\psi,\psi^{+}]\label{Eq.SimplifnResult3BNC}\end{eqnarray}
Using the result (\ref{Eq.SimplifnResult3BNC}) we find that\begin{eqnarray}
 &  & P[\psi(\mathbf{r}),\psi^{+}(\mathbf{r})]\nonumber \\
 & \rightarrow & \frac{\hbar^{2}}{2m}\sum\limits _{\mu}\int d\mathbf{s\,}\left\{ \left(\partial_{\mu}\psi(\mathbf{s})\right)\left(\partial_{\mu}\psi^{+}(\mathbf{s})\right)\right\} P[\psi,\psi^{+}]\nonumber \\
 &  & +\frac{\hbar^{2}}{2m}\sum\limits _{\mu}\int d\mathbf{s}\left\{ \left(\frac{\delta}{\delta\psi^{+}(\mathbf{s})}\right)\left(\partial_{\mu}^{2}\psi^{+}(\mathbf{s})\right)\right\} P[\psi,\psi^{+}]\label{Eq.RhoTResultNC}\end{eqnarray}

We now combine the contributions so that when \begin{equation}
\widehat{\rho}\rightarrow\lbrack\widehat{T},\widehat{\rho}]=[\frac{\hbar^{2}}{2m}\sum\limits _{\mu}\int d\mathbf{s(}\partial_{\mu}\widehat{\Psi}(\mathbf{s})^{\dagger}\,\partial_{\mu}\widehat{\Psi}(\mathbf{s})),\widehat{\rho}]\end{equation}
then

\begin{equation}
P[\psi(\mathbf{r}),\psi^{+}(\mathbf{r})]\rightarrow P^{1}\end{equation}
where\begin{eqnarray}
P^{1} & = & -\int d\mathbf{s}\left\{ \frac{\delta}{\delta\psi^{+}(\mathbf{s})}\left(\sum\limits _{\mu}\frac{\hbar^{2}}{2m}\partial_{\mu}^{2}\psi^{+}(\mathbf{s})\right))P[\psi,\psi^{+}]\right\} \nonumber \\
 &  & +\int d\mathbf{s}\left\{ \frac{\delta}{\delta\psi(\mathbf{s})}\left(\sum\limits _{\mu}\frac{\hbar^{2}}{2m}\partial_{\mu}^{2}\psi(\mathbf{s})\right)P[\psi,\psi^{+}])\right\} \nonumber \\
 &  & \,\end{eqnarray}
where the $\left(\partial_{\mu}\psi(\mathbf{s})\right)\left(\partial_{\mu}\psi^{+}(\mathbf{s})\right)$
terms cancel. Thus only a first order functional derivative term occurs.

Overall, the contribution to the functional Fokker-Planck equation
from the kinetic energy term is given by\begin{eqnarray}
 &  & \left(\frac{\partial}{\partial t}P[\psi,\psi^{+}]\right)_{K}\nonumber \\
 & = & \frac{-i}{\hbar}\left\{ -\int d\mathbf{s}\left\{ \frac{\delta}{\delta\psi^{+}(\mathbf{s})}\left(\sum\limits _{\mu}\frac{\hbar^{2}}{2m}\partial_{\mu}^{2}\psi^{+}(\mathbf{s})\right)P[\psi,\psi^{+}]\right\} \right\} \nonumber \\
 &  & +\frac{-i}{\hbar}\left\{ +\int d\mathbf{s}\left\{ \frac{\delta}{\delta\psi(\mathbf{s})}\left(\sum\limits _{\mu}\frac{\hbar^{2}}{2m}\partial_{\mu}^{2}\psi(\mathbf{s})\right)P[\psi,\psi^{+}]\right\} \right\} \nonumber \\
 &  & \,\end{eqnarray}

Reverting to the original notation, the contribution to the functional
Fokker-Planck equation from the non-condensate kinetic energy term
is given by\begin{eqnarray}
 &  & \left(\frac{\partial}{\partial t}P[\underrightarrow{\psi}(\mathbf{r}),\underrightarrow{\psi^{\ast}}(\mathbf{r})]\right)_{K}\nonumber \\
 & = & \frac{-i}{\hbar}\left\{ -\int d\mathbf{s}\left\{ \frac{\delta}{\delta\psi_{NC}^{+}(\mathbf{s})}\left(\sum\limits _{\mu}\frac{\hbar^{2}}{2m}\partial_{\mu}^{2}\psi_{NC}^{+}(\mathbf{s})\right)P[\underrightarrow{\psi}(\mathbf{r}),\underrightarrow{\psi^{\ast}}(\mathbf{r})]\right\} \right\} \nonumber \\
 &  & +\frac{-i}{\hbar}\left\{ +\int d\mathbf{s}\left\{ \frac{\delta}{\delta\psi_{NC}(\mathbf{s})}\left(\sum\limits _{\mu}\frac{\hbar^{2}}{2m}\partial_{\mu}^{2}\psi_{NC}(\mathbf{s})\right)P[\underrightarrow{\psi}(\mathbf{r}),\underrightarrow{\psi^{\ast}}(\mathbf{r})]\right\} \right\} \nonumber \\
 &  & \,\label{eq:FFPENonCondKE}\end{eqnarray}

\subsection{Non-Condensate Trap Potential Terms}

We write the trap potential as \begin{equation}
\widehat{V}=\int d\mathbf{s(}\widehat{\Psi}(\mathbf{s})^{\dagger}V\widehat{\Psi}(\mathbf{s}))\end{equation}

Now if \begin{equation}
\widehat{\rho}\rightarrow\widehat{V}\widehat{\rho}=\int d\mathbf{s(}\widehat{\Psi}(\mathbf{s})^{\dagger}V\widehat{\Psi}(\mathbf{s}))\widehat{\rho}\end{equation}
then\begin{eqnarray}
 &  & P[\psi(\mathbf{r}),\psi^{+}(\mathbf{r})]\nonumber \\
 & \rightarrow & \int d\mathbf{s}\left\{ \left(\psi^{+}(\mathbf{s})-\frac{\delta}{\delta\psi(\mathbf{s})}\right)V(\mathbf{s})\left(\psi(\mathbf{s})\right)\right\} P[\psi,\psi^{+}]\nonumber \\
 &  & \,\end{eqnarray}

After expanding we find that\begin{eqnarray}
 &  & P[\psi(\mathbf{r}),\psi^{+}(\mathbf{r})]\nonumber \\
 & \rightarrow & \int d\mathbf{s}\left\{ \psi(\mathbf{s})V(\mathbf{s})\psi^{+}(\mathbf{s})\right\} P[\psi,\psi^{+}]-\int d\mathbf{s}\left\{ \frac{\delta}{\delta\psi(\mathbf{s})}V(\mathbf{s})\psi(\mathbf{s})\right\} P[\psi,\psi^{+}]\nonumber \\
 &  & \,\end{eqnarray}
where we have re-ordered the $\psi^{+}(\mathbf{s})V(\mathbf{s})\psi(\mathbf{s})$
factor in the first term.

Now if \begin{equation}
\widehat{\rho}\rightarrow\widehat{\rho}\widehat{V}=\int d\mathbf{s}\widehat{\mathbf{\rho}}\widehat{\Psi}(\mathbf{s})^{\dagger}V\widehat{\Psi}(\mathbf{s})\end{equation}
then\begin{eqnarray}
 &  & P[\psi(\mathbf{r}),\psi^{+}(\mathbf{r})]\nonumber \\
 & \rightarrow & \int d\mathbf{s}\left(\psi(\mathbf{s})-\frac{\delta}{\delta\psi^{+}(\mathbf{s})}\right)V(\mathbf{s})\left(\psi^{+}(\mathbf{s})\right)P[\psi,\psi^{+}]\nonumber \\
 &  & \,\end{eqnarray}
After expanding we have\begin{eqnarray}
 &  & P[\psi(\mathbf{r}),\psi^{+}(\mathbf{r})]\nonumber \\
 & \rightarrow & \int d\mathbf{s\{}\psi(\mathbf{s})V(\mathbf{s})\psi^{+}(\mathbf{s})\}P[\psi,\psi^{+}]-\int d\mathbf{s}\left\{ \frac{\delta}{\delta\psi^{+}(\mathbf{s})}V(\mathbf{s})\psi^{+}(\mathbf{s})\right\} P[\psi,\psi^{+}]\nonumber \\
 &  & \,\end{eqnarray}

We now combine the contributions so that when \begin{equation}
\widehat{\rho}\rightarrow\lbrack\widehat{V},\widehat{\rho}]=[\int d\mathbf{s}\,\,\widehat{\Psi}(\mathbf{s})^{\dagger}V(\mathbf{s})\widehat{\Psi}(\mathbf{s}),\widehat{\rho}]\end{equation}
then

\begin{align}
P[\psi(\mathbf{r}),\psi^{+}(\mathbf{r})] & \rightarrow P^{1}\end{align}
which only involves a first order derivative since the zero order
terms cancel.\begin{eqnarray}
 &  & P^{1}\nonumber \\
 & = & -\int d\mathbf{s}\left\{ \frac{\delta}{\delta\psi(\mathbf{s})}\{V(\mathbf{s})\psi(\mathbf{s})\}\right\} P[\psi,\psi^{+}]+\int d\mathbf{s}\left\{ \frac{\delta}{\delta\psi^{+}(\mathbf{s})}V(\mathbf{s})\psi^{+}(\mathbf{s})\right\} P[\psi,\psi^{+}]\nonumber \\
 &  & \,\end{eqnarray}

Overall, the contribution to the functional Fokker-Planck equation
from the trap potential term is given by\begin{eqnarray}
 &  & \left(\frac{\partial}{\partial t}P[\psi,\psi^{+}]\right)_{V}\nonumber \\
 & = & \frac{-i}{\hbar}\left\{ -\int d\mathbf{s}\left\{ \frac{\delta}{\delta\psi(\mathbf{s})}\{V(\mathbf{s})\psi(\mathbf{s})\}\right\} P[\psi,\psi^{+}]+\int d\mathbf{s}\left\{ \frac{\delta}{\delta\psi^{+}(\mathbf{s})}V(\mathbf{s})\psi^{+}(\mathbf{s})\right\} P[\psi,\psi^{+}]\right\} \nonumber \\
 &  & \,\end{eqnarray}
which only involves first order functional derivatives.

Reverting to the original notation, the contribution to the functional
Fokker-Planck equation from the non-condensate trap potential term
is given by\begin{eqnarray}
 &  & \left(\frac{\partial}{\partial t}P[\underrightarrow{\psi}(\mathbf{r}),\underrightarrow{\psi^{\ast}}(\mathbf{r})]\right)_{V}\nonumber \\
 & = & \frac{-i}{\hbar}\left\{ -\int d\mathbf{s}\left\{ \frac{\delta}{\delta\psi_{NC}(\mathbf{s})}\{V(\mathbf{s})\psi_{NC}(\mathbf{s})\}\right\} P[\underrightarrow{\psi}(\mathbf{r}),\underrightarrow{\psi^{\ast}}(\mathbf{r})]\right\} \nonumber \\
 &  & \frac{-i}{\hbar}\left\{ +\int d\mathbf{s}\left\{ \frac{\delta}{\delta\psi_{NC}^{+}(\mathbf{s})}V(\mathbf{s})\psi_{NC}^{+}(\mathbf{s})\right\} P[\underrightarrow{\psi}(\mathbf{r}),\underrightarrow{\psi^{\ast}}(\mathbf{r})]\right\} \label{Eq.FFPENonCondTrap}\end{eqnarray}

\subsection{Non-Condensate Boson-Boson Interaction Terms}

For the Bogoliubov Hamiltonian for which we derive the functional
Fokker-Planck equation this boson-boson interaction in the non-condensate
Hamiltonian $\widehat{H}_{NC}$ is discarded, but for completeness
we treat it here. We write the boson-boson interaction potential as
\begin{align}
\widehat{U} & =\frac{g}{2}\int d\mathbf{s}\widehat{\Psi}(\mathbf{s})^{\dagger}\widehat{\Psi}(\mathbf{s})^{\dagger}\widehat{\Psi}(\mathbf{s})\widehat{\Psi}(\mathbf{s})\nonumber \\
 & \,\end{align}

Now if \begin{align}
\widehat{\rho} & \rightarrow\widehat{U}\widehat{\rho}=\frac{g}{2}\int d\mathbf{s}\widehat{\Psi}(\mathbf{s})^{\dagger}\widehat{\Psi}(\mathbf{s})^{\dagger}\widehat{\Psi}(\mathbf{s})\widehat{\Psi}(\mathbf{s})\widehat{\rho}\nonumber \\
 & \,\end{align}

\begin{eqnarray}
 &  & P[\psi(\mathbf{r}),\psi^{+}(\mathbf{r})]\nonumber \\
 & \rightarrow & \frac{g}{2}\int d\mathbf{s}\left(\psi^{+}(\mathbf{s})-\frac{\delta}{\delta\psi(\mathbf{s})}\right)\left(\psi^{+}(\mathbf{s})-\frac{\delta}{\delta\psi(\mathbf{s})}\right)\left(\psi(\mathbf{s})\right)\left(\psi(\mathbf{s})\right)P[\psi,\psi^{+}]\nonumber \\
 & = & \frac{g}{2}\int d\mathbf{s}\left\{ \psi^{+}(\mathbf{s})\psi^{+}(\mathbf{s})\psi(\mathbf{s})\psi(\mathbf{s})\right\} P[\psi,\psi^{+}]\nonumber \\
 &  & +\frac{g}{2}\int d\mathbf{s}\left\{ -\psi^{+}(\mathbf{s})\frac{\delta}{\delta\psi(\mathbf{s})}\psi(\mathbf{s})\psi(\mathbf{s})\right\} P[\psi,\psi^{+}]\nonumber \\
 &  & +\frac{g}{2}\int d\mathbf{s}\left\{ -\frac{\delta}{\delta\psi(\mathbf{s})}\psi^{+}(\mathbf{s})\psi(\mathbf{s})\psi(\mathbf{s})\right\} P[\psi,\psi^{+}]\nonumber \\
 &  & +\frac{g}{2}\int d\mathbf{s}\left\{ \frac{\delta}{\delta\psi(\mathbf{s})}\frac{\delta}{\delta\psi(\mathbf{s})}\psi(\mathbf{s})\psi(\mathbf{s})\right\} P[\psi,\psi^{+}]\nonumber \\
 &  & \,\end{eqnarray}
After expanding we get

\begin{eqnarray}
P[\psi(\mathbf{r}),\psi^{+}(\mathbf{r})] & \rightarrow & \frac{g}{2}\int d\mathbf{s}\left\{ \psi^{+}(\mathbf{s})\psi^{+}(\mathbf{s})\psi(\mathbf{s})\psi(\mathbf{s})\right\} P[\psi,\psi^{+}]\nonumber \\
 &  & -\frac{g}{2}\int d\mathbf{s}\left\{ \psi^{+}(\mathbf{s})\frac{\delta}{\delta\psi(\mathbf{s})}\psi(\mathbf{s})\psi(\mathbf{s})\right\} P[\psi,\psi^{+}]\nonumber \\
 &  & -\frac{g}{2}\int d\mathbf{s}\left\{ \frac{\delta}{\delta\psi(\mathbf{s})}\psi^{+}(\mathbf{s})\psi(\mathbf{s})\psi(\mathbf{s})\right\} P[\psi,\psi^{+}]\nonumber \\
 &  & +\frac{g}{2}\int d\mathbf{s}\left\{ \frac{\delta}{\delta\psi(\mathbf{s})}\frac{\delta}{\delta\psi(\mathbf{s})}\psi(\mathbf{s})\psi(\mathbf{s})\right\} P[\psi,\psi^{+}]\nonumber \\
 &  & \,\end{eqnarray}

We can now use the product rule for functional derivatives (\ref{eq:ProdRuleFnalDeriv-1})
together with (\ref{Eq.FuncDerivativeRule1-1}) and (\ref{Eq.FuncDerivativeRule2-1})
to place all the derivatives on the left of the expression and obtain\begin{eqnarray}
 &  & P[\psi(\mathbf{r}),\psi^{+}(\mathbf{r})]\nonumber \\
 & \rightarrow & \frac{g}{2}\int d\mathbf{s}\left\{ \psi(\mathbf{s})\psi(\mathbf{s})\psi^{+}(\mathbf{s})\psi^{+}(\mathbf{s})\right\} P[\psi,\psi^{+}]\nonumber \\
 &  & -\frac{g}{2}\int d\mathbf{s}\left\{ \frac{\delta}{\delta\psi(\mathbf{s})}\psi^{+}(\mathbf{s})\psi(\mathbf{s})\psi(\mathbf{s})\right\} P[\psi,\psi^{+}]\qquad T2\nonumber \\
 &  & -\frac{g}{2}\int d\mathbf{s}\left\{ \frac{\delta}{\delta\psi(\mathbf{s})}\psi^{+}(\mathbf{s})\psi(\mathbf{s})\psi(\mathbf{s})\right\} P[\psi,\psi^{+}]\nonumber \\
 &  & +\frac{g}{2}\int d\mathbf{s}\left\{ \frac{\delta}{\delta\psi(\mathbf{s})}\frac{\delta}{\delta\psi(\mathbf{s})}\psi(\mathbf{s})\psi(\mathbf{s})\right\} W[\psi,\psi^{+}]\nonumber \\
 & = & \frac{g}{2}\int d\mathbf{s}\left\{ \psi(\mathbf{s})\psi(\mathbf{s})\psi^{+}(\mathbf{s})\psi^{+}(\mathbf{s})\right\} P[\psi,\psi^{+}]\nonumber \\
 &  & -g\int d\mathbf{s}\left\{ \frac{\delta}{\delta\psi(\mathbf{s})}\psi^{+}(\mathbf{s})\psi(\mathbf{s})\psi(\mathbf{s})\right\} P[\psi,\psi^{+}]\nonumber \\
 &  & +\frac{g}{2}\int d\mathbf{s}\left\{ \frac{\delta}{\delta\psi(\mathbf{s})}\frac{\delta}{\delta\psi(\mathbf{s})}\psi(\mathbf{s})\psi(\mathbf{s})\right\} W[\psi,\psi^{+}]\nonumber \\
 &  & \,\end{eqnarray}
where we have also rearranged the order of the factors in $\psi^{+}(\mathbf{s})\psi^{+}(\mathbf{s})\psi(\mathbf{s})\psi(\mathbf{s})$.

Details include

T2

\begin{eqnarray*}
 &  & \frac{\delta}{\delta\psi(\mathbf{s})}\{\psi^{+}(\mathbf{s})\psi(\mathbf{s})\psi(\mathbf{s})P\}\\
 & = & \frac{\delta}{\delta\psi(\mathbf{s})}\{\psi^{+}(\mathbf{s})\}\psi(\mathbf{s})\psi(\mathbf{s})P+\psi^{+}(\mathbf{s})\{\frac{\delta}{\delta\psi(\mathbf{s})}\psi(\mathbf{s})\psi(\mathbf{s})P\}\\
 & = & \psi^{+}(\mathbf{s})\{\frac{\delta}{\delta\psi(\mathbf{s})}\psi(\mathbf{s})\psi(\mathbf{s})P\}\\
 &  & \psi^{+}(\mathbf{s})\{\frac{\delta}{\delta\psi(\mathbf{s})}\psi(\mathbf{s})\psi(\mathbf{s})P[\psi,\psi^{+}]\}\\
 & = & \frac{\delta}{\delta\psi(\mathbf{s})}\{\psi^{+}(\mathbf{s})\psi(\mathbf{s})\psi(\mathbf{s})P[\psi,\psi^{+}]\}\end{eqnarray*}

Now if \begin{align}
\widehat{\rho} & \rightarrow\widehat{\rho}\widehat{U}=\frac{g}{2}\int d\mathbf{s}\widehat{\rho}\widehat{\Psi}(\mathbf{s})^{\dagger}\widehat{\Psi}(\mathbf{s})^{\dagger}\widehat{\Psi}(\mathbf{s})\widehat{\Psi}(\mathbf{s})\nonumber \\
 & \,\end{align}
then\begin{eqnarray}
 &  & P[\psi(\mathbf{r}),\psi^{+}(\mathbf{r})]\nonumber \\
 & \rightarrow & \frac{g}{2}\int d\mathbf{s}\left(\psi(\mathbf{s})-\frac{\delta}{\delta\psi^{+}(\mathbf{s})}\right)\left(\psi(\mathbf{s})-\frac{\delta}{\delta\psi^{+}(\mathbf{s})}\right)\left(\psi^{+}(\mathbf{s})\right)\left(\psi^{+}(\mathbf{s})\right)P[\psi,\psi^{+}]\nonumber \\
 & = & \frac{g}{2}\int d\mathbf{s}\left\{ \psi(\mathbf{s})\psi(\mathbf{s})\psi^{+}(\mathbf{s})\psi^{+}(\mathbf{s})\right\} P[\psi,\psi^{+}]\nonumber \\
 &  & +\frac{g}{2}\int d\mathbf{s}\left\{ -\psi(\mathbf{s})\frac{\delta}{\delta\psi^{+}(\mathbf{s})}\psi^{+}(\mathbf{s})\psi^{+}(\mathbf{s})\right\} P[\psi,\psi^{+}]\nonumber \\
 &  & +\frac{g}{2}\int d\mathbf{s}\left\{ -\frac{\delta}{\delta\psi^{+}(\mathbf{s})}\psi(\mathbf{s})\psi^{+}(\mathbf{s})\psi^{+}(\mathbf{s})\right\} P[\psi,\psi^{+}]\nonumber \\
 &  & +\frac{g}{2}\int d\mathbf{s}\left\{ \frac{\delta}{\delta\psi^{+}(\mathbf{s})}\frac{\delta}{\delta\psi^{+}(\mathbf{s})}\psi^{+}(\mathbf{s})\psi^{+}(\mathbf{s})\right\} P[\psi,\psi^{+}]\nonumber \\
 &  & \,\end{eqnarray}
After expanding we obtain

\begin{eqnarray}
P[\psi(\mathbf{r}),\psi^{+}(\mathbf{r})] & \rightarrow & \frac{g}{2}\int d\mathbf{s}\left\{ \psi(\mathbf{s})\psi(\mathbf{s})\psi^{+}(\mathbf{s})\psi^{+}(\mathbf{s})\right\} P[\psi,\psi^{+}]\nonumber \\
 &  & -\frac{g}{2}\int d\mathbf{s}\left\{ \psi(\mathbf{s})\frac{\delta}{\delta\psi^{+}(\mathbf{s})}\psi^{+}(\mathbf{s})\psi^{+}(\mathbf{s})\right\} P[\psi,\psi^{+}]\nonumber \\
 &  & -\frac{g}{2}\int d\mathbf{s}\left\{ \frac{\delta}{\delta\psi^{+}(\mathbf{s})}\psi(\mathbf{s})\psi^{+}(\mathbf{s})\psi^{+}(\mathbf{s})\right\} P[\psi,\psi^{+}]\nonumber \\
 &  & +\frac{g}{2}\int d\mathbf{s}\left\{ \frac{\delta}{\delta\psi^{+}(\mathbf{s})}\frac{\delta}{\delta\psi^{+}(\mathbf{s})}\psi^{+}(\mathbf{s})\psi^{+}(\mathbf{s})\right\} P[\psi,\psi^{+}]\nonumber \\
 &  & \,\end{eqnarray}

We can now use the product rule for functional derivatives (\ref{eq:ProdRuleFnalDeriv-1})
together with (\ref{Eq.FuncDerivativeRule1-1}) and (\ref{Eq.FuncDerivativeRule2-1})
to place all the derivatives on the left of the expression. However
the results can more easily be obtained by noticing that the $\widehat{\rho}\widehat{U}$
is the same as the $\widehat{U}\widehat{\rho}$ if we interchange
$\psi(\mathbf{s})$ and $\psi^{+}(\mathbf{s})$ everywhere. Hence

\begin{eqnarray}
 &  & P[\psi(\mathbf{r}),\psi^{+}(\mathbf{r})]\nonumber \\
 & \rightarrow & \frac{g}{2}\int d\mathbf{s}\left\{ \psi(\mathbf{s})\psi(\mathbf{s})\psi^{+}(\mathbf{s})\psi^{+}(\mathbf{s})\right\} P[\psi,\psi^{+}]\nonumber \\
 &  & -\frac{g}{2}\int d\mathbf{s}\left\{ \frac{\delta}{\delta\psi^{+}(\mathbf{s})}\psi(\mathbf{s})\psi^{+}(\mathbf{s})\psi^{+}(\mathbf{s})\right\} P[\psi,\psi^{+}]\nonumber \\
 &  & -\frac{g}{2}\int d\mathbf{s}\left\{ \frac{\delta}{\delta\psi^{+}(\mathbf{s})}\psi(\mathbf{s})\psi^{+}(\mathbf{s})\psi^{+}(\mathbf{s})\right\} P[\psi,\psi^{+}]\nonumber \\
 &  & +\frac{g}{2}\int d\mathbf{s}\left\{ \frac{\delta}{\delta\psi^{+}(\mathbf{s})}\frac{\delta}{\delta\psi^{+}(\mathbf{s})}\psi^{+}(\mathbf{s})\psi^{+}(\mathbf{s})\right\} P[\psi,\psi^{+}]\nonumber \\
 & = & \frac{g}{2}\int d\mathbf{s}\left\{ \psi(\mathbf{s})\psi(\mathbf{s})\psi^{+}(\mathbf{s})\psi^{+}(\mathbf{s})\right\} P[\psi,\psi^{+}]\nonumber \\
 &  & -g\int d\mathbf{s}\left\{ \frac{\delta}{\delta\psi^{+}(\mathbf{s})}\psi(\mathbf{s})\psi^{+}(\mathbf{s})\psi^{+}(\mathbf{s})\right\} P[\psi,\psi^{+}]\nonumber \\
 &  & +\frac{g}{2}\int d\mathbf{s}\left\{ \frac{\delta}{\delta\psi^{+}(\mathbf{s})}\frac{\delta}{\delta\psi^{+}(\mathbf{s})}\psi^{+}(\mathbf{s})\psi^{+}(\mathbf{s})\right\} P[\psi,\psi^{+}]\nonumber \\
 &  & \,\end{eqnarray}

We now combine the contributions so that when \begin{align}
\widehat{\rho} & \rightarrow\lbrack\widehat{U},\widehat{\rho}]=[\frac{g}{2}\int d\mathbf{s}\,\,\widehat{\Psi}(\mathbf{s})^{\dagger}\widehat{\Psi}(\mathbf{s})^{\dagger}\widehat{\Psi}(\mathbf{s})\widehat{\Psi}(\mathbf{s}),\widehat{\rho}]\nonumber \\
 & \,\end{align}
then

\begin{align}
P[\psi(\mathbf{r}),\psi^{+}(\mathbf{r})] & \rightarrow P^{1}+P^{2}\nonumber \\
 & \,\end{align}
where the terms are listed via the order of derivatives that occur\begin{eqnarray}
 &  & P^{1}\nonumber \\
 & = & +g\int d\mathbf{s}\left\{ \frac{\delta}{\delta\psi^{+}(\mathbf{s})}\psi^{+}(\mathbf{s})\psi^{+}(\mathbf{s})\psi(\mathbf{s})\right\} P[\psi,\psi^{+}]\nonumber \\
 &  & -g\int d\mathbf{s}\left\{ \frac{\delta}{\delta\psi(\mathbf{s})}\psi(\mathbf{s})\psi(\mathbf{s})\psi^{+}(\mathbf{s})\right\} P[\psi,\psi^{+}]\nonumber \\
 &  & \,\end{eqnarray}
\begin{eqnarray}
 &  & P^{2}\nonumber \\
 & = & -\frac{g}{2}\int d\mathbf{s}\left\{ \frac{\delta}{\delta\psi^{+}(\mathbf{s})}\frac{\delta}{\delta\psi^{+}(\mathbf{s})}\psi^{+}(\mathbf{s})\psi^{+}(\mathbf{s})\right\} P[\psi,\psi^{+}]\nonumber \\
 &  & +\frac{g}{2}\int d\mathbf{s}\left\{ \frac{\delta}{\delta\psi(\mathbf{s})}\frac{\delta}{\delta\psi(\mathbf{s})}\psi(\mathbf{s})\psi(\mathbf{s})\right\} P[\psi,\psi^{+}]\nonumber \\
 &  & \,\end{eqnarray}

Overall, the contribution to the functional Fokker-Planck equation
from the boson-boson interaction is given by\begin{eqnarray}
 &  & \left(\frac{\partial}{\partial t}P[\psi,\psi^{+}]\right)_{U}\nonumber \\
 & = & \frac{-i}{\hbar}\left\{ -g\int d\mathbf{s}\frac{\delta}{\delta\psi(\mathbf{s})}\left\{ (\psi^{+}(\mathbf{s})\psi(\mathbf{s})\}\psi(\mathbf{s})\right\} P[\psi,\psi^{+}]\right\} \nonumber \\
 &  & +\frac{-i}{\hbar}\left\{ +g\int d\mathbf{s}\frac{\delta}{\delta\psi^{+}(\mathbf{s})}\left\{ (\psi^{+}(\mathbf{s})\psi(\mathbf{s}))\psi^{+}(\mathbf{s})\right\} P[\psi,\psi^{+}]\right\} \nonumber \\
 &  & +\frac{-i}{\hbar}\left\{ \frac{g}{2}\int d\mathbf{s}\frac{\delta}{\delta\psi(\mathbf{s})}\frac{\delta}{\delta\psi(\mathbf{s})}\{\psi(\mathbf{s})\psi(\mathbf{s})\}P[\psi,\psi^{+}]\right\} \nonumber \\
 &  & +\frac{-i}{\hbar}\left\{ -\frac{g}{2}\int d\mathbf{s}\frac{\delta}{\delta\psi^{+}(\mathbf{s})}\frac{\delta}{\delta\psi^{+}(\mathbf{s})}\{\psi^{+}(\mathbf{s})\psi^{+}(\mathbf{s})\}P[\psi,\psi^{+}]\right\} \nonumber \\
 &  & \,\end{eqnarray}
which involves first order and second order functional derivatives.

Reverting to the original notation the contribution to the functional
Fokker-Planck equation from the non-condensate boson-boson interaction
term is given by \begin{eqnarray}
 &  & \left(\frac{\partial}{\partial t}P[\underrightarrow{\psi}(\mathbf{r}),\underrightarrow{\psi^{\ast}}(\mathbf{r})]\right)_{U}\nonumber \\
 & = & \frac{-i}{\hbar}\left\{ -g\int d\mathbf{s}\frac{\delta}{\delta\psi_{NC}(\mathbf{s})}\left\{ (\psi_{NC}^{+}(\mathbf{s})\psi_{NC}(\mathbf{s}))\psi_{NC}(\mathbf{s})\right\} P[\underrightarrow{\psi}(\mathbf{r}),\underrightarrow{\psi^{\ast}}(\mathbf{r})]\right\} \nonumber \\
 &  & \frac{-i}{\hbar}\left\{ +g\int d\mathbf{s}\frac{\delta}{\delta\psi_{NC}^{+}(\mathbf{s})}\left\{ (\psi_{NC}^{+}(\mathbf{s})\psi_{NC}(\mathbf{s}))\psi_{NC}^{+}(\mathbf{s})\right\} P[\underrightarrow{\psi}(\mathbf{r}),\underrightarrow{\psi^{\ast}}(\mathbf{r})]\right\} \nonumber \\
 &  & \frac{-i}{\hbar}\left\{ g\int d\mathbf{s}\frac{\delta}{\delta\psi_{NC}(\mathbf{s})}\frac{\delta}{\delta\psi_{NC}(\mathbf{s})}\{\frac{1}{2}\psi_{NC}(\mathbf{s})\psi_{NC}(\mathbf{s})\}P[\underrightarrow{\psi}(\mathbf{r}),\underrightarrow{\psi^{\ast}}(\mathbf{r})]\right\} \nonumber \\
 &  & \frac{-i}{\hbar}\left\{ -g\int d\mathbf{s}\frac{\delta}{\delta\psi_{NC}^{+}(\mathbf{s})}\frac{\delta}{\delta\psi_{NC}^{+}(\mathbf{s})}\{\frac{1}{2}\psi_{NC}^{+}(\mathbf{s})\psi_{NC}^{+}(\mathbf{s})\}P[\underrightarrow{\psi}(\mathbf{r}),\underrightarrow{\psi^{\ast}}(\mathbf{r})]\right\} \nonumber \\
 &  & \,\label{eq:FFPENonCondIntn}\end{eqnarray}
Note that this term is not included in the final functional Fokker-Planck
equation for the Bogoliubov Hamiltonian.

Similar expressions for the functional Fokker-Planck equation in the
case of a pure P representation (but not involving a doubled phase
space) are given in the paper by Steel et al {[}\citep{Steel98b}{]}
(see Eq. (17)). Comparisons can be made by substituting $\psi_{NC}^{+}(\mathbf{s})$
with $\psi_{NC}^{\ast}(\mathbf{s})$. As in the present result, no
restricted delta function $\delta_{C}(\mathbf{s},\mathbf{s})$ term
in the interaction contribution appears in a P representation approach.

\subsection{Condensate - Non-Condensate Interaction - First Order in Non-Condensate}

The first order term in the interaction between the condensate and
the non-condensate is\begin{eqnarray}
\widehat{V}_{1} & = & g\int d\mathbf{r\,}\widehat{\Psi}_{NC}^{\dagger}(\mathbf{r})\{\widehat{\Psi}_{C}^{\dagger}(\mathbf{r})\,\widehat{\Psi}_{C}(\mathbf{r})\}\widehat{\Psi}_{C}(\mathbf{r})-g\int\int d\mathbf{r\,}d\mathbf{s\,}F(\mathbf{r},\mathbf{s})\widehat{\Psi}_{NC}(\mathbf{r})^{\dagger}\,\widehat{\Psi}_{C}(\mathbf{s})\nonumber \\
 &  & +g\int d\mathbf{r\,}\widehat{\Psi}_{C}^{\dagger}(\mathbf{r})\{\widehat{\Psi}_{C}^{\dagger}(\mathbf{r})\,\widehat{\Psi}_{C}(\mathbf{r})\}\widehat{\Psi}_{NC}(\mathbf{r})-g\int\int d\mathbf{r\,}d\mathbf{s\,}F^{\ast}(\mathbf{s},\mathbf{r})\widehat{\Psi}_{C}(\mathbf{r})^{\dagger}\widehat{\Psi}_{NC}(\mathbf{s})\,\nonumber \\
 &  & \,\end{eqnarray}

This is the sum of two terms, one fourth order in the field operators,
the other second order.\begin{eqnarray}
\widehat{V}_{14} & = & g\int d\mathbf{r\,}\widehat{\Psi}_{NC}^{\dagger}(\mathbf{r})\widehat{\Psi}_{C}^{\dagger}(\mathbf{r})\,\widehat{\Psi}_{C}(\mathbf{r})\widehat{\Psi}_{C}(\mathbf{r})+g\int d\mathbf{r\,}\widehat{\Psi}_{C}^{\dagger}(\mathbf{r})\widehat{\Psi}_{C}^{\dagger}(\mathbf{r})\,\widehat{\Psi}_{C}(\mathbf{r})\widehat{\Psi}_{NC}(\mathbf{r})\nonumber \\
 &  & \,\\
\widehat{V}_{12} & = & -g\int\int d\mathbf{r\,}d\mathbf{s\,}F(\mathbf{r},\mathbf{s})\widehat{\Psi}_{NC}(\mathbf{r})^{\dagger}\,\widehat{\Psi}_{C}(\mathbf{s})-g\int\int d\mathbf{r\,}d\mathbf{s\,}F^{\ast}(\mathbf{s},\mathbf{r})\widehat{\Psi}_{C}(\mathbf{r})^{\dagger}\widehat{\Psi}_{NC}(\mathbf{s})\,\nonumber \\
 &  & \,\end{eqnarray}

\subsubsection{Fourth Order Term}

Now if \begin{align}
\widehat{\rho} & \rightarrow\widehat{V}_{14}\,\widehat{\rho}=g\int d\mathbf{s\,(}\widehat{\Psi}_{NC}^{\dagger}(\mathbf{s})\widehat{\Psi}_{C}^{\dagger}(\mathbf{s})\,\widehat{\Psi}_{C}(\mathbf{s})\widehat{\Psi}_{C}(\mathbf{s}))+\widehat{\Psi}_{C}^{\dagger}(\mathbf{s})\widehat{\Psi}_{C}^{\dagger}(\mathbf{s})\,\widehat{\Psi}_{C}(\mathbf{s})\widehat{\Psi}_{NC}(\mathbf{s}))\widehat{\rho}\nonumber \\
 & \,\end{align}
then\begin{eqnarray}
 &  & WP[\psi(\mathbf{r}),\psi^{+}(\mathbf{r}),\phi(\mathbf{r}),\phi^{+}(\mathbf{r})]\nonumber \\
 & \rightarrow & g\int d\mathbf{s\,}\left(\phi^{+}(\mathbf{s})-\frac{\delta}{\delta\phi(\mathbf{s})}\right)\left(\psi^{+}(\mathbf{s})-\frac{1}{2}\frac{\delta}{\delta\psi(\mathbf{s})}\right)\left(\psi(\mathbf{s})+\frac{1}{2}\frac{\delta}{\delta\psi^{+}(\mathbf{s})}\right)\nonumber \\
 &  & \times\left(\psi(\mathbf{s})+\frac{\textrm{1}}{2}\frac{\delta}{\delta\psi^{+}(\mathbf{s})}\right)WP\nonumber \\
 &  & +g\int d\mathbf{s\,}\mathbf{\left(\psi^{+}(\mathbf{s})-\textrm{\ensuremath{\frac{1}{2}}}\frac{\delta}{\delta\psi(\mathbf{s})}\right)\left(\psi^{+}(\mathbf{s})-\textrm{\ensuremath{\frac{1}{2}}}\frac{\delta}{\delta\psi(\mathbf{s})}\right)\left(\psi(\mathbf{s})+\frac{\textrm{1}}{\textrm{2}}\frac{\delta}{\delta\psi^{+}(\mathbf{s})}\right)}\nonumber \\
 &  & \times\left(\mathbf{\phi(\mathbf{s})}\right)WP[\psi,\psi^{+},\phi,\phi^{+}]\nonumber \\
 &  & \,\end{eqnarray}

Expanding out the terms gives\begin{eqnarray}
 &  & WP[\psi(\mathbf{r}),\psi^{+}(\mathbf{r}),\phi(\mathbf{r}),\phi^{+}(\mathbf{r})]\nonumber \\
 & \rightarrow & WP[\psi(\mathbf{r}),\psi^{+}(\mathbf{r}),\phi(\mathbf{r}),\phi^{+}(\mathbf{r})]_{1-16}+WP[\psi(\mathbf{r}),\psi^{+}(\mathbf{r}),\phi(\mathbf{r}),\phi^{+}(\mathbf{r})]_{17-24}\nonumber \\
 &  & \,\end{eqnarray}
where\begin{eqnarray*}
 &  & WP[\psi(\mathbf{r}),\psi^{+}(\mathbf{r}),\phi(\mathbf{r}),\phi^{+}(\mathbf{r})]_{1-16}\\
 & = & g\int d\mathbf{s\,}\left\{ \left(\phi^{+}(\mathbf{s})\right)\left(\psi^{+}(\mathbf{s})\right)\left(\psi(\mathbf{s})\right)\left(\psi(\mathbf{s})\right)\right\} WP\\
 &  & +g\int d\mathbf{s\,}\left\{ \left(\phi^{+}(\mathbf{s})\right)\left(\psi^{+}(\mathbf{s})\right)\left(\psi(\mathbf{s})\right)\left(\frac{1}{2}\frac{\delta}{\delta\psi^{+}(\mathbf{s})}\right)\right\} WP\\
 &  & +g\int d\mathbf{s\,}\left\{ \left(\phi^{+}(\mathbf{s})\right)\left(\psi^{+}(\mathbf{s})\right)\left(\frac{1}{2}\frac{\delta}{\delta\psi^{+}(\mathbf{s})}\right)\left(\psi(\mathbf{s})\right)\right\} WP\\
 &  & +g\int d\mathbf{s\,}\left\{ \left(\phi^{+}(\mathbf{s})\right)\left(\psi^{+}(\mathbf{s})\right)\left(\frac{1}{2}\frac{\delta}{\delta\psi^{+}(\mathbf{s})}\right)\left(\frac{1}{2}\frac{\delta}{\delta\psi^{+}(\mathbf{s})}\right)\right\} WP\\
 &  & +g\int d\mathbf{s\,}\left\{ \left(\phi^{+}(\mathbf{s})\right)\left(-\frac{1}{2}\frac{\delta}{\delta\psi(\mathbf{s})}\right)\left(\psi(\mathbf{s})\right)\left(\psi(\mathbf{s})\right)\right\} WP\\
 &  & +g\int d\mathbf{s\,}\left\{ \left(\phi^{+}(\mathbf{s})\right)\left(-\frac{1}{2}\frac{\delta}{\delta\psi(\mathbf{s})}\right)\left(\psi(\mathbf{s})\right)\left(\frac{1}{2}\frac{\delta}{\delta\psi^{+}(\mathbf{s})}\right)\right\} WP\\
 &  & +g\int d\mathbf{s\,}\left\{ \left(\phi^{+}(\mathbf{s})\right)\left(-\frac{1}{2}\frac{\delta}{\delta\psi(\mathbf{s})}\right)\left(\frac{1}{2}\frac{\delta}{\delta\psi^{+}(\mathbf{s})}\right)\left(\psi(\mathbf{s})\right)\right\} WP\\
 &  & +g\int d\mathbf{s\,}\left\{ \left(\phi^{+}(\mathbf{s})\right)\left(-\frac{1}{2}\frac{\delta}{\delta\psi(\mathbf{s})}\right)\left(\frac{1}{2}\frac{\delta}{\delta\psi^{+}(\mathbf{s})}\right)\left(\frac{1}{2}\frac{\delta}{\delta\psi^{+}(\mathbf{s})}\right)\right\} WP\\
 &  & +g\int d\mathbf{s\,}\left\{ \left(-\frac{\delta}{\delta\phi(\mathbf{s})}\right)\left(\psi^{+}(\mathbf{s})\right)\left(\psi(\mathbf{s})\right)\left(\psi(\mathbf{s})\right)\right\} WP\\
 &  & +g\int d\mathbf{s\,}\left\{ \left(-\frac{\delta}{\delta\phi(\mathbf{s})}\right)\left(\psi^{+}(\mathbf{s})\right)\left(\psi(\mathbf{s})\right)\left(\frac{1}{2}\frac{\delta}{\delta\psi^{+}(\mathbf{s})}\right)\right\} WP\\
 &  & +g\int d\mathbf{s\,}\left\{ \left(-\frac{\delta}{\delta\phi(\mathbf{s})}\right)\left(\psi^{+}(\mathbf{s})\right)\left(\frac{1}{2}\frac{\delta}{\delta\psi^{+}(\mathbf{s})}\right)\left(\psi(\mathbf{s})\right)\right\} WP\\
 &  & +g\int d\mathbf{s\,}\left\{ \left(-\frac{\delta}{\delta\phi(\mathbf{s})}\right)\left(\psi^{+}(\mathbf{s})\right)\left(\frac{1}{2}\frac{\delta}{\delta\psi^{+}(\mathbf{s})}\right)\left(\frac{1}{2}\frac{\delta}{\delta\psi^{+}(\mathbf{s})}\right)\right\} WP\\
 &  & +g\int d\mathbf{s\,}\left\{ \left(-\frac{\delta}{\delta\phi(\mathbf{s})}\right)\left(-\frac{1}{2}\frac{\delta}{\delta\psi(\mathbf{s})}\right)\left(\psi(\mathbf{s})\right)\left(\psi(\mathbf{s})\right)\right\} WP\\
 &  & +g\int d\mathbf{s\,}\left\{ \left(-\frac{\delta}{\delta\phi(\mathbf{s})}\right)\left(-\frac{1}{2}\frac{\delta}{\delta\psi(\mathbf{s})}\right)\left(\psi(\mathbf{s})\right)\left(\frac{1}{2}\frac{\delta}{\delta\psi^{+}(\mathbf{s})}\right)\right\} WP\\
 &  & +g\int d\mathbf{s\,}\left\{ \left(-\frac{\delta}{\delta\phi(\mathbf{s})}\right)\left(-\frac{1}{2}\frac{\delta}{\delta\psi(\mathbf{s})}\right)\left(\frac{1}{2}\frac{\delta}{\delta\psi^{+}(\mathbf{s})}\right)\left(\psi(\mathbf{s})\right)\right\} WP\\
 &  & +g\int d\mathbf{s\,}\left\{ \left(-\frac{\delta}{\delta\phi(\mathbf{s})}\right)\left(-\frac{1}{2}\frac{\delta}{\delta\psi(\mathbf{s})}\right)\left(\frac{1}{2}\frac{\delta}{\delta\psi^{+}(\mathbf{s})}\right)\left(\frac{1}{2}\frac{\delta}{\delta\psi^{+}(\mathbf{s})}\right)\right\} WP\end{eqnarray*}

and\begin{eqnarray*}
 &  & WP[\psi(\mathbf{r}),\psi^{+}(\mathbf{r}),\phi(\mathbf{r}),\phi^{+}(\mathbf{r})]_{17-24}\\
 & = & +g\int d\mathbf{s\,}\left\{ \mathbf{\left(\psi^{+}(\mathbf{s})\right)\left(\psi^{+}(\mathbf{s})\right)\left(\psi(\mathbf{s})\right)}\left(\mathbf{\phi(\mathbf{s})}\right)\right\} WP\\
 &  & +g\int d\mathbf{s\,}\left\{ \mathbf{\left(\psi^{+}(\mathbf{s})\right)\left(\psi^{+}(\mathbf{s})\right)\left(\frac{1}{2}\frac{\delta}{\delta\psi^{+}(\mathbf{s})}\right)}\left(\mathbf{\phi(\mathbf{s})}\right)\right\} WP\\
 &  & +g\int d\mathbf{s\,}\left\{ \mathbf{\left(\psi^{+}(\mathbf{s})\right)\left(-\frac{1}{2}\frac{\delta}{\delta\psi(\mathbf{s})}\right)\left(\psi(\mathbf{s})\right)}\left(\mathbf{\phi(\mathbf{s})}\right)\right\} WP\\
 &  & +g\int d\mathbf{s\,}\left\{ \mathbf{\left(\psi^{+}(\mathbf{s})\right)\left(-\frac{1}{2}\frac{\delta}{\delta\psi(\mathbf{s})}\right)\left(\frac{1}{2}\frac{\delta}{\delta\psi^{+}(\mathbf{s})}\right)}\left(\mathbf{\phi(\mathbf{s})}\right)\right\} WP\\
 &  & +g\int d\mathbf{s\,}\left\{ \mathbf{\left(-\frac{1}{2}\frac{\delta}{\delta\psi(\mathbf{s})}\right)\left(\psi^{+}(\mathbf{s})\right)\left(\psi(\mathbf{s})\right)}\left(\mathbf{\phi(\mathbf{s})}\right)\right\} WP\\
 &  & +g\int d\mathbf{s\,}\left\{ \mathbf{\left(-\frac{1}{2}\frac{\delta}{\delta\psi(\mathbf{s})}\right)\left(\psi^{+}(\mathbf{s})\right)\left(\frac{1}{2}\frac{\delta}{\delta\psi^{+}(\mathbf{s})}\right)}\left(\mathbf{\phi(\mathbf{s})}\right)\right\} WP\\
 &  & +g\int d\mathbf{s\,}\left\{ \mathbf{\left(-\frac{1}{2}\frac{\delta}{\delta\psi(\mathbf{s})}\right)\left(-\frac{1}{2}\frac{\delta}{\delta\psi(\mathbf{s})}\right)\left(\psi(\mathbf{s})\right)}\left(\mathbf{\phi(\mathbf{s})}\right)\right\} WP\\
 &  & +g\int d\mathbf{s\,}\left\{ \mathbf{\left(-\frac{1}{2}\frac{\delta}{\delta\psi(\mathbf{s})}\right)\left(-\frac{1}{2}\frac{\delta}{\delta\psi(\mathbf{s})}\right)\left(\frac{1}{2}\frac{\delta}{\delta\psi^{+}(\mathbf{s})}\right)}\left(\mathbf{\phi(\mathbf{s})}\right)\right\} WP\end{eqnarray*}
or on further simplification

\begin{eqnarray}
 &  & WP[\psi(\mathbf{r}),\psi^{+}(\mathbf{r}),\phi(\mathbf{r}),\phi^{+}(\mathbf{r})]_{1-16}\nonumber \\
 & = & g\int d\mathbf{s\,}\left\{ \left(\phi^{+}(\mathbf{s})\right)\left(\psi^{+}(\mathbf{s})\right)\left(\psi(\mathbf{s})\right)\left(\psi(\mathbf{s})\right)\right\} WP\qquad\emph{T1}\nonumber \\
 &  & +g\int d\mathbf{s\,}\frac{1}{2}\left\{ \left(\phi^{+}(\mathbf{s})\right)\left(\psi^{+}(\mathbf{s})\right)\left(\psi(\mathbf{s})\right)\left(\frac{\delta}{\delta\psi^{+}(\mathbf{s})}\right)\right\} WP\qquad\emph{T2}\nonumber \\
 &  & +g\int d\mathbf{s\,}\frac{1}{2}\left\{ \left(\phi^{+}(\mathbf{s})\right)\left(\psi^{+}(\mathbf{s})\right)\left(\frac{\delta}{\delta\psi^{+}(\mathbf{s})}\right)\left(\psi(\mathbf{s})\right)\right\} WP\qquad\emph{T3}\nonumber \\
 &  & +g\int d\mathbf{s\,}\frac{1}{4}\left\{ \left(\phi^{+}(\mathbf{s})\right)\left(\psi^{+}(\mathbf{s})\right)\left(\frac{\delta}{\delta\psi^{+}(\mathbf{s})}\right)\left(\frac{\delta}{\delta\psi^{+}(\mathbf{s})}\right)\right\} WP\qquad\emph{T4}\nonumber \\
 &  & -g\int d\mathbf{s\,}\frac{1}{2}\left\{ \left(\phi^{+}(\mathbf{s})\right)\left(\frac{\delta}{\delta\psi(\mathbf{s})}\right)\left(\psi(\mathbf{s})\right)\left(\psi(\mathbf{s})\right)\right\} WP\qquad\emph{T5}\nonumber \\
 &  & -g\int d\mathbf{s\,}\frac{1}{4}\left\{ \left(\phi^{+}(\mathbf{s})\right)\left(\frac{\delta}{\delta\psi(\mathbf{s})}\right)\left(\psi(\mathbf{s})\right)\left(\frac{\delta}{\delta\psi^{+}(\mathbf{s})}\right)\right\} WP\qquad\emph{T6}\nonumber \\
 &  & -g\int d\mathbf{s\,}\frac{1}{4}\left\{ \left(\phi^{+}(\mathbf{s})\right)\left(\frac{\delta}{\delta\psi(\mathbf{s})}\right)\left(\frac{\delta}{\delta\psi^{+}(\mathbf{s})}\right)\left(\psi(\mathbf{s})\right)\right\} WP\qquad\emph{T7}\nonumber \\
 &  & -g\int d\mathbf{s\,}\frac{1}{8}\left\{ \left(\phi^{+}(\mathbf{s})\right)\left(\frac{\delta}{\delta\psi(\mathbf{s})}\right)\left(\frac{\delta}{\delta\psi^{+}(\mathbf{s})}\right)\left(\frac{\delta}{\delta\psi^{+}(\mathbf{s})}\right)\right\} WP\qquad\emph{T8}\nonumber \\
 &  & -g\int d\mathbf{s\,}\left\{ \left(\frac{\delta}{\delta\phi(\mathbf{s})}\right)\left(\psi^{+}(\mathbf{s})\right)\left(\psi(\mathbf{s})\right)\left(\psi(\mathbf{s})\right)\right\} WP\qquad\emph{T9}\nonumber \\
 &  & -g\int d\mathbf{s\,}\frac{1}{2}\left\{ \left(\frac{\delta}{\delta\phi(\mathbf{s})}\right)\left(\psi^{+}(\mathbf{s})\right)\left(\psi(\mathbf{s})\right)\left(\frac{\delta}{\delta\psi^{+}(\mathbf{s})}\right)\right\} WP\qquad\emph{T10}\nonumber \\
 &  & -g\int d\mathbf{s\,}\frac{1}{2}\left\{ \left(\frac{\delta}{\delta\phi(\mathbf{s})}\right)\left(\psi^{+}(\mathbf{s})\right)\left(\frac{\delta}{\delta\psi^{+}(\mathbf{s})}\right)\left(\psi(\mathbf{s})\right)\right\} WP\qquad\emph{T11}\nonumber \\
 &  & -g\int d\mathbf{s\,}\frac{1}{4}\left\{ \left(\frac{\delta}{\delta\phi(\mathbf{s})}\right)\left(\psi^{+}(\mathbf{s})\right)\left(\frac{\delta}{\delta\psi^{+}(\mathbf{s})}\right)\left(\frac{\delta}{\delta\psi^{+}(\mathbf{s})}\right)\right\} WP\qquad\emph{T12}\nonumber \\
 &  & +g\int d\mathbf{s\,}\frac{1}{2}\left\{ \left(\frac{\delta}{\delta\phi(\mathbf{s})}\right)\left(\frac{\delta}{\delta\psi(\mathbf{s})}\right)\left(\psi(\mathbf{s})\right)\left(\psi(\mathbf{s})\right)\right\} WP\qquad\emph{T13}\nonumber \\
 &  & +g\int d\mathbf{s\,}\frac{1}{4}\left\{ \left(\frac{\delta}{\delta\phi(\mathbf{s})}\right)\left(\frac{\delta}{\delta\psi(\mathbf{s})}\right)\left(\psi(\mathbf{s})\right)\left(\frac{\delta}{\delta\psi^{+}(\mathbf{s})}\right)\right\} WP\qquad\emph{T14}\nonumber \\
 &  & +g\int d\mathbf{s\,}\frac{1}{4}\left\{ \left(\frac{\delta}{\delta\phi(\mathbf{s})}\right)\left(\frac{\delta}{\delta\psi(\mathbf{s})}\right)\left(\frac{\delta}{\delta\psi^{+}(\mathbf{s})}\right)\left(\psi(\mathbf{s})\right)\right\} WP\qquad\emph{T15}\nonumber \\
 &  & +g\int d\mathbf{s\,}\frac{1}{8}\left\{ \left(\frac{\delta}{\delta\phi(\mathbf{s})}\right)\left(\frac{\delta}{\delta\psi(\mathbf{s})}\right)\left(\frac{\delta}{\delta\psi^{+}(\mathbf{s})}\right)\left(\frac{\delta}{\delta\psi^{+}(\mathbf{s})}\right)\right\} WP\qquad\emph{T16 }\nonumber \\
 &  & \,\end{eqnarray}
and\begin{eqnarray}
 &  & WP[\psi(\mathbf{r}),\psi^{+}(\mathbf{r}),\phi(\mathbf{r}),\phi^{+}(\mathbf{r})]_{17-24}\nonumber \\
 & = & +g\int d\mathbf{s\,}\left\{ \mathbf{\left(\psi^{+}(\mathbf{s})\right)\left(\psi^{+}(\mathbf{s})\right)\left(\psi(\mathbf{s})\right)}\left(\mathbf{\phi(\mathbf{s})}\right)\right\} WP\qquad\emph{T17}\nonumber \\
 &  & +g\int d\mathbf{s\,}\frac{1}{2}\left\{ \mathbf{\left(\psi^{+}(\mathbf{s})\right)\left(\psi^{+}(\mathbf{s})\right)\left(\frac{\delta}{\delta\psi^{+}(\mathbf{s})}\right)}\left(\mathbf{\phi(\mathbf{s})}\right)\right\} WP\qquad\emph{T18}\nonumber \\
 &  & -g\int d\mathbf{s\,}\frac{1}{2}\left\{ \mathbf{\left(\psi^{+}(\mathbf{s})\right)\left(\frac{\delta}{\delta\psi(\mathbf{s})}\right)\left(\psi(\mathbf{s})\right)}\left(\mathbf{\phi(\mathbf{s})}\right)\right\} WP\qquad\emph{T19}\nonumber \\
 &  & -g\int d\mathbf{s\,}\frac{1}{4}\left\{ \mathbf{\left(\psi^{+}(\mathbf{s})\right)\left(\frac{\delta}{\delta\psi(\mathbf{s})}\right)\left(\frac{\delta}{\delta\psi^{+}(\mathbf{s})}\right)}\left(\mathbf{\phi(\mathbf{s})}\right)\right\} WP\qquad\emph{T20}\nonumber \\
 &  & -g\int d\mathbf{s\,}\frac{1}{2}\left\{ \mathbf{\left(\frac{\delta}{\delta\psi(\mathbf{s})}\right)\left(\psi^{+}(\mathbf{s})\right)\left(\psi(\mathbf{s})\right)}\left(\mathbf{\phi(\mathbf{s})}\right)\right\} WP\qquad\emph{T21}\nonumber \\
 &  & -g\int d\mathbf{s\,}\frac{1}{4}\left\{ \mathbf{\left(\frac{\delta}{\delta\psi(\mathbf{s})}\right)\left(\psi^{+}(\mathbf{s})\right)\left(\frac{\delta}{\delta\psi^{+}(\mathbf{s})}\right)}\left(\mathbf{\phi(\mathbf{s})}\right)\right\} WP\qquad\emph{T22}\nonumber \\
 &  & +g\int d\mathbf{s\,}\frac{1}{4}\left\{ \mathbf{\left(\frac{\delta}{\delta\psi(\mathbf{s})}\right)\left(\frac{\delta}{\delta\psi(\mathbf{s})}\right)\left(\psi(\mathbf{s})\right)}\left(\mathbf{\phi(\mathbf{s})}\right)\right\} WP\qquad\emph{T23}\nonumber \\
 &  & +g\int d\mathbf{s\,}\frac{1}{8}\left\{ \mathbf{\left(\frac{\delta}{\delta\psi(\mathbf{s})}\right)\left(\frac{\delta}{\delta\psi(\mathbf{s})}\right)\left(\frac{\delta}{\delta\psi^{+}(\mathbf{s})}\right)}\left(\mathbf{\phi(\mathbf{s})}\right)\right\} WP\qquad\emph{T24}\nonumber \\
 &  & \,\end{eqnarray}

These terms can be simplified in terms of placing all the functional
derivatives on the left by using the product rule (\ref{eq:ProdRuleFnalDeriv-1})
together with (\ref{Eq.FuncDerivativeRule1-1}) and (\ref{Eq.FuncDerivativeRule2-1})
for functional differentiation and noting that many functional derivatives
are zero.

For the T2 term\begin{eqnarray*}
 &  & \int d\mathbf{s\,}\frac{1}{2}\left\{ \left(\phi^{+}(\mathbf{s})\right)\left(\psi^{+}(\mathbf{s})\right)\left(\psi(\mathbf{s})\right)\left(\frac{\delta}{\delta\psi^{+}(\mathbf{s})}\right)\right\} WP\\
 & = & \int d\mathbf{s\,}\frac{1}{2}\left\{ \left(\frac{\delta}{\delta\psi^{+}(\mathbf{s})}\right)\{\phi^{+}(\mathbf{s})\psi^{+}(\mathbf{s})\psi(\mathbf{s})\}-\{\delta_{C}(\mathbf{s},\mathbf{s})\phi^{+}(\mathbf{s})\psi(\mathbf{s})\}\right\} WP\\
 & = & \int d\mathbf{s\,}\frac{1}{2}\left\{ \left(\frac{\delta}{\delta\psi^{+}(\mathbf{s})}\right)\{\phi^{+}(\mathbf{s})\psi^{+}(\mathbf{s})\psi(\mathbf{s})\}\right\} WP\\
 &  & -\int d\mathbf{s\,}\frac{1}{2}\{\delta_{C}(\mathbf{s},\mathbf{s})\phi^{+}(\mathbf{s})\psi(\mathbf{s})\}WP\end{eqnarray*}

For the T3 term\begin{eqnarray*}
 &  & \int d\mathbf{s\,}\frac{1}{2}\left\{ \left(\phi^{+}(\mathbf{s})\right)\left(\psi^{+}(\mathbf{s})\right)\left(\frac{\delta}{\delta\psi^{+}(\mathbf{s})}\right)\left(\psi(\mathbf{s})\right)\right\} WP\\
 & = & \int d\mathbf{s\,}\frac{1}{2}\left\{ \left(\frac{\delta}{\delta\psi^{+}(\mathbf{s})}\right)\{\phi^{+}(\mathbf{s})\psi^{+}(\mathbf{s})\psi(\mathbf{s})\}-\{\delta_{C}(\mathbf{s},\mathbf{s})\phi^{+}(\mathbf{s})\psi(\mathbf{s})\}\right\} WP\\
 & = & \int d\mathbf{s\,}\frac{1}{2}\left\{ \left(\frac{\delta}{\delta\psi^{+}(\mathbf{s})}\right)\{\phi^{+}(\mathbf{s})\psi^{+}(\mathbf{s})\psi(\mathbf{s})\}\right\} WP\\
 &  & -\int d\mathbf{s\,}\frac{1}{2}\{\delta_{C}(\mathbf{s},\mathbf{s})\phi^{+}(\mathbf{s})\psi(\mathbf{s})\}WP\end{eqnarray*}

For the T4 term\begin{eqnarray*}
 &  & \int d\mathbf{s\,}\frac{1}{4}\left\{ \left(\phi^{+}(\mathbf{s})\right)\left(\psi^{+}(\mathbf{s})\right)\left(\frac{\delta}{\delta\psi^{+}(\mathbf{s})}\right)\left(\frac{\delta}{\delta\psi^{+}(\mathbf{s})}\right)\right\} WP\\
 & = & \int d\mathbf{s\,}\frac{1}{4}\left\{ \left(\frac{\delta}{\delta\psi^{+}(\mathbf{s})}\right)\{\phi^{+}(\mathbf{s})\psi^{+}(\mathbf{s})\}-\{\phi^{+}(\mathbf{s})\delta_{C}(\mathbf{s},\mathbf{s})\}\right\} \left(\frac{\delta}{\delta\psi^{+}(\mathbf{s})}\right)WP\\
 & = & \int d\mathbf{s\,}\frac{1}{4}\left\{ \left(\frac{\delta}{\delta\psi^{+}(\mathbf{s})}\right)\left(\frac{\delta}{\delta\psi^{+}(\mathbf{s})}\right)\{\phi^{+}(\mathbf{s})\psi^{+}(\mathbf{s})\}\right\} WP\\
 &  & -\int d\mathbf{s\,}\frac{1}{4}\left\{ \left(\frac{\delta}{\delta\psi^{+}(\mathbf{s})}\right)\{\phi^{+}(\mathbf{s})\delta_{C}(\mathbf{s},\mathbf{s})\}\right\} WP\\
 &  & -\int d\mathbf{s\,}\frac{1}{4}\left(\frac{\delta}{\delta\psi^{+}(\mathbf{s})}\{\phi^{+}(\mathbf{s})\delta_{C}(\mathbf{s},\mathbf{s})\}\right)WP\\
 & = & \int d\mathbf{s\,}\frac{1}{4}\left\{ \left(\frac{\delta}{\delta\psi^{+}(\mathbf{s})}\right)\left(\frac{\delta}{\delta\psi^{+}(\mathbf{s})}\right)\{\phi^{+}(\mathbf{s})\psi^{+}(\mathbf{s})\}\right\} WP\\
 &  & -\int d\mathbf{s\,}\frac{1}{2}\left\{ \left(\frac{\delta}{\delta\psi^{+}(\mathbf{s})}\right)\{\phi^{+}(\mathbf{s})\delta_{C}(\mathbf{s},\mathbf{s})\}\right\} WP\end{eqnarray*}

For the T5 term\begin{eqnarray*}
 &  & \int d\mathbf{s\,}\frac{1}{2}\left\{ \left(\phi^{+}(\mathbf{s})\right)\left(\frac{\delta}{\delta\psi(\mathbf{s})}\right)\left(\psi(\mathbf{s})\right)\left(\psi(\mathbf{s})\right)\right\} WP\\
 & = & \int d\mathbf{s\,}\frac{1}{2}\left\{ \left(\frac{\delta}{\delta\psi(\mathbf{s})}\right)\{\phi^{+}(\mathbf{s})\psi(\mathbf{s})\psi(\mathbf{s})\}\right\} WP\end{eqnarray*}

For the T6 term\begin{eqnarray*}
 &  & \int d\mathbf{s\,}\frac{1}{4}\left\{ \left(\phi^{+}(\mathbf{s})\right)\left(\frac{\delta}{\delta\psi(\mathbf{s})}\right)\left(\psi(\mathbf{s})\right)\left(\frac{\delta}{\delta\psi^{+}(\mathbf{s})}\right)\right\} WP\\
 & = & \int d\mathbf{s\,}\frac{1}{4}\left\{ \left(\frac{\delta}{\delta\psi(\mathbf{s})}\right)\left(\frac{\delta}{\delta\psi^{+}(\mathbf{s})}\right)\{\phi^{+}(\mathbf{s})\psi(\mathbf{s})\}\right\} WP\end{eqnarray*}

For the T7 term\begin{eqnarray*}
 &  & \int d\mathbf{s\,}\frac{1}{4}\left\{ \left(\phi^{+}(\mathbf{s})\right)\left(\frac{\delta}{\delta\psi(\mathbf{s})}\right)\left(\frac{\delta}{\delta\psi^{+}(\mathbf{s})}\right)\left(\psi(\mathbf{s})\right)\right\} WP\\
 & = & \int d\mathbf{s\,}\frac{1}{4}\left\{ \left(\frac{\delta}{\delta\psi(\mathbf{s})}\right)\left(\frac{\delta}{\delta\psi^{+}(\mathbf{s})}\right)\{\phi^{+}(\mathbf{s})\psi(\mathbf{s})\}\right\} WP\end{eqnarray*}

For the T8 term\begin{eqnarray*}
 &  & \int d\mathbf{s\,}\frac{1}{8}\left\{ \left(\phi^{+}(\mathbf{s})\right)\left(\frac{\delta}{\delta\psi(\mathbf{s})}\right)\left(\frac{\delta}{\delta\psi^{+}(\mathbf{s})}\right)\left(\frac{\delta}{\delta\psi^{+}(\mathbf{s})}\right)\right\} WP\\
 & = & \int d\mathbf{s\,}\frac{1}{8}\left\{ \left(\frac{\delta}{\delta\psi(\mathbf{s})}\right)\left(\frac{\delta}{\delta\psi^{+}(\mathbf{s})}\right)\left(\frac{\delta}{\delta\psi^{+}(\mathbf{s})}\right)\{\phi^{+}(\mathbf{s})\}\right\} WP\end{eqnarray*}

For the T9 term\begin{eqnarray*}
 &  & \int d\mathbf{s\,}\left\{ \left(\frac{\delta}{\delta\phi(\mathbf{s})}\right)\left(\psi^{+}(\mathbf{s})\right)\left(\psi(\mathbf{s})\right)\left(\psi(\mathbf{s})\right)\right\} WP\\
 & = & \int d\mathbf{s\,}\left\{ \left(\frac{\delta}{\delta\phi(\mathbf{s})}\right)\{\psi^{+}(\mathbf{s})\psi(\mathbf{s})\psi(\mathbf{s})\}\right\} WP\end{eqnarray*}

For the T10 term\begin{eqnarray*}
 &  & \int d\mathbf{s\,}\frac{1}{2}\left\{ \left(\frac{\delta}{\delta\phi(\mathbf{s})}\right)\left(\psi^{+}(\mathbf{s})\right)\left(\psi(\mathbf{s})\right)\left(\frac{\delta}{\delta\psi^{+}(\mathbf{s})}\right)\right\} WP\\
 & = & \int d\mathbf{s\,}\frac{1}{2}\left\{ \left(\frac{\delta}{\delta\phi(\mathbf{s})}\right)\left(\frac{\delta}{\delta\psi^{+}(\mathbf{s})}\right)\{\psi^{+}(\mathbf{s})\psi(\mathbf{s})\}\right\} WP\\
 &  & -\int d\mathbf{s\,}\frac{1}{2}\left\{ \left(\frac{\delta}{\delta\phi(\mathbf{s})}\right)\{\delta_{C}(\mathbf{s},\mathbf{s})\psi(\mathbf{s})\}\right\} WP\end{eqnarray*}

For the T11 term\begin{eqnarray*}
 &  & \int d\mathbf{s\,}\frac{1}{2}\left\{ \left(\frac{\delta}{\delta\phi(\mathbf{s})}\right)\left(\psi^{+}(\mathbf{s})\right)\left(\frac{\delta}{\delta\psi^{+}(\mathbf{s})}\right)\left(\psi(\mathbf{s})\right)\right\} WP\\
 & = & \int d\mathbf{s\,}\frac{1}{2}\left\{ \left(\frac{\delta}{\delta\phi(\mathbf{s})}\right)\left(\frac{\delta}{\delta\psi^{+}(\mathbf{s})}\right)\{\psi^{+}(\mathbf{s})\psi(\mathbf{s})\}\right\} WP\\
 &  & -\int d\mathbf{s\,}\frac{1}{2}\left\{ \left(\frac{\delta}{\delta\phi(\mathbf{s})}\right)\{\delta_{C}(\mathbf{s},\mathbf{s})\psi(\mathbf{s})\}\right\} WP\end{eqnarray*}

For the T12 term\begin{eqnarray*}
 &  & \int d\mathbf{s\,}\frac{1}{4}\left\{ \left(\frac{\delta}{\delta\phi(\mathbf{s})}\right)\left(\psi^{+}(\mathbf{s})\right)\left(\frac{\delta}{\delta\psi^{+}(\mathbf{s})}\right)\left(\frac{\delta}{\delta\psi^{+}(\mathbf{s})}\right)\right\} WP\\
 & = & \int d\mathbf{s\,}\frac{1}{4}\left(\frac{\delta}{\delta\phi(\mathbf{s})}\right)\left\{ \left(\frac{\delta}{\delta\psi^{+}(\mathbf{s})}\right)\left(\psi^{+}(\mathbf{s})\right)\left(\frac{\delta}{\delta\psi^{+}(\mathbf{s})}\right)\right\} WP\\
 &  & -\int d\mathbf{s\,}\frac{1}{4}\left(\frac{\delta}{\delta\phi(\mathbf{s})}\right)\left\{ \delta_{C}(\mathbf{s},\mathbf{s})\left(\frac{\delta}{\delta\psi^{+}(\mathbf{s})}\right)\right\} WP\\
 & = & \int d\mathbf{s\,}\frac{1}{4}\left(\frac{\delta}{\delta\phi(\mathbf{s})}\right)\left\{ \left(\frac{\delta}{\delta\psi^{+}(\mathbf{s})}\right)\left(\frac{\delta}{\delta\psi^{+}(\mathbf{s})}\right)\{\psi^{+}(\mathbf{s})\}\right\} WP\\
 &  & -\int d\mathbf{s\,}\frac{1}{4}\left(\frac{\delta}{\delta\phi(\mathbf{s})}\right)\left\{ \left(\frac{\delta}{\delta\psi^{+}(\mathbf{s})}\right)\left\{ \delta_{C}(\mathbf{s},\mathbf{s})\right\} \right\} WP\\
 &  & -\int d\mathbf{s\,}\frac{1}{4}\left\{ \left(\frac{\delta}{\delta\phi(\mathbf{s})}\right)\left(\frac{\delta}{\delta\psi^{+}(\mathbf{s})}\right)\left\{ \delta_{C}(\mathbf{s},\mathbf{s})\right\} \right\} WP\\
 & = & \int d\mathbf{s\,}\frac{1}{4}\left\{ \left(\frac{\delta}{\delta\phi(\mathbf{s})}\right)\left(\frac{\delta}{\delta\psi^{+}(\mathbf{s})}\right)\left(\frac{\delta}{\delta\psi^{+}(\mathbf{s})}\right)\{\psi^{+}(\mathbf{s})\}\right\} WP\\
 &  & -\int d\mathbf{s\,}\frac{1}{2}\left\{ \left(\frac{\delta}{\delta\phi(\mathbf{s})}\right)\left(\frac{\delta}{\delta\psi^{+}(\mathbf{s})}\right)\left\{ \delta_{C}(\mathbf{s},\mathbf{s})\right\} \right\} WP\end{eqnarray*}

For the T13 term\begin{eqnarray*}
 &  & \int d\mathbf{s\,}\frac{1}{2}\left\{ \left(\frac{\delta}{\delta\phi(\mathbf{s})}\right)\left(\frac{\delta}{\delta\psi(\mathbf{s})}\right)\left(\psi(\mathbf{s})\right)\left(\psi(\mathbf{s})\right)\right\} WP\\
 & = & \int d\mathbf{s\,}\frac{1}{2}\left\{ \left(\frac{\delta}{\delta\phi(\mathbf{s})}\right)\left(\frac{\delta}{\delta\psi(\mathbf{s})}\right)\{\psi(\mathbf{s})\psi(\mathbf{s})\}\right\} WP\end{eqnarray*}

For the T14 term\begin{eqnarray*}
 &  & \int d\mathbf{s\,}\frac{1}{4}\left\{ \left(\frac{\delta}{\delta\phi(\mathbf{s})}\right)\left(\frac{\delta}{\delta\psi(\mathbf{s})}\right)\left(\psi(\mathbf{s})\right)\left(\frac{\delta}{\delta\psi^{+}(\mathbf{s})}\right)\right\} WP\\
 & = & \int d\mathbf{s\,}\frac{1}{4}\left\{ \left(\frac{\delta}{\delta\phi(\mathbf{s})}\right)\left(\frac{\delta}{\delta\psi(\mathbf{s})}\right)\left(\frac{\delta}{\delta\psi^{+}(\mathbf{s})}\right)\{\psi(\mathbf{s})\}\right\} WP\end{eqnarray*}

For the T15 term\begin{eqnarray*}
 &  & \int d\mathbf{s\,}\frac{1}{4}\left\{ \left(\frac{\delta}{\delta\phi(\mathbf{s})}\right)\left(\frac{\delta}{\delta\psi(\mathbf{s})}\right)\left(\frac{\delta}{\delta\psi^{+}(\mathbf{s})}\right)\left(\psi(\mathbf{s})\right)\right\} WP\\
 & = & \int d\mathbf{s\,}\frac{1}{4}\left\{ \left(\frac{\delta}{\delta\phi(\mathbf{s})}\right)\left(\frac{\delta}{\delta\psi(\mathbf{s})}\right)\left(\frac{\delta}{\delta\psi^{+}(\mathbf{s})}\right)\{\psi(\mathbf{s})\}\right\} WP\end{eqnarray*}

For the T18 term\begin{eqnarray*}
 &  & \int d\mathbf{s\,}\frac{1}{2}\left\{ \mathbf{\left(\psi^{+}(\mathbf{s})\right)\left(\psi^{+}(\mathbf{s})\right)\left(\frac{\delta}{\delta\psi^{+}(\mathbf{s})}\right)}\left(\mathbf{\phi(\mathbf{s})}\right)\right\} WP\\
 & = & \int d\mathbf{s\,}\frac{1}{2}\left\{ \mathbf{\left(\frac{\delta}{\delta\psi^{+}(\mathbf{s})}\right)\{\psi^{+}(\mathbf{s})\psi^{+}(\mathbf{s})\phi(\mathbf{s})\}}\right\} WP\\
 &  & -\int d\mathbf{s\,}\frac{1}{2}\left\{ \mathbf{2\delta_{C}(\mathbf{s},\mathbf{s})\psi^{+}(\mathbf{s})\phi(\mathbf{s})}\right\} WP\end{eqnarray*}

For the T19 term\begin{eqnarray*}
 &  & \int d\mathbf{s\,}\frac{1}{2}\left\{ \mathbf{\left(\psi^{+}(\mathbf{s})\right)\left(\frac{\delta}{\delta\psi(\mathbf{s})}\right)\left(\psi(\mathbf{s})\right)}\left(\mathbf{\phi(\mathbf{s})}\right)\right\} WP\\
 & = & \int d\mathbf{s\,}\frac{1}{2}\left\{ \mathbf{\left(\frac{\delta}{\delta\psi(\mathbf{s})}\right)\{\psi^{+}(\mathbf{s})\psi(\mathbf{s})\phi(\mathbf{s})\}}\right\} WP\end{eqnarray*}

For the T20 term\begin{eqnarray*}
 &  & \int d\mathbf{s\,}\frac{1}{4}\left\{ \mathbf{\left(\psi^{+}(\mathbf{s})\right)\left(\frac{\delta}{\delta\psi(\mathbf{s})}\right)\left(\frac{\delta}{\delta\psi^{+}(\mathbf{s})}\right)}\left(\mathbf{\phi(\mathbf{s})}\right)\right\} WP\\
 & = & \int d\mathbf{s\,}\frac{1}{4}\left\{ \mathbf{\left(\frac{\delta}{\delta\psi(\mathbf{s})}\right)\left(\psi^{+}(\mathbf{s})\right)\left(\frac{\delta}{\delta\psi^{+}(\mathbf{s})}\right)}\left(\mathbf{\phi(\mathbf{s})}\right)\right\} WP\\
 & = & \int d\mathbf{s\,}\frac{1}{4}\left\{ \mathbf{\left(\frac{\delta}{\delta\psi(\mathbf{s})}\right)\left(\frac{\delta}{\delta\psi^{+}(\mathbf{s})}\right)\{\psi^{+}(\mathbf{s})\phi(\mathbf{s})\}}\right\} WP\\
 &  & -\int d\mathbf{s\,}\frac{1}{4}\left\{ \mathbf{\left(\frac{\delta}{\delta\psi(\mathbf{s})}\right)\{\delta_{C}(\mathbf{s},\mathbf{s})\phi(\mathbf{s})\}}\right\} WP\end{eqnarray*}

For the T21 term\begin{eqnarray*}
 &  & \int d\mathbf{s\,}\frac{1}{2}\left\{ \mathbf{\left(\frac{\delta}{\delta\psi(\mathbf{s})}\right)\left(\psi^{+}(\mathbf{s})\right)\left(\psi(\mathbf{s})\right)}\left(\mathbf{\phi(\mathbf{s})}\right)\right\} WP\\
 & = & \int d\mathbf{s\,}\frac{1}{2}\left\{ \mathbf{\left(\frac{\delta}{\delta\psi(\mathbf{s})}\right)\{\psi^{+}(\mathbf{s})\psi(\mathbf{s})\phi(\mathbf{s})\}}\right\} WP\end{eqnarray*}

For the T22 term\begin{eqnarray*}
 &  & \int d\mathbf{s\,}\frac{1}{4}\left\{ \mathbf{\left(\frac{\delta}{\delta\psi(\mathbf{s})}\right)\left(\psi^{+}(\mathbf{s})\right)\left(\frac{\delta}{\delta\psi^{+}(\mathbf{s})}\right)}\left(\mathbf{\phi(\mathbf{s})}\right)\right\} WP\\
 & = & \int d\mathbf{s\,}\frac{1}{4}\left\{ \mathbf{\left(\frac{\delta}{\delta\psi(\mathbf{s})}\right)\left(\frac{\delta}{\delta\psi^{+}(\mathbf{s})}\right)\{\psi^{+}(\mathbf{s})\phi(\mathbf{s})\}}\right\} WP\\
 &  & -\int d\mathbf{s\,}\frac{1}{4}\left\{ \mathbf{\left(\frac{\delta}{\delta\psi(\mathbf{s})}\right)\{\delta_{C}(\mathbf{s},\mathbf{s})\phi(\mathbf{s})\}}\right\} WP\end{eqnarray*}

For the T23 term\[
\int d\mathbf{s\,}\frac{1}{4}\left\{ \mathbf{\left(\frac{\delta}{\delta\psi(\mathbf{s})}\right)\left(\frac{\delta}{\delta\psi(\mathbf{s})}\right)\left(\psi(\mathbf{s})\right)}\left(\mathbf{\phi(\mathbf{s})}\right)\right\} WP\]

For the T24 term\begin{eqnarray*}
 &  & \int d\mathbf{s\,}\frac{1}{8}\left\{ \mathbf{\left(\frac{\delta}{\delta\psi(\mathbf{s})}\right)\left(\frac{\delta}{\delta\psi(\mathbf{s})}\right)\left(\frac{\delta}{\delta\psi^{+}(\mathbf{s})}\right)}\left(\mathbf{\phi(\mathbf{s})}\right)\right\} WP\\
 & = & \int d\mathbf{s\,}\frac{1}{8}\left\{ \mathbf{\left(\frac{\delta}{\delta\psi(\mathbf{s})}\right)\left(\frac{\delta}{\delta\psi(\mathbf{s})}\right)\left(\frac{\delta}{\delta\psi^{+}(\mathbf{s})}\right)\{\phi(\mathbf{s})\}}\right\} WP\end{eqnarray*}

Hence substituting these results we have

\begin{eqnarray}
 &  & WP[\psi(\mathbf{r}),\psi^{+}(\mathbf{r}),\phi(\mathbf{r}),\phi^{+}(\mathbf{r})]_{1-16}\nonumber \\
 & = & g\int d\mathbf{s\,}\left\{ \phi^{+}(\mathbf{s})\psi^{+}(\mathbf{s})\psi(\mathbf{s})\psi(\mathbf{s})\right\} WP\qquad\emph{F1}\nonumber \\
 &  & +g\int d\mathbf{s\,}\frac{1}{2}\left\{ \left(\frac{\delta}{\delta\psi^{+}(\mathbf{s})}\right)\{\phi^{+}(\mathbf{s})\psi^{+}(\mathbf{s})\psi(\mathbf{s})\}\right\} WP\qquad\emph{F2.1}\nonumber \\
 &  & -g\int d\mathbf{s\,}\frac{1}{2}\{\delta_{C}(\mathbf{s},\mathbf{s})\phi^{+}(\mathbf{s})\psi(\mathbf{s})\}WP\qquad\emph{F2.2}\nonumber \\
 &  & +g\int d\mathbf{s\,}\frac{1}{2}\left\{ \left(\frac{\delta}{\delta\psi^{+}(\mathbf{s})}\right)\{\phi^{+}(\mathbf{s})\psi^{+}(\mathbf{s})\psi(\mathbf{s})\}\right\} WP\qquad\emph{F3.1}\nonumber \\
 &  & -g\int d\mathbf{s\,}\frac{1}{2}\{\delta_{C}(\mathbf{s},\mathbf{s})\phi^{+}(\mathbf{s})\psi(\mathbf{s})\}WP\qquad\emph{F3.2}\nonumber \\
 &  & +g\int d\mathbf{s\,}\frac{1}{4}\left\{ \left(\frac{\delta}{\delta\psi^{+}(\mathbf{s})}\right)\left(\frac{\delta}{\delta\psi^{+}(\mathbf{s})}\right)\{\phi^{+}(\mathbf{s})\psi^{+}(\mathbf{s})\}\right\} WP\qquad\emph{F4.1}\nonumber \\
 &  & -g\int d\mathbf{s\,}\frac{1}{2}\left\{ \left(\frac{\delta}{\delta\psi^{+}(\mathbf{s})}\right)\{\phi^{+}(\mathbf{s})\delta_{C}(\mathbf{s},\mathbf{s})\}\right\} WP\qquad\emph{F4.2}\nonumber \\
 &  & -g\int d\mathbf{s\,}\frac{1}{2}\left\{ \left(\frac{\delta}{\delta\psi(\mathbf{s})}\right)\{\phi^{+}(\mathbf{s})\psi(\mathbf{s})\psi(\mathbf{s})\}\right\} WP\qquad\emph{F5}\nonumber \\
 &  & -g\int d\mathbf{s\,}\frac{1}{4}\left\{ \left(\frac{\delta}{\delta\psi(\mathbf{s})}\right)\left(\frac{\delta}{\delta\psi^{+}(\mathbf{s})}\right)\{\phi^{+}(\mathbf{s})\psi(\mathbf{s})\}\right\} WP\qquad\emph{F6}\nonumber \\
 &  & -g\int d\mathbf{s\,}\frac{1}{4}\left\{ \left(\frac{\delta}{\delta\psi(\mathbf{s})}\right)\left(\frac{\delta}{\delta\psi^{+}(\mathbf{s})}\right)\{\phi^{+}(\mathbf{s})\psi(\mathbf{s})\}\right\} WP\qquad\emph{F7}\nonumber \\
 &  & -g\int d\mathbf{s\,}\frac{1}{8}\left\{ \left(\frac{\delta}{\delta\psi(\mathbf{s})}\right)\left(\frac{\delta}{\delta\psi^{+}(\mathbf{s})}\right)\left(\frac{\delta}{\delta\psi^{+}(\mathbf{s})}\right)\{\phi^{+}(\mathbf{s})\}\right\} WP\qquad\emph{F8}\nonumber \\
 &  & -g\int d\mathbf{s\,}\left\{ \left(\frac{\delta}{\delta\phi(\mathbf{s})}\right)\{\psi^{+}(\mathbf{s})\psi(\mathbf{s})\psi(\mathbf{s})\}\right\} WP\qquad\emph{F9}\nonumber \\
 &  & -g\int d\mathbf{s\,}\frac{1}{2}\left\{ \left(\frac{\delta}{\delta\phi(\mathbf{s})}\right)\left(\frac{\delta}{\delta\psi^{+}(\mathbf{s})}\right)\{\psi^{+}(\mathbf{s})\psi(\mathbf{s})\}\right\} WP\qquad\emph{F10.1}\nonumber \\
 &  & +g\int d\mathbf{s\,}\frac{1}{2}\left\{ \left(\frac{\delta}{\delta\phi(\mathbf{s})}\right)\{\delta_{C}(\mathbf{s},\mathbf{s})\psi(\mathbf{s})\}\right\} WP\qquad\emph{F10.2}\nonumber \\
 &  & -g\int d\mathbf{s\,}\frac{1}{2}\left\{ \left(\frac{\delta}{\delta\phi(\mathbf{s})}\right)\left(\frac{\delta}{\delta\psi^{+}(\mathbf{s})}\right)\{\psi^{+}(\mathbf{s})\psi(\mathbf{s})\}\right\} WP\qquad\emph{F11.1}\nonumber \\
 &  & +g\int d\mathbf{s\,}\frac{1}{2}\left\{ \left(\frac{\delta}{\delta\phi(\mathbf{s})}\right)\{\delta_{C}(\mathbf{s},\mathbf{s})\psi(\mathbf{s})\}\right\} WP\qquad\emph{F11.2}\nonumber \\
 &  & -g\int d\mathbf{s\,}\frac{1}{4}\left\{ \left(\frac{\delta}{\delta\phi(\mathbf{s})}\right)\left(\frac{\delta}{\delta\psi^{+}(\mathbf{s})}\right)\left(\frac{\delta}{\delta\psi^{+}(\mathbf{s})}\right)\{\psi^{+}(\mathbf{s})\}\right\} WP\qquad\emph{F12.1}\nonumber \\
 &  & +g\int d\mathbf{s\,}\frac{1}{2}\left\{ \left(\frac{\delta}{\delta\phi(\mathbf{s})}\right)\left(\frac{\delta}{\delta\psi^{+}(\mathbf{s})}\right)\left\{ \delta_{C}(\mathbf{s},\mathbf{s})\right\} \right\} WP\qquad\emph{F12.2}\nonumber \\
 &  & +g\int d\mathbf{s\,}\frac{1}{2}\left\{ \left(\frac{\delta}{\delta\phi(\mathbf{s})}\right)\left(\frac{\delta}{\delta\psi(\mathbf{s})}\right)\{\psi(\mathbf{s})\psi(\mathbf{s})\}\right\} WP\qquad\emph{F13}\nonumber \\
 &  & +g\int d\mathbf{s\,}\frac{1}{4}\left\{ \left(\frac{\delta}{\delta\phi(\mathbf{s})}\right)\left(\frac{\delta}{\delta\psi(\mathbf{s})}\right)\left(\frac{\delta}{\delta\psi^{+}(\mathbf{s})}\right)\{\psi(\mathbf{s})\}\right\} WP\qquad\emph{F14}\nonumber \\
 &  & +g\int d\mathbf{s\,}\frac{1}{4}\left\{ \left(\frac{\delta}{\delta\phi(\mathbf{s})}\right)\left(\frac{\delta}{\delta\psi(\mathbf{s})}\right)\left(\frac{\delta}{\delta\psi^{+}(\mathbf{s})}\right)\{\psi(\mathbf{s})\}\right\} WP\qquad\emph{F15}\nonumber \\
 &  & +g\int d\mathbf{s\,}\frac{1}{8}\left\{ \left(\frac{\delta}{\delta\phi(\mathbf{s})}\right)\left(\frac{\delta}{\delta\psi(\mathbf{s})}\right)\left(\frac{\delta}{\delta\psi^{+}(\mathbf{s})}\right)\left(\frac{\delta}{\delta\psi^{+}(\mathbf{s})}\right)\right\} WP\qquad\emph{F16}\nonumber \\
 &  & \,\end{eqnarray}
and\begin{eqnarray}
 &  & WP[\psi(\mathbf{r}),\psi^{+}(\mathbf{r}),\phi(\mathbf{r}),\phi^{+}(\mathbf{r})]_{17-24}\nonumber \\
 & = & +g\int d\mathbf{s\,}\left\{ \mathbf{\psi^{+}(\mathbf{s})\psi^{+}(\mathbf{s})\psi(\mathbf{s})\phi(\mathbf{s})}\right\} WP\qquad\emph{F17}\nonumber \\
 &  & +g\int d\mathbf{s\,}\frac{1}{2}\left\{ \mathbf{\left(\frac{\delta}{\delta\psi^{+}(\mathbf{s})}\right)\{\psi^{+}(\mathbf{s})\psi^{+}(\mathbf{s})\phi(\mathbf{s})\}}\right\} WP\qquad\emph{F18.1}\nonumber \\
 &  & -g\int d\mathbf{s\,}\frac{1}{2}\left\{ \mathbf{2\delta_{C}(\mathbf{s},\mathbf{s})\psi^{+}(\mathbf{s})\phi(\mathbf{s})}\right\} WP\qquad\emph{F18.2}\nonumber \\
 &  & -g\int d\mathbf{s\,}\frac{1}{2}\left\{ \mathbf{\left(\frac{\delta}{\delta\psi(\mathbf{s})}\right)\{\psi^{+}(\mathbf{s})\psi(\mathbf{s})\phi(\mathbf{s})\}}\right\} WP\qquad\emph{F19}\nonumber \\
 &  & -g\int d\mathbf{s\,}\frac{1}{4}\left\{ \mathbf{\left(\frac{\delta}{\delta\psi(\mathbf{s})}\right)\left(\frac{\delta}{\delta\psi^{+}(\mathbf{s})}\right)\{\psi^{+}(\mathbf{s})\phi(\mathbf{s})\}}\right\} WP\qquad\emph{F20.1}\nonumber \\
 &  & +g\int d\mathbf{s\,}\frac{1}{4}\left\{ \mathbf{\left(\frac{\delta}{\delta\psi(\mathbf{s})}\right)\{\delta_{C}(\mathbf{s},\mathbf{s})\phi(\mathbf{s})\}}\right\} WP\qquad\emph{F20.2}\nonumber \\
 &  & -g\int d\mathbf{s\,}\frac{1}{2}\left\{ \mathbf{\left(\frac{\delta}{\delta\psi(\mathbf{s})}\right)\{\psi^{+}(\mathbf{s})\psi(\mathbf{s})\phi(\mathbf{s})\}}\right\} WP\qquad\emph{F21}\nonumber \\
 &  & -g\int d\mathbf{s\,}\frac{1}{4}\left\{ \mathbf{\left(\frac{\delta}{\delta\psi(\mathbf{s})}\right)\left(\frac{\delta}{\delta\psi^{+}(\mathbf{s})}\right)\{\psi^{+}(\mathbf{s})\phi(\mathbf{s})\}}\right\} WP\qquad\emph{F22.1}\nonumber \\
 &  & +g\int d\mathbf{s\,}\frac{1}{4}\left\{ \mathbf{\left(\frac{\delta}{\delta\psi(\mathbf{s})}\right)\{\delta_{C}(\mathbf{s},\mathbf{s})\phi(\mathbf{s})\}}\right\} WP\qquad\emph{F22.2}\nonumber \\
 &  & +g\int d\mathbf{s\,}\frac{1}{4}\left\{ \mathbf{\left(\frac{\delta}{\delta\psi(\mathbf{s})}\right)\left(\frac{\delta}{\delta\psi(\mathbf{s})}\right)\left(\psi(\mathbf{s})\right)}\left(\mathbf{\phi(\mathbf{s})}\right)\right\} WP\qquad\emph{F23}\nonumber \\
 &  & +g\int d\mathbf{s\,}\frac{1}{8}\left\{ \mathbf{\left(\frac{\delta}{\delta\psi(\mathbf{s})}\right)\left(\frac{\delta}{\delta\psi(\mathbf{s})}\right)\left(\frac{\delta}{\delta\psi^{+}(\mathbf{s})}\right)\{\phi(\mathbf{s})\}}\right\} WP\qquad\emph{F24 }\nonumber \\
 &  & \,\end{eqnarray}

Collecting terms with the same order of functional derivatives we
have\begin{eqnarray}
 &  & WP[\psi(\mathbf{r}),\psi^{+}(\mathbf{r}),\phi(\mathbf{r}),\phi^{+}(\mathbf{r})]\nonumber \\
 & \rightarrow & WP^{0}+WP^{1}+WP^{2}+WP^{3}+WP^{4}\nonumber \\
 &  & \,\label{eq:V14RhoResult1}\end{eqnarray}
where we have used upper subscripts for the $\,\widehat{V}_{14}\widehat{\rho}$
contributions and\begin{eqnarray}
 &  & WP^{0}\nonumber \\
 & = & g\int d\mathbf{s\,}\left\{ \phi^{+}(\mathbf{s})\psi^{+}(\mathbf{s})\psi(\mathbf{s})\psi(\mathbf{s})\right\} WP\qquad\emph{F1}\nonumber \\
 &  & +g\int d\mathbf{s\,}\frac{1}{2}\left\{ -\{\delta_{C}(\mathbf{s},\mathbf{s})\phi^{+}(\mathbf{s})\psi(\mathbf{s})\}\right\} WP\qquad\emph{F2.2}\nonumber \\
 &  & +g\int d\mathbf{s\,}\frac{1}{2}\left\{ -\{\delta_{C}(\mathbf{s},\mathbf{s})\phi^{+}(\mathbf{s})\psi(\mathbf{s})\}\right\} WP\qquad\emph{F3.2}\nonumber \\
 &  & +g\int d\mathbf{s\,}\left\{ \mathbf{\psi^{+}(\mathbf{s})\psi^{+}(\mathbf{s})\psi(\mathbf{s})\phi(\mathbf{s})}\right\} WP\qquad\emph{F17}\nonumber \\
 &  & -g\int d\mathbf{s\,}\frac{1}{2}\left\{ \mathbf{2\delta_{C}(\mathbf{s},\mathbf{s})\psi^{+}(\mathbf{s})\phi(\mathbf{s})}\right\} WP\qquad\emph{F18}\nonumber \\
 & = & g\int d\mathbf{s\,}\left\{ \phi^{+}(\mathbf{s})\psi^{+}(\mathbf{s})\psi(\mathbf{s})\psi(\mathbf{s})\right\} WP[\psi,\psi^{+},\phi,\phi^{+}]\nonumber \\
 &  & +g\int d\mathbf{s\,}\left\{ \mathbf{\psi^{+}(\mathbf{s})\psi^{+}(\mathbf{s})\psi(\mathbf{s})\phi(\mathbf{s})}\right\} WP[\psi,\psi^{+},\phi,\phi^{+}]\nonumber \\
 &  & -g\int d\mathbf{s\,}\{\delta_{C}(\mathbf{s},\mathbf{s})\phi^{+}(\mathbf{s})\psi(\mathbf{s})\}WP[\psi,\psi^{+},\phi,\phi^{+}]\nonumber \\
 &  & -g\int d\mathbf{s\,}\{\delta_{C}(\mathbf{s},\mathbf{s})\psi^{+}(\mathbf{s})\phi(\mathbf{s})\}WP[\psi,\psi^{+},\phi,\phi^{+}]\nonumber \\
 &  & \,\end{eqnarray}
\begin{eqnarray}
 &  & WP^{1}\nonumber \\
 & = & +g\int d\mathbf{s\,}\frac{1}{2}\left\{ \left(\frac{\delta}{\delta\psi^{+}(\mathbf{s})}\right)\{\phi^{+}(\mathbf{s})\psi^{+}(\mathbf{s})\psi(\mathbf{s})\}\right\} WP\qquad\emph{F2.1}\nonumber \\
 &  & +g\int d\mathbf{s\,}\frac{1}{2}\left\{ \left(\frac{\delta}{\delta\psi^{+}(\mathbf{s})}\right)\{\phi^{+}(\mathbf{s})\psi^{+}(\mathbf{s})\psi(\mathbf{s})\}\right\} WP\qquad\emph{F3.1}\nonumber \\
 &  & -g\int d\mathbf{s\,}\frac{1}{2}\left\{ \left(\frac{\delta}{\delta\psi^{+}(\mathbf{s})}\right)\{\phi^{+}(\mathbf{s})\delta_{C}(\mathbf{s},\mathbf{s})\}\right\} WP\qquad\emph{F4.2}\nonumber \\
 &  & -g\int d\mathbf{s\,}\frac{1}{2}\left\{ \left(\frac{\delta}{\delta\psi(\mathbf{s})}\right)\{\phi^{+}(\mathbf{s})\psi(\mathbf{s})\psi(\mathbf{s})\}\right\} WP\qquad\emph{F5}\nonumber \\
 &  & -g\int d\mathbf{s\,}\left\{ \left(\frac{\delta}{\delta\phi(\mathbf{s})}\right)\{\psi^{+}(\mathbf{s})\psi(\mathbf{s})\psi(\mathbf{s})\}\right\} WP\qquad\emph{F9}\nonumber \\
 &  & +g\int d\mathbf{s\,}\frac{1}{2}\left\{ \left(\frac{\delta}{\delta\phi(\mathbf{s})}\right)\{\delta_{C}(\mathbf{s},\mathbf{s})\psi(\mathbf{s})\}\right\} WP\qquad\emph{F10.2}\nonumber \\
 &  & +g\int d\mathbf{s\,}\frac{1}{2}\left\{ \left(\frac{\delta}{\delta\phi(\mathbf{s})}\right)\{\delta_{C}(\mathbf{s},\mathbf{s})\psi(\mathbf{s})\}\right\} WP\qquad\emph{F11.2}\nonumber \\
 &  & +g\int d\mathbf{s\,}\frac{1}{2}\left\{ \mathbf{\left(\frac{\delta}{\delta\psi^{+}(\mathbf{s})}\right)\{\psi^{+}(\mathbf{s})\psi^{+}(\mathbf{s})\phi(\mathbf{s})\}}\right\} WP\qquad\emph{F18.1}\nonumber \\
 &  & -g\int d\mathbf{s\,}\frac{1}{2}\left\{ \mathbf{\left(\frac{\delta}{\delta\psi(\mathbf{s})}\right)\{\psi^{+}(\mathbf{s})\psi(\mathbf{s})\phi(\mathbf{s})\}}\right\} WP\qquad\emph{F19}\nonumber \\
 &  & +g\int d\mathbf{s\,}\frac{1}{4}\left\{ \mathbf{\left(\frac{\delta}{\delta\psi(\mathbf{s})}\right)\{\delta_{C}(\mathbf{s},\mathbf{s})\phi(\mathbf{s})\}}\right\} WP\qquad\emph{F20.2}\nonumber \\
 &  & -g\int d\mathbf{s\,}\frac{1}{2}\left\{ \mathbf{\left(\frac{\delta}{\delta\psi(\mathbf{s})}\right)\{\psi^{+}(\mathbf{s})\psi(\mathbf{s})\phi(\mathbf{s})\}}\right\} WP\qquad\emph{F21}\nonumber \\
 &  & +g\int d\mathbf{s\,}\frac{1}{4}\left\{ \mathbf{\left(\frac{\delta}{\delta\psi(\mathbf{s})}\right)\{\delta_{C}(\mathbf{s},\mathbf{s})\phi(\mathbf{s})\}}\right\} WP\qquad\emph{F22.2}\nonumber \\
 & = & +g\int d\mathbf{s\,}\left\{ \left(\frac{\delta}{\delta\psi^{+}(\mathbf{s})}\right)\{\phi^{+}(\mathbf{s})\psi^{+}(\mathbf{s})\psi(\mathbf{s})\}\right\} WP[\psi,\psi^{+},\phi,\phi^{+}]\nonumber \\
 &  & +g\int d\mathbf{s\,}\frac{1}{2}\left\{ \left(\frac{\delta}{\delta\psi^{+}(\mathbf{s})}\right)\{\psi^{+}(\mathbf{s})\psi^{+}(\mathbf{s})\phi(\mathbf{s})\}\right\} WP[\psi,\psi^{+},\phi,\phi^{+}]\nonumber \\
 &  & -g\int d\mathbf{s\,}\frac{1}{2}\left\{ \left(\frac{\delta}{\delta\psi^{+}(\mathbf{s})}\right)\{\phi^{+}(\mathbf{s})\delta_{C}(\mathbf{s},\mathbf{s})\}\right\} WP[\psi,\psi^{+},\phi,\phi^{+}]\nonumber \\
 &  & -g\int d\mathbf{s\,}\left\{ \left(\frac{\delta}{\delta\psi(\mathbf{s})}\right)\{\psi^{+}(\mathbf{s})\psi(\mathbf{s})\phi(\mathbf{s})\}\right\} WP[\psi,\psi^{+},\phi,\phi^{+}]\nonumber \\
 &  & -g\int d\mathbf{s\,}\frac{1}{2}\left\{ \left(\frac{\delta}{\delta\psi(\mathbf{s})}\right)\{\phi^{+}(\mathbf{s})\psi(\mathbf{s})\psi(\mathbf{s})\}\right\} WP[\psi,\psi^{+},\phi,\phi^{+}]\nonumber \\
 &  & +g\int d\mathbf{s\,}\frac{1}{2}\left\{ \left(\frac{\delta}{\delta\psi(\mathbf{s})}\right)\{\delta_{C}(\mathbf{s},\mathbf{s})\phi(\mathbf{s})\}\right\} WP[\psi,\psi^{+},\phi,\phi^{+}]\nonumber \\
 &  & -g\int d\mathbf{s\,}\left\{ \left(\frac{\delta}{\delta\phi(\mathbf{s})}\right)\{\psi^{+}(\mathbf{s})\psi(\mathbf{s})\psi(\mathbf{s})\}\right\} WP[\psi,\psi^{+},\phi,\phi^{+}]\nonumber \\
 &  & +g\int d\mathbf{s\,}\left\{ \left(\frac{\delta}{\delta\phi(\mathbf{s})}\right)\{\delta_{C}(\mathbf{s},\mathbf{s})\psi(\mathbf{s})\}\right\} WP[\psi,\psi^{+},\phi,\phi^{+}]\label{Eq.V14RhoResult3}\end{eqnarray}

\begin{eqnarray}
 &  & WP^{2}\nonumber \\
 & = & +g\int d\mathbf{s\,}\frac{1}{4}\left\{ \left(\frac{\delta}{\delta\psi^{+}(\mathbf{s})}\right)\left(\frac{\delta}{\delta\psi^{+}(\mathbf{s})}\right)\{\phi^{+}(\mathbf{s})\psi^{+}(\mathbf{s})\}\right\} WP\qquad\emph{F4.1}\nonumber \\
 &  & -g\int d\mathbf{s\,}\frac{1}{4}\left\{ \left(\frac{\delta}{\delta\psi(\mathbf{s})}\right)\left(\frac{\delta}{\delta\psi^{+}(\mathbf{s})}\right)\{\phi^{+}(\mathbf{s})\psi(\mathbf{s})\}\right\} WP\qquad\emph{F6}\nonumber \\
 &  & -g\int d\mathbf{s\,}\frac{1}{4}\left\{ \left(\frac{\delta}{\delta\psi(\mathbf{s})}\right)\left(\frac{\delta}{\delta\psi^{+}(\mathbf{s})}\right)\{\phi^{+}(\mathbf{s})\psi(\mathbf{s})\}\right\} WP\qquad\emph{F7}\nonumber \\
 &  & -g\int d\mathbf{s\,}\frac{1}{2}\left\{ \left(\frac{\delta}{\delta\phi(\mathbf{s})}\right)\left(\frac{\delta}{\delta\psi^{+}(\mathbf{s})}\right)\{\psi^{+}(\mathbf{s})\psi(\mathbf{s})\}\right\} WP\qquad\emph{F10.1}\nonumber \\
 &  & -g\int d\mathbf{s\,}\frac{1}{2}\left\{ \left(\frac{\delta}{\delta\phi(\mathbf{s})}\right)\left(\frac{\delta}{\delta\psi^{+}(\mathbf{s})}\right)\{\psi^{+}(\mathbf{s})\psi(\mathbf{s})\}\right\} WP\qquad\emph{F11.1}\nonumber \\
 &  & +g\int d\mathbf{s\,}\frac{1}{2}\left\{ \left(\frac{\delta}{\delta\phi(\mathbf{s})}\right)\left(\frac{\delta}{\delta\psi^{+}(\mathbf{s})}\right)\left\{ \delta_{C}(\mathbf{s},\mathbf{s})\right\} \right\} WP\qquad\emph{F12.2}\nonumber \\
 &  & +g\int d\mathbf{s\,}\frac{1}{2}\left\{ \left(\frac{\delta}{\delta\phi(\mathbf{s})}\right)\left(\frac{\delta}{\delta\psi(\mathbf{s})}\right)\{\psi(\mathbf{s})\psi(\mathbf{s})\}\right\} WP\qquad\emph{F13}\nonumber \\
 &  & -g\int d\mathbf{s\,}\frac{1}{4}\left\{ \left(\frac{\delta}{\delta\psi(\mathbf{s})}\right)\left(\frac{\delta}{\delta\psi^{+}(\mathbf{s})}\right)\{\psi^{+}(\mathbf{s})\phi(\mathbf{s})\}\right\} WP\qquad\emph{F20.1}\nonumber \\
 &  & -g\int d\mathbf{s\,}\frac{1}{4}\left\{ \left(\frac{\delta}{\delta\psi(\mathbf{s})}\right)\left(\frac{\delta}{\delta\psi^{+}(\mathbf{s})}\right)\{\psi^{+}(\mathbf{s})\phi(\mathbf{s})\}\right\} WP\qquad\emph{F22.1}\nonumber \\
 &  & +g\int d\mathbf{s\,}\frac{1}{4}\left\{ \left(\frac{\delta}{\delta\psi(\mathbf{s})}\right)\left(\frac{\delta}{\delta\psi(\mathbf{s})}\right)\left(\psi(\mathbf{s})\right)\left(\mathbf{\phi(\mathbf{s})}\right)\right\} WP\qquad\emph{F23}\nonumber \\
 & = & +g\int d\mathbf{s\,}\frac{1}{4}\left\{ \left(\frac{\delta}{\delta\psi^{+}(\mathbf{s})}\right)\left(\frac{\delta}{\delta\psi^{+}(\mathbf{s})}\right)\{\phi^{+}(\mathbf{s})\psi^{+}(\mathbf{s})\}\right\} WP[\psi,\psi^{+},\phi,\phi^{+}]\nonumber \\
 &  & -g\int d\mathbf{s\,}\frac{1}{2}\left\{ \left(\frac{\delta}{\delta\psi^{+}(\mathbf{s})}\right)\left(\frac{\delta}{\delta\psi(\mathbf{s})}\right)\{\phi^{+}(\mathbf{s})\psi(\mathbf{s})\}\right\} WP[\psi,\psi^{+},\phi,\phi^{+}]\nonumber \\
 &  & -g\int d\mathbf{s\,}\left\{ \left(\frac{\delta}{\delta\psi^{+}(\mathbf{s})}\right)\left(\frac{\delta}{\delta\phi(\mathbf{s})}\right)\{\psi^{+}(\mathbf{s})\psi(\mathbf{s})\}\right\} WP[\psi,\psi^{+},\phi,\phi^{+}]\nonumber \\
 &  & +g\int d\mathbf{s\,}\frac{1}{2}\left\{ \left(\frac{\delta}{\delta\psi^{+}(\mathbf{s})}\right)\left(\frac{\delta}{\delta\phi(\mathbf{s})}\right)\left\{ \delta_{C}(\mathbf{s},\mathbf{s})\right\} \right\} WP[\psi,\psi^{+},\phi,\phi^{+}]\nonumber \\
 &  & +g\int d\mathbf{s\,}\frac{1}{2}\left\{ \left(\frac{\delta}{\delta\psi(\mathbf{s})}\right)\left(\frac{\delta}{\delta\phi(\mathbf{s})}\right)\{\psi(\mathbf{s})\psi(\mathbf{s})\}\right\} WP[\psi,\psi^{+},\phi,\phi^{+}]\nonumber \\
 &  & -g\int d\mathbf{s\,}\frac{1}{2}\left\{ \left(\frac{\delta}{\delta\psi^{+}(\mathbf{s})}\right)\left(\frac{\delta}{\delta\psi(\mathbf{s})}\right)\{\psi^{+}(\mathbf{s})\phi(\mathbf{s})\}\right\} WP[\psi,\psi^{+},\phi,\phi^{+}]\nonumber \\
 &  & +g\int d\mathbf{s\,}\frac{1}{4}\left\{ \left(\frac{\delta}{\delta\psi(\mathbf{s})}\right)\left(\frac{\delta}{\delta\psi(\mathbf{s})}\right)\{\psi(\mathbf{s})\phi(\mathbf{s})\}\right\} WP[\psi,\psi^{+},\phi,\phi^{+}]\nonumber \\
 &  & \,\label{eq:V14RhoResult4}\end{eqnarray}

\begin{eqnarray}
 &  & WP^{3}\nonumber \\
 & = & -g\int d\mathbf{s\,}\frac{1}{8}\left\{ \left(\frac{\delta}{\delta\psi^{+}(\mathbf{s})}\right)\left(\frac{\delta}{\delta\psi^{+}(\mathbf{s})}\right)\left(\frac{\delta}{\delta\psi(\mathbf{s})}\right)\{\phi^{+}(\mathbf{s})\}\right\} WP\qquad\emph{F8}\nonumber \\
 &  & -g\int d\mathbf{s\,}\frac{1}{4}\left\{ \left(\frac{\delta}{\delta\phi(\mathbf{s})}\right)\left(\frac{\delta}{\delta\psi^{+}(\mathbf{s})}\right)\left(\frac{\delta}{\delta\psi^{+}(\mathbf{s})}\right)\{\psi^{+}(\mathbf{s})\}\right\} WP\qquad\emph{F12.1}\nonumber \\
 &  & +g\int d\mathbf{s\,}\frac{1}{4}\left\{ \left(\frac{\delta}{\delta\psi^{+}(\mathbf{s})}\right)\left(\frac{\delta}{\delta\psi(\mathbf{s})}\right)\left(\frac{\delta}{\delta\phi(\mathbf{s})}\right)\{\psi(\mathbf{s})\}\right\} WP\qquad\emph{F14}\nonumber \\
 &  & +g\int d\mathbf{s\,}\frac{1}{4}\left\{ \left(\frac{\delta}{\delta\psi^{+}(\mathbf{s})}\right)\left(\frac{\delta}{\delta\psi(\mathbf{s})}\right)\left(\frac{\delta}{\delta\phi(\mathbf{s})}\right)\{\psi(\mathbf{s})\}\right\} WP\qquad\emph{F15}\nonumber \\
 &  & +g\int d\mathbf{s\,}\frac{1}{8}\left\{ \mathbf{\left(\frac{\delta}{\delta\psi^{+}(\mathbf{s})}\right)\left(\frac{\delta}{\delta\psi(\mathbf{s})}\right)\left(\frac{\delta}{\delta\psi(\mathbf{s})}\right)\{\phi(\mathbf{s})\}}\right\} WP\qquad\emph{F24}\nonumber \\
 & = & -g\int d\mathbf{s\,}\frac{1}{8}\left\{ \left(\frac{\delta}{\delta\psi^{+}(\mathbf{s})}\right)\left(\frac{\delta}{\delta\psi^{+}(\mathbf{s})}\right)\left(\frac{\delta}{\delta\psi(\mathbf{s})}\right)\{\phi^{+}(\mathbf{s})\}\right\} WP[\psi,\psi^{+},\phi,\phi^{+}]\nonumber \\
 &  & -g\int d\mathbf{s\,}\frac{1}{4}\left\{ \left(\frac{\delta}{\delta\psi^{+}(\mathbf{s})}\right)\left(\frac{\delta}{\delta\psi^{+}(\mathbf{s})}\right)\left(\frac{\delta}{\delta\phi(\mathbf{s})}\right)\{\psi^{+}(\mathbf{s})\}\right\} WP[\psi,\psi^{+},\phi,\phi^{+}]\nonumber \\
 &  & +g\int d\mathbf{s\,}\frac{1}{2}\left\{ \left(\frac{\delta}{\delta\psi^{+}(\mathbf{s})}\right)\left(\frac{\delta}{\delta\psi(\mathbf{s})}\right)\left(\frac{\delta}{\delta\phi(\mathbf{s})}\right)\{\psi(\mathbf{s})\}\right\} WP[\psi,\psi^{+},\phi,\phi^{+}]\nonumber \\
 &  & +g\int d\mathbf{s\,}\frac{1}{8}\left\{ \mathbf{\left(\frac{\delta}{\delta\psi^{+}(\mathbf{s})}\right)\left(\frac{\delta}{\delta\psi(\mathbf{s})}\right)\left(\frac{\delta}{\delta\psi(\mathbf{s})}\right)\{\phi(\mathbf{s})\}}\right\} WP[\psi,\psi^{+},\phi,\phi^{+}]\nonumber \\
 &  & \,\label{eq:V14RhoResult5}\end{eqnarray}
\begin{eqnarray}
 &  & WP^{4}\nonumber \\
 & = & +g\int d\mathbf{s\,}\frac{1}{8}\left\{ \left(\frac{\delta}{\delta\phi(\mathbf{s})}\right)\left(\frac{\delta}{\delta\psi(\mathbf{s})}\right)\left(\frac{\delta}{\delta\psi^{+}(\mathbf{s})}\right)\left(\frac{\delta}{\delta\psi^{+}(\mathbf{s})}\right)\right\} WP\qquad\emph{F16}\nonumber \\
 & = & +g\int d\mathbf{s\,}\frac{1}{8}\left\{ \left(\frac{\delta}{\delta\psi^{+}(\mathbf{s})}\right)\left(\frac{\delta}{\delta\psi^{+}(\mathbf{s})}\right)\left(\frac{\delta}{\delta\psi(\mathbf{s})}\right)\left(\frac{\delta}{\delta\phi(\mathbf{s})}\right)\right\} WP[\psi,\psi^{+},\phi,\phi^{+}]\nonumber \\
 &  & \,\label{eq:V14RhoResult6}\end{eqnarray}

Now if \begin{align}
\widehat{\rho} & \rightarrow\widehat{\rho}\,\widehat{V}_{14}\,=g\int d\mathbf{s\,\widehat{\rho}\,(}\widehat{\Psi}_{NC}^{\dagger}(\mathbf{s})\widehat{\Psi}_{C}^{\dagger}(\mathbf{s})\,\widehat{\Psi}_{C}(\mathbf{s})\widehat{\Psi}_{C}(\mathbf{s}))\nonumber \\
 & +g\int d\mathbf{s\,\widehat{\rho}\,(}\widehat{\Psi}_{C}^{\dagger}(\mathbf{s})\widehat{\Psi}_{C}^{\dagger}(\mathbf{s})\,\widehat{\Psi}_{C}(\mathbf{s})\widehat{\Psi}_{NC}(\mathbf{s}))\nonumber \\
 & \,\end{align}
then\begin{eqnarray}
 &  & WP[\psi(\mathbf{r}),\psi^{+}(\mathbf{r}),\phi(\mathbf{r}),\phi^{+}(\mathbf{r})]\nonumber \\
 & \rightarrow & g\int d\mathbf{s}\left(\psi(\mathbf{s})-\frac{1}{2}\frac{\delta}{\delta\psi^{+}(\mathbf{s})}\right)\left(\psi(\mathbf{s})-\frac{1}{2}\frac{\delta}{\delta\psi^{+}(\mathbf{s})}\right)\left(\psi^{+}(\mathbf{s})+\frac{1}{2}\frac{\delta}{\delta\psi(\mathbf{s})}\right)\nonumber \\
 &  & \times\left(\phi^{+}(\mathbf{s})\right)WP\nonumber \\
 &  & +g\int d\mathbf{s}\left(\phi(\mathbf{s})-\frac{\delta}{\delta\phi^{+}(\mathbf{s})}\right)\left(\psi(\mathbf{s})-\frac{1}{2}\frac{\delta}{\delta\psi^{+}(\mathbf{s})}\right)\left(\psi^{+}(\mathbf{s})+\frac{1}{2}\frac{\delta}{\delta\psi(\mathbf{s})}\right)\nonumber \\
 &  & \times\left(\psi^{+}(\mathbf{s})+\frac{1}{2}\frac{\delta}{\delta\psi(\mathbf{s})}\right)WP\nonumber \\
 &  & \,\end{eqnarray}

Expanding this expression gives\begin{eqnarray}
 &  & WP[\psi(\mathbf{r}),\psi^{+}(\mathbf{r}),\phi(\mathbf{r}),\phi^{+}(\mathbf{r})]\nonumber \\
 & \rightarrow & WP[\psi(\mathbf{r}),\psi^{+}(\mathbf{r}),\phi(\mathbf{r}),\phi^{+}(\mathbf{r})]_{1-8}+WP[\psi(\mathbf{r}),\psi^{+}(\mathbf{r}),\phi(\mathbf{r}),\phi^{+}(\mathbf{r})]_{9-24}\nonumber \\
 &  & \,\end{eqnarray}
where\begin{eqnarray*}
 &  & WP[\psi(\mathbf{r}),\psi^{+}(\mathbf{r}),\phi(\mathbf{r}),\phi^{+}(\mathbf{r})]_{1-8}\\
 & = & g\int d\mathbf{s}\left\{ \left(\psi(\mathbf{s})\right)\left(\psi(\mathbf{s})\right)\left(\psi^{+}(\mathbf{s})\right)\left(\phi^{+}(\mathbf{s})\right)\right\} WP\\
 &  & +g\int d\mathbf{s}\left\{ \left(\psi(\mathbf{s})\right)\left(\psi(\mathbf{s})\right)\left(\frac{1}{2}\frac{\delta}{\delta\psi(\mathbf{s})}\right)\left(\phi^{+}(\mathbf{s})\right)\right\} WP\\
 &  & +g\int d\mathbf{s}\left\{ \left(\psi(\mathbf{s})\right)\left(-\frac{1}{2}\frac{\delta}{\delta\psi^{+}(\mathbf{s})}\right)\left(\psi^{+}(\mathbf{s})\right)\left(\phi^{+}(\mathbf{s})\right)\right\} WP\\
 &  & +g\int d\mathbf{s}\left\{ \left(\psi(\mathbf{s})\right)\left(-\frac{1}{2}\frac{\delta}{\delta\psi^{+}(\mathbf{s})}\right)\left(\frac{1}{2}\frac{\delta}{\delta\psi(\mathbf{s})}\right)\left(\phi^{+}(\mathbf{s})\right)\right\} WP\\
 &  & +g\int d\mathbf{s}\left\{ \left(-\frac{1}{2}\frac{\delta}{\delta\psi^{+}(\mathbf{s})}\right)\left(\psi(\mathbf{s})\right)\left(\psi^{+}(\mathbf{s})\right)\left(\phi^{+}(\mathbf{s})\right)\right\} WP\\
 &  & +g\int d\mathbf{s}\left\{ \left(-\frac{1}{2}\frac{\delta}{\delta\psi^{+}(\mathbf{s})}\right)\left(\psi(\mathbf{s})\right)\left(\frac{1}{2}\frac{\delta}{\delta\psi(\mathbf{s})}\right)\left(\phi^{+}(\mathbf{s})\right)\right\} WP\\
 &  & +g\int d\mathbf{s}\left\{ \left(-\frac{1}{2}\frac{\delta}{\delta\psi^{+}(\mathbf{s})}\right)\left(-\frac{1}{2}\frac{\delta}{\delta\psi^{+}(\mathbf{s})}\right)\left(\psi^{+}(\mathbf{s})\right)\left(\phi^{+}(\mathbf{s})\right)\right\} WP\\
 &  & +g\int d\mathbf{s}\left\{ \left(-\frac{1}{2}\frac{\delta}{\delta\psi^{+}(\mathbf{s})}\right)\left(-\frac{1}{2}\frac{\delta}{\delta\psi^{+}(\mathbf{s})}\right)\left(\frac{1}{2}\frac{\delta}{\delta\psi(\mathbf{s})}\right)\left(\phi^{+}(\mathbf{s})\right)\right\} WP\end{eqnarray*}
and\begin{eqnarray*}
 &  & WP[\psi(\mathbf{r}),\psi^{+}(\mathbf{r}),\phi(\mathbf{r}),\phi^{+}(\mathbf{r})]_{9-24}\\
 & = & +g\int d\mathbf{s}\left(\phi(\mathbf{s})\right)\left(\psi(\mathbf{s})\right)\left(\psi^{+}(\mathbf{s})\right)\left(\psi^{+}(\mathbf{s})\right)WP\\
 &  & +g\int d\mathbf{s}\left(\phi(\mathbf{s})\right)\left(\psi(\mathbf{s})\right)\left(\psi^{+}(\mathbf{s})\right)\left(\frac{1}{2}\frac{\delta}{\delta\psi(\mathbf{s})}\right)WP\\
 &  & +g\int d\mathbf{s}\left(\phi(\mathbf{s})\right)\left(\psi(\mathbf{s})\right)\left(\frac{1}{2}\frac{\delta}{\delta\psi(\mathbf{s})}\right)\left(\psi^{+}(\mathbf{s})\right)WP\\
 &  & +g\int d\mathbf{s}\left(\phi(\mathbf{s})\right)\left(\psi(\mathbf{s})\right)\left(\frac{1}{2}\frac{\delta}{\delta\psi(\mathbf{s})}\right)\left(\frac{1}{2}\frac{\delta}{\delta\psi(\mathbf{s})}\right)WP\\
 &  & +g\int d\mathbf{s}\left(\phi(\mathbf{s})\right)\left(-\frac{1}{2}\frac{\delta}{\delta\psi^{+}(\mathbf{s})}\right)\left(\psi^{+}(\mathbf{s})\right)\left(\psi^{+}(\mathbf{s})\right)WP\\
 &  & +g\int d\mathbf{s}\left(\phi(\mathbf{s})\right)\left(-\frac{1}{2}\frac{\delta}{\delta\psi^{+}(\mathbf{s})}\right)\left(\psi^{+}(\mathbf{s})\right)\left(\frac{1}{2}\frac{\delta}{\delta\psi(\mathbf{s})}\right)WP\\
 &  & +g\int d\mathbf{s}\left(\phi(\mathbf{s})\right)\left(-\frac{1}{2}\frac{\delta}{\delta\psi^{+}(\mathbf{s})}\right)\left(\frac{1}{2}\frac{\delta}{\delta\psi(\mathbf{s})}\right)\left(\psi^{+}(\mathbf{s})\right)WP\\
 &  & +g\int d\mathbf{s}\left(\phi(\mathbf{s})\right)\left(-\frac{1}{2}\frac{\delta}{\delta\psi^{+}(\mathbf{s})}\right)\left(\frac{1}{2}\frac{\delta}{\delta\psi(\mathbf{s})}\right)\left(\frac{1}{2}\frac{\delta}{\delta\psi(\mathbf{s})}\right)WP\\
 &  & +g\int d\mathbf{s}\left(-\frac{\delta}{\delta\phi^{+}(\mathbf{s})}\right)\left(\psi(\mathbf{s})\right)\left(\psi^{+}(\mathbf{s})\right)\left(\psi^{+}(\mathbf{s})\right)WP\\
 &  & +g\int d\mathbf{s}\left(-\frac{\delta}{\delta\phi^{+}(\mathbf{s})}\right)\left(\psi(\mathbf{s})\right)\left(\psi^{+}(\mathbf{s})\right)\left(\frac{1}{2}\frac{\delta}{\delta\psi(\mathbf{s})}\right)WP\\
 &  & +g\int d\mathbf{s}\left(-\frac{\delta}{\delta\phi^{+}(\mathbf{s})}\right)\left(\psi(\mathbf{s})\right)\left(\frac{1}{2}\frac{\delta}{\delta\psi(\mathbf{s})}\right)\left(\psi^{+}(\mathbf{s})\right)WP\\
 &  & +g\int d\mathbf{s}\left(-\frac{\delta}{\delta\phi^{+}(\mathbf{s})}\right)\left(\psi(\mathbf{s})\right)\left(\frac{1}{2}\frac{\delta}{\delta\psi(\mathbf{s})}\right)\left(\frac{1}{2}\frac{\delta}{\delta\psi(\mathbf{s})}\right)WP\\
 &  & +g\int d\mathbf{s}\left(-\frac{\delta}{\delta\phi^{+}(\mathbf{s})}\right)\left(-\frac{1}{2}\frac{\delta}{\delta\psi^{+}(\mathbf{s})}\right)\left(\psi^{+}(\mathbf{s})\right)\left(\psi^{+}(\mathbf{s})\right)WP\\
 &  & +g\int d\mathbf{s}\left(-\frac{\delta}{\delta\phi^{+}(\mathbf{s})}\right)\left(-\frac{1}{2}\frac{\delta}{\delta\psi^{+}(\mathbf{s})}\right)\left(\psi^{+}(\mathbf{s})\right)\left(\frac{1}{2}\frac{\delta}{\delta\psi(\mathbf{s})}\right)WP\\
 &  & +g\int d\mathbf{s}\left(-\frac{\delta}{\delta\phi^{+}(\mathbf{s})}\right)\left(-\frac{1}{2}\frac{\delta}{\delta\psi^{+}(\mathbf{s})}\right)\left(\frac{1}{2}\frac{\delta}{\delta\psi(\mathbf{s})}\right)\left(\psi^{+}(\mathbf{s})\right)WP\\
 &  & +g\int d\mathbf{s}\left(-\frac{\delta}{\delta\phi^{+}(\mathbf{s})}\right)\left(-\frac{1}{2}\frac{\delta}{\delta\psi^{+}(\mathbf{s})}\right)\left(\frac{1}{2}\frac{\delta}{\delta\psi(\mathbf{s})}\right)\left(\frac{1}{2}\frac{\delta}{\delta\psi(\mathbf{s})}\right)WP\end{eqnarray*}
Collecting terms gives\begin{eqnarray}
 &  & WP[\psi(\mathbf{r}),\psi^{+}(\mathbf{r}),\phi(\mathbf{r}),\phi^{+}(\mathbf{r})]_{1-8}\nonumber \\
 & = & g\int d\mathbf{s}\left\{ \left(\psi(\mathbf{s})\right)\left(\psi(\mathbf{s})\right)\left(\psi^{+}(\mathbf{s})\right)\left(\phi^{+}(\mathbf{s})\right)\right\} WP\qquad\emph{S1}\nonumber \\
 &  & +g\int d\mathbf{s}\frac{1}{2}\left\{ \left(\psi(\mathbf{s})\right)\left(\psi(\mathbf{s})\right)\left(\frac{\delta}{\delta\psi(\mathbf{s})}\right)\left(\phi^{+}(\mathbf{s})\right)\right\} WP\qquad\emph{S2}\nonumber \\
 &  & -g\int d\mathbf{s}\frac{1}{2}\left\{ \left(\psi(\mathbf{s})\right)\left(\frac{\delta}{\delta\psi^{+}(\mathbf{s})}\right)\left(\psi^{+}(\mathbf{s})\right)\left(\phi^{+}(\mathbf{s})\right)\right\} WP\qquad\emph{S3}\nonumber \\
 &  & -g\int d\mathbf{s}\frac{1}{4}\left\{ \left(\psi(\mathbf{s})\right)\left(\frac{\delta}{\delta\psi^{+}(\mathbf{s})}\right)\left(\frac{\delta}{\delta\psi(\mathbf{s})}\right)\left(\phi^{+}(\mathbf{s})\right)\right\} WP\qquad\mathcal{S}\emph{4}\nonumber \\
 &  & -g\int d\mathbf{s}\frac{1}{2}\left\{ \left(\frac{\delta}{\delta\psi^{+}(\mathbf{s})}\right)\left(\psi(\mathbf{s})\right)\left(\psi^{+}(\mathbf{s})\right)\left(\phi^{+}(\mathbf{s})\right)\right\} WP\qquad\emph{S5}\nonumber \\
 &  & -g\int d\mathbf{s}\frac{1}{4}\left\{ \left(\frac{\delta}{\delta\psi^{+}(\mathbf{s})}\right)\left(\psi(\mathbf{s})\right)\left(\frac{\delta}{\delta\psi(\mathbf{s})}\right)\left(\phi^{+}(\mathbf{s})\right)\right\} WP\qquad\emph{S6}\nonumber \\
 &  & +g\int d\mathbf{s}\frac{1}{4}\left\{ \left(\frac{\delta}{\delta\psi^{+}(\mathbf{s})}\right)\left(\frac{\delta}{\delta\psi^{+}(\mathbf{s})}\right)\left(\psi^{+}(\mathbf{s})\right)\left(\phi^{+}(\mathbf{s})\right)\right\} WP\qquad\emph{S7}\nonumber \\
 &  & +g\int d\mathbf{s}\frac{1}{8}\left\{ \left(\frac{\delta}{\delta\psi^{+}(\mathbf{s})}\right)\left(\frac{\delta}{\delta\psi^{+}(\mathbf{s})}\right)\left(\frac{\delta}{\delta\psi(\mathbf{s})}\right)\left(\phi^{+}(\mathbf{s})\right)\right\} WP\qquad\emph{S8}\nonumber \\
 &  & \,\end{eqnarray}
and \begin{eqnarray}
 &  & WP[\psi(\mathbf{r}),\psi^{+}(\mathbf{r}),\phi(\mathbf{r}),\phi^{+}(\mathbf{r})]_{9-24}\nonumber \\
 & = & +g\int d\mathbf{s}\left\{ \left(\phi(\mathbf{s})\right)\left(\psi(\mathbf{s})\right)\left(\psi^{+}(\mathbf{s})\right)\left(\psi^{+}(\mathbf{s})\right)\right\} WP\qquad\emph{S9}\nonumber \\
 &  & +g\int d\mathbf{s}\frac{1}{2}\left\{ \left(\phi(\mathbf{s})\right)\left(\psi(\mathbf{s})\right)\left(\psi^{+}(\mathbf{s})\right)\left(\frac{\delta}{\delta\psi(\mathbf{s})}\right)\right\} WP\qquad\emph{S10}\nonumber \\
 &  & +g\int d\mathbf{s}\frac{1}{2}\left\{ \left(\phi(\mathbf{s})\right)\left(\psi(\mathbf{s})\right)\left(\frac{\delta}{\delta\psi(\mathbf{s})}\right)\left(\psi^{+}(\mathbf{s})\right)\right\} WP\qquad\emph{S11}\nonumber \\
 &  & +g\int d\mathbf{s}\frac{1}{4}\left\{ \left(\phi(\mathbf{s})\right)\left(\psi(\mathbf{s})\right)\left(\frac{\delta}{\delta\psi(\mathbf{s})}\right)\left(\frac{\delta}{\delta\psi(\mathbf{s})}\right)\right\} WP\qquad\emph{S12}\nonumber \\
 &  & -g\int d\mathbf{s}\frac{1}{2}\left\{ \left(\phi(\mathbf{s})\right)\left(\frac{\delta}{\delta\psi^{+}(\mathbf{s})}\right)\left(\psi^{+}(\mathbf{s})\right)\left(\psi^{+}(\mathbf{s})\right)\right\} WP\qquad\emph{S13}\nonumber \\
 &  & -g\int d\mathbf{s}\frac{1}{4}\left\{ \left(\phi(\mathbf{s})\right)\left(\frac{\delta}{\delta\psi^{+}(\mathbf{s})}\right)\left(\psi^{+}(\mathbf{s})\right)\left(\frac{\delta}{\delta\psi(\mathbf{s})}\right)\right\} WP\qquad\emph{S14}\nonumber \\
 &  & -g\int d\mathbf{s}\frac{1}{4}\left\{ \left(\phi(\mathbf{s})\right)\left(\frac{\delta}{\delta\psi^{+}(\mathbf{s})}\right)\left(\frac{\delta}{\delta\psi(\mathbf{s})}\right)\left(\psi^{+}(\mathbf{s})\right)\right\} WP\qquad\emph{S15}\nonumber \\
 &  & -g\int d\mathbf{s}\frac{1}{8}\left\{ \left(\phi(\mathbf{s})\right)\left(\frac{\delta}{\delta\psi^{+}(\mathbf{s})}\right)\left(\frac{\delta}{\delta\psi(\mathbf{s})}\right)\left(\frac{\delta}{\delta\psi(\mathbf{s})}\right)\right\} WP\qquad\emph{S16}\nonumber \\
 &  & -g\int d\mathbf{s}\left\{ \left(\frac{\delta}{\delta\phi^{+}(\mathbf{s})}\right)\left(\psi(\mathbf{s})\right)\left(\psi^{+}(\mathbf{s})\right)\left(\psi^{+}(\mathbf{s})\right)\right\} WP\qquad\emph{S17}\nonumber \\
 &  & -g\int d\mathbf{s}\frac{1}{2}\left\{ \left(\frac{\delta}{\delta\phi^{+}(\mathbf{s})}\right)\left(\psi(\mathbf{s})\right)\left(\psi^{+}(\mathbf{s})\right)\left(\frac{\delta}{\delta\psi(\mathbf{s})}\right)\right\} WP\qquad\emph{S18}\nonumber \\
 &  & -g\int d\mathbf{s}\frac{1}{2}\left\{ \left(\frac{\delta}{\delta\phi^{+}(\mathbf{s})}\right)\left(\psi(\mathbf{s})\right)\left(\frac{\delta}{\delta\psi(\mathbf{s})}\right)\left(\psi^{+}(\mathbf{s})\right)\right\} WP\qquad\emph{S19}\nonumber \\
 &  & -g\int d\mathbf{s}\frac{1}{4}\left\{ \left(\frac{\delta}{\delta\phi^{+}(\mathbf{s})}\right)\left(\psi(\mathbf{s})\right)\left(\frac{\delta}{\delta\psi(\mathbf{s})}\right)\left(\frac{\delta}{\delta\psi(\mathbf{s})}\right)\right\} WP\qquad\emph{S20}\nonumber \\
 &  & +g\int d\mathbf{s}\frac{1}{2}\left\{ \left(\frac{\delta}{\delta\phi^{+}(\mathbf{s})}\right)\left(\frac{\delta}{\delta\psi^{+}(\mathbf{s})}\right)\left(\psi^{+}(\mathbf{s})\right)\left(\psi^{+}(\mathbf{s})\right)\right\} WP\qquad\emph{S21}\nonumber \\
 &  & +g\int d\mathbf{s}\frac{1}{4}\left\{ \left(\frac{\delta}{\delta\phi^{+}(\mathbf{s})}\right)\left(\frac{\delta}{\delta\psi^{+}(\mathbf{s})}\right)\left(\psi^{+}(\mathbf{s})\right)\left(\frac{\delta}{\delta\psi(\mathbf{s})}\right)\right\} WP\qquad\emph{S22}\nonumber \\
 &  & +g\int d\mathbf{s}\frac{1}{4}\left\{ \left(\frac{\delta}{\delta\phi^{+}(\mathbf{s})}\right)\left(\frac{\delta}{\delta\psi^{+}(\mathbf{s})}\right)\left(\frac{\delta}{\delta\psi(\mathbf{s})}\right)\left(\psi^{+}(\mathbf{s})\right)\right\} WP\qquad\emph{S23}\nonumber \\
 &  & +g\int d\mathbf{s}\frac{1}{8}\left\{ \left(\frac{\delta}{\delta\phi^{+}(\mathbf{s})}\right)\left(\frac{\delta}{\delta\psi^{+}(\mathbf{s})}\right)\left(\frac{\delta}{\delta\psi(\mathbf{s})}\right)\left(\frac{\delta}{\delta\psi(\mathbf{s})}\right)\right\} WP\qquad\emph{S24}\nonumber \\
 &  & \,\end{eqnarray}

Again we use the product rule (\ref{eq:ProdRuleFnalDeriv-1}) together
with (\ref{Eq.FuncDerivativeRule1-1}) and (\ref{Eq.FuncDerivativeRule2-1})
to place all the functional derivatives on the left.

For the \emph{S2}T18 term\begin{eqnarray*}
 &  & \int d\mathbf{s\,}\frac{1}{2}\left\{ \mathbf{\left(\psi(\mathbf{s})\right)\left(\psi(\mathbf{s})\right)\left(\frac{\delta}{\delta\psi(\mathbf{s})}\right)}\left(\mathbf{\phi}^{+}\mathbf{(\mathbf{s})}\right)\right\} WP\\
 & = & \int d\mathbf{s\,}\frac{1}{2}\left\{ \mathbf{\left(\frac{\delta}{\delta\psi(\mathbf{s})}\right)\{\psi(\mathbf{s})\psi(\mathbf{s})\mathbf{\phi}^{+}\mathbf{(\mathbf{s})}\}}\right\} WP\\
 &  & -\int d\mathbf{s\,}\frac{1}{2}\left\{ \mathbf{2\delta_{C}(\mathbf{s},\mathbf{s})\psi(\mathbf{s})\mathbf{\phi}^{+}\mathbf{(\mathbf{s})}}\right\} WP\end{eqnarray*}

For the \emph{S3}T19 term\begin{eqnarray*}
 &  & \int d\mathbf{s\,}\frac{1}{2}\left\{ \mathbf{\left(\psi(\mathbf{s})\right)\left(\frac{\delta}{\delta\psi^{+}(\mathbf{s})}\right)\left(\psi^{+}(\mathbf{s})\right)}\left(\mathbf{\phi}^{+}\mathbf{(\mathbf{s})}\right)\right\} WP\\
 & = & \int d\mathbf{s\,}\frac{1}{2}\left\{ \mathbf{\left(\frac{\delta}{\delta\psi^{+}(\mathbf{s})}\right)\{\psi(\mathbf{s})\psi^{+}(\mathbf{s})\mathbf{\phi}^{+}\mathbf{(\mathbf{s})}\phi\}}\right\} WP\end{eqnarray*}

For the \emph{S4}T20 term\begin{eqnarray*}
 &  & \int d\mathbf{s\,}\frac{1}{4}\left\{ \mathbf{\left(\psi(\mathbf{s})\right)\left(\frac{\delta}{\delta\psi^{+}(\mathbf{s})}\right)\left(\frac{\delta}{\delta\psi(\mathbf{s})}\right)}\left(\mathbf{\phi}^{+}\mathbf{(\mathbf{s})}\right)\right\} WP\\
 & = & \int d\mathbf{s\,}\frac{1}{4}\left\{ \mathbf{\left(\frac{\delta}{\delta\psi^{+}(\mathbf{s})}\right)\left(\psi(\mathbf{s})\right)\left(\frac{\delta}{\delta\psi(\mathbf{s})}\right)}\left(\mathbf{\phi}^{+}\mathbf{(\mathbf{s})}\right)\right\} WP\\
 & = & \int d\mathbf{s\,}\frac{1}{4}\left\{ \mathbf{\left(\frac{\delta}{\delta\psi^{+}(\mathbf{s})}\right)\left(\frac{\delta}{\delta\psi(\mathbf{s})}\right)\{\psi(\mathbf{s})\mathbf{\phi}^{+}\mathbf{(\mathbf{s})}\}}\right\} WP\\
 &  & -\int d\mathbf{s\,}\frac{1}{4}\left\{ \mathbf{\left(\frac{\delta}{\delta\psi^{+}(\mathbf{s})}\right)\{\delta_{C}(\mathbf{s},\mathbf{s})\mathbf{\phi}^{+}\mathbf{(\mathbf{s})}\}}\right\} WP\end{eqnarray*}

For the \emph{S5}T21 term\begin{eqnarray*}
 &  & \int d\mathbf{s\,}\frac{1}{2}\left\{ \mathbf{\left(\frac{\delta}{\delta\psi^{+}(\mathbf{s})}\right)\left(\psi(\mathbf{s})\right)\left(\psi^{+}(\mathbf{s})\right)}\left(\mathbf{\phi}^{+}\mathbf{(\mathbf{s})}\right)\right\} WP\\
 & = & \int d\mathbf{s\,}\frac{1}{2}\left\{ \mathbf{\left(\frac{\delta}{\delta\psi^{+}(\mathbf{s})}\right)\{\psi(\mathbf{s})\psi^{+}(\mathbf{s})\mathbf{\phi}^{+}\mathbf{(\mathbf{s})}\}}\right\} WP\end{eqnarray*}

For the \emph{S6}T22 term\begin{eqnarray*}
 &  & \int d\mathbf{s\,}\frac{1}{4}\left\{ \mathbf{\left(\frac{\delta}{\delta\psi^{+}(\mathbf{s})}\right)\left(\psi(\mathbf{s})\right)\left(\frac{\delta}{\delta\psi(\mathbf{s})}\right)}\left(\mathbf{\phi}^{+}\mathbf{(\mathbf{s})}\right)\right\} WP\\
 & = & \int d\mathbf{s\,}\frac{1}{4}\left\{ \mathbf{\left(\frac{\delta}{\delta\psi^{+}(\mathbf{s})}\right)\left(\frac{\delta}{\delta\psi(\mathbf{s})}\right)\{\psi(\mathbf{s})\mathbf{\phi}^{+}\mathbf{(\mathbf{s})}\}}\right\} WP\\
 &  & -\int d\mathbf{s\,}\frac{1}{4}\left\{ \mathbf{\left(\frac{\delta}{\delta\psi^{+}(\mathbf{s})}\right)\{\delta_{C}(\mathbf{s},\mathbf{s})\mathbf{\phi}^{+}\mathbf{(\mathbf{s})}\}}\right\} WP\end{eqnarray*}

For the \emph{S7}T23 term\[
\int d\mathbf{s\,}\frac{1}{4}\left\{ \mathbf{\left(\frac{\delta}{\delta\psi^{+}(\mathbf{s})}\right)\left(\frac{\delta}{\delta\psi^{+}(\mathbf{s})}\right)\left(\psi^{+}(\mathbf{s})\right)}\left(\mathbf{\phi}^{+}\mathbf{(\mathbf{s})}\right)\right\} WP\]

For the \emph{S8}T24 term\begin{eqnarray*}
 &  & \int d\mathbf{s\,}\frac{1}{8}\left\{ \mathbf{\left(\frac{\delta}{\delta\psi^{+}(\mathbf{s})}\right)\left(\frac{\delta}{\delta\psi^{+}(\mathbf{s})}\right)\left(\frac{\delta}{\delta\psi(\mathbf{s})}\right)}\left(\mathbf{\phi}^{+}\mathbf{(\mathbf{s})}\right)\right\} WP\\
 & = & \int d\mathbf{s\,}\frac{1}{8}\left\{ \mathbf{\left(\frac{\delta}{\delta\psi^{+}(\mathbf{s})}\right)\left(\frac{\delta}{\delta\psi^{+}(\mathbf{s})}\right)\left(\frac{\delta}{\delta\psi(\mathbf{s})}\right)\{\mathbf{\phi}^{+}\mathbf{(\mathbf{s})}\}}\right\} WP\end{eqnarray*}

For the \emph{S10}T2 term\begin{eqnarray*}
 &  & \int d\mathbf{s\,}\frac{1}{2}\left\{ \left(\phi(\mathbf{s})\right)\left(\psi(\mathbf{s})\right)\left(\psi^{+}(\mathbf{s})\right)\left(\frac{\delta}{\delta\psi(\mathbf{s})}\right)\right\} WP\\
 & = & \int d\mathbf{s\,}\frac{1}{2}\left\{ \left(\frac{\delta}{\delta\psi(\mathbf{s})}\right)\{\phi(\mathbf{s})\psi(\mathbf{s})\psi^{+}(\mathbf{s})\}-\{\delta_{C}(\mathbf{s},\mathbf{s})\phi(\mathbf{s})\psi^{+}(\mathbf{s})\}\right\} WP\\
 & = & \int d\mathbf{s\,}\frac{1}{2}\left\{ \left(\frac{\delta}{\delta\psi(\mathbf{s})}\right)\{\phi(\mathbf{s})\psi(\mathbf{s})\psi^{+}(\mathbf{s})\}\right\} WP\\
 &  & -\int d\mathbf{s\,}\frac{1}{2}\{\delta_{C}(\mathbf{s},\mathbf{s})\phi(\mathbf{s})\psi^{+}(\mathbf{s})\}WP\end{eqnarray*}

For the \emph{S11}T3 term\begin{eqnarray*}
 &  & \int d\mathbf{s\,}\frac{1}{2}\left\{ \left(\phi(\mathbf{s})\right)\left(\psi(\mathbf{s})\right)\left(\frac{\delta}{\delta\psi(\mathbf{s})}\right)\left(\psi^{+}(\mathbf{s})\right)\right\} WP\\
 & = & \int d\mathbf{s\,}\frac{1}{2}\left\{ \left(\frac{\delta}{\delta\psi(\mathbf{s})}\right)\{\phi(\mathbf{s})\psi(\mathbf{s})\psi^{+}(\mathbf{s})\}-\{\delta_{C}(\mathbf{s},\mathbf{s})\phi(\mathbf{s})\psi^{+}(\mathbf{s})\}\right\} WP\\
 & = & \int d\mathbf{s\,}\frac{1}{2}\left\{ \left(\frac{\delta}{\delta\psi(\mathbf{s})}\right)\{\phi(\mathbf{s})\psi(\mathbf{s})\psi^{+}(\mathbf{s})\}\right\} WP\\
 &  & -\int d\mathbf{s\,}\frac{1}{2}\{\delta_{C}(\mathbf{s},\mathbf{s})\phi(\mathbf{s})\psi^{+}(\mathbf{s})\}WP\end{eqnarray*}

For the \emph{S12}T4 term\begin{eqnarray*}
 &  & \int d\mathbf{s\,}\frac{1}{4}\left\{ \left(\phi(\mathbf{s})\right)\left(\psi(\mathbf{s})\right)\left(\frac{\delta}{\delta\psi(\mathbf{s})}\right)\left(\frac{\delta}{\delta\psi(\mathbf{s})}\right)\right\} WP\\
 & = & \int d\mathbf{s\,}\frac{1}{4}\left\{ \left(\frac{\delta}{\delta\psi(\mathbf{s})}\right)\{\phi(\mathbf{s})\psi(\mathbf{s})\}-\{\phi(\mathbf{s})\delta_{C}(\mathbf{s},\mathbf{s})\}\right\} \left(\frac{\delta}{\delta\psi(\mathbf{s})}\right)WP\\
 & = & \int d\mathbf{s\,}\frac{1}{4}\left\{ \left(\frac{\delta}{\delta\psi(\mathbf{s})}\right)\left(\frac{\delta}{\delta\psi(\mathbf{s})}\right)\{\phi(\mathbf{s})\psi(\mathbf{s})\}\right\} WP\\
 &  & -\int d\mathbf{s\,}\frac{1}{4}\left\{ \left(\frac{\delta}{\delta\psi(\mathbf{s})}\right)\{\phi(\mathbf{s})\delta_{C}(\mathbf{s},\mathbf{s})\}\right\} WP\\
 &  & -\int d\mathbf{s\,}\frac{1}{4}\left(\frac{\delta}{\delta\psi(\mathbf{s})}\{\phi(\mathbf{s})\delta_{C}(\mathbf{s},\mathbf{s})\}\right)WP\\
 & = & \int d\mathbf{s\,}\frac{1}{4}\left\{ \left(\frac{\delta}{\delta\psi(\mathbf{s})}\right)\left(\frac{\delta}{\delta\psi(\mathbf{s})}\right)\{\phi(\mathbf{s})\psi(\mathbf{s})\}\right\} WP\\
 &  & -\int d\mathbf{s\,}\frac{1}{2}\left\{ \left(\frac{\delta}{\delta\psi(\mathbf{s})}\right)\{\phi(\mathbf{s})\delta_{C}(\mathbf{s},\mathbf{s})\}\right\} WP\end{eqnarray*}

For the \emph{S13}T5 term\begin{eqnarray*}
 &  & \int d\mathbf{s\,}\frac{1}{2}\left\{ \left(\phi(\mathbf{s})\right)\left(\frac{\delta}{\delta\psi^{+}(\mathbf{s})}\right)\left(\psi^{+}(\mathbf{s})\right)\left(\psi^{+}(\mathbf{s})\right)\right\} WP\\
 & = & \int d\mathbf{s\,}\frac{1}{2}\left\{ \left(\frac{\delta}{\delta\psi^{+}(\mathbf{s})}\right)\{\phi(\mathbf{s})\psi^{+}(\mathbf{s})\psi^{+}(\mathbf{s})\}\right\} WP\end{eqnarray*}

For the \emph{S14}T6 term\begin{eqnarray*}
 &  & \int d\mathbf{s\,}\frac{1}{4}\left\{ \left(\phi(\mathbf{s})\right)\left(\frac{\delta}{\delta\psi^{+}(\mathbf{s})}\right)\left(\psi^{+}(\mathbf{s})\right)\left(\frac{\delta}{\delta\psi(\mathbf{s})}\right)\right\} WP\\
 & = & \int d\mathbf{s\,}\frac{1}{4}\left\{ \left(\frac{\delta}{\delta\psi^{+}(\mathbf{s})}\right)\left(\frac{\delta}{\delta\psi(\mathbf{s})}\right)\{\phi(\mathbf{s})\psi^{+}(\mathbf{s})\}\right\} WP\end{eqnarray*}

For the \emph{S15}T7 term\begin{eqnarray*}
 &  & \int d\mathbf{s\,}\frac{1}{4}\left\{ \left(\phi(\mathbf{s})\right)\left(\frac{\delta}{\delta\psi^{+}(\mathbf{s})}\right)\left(\frac{\delta}{\delta\psi(\mathbf{s})}\right)\left(\psi^{+}(\mathbf{s})\right)\right\} WP\\
 & = & \int d\mathbf{s\,}\frac{1}{4}\left\{ \left(\frac{\delta}{\delta\psi^{+}(\mathbf{s})}\right)\left(\frac{\delta}{\delta\psi(\mathbf{s})}\right)\{\phi(\mathbf{s})\psi^{+}(\mathbf{s})\}\right\} WP\end{eqnarray*}

For the \emph{S16}T8 term\begin{eqnarray*}
 &  & \int d\mathbf{s\,}\frac{1}{8}\left\{ \left(\phi(\mathbf{s})\right)\left(\frac{\delta}{\delta\psi^{+}(\mathbf{s})}\right)\left(\frac{\delta}{\delta\psi(\mathbf{s})}\right)\left(\frac{\delta}{\delta\psi(\mathbf{s})}\right)\right\} WP\\
 & = & \int d\mathbf{s\,}\frac{1}{8}\left\{ \left(\frac{\delta}{\delta\psi^{+}(\mathbf{s})}\right)\left(\frac{\delta}{\delta\psi(\mathbf{s})}\right)\left(\frac{\delta}{\delta\psi(\mathbf{s})}\right)\{\phi(\mathbf{s})\}\right\} WP\end{eqnarray*}

For the \emph{S17}T9 term\begin{eqnarray*}
 &  & \int d\mathbf{s\,}\left\{ \left(\frac{\delta}{\delta\phi^{+}(\mathbf{s})}\right)\left(\psi(\mathbf{s})\right)\left(\psi^{+}(\mathbf{s})\right)\left(\psi^{+}(\mathbf{s})\right)\right\} WP\\
 & = & \int d\mathbf{s\,}\left\{ \left(\frac{\delta}{\delta\phi^{+}(\mathbf{s})}\right)\{\psi(\mathbf{s})\psi^{+}(\mathbf{s})\psi^{+}(\mathbf{s})\}\right\} WP\end{eqnarray*}

For the \emph{S18}T10 term\begin{eqnarray*}
 &  & \int d\mathbf{s\,}\frac{1}{2}\left\{ \left(\frac{\delta}{\delta\phi^{+}(\mathbf{s})}\right)\left(\psi(\mathbf{s})\right)\left(\psi^{+}(\mathbf{s})\right)\left(\frac{\delta}{\delta\psi(\mathbf{s})}\right)\right\} WP\\
 & = & \int d\mathbf{s\,}\frac{1}{2}\left\{ \left(\frac{\delta}{\delta\phi^{+}(\mathbf{s})}\right)\left(\frac{\delta}{\delta\psi(\mathbf{s})}\right)\{\psi(\mathbf{s})\psi^{+}(\mathbf{s})\}\right\} WP\\
 &  & -\int d\mathbf{s\,}\frac{1}{2}\left\{ \left(\frac{\delta}{\delta\phi^{+}(\mathbf{s})}\right)\{\delta_{C}(\mathbf{s},\mathbf{s})\psi^{+}(\mathbf{s})\}\right\} WP\end{eqnarray*}

For the \emph{S19}T11 term\begin{eqnarray*}
 &  & \int d\mathbf{s\,}\frac{1}{2}\left\{ \left(\frac{\delta}{\delta\phi^{+}(\mathbf{s})}\right)\left(\psi(\mathbf{s})\right)\left(\frac{\delta}{\delta\psi(\mathbf{s})}\right)\left(\psi^{+}(\mathbf{s})\right)\right\} WP\\
 & = & \int d\mathbf{s\,}\frac{1}{2}\left\{ \left(\frac{\delta}{\delta\phi^{+}(\mathbf{s})}\right)\left(\frac{\delta}{\delta\psi(\mathbf{s})}\right)\{\psi(\mathbf{s})\psi^{+}(\mathbf{s})\}\right\} WP\\
 &  & -\int d\mathbf{s\,}\frac{1}{2}\left\{ \left(\frac{\delta}{\delta\phi^{+}(\mathbf{s})}\right)\{\delta_{C}(\mathbf{s},\mathbf{s})\psi^{+}(\mathbf{s})\}\right\} WP\end{eqnarray*}

For the \emph{S20}T12 term\begin{eqnarray*}
 &  & \int d\mathbf{s\,}\frac{1}{4}\left\{ \left(\frac{\delta}{\delta\phi^{+}(\mathbf{s})}\right)\left(\psi(\mathbf{s})\right)\left(\frac{\delta}{\delta\psi(\mathbf{s})}\right)\left(\frac{\delta}{\delta\psi(\mathbf{s})}\right)\right\} WP\\
 & = & \int d\mathbf{s\,}\frac{1}{4}\left(\frac{\delta}{\delta\phi^{+}(\mathbf{s})}\right)\left\{ \left(\frac{\delta}{\delta\psi(\mathbf{s})}\right)\left(\psi(\mathbf{s})\right)\left(\frac{\delta}{\delta\psi(\mathbf{s})}\right)\right\} WP\\
 &  & -\int d\mathbf{s\,}\frac{1}{4}\left(\frac{\delta}{\delta\phi^{+}(\mathbf{s})}\right)\left\{ \delta_{C}(\mathbf{s},\mathbf{s})\left(\frac{\delta}{\delta\psi(\mathbf{s})}\right)\right\} WP\\
 & = & \int d\mathbf{s\,}\frac{1}{4}\left(\frac{\delta}{\delta\phi^{+}(\mathbf{s})}\right)\left\{ \left(\frac{\delta}{\delta\psi(\mathbf{s})}\right)\left(\frac{\delta}{\delta\psi(\mathbf{s})}\right)\{\psi(\mathbf{s})\}\right\} WP\\
 &  & -\int d\mathbf{s\,}\frac{1}{4}\left(\frac{\delta}{\delta\phi^{+}(\mathbf{s})}\right)\left\{ \left(\frac{\delta}{\delta\psi(\mathbf{s})}\right)\left\{ \delta_{C}(\mathbf{s},\mathbf{s})\right\} \right\} WP\\
 &  & -\int d\mathbf{s\,}\frac{1}{4}\left\{ \left(\frac{\delta}{\delta\phi^{+}(\mathbf{s})}\right)\left(\frac{\delta}{\delta\psi(\mathbf{s})}\right)\left\{ \delta_{C}(\mathbf{s},\mathbf{s})\right\} \right\} WP\\
 & = & \int d\mathbf{s\,}\frac{1}{4}\left\{ \left(\frac{\delta}{\delta\phi^{+}(\mathbf{s})}\right)\left(\frac{\delta}{\delta\psi(\mathbf{s})}\right)\left(\frac{\delta}{\delta\psi(\mathbf{s})}\right)\{\psi(\mathbf{s})\}\right\} WP\\
 &  & -\int d\mathbf{s\,}\frac{1}{2}\left\{ \left(\frac{\delta}{\delta\phi^{+}(\mathbf{s})}\right)\left(\frac{\delta}{\delta\psi(\mathbf{s})}\right)\left\{ \delta_{C}(\mathbf{s},\mathbf{s})\right\} \right\} WP\end{eqnarray*}

For the \emph{S21}T13 term\begin{eqnarray*}
 &  & \int d\mathbf{s\,}\frac{1}{2}\left\{ \left(\frac{\delta}{\delta\phi^{+}(\mathbf{s})}\right)\left(\frac{\delta}{\delta\psi^{+}(\mathbf{s})}\right)\left(\psi^{+}(\mathbf{s})\right)\left(\psi^{+}(\mathbf{s})\right)\right\} WP\\
 & = & \int d\mathbf{s\,}\frac{1}{2}\left\{ \left(\frac{\delta}{\delta\phi^{+}(\mathbf{s})}\right)\left(\frac{\delta}{\delta\psi^{+}(\mathbf{s})}\right)\{\psi^{+}(\mathbf{s})\psi^{+}(\mathbf{s})\}\right\} WP\end{eqnarray*}

For the \emph{S22}T14 term\begin{eqnarray*}
 &  & \int d\mathbf{s\,}\frac{1}{4}\left\{ \left(\frac{\delta}{\delta\phi^{+}(\mathbf{s})}\right)\left(\frac{\delta}{\delta\psi^{+}(\mathbf{s})}\right)\left(\psi^{+}(\mathbf{s})\right)\left(\frac{\delta}{\delta\psi(\mathbf{s})}\right)\right\} WP\\
 & = & \int d\mathbf{s\,}\frac{1}{4}\left\{ \left(\frac{\delta}{\delta\phi^{+}(\mathbf{s})}\right)\left(\frac{\delta}{\delta\psi^{+}(\mathbf{s})}\right)\left(\frac{\delta}{\delta\psi(\mathbf{s})}\right)\{\psi^{+}(\mathbf{s})\}\right\} WP\end{eqnarray*}

For the \emph{S23}T15 term\begin{eqnarray*}
 &  & \int d\mathbf{s\,}\frac{1}{4}\left\{ \left(\frac{\delta}{\delta\phi^{+}(\mathbf{s})}\right)\left(\frac{\delta}{\delta\psi^{+}(\mathbf{s})}\right)\left(\frac{\delta}{\delta\psi(\mathbf{s})}\right)\left(\psi^{+}(\mathbf{s})\right)\right\} WP\\
 & = & \int d\mathbf{s\,}\frac{1}{4}\left\{ \left(\frac{\delta}{\delta\phi^{+}(\mathbf{s})}\right)\left(\frac{\delta}{\delta\psi^{+}(\mathbf{s})}\right)\left(\frac{\delta}{\delta\psi(\mathbf{s})}\right)\{\psi^{+}(\mathbf{s})\}\right\} WP\end{eqnarray*}

Substituting these results gives\begin{eqnarray}
 &  & WP[\psi(\mathbf{r}),\psi^{+}(\mathbf{r}),\phi(\mathbf{r}),\phi^{+}(\mathbf{r})]_{1-8}\nonumber \\
 & = & g\int d\mathbf{s}\left\{ \psi(\mathbf{s})\psi(\mathbf{s})\psi^{+}(\mathbf{s})\phi^{+}(\mathbf{s})\right\} WP\qquad\emph{G1}\nonumber \\
 &  & +g\int d\mathbf{s\,}\frac{1}{2}\left\{ \mathbf{\left(\frac{\delta}{\delta\psi(\mathbf{s})}\right)\{\psi(\mathbf{s})\psi(\mathbf{s})\mathbf{\phi}^{+}\mathbf{(\mathbf{s})}\}}\right\} WP\qquad\emph{G2.1}\nonumber \\
 &  & -g\int d\mathbf{s\,}\frac{1}{2}\left\{ \mathbf{2\delta_{C}(\mathbf{s},\mathbf{s})\psi(\mathbf{s})\mathbf{\phi}^{+}\mathbf{(\mathbf{s})}}\right\} WP\qquad\emph{G2.2}\nonumber \\
 &  & -g\int d\mathbf{s\,}\frac{1}{2}\left\{ \mathbf{\left(\frac{\delta}{\delta\psi^{+}(\mathbf{s})}\right)\{\psi(\mathbf{s})\psi^{+}(\mathbf{s})\mathbf{\phi}^{+}\mathbf{(\mathbf{s})}\}}\right\} WP\qquad\emph{G3}\nonumber \\
 &  & -g\int d\mathbf{s\,}\frac{1}{4}\left\{ \mathbf{\left(\frac{\delta}{\delta\psi^{+}(\mathbf{s})}\right)\left(\frac{\delta}{\delta\psi(\mathbf{s})}\right)\{\psi(\mathbf{s})\mathbf{\phi}^{+}\mathbf{(\mathbf{s})}\}}\right\} WP\qquad\emph{G4.1}\nonumber \\
 &  & +g\int d\mathbf{s\,}\frac{1}{4}\left\{ \mathbf{\left(\frac{\delta}{\delta\psi^{+}(\mathbf{s})}\right)\{\delta_{C}(\mathbf{s},\mathbf{s})\mathbf{\phi}^{+}\mathbf{(\mathbf{s})}\}}\right\} WP\qquad\emph{G4.2}\nonumber \\
 &  & -g\int d\mathbf{s\,}\frac{1}{2}\left\{ \mathbf{\left(\frac{\delta}{\delta\psi^{+}(\mathbf{s})}\right)\{\psi(\mathbf{s})\psi^{+}(\mathbf{s})\mathbf{\phi}^{+}\mathbf{(\mathbf{s})}\}}\right\} WP\qquad\emph{G5}\nonumber \\
 &  & -g\int d\mathbf{s\,}\frac{1}{4}\left\{ \mathbf{\left(\frac{\delta}{\delta\psi^{+}(\mathbf{s})}\right)\left(\frac{\delta}{\delta\psi(\mathbf{s})}\right)\{\psi(\mathbf{s})\mathbf{\phi}^{+}\mathbf{(\mathbf{s})}\}}\right\} WP\qquad\emph{G6.1}\nonumber \\
 &  & +g\int d\mathbf{s\,}\frac{1}{4}\left\{ \mathbf{\left(\frac{\delta}{\delta\psi^{+}(\mathbf{s})}\right)\{\delta_{C}(\mathbf{s},\mathbf{s})\mathbf{\phi}^{+}\mathbf{(\mathbf{s})}\}}\right\} WP\qquad\emph{G6.2}\nonumber \\
 &  & +g\int d\mathbf{s}\frac{1}{4}\left\{ \left(\frac{\delta}{\delta\psi^{+}(\mathbf{s})}\right)\left(\frac{\delta}{\delta\psi^{+}(\mathbf{s})}\right)\{\psi^{+}(\mathbf{s})\phi^{+}(\mathbf{s})\}\right\} WP\qquad\emph{G7}\nonumber \\
 &  & +g\int d\mathbf{s\,}\frac{1}{8}\left\{ \mathbf{\left(\frac{\delta}{\delta\psi^{+}(\mathbf{s})}\right)\left(\frac{\delta}{\delta\psi^{+}(\mathbf{s})}\right)\left(\frac{\delta}{\delta\psi(\mathbf{s})}\right)\{\mathbf{\phi}^{+}\mathbf{(\mathbf{s})}\}}\right\} WP\qquad\emph{G8 }\nonumber \\
 &  & \,\end{eqnarray}
and

\begin{eqnarray}
 &  & WP[\psi(\mathbf{r}),\psi^{+}(\mathbf{r}),\phi(\mathbf{r}),\phi^{+}(\mathbf{r})]_{9-24}\nonumber \\
 & = & +g\int d\mathbf{s}\left\{ \phi(\mathbf{s})\psi(\mathbf{s})\psi^{+}(\mathbf{s})\psi^{+}(\mathbf{s})\right\} WP[\psi,\psi^{+},\phi,\phi^{+}]\qquad\emph{G9}\nonumber \\
 &  & +g\int d\mathbf{s\,}\frac{1}{2}\left\{ \left(\frac{\delta}{\delta\psi(\mathbf{s})}\right)\{\phi(\mathbf{s})\psi(\mathbf{s})\psi^{+}(\mathbf{s})\}\right\} WP\qquad\emph{G10.1}\nonumber \\
 &  & -g\int d\mathbf{s\,}\frac{1}{2}\{\delta_{C}(\mathbf{s},\mathbf{s})\phi(\mathbf{s})\psi^{+}(\mathbf{s})\}WP\qquad\emph{G10.2}\nonumber \\
 &  & +g\int d\mathbf{s\,}\frac{1}{2}\left\{ \left(\frac{\delta}{\delta\psi(\mathbf{s})}\right)\{\phi(\mathbf{s})\psi(\mathbf{s})\psi^{+}(\mathbf{s})\}\right\} WP\qquad\emph{G11.1}\nonumber \\
 &  & -g\int d\mathbf{s\,}\frac{1}{2}\{\delta_{C}(\mathbf{s},\mathbf{s})\phi(\mathbf{s})\psi^{+}(\mathbf{s})\}WP\qquad\emph{G11.2}\nonumber \\
 &  & +g\int d\mathbf{s\,}\frac{1}{4}\left\{ \left(\frac{\delta}{\delta\psi(\mathbf{s})}\right)\left(\frac{\delta}{\delta\psi(\mathbf{s})}\right)\{\phi(\mathbf{s})\psi(\mathbf{s})\}\right\} WP\qquad\emph{G12.1}\nonumber \\
 &  & -g\int d\mathbf{s\,}\frac{1}{2}\left\{ \left(\frac{\delta}{\delta\psi(\mathbf{s})}\right)\{\phi(\mathbf{s})\delta_{C}(\mathbf{s},\mathbf{s})\}\right\} WP\qquad\emph{G12.2}\nonumber \\
 &  & -g\int d\mathbf{s\,}\frac{1}{2}\left\{ \left(\frac{\delta}{\delta\psi^{+}(\mathbf{s})}\right)\{\phi(\mathbf{s})\psi^{+}(\mathbf{s})\psi^{+}(\mathbf{s})\}\right\} WP\qquad\emph{G13}\nonumber \\
 &  & -g\int d\mathbf{s\,}\frac{1}{4}\left\{ \left(\frac{\delta}{\delta\psi^{+}(\mathbf{s})}\right)\left(\frac{\delta}{\delta\psi(\mathbf{s})}\right)\{\phi(\mathbf{s})\psi^{+}(\mathbf{s})\}\right\} WP\qquad\emph{G14}\nonumber \\
 &  & -g\int d\mathbf{s\,}\frac{1}{4}\left\{ \left(\frac{\delta}{\delta\psi^{+}(\mathbf{s})}\right)\left(\frac{\delta}{\delta\psi(\mathbf{s})}\right)\{\phi(\mathbf{s})\psi^{+}(\mathbf{s})\}\right\} WP\qquad\emph{G15}\nonumber \\
 &  & -g\int d\mathbf{s\,}\frac{1}{8}\left\{ \left(\frac{\delta}{\delta\psi^{+}(\mathbf{s})}\right)\left(\frac{\delta}{\delta\psi(\mathbf{s})}\right)\left(\frac{\delta}{\delta\psi(\mathbf{s})}\right)\{\phi(\mathbf{s})\}\right\} WP\qquad\emph{G16}\nonumber \\
 &  & -g\int d\mathbf{s\,}\left\{ \left(\frac{\delta}{\delta\phi^{+}(\mathbf{s})}\right)\{\psi(\mathbf{s})\psi^{+}(\mathbf{s})\psi^{+}(\mathbf{s})\}\right\} WP\qquad\emph{G17}\nonumber \\
 &  & -g\int d\mathbf{s\,}\frac{1}{2}\left\{ \left(\frac{\delta}{\delta\phi^{+}(\mathbf{s})}\right)\left(\frac{\delta}{\delta\psi(\mathbf{s})}\right)\{\psi(\mathbf{s})\psi^{+}(\mathbf{s})\}\right\} WP\qquad\emph{G18.1}\nonumber \\
 &  & +g\int d\mathbf{s\,}\frac{1}{2}\left\{ \left(\frac{\delta}{\delta\phi^{+}(\mathbf{s})}\right)\{\delta_{C}(\mathbf{s},\mathbf{s})\psi^{+}(\mathbf{s})\}\right\} WP\qquad\emph{G18.2}\nonumber \\
 &  & -g\int d\mathbf{s\,}\frac{1}{2}\left\{ \left(\frac{\delta}{\delta\phi^{+}(\mathbf{s})}\right)\left(\frac{\delta}{\delta\psi(\mathbf{s})}\right)\{\psi(\mathbf{s})\psi^{+}(\mathbf{s})\}\right\} WP\qquad\emph{G19.1}\nonumber \\
 &  & +g\int d\mathbf{s\,}\frac{1}{2}\left\{ \left(\frac{\delta}{\delta\phi^{+}(\mathbf{s})}\right)\{\delta_{C}(\mathbf{s},\mathbf{s})\psi^{+}(\mathbf{s})\}\right\} WP\qquad\emph{G19.2}\nonumber \\
 &  & -g\int d\mathbf{s\,}\frac{1}{4}\left\{ \left(\frac{\delta}{\delta\phi^{+}(\mathbf{s})}\right)\left(\frac{\delta}{\delta\psi(\mathbf{s})}\right)\left(\frac{\delta}{\delta\psi(\mathbf{s})}\right)\{\psi(\mathbf{s})\}\right\} WP\qquad\emph{G20.1}\nonumber \\
 &  & +g\int d\mathbf{s\,}\frac{1}{2}\left\{ \left(\frac{\delta}{\delta\phi^{+}(\mathbf{s})}\right)\left(\frac{\delta}{\delta\psi(\mathbf{s})}\right)\left\{ \delta_{C}(\mathbf{s},\mathbf{s})\right\} \right\} WP\qquad\emph{G20.2}\nonumber \\
 &  & +g\int d\mathbf{s\,}\frac{1}{2}\left\{ \left(\frac{\delta}{\delta\phi^{+}(\mathbf{s})}\right)\left(\frac{\delta}{\delta\psi^{+}(\mathbf{s})}\right)\{\psi^{+}(\mathbf{s})\psi^{+}(\mathbf{s})\}\right\} WP\qquad\emph{G21}\nonumber \\
 &  & +g\int d\mathbf{s\,}\frac{1}{4}\left\{ \left(\frac{\delta}{\delta\phi^{+}(\mathbf{s})}\right)\left(\frac{\delta}{\delta\psi^{+}(\mathbf{s})}\right)\left(\frac{\delta}{\delta\psi(\mathbf{s})}\right)\{\psi^{+}(\mathbf{s})\}\right\} WP\qquad\emph{G22}\nonumber \\
 &  & +g\int d\mathbf{s\,}\frac{1}{4}\left\{ \left(\frac{\delta}{\delta\phi^{+}(\mathbf{s})}\right)\left(\frac{\delta}{\delta\psi^{+}(\mathbf{s})}\right)\left(\frac{\delta}{\delta\psi(\mathbf{s})}\right)\{\psi^{+}(\mathbf{s})\}\right\} WP\qquad\emph{G23}\nonumber \\
 &  & +g\int d\mathbf{s}\frac{1}{8}\left\{ \left(\frac{\delta}{\delta\phi^{+}(\mathbf{s})}\right)\left(\frac{\delta}{\delta\psi^{+}(\mathbf{s})}\right)\left(\frac{\delta}{\delta\psi(\mathbf{s})}\right)\left(\frac{\delta}{\delta\psi(\mathbf{s})}\right)\right\} WP\qquad G\mathit{24}\nonumber \\
 &  & \emph{\,}\end{eqnarray}

Collecting terms with the same order of functional derivatives we
have\begin{eqnarray}
 &  & WP[\psi(\mathbf{r}),\psi^{+}(\mathbf{r}),\phi(\mathbf{r}),\phi^{+}(\mathbf{r})]\nonumber \\
 & \rightarrow & WP_{0}+WP_{1}+WP_{2}+WP_{3}+WP_{4}\label{Eq.RhoV4Result1}\end{eqnarray}
where we have used lower subscripts for the $\widehat{\rho}\,\widehat{V}_{14}$
contributions and \begin{eqnarray}
 &  & WP_{0}\nonumber \\
 & = & g\int d\mathbf{s}\left\{ \psi(\mathbf{s})\psi(\mathbf{s})\psi^{+}(\mathbf{s})\phi^{+}(\mathbf{s})\right\} WP\qquad\emph{G1}\nonumber \\
 &  & -g\int d\mathbf{s\,}\frac{1}{2}\left\{ \mathbf{2\delta_{C}(\mathbf{s},\mathbf{s})\psi(\mathbf{s})\mathbf{\phi}^{+}\mathbf{(\mathbf{s})}}\right\} WP\qquad\emph{G2.2}\nonumber \\
 &  & +g\int d\mathbf{s}\left\{ \phi(\mathbf{s})\psi(\mathbf{s})\psi^{+}(\mathbf{s})\psi^{+}(\mathbf{s})\right\} WP\qquad\emph{G9}\nonumber \\
 &  & -g\int d\mathbf{s\,}\frac{1}{2}\{\delta_{C}(\mathbf{s},\mathbf{s})\phi(\mathbf{s})\psi^{+}(\mathbf{s})\}WP\qquad\emph{G10.2}\nonumber \\
 &  & -g\int d\mathbf{s\,}\frac{1}{2}\{\delta_{C}(\mathbf{s},\mathbf{s})\phi(\mathbf{s})\psi^{+}(\mathbf{s})\}WP\qquad\emph{G11.2}\nonumber \\
 & = & g\int d\mathbf{s}\left\{ \psi(\mathbf{s})\psi(\mathbf{s})\psi^{+}(\mathbf{s})\phi^{+}(\mathbf{s})\right\} WP[\psi,\psi^{+},\phi,\phi^{+}]\nonumber \\
 &  & +g\int d\mathbf{s}\left\{ \phi(\mathbf{s})\psi(\mathbf{s})\psi^{+}(\mathbf{s})\psi^{+}(\mathbf{s})\right\} WP[\psi,\psi^{+},\phi,\phi^{+}]\nonumber \\
 &  & -g\int d\mathbf{s\,}\left\{ \mathbf{\delta_{C}(\mathbf{s},\mathbf{s})\psi(\mathbf{s})\mathbf{\phi}^{+}\mathbf{(\mathbf{s})}}\right\} WP[\psi,\psi^{+},\phi,\phi^{+}]\nonumber \\
 &  & -g\int d\mathbf{s\,}\{\delta_{C}(\mathbf{s},\mathbf{s})\phi(\mathbf{s})\psi^{+}(\mathbf{s})\}WP[\psi,\psi^{+},\phi,\phi^{+}]\label{Eq.RhoV14Result2}\end{eqnarray}

\begin{eqnarray}
 &  & WP_{1}\nonumber \\
 & = & +g\int d\mathbf{s\,}\frac{1}{2}\left\{ \mathbf{\left(\frac{\delta}{\delta\psi(\mathbf{s})}\right)\{\psi(\mathbf{s})\psi(\mathbf{s})\mathbf{\phi}^{+}\mathbf{(\mathbf{s})}\}}\right\} WP\qquad\emph{G2.1}\nonumber \\
 &  & -g\int d\mathbf{s\,}\frac{1}{2}\left\{ \mathbf{\left(\frac{\delta}{\delta\psi^{+}(\mathbf{s})}\right)\{\psi(\mathbf{s})\psi^{+}(\mathbf{s})\mathbf{\phi}^{+}\mathbf{(\mathbf{s})}\}}\right\} WP\qquad\emph{G3}\nonumber \\
 &  & +g\int d\mathbf{s\,}\frac{1}{4}\left\{ \mathbf{\left(\frac{\delta}{\delta\psi^{+}(\mathbf{s})}\right)\{\delta_{C}(\mathbf{s},\mathbf{s})\mathbf{\phi}^{+}\mathbf{(\mathbf{s})}\}}\right\} WP\qquad\emph{G4.2}\nonumber \\
 &  & -g\int d\mathbf{s\,}\frac{1}{2}\left\{ \mathbf{\left(\frac{\delta}{\delta\psi^{+}(\mathbf{s})}\right)\{\psi(\mathbf{s})\psi^{+}(\mathbf{s})\mathbf{\phi}^{+}\mathbf{(\mathbf{s})}\}}\right\} WP\qquad\emph{G5}\nonumber \\
 &  & +g\int d\mathbf{s\,}\frac{1}{4}\left\{ \mathbf{\left(\frac{\delta}{\delta\psi^{+}(\mathbf{s})}\right)\{\delta_{C}(\mathbf{s},\mathbf{s})\mathbf{\phi}^{+}\mathbf{(\mathbf{s})}\}}\right\} WP\qquad\emph{G6.2}\nonumber \\
 &  & +g\int d\mathbf{s\,}\frac{1}{2}\left\{ \left(\frac{\delta}{\delta\psi(\mathbf{s})}\right)\{\phi(\mathbf{s})\psi(\mathbf{s})\psi^{+}(\mathbf{s})\}\right\} WP\qquad\emph{G10.1}\nonumber \\
 &  & +g\int d\mathbf{s\,}\frac{1}{2}\left\{ \left(\frac{\delta}{\delta\psi(\mathbf{s})}\right)\{\phi(\mathbf{s})\psi(\mathbf{s})\psi^{+}(\mathbf{s})\}\right\} WP\qquad\emph{G11.1}\nonumber \\
 &  & -g\int d\mathbf{s\,}\frac{1}{2}\left\{ \left(\frac{\delta}{\delta\psi(\mathbf{s})}\right)\{\phi(\mathbf{s})\delta_{C}(\mathbf{s},\mathbf{s})\}\right\} WP\qquad\emph{G12.2}\nonumber \\
 &  & -g\int d\mathbf{s\,}\frac{1}{2}\left\{ \left(\frac{\delta}{\delta\psi^{+}(\mathbf{s})}\right)\{\phi(\mathbf{s})\psi^{+}(\mathbf{s})\psi^{+}(\mathbf{s})\}\right\} WP\qquad\emph{G13}\nonumber \\
 &  & -g\int d\mathbf{s\,}\left\{ \left(\frac{\delta}{\delta\phi^{+}(\mathbf{s})}\right)\{\psi(\mathbf{s})\psi^{+}(\mathbf{s})\psi^{+}(\mathbf{s})\}\right\} WP\qquad\emph{G17}\nonumber \\
 &  & +g\int d\mathbf{s\,}\frac{1}{2}\left\{ \left(\frac{\delta}{\delta\phi^{+}(\mathbf{s})}\right)\{\delta_{C}(\mathbf{s},\mathbf{s})\psi^{+}(\mathbf{s})\}\right\} WP\qquad\emph{G18.2}\nonumber \\
 &  & +g\int d\mathbf{s\,}\frac{1}{2}\left\{ \left(\frac{\delta}{\delta\phi^{+}(\mathbf{s})}\right)\{\delta_{C}(\mathbf{s},\mathbf{s})\psi^{+}(\mathbf{s})\}\right\} WP\qquad\emph{G19.2}\nonumber \\
 & = & +g\int d\mathbf{s\,}\left\{ \left(\frac{\delta}{\delta\psi(\mathbf{s})}\right)\{\phi(\mathbf{s})\psi(\mathbf{s})\psi^{+}(\mathbf{s})\}\right\} WP[\psi,\psi^{+},\phi,\phi^{+}]\qquad\emph{G10.1,G11.1}\nonumber \\
 &  & +g\int d\mathbf{s\,}\frac{1}{2}\left\{ \mathbf{\left(\frac{\delta}{\delta\psi(\mathbf{s})}\right)\{\psi(\mathbf{s})\psi(\mathbf{s})\mathbf{\phi}^{+}\mathbf{(\mathbf{s})}\}}\right\} WP[\psi,\psi^{+},\phi,\phi^{+}]\qquad\emph{G2.1}\nonumber \\
 &  & -g\int d\mathbf{s\,}\frac{1}{2}\left\{ \left(\frac{\delta}{\delta\psi(\mathbf{s})}\right)\{\phi(\mathbf{s})\delta_{C}(\mathbf{s},\mathbf{s})\}\right\} WP[\psi,\psi^{+},\phi,\phi^{+}]\qquad\emph{G12.2}\nonumber \\
 &  & -g\int d\mathbf{s\,}\left\{ \mathbf{\left(\frac{\delta}{\delta\psi^{+}(\mathbf{s})}\right)\{\psi(\mathbf{s})\psi^{+}(\mathbf{s})\mathbf{\phi}^{+}\mathbf{(\mathbf{s})}\}}\right\} WP[\psi,\psi^{+},\phi,\phi^{+}]\qquad\emph{G3,G5}\nonumber \\
 &  & -g\int d\mathbf{s\,}\frac{1}{2}\left\{ \left(\frac{\delta}{\delta\psi^{+}(\mathbf{s})}\right)\{\phi(\mathbf{s})\psi^{+}(\mathbf{s})\psi^{+}(\mathbf{s})\}\right\} WP[\psi,\psi^{+},\phi,\phi^{+}]\qquad\emph{G13}\nonumber \\
 &  & +g\int d\mathbf{s\,}\frac{1}{2}\left\{ \mathbf{\left(\frac{\delta}{\delta\psi^{+}(\mathbf{s})}\right)\{\delta_{C}(\mathbf{s},\mathbf{s})\mathbf{\phi}^{+}\mathbf{(\mathbf{s})}\}}\right\} WP[\psi,\psi^{+},\phi,\phi^{+}]\qquad\emph{G4.2,G6.2}\nonumber \\
 &  & -g\int d\mathbf{s\,}\left\{ \left(\frac{\delta}{\delta\phi^{+}(\mathbf{s})}\right)\{\psi(\mathbf{s})\psi^{+}(\mathbf{s})\psi^{+}(\mathbf{s})\}\right\} WP[\psi,\psi^{+},\phi,\phi^{+}]\qquad\emph{G17}\nonumber \\
 &  & +g\int d\mathbf{s\,}\left\{ \left(\frac{\delta}{\delta\phi^{+}(\mathbf{s})}\right)\{\delta_{C}(\mathbf{s},\mathbf{s})\psi^{+}(\mathbf{s})\}\right\} WP[\psi,\psi^{+},\phi,\phi^{+}]\qquad\emph{G18.2,G19.2}\nonumber \\
 &  & \,\label{eq:RhoV14Result3}\end{eqnarray}

\begin{eqnarray}
 &  & WP_{2}\nonumber \\
 & = & -g\int d\mathbf{s\,}\frac{1}{4}\left\{ \mathbf{\left(\frac{\delta}{\delta\psi^{+}(\mathbf{s})}\right)\left(\frac{\delta}{\delta\psi(\mathbf{s})}\right)\{\psi(\mathbf{s})\mathbf{\phi}^{+}\mathbf{(\mathbf{s})}\}}\right\} WP\qquad\emph{G4.1}\nonumber \\
 &  & -g\int d\mathbf{s\,}\frac{1}{4}\left\{ \mathbf{\left(\frac{\delta}{\delta\psi^{+}(\mathbf{s})}\right)\left(\frac{\delta}{\delta\psi(\mathbf{s})}\right)\{\psi(\mathbf{s})\mathbf{\phi}^{+}\mathbf{(\mathbf{s})}\}}\right\} WP\qquad\emph{G6.1}\nonumber \\
 &  & +g\int d\mathbf{s}\frac{1}{4}\left\{ \left(\frac{\delta}{\delta\psi^{+}(\mathbf{s})}\right)\left(\frac{\delta}{\delta\psi^{+}(\mathbf{s})}\right)\{\psi^{+}(\mathbf{s})\phi^{+}(\mathbf{s})\}\right\} WP\qquad\emph{G7}\nonumber \\
 &  & +g\int d\mathbf{s\,}\frac{1}{4}\left\{ \left(\frac{\delta}{\delta\psi(\mathbf{s})}\right)\left(\frac{\delta}{\delta\psi(\mathbf{s})}\right)\{\phi(\mathbf{s})\psi(\mathbf{s})\}\right\} WP\qquad\emph{G12.1}\nonumber \\
 &  & -g\int d\mathbf{s\,}\frac{1}{4}\left\{ \left(\frac{\delta}{\delta\psi^{+}(\mathbf{s})}\right)\left(\frac{\delta}{\delta\psi(\mathbf{s})}\right)\{\phi(\mathbf{s})\psi^{+}(\mathbf{s})\}\right\} WP\qquad\emph{G14}\nonumber \\
 &  & -g\int d\mathbf{s\,}\frac{1}{4}\left\{ \left(\frac{\delta}{\delta\psi^{+}(\mathbf{s})}\right)\left(\frac{\delta}{\delta\psi(\mathbf{s})}\right)\{\phi(\mathbf{s})\psi^{+}(\mathbf{s})\}\right\} WP\qquad\emph{G15}\nonumber \\
 &  & -g\int d\mathbf{s\,}\frac{1}{2}\left\{ \left(\frac{\delta}{\delta\phi^{+}(\mathbf{s})}\right)\left(\frac{\delta}{\delta\psi(\mathbf{s})}\right)\{\psi(\mathbf{s})\psi^{+}(\mathbf{s})\}\right\} WP\qquad\emph{G18.1}\nonumber \\
 &  & -g\int d\mathbf{s\,}\frac{1}{2}\left\{ \left(\frac{\delta}{\delta\phi^{+}(\mathbf{s})}\right)\left(\frac{\delta}{\delta\psi(\mathbf{s})}\right)\{\psi(\mathbf{s})\psi^{+}(\mathbf{s})\}\right\} WP\qquad\emph{G19.1}\nonumber \\
 &  & +g\int d\mathbf{s\,}\frac{1}{2}\left\{ \left(\frac{\delta}{\delta\phi^{+}(\mathbf{s})}\right)\left(\frac{\delta}{\delta\psi(\mathbf{s})}\right)\left\{ \delta_{C}(\mathbf{s},\mathbf{s})\right\} \right\} WP\qquad\emph{G20.2}\nonumber \\
 &  & +g\int d\mathbf{s\,}\frac{1}{2}\left\{ \left(\frac{\delta}{\delta\phi^{+}(\mathbf{s})}\right)\left(\frac{\delta}{\delta\psi^{+}(\mathbf{s})}\right)\{\psi^{+}(\mathbf{s})\psi^{+}(\mathbf{s})\}\right\} WP\qquad\emph{G21}\nonumber \\
 & = & +g\int d\mathbf{s\,}\frac{1}{4}\left\{ \left(\frac{\delta}{\delta\psi(\mathbf{s})}\right)\left(\frac{\delta}{\delta\psi(\mathbf{s})}\right)\{\phi(\mathbf{s})\psi(\mathbf{s})\}\right\} WP[\psi,\psi^{+},\phi,\phi^{+}]\qquad\emph{G12.1}\nonumber \\
 &  & -g\int d\mathbf{s\,}\frac{1}{2}\left\{ \left(\frac{\delta}{\delta\psi(\mathbf{s})}\right)\left(\frac{\delta}{\delta\psi^{+}(\mathbf{s})}\right)\{\phi(\mathbf{s})\psi^{+}(\mathbf{s})\}\right\} WP[\psi,\psi^{+},\phi,\phi^{+}]\qquad\emph{G14,G15}\nonumber \\
 &  & -g\int d\mathbf{s\,}\left\{ \left(\frac{\delta}{\delta\psi(\mathbf{s})}\right)\left(\frac{\delta}{\delta\phi^{+}(\mathbf{s})}\right)\{\psi(\mathbf{s})\psi^{+}(\mathbf{s})\}\right\} WP[\psi,\psi^{+},\phi,\phi^{+}]\qquad\emph{G18.1,19.1}\nonumber \\
 &  & +g\int d\mathbf{s\,}\frac{1}{2}\left\{ \left(\frac{\delta}{\delta\psi(\mathbf{s})}\right)\left(\frac{\delta}{\delta\phi^{+}(\mathbf{s})}\right)\left\{ \delta_{C}(\mathbf{s},\mathbf{s})\right\} \right\} WP[\psi,\psi^{+},\phi,\phi^{+}]\qquad\emph{G20.2}\nonumber \\
 &  & +g\int d\mathbf{s\,}\frac{1}{2}\left\{ \left(\frac{\delta}{\delta\psi^{+}(\mathbf{s})}\right)\left(\frac{\delta}{\delta\phi^{+}(\mathbf{s})}\right)\{\psi^{+}(\mathbf{s})\psi^{+}(\mathbf{s})\}\right\} WP[\psi,\psi^{+},\phi,\phi^{+}]\qquad\emph{G21}\nonumber \\
 &  & -g\int d\mathbf{s\,}\frac{1}{2}\left\{ \mathbf{\left(\frac{\delta}{\delta\psi(\mathbf{s})}\right)\left(\frac{\delta}{\delta\psi^{+}(\mathbf{s})}\right)\{\psi(\mathbf{s})\mathbf{\phi}^{+}\mathbf{(\mathbf{s})}\}}\right\} WP[\psi,\psi^{+},\phi,\phi^{+}]\qquad\emph{G4.1,6.1}\nonumber \\
 &  & +g\int d\mathbf{s}\frac{1}{4}\left\{ \left(\frac{\delta}{\delta\psi^{+}(\mathbf{s})}\right)\left(\frac{\delta}{\delta\psi^{+}(\mathbf{s})}\right)\{\psi^{+}(\mathbf{s})\phi^{+}(\mathbf{s})\}\right\} WP[\psi,\psi^{+},\phi,\phi^{+}]\qquad\emph{G7}\nonumber \\
 &  & \,\label{eq:RhoV14Result4}\end{eqnarray}

\begin{eqnarray}
 &  & WP_{3}\nonumber \\
 & = & +g\int d\mathbf{s\,}\frac{1}{8}\left\{ \mathbf{\left(\frac{\delta}{\delta\psi^{+}(\mathbf{s})}\right)\left(\frac{\delta}{\delta\psi^{+}(\mathbf{s})}\right)\left(\frac{\delta}{\delta\psi(\mathbf{s})}\right)\{\mathbf{\phi}^{+}\mathbf{(\mathbf{s})}\}}\right\} WP\qquad\emph{G8}\nonumber \\
 &  & -g\int d\mathbf{s\,}\frac{1}{8}\left\{ \left(\frac{\delta}{\delta\psi^{+}(\mathbf{s})}\right)\left(\frac{\delta}{\delta\psi(\mathbf{s})}\right)\left(\frac{\delta}{\delta\psi(\mathbf{s})}\right)\{\phi(\mathbf{s})\}\right\} WP\qquad\emph{G16}\nonumber \\
 &  & -g\int d\mathbf{s\,}\frac{1}{4}\left\{ \left(\frac{\delta}{\delta\phi^{+}(\mathbf{s})}\right)\left(\frac{\delta}{\delta\psi(\mathbf{s})}\right)\left(\frac{\delta}{\delta\psi(\mathbf{s})}\right)\{\psi(\mathbf{s})\}\right\} WP\qquad\emph{G20.1}\nonumber \\
 &  & +g\int d\mathbf{s\,}\frac{1}{4}\left\{ \left(\frac{\delta}{\delta\phi^{+}(\mathbf{s})}\right)\left(\frac{\delta}{\delta\psi^{+}(\mathbf{s})}\right)\left(\frac{\delta}{\delta\psi(\mathbf{s})}\right)\{\psi^{+}(\mathbf{s})\}\right\} WP\qquad\emph{G22}\nonumber \\
 &  & +g\int d\mathbf{s\,}\frac{1}{4}\left\{ \left(\frac{\delta}{\delta\phi^{+}(\mathbf{s})}\right)\left(\frac{\delta}{\delta\psi^{+}(\mathbf{s})}\right)\left(\frac{\delta}{\delta\psi(\mathbf{s})}\right)\{\psi^{+}(\mathbf{s})\}\right\} WP\qquad\emph{G23}\nonumber \\
 & = & -g\int d\mathbf{s\,}\frac{1}{8}\left\{ \left(\frac{\delta}{\delta\psi(\mathbf{s})}\right)\left(\frac{\delta}{\delta\psi(\mathbf{s})}\right)\left(\frac{\delta}{\delta\psi^{+}(\mathbf{s})}\right)\{\phi(\mathbf{s})\}\right\} WP[\psi,\psi^{+},\phi,\phi^{+}]\qquad\emph{G16}\nonumber \\
 &  & -g\int d\mathbf{s\,}\frac{1}{4}\left\{ \left(\frac{\delta}{\delta\psi(\mathbf{s})}\right)\left(\frac{\delta}{\delta\psi(\mathbf{s})}\right)\left(\frac{\delta}{\delta\phi^{+}(\mathbf{s})}\right)\{\psi(\mathbf{s})\}\right\} WP[\psi,\psi^{+},\phi,\phi^{+}]\qquad\emph{G20.1}\nonumber \\
 &  & +g\int d\mathbf{s\,}\frac{1}{2}\left\{ \left(\frac{\delta}{\delta\psi(\mathbf{s})}\right)\left(\frac{\delta}{\delta\psi^{+}(\mathbf{s})}\right)\left(\frac{\delta}{\delta\phi^{+}(\mathbf{s})}\right)\{\psi^{+}(\mathbf{s})\}\right\} WP[\psi,\psi^{+},\phi,\phi^{+}]\qquad\emph{G22,G23}\nonumber \\
 &  & +g\int d\mathbf{s\,}\frac{1}{8}\left\{ \mathbf{\left(\frac{\delta}{\delta\psi(\mathbf{s})}\right)\left(\frac{\delta}{\delta\psi^{+}(\mathbf{s})}\right)\left(\frac{\delta}{\delta\psi^{+}(\mathbf{s})}\right)\{\mathbf{\phi}^{+}\mathbf{(\mathbf{s})}\}}\right\} WP[\psi,\psi^{+},\phi,\phi^{+}]\qquad\emph{G8}\nonumber \\
 &  & \,\label{eq:RhoV14Result5}\end{eqnarray}

\begin{eqnarray}
 &  & WP_{4}\nonumber \\
 & = & +g\int d\mathbf{s}\frac{1}{8}\left\{ \left(\frac{\delta}{\delta\psi(\mathbf{s})}\right)\left(\frac{\delta}{\delta\psi(\mathbf{s})}\right)\left(\frac{\delta}{\delta\psi^{+}(\mathbf{s})}\right)\left(\frac{\delta}{\delta\phi^{+}(\mathbf{s})}\right)\right\} WP[\psi,\psi^{+},\phi,\phi^{+}]\nonumber \\
 &  & \,\label{eq:RhoV14Result6}\end{eqnarray}

Now if \begin{align}
\widehat{\rho} & \rightarrow\lbrack\widehat{V}_{14}\,,\widehat{\rho}]\nonumber \\
= & [g\int d\mathbf{s\,(}\widehat{\Psi}_{NC}^{\dagger}(\mathbf{s})\widehat{\Psi}_{C}^{\dagger}(\mathbf{s})\,\widehat{\Psi}_{C}(\mathbf{s})\widehat{\Psi}_{C}(\mathbf{s}))+\widehat{\Psi}_{C}^{\dagger}(\mathbf{s})\widehat{\Psi}_{C}^{\dagger}(\mathbf{s})\,\widehat{\Psi}_{C}(\mathbf{s})\widehat{\Psi}_{NC}(\mathbf{s}))\,,\widehat{\rho}]\nonumber \\
 & \,\end{align}
then \begin{eqnarray}
 &  & WP[\psi(\mathbf{r}),\psi^{+}(\mathbf{r}),\phi(\mathbf{r}),\phi^{+}(\mathbf{r})]\nonumber \\
 & \rightarrow & WP_{T}^{0}+WP_{T}^{1}+WP_{T}^{2}+WP_{T}^{3}+WP_{T}^{4}\label{Eq.V14CommutRhoResult1}\end{eqnarray}
where the $WP_{T}^{n}$ are obtained by subtracting the results for
$\widehat{\rho}\widehat{\, V}_{14}$ from those for $\widehat{V}_{14}\,\widehat{\rho}$

Collecting terms gives\textbf{\ } \begin{eqnarray}
 &  & WP_{T}^{0}\nonumber \\
 & = & g\int d\mathbf{s\,}\left\{ \phi^{+}(\mathbf{s})\psi^{+}(\mathbf{s})\psi(\mathbf{s})\psi(\mathbf{s})\right\} WP+g\int d\mathbf{s\,}\left\{ \mathbf{\psi^{+}(\mathbf{s})\psi^{+}(\mathbf{s})\psi(\mathbf{s})\phi(\mathbf{s})}\right\} WP\nonumber \\
 &  & -g\int d\mathbf{s\,}\{\delta_{C}(\mathbf{s},\mathbf{s})\phi^{+}(\mathbf{s})\psi(\mathbf{s})\}WP-g\int d\mathbf{s\,}\{\delta_{C}(\mathbf{s},\mathbf{s})\psi^{+}(\mathbf{s})\phi(\mathbf{s})\}WP\nonumber \\
 &  & -g\int d\mathbf{s}\left\{ \psi(\mathbf{s})\psi(\mathbf{s})\psi^{+}(\mathbf{s})\phi^{+}(\mathbf{s})\right\} WP-g\int d\mathbf{s}\left\{ \phi(\mathbf{s})\psi(\mathbf{s})\psi^{+}(\mathbf{s})\psi^{+}(\mathbf{s})\right\} WP\nonumber \\
 &  & +g\int d\mathbf{s\,}\left\{ \mathbf{\delta_{C}(\mathbf{s},\mathbf{s})\psi(\mathbf{s})\mathbf{\phi}^{+}\mathbf{(\mathbf{s})}}\right\} WP+g\int d\mathbf{s\,}\{\delta_{C}(\mathbf{s},\mathbf{s})\phi(\mathbf{s})\psi^{+}(\mathbf{s})\}WP\nonumber \\
 & = & 0\label{Eq.V14CommutRhoResult2}\end{eqnarray}

\begin{eqnarray}
 &  & WP_{T}^{1}\nonumber \\
 & = & +g\int d\mathbf{s\,}\left\{ \left(\frac{\delta}{\delta\psi^{+}(\mathbf{s})}\right)\{2\phi^{+}(\mathbf{s})\psi^{+}(\mathbf{s})\psi(\mathbf{s})\}\right\} WP\nonumber \\
 &  & +g\int d\mathbf{s\,}\left\{ \mathbf{\left(\frac{\delta}{\delta\psi^{+}(\mathbf{s})}\right)\{\psi^{+}(\mathbf{s})\psi^{+}(\mathbf{s})\phi(\mathbf{s})\}}\right\} WP\nonumber \\
 &  & -g\int d\mathbf{s\,}\left\{ \left(\frac{\delta}{\delta\psi^{+}(\mathbf{s})}\right)\{\phi^{+}(\mathbf{s})\delta_{C}(\mathbf{s},\mathbf{s})\}\right\} WP\nonumber \\
 &  & -g\int d\mathbf{s\,}\left\{ \mathbf{\left(\frac{\delta}{\delta\psi(\mathbf{s})}\right)\{}2\psi\mathbf{^{+}(\mathbf{s})}\psi\mathbf{(\mathbf{s})}\phi\mathbf{(\mathbf{s})\}}\right\} WP\nonumber \\
 &  & -g\int d\mathbf{s\,}\left\{ \left(\frac{\delta}{\delta\psi(\mathbf{s})}\right)\{\phi^{+}(\mathbf{s})\psi(\mathbf{s})\psi(\mathbf{s})\}\right\} WP\nonumber \\
 &  & +g\int d\mathbf{s\,}\left\{ \mathbf{\left(\frac{\delta}{\delta\psi(\mathbf{s})}\right)\{\delta_{C}(\mathbf{s},\mathbf{s})\phi(\mathbf{s})\}}\right\} WP\nonumber \\
 &  & -g\int d\mathbf{s\,}\left\{ \left(\frac{\delta}{\delta\phi(\mathbf{s})}\right)\{\psi^{+}(\mathbf{s})\psi(\mathbf{s})\psi(\mathbf{s})\}\right\} WP\nonumber \\
 &  & +g\int d\mathbf{s\,}\left\{ \left(\frac{\delta}{\delta\phi(\mathbf{s})}\right)\{\delta_{C}(\mathbf{s},\mathbf{s})\psi(\mathbf{s})\}\right\} WP\nonumber \\
 &  & +g\int d\mathbf{s\,}\left\{ \left(\frac{\delta}{\delta\phi^{+}(\mathbf{s})}\right)\{\psi(\mathbf{s})\psi^{+}(\mathbf{s})\psi^{+}(\mathbf{s})\}\right\} WP\nonumber \\
 &  & -g\int d\mathbf{s\,}\left\{ \left(\frac{\delta}{\delta\phi^{+}(\mathbf{s})}\right)\{\delta_{C}(\mathbf{s},\mathbf{s})\psi^{+}(\mathbf{s})\}\right\} WP\nonumber \\
 &  & \,\label{eq:V14RhoCommResult3}\end{eqnarray}
\begin{eqnarray*}
 &  & WP_{T}^{2}\\
 & = & +g\int d\mathbf{s\,}\frac{1}{4}\left\{ \left(\frac{\delta}{\delta\psi^{+}(\mathbf{s})}\right)\left(\frac{\delta}{\delta\psi^{+}(\mathbf{s})}\right)\{\phi^{+}(\mathbf{s})\psi^{+}(\mathbf{s})\}\right\} WP\\
 &  & -g\int d\mathbf{s\,}\frac{1}{2}\left\{ \left(\frac{\delta}{\delta\psi^{+}(\mathbf{s})}\right)\left(\frac{\delta}{\delta\psi(\mathbf{s})}\right)\{\phi^{+}(\mathbf{s})\psi(\mathbf{s})\}\right\} WP\\
 &  & -g\int d\mathbf{s\,}\left\{ \left(\frac{\delta}{\delta\psi^{+}(\mathbf{s})}\right)\left(\frac{\delta}{\delta\phi(\mathbf{s})}\right)\{\psi^{+}(\mathbf{s})\psi(\mathbf{s})\}\right\} WP\\
 &  & +g\int d\mathbf{s\,}\frac{1}{2}\left\{ \left(\frac{\delta}{\delta\psi^{+}(\mathbf{s})}\right)\left(\frac{\delta}{\delta\phi(\mathbf{s})}\right)\left\{ \delta_{C}(\mathbf{s},\mathbf{s})\right\} \right\} WP\\
 &  & +g\int d\mathbf{s\,}\frac{1}{2}\left\{ \left(\frac{\delta}{\delta\psi(\mathbf{s})}\right)\left(\frac{\delta}{\delta\phi(\mathbf{s})}\right)\{\psi(\mathbf{s})\psi(\mathbf{s})\}\right\} WP\\
 &  & -g\int d\mathbf{s\,}\frac{1}{2}\left\{ \mathbf{\left(\frac{\delta}{\delta\psi^{+}(\mathbf{s})}\right)\left(\frac{\delta}{\delta\psi(\mathbf{s})}\right)\{\psi^{+}(\mathbf{s})\phi(\mathbf{s})\}}\right\} WP\\
 &  & +g\int d\mathbf{s\,}\frac{1}{4}\left\{ \mathbf{\left(\frac{\delta}{\delta\psi(\mathbf{s})}\right)\left(\frac{\delta}{\delta\psi(\mathbf{s})}\right)\{\psi(\mathbf{s})\phi(\mathbf{s})\}}\right\} WP\\
 &  & -g\int d\mathbf{s\,}\frac{1}{4}\left\{ \left(\frac{\delta}{\delta\psi(\mathbf{s})}\right)\left(\frac{\delta}{\delta\psi(\mathbf{s})}\right)\{\phi(\mathbf{s})\psi(\mathbf{s})\}\right\} WP\\
 &  & +g\int d\mathbf{s\,}\frac{1}{2}\left\{ \left(\frac{\delta}{\delta\psi(\mathbf{s})}\right)\left(\frac{\delta}{\delta\psi^{+}(\mathbf{s})}\right)\{\phi(\mathbf{s})\psi^{+}(\mathbf{s})\}\right\} WP\\
 &  & +g\int d\mathbf{s\,}\left\{ \left(\frac{\delta}{\delta\psi(\mathbf{s})}\right)\left(\frac{\delta}{\delta\phi^{+}(\mathbf{s})}\right)\{\psi(\mathbf{s})\psi^{+}(\mathbf{s})\}\right\} WP\\
 &  & -g\int d\mathbf{s\,}\frac{1}{2}\left\{ \left(\frac{\delta}{\delta\psi(\mathbf{s})}\right)\left(\frac{\delta}{\delta\phi^{+}(\mathbf{s})}\right)\left\{ \delta_{C}(\mathbf{s},\mathbf{s})\right\} \right\} WP\\
 &  & -g\int d\mathbf{s\,}\frac{1}{2}\left\{ \left(\frac{\delta}{\delta\psi^{+}(\mathbf{s})}\right)\left(\frac{\delta}{\delta\phi^{+}(\mathbf{s})}\right)\{\psi^{+}(\mathbf{s})\psi^{+}(\mathbf{s})\}\right\} WP\\
 &  & +g\int d\mathbf{s\,}\frac{1}{2}\left\{ \mathbf{\left(\frac{\delta}{\delta\psi(\mathbf{s})}\right)\left(\frac{\delta}{\delta\psi^{+}(\mathbf{s})}\right)\{\psi(\mathbf{s})\mathbf{\phi}^{+}\mathbf{(\mathbf{s})}\}}\right\} WP\\
 &  & -g\int d\mathbf{s}\frac{1}{4}\left\{ \left(\frac{\delta}{\delta\psi^{+}(\mathbf{s})}\right)\left(\frac{\delta}{\delta\psi^{+}(\mathbf{s})}\right)\{\psi^{+}(\mathbf{s})\phi^{+}(\mathbf{s})\}\right\} WP\end{eqnarray*}
Collecting all the terms gives\begin{eqnarray}
 &  & WP_{T}^{2}\nonumber \\
 & = & +g\int d\mathbf{s\,}\frac{1}{4}\left\{ \left(\frac{\delta}{\delta\psi^{+}(\mathbf{s})}\right)\left(\frac{\delta}{\delta\psi^{+}(\mathbf{s})}\right)\{\phi^{+}(\mathbf{s})\psi^{+}(\mathbf{s})\}\right\} WP\qquad\emph{Cancel}\nonumber \\
 &  & -g\int d\mathbf{s}\frac{1}{4}\left\{ \left(\frac{\delta}{\delta\psi^{+}(\mathbf{s})}\right)\left(\frac{\delta}{\delta\psi^{+}(\mathbf{s})}\right)\{\psi^{+}(\mathbf{s})\phi^{+}(\mathbf{s})\}\right\} WP\qquad\emph{Cancel}\nonumber \\
 &  & -g\int d\mathbf{s\,}\frac{1}{2}\left\{ \left(\frac{\delta}{\delta\psi^{+}(\mathbf{s})}\right)\left(\frac{\delta}{\delta\psi(\mathbf{s})}\right)\{\phi^{+}(\mathbf{s})\psi(\mathbf{s})\}\right\} WP\qquad\emph{Cancel}\nonumber \\
 &  & +g\int d\mathbf{s\,}\frac{1}{2}\left\{ \mathbf{\left(\frac{\delta}{\delta\psi(\mathbf{s})}\right)\left(\frac{\delta}{\delta\psi^{+}(\mathbf{s})}\right)\{\psi(\mathbf{s})\mathbf{\phi}^{+}\mathbf{(\mathbf{s})}\}}\right\} WP\qquad\emph{Cancel}\nonumber \\
 &  & -g\int d\mathbf{s\,}\left\{ \left(\frac{\delta}{\delta\psi^{+}(\mathbf{s})}\right)\left(\frac{\delta}{\delta\phi(\mathbf{s})}\right)\{\psi^{+}(\mathbf{s})\psi(\mathbf{s})\}\right\} WP\nonumber \\
 &  & +g\int d\mathbf{s\,}\left\{ \left(\frac{\delta}{\delta\psi(\mathbf{s})}\right)\left(\frac{\delta}{\delta\phi^{+}(\mathbf{s})}\right)\{\psi(\mathbf{s})\psi^{+}(\mathbf{s})\}\right\} WP\nonumber \\
 &  & +g\int d\mathbf{s\,}\frac{1}{2}\left\{ \left(\frac{\delta}{\delta\psi^{+}(\mathbf{s})}\right)\left(\frac{\delta}{\delta\phi(\mathbf{s})}\right)\left\{ \delta_{C}(\mathbf{s},\mathbf{s})\right\} \right\} WP\nonumber \\
 &  & -g\int d\mathbf{s\,}\frac{1}{2}\left\{ \left(\frac{\delta}{\delta\psi(\mathbf{s})}\right)\left(\frac{\delta}{\delta\phi^{+}(\mathbf{s})}\right)\left\{ \delta_{C}(\mathbf{s},\mathbf{s})\right\} \right\} WP\nonumber \\
 &  & +g\int d\mathbf{s\,}\frac{1}{2}\left\{ \left(\frac{\delta}{\delta\psi(\mathbf{s})}\right)\left(\frac{\delta}{\delta\phi(\mathbf{s})}\right)\{\psi(\mathbf{s})\psi(\mathbf{s})\}\right\} WP\nonumber \\
 &  & -g\int d\mathbf{s\,}\frac{1}{2}\left\{ \left(\frac{\delta}{\delta\psi^{+}(\mathbf{s})}\right)\left(\frac{\delta}{\delta\phi^{+}(\mathbf{s})}\right)\{\psi^{+}(\mathbf{s})\psi^{+}(\mathbf{s})\}\right\} WP\nonumber \\
 &  & -g\int d\mathbf{s\,}\frac{1}{2}\left\{ \mathbf{\left(\frac{\delta}{\delta\psi^{+}(\mathbf{s})}\right)\left(\frac{\delta}{\delta\psi(\mathbf{s})}\right)\{\psi^{+}(\mathbf{s})\phi(\mathbf{s})\}}\right\} WP\qquad\emph{Cancel}\nonumber \\
 &  & +g\int d\mathbf{s\,}\frac{1}{2}\left\{ \left(\frac{\delta}{\delta\psi(\mathbf{s})}\right)\left(\frac{\delta}{\delta\psi^{+}(\mathbf{s})}\right)\{\phi(\mathbf{s})\psi^{+}(\mathbf{s})\}\right\} WP\qquad\emph{Cancel}\nonumber \\
 &  & +g\int d\mathbf{s\,}\frac{1}{4}\left\{ \mathbf{\left(\frac{\delta}{\delta\psi(\mathbf{s})}\right)\left(\frac{\delta}{\delta\psi(\mathbf{s})}\right)\{\psi(\mathbf{s})\phi(\mathbf{s})\}}\right\} WP\qquad\emph{Cancel}\nonumber \\
 &  & -g\int d\mathbf{s\,}\frac{1}{4}\left\{ \left(\frac{\delta}{\delta\psi(\mathbf{s})}\right)\left(\frac{\delta}{\delta\psi(\mathbf{s})}\right)\{\phi(\mathbf{s})\psi(\mathbf{s})\}\right\} WP\qquad\emph{Cancel}\nonumber \\
 & = & -g\int d\mathbf{s\,}\left\{ \left(\frac{\delta}{\delta\psi^{+}(\mathbf{s})}\right)\left(\frac{\delta}{\delta\phi(\mathbf{s})}\right)\{\psi^{+}(\mathbf{s})\psi(\mathbf{s})\}\right\} WP\nonumber \\
 &  & +g\int d\mathbf{s\,}\left\{ \left(\frac{\delta}{\delta\psi(\mathbf{s})}\right)\left(\frac{\delta}{\delta\phi^{+}(\mathbf{s})}\right)\{\psi(\mathbf{s})\psi^{+}(\mathbf{s})\}\right\} WP\nonumber \\
 &  & +g\int d\mathbf{s}\left\{ \left(\frac{\delta}{\delta\psi^{+}(\mathbf{s})}\right)\left(\frac{\delta}{\delta\phi(\mathbf{s})}\right)\left\{ \mathbf{\,}\frac{1}{2}\delta_{C}(\mathbf{s},\mathbf{s})\right\} \right\} WP\nonumber \\
 &  & -g\int d\mathbf{s\,}\left\{ \left(\frac{\delta}{\delta\psi(\mathbf{s})}\right)\left(\frac{\delta}{\delta\phi^{+}(\mathbf{s})}\right)\left\{ \frac{1}{2}\delta_{C}(\mathbf{s},\mathbf{s})\right\} \right\} WP\nonumber \\
 &  & +g\int d\mathbf{s\,}\left\{ \left(\frac{\delta}{\delta\psi(\mathbf{s})}\right)\left(\frac{\delta}{\delta\phi(\mathbf{s})}\right)\{\frac{1}{2}\psi(\mathbf{s})\psi(\mathbf{s})\}\right\} WP\nonumber \\
 &  & -g\int d\mathbf{s\,}\left\{ \left(\frac{\delta}{\delta\psi^{+}(\mathbf{s})}\right)\left(\frac{\delta}{\delta\phi^{+}(\mathbf{s})}\right)\{\frac{1}{2}\psi^{+}(\mathbf{s})\psi^{+}(\mathbf{s})\}\right\} WP\nonumber \\
 &  & \,\label{eq:V14RhoCommResult4}\end{eqnarray}

\begin{eqnarray*}
 &  & WP_{T}^{3}\\
 & = & -g\int d\mathbf{s\,}\frac{1}{8}\left\{ \left(\frac{\delta}{\delta\psi^{+}(\mathbf{s})}\right)\left(\frac{\delta}{\delta\psi^{+}(\mathbf{s})}\right)\left(\frac{\delta}{\delta\psi(\mathbf{s})}\right)\{\phi^{+}(\mathbf{s})\}\right\} WP\\
 &  & -g\int d\mathbf{s\,}\frac{1}{4}\left\{ \left(\frac{\delta}{\delta\psi^{+}(\mathbf{s})}\right)\left(\frac{\delta}{\delta\psi^{+}(\mathbf{s})}\right)\left(\frac{\delta}{\delta\phi(\mathbf{s})}\right)\{\psi^{+}(\mathbf{s})\}\right\} WP\\
 &  & +g\int d\mathbf{s\,}\frac{1}{2}\left\{ \left(\frac{\delta}{\delta\psi^{+}(\mathbf{s})}\right)\left(\frac{\delta}{\delta\psi(\mathbf{s})}\right)\left(\frac{\delta}{\delta\phi(\mathbf{s})}\right)\{\psi(\mathbf{s})\}\right\} WP\\
 &  & +g\int d\mathbf{s\,}\frac{1}{8}\left\{ \mathbf{\left(\frac{\delta}{\delta\psi^{+}(\mathbf{s})}\right)\left(\frac{\delta}{\delta\psi(\mathbf{s})}\right)\left(\frac{\delta}{\delta\psi(\mathbf{s})}\right)\{\phi(\mathbf{s})\}}\right\} WP\\
 &  & +g\int d\mathbf{s\,}\frac{1}{8}\left\{ \left(\frac{\delta}{\delta\psi(\mathbf{s})}\right)\left(\frac{\delta}{\delta\psi(\mathbf{s})}\right)\left(\frac{\delta}{\delta\psi^{+}(\mathbf{s})}\right)\{\phi(\mathbf{s})\}\right\} WP\\
 &  & +g\int d\mathbf{s\,}\frac{1}{4}\left\{ \left(\frac{\delta}{\delta\psi(\mathbf{s})}\right)\left(\frac{\delta}{\delta\psi(\mathbf{s})}\right)\left(\frac{\delta}{\delta\phi^{+}(\mathbf{s})}\right)\{\psi(\mathbf{s})\}\right\} WP\\
 &  & -g\int d\mathbf{s\,}\frac{1}{2}\left\{ \left(\frac{\delta}{\delta\psi(\mathbf{s})}\right)\left(\frac{\delta}{\delta\psi^{+}(\mathbf{s})}\right)\left(\frac{\delta}{\delta\phi^{+}(\mathbf{s})}\right)\{\psi^{+}(\mathbf{s})\}\right\} WP\\
 &  & -g\int d\mathbf{s\,}\frac{1}{8}\left\{ \mathbf{\left(\frac{\delta}{\delta\psi(\mathbf{s})}\right)\left(\frac{\delta}{\delta\psi^{+}(\mathbf{s})}\right)\left(\frac{\delta}{\delta\psi^{+}(\mathbf{s})}\right)\{\mathbf{\phi}^{+}\mathbf{(\mathbf{s})}\}}\right\} WP\end{eqnarray*}
Collecting the terms gives\begin{eqnarray}
 &  & WP_{T}^{3}\nonumber \\
 & = & -g\int d\mathbf{s\,}\frac{1}{4}\left\{ \left(\frac{\delta}{\delta\psi^{+}(\mathbf{s})}\right)\left(\frac{\delta}{\delta\psi^{+}(\mathbf{s})}\right)\left(\frac{\delta}{\delta\psi(\mathbf{s})}\right)\{\phi^{+}(\mathbf{s})\}\right\} WP\nonumber \\
 &  & -g\int d\mathbf{s\,}\frac{1}{4}\left\{ \left(\frac{\delta}{\delta\psi^{+}(\mathbf{s})}\right)\left(\frac{\delta}{\delta\psi^{+}(\mathbf{s})}\right)\left(\frac{\delta}{\delta\phi(\mathbf{s})}\right)\{\psi^{+}(\mathbf{s})\}\right\} WP\nonumber \\
 &  & +g\int d\mathbf{s\,}\frac{1}{4}\left\{ \left(\frac{\delta}{\delta\psi(\mathbf{s})}\right)\left(\frac{\delta}{\delta\psi(\mathbf{s})}\right)\left(\frac{\delta}{\delta\phi^{+}(\mathbf{s})}\right)\{\psi(\mathbf{s})\}\right\} WP\nonumber \\
 &  & +g\int d\mathbf{s\,}\frac{1}{2}\left\{ \left(\frac{\delta}{\delta\psi^{+}(\mathbf{s})}\right)\left(\frac{\delta}{\delta\psi(\mathbf{s})}\right)\left(\frac{\delta}{\delta\phi(\mathbf{s})}\right)\{\psi(\mathbf{s})\}\right\} WP\nonumber \\
 &  & -g\int d\mathbf{s\,}\frac{1}{2}\left\{ \left(\frac{\delta}{\delta\psi(\mathbf{s})}\right)\left(\frac{\delta}{\delta\psi^{+}(\mathbf{s})}\right)\left(\frac{\delta}{\delta\phi^{+}(\mathbf{s})}\right)\{\psi^{+}(\mathbf{s})\}\right\} WP\nonumber \\
 &  & +g\int d\mathbf{s\,}\frac{1}{4}\left\{ \left(\frac{\delta}{\delta\psi(\mathbf{s})}\right)\left(\frac{\delta}{\delta\psi(\mathbf{s})}\right)\left(\frac{\delta}{\delta\psi^{+}(\mathbf{s})}\right)\{\phi(\mathbf{s})\}\right\} WP\nonumber \\
 &  & \,\label{eq:V14RhoCommResult5}\end{eqnarray}
\begin{eqnarray}
 &  & WP_{T}^{4}\nonumber \\
 & = & +g\int d\mathbf{s\,}\frac{1}{8}\left\{ \left(\frac{\delta}{\delta\psi^{+}(\mathbf{s})}\right)\left(\frac{\delta}{\delta\psi^{+}(\mathbf{s})}\right)\left(\frac{\delta}{\delta\psi(\mathbf{s})}\right)\left(\frac{\delta}{\delta\phi(\mathbf{s})}\right)\right\} WP\nonumber \\
 &  & -g\int d\mathbf{s}\frac{1}{8}\left\{ \left(\frac{\delta}{\delta\psi(\mathbf{s})}\right)\left(\frac{\delta}{\delta\psi(\mathbf{s})}\right)\left(\frac{\delta}{\delta\psi^{+}(\mathbf{s})}\right)\left(\frac{\delta}{\delta\phi^{+}(\mathbf{s})}\right)\right\} WP\nonumber \\
 &  & \,\label{eq:V14RhoCommResult6}\end{eqnarray}

Thus we see that the $\widehat{\, V}_{14}$ term produces functional
derivatives of orders one, two, three and four. We may write the contributions
to the functional Fokker-Planck equation in the form\begin{eqnarray}
 &  & \left(\frac{\partial}{\partial t}WP[\psi,\psi^{+},\phi,\phi^{+}]\right)_{V14}\nonumber \\
 & = & \left(\frac{\partial}{\partial t}WP[\psi,\psi^{+},\phi,\phi^{+}]\right)_{V14}^{1}+\left(\frac{\partial}{\partial t}WP[\psi,\psi^{+},\phi,\phi^{+}]\right)_{V14}^{2}\nonumber \\
 &  & +\left(\frac{\partial}{\partial t}WP[\psi,\psi^{+},\phi,\phi^{+}]\right)_{V14}^{3}+\left(\frac{\partial}{\partial t}WP[\psi,\psi^{+},\phi,\phi^{+}]\right)_{V14}^{4}\nonumber \\
 &  & \,\label{eq:FnalFPV14}\end{eqnarray}
where\begin{eqnarray}
 &  & \left(\frac{\partial}{\partial t}WP[\psi,\psi^{+},\phi,\phi^{+}]\right)_{V14}^{1}\nonumber \\
 & = & \frac{-i}{\hbar}\left\{ +g\int d\mathbf{s\,}\left\{ \left(\frac{\delta}{\delta\psi^{+}(\mathbf{s})}\right)\{2\phi^{+}(\mathbf{s})\psi^{+}(\mathbf{s})\psi(\mathbf{s})\}\right\} WP[\psi,\psi^{+},\phi,\phi^{+}]\right\} \nonumber \\
 &  & \frac{-i}{\hbar}\left\{ +g\int d\mathbf{s\,}\left\{ \mathbf{\left(\frac{\delta}{\delta\psi^{+}(\mathbf{s})}\right)\{\psi^{+}(\mathbf{s})\psi^{+}(\mathbf{s})\phi(\mathbf{s})\}}\right\} WP[\psi,\psi^{+},\phi,\phi^{+}]\right\} \nonumber \\
 &  & \frac{-i}{\hbar}\left\{ -g\int d\mathbf{s\,}\left\{ \left(\frac{\delta}{\delta\psi^{+}(\mathbf{s})}\right)\{\phi^{+}(\mathbf{s})\delta_{C}(\mathbf{s},\mathbf{s})\}\right\} WP[\psi,\psi^{+},\phi,\phi^{+}]\right\} \nonumber \\
 &  & \frac{-i}{\hbar}\left\{ -g\int d\mathbf{s\,}\left\{ \mathbf{\left(\frac{\delta}{\delta\psi(\mathbf{s})}\right)\{}2\psi\mathbf{^{+}(\mathbf{s})}\psi\mathbf{(\mathbf{s})\phi(\mathbf{s})\}}\right\} WP[\psi,\psi^{+},\phi,\phi^{+}]\right\} \nonumber \\
 &  & \frac{-i}{\hbar}\left\{ -g\int d\mathbf{s\,}\left\{ \left(\frac{\delta}{\delta\psi(\mathbf{s})}\right)\{\phi^{+}(\mathbf{s})\psi(\mathbf{s})\psi(\mathbf{s})\}\right\} WP[\psi,\psi^{+},\phi,\phi^{+}]\right\} \nonumber \\
 &  & \frac{-i}{\hbar}\left\{ +g\int d\mathbf{s\,}\left\{ \mathbf{\left(\frac{\delta}{\delta\psi(\mathbf{s})}\right)\{\delta_{C}(\mathbf{s},\mathbf{s})\phi(\mathbf{s})\}}\right\} WP[\psi,\psi^{+},\phi,\phi^{+}]\right\} \nonumber \\
 &  & \frac{-i}{\hbar}\left\{ -g\int d\mathbf{s\,}\left\{ \left(\frac{\delta}{\delta\phi(\mathbf{s})}\right)\{\psi^{+}(\mathbf{s})\psi(\mathbf{s})\psi(\mathbf{s})\}\right\} WP[\psi,\psi^{+},\phi,\phi^{+}]\right\} \nonumber \\
 &  & \frac{-i}{\hbar}\left\{ +g\int d\mathbf{s\,}\left\{ \left(\frac{\delta}{\delta\phi(\mathbf{s})}\right)\{\delta_{C}(\mathbf{s},\mathbf{s})\psi(\mathbf{s})\}\right\} WP[\psi,\psi^{+},\phi,\phi^{+}]\right\} \nonumber \\
 &  & \frac{-i}{\hbar}\left\{ +g\int d\mathbf{s\,}\left\{ \left(\frac{\delta}{\delta\phi^{+}(\mathbf{s})}\right)\{\psi(\mathbf{s})\psi^{+}(\mathbf{s})\psi^{+}(\mathbf{s})\}\right\} WP[\psi,\psi^{+},\phi,\phi^{+}]\right\} \nonumber \\
 &  & \frac{-i}{\hbar}\left\{ -g\int d\mathbf{s\,}\left\{ \left(\frac{\delta}{\delta\phi^{+}(\mathbf{s})}\right)\{\delta_{C}(\mathbf{s},\mathbf{s})\psi^{+}(\mathbf{s})\}\right\} WP[\psi,\psi^{+},\phi,\phi^{+}]\right\} \nonumber \\
 &  & \,\label{eq:FnalFPV14Linear}\end{eqnarray}
\begin{eqnarray}
 &  & \left(\frac{\partial}{\partial t}WP[\psi,\psi^{+},\phi,\phi^{+}]\right)_{V14}^{2}\nonumber \\
 & = & \frac{-i}{\hbar}\left\{ -g\int d\mathbf{s\,}\left\{ \left(\frac{\delta}{\delta\psi^{+}(\mathbf{s})}\right)\left(\frac{\delta}{\delta\phi(\mathbf{s})}\right)\{\psi^{+}(\mathbf{s})\psi(\mathbf{s})\}\right\} WP[\psi,\psi^{+},\phi,\phi^{+}]\right\} \nonumber \\
 &  & \frac{-i}{\hbar}\left\{ +g\int d\mathbf{s\,}\left\{ \left(\frac{\delta}{\delta\psi(\mathbf{s})}\right)\left(\frac{\delta}{\delta\phi^{+}(\mathbf{s})}\right)\{\psi(\mathbf{s})\psi^{+}(\mathbf{s})\}\right\} WP[\psi,\psi^{+},\phi,\phi^{+}]\right\} \nonumber \\
 &  & \frac{-i}{\hbar}\left\{ +g\int d\mathbf{s\,}\left\{ \left(\frac{\delta}{\delta\psi^{+}(\mathbf{s})}\right)\left(\frac{\delta}{\delta\phi(\mathbf{s})}\right)\left\{ \frac{1}{2}\delta_{C}(\mathbf{s},\mathbf{s})\right\} \right\} WP[\psi,\psi^{+},\phi,\phi^{+}]\right\} \nonumber \\
 &  & \frac{-i}{\hbar}\left\{ -g\int d\mathbf{s\,}\left\{ \left(\frac{\delta}{\delta\psi(\mathbf{s})}\right)\left(\frac{\delta}{\delta\phi^{+}(\mathbf{s})}\right)\left\{ \frac{1}{2}\delta_{C}(\mathbf{s},\mathbf{s})\right\} \right\} WP[\psi,\psi^{+},\phi,\phi^{+}]\right\} \nonumber \\
 &  & \frac{-i}{\hbar}\left\{ +g\int d\mathbf{s\,}\left\{ \left(\frac{\delta}{\delta\psi(\mathbf{s})}\right)\left(\frac{\delta}{\delta\phi(\mathbf{s})}\right)\{\frac{1}{2}\psi(\mathbf{s})\psi(\mathbf{s})\}\right\} WP[\psi,\psi^{+},\phi,\phi^{+}]\right\} \nonumber \\
 &  & \frac{-i}{\hbar}\left\{ -g\int d\mathbf{s\,}\left\{ \left(\frac{\delta}{\delta\psi^{+}(\mathbf{s})}\right)\left(\frac{\delta}{\delta\phi^{+}(\mathbf{s})}\right)\{\frac{1}{2}\psi^{+}(\mathbf{s})\psi^{+}(\mathbf{s})\}\right\} WP[\psi,\psi^{+},\phi,\phi^{+}]\right\} \nonumber \\
 &  & \,\label{eq:FnalFPV14Quadratic}\end{eqnarray}

\begin{eqnarray}
 &  & \left(\frac{\partial}{\partial t}WP[\psi,\psi^{+},\phi,\phi^{+}]\right)_{V14}^{3}\nonumber \\
 & = & \frac{-i}{\hbar}\left\{ -g\int d\mathbf{s\,}\left\{ \left(\frac{\delta}{\delta\psi^{+}(\mathbf{s})}\right)\left(\frac{\delta}{\delta\psi^{+}(\mathbf{s})}\right)\left(\frac{\delta}{\delta\psi(\mathbf{s})}\right)\{\frac{1}{4}\phi^{+}(\mathbf{s})\}\right\} WP[\psi,\psi^{+},\phi,\phi^{+}]\right\} \nonumber \\
 &  & \frac{-i}{\hbar}\left\{ +g\int d\mathbf{s\,}\left\{ \left(\frac{\delta}{\delta\psi(\mathbf{s})}\right)\left(\frac{\delta}{\delta\psi(\mathbf{s})}\right)\left(\frac{\delta}{\delta\psi^{+}(\mathbf{s})}\right)\{\frac{1}{4}\phi(\mathbf{s})\}\right\} WP[\psi,\psi^{+},\phi,\phi^{+}]\right\} \nonumber \\
 &  & \frac{-i}{\hbar}\left\{ -g\int d\mathbf{s\,}\left\{ \left(\frac{\delta}{\delta\psi^{+}(\mathbf{s})}\right)\left(\frac{\delta}{\delta\psi^{+}(\mathbf{s})}\right)\left(\frac{\delta}{\delta\phi(\mathbf{s})}\right)\{\frac{1}{4}\psi^{+}(\mathbf{s})\}\right\} WP[\psi,\psi^{+},\phi,\phi^{+}]\right\} \nonumber \\
 &  & \frac{-i}{\hbar}\left\{ +g\int d\mathbf{s\,}\left\{ \left(\frac{\delta}{\delta\psi(\mathbf{s})}\right)\left(\frac{\delta}{\delta\psi(\mathbf{s})}\right)\left(\frac{\delta}{\delta\phi^{+}(\mathbf{s})}\right)\{\frac{1}{4}\psi(\mathbf{s})\}\right\} WP[\psi,\psi^{+},\phi,\phi^{+}]\right\} \nonumber \\
 &  & \frac{-i}{\hbar}\left\{ +g\int d\mathbf{s\,}\left\{ \left(\frac{\delta}{\delta\psi(\mathbf{s})}\right)\left(\frac{\delta}{\delta\psi^{+}(\mathbf{s})}\right)\left(\frac{\delta}{\delta\phi(\mathbf{s})}\right)\{\frac{1}{2}\psi(\mathbf{s})\}\right\} WP[\psi,\psi^{+},\phi,\phi^{+}]\right\} \nonumber \\
 &  & \frac{-i}{\hbar}\left\{ -g\int d\mathbf{s\,}\left\{ \left(\frac{\delta}{\delta\psi(\mathbf{s})}\right)\left(\frac{\delta}{\delta\psi^{+}(\mathbf{s})}\right)\left(\frac{\delta}{\delta\phi^{+}(\mathbf{s})}\right)\{\frac{1}{2}\psi^{+}(\mathbf{s})\}\right\} WP[\psi,\psi^{+},\phi,\phi^{+}]\right\} \nonumber \\
 &  & \,\label{eq:FnalFPV14Cubic}\end{eqnarray}
\begin{eqnarray}
 &  & \left(\frac{\partial}{\partial t}WP[\psi,\psi^{+},\phi,\phi^{+}]\right)_{V14}^{4}\nonumber \\
 & = & \frac{-i}{\hbar}\left\{ g\int d\mathbf{s\,}\left\{ \left(\frac{\delta}{\delta\psi^{+}(\mathbf{s})}\right)\left(\frac{\delta}{\delta\psi^{+}(\mathbf{s})}\right)\left(\frac{\delta}{\delta\psi(\mathbf{s})}\right)\left(\frac{\delta}{\delta\phi(\mathbf{s})}\right)\{\frac{1}{8}\}\right\} WP[\psi,\psi^{+},\phi,\phi^{+}]\right\} \nonumber \\
 &  & \frac{-i}{\hbar}\left\{ -g\int d\mathbf{s}\left\{ \left(\frac{\delta}{\delta\psi(\mathbf{s})}\right)\left(\frac{\delta}{\delta\psi(\mathbf{s})}\right)\left(\frac{\delta}{\delta\psi^{+}(\mathbf{s})}\right)\left(\frac{\delta}{\delta\phi^{+}(\mathbf{s})}\right)\{\frac{1}{8}\}\right\} WP[\psi,\psi^{+},\phi,\phi^{+}]\right\} \nonumber \\
 &  & \,\label{eq:FnalFPV14Quartic}\end{eqnarray}

Reverting to the original notation we have\begin{eqnarray}
 &  & \left(\frac{\partial}{\partial t}P[\underrightarrow{\psi}(\mathbf{r}),\underrightarrow{\psi^{\ast}}(\mathbf{r})]\right)_{V14}^{1}\nonumber \\
 & = & \frac{-i}{\hbar}\left\{ +g\int d\mathbf{s\,}\left\{ \left(\frac{\delta}{\delta\psi_{C}^{+}(\mathbf{s})}\right)\{[2\psi_{C}^{+}(\mathbf{s})\psi_{C}(\mathbf{s})-\delta_{C}(\mathbf{s},\mathbf{s})]\psi_{NC}^{+}(\mathbf{s})\}\right\} P[\underrightarrow{\psi}(\mathbf{r}),\underrightarrow{\psi^{\ast}}(\mathbf{r})]\right\} \nonumber \\
 &  & \frac{-i}{\hbar}\left\{ +g\int d\mathbf{s\,}\left\{ \mathbf{\left(\frac{\delta}{\delta\psi_{C}^{+}(\mathbf{s})}\right)\{[}\psi_{C}^{+}\mathbf{(\mathbf{s})}\psi_{C}^{+}\mathbf{(\mathbf{s})]}\psi_{NC}\mathbf{(\mathbf{s})\}}\right\} P[\underrightarrow{\psi}(\mathbf{r}),\underrightarrow{\psi^{\ast}}(\mathbf{r})]\right\} \nonumber \\
 &  & \frac{-i}{\hbar}\left\{ -g\int d\mathbf{s\,}\left\{ \mathbf{\left(\frac{\delta}{\delta\psi_{C}(\mathbf{s})}\right)\{[}2\psi\mathbf{_{C}(\mathbf{s})}\psi_{C}^{+}\mathbf{(\mathbf{s})-\delta_{C}(\mathbf{s},\mathbf{s})]\psi}_{NC}\mathbf{(\mathbf{s})\}}\right\} P[\underrightarrow{\psi}(\mathbf{r}),\underrightarrow{\psi^{\ast}}(\mathbf{r})]\right\} \nonumber \\
 &  & \frac{-i}{\hbar}\left\{ -g\int d\mathbf{s\,}\left\{ \left(\frac{\delta}{\delta\psi_{C}(\mathbf{s})}\right)\{[\psi_{C}(\mathbf{s})\psi_{C}(\mathbf{s})]\psi_{NC}^{+}(\mathbf{s})\}\right\} P[\underrightarrow{\psi}(\mathbf{r}),\underrightarrow{\psi^{\ast}}(\mathbf{r})]\right\} \nonumber \\
 &  & \frac{-i}{\hbar}\left\{ -g\int d\mathbf{s\,}\left\{ \left(\frac{\delta}{\delta\psi_{NC}(\mathbf{s})}\right)\{[\psi_{C}^{+}(\mathbf{s})\psi_{C}(\mathbf{s})-\delta_{C}(\mathbf{s},\mathbf{s)}]\psi_{C}(\mathbf{s})\}\right\} P[\underrightarrow{\psi}(\mathbf{r}),\underrightarrow{\psi^{\ast}}(\mathbf{r})]\right\} \nonumber \\
 &  & \frac{-i}{\hbar}\left\{ +g\int d\mathbf{s\,}\left\{ \left(\frac{\delta}{\delta\psi_{NC}^{+}(\mathbf{s})}\right)\{[\psi_{C}(\mathbf{s})\psi_{C}^{+}(\mathbf{s})-\delta_{C}(\mathbf{s},\mathbf{s})]\psi_{C}^{+}(\mathbf{s})\}\right\} P[\underrightarrow{\psi}(\mathbf{r}),\underrightarrow{\psi^{\ast}}(\mathbf{r})]\right\} \nonumber \\
 &  & \,\label{eq:FFPECondNonCondV14Linear}\end{eqnarray}
\begin{eqnarray}
 &  & \left(\frac{\partial}{\partial t}P[\underrightarrow{\psi}(\mathbf{r}),\underrightarrow{\psi^{\ast}}(\mathbf{r})]\right)_{V14}^{2}\nonumber \\
 & = & \frac{-i}{\hbar}\left\{ -g\int d\mathbf{s\,}\left\{ \left(\frac{\delta}{\delta\psi_{C}^{+}(\mathbf{s})}\right)\left(\frac{\delta}{\delta\psi_{NC}(\mathbf{s})}\right)\{\psi_{C}^{+}(\mathbf{s})\psi_{C}(\mathbf{s})\}\right\} P[\underrightarrow{\psi}(\mathbf{r}),\underrightarrow{\psi^{\ast}}(\mathbf{r})]\right\} \nonumber \\
 &  & \frac{-i}{\hbar}\left\{ +g\int d\mathbf{s\,}\left\{ \left(\frac{\delta}{\delta\psi_{C}(\mathbf{s})}\right)\left(\frac{\delta}{\delta\psi_{NC}^{+}(\mathbf{s})}\right)\{\psi_{C}(\mathbf{s})\psi_{C}^{+}(\mathbf{s})\}\right\} P[\underrightarrow{\psi}(\mathbf{r}),\underrightarrow{\psi^{\ast}}(\mathbf{r})]\right\} \nonumber \\
 &  & \frac{-i}{\hbar}\left\{ +g\int d\mathbf{s\,}\left\{ \left(\frac{\delta}{\delta\psi_{C}^{+}(\mathbf{s})}\right)\left(\frac{\delta}{\delta\psi_{NC}(\mathbf{s})}\right)\left\{ \frac{1}{2}\delta_{C}(\mathbf{s},\mathbf{s})\right\} \right\} P[\underrightarrow{\psi}(\mathbf{r}),\underrightarrow{\psi^{\ast}}(\mathbf{r})]\right\} \nonumber \\
 &  & \frac{-i}{\hbar}\left\{ -g\int d\mathbf{s\,}\left\{ \left(\frac{\delta}{\delta\psi_{C}(\mathbf{s})}\right)\left(\frac{\delta}{\delta\psi_{NC}^{+}(\mathbf{s})}\right)\left\{ \frac{1}{2}\delta_{C}(\mathbf{s},\mathbf{s})\right\} \right\} P[\underrightarrow{\psi}(\mathbf{r}),\underrightarrow{\psi^{\ast}}(\mathbf{r})]\right\} \nonumber \\
 &  & \frac{-i}{\hbar}\left\{ +g\int d\mathbf{s\,}\left\{ \left(\frac{\delta}{\delta\psi_{C}(\mathbf{s})}\right)\left(\frac{\delta}{\delta\psi_{NC}(\mathbf{s})}\right)\{\frac{1}{2}\psi_{C}(\mathbf{s})\psi_{C}(\mathbf{s})\}\right\} P[\underrightarrow{\psi}(\mathbf{r}),\underrightarrow{\psi^{\ast}}(\mathbf{r})]\right\} \nonumber \\
 &  & \frac{-i}{\hbar}\left\{ -g\int d\mathbf{s\,}\left\{ \left(\frac{\delta}{\delta\psi_{C}^{+}(\mathbf{s})}\right)\left(\frac{\delta}{\delta\psi_{NC}^{+}(\mathbf{s})}\right)\{\frac{1}{2}\psi_{C}^{+}(\mathbf{s})\psi_{C}^{+}(\mathbf{s})\}\right\} P[\underrightarrow{\psi}(\mathbf{r}),\underrightarrow{\psi^{\ast}}(\mathbf{r})]\right\} \nonumber \\
 &  & \,\label{eq:FFPECondNonCondV14Quad}\end{eqnarray}

\begin{eqnarray}
 &  & \left(\frac{\partial}{\partial t}P[\underrightarrow{\psi}(\mathbf{r}),\underrightarrow{\psi^{\ast}}(\mathbf{r})]\right)_{V14}^{3}\nonumber \\
 & = & \frac{-i}{\hbar}\left\{ -g\int d\mathbf{s\,}\left\{ \left(\frac{\delta}{\delta\psi_{C}^{+}(\mathbf{s})}\right)\left(\frac{\delta}{\delta\psi_{C}^{+}(\mathbf{s})}\right)\left(\frac{\delta}{\delta\psi_{C}(\mathbf{s})}\right)\{\frac{1}{4}\psi_{NC}^{+}(\mathbf{s})\}\right\} P[\underrightarrow{\psi}(\mathbf{r}),\underrightarrow{\psi^{\ast}}(\mathbf{r})]\right\} \nonumber \\
 &  & \frac{-i}{\hbar}\left\{ +g\int d\mathbf{s\,}\left\{ \left(\frac{\delta}{\delta\psi_{C}(\mathbf{s})}\right)\left(\frac{\delta}{\delta\psi_{C}(\mathbf{s})}\right)\left(\frac{\delta}{\delta\psi_{C}^{+}(\mathbf{s})}\right)\{\frac{1}{4}\psi_{NC}(\mathbf{s})\}\right\} P[\underrightarrow{\psi}(\mathbf{r}),\underrightarrow{\psi^{\ast}}(\mathbf{r})]\right\} \nonumber \\
 &  & \frac{-i}{\hbar}\left\{ -g\int d\mathbf{s\,}\left\{ \left(\frac{\delta}{\delta\psi_{C}^{+}(\mathbf{s})}\right)\left(\frac{\delta}{\delta\psi_{C}^{+}(\mathbf{s})}\right)\left(\frac{\delta}{\delta\psi_{NC}(\mathbf{s})}\right)\{\frac{1}{4}\psi_{C}^{+}(\mathbf{s})\}\right\} P[\underrightarrow{\psi}(\mathbf{r}),\underrightarrow{\psi^{\ast}}(\mathbf{r})]\right\} \nonumber \\
 &  & \frac{-i}{\hbar}\left\{ +g\int d\mathbf{s\,}\left\{ \left(\frac{\delta}{\delta\psi_{C}(\mathbf{s})}\right)\left(\frac{\delta}{\delta\psi_{C}(\mathbf{s})}\right)\left(\frac{\delta}{\delta\psi_{NC}^{+}(\mathbf{s})}\right)\{\frac{1}{4}\psi_{C}(\mathbf{s})\}\right\} P[\underrightarrow{\psi}(\mathbf{r}),\underrightarrow{\psi^{\ast}}(\mathbf{r})]\right\} \nonumber \\
 &  & \frac{-i}{\hbar}\left\{ +g\int d\mathbf{s\,}\left\{ \left(\frac{\delta}{\delta\psi_{C}^{+}(\mathbf{s})}\right)\left(\frac{\delta}{\delta\psi_{C}(\mathbf{s})}\right)\left(\frac{\delta}{\delta\psi_{NC}(\mathbf{s})}\right)\{\frac{1}{2}\psi_{C}(\mathbf{s})\}\right\} P[\underrightarrow{\psi}(\mathbf{r}),\underrightarrow{\psi^{\ast}}(\mathbf{r})]\right\} \nonumber \\
 &  & \frac{-i}{\hbar}\left\{ -g\int d\mathbf{s\,}\left\{ \left(\frac{\delta}{\delta\psi_{C}(\mathbf{s})}\right)\left(\frac{\delta}{\delta\psi_{C}^{+}(\mathbf{s})}\right)\left(\frac{\delta}{\delta\psi_{NC}^{+}(\mathbf{s})}\right)\{\frac{1}{2}\psi_{C}^{+}(\mathbf{s})\}\right\} P[\underrightarrow{\psi}(\mathbf{r}),\underrightarrow{\psi^{\ast}}(\mathbf{r})]\right\} \nonumber \\
 &  & \,\label{eq:FFPECondNonCondV14Cubic}\end{eqnarray}
\begin{eqnarray}
 &  & \left(\frac{\partial}{\partial t}P[\underrightarrow{\psi}(\mathbf{r}),\underrightarrow{\psi^{\ast}}(\mathbf{r})]\right)_{V14}^{4}\nonumber \\
 & = & \frac{-i}{\hbar}\left\{ g\int d\mathbf{s\,}\left\{ \left(\frac{\delta}{\delta\psi_{C}^{+}(\mathbf{s})}\right)\left(\frac{\delta}{\delta\psi_{C}^{+}(\mathbf{s})}\right)\left(\frac{\delta}{\delta\psi_{C}(\mathbf{s})}\right)\left(\frac{\delta}{\delta\psi_{NC}(\mathbf{s})}\right)\{\frac{1}{8}\}\right\} P[\underrightarrow{\psi}(\mathbf{r}),\underrightarrow{\psi^{\ast}}(\mathbf{r})]\right\} \nonumber \\
 &  & \frac{-i}{\hbar}\left\{ -g\int d\mathbf{s}\left\{ \left(\frac{\delta}{\delta\psi_{C}(\mathbf{s})}\right)\left(\frac{\delta}{\delta\psi_{C}(\mathbf{s})}\right)\left(\frac{\delta}{\delta\psi_{C}^{+}(\mathbf{s})}\right)\left(\frac{\delta}{\delta\psi_{NC}^{+}(\mathbf{s})}\right)\{\frac{1}{8}\}\right\} P[\underrightarrow{\psi}(\mathbf{r}),\underrightarrow{\psi^{\ast}}(\mathbf{r})]\right\} \nonumber \\
 &  & \,\label{eq:FFPECondNonCondV14Quartic}\end{eqnarray}

\subsubsection{Second Order Term}

\begin{equation}
\widehat{V}_{12}=-g\int\int d\mathbf{r\,}d\mathbf{s\,}F(\mathbf{r},\mathbf{s})\widehat{\Psi}_{NC}(\mathbf{r})^{\dagger}\,\widehat{\Psi}_{C}(\mathbf{s})-g\int\int d\mathbf{r\,}d\mathbf{s\,}F^{\ast}(\mathbf{s},\mathbf{r})\widehat{\Psi}_{C}(\mathbf{r})^{\dagger}\widehat{\Psi}_{NC}(\mathbf{s})\,\end{equation}

Now if \begin{align}
\widehat{\rho} & \rightarrow\widehat{V}_{12}\,\widehat{\rho}=\int\int d\mathbf{s\,}d\mathbf{u\,(}\widehat{\Psi}_{NC}(\mathbf{s})^{\dagger}\triangle V(\mathbf{s},\mathbf{u})\widehat{\Psi}_{C}(\mathbf{u})+\widehat{\Psi}_{C}(\mathbf{s})^{\dagger}\triangle V(\mathbf{u},\mathbf{s})^{\ast}\widehat{\Psi}_{NC}(\mathbf{u}))\,\widehat{\rho}\nonumber \\
 & \,\end{align}
where we write $\triangle V(\mathbf{s},\mathbf{u})=-gF(\mathbf{s},\mathbf{u})$
for short, then \begin{eqnarray}
 &  & WP[\psi(\mathbf{r}),\psi^{+}(\mathbf{r}),\phi(\mathbf{r}),\phi^{+}(\mathbf{r})]\nonumber \\
 & \rightarrow & \int\int d\mathbf{s\,}d\mathbf{u\,}\left\{ \left(\phi^{+}(\mathbf{s})-\frac{\delta}{\delta\phi(\mathbf{s})}\right)\triangle V(\mathbf{s},\mathbf{u})\left(\psi(\mathbf{u})+\frac{1}{2}\frac{\delta}{\delta\psi^{+}(\mathbf{u})}\right)\right\} \, WP[\psi,\psi^{+},\phi,\phi^{+}]\nonumber \\
 &  & +\int\int d\mathbf{s\,}d\mathbf{u\,}\left\{ \left(\psi^{+}(\mathbf{s})-\frac{1}{2}\frac{\delta}{\delta\psi(\mathbf{s})}\right)\triangle V(\mathbf{u},\mathbf{s})^{\ast}\left(\phi(\mathbf{u})\right)\right\} \, WP[\psi,\psi^{+},\phi,\phi^{+}]\nonumber \\
 &  & \,\end{eqnarray}

Expanding we get\begin{eqnarray}
 &  & WP[\psi(\mathbf{r}),\psi^{+}(\mathbf{r}),\phi(\mathbf{r}),\phi^{+}(\mathbf{r})]\nonumber \\
 & \rightarrow & \int\int d\mathbf{s\,}d\mathbf{u\,}\left\{ \left(\phi^{+}(\mathbf{s})\right)\triangle V(\mathbf{s},\mathbf{u})\left(\psi(\mathbf{u})\right)\right\} \, WP[\psi,\psi^{+},\phi,\phi^{+}]\nonumber \\
 &  & +\int\int d\mathbf{s\,}d\mathbf{u\,}\frac{1}{2}\left\{ \left(\phi^{+}(\mathbf{s})\right)\triangle V(\mathbf{s},\mathbf{u})\left(\frac{\delta}{\delta\psi^{+}(\mathbf{u})}\right)\right\} \, WP[\psi,\psi^{+},\phi,\phi^{+}]\nonumber \\
 &  & -\int\int d\mathbf{s\,}d\mathbf{u\,}\left\{ \left(\frac{\delta}{\delta\phi(\mathbf{s})}\right)\triangle V(\mathbf{s},\mathbf{u})\left(\psi(\mathbf{u})\right)\right\} \, WP[\psi,\psi^{+},\phi,\phi^{+}]\nonumber \\
 &  & -\int\int d\mathbf{s\,}d\mathbf{u\,}\frac{1}{2}\left\{ \left(\frac{\delta}{\delta\phi(\mathbf{s})}\right)\triangle V(\mathbf{s},\mathbf{u})\left(\frac{\delta}{\delta\psi^{+}(\mathbf{u})}\right)\right\} \, WP[\psi,\psi^{+},\phi,\phi^{+}]\nonumber \\
 &  & +\int\int d\mathbf{s\,}d\mathbf{u\,}\left\{ \left(\psi^{+}(\mathbf{s})\right)\triangle V(\mathbf{u},\mathbf{s})^{\ast}\left(\phi(\mathbf{u})\right)\right\} \, WP[\psi,\psi^{+},\phi,\phi^{+}]\nonumber \\
 &  & -\int\int d\mathbf{s\,}d\mathbf{u\,}\frac{1}{2}\left\{ \left(\frac{\delta}{\delta\psi(\mathbf{s})}\right)\triangle V(\mathbf{u},\mathbf{s})^{\ast}\left(\phi(\mathbf{u})\right)\right\} \, WP[\psi,\psi^{+},\phi,\phi^{+}]\nonumber \\
 &  & \,\end{eqnarray}

The first term is \begin{eqnarray}
 &  & \int\int d\mathbf{s\,}d\mathbf{u\,}\left\{ \left(\phi^{+}(\mathbf{s})\right)\triangle V(\mathbf{s},\mathbf{u})\left(\psi(\mathbf{u})\right)\right\} \, WP[\psi,\psi^{+},\phi,\phi^{+}]\nonumber \\
 & = & \int\int d\mathbf{s\,}d\mathbf{u\,}\left\{ \left(\psi(\mathbf{u})\right)\triangle V(\mathbf{s},\mathbf{u})\left(\phi^{+}(\mathbf{s})\right)\right\} \, WP[\psi,\psi^{+},\phi,\phi^{+}]\nonumber \\
 &  & \,\label{eq:SimplifnResult14}\end{eqnarray}

Using the product rule and the second term becomes\begin{eqnarray}
 &  & \int\int d\mathbf{s\,}d\mathbf{u\,}\frac{1}{2}\left\{ \left(\phi^{+}(\mathbf{s})\right)\triangle V(\mathbf{s},\mathbf{u})\left(\frac{\delta}{\delta\psi^{+}(\mathbf{u})}\right)\right\} \, WP[\psi,\psi^{+},\phi,\phi^{+}]\nonumber \\
 & = & \int\int d\mathbf{s\,}d\mathbf{u\,}\frac{1}{2}\left\{ \left(\frac{\delta}{\delta\psi^{+}(\mathbf{u})}\right)\left(\triangle V(\mathbf{s},\mathbf{u})\phi^{+}(\mathbf{s})\right)\right\} \, WP[\psi,\psi^{+},\phi,\phi^{+}]\nonumber \\
 &  & \,\label{eq:SimplifnResult15}\end{eqnarray}

The third term is \begin{eqnarray}
 &  & \int\int d\mathbf{s\,}d\mathbf{u}\left\{ \left(\frac{\delta}{\delta\phi(\mathbf{s})}\right)\triangle V(\mathbf{s},\mathbf{u})\left(\psi(\mathbf{u})\right)\right\} \, WP(\psi,\psi^{+},\phi,\phi^{+}]\nonumber \\
 & = & \int\int d\mathbf{s\,}d\mathbf{u}\left\{ \left(\frac{\delta}{\delta\phi(\mathbf{s})}\right)\left(\triangle V(\mathbf{s},\mathbf{u})\psi(\mathbf{u})\right)\right\} \, WP(\psi,\psi^{+},\phi,\phi^{+}]\nonumber \\
 &  & \,\label{eq:SimplifnResult16}\end{eqnarray}

Using the result that the functional derivatives can be performed
in either order the fourth term is\begin{eqnarray}
 &  & \int\int d\mathbf{s\,}d\mathbf{u}\frac{1}{2}\left\{ \left(\frac{\delta}{\delta\phi(\mathbf{s})}\right)\triangle V(\mathbf{s},\mathbf{u})\left(\frac{\delta}{\delta\psi^{+}(\mathbf{u})}\right)\right\} \, WP[\psi,\psi^{+},\phi,\phi^{+}]\nonumber \\
 & = & \int\int d\mathbf{s\,}d\mathbf{u}\frac{1}{2}\left\{ \left(\frac{\delta}{\delta\phi(\mathbf{s})}\right)\left(\frac{\delta}{\delta\psi^{+}(\mathbf{u})}\right)\triangle V(\mathbf{s},\mathbf{u})\right\} \, WP[\psi,\psi^{+},\phi,\phi^{+}]\nonumber \\
 &  & \,\label{eq:SimplifnResult17}\end{eqnarray}

Combining these results we find that\begin{eqnarray}
 &  & WP[\psi(\mathbf{r}),\psi^{+}(\mathbf{r}),\phi(\mathbf{r}),\phi^{+}(\mathbf{r})]\nonumber \\
 & \rightarrow & \int\int d\mathbf{s\,}d\mathbf{u}\left\{ \phi^{+}(\mathbf{s})\triangle V(\mathbf{s},\mathbf{u})\,\psi(\mathbf{u})\right\} \, WP[\psi,\psi^{+},\phi,\phi^{+}]\nonumber \\
 &  & +\int\int d\mathbf{s\,}d\mathbf{u}\frac{1}{2}\left\{ \left(\frac{\delta}{\delta\psi^{+}(\mathbf{u})}\right)\{\triangle V(\mathbf{s},\mathbf{u})\,\phi^{+}(\mathbf{s})\}\right\} \, WP[\psi,\psi^{+},\phi,\phi^{+}]\nonumber \\
 &  & -\int\int d\mathbf{s\,}d\mathbf{u}\left\{ \left(\frac{\delta}{\delta\phi(\mathbf{s})}\right)\{\triangle V(\mathbf{s},\mathbf{u})\,\psi(\mathbf{u})\}\right\} \, WP[\psi,\psi^{+},\phi,\phi^{+}]\nonumber \\
 &  & -\int\int d\mathbf{s\,}d\mathbf{u}\frac{1}{2}\left\{ \left(\frac{\delta}{\delta\phi(\mathbf{s})}\right)\left(\frac{\delta}{\delta\psi^{+}(\mathbf{u})}\right)\{\triangle V(\mathbf{s},\mathbf{u})\}\right\} \, WP[\psi,\psi^{+},\phi,\phi^{+}]\nonumber \\
 &  & +\int\int d\mathbf{s\,}d\mathbf{u}\left\{ \psi^{+}(\mathbf{s})\triangle V(\mathbf{u},\mathbf{s})^{\ast}\,\phi(\mathbf{u})\right\} \, WP[\psi,\psi^{+},\phi,\phi^{+}]\nonumber \\
 &  & -\int\int d\mathbf{s\,}d\mathbf{u}\frac{1}{2}\left\{ \left(\frac{\delta}{\delta\psi(\mathbf{s})}\right)\{\triangle V(\mathbf{u},\mathbf{s})^{\ast}\,\phi(\mathbf{u})\}\right\} \, WP[\psi,\psi^{+},\phi,\phi^{+}]\nonumber \\
 &  & \,\label{eq:NewV12RhoResult}\end{eqnarray}

Now if \begin{align}
\widehat{\rho} & \rightarrow\widehat{\rho}\,\widehat{V}_{12}\,\nonumber \\
= & \int\int d\mathbf{s\, d\mathbf{u\,}\widehat{\rho}\mathbf{(}\widehat{\Psi}_{NC}(\mathbf{s})^{\dagger}\triangle V(\mathbf{s},\mathbf{u})\widehat{\Psi}_{C}(\mathbf{u})+\widehat{\Psi}_{C}(\mathbf{s})^{\dagger}\triangle V(\mathbf{u},\mathbf{s})^{\ast}\widehat{\Psi}_{NC}(\mathbf{u}))}\,\nonumber \\
 & \,\end{align}
then \begin{eqnarray}
 &  & WP[\psi(\mathbf{r}),\psi^{+}(\mathbf{r}),\phi(\mathbf{r}),\phi^{+}(\mathbf{r})]\nonumber \\
 & \rightarrow & \int d\mathbf{s\,}\left\{ \left(\psi(\mathbf{u})-\frac{1}{2}\frac{\delta}{\delta\psi^{+}(\mathbf{u})}\right)\mathbf{\triangle V(\mathbf{s},\mathbf{u})}\left(\phi^{+}(\mathbf{s})\right)\right\} \, WP[\psi,\psi^{+},\phi,\phi^{+}]\nonumber \\
 &  & +\int d\mathbf{s\,}\left\{ \left(\phi(\mathbf{u})-\frac{\delta}{\delta\phi^{+}(\mathbf{u})}\right)\mathbf{\triangle V(\mathbf{u},\mathbf{s})^{\ast}}\left(\psi^{+}(\mathbf{s})+\frac{1}{2}\frac{\delta}{\delta\psi(\mathbf{s})}\right)\right\} \, WP[\psi,\psi^{+},\phi,\phi^{+}]\nonumber \\
 &  & \,\end{eqnarray}

Expanding out gives\begin{eqnarray*}
 &  & WP[\psi(\mathbf{r}),\psi^{+}(\mathbf{r}),\phi(\mathbf{r}),\phi^{+}(\mathbf{r})]\\
 & \rightarrow & \int\int d\mathbf{s\, d\mathbf{u\,}}\left\{ \left(\psi(\mathbf{u})\right)\mathbf{\triangle V(\mathbf{s},\mathbf{u})}\left(\phi^{+}(\mathbf{s})\right)\right\} \, WP[\psi,\psi^{+},\phi,\phi^{+}]\\
 &  & -\int\int d\mathbf{s\, d\mathbf{u\,}}\frac{1}{2}\left\{ \left(\frac{\delta}{\delta\psi^{+}(\mathbf{u})}\right)\mathbf{\triangle V(\mathbf{s},\mathbf{u})}\left(\phi^{+}(\mathbf{s})\right)\right\} \, WP[\psi,\psi^{+},\phi,\phi^{+}]\\
 &  & +\int\int d\mathbf{s\, d\mathbf{u\,}}\left\{ \left(\phi(\mathbf{u})\right)\mathbf{\triangle V(\mathbf{u},\mathbf{s})^{\ast}}\left(\psi^{+}(\mathbf{s})\right)\right\} \, WP[\psi,\psi^{+},\phi,\phi^{+}]\\
 &  & +\int\int d\mathbf{s\, d\mathbf{u\,}}\frac{1}{2}\left\{ \left(\phi(\mathbf{u})\right)\mathbf{\triangle V(\mathbf{u},\mathbf{s})^{\ast}}\left(\frac{\delta}{\delta\psi(\mathbf{s})}\right)\right\} \, WP[\psi,\psi^{+},\phi,\phi^{+}]\\
 &  & -\int\int d\mathbf{s\, d\mathbf{u\,}}\left\{ \left(\frac{\delta}{\delta\phi^{+}(\mathbf{u})}\right)\mathbf{\triangle V(\mathbf{u},\mathbf{s})^{\ast}}\left(\psi^{+}(\mathbf{s})\right)\right\} \, WP[\psi,\psi^{+},\phi,\phi^{+}]\\
 &  & -\int\int d\mathbf{s\, d\mathbf{u\,}}\frac{1}{2}\left\{ \left(\frac{\delta}{\delta\phi^{+}(\mathbf{u})}\right)\mathbf{\triangle V(\mathbf{u},\mathbf{s})^{\ast}}\left(\frac{\delta}{\delta\psi(\mathbf{s})}\right)\right\} \, WP[\psi,\psi^{+},\phi,\phi^{+}]\end{eqnarray*}

Using a similar approach to that above we find that\begin{eqnarray}
 &  & WP[\psi(\mathbf{r}),\psi^{+}(\mathbf{r}),\phi(\mathbf{r}),\phi^{+}(\mathbf{r})]\nonumber \\
 & \rightarrow & \int\int d\mathbf{s\, d\mathbf{u\,}}\left\{ \psi(\mathbf{u})\mathbf{\triangle V(\mathbf{s},\mathbf{u})}\,\phi^{+}(\mathbf{s})\right\} \, WP[\psi,\psi^{+},\phi,\phi^{+}]\nonumber \\
 &  & -\int\int d\mathbf{s\, d\mathbf{u\,}}\frac{1}{2}\left\{ \left(\frac{\delta}{\delta\psi^{+}(\mathbf{u})}\right)\{\mathbf{\triangle V(\mathbf{s},\mathbf{u})}\phi^{+}(\mathbf{s})\}\right\} \, WP[\psi,\psi^{+},\phi,\phi^{+}]\nonumber \\
 &  & +\int\int d\mathbf{s\, d\mathbf{u\,}}\left\{ \phi(\mathbf{u})\mathbf{\triangle V(\mathbf{u},\mathbf{s})^{\ast}}\,\psi^{+}(\mathbf{s})\right\} \, WP[\psi,\psi^{+},\phi,\phi^{+}]\nonumber \\
 &  & +\int\int d\mathbf{s\, d\mathbf{u\,}}\frac{1}{2}\left\{ \left(\frac{\delta}{\delta\psi(\mathbf{s})}\right)\{\phi(\mathbf{u})\mathbf{\triangle V(\mathbf{u},\mathbf{s})^{\ast}}\}\right\} \, WP[\psi,\psi^{+},\phi,\phi^{+}]\nonumber \\
 &  & -\int\int d\mathbf{s\, d\mathbf{u\,}}\left\{ \left(\frac{\delta}{\delta\phi^{+}(\mathbf{u})}\right)\{\mathbf{\triangle V(\mathbf{u},\mathbf{s})^{\ast}}\,\psi^{+}(\mathbf{s})\}\right\} \, WP[\psi,\psi^{+},\phi,\phi^{+}]\nonumber \\
 &  & -\int\int d\mathbf{s\, d\mathbf{u\,}}\frac{1}{2}\left\{ \left(\frac{\delta}{\delta\phi^{+}(\mathbf{u})}\right)\left(\frac{\delta}{\delta\psi(\mathbf{s})}\right)\{\mathbf{\triangle V(\mathbf{u},\mathbf{s})^{\ast}}\}\right\} \, WP[\psi,\psi^{+},\phi,\phi^{+}]\nonumber \\
 &  & \,\label{eq:NewRhoV12Result}\end{eqnarray}

Now if \begin{align}
\widehat{\rho} & \rightarrow\lbrack\widehat{V}_{12}\,,\widehat{\rho}]\nonumber \\
= & [\int\int d\mathbf{s\, d\mathbf{u\,\mathbf{(}\widehat{\Psi}_{NC}(\mathbf{s})^{\dagger}\triangle V(\mathbf{s},\mathbf{u})\widehat{\Psi}_{C}(\mathbf{u})+\widehat{\Psi}_{C}(\mathbf{s})^{\dagger}\triangle V(\mathbf{u},\mathbf{s})^{\ast}\widehat{\Psi}_{NC}(\mathbf{u}))}}\,,\widehat{\rho}]\nonumber \\
 & \,\end{align}
then\begin{eqnarray}
 &  & WP[\psi(\mathbf{r}),\psi^{+}(\mathbf{r}),\phi(\mathbf{r}),\phi^{+}(\mathbf{r})]\nonumber \\
 & \rightarrow & \int\int d\mathbf{s\,}d\mathbf{u}\left\{ \phi^{+}(\mathbf{s})\triangle V(\mathbf{s},\mathbf{u})\,\psi(\mathbf{u})\right\} \, WP\nonumber \\
 &  & -\int\int d\mathbf{s\, d\mathbf{u\,}}\left\{ \psi(\mathbf{u})\mathbf{\triangle V(\mathbf{s},\mathbf{u})}\,\phi^{+}(\mathbf{s})\right\} \, WP\nonumber \\
 &  & +\int\int d\mathbf{s\,}d\mathbf{u}\frac{1}{2}\left\{ \left(\frac{\delta}{\delta\psi^{+}(\mathbf{u})}\right)\{\triangle V(\mathbf{s},\mathbf{u})\,\phi^{+}(\mathbf{s})\}\right\} \, WP\nonumber \\
 &  & +\int\int d\mathbf{s\, d\mathbf{u\,}}\frac{1}{2}\left\{ \left(\frac{\delta}{\delta\psi^{+}(\mathbf{u})}\right)\{\mathbf{\triangle V(\mathbf{s},\mathbf{u})}\phi^{+}(\mathbf{s})\}\right\} \, WP\nonumber \\
 &  & -\int\int d\mathbf{s\,}d\mathbf{u}\left\{ \left(\frac{\delta}{\delta\phi(\mathbf{s})}\right)\{\triangle V(\mathbf{s},\mathbf{u})\,\psi(\mathbf{u})\}\right\} \, WP\nonumber \\
 &  & +\int\int d\mathbf{s\, d\mathbf{u\,}}\left\{ \left(\frac{\delta}{\delta\phi^{+}(\mathbf{u})}\right)\{\mathbf{\triangle V(\mathbf{u},\mathbf{s})^{\ast}}\,\psi^{+}(\mathbf{s})\}\right\} \, WP\nonumber \\
 &  & -\int\int d\mathbf{s\,}d\mathbf{u}\frac{1}{2}\left\{ \left(\frac{\delta}{\delta\phi(\mathbf{s})}\right)\left(\frac{\delta}{\delta\psi^{+}(\mathbf{u})}\right)\{\triangle V(\mathbf{s},\mathbf{u})\}\right\} \, WP\nonumber \\
 &  & +\int\int d\mathbf{s\, d\mathbf{u\,}}\frac{1}{2}\left\{ \left(\frac{\delta}{\delta\phi^{+}(\mathbf{u})}\right)\left(\frac{\delta}{\delta\psi(\mathbf{s})}\right)\{\mathbf{\triangle V(\mathbf{u},\mathbf{s})^{\ast}}\}\right\} \, WP\nonumber \\
 &  & +\int\int d\mathbf{s\,}d\mathbf{u}\left\{ \psi^{+}(\mathbf{s})\triangle V(\mathbf{u},\mathbf{s})^{\ast}\,\phi(\mathbf{u})\right\} \, WP\nonumber \\
 &  & -\int\int d\mathbf{s\, d\mathbf{u\,}}\left\{ \phi(\mathbf{u})\mathbf{\triangle V(\mathbf{u},\mathbf{s})^{\ast}}\,\psi^{+}(\mathbf{s})\right\} \, WP\nonumber \\
 &  & -\int\int d\mathbf{s\,}d\mathbf{u}\frac{1}{2}\left\{ \left(\frac{\delta}{\delta\psi(\mathbf{s})}\right)\{\triangle V(\mathbf{u},\mathbf{s})^{\ast}\,\phi(\mathbf{u})\}\right\} \, WP\nonumber \\
 &  & -\int\int d\mathbf{s\, d\mathbf{u\,}}\frac{1}{2}\left\{ \left(\frac{\delta}{\delta\psi(\mathbf{s})}\right)\{\phi(\mathbf{u})\mathbf{\triangle V(\mathbf{u},\mathbf{s})^{\ast}}\}\right\} \, WP\nonumber \\
 &  & \,\end{eqnarray}
\begin{eqnarray}
 &  & WP[\psi(\mathbf{r}),\psi^{+}(\mathbf{r}),\phi(\mathbf{r}),\phi^{+}(\mathbf{r})]\nonumber \\
 & \rightarrow & \int\int d\mathbf{s\,}d\mathbf{u}\left\{ \left(\frac{\delta}{\delta\psi^{+}(\mathbf{u})}\right)\{\triangle V(\mathbf{s},\mathbf{u})\,\phi^{+}(\mathbf{s})\}\right\} \, WP[\psi,\psi^{+},\phi,\phi^{+}]\nonumber \\
 &  & -\int\int d\mathbf{s\,}d\mathbf{u}\left\{ \left(\frac{\delta}{\delta\psi(\mathbf{s})}\right)\{\triangle V(\mathbf{u},\mathbf{s})^{\ast}\,\phi(\mathbf{u})\}\right\} \, WP[\psi,\psi^{+},\phi,\phi^{+}]\nonumber \\
 &  & -\int\int d\mathbf{s\,}d\mathbf{u}\left\{ \left(\frac{\delta}{\delta\phi(\mathbf{s})}\right)\{\triangle V(\mathbf{s},\mathbf{u})\,\psi(\mathbf{u})\}\right\} \, WP[\psi,\psi^{+},\phi,\phi^{+}]\nonumber \\
 &  & +\int\int d\mathbf{s\, d\mathbf{u\,}}\left\{ \left(\frac{\delta}{\delta\phi^{+}(\mathbf{u})}\right)\{\mathbf{\triangle V(\mathbf{u},\mathbf{s})^{\ast}}\,\psi^{+}(\mathbf{s})\}\right\} \, WP[\psi,\psi^{+},\phi,\phi^{+}]\nonumber \\
 &  & -\int\int d\mathbf{s\,}d\mathbf{u}\left\{ \left(\frac{\delta}{\delta\phi(\mathbf{s})}\right)\left(\frac{\delta}{\delta\psi^{+}(\mathbf{u})}\right)\{\frac{1}{2}\triangle V(\mathbf{s},\mathbf{u})\}\right\} \, WP[\psi,\psi^{+},\phi,\phi^{+}]\nonumber \\
 &  & +\int\int d\mathbf{s\, d\mathbf{u\,}}\left\{ \left(\frac{\delta}{\delta\phi^{+}(\mathbf{u})}\right)\left(\frac{\delta}{\delta\psi(\mathbf{s})}\right)\{\frac{1}{2}\mathbf{\triangle V(\mathbf{u},\mathbf{s})^{\ast}}\}\right\} \, WP[\psi,\psi^{+},\phi,\phi^{+}]\nonumber \\
 &  & \,\end{eqnarray}

Thus we see that the $\widehat{\, V}_{12}$ term produces functional
derivatives of orders one and two. We may write the contributions
to the functional Fokker-Planck equation in the form\begin{eqnarray}
 &  & \left(\frac{\partial}{\partial t}WP[\psi,\psi^{+},\phi,\phi^{+}]\right)_{V12}\nonumber \\
 & = & \left(\frac{\partial}{\partial t}WP[\psi,\psi^{+},\phi,\phi^{+}]\right)_{V12}^{1}+\left(\frac{\partial}{\partial t}WP[\psi,\psi^{+},\phi,\phi^{+}]\right)_{V12}^{2}\nonumber \\
 &  & \,\label{eq:NewFnalFPV12}\end{eqnarray}
where\begin{eqnarray}
 &  & \left(\frac{\partial}{\partial t}WP[\psi,\psi^{+},\phi,\phi^{+}]\right)_{V12}^{1}\nonumber \\
 & = & \frac{-i}{\hbar}\left\{ \int\int d\mathbf{s\,}d\mathbf{u}\left\{ \left(\frac{\delta}{\delta\psi^{+}(\mathbf{u})}\right)\{\triangle V(\mathbf{s},\mathbf{u})\,\phi^{+}(\mathbf{s})\}\right\} \, WP[\psi,\psi^{+},\phi,\phi^{+}]\right\} \nonumber \\
 &  & +\frac{-i}{\hbar}\left\{ -\int\int d\mathbf{s\,}d\mathbf{u}\left\{ \left(\frac{\delta}{\delta\psi(\mathbf{s})}\right)\{\triangle V(\mathbf{u},\mathbf{s})^{\ast}\,\phi(\mathbf{u})\}\right\} \, WP[\psi,\psi^{+},\phi,\phi^{+}]\right\} \nonumber \\
 &  & +\frac{-i}{\hbar}\left\{ \int\int d\mathbf{s\, d\mathbf{u\,}}\left\{ \left(\frac{\delta}{\delta\phi^{+}(\mathbf{u})}\right)\{\mathbf{\triangle V(\mathbf{u},\mathbf{s})^{\ast}}\,\psi^{+}(\mathbf{s})\}\right\} \, WP[\psi,\psi^{+},\phi,\phi^{+}]\right\} \nonumber \\
 &  & +\frac{-i}{\hbar}\left\{ -\int\int d\mathbf{s\,}d\mathbf{u}\left\{ \left(\frac{\delta}{\delta\phi(\mathbf{s})}\right)\{\triangle V(\mathbf{s},\mathbf{u})\,\psi(\mathbf{u})\}\right\} \, WP[\psi,\psi^{+},\phi,\phi^{+}]\right\} \nonumber \\
 &  & \,\label{eq:NewFnalFPV12Linear}\end{eqnarray}
\begin{eqnarray}
 &  & \left(\frac{\partial}{\partial t}WP[\psi,\psi^{+},\phi,\phi^{+}]\right)_{V12}^{2}\nonumber \\
 & = & \frac{-i}{\hbar}\left\{ -\int\int d\mathbf{s\,}d\mathbf{u}\left\{ \left(\frac{\delta}{\delta\psi^{+}(\mathbf{u})}\right)\left(\frac{\delta}{\delta\phi(\mathbf{s})}\right)\{\frac{1}{2}\triangle V(\mathbf{s},\mathbf{u})\}\right\} \, WP[\psi,\psi^{+},\phi,\phi^{+}]\right\} \nonumber \\
 &  & +\frac{-i}{\hbar}\left\{ \int\int d\mathbf{s\, d\mathbf{u\,}}\left\{ \left(\frac{\delta}{\delta\psi(\mathbf{s})}\right)\left(\frac{\delta}{\delta\phi^{+}(\mathbf{u})}\right)\{\frac{1}{2}\mathbf{\triangle V(\mathbf{u},\mathbf{s})^{\ast}}\}\right\} \, WP[\psi,\psi^{+},\phi,\phi^{+}]\right\} \nonumber \\
 &  & \,\label{eq:NewFnalFPV12Quadratic}\end{eqnarray}

For the single condensate mode case the result is simpler and can
be obtained via the substitution $\triangle V(\mathbf{s},\mathbf{u})=\triangle V(\mathbf{s)}\delta(\mathbf{u}-\mathbf{s})$
with $\triangle V(\mathbf{s)}=-g\left\langle \widehat{\Psi}_{C}(\mathbf{s})^{\dagger}\widehat{\Psi}_{C}(\mathbf{s})\right\rangle $
and is given by\begin{eqnarray}
 &  & \left(\frac{\partial}{\partial t}WP[\psi,\psi^{+},\phi,\phi^{+}]\right)_{V12}^{1}\nonumber \\
 & = & \frac{-i}{\hbar}\left\{ +\int d\mathbf{s\,}\left\{ \left(\frac{\delta}{\delta\psi^{+}(\mathbf{s})}\right)\{\triangle V\,\phi^{+}(\mathbf{s})\}\right\} \, WP[\psi,\psi^{+},\phi,\phi^{+}]\right\} \nonumber \\
 &  & \frac{-i}{\hbar}\left\{ -\int d\mathbf{s\,}\left\{ \left(\frac{\delta}{\delta\psi(\mathbf{s})}\right)\{\triangle V\,\phi(\mathbf{s})\}\right\} \, WP[\psi,\psi^{+},\phi,\phi^{+}]\right\} \nonumber \\
 &  & \frac{-i}{\hbar}\left\{ +\int d\mathbf{s\,}\left\{ \left(\frac{\delta}{\delta\phi^{+}(\mathbf{s})}\right)\{\triangle V\,\psi^{+}(\mathbf{s})\}\right\} \, WP[\psi,\psi^{+},\phi,\phi^{+}]\right\} \nonumber \\
 &  & \frac{-i}{\hbar}\left\{ -\int d\mathbf{s\,}\left\{ \left(\frac{\delta}{\delta\phi(\mathbf{s})}\right)\{\triangle V\,\psi(\mathbf{s})\}\right\} \, WP[\psi,\psi^{+},\phi,\phi^{+}]\right\} \nonumber \\
 &  & \,\label{eq:FnalFPV12Linear}\end{eqnarray}
\begin{eqnarray}
 &  & \left(\frac{\partial}{\partial t}WP[\psi,\psi^{+},\phi,\phi^{+}]\right)_{V12}^{2}\nonumber \\
 & = & \frac{-i}{\hbar}\left\{ -\int d\mathbf{s\,}\left\{ \left(\frac{\delta}{\delta\psi^{+}(\mathbf{s})}\right)\left(\frac{\delta}{\delta\phi(\mathbf{s})}\right)\{\frac{1}{2}\triangle V\}\right\} \, WP[\psi,\psi^{+},\phi,\phi^{+}]\right\} \nonumber \\
 &  & \frac{-i}{\hbar}\left\{ +\int d\mathbf{s\,}\left\{ \left(\frac{\delta}{\delta\psi(\mathbf{s})}\right)\left(\frac{\delta}{\delta\phi^{+}(\mathbf{s})}\right)\{\frac{1}{2}\triangle V\}\right\} \, WP[\psi,\psi^{+},\phi,\phi^{+}]\right\} \nonumber \\
 &  & \,\label{eq:FnalFPV12Quadratic}\end{eqnarray}

Reverting to the original notation and replacing $\triangle V(\mathbf{s},\mathbf{u})=-gF(\mathbf{s},\mathbf{u})$
we have for the two mode condensate case\begin{eqnarray}
 &  & \left(\frac{\partial}{\partial t}P[\underrightarrow{\psi}(\mathbf{r}),\underrightarrow{\psi^{\ast}}(\mathbf{r})]\right)_{V12}^{1}\nonumber \\
 & = & \frac{-i}{\hbar}\left\{ -g\int\int d\mathbf{s\,}d\mathbf{u}\left\{ \left(\frac{\delta}{\delta\psi_{C}^{+}(\mathbf{u})}\right)\{F(\mathbf{s},\mathbf{u})\,\psi_{NC}^{+}(\mathbf{s})\}\right\} \, P[\underrightarrow{\psi}(\mathbf{r}),\underrightarrow{\psi^{\ast}}(\mathbf{r})]\right\} \nonumber \\
 &  & +\frac{-i}{\hbar}\left\{ +g\int\int d\mathbf{s\,}d\mathbf{u}\left\{ \left(\frac{\delta}{\delta\psi_{C}(\mathbf{s})}\right)\{F(\mathbf{u},\mathbf{s})^{\ast}\,\psi_{NC}(\mathbf{u})\}\right\} \, P[\underrightarrow{\psi}(\mathbf{r}),\underrightarrow{\psi^{\ast}}(\mathbf{r})]\right\} \nonumber \\
 &  & +\frac{-i}{\hbar}\left\{ -g\int\int d\mathbf{s\, d\mathbf{u\,}}\left\{ \left(\frac{\delta}{\delta\psi_{NC}^{+}(\mathbf{u})}\right)\{F\mathbf{(\mathbf{u},\mathbf{s})^{\ast}}\,\psi_{C}^{+}(\mathbf{s})\}\right\} \, P[\underrightarrow{\psi}(\mathbf{r}),\underrightarrow{\psi^{\ast}}(\mathbf{r})]\right\} \nonumber \\
 &  & +\frac{-i}{\hbar}\left\{ +g\int\int d\mathbf{s\,}d\mathbf{u}\left\{ \left(\frac{\delta}{\delta\psi_{NC}(\mathbf{s})}\right)\{F(\mathbf{s},\mathbf{u})\,\psi_{C}(\mathbf{u})\}\right\} \, P[\underrightarrow{\psi}(\mathbf{r}),\underrightarrow{\psi^{\ast}}(\mathbf{r})]\right\} \nonumber \\
 &  & \,\label{eq:FFPECondNonCondV12Linear}\end{eqnarray}
\begin{eqnarray}
 &  & \left(\frac{\partial}{\partial t}P[\underrightarrow{\psi}(\mathbf{r}),\underrightarrow{\psi^{\ast}}(\mathbf{r})]\right)_{V12}^{2}\nonumber \\
 & = & \frac{-i}{\hbar}\left\{ +g\int\int d\mathbf{s\,}d\mathbf{u}\left\{ \left(\frac{\delta}{\delta\psi_{C}^{+}(\mathbf{u})}\right)\left(\frac{\delta}{\delta\psi_{NC}(\mathbf{s})}\right)\{\frac{1}{2}F(\mathbf{s},\mathbf{u})\}\right\} \, P[\underrightarrow{\psi}(\mathbf{r}),\underrightarrow{\psi^{\ast}}(\mathbf{r})]\right\} \nonumber \\
 &  & +\frac{-i}{\hbar}\left\{ -g\int\int d\mathbf{s\, d\mathbf{u\,}}\left\{ \left(\frac{\delta}{\delta\psi_{C}(\mathbf{s})}\right)\left(\frac{\delta}{\delta\psi_{NC}^{+}(\mathbf{u})}\right)\{\frac{1}{2}\mathbf{F(\mathbf{u},\mathbf{s})^{\ast}}\}\right\} \, P[\underrightarrow{\psi}(\mathbf{r}),\underrightarrow{\psi^{\ast}}(\mathbf{r})]\right\} \nonumber \\
 &  & \,\label{eq:FFPECondNonCondV12Quadratic}\end{eqnarray}

For the case of the single mode condensate with $\triangle V(\mathbf{s)}=-g\left\langle \widehat{\Psi}_{C}(\mathbf{s})^{\dagger}\widehat{\Psi}_{C}(\mathbf{s})\right\rangle $\begin{eqnarray}
 &  & \left(\frac{\partial}{\partial t}P[\underrightarrow{\psi}(\mathbf{r}),\underrightarrow{\psi^{\ast}}(\mathbf{r})]\right)_{V12}^{1}\nonumber \\
 & = & \frac{-i}{\hbar}\left\{ -g\int d\mathbf{s\,}\left\{ \left(\frac{\delta}{\delta\psi_{C}^{+}(\mathbf{s})}\right)\{\left\langle \widehat{\Psi}_{C}(\mathbf{s})^{\dagger}\widehat{\Psi}_{C}(\mathbf{s})\right\rangle \,\psi_{NC}^{+}(\mathbf{s})\}\right\} \, P[\underrightarrow{\psi}(\mathbf{r}),\underrightarrow{\psi^{\ast}}(\mathbf{r})]\right\} \nonumber \\
 &  & \frac{-i}{\hbar}\left\{ +g\int d\mathbf{s\,}\left\{ \left(\frac{\delta}{\delta\psi_{C}(\mathbf{s})}\right)\{\left\langle \widehat{\Psi}_{C}(\mathbf{s})^{\dagger}\widehat{\Psi}_{C}(\mathbf{s})\right\rangle \,\psi_{NC}(\mathbf{s})\}\right\} \, P[\underrightarrow{\psi}(\mathbf{r}),\underrightarrow{\psi^{\ast}}(\mathbf{r})]\right\} \nonumber \\
 &  & \frac{-i}{\hbar}\left\{ -g\int d\mathbf{s\,}\left\{ \left(\frac{\delta}{\delta\psi_{NC}^{+}(\mathbf{s})}\right)\{\left\langle \widehat{\Psi}_{C}(\mathbf{s})^{\dagger}\widehat{\Psi}_{C}(\mathbf{s})\right\rangle \,\psi_{C}^{+}(\mathbf{s})\}\right\} \, P[\underrightarrow{\psi}(\mathbf{r}),\underrightarrow{\psi^{\ast}}(\mathbf{r})]\right\} \nonumber \\
 &  & \frac{-i}{\hbar}\left\{ +g\int d\mathbf{s\,}\left\{ \left(\frac{\delta}{\delta\psi_{NC}(\mathbf{s})}\right)\{\left\langle \widehat{\Psi}_{C}(\mathbf{s})^{\dagger}\widehat{\Psi}_{C}(\mathbf{s})\right\rangle \,\psi_{C}(\mathbf{s})\}\right\} \, P[\underrightarrow{\psi}(\mathbf{r}),\underrightarrow{\psi^{\ast}}(\mathbf{r})]\right\} \nonumber \\
 &  & \,\label{eq:FFPECondNonCondV12LinearSingle}\end{eqnarray}
\begin{eqnarray}
 &  & \left(\frac{\partial}{\partial t}P[\underrightarrow{\psi}(\mathbf{r}),\underrightarrow{\psi^{\ast}}(\mathbf{r})]\right)_{V12}^{2}\nonumber \\
 & = & \frac{-i}{\hbar}\left\{ +g\int d\mathbf{s\,}\left\{ \left(\frac{\delta}{\delta\psi_{C}^{+}(\mathbf{s})}\right)\left(\frac{\delta}{\delta\psi_{NC}(\mathbf{s})}\right)\{\frac{1}{2}\left\langle \widehat{\Psi}_{C}(\mathbf{s})^{\dagger}\widehat{\Psi}_{C}(\mathbf{s})\right\rangle \}\right\} \, P[\underrightarrow{\psi}(\mathbf{r}),\underrightarrow{\psi^{\ast}}(\mathbf{r})]\right\} \nonumber \\
 &  & \frac{-i}{\hbar}\left\{ -g\int d\mathbf{s\,}\left\{ \left(\frac{\delta}{\delta\psi_{C}(\mathbf{s})}\right)\left(\frac{\delta}{\delta\psi_{NC}^{+}(\mathbf{s})}\right)\{\frac{1}{2}\left\langle \widehat{\Psi}_{C}(\mathbf{s})^{\dagger}\widehat{\Psi}_{C}(\mathbf{s})\right\rangle \}\right\} \, P[\underrightarrow{\psi}(\mathbf{r}),\underrightarrow{\psi^{\ast}}(\mathbf{r})]\right\} \nonumber \\
 &  & \,\label{eq:FFPECondNonCondV12QuadraticSingle}\end{eqnarray}

We can show using the special form of $F(\mathbf{r},\mathbf{s})$
for a single mode condensate, that the Fokker-Planck equation terms
for $\widehat{V}_{12}$ can be obtained from the two mode case. We
have \begin{equation}
F(\mathbf{r},\mathbf{s})=(N-1)\phi_{1}^{\ast}(\mathbf{r})\phi_{1}(\mathbf{r})\phi_{1}(\mathbf{r})\,\phi_{1}^{\ast}(\mathbf{s\,)}\end{equation}
we can use the forms (\ref{eq:RestrictFnalDerivIdent1-1}) for the
functional derivatives involving the expansion coefficients\begin{eqnarray}
\frac{\delta}{\delta\psi_{C}(\mathbf{s})} & \equiv & \phi_{1}^{\ast}(\mathbf{s\,)}\frac{\partial}{\partial\alpha_{1}}\qquad\frac{\delta}{\delta\psi_{C}^{+}(\mathbf{s})}\equiv\phi_{1}(\mathbf{s\,)}\frac{\partial}{\partial\alpha_{1}^{+}}\nonumber \\
\frac{\delta}{\delta\psi_{NC}(\mathbf{s})} & \equiv & \sum_{k\neq1}\phi_{k}^{\ast}(\mathbf{s\,)}\frac{\partial}{\partial\alpha_{k}}\qquad\frac{\delta}{\delta\psi_{NC}^{+}(\mathbf{s})}\equiv\sum_{k\neq1}\phi_{k}(\mathbf{s\,)}\frac{\partial}{\partial\alpha_{k}^{+}}\nonumber \\
P[\underrightarrow{\psi}(\mathbf{r}),\underrightarrow{\psi^{\ast}}(\mathbf{r})] & \equiv & P_{b}(\underrightarrow{\alpha},\underrightarrow{\alpha^{\ast}})\end{eqnarray}
to show that this is the case.

Considering the first order functional derivative terms we see that\begin{eqnarray}
 &  & \left(\frac{\partial}{\partial t}P[\underrightarrow{\psi}(\mathbf{r}),\underrightarrow{\psi^{\ast}}(\mathbf{r})]\right)_{V12}^{1}\nonumber \\
 & = & \frac{-i}{\hbar}\left\{ -g\int\int d\mathbf{s\,}d\mathbf{u}\left\{ \left(\frac{\delta}{\delta\psi_{C}^{+}(\mathbf{u})}\right)\{F(\mathbf{s},\mathbf{u})\,\psi_{NC}^{+}(\mathbf{s})\}\right\} \, P[\underrightarrow{\psi}(\mathbf{r}),\underrightarrow{\psi^{\ast}}(\mathbf{r})]\right\} \nonumber \\
 &  & +\frac{-i}{\hbar}\left\{ +g\int\int d\mathbf{s\,}d\mathbf{u}\left\{ \left(\frac{\delta}{\delta\psi_{C}(\mathbf{s})}\right)\{F(\mathbf{u},\mathbf{s})^{\ast}\,\psi_{NC}(\mathbf{u})\}\right\} \, P[\underrightarrow{\psi}(\mathbf{r}),\underrightarrow{\psi^{\ast}}(\mathbf{r})]\right\} \nonumber \\
 &  & +\frac{-i}{\hbar}\left\{ -g\int\int d\mathbf{s\, d\mathbf{u\,}}\left\{ \left(\frac{\delta}{\delta\psi_{NC}^{+}(\mathbf{u})}\right)\{F\mathbf{(\mathbf{u},\mathbf{s})^{\ast}}\,\psi_{C}^{+}(\mathbf{s})\}\right\} \, P[\underrightarrow{\psi}(\mathbf{r}),\underrightarrow{\psi^{\ast}}(\mathbf{r})]\right\} \nonumber \\
 &  & +\frac{-i}{\hbar}\left\{ +g\int\int d\mathbf{s\,}d\mathbf{u}\left\{ \left(\frac{\delta}{\delta\psi_{NC}(\mathbf{s})}\right)\{F(\mathbf{s},\mathbf{u})\,\psi_{C}(\mathbf{u})\}\right\} \, P[\underrightarrow{\psi}(\mathbf{r}),\underrightarrow{\psi^{\ast}}(\mathbf{r})]\right\} \nonumber \\
 & = & \frac{-i}{\hbar}\left\{ -g\int\int d\mathbf{s\,}d\mathbf{u}\left\{ \left(\phi_{1}(\mathbf{u)}\frac{\partial}{\partial\alpha_{1}^{+}}\right)\{(N-1)\phi_{1}^{\ast}(\mathbf{s})\phi_{1}(\mathbf{s})\phi_{1}(\mathbf{s})\,\phi_{1}^{\ast}(\mathbf{u)}\,\psi_{NC}^{+}(\mathbf{s})\}\right\} \, P_{b}(\underrightarrow{\alpha},\underrightarrow{\alpha^{\ast}})\right\} \nonumber \\
 &  & +\frac{-i}{\hbar}\left\{ +g\int\int d\mathbf{s\,}d\mathbf{u}\left\{ \left(\phi_{1}^{\ast}(\mathbf{s)}\frac{\partial}{\partial\alpha_{1}}\right)\{(N-1)\phi_{1}(\mathbf{u})\phi_{1}^{\ast}(\mathbf{u})\phi_{1}^{\ast}(\mathbf{u})\,\phi_{1}(\mathbf{s)}\,\psi_{NC}(\mathbf{u})\}\right\} \, P_{b}(\underrightarrow{\alpha},\underrightarrow{\alpha^{\ast}})\right\} \nonumber \\
 &  & +\frac{-i}{\hbar}\left\{ -g\int\int d\mathbf{s\, d\mathbf{u\,}}\left\{ \left(\sum_{k\neq1}\phi_{k}(\mathbf{u)}\frac{\partial}{\partial\alpha_{k}^{+}}\right)\{(N-1)\phi_{1}(\mathbf{u})\phi_{1}^{\ast}(\mathbf{u})\phi_{1}^{\ast}(\mathbf{u})\,\phi_{1}(\mathbf{s)}\,\psi_{C}^{+}(\mathbf{s})\}\right\} \, P_{b}(\underrightarrow{\alpha},\underrightarrow{\alpha^{\ast}})\right\} \nonumber \\
 &  & +\frac{-i}{\hbar}\left\{ +g\int\int d\mathbf{s\,}d\mathbf{u}\left\{ \left(\sum_{k\neq1}\phi_{k}^{\ast}(\mathbf{s)}\frac{\partial}{\partial\alpha_{k}}\right)\{(N-1)\phi_{1}^{\ast}(\mathbf{s})\phi_{1}(\mathbf{s})\phi_{1}(\mathbf{s})\,\phi_{1}^{\ast}(\mathbf{u)}\,\psi_{C}(\mathbf{u})\}\right\} P_{b}(\underrightarrow{\alpha},\underrightarrow{\alpha^{\ast}})\,\right\} \nonumber \\
 & = & \frac{-i}{\hbar}\left\{ -g\int d\mathbf{s\,}\left\{ \left(\phi_{1}(\mathbf{s})\frac{\partial}{\partial\alpha_{1}^{+}}\right)\{(N-1)\phi_{1}^{\ast}(\mathbf{s})\phi_{1}(\mathbf{s})\,\psi_{NC}^{+}(\mathbf{s})\}\right\} P_{b}(\underrightarrow{\alpha},\underrightarrow{\alpha^{\ast}})\right\} \nonumber \\
 &  & +\frac{-i}{\hbar}\left\{ +g\int\mathbf{\,}d\mathbf{u}\left\{ \left(\phi_{1}^{\ast}(\mathbf{u})\frac{\partial}{\partial\alpha_{1}}\right)\{(N-1)\phi_{1}(\mathbf{u})\phi_{1}^{\ast}(\mathbf{u})\,\psi_{NC}(\mathbf{u})\}\right\} \, P_{b}(\underrightarrow{\alpha},\underrightarrow{\alpha^{\ast}})\right\} \nonumber \\
 &  & +\frac{-i}{\hbar}\left\{ -g\int\mathbf{\, d\mathbf{u\,}}\left\{ \left(\sum_{k\neq1}\phi_{k}(\mathbf{u\,)}\frac{\partial}{\partial\alpha_{k}^{+}}\right)\{(N-1)\phi_{1}(\mathbf{u})\phi_{1}^{\ast}(\mathbf{u})\,\alpha_{1}^{+}\phi_{1}^{\ast}(\mathbf{u})\}\right\} \, P_{b}(\underrightarrow{\alpha},\underrightarrow{\alpha^{\ast}})\right\} \nonumber \\
 &  & +\frac{-i}{\hbar}\left\{ +g\int d\mathbf{s\,}\left\{ \left(\sum_{k\neq1}\phi_{k}^{\ast}(\mathbf{s\,)}\frac{\partial}{\partial\alpha_{k}}\right)\{(N-1)\phi_{1}^{\ast}(\mathbf{s})\phi_{1}(\mathbf{s})\,\alpha_{1}\phi_{1}(\mathbf{s})\}\right\} \, P_{b}(\underrightarrow{\alpha},\underrightarrow{\alpha^{\ast}})\right\} \nonumber \\
 & = & \frac{-i}{\hbar}\left\{ -g\int d\mathbf{s\,}\left\{ \left(\frac{\delta}{\delta\psi_{C}^{+}(\mathbf{s})}\right)\{\left\langle \widehat{\Psi}_{C}(\mathbf{s})^{\dagger}\widehat{\Psi}_{C}(\mathbf{s})\right\rangle \,\psi_{NC}^{+}(\mathbf{s})\}\right\} \, P[\underrightarrow{\psi}(\mathbf{r}),\underrightarrow{\psi^{\ast}}(\mathbf{r})]\right\} \nonumber \\
 &  & +\frac{-i}{\hbar}\left\{ +g\int\mathbf{\,}d\mathbf{s}\left\{ \left(\frac{\delta}{\delta\psi_{C}(\mathbf{s})}\right)\{\left\langle \widehat{\Psi}_{C}(\mathbf{s})^{\dagger}\widehat{\Psi}_{C}(\mathbf{s})\right\rangle \,\psi_{NC}(\mathbf{s})\}\right\} \, P[\underrightarrow{\psi}(\mathbf{r}),\underrightarrow{\psi^{\ast}}(\mathbf{r})]\right\} \nonumber \\
 &  & +\frac{-i}{\hbar}\left\{ -g\int\mathbf{\, d}s\mathbf{\mathbf{\,}}\left\{ \left(\frac{\delta}{\delta\psi_{NC}^{+}(\mathbf{s})}\right)\{\left\langle \widehat{\Psi}_{C}(\mathbf{s})^{\dagger}\widehat{\Psi}_{C}(\mathbf{s})\right\rangle \,\psi_{C}^{+}(\mathbf{s})\}\right\} \, P[\underrightarrow{\psi}(\mathbf{r}),\underrightarrow{\psi^{\ast}}(\mathbf{r})]\right\} \nonumber \\
 &  & +\frac{-i}{\hbar}\left\{ +g\int d\mathbf{s\,}\left\{ \left(\frac{\delta}{\delta\psi_{NC}(\mathbf{s})}\right)\{\left\langle \widehat{\Psi}_{C}(\mathbf{s})^{\dagger}\widehat{\Psi}_{C}(\mathbf{s})\right\rangle \,\psi_{C}(\mathbf{s})\}\right\} \, P[\underrightarrow{\psi}(\mathbf{r}),\underrightarrow{\psi^{\ast}}(\mathbf{r})]\right\} \end{eqnarray}
which is the same as the single mode condensate result. We have used
the results\[
\int d\mathbf{u\,}\phi_{1}(\mathbf{u\,)}\phi_{1}^{\ast}(\mathbf{u\,)=}\int d\mathbf{s\,}\phi_{1}(\mathbf{s\,)}\phi_{1}^{\ast}(\mathbf{s\,)=}1\]
in the first and second terms and \begin{eqnarray*}
\int d\mathbf{s\,}\phi_{1}(\mathbf{s\,)}\,\psi_{C}^{+}(\mathbf{s}) & = & \alpha_{1}^{+}\\
\int d\mathbf{u\,}\phi_{1}^{\ast}(\mathbf{u\,)}\,\psi_{C}(\mathbf{u}) & = & \alpha_{1}\end{eqnarray*}
in the third and fourth terms, changed dummies of integration and
recalled the notation $\left\langle \widehat{\Psi}_{C}(\mathbf{s})^{\dagger}\widehat{\Psi}_{C}(\mathbf{s})\right\rangle =({\small N-1})\left\vert \phi_{1}(\mathbf{s})\right\vert ^{2}$.

For the second order functional derivative terms\begin{eqnarray}
 &  & \left(\frac{\partial}{\partial t}P[\underrightarrow{\psi}(\mathbf{r}),\underrightarrow{\psi^{\ast}}(\mathbf{r})]\right)_{V12}^{2}\nonumber \\
 & = & \frac{-i}{\hbar}\left\{ +g\int\int d\mathbf{s\,}d\mathbf{u}\,\frac{1}{2}\left\{ \left(\frac{\delta}{\delta\psi_{C}^{+}(\mathbf{u})}\right)\left(\frac{\delta}{\delta\psi_{NC}(\mathbf{s})}\right)\{F(\mathbf{s},\mathbf{u})\}\right\} \, P[\underrightarrow{\psi}(\mathbf{r}),\underrightarrow{\psi^{\ast}}(\mathbf{r})]\right\} \nonumber \\
 &  & +\frac{-i}{\hbar}\left\{ -g\int\int d\mathbf{s\, d\mathbf{u\,}}\frac{1}{2}\left\{ \left(\frac{\delta}{\delta\psi_{C}(\mathbf{s})}\right)\left(\frac{\delta}{\delta\psi_{NC}^{+}(\mathbf{u})}\right)\{\mathbf{F(\mathbf{u},\mathbf{s})^{\ast}}\}\right\} \, P[\underrightarrow{\psi}(\mathbf{r}),\underrightarrow{\psi^{\ast}}(\mathbf{r})]\right\} \nonumber \\
 & = & \frac{-i}{\hbar}\left\{ +g\int\int d\mathbf{s\,}d\mathbf{u}\,\frac{1}{2}\left\{ \left(\phi_{1}(\mathbf{u)}\frac{\partial}{\partial\alpha_{1}^{+}}\right)\left(\sum_{k\neq1}\phi_{k}^{\ast}(\mathbf{s)}\frac{\partial}{\partial\alpha_{k}}\right)\{(N-1)\phi_{1}^{\ast}(\mathbf{s})\phi_{1}(\mathbf{s})\phi_{1}(\mathbf{s})\phi_{1}^{\ast}(\mathbf{u})\}\right\} P_{b}(\underrightarrow{\alpha},\underrightarrow{\alpha^{\ast}})\right\} \nonumber \\
 &  & \frac{-i}{\hbar}\left\{ -g\int\int d\mathbf{s\, d\mathbf{u\,}}\frac{1}{2}\left\{ \left(\phi_{1}^{\ast}(\mathbf{s)}\frac{\partial}{\partial\alpha_{1}}\right)\left(\sum_{k\neq1}\phi_{k}(\mathbf{u)}\frac{\partial}{\partial\alpha_{k}^{+}}\right)\{(N-1)\phi_{1}(\mathbf{u})\phi_{1}^{\ast}(\mathbf{u})\phi_{1}^{\ast}(\mathbf{u})\phi_{1}(\mathbf{s)}\}\right\} P_{b}(\underrightarrow{\alpha},\underrightarrow{\alpha^{\ast}})\right\} \nonumber \\
 & = & \frac{-i}{\hbar}\left\{ +g\int d\mathbf{s\,}\frac{1}{2}\left\{ \left(\phi_{1}(\mathbf{s})\,\frac{\partial}{\partial\alpha_{1}^{+}}\right)\left(\sum_{k\neq1}\phi_{k}^{\ast}(\mathbf{s\,)}\frac{\partial}{\partial\alpha_{k}}\right)\{(N-1)\phi_{1}^{\ast}(\mathbf{s})\phi_{1}(\mathbf{s})\}\right\} P_{b}(\underrightarrow{\alpha},\underrightarrow{\alpha^{\ast}})\right\} \nonumber \\
 &  & +\frac{-i}{\hbar}\left\{ -g\int\mathbf{\, d\mathbf{u\,}}\frac{1}{2}\left\{ \left(\phi_{1}^{\ast}(\mathbf{u})\,\frac{\partial}{\partial\alpha_{1}}\right)\left(\sum_{k\neq1}\phi_{k}(\mathbf{u\,)}\frac{\partial}{\partial\alpha_{k}^{+}}\right)\{(N-1)\phi_{1}(\mathbf{u})\phi_{1}^{\ast}(\mathbf{u})\}\right\} P_{b}(\underrightarrow{\alpha},\underrightarrow{\alpha^{\ast}})\right\} \nonumber \\
 & = & \frac{-i}{\hbar}\left\{ +g\int d\mathbf{s\,}\frac{1}{2}\left\{ \left(\frac{\delta}{\delta\psi_{C}^{+}(\mathbf{s})}\right)\left(\frac{\delta}{\delta\psi_{NC}(\mathbf{s})}\right)\{\left\langle \widehat{\Psi}_{C}(\mathbf{s})^{\dagger}\widehat{\Psi}_{C}(\mathbf{s})\right\rangle \}\right\} \, P[\underrightarrow{\psi}(\mathbf{r}),\underrightarrow{\psi^{\ast}}(\mathbf{r})]\right\} \nonumber \\
 &  & +\frac{-i}{\hbar}\left\{ -g\int\mathbf{\, d}s\mathbf{\mathbf{\,}}\frac{1}{2}\left\{ \left(\frac{\delta}{\delta\psi_{C}(\mathbf{s})}\right)\left(\frac{\delta}{\delta\psi_{NC}^{+}(\mathbf{s})}\right)\{\left\langle \widehat{\Psi}_{C}(\mathbf{s})^{\dagger}\widehat{\Psi}_{C}(\mathbf{s})\right\rangle \}\right\} \, P[\underrightarrow{\psi}(\mathbf{r}),\underrightarrow{\psi^{\ast}}(\mathbf{r})]\right\} \end{eqnarray}
which is the same as the single condensate result. Again the results
$\int d\mathbf{u\,}\phi_{1}(\mathbf{u\,)}\phi_{1}^{\ast}(\mathbf{u\,)=}\int d\mathbf{s\,}\phi_{1}(\mathbf{s\,)}\phi_{1}^{\ast}(\mathbf{s\,)=}1$
are used.

\subsection{Condensate - Non-Condensate Interaction - Second Order in Non-Condensate}

The second order term in the interaction between the condensate and
the non-condensate is\begin{eqnarray}
\widehat{V}_{2} & = & \frac{g}{2}\int d\mathbf{s\{}\widehat{\Psi}_{NC}(\mathbf{s})^{\dagger}\widehat{\Psi}_{NC}(\mathbf{s})^{\dagger}\widehat{\Psi}_{C}(\mathbf{s})\widehat{\Psi}_{C}(\mathbf{s})+\widehat{\Psi}_{C}(\mathbf{s})^{\dagger}\widehat{\Psi}_{C}(\mathbf{s})^{\dagger}\widehat{\Psi}_{NC}(\mathbf{s})\widehat{\Psi}_{NC}(\mathbf{s})\nonumber \\
 &  & +4\widehat{\Psi}_{NC}(\mathbf{s})^{\dagger}\widehat{\Psi}_{C}(\mathbf{s})^{\dagger}\widehat{\Psi}_{NC}(\mathbf{s})\widehat{\Psi}_{C}(\mathbf{s})\}\nonumber \\
 &  & \,\end{eqnarray}
This term is due to the boson-boson interaction.

Now if \begin{eqnarray}
\widehat{\rho} & \rightarrow & \widehat{V}_{2}\,\widehat{\rho}\nonumber \\
 & = & \frac{g}{2}\int d\mathbf{s(}\widehat{\Psi}_{NC}(\mathbf{s})^{\dagger}\widehat{\Psi}_{NC}(\mathbf{s})^{\dagger}\widehat{\Psi}_{C}(\mathbf{s})\widehat{\Psi}_{C}(\mathbf{s})+\widehat{\Psi}_{C}(\mathbf{s})^{\dagger}\widehat{\Psi}_{C}(\mathbf{s})^{\dagger}\widehat{\Psi}_{NC}(\mathbf{s})\widehat{\Psi}_{NC}(\mathbf{s}))\widehat{\rho}\nonumber \\
 &  & +\frac{g}{2}\int d\mathbf{s(}4\widehat{\Psi}_{NC}(\mathbf{s})^{\dagger}\widehat{\Psi}_{C}(\mathbf{s})^{\dagger}\widehat{\Psi}_{NC}(\mathbf{s})\widehat{\Psi}_{C}(\mathbf{s}))\widehat{\rho}\nonumber \\
 &  & \,\end{eqnarray}
then\begin{eqnarray}
 &  & WP[\psi(\mathbf{r}),\psi^{+}(\mathbf{r}),\phi(\mathbf{r}),\phi^{+}(\mathbf{r})]\nonumber \\
 & \rightarrow & \frac{g}{2}\int d\mathbf{s}\left(\phi^{+}(\mathbf{s})-\frac{\delta}{\delta\phi(\mathbf{s})}\right)\left(\phi^{+}(\mathbf{s})-\frac{\delta}{\delta\phi(\mathbf{s})}\right)\left(\psi(\mathbf{s})+\frac{1}{2}\frac{\delta}{\delta\psi^{+}(\mathbf{s})}\right)\nonumber \\
 &  & \times\left(\psi(\mathbf{s})+\frac{1}{2}\frac{\delta}{\delta\psi^{+}(\mathbf{s})}\right)WP[\psi,\psi^{+},\phi,\phi^{+}]\nonumber \\
 &  & +\frac{g}{2}\int d\mathbf{s}\left(\psi^{+}(\mathbf{s})-\frac{1}{2}\frac{\delta}{\delta\psi(\mathbf{s})}\right)\left(\psi^{+}(\mathbf{s})-\frac{1}{2}\frac{\delta}{\delta\psi(\mathbf{s})}\right)\left(\phi(\mathbf{s})\right)\nonumber \\
 &  & \times\left(\phi(\mathbf{s})\right)WP[\psi,\psi^{+},\phi,\phi^{+}]\nonumber \\
 &  & +2g\int d\mathbf{s}\left(\phi^{+}(\mathbf{s})-\frac{\delta}{\delta\phi(\mathbf{s})}\right)\left(\psi^{+}(\mathbf{s})-\frac{1}{2}\frac{\delta}{\delta\psi(\mathbf{s})}\right)\left(\phi(\mathbf{s})\right)\nonumber \\
 &  & \times\left(\psi(\mathbf{s})+\frac{1}{2}\frac{\delta}{\delta\psi^{+}(\mathbf{s})}\right)WP[\psi,\psi^{+},\phi,\phi^{+}]\nonumber \\
 &  & \,\end{eqnarray}

Expanding gives\begin{eqnarray}
 &  & WP[\psi(\mathbf{r}),\psi^{+}(\mathbf{r}),\phi(\mathbf{r}),\phi^{+}(\mathbf{r})]\nonumber \\
 & = & WP[\psi,\psi^{+},\phi,\phi^{+}]_{1-16}+WP[\psi,\psi^{+},\phi,\phi^{+}]_{17-20}+WP[\psi,\psi^{+},\phi,\phi^{+}]_{21-28}\nonumber \\
 &  & \,\end{eqnarray}
\begin{eqnarray}
 &  & WP[\psi(\mathbf{r}),\psi^{+}(\mathbf{r}),\phi(\mathbf{r}),\phi^{+}(\mathbf{r})]_{1-16}\nonumber \\
 & = & \frac{g}{2}\int d\mathbf{s}\left(\phi^{+}(\mathbf{s})\right)\left(\phi^{+}(\mathbf{s})\right)\left(\psi(\mathbf{s})\right)\left(\psi(\mathbf{s})\right)WP\nonumber \\
 &  & +\frac{g}{2}\int d\mathbf{s}\left(\phi^{+}(\mathbf{s})\right)\left(\phi^{+}(\mathbf{s})\right)\left(\psi(\mathbf{s})\right)\left(\frac{1}{2}\frac{\delta}{\delta\psi^{+}(\mathbf{s})}\right)WP\nonumber \\
 &  & +\frac{g}{2}\int d\mathbf{s}\left(\phi^{+}(\mathbf{s})\right)\left(\phi^{+}(\mathbf{s})\right)\left(\frac{1}{2}\frac{\delta}{\delta\psi^{+}(\mathbf{s})}\right)\left(\psi(\mathbf{s})\right)WP\nonumber \\
 &  & +\frac{g}{2}\int d\mathbf{s}\left(\phi^{+}(\mathbf{s})\right)\left(\phi^{+}(\mathbf{s})\right)\left(\frac{1}{2}\frac{\delta}{\delta\psi^{+}(\mathbf{s})}\right)\left(\frac{1}{2}\frac{\delta}{\delta\psi^{+}(\mathbf{s})}\right)WP\nonumber \\
 &  & +\frac{g}{2}\int d\mathbf{s}\left(\phi^{+}(\mathbf{s})\right)\left(-\frac{\delta}{\delta\phi(\mathbf{s})}\right)\left(\psi(\mathbf{s})\right)\left(\psi(\mathbf{s})\right)WP\nonumber \\
 &  & +\frac{g}{2}\int d\mathbf{s}\left(\phi^{+}(\mathbf{s})\right)\left(-\frac{\delta}{\delta\phi(\mathbf{s})}\right)\left(\psi(\mathbf{s})\right)\left(\frac{1}{2}\frac{\delta}{\delta\psi^{+}(\mathbf{s})}\right)WP\nonumber \\
 &  & +\frac{g}{2}\int d\mathbf{s}\left(\phi^{+}(\mathbf{s})\right)\left(-\frac{\delta}{\delta\phi(\mathbf{s})}\right)\left(\frac{1}{2}\frac{\delta}{\delta\psi^{+}(\mathbf{s})}\right)\left(\psi(\mathbf{s})\right)WP\nonumber \\
 &  & +\frac{g}{2}\int d\mathbf{s}\left(\phi^{+}(\mathbf{s})\right)\left(-\frac{\delta}{\delta\phi(\mathbf{s})}\right)\left(\frac{1}{2}\frac{\delta}{\delta\psi^{+}(\mathbf{s})}\right)\left(\frac{1}{2}\frac{\delta}{\delta\psi^{+}(\mathbf{s})}\right)WP\nonumber \\
 &  & +\frac{g}{2}\int d\mathbf{s}\left(-\frac{\delta}{\delta\phi(\mathbf{s})}\right)\left(\phi^{+}(\mathbf{s})\right)\left(\psi(\mathbf{s})\right)\left(\psi(\mathbf{s})\right)WP\nonumber \\
 &  & +\frac{g}{2}\int d\mathbf{s}\left(-\frac{\delta}{\delta\phi(\mathbf{s})}\right)\left(\phi^{+}(\mathbf{s})\right)\left(\psi(\mathbf{s})\right)\left(\frac{1}{2}\frac{\delta}{\delta\psi^{+}(\mathbf{s})}\right)WP\nonumber \\
 &  & +\frac{g}{2}\int d\mathbf{s}\left(-\frac{\delta}{\delta\phi(\mathbf{s})}\right)\left(\phi^{+}(\mathbf{s})\right)\left(\frac{1}{2}\frac{\delta}{\delta\psi^{+}(\mathbf{s})}\right)\left(\psi(\mathbf{s})\right)WP\nonumber \\
 &  & +\frac{g}{2}\int d\mathbf{s}\left(-\frac{\delta}{\delta\phi(\mathbf{s})}\right)\left(\phi^{+}(\mathbf{s})\right)\left(\frac{1}{2}\frac{\delta}{\delta\psi^{+}(\mathbf{s})}\right)\left(\frac{1}{2}\frac{\delta}{\delta\psi^{+}(\mathbf{s})}\right)WP\nonumber \\
 &  & +\frac{g}{2}\int d\mathbf{s}\left(-\frac{\delta}{\delta\phi(\mathbf{s})}\right)\left(-\frac{\delta}{\delta\phi(\mathbf{s})}\right)\left(\psi(\mathbf{s})\right)\left(\psi(\mathbf{s})\right)WP\nonumber \\
 &  & +\frac{g}{2}\int d\mathbf{s}\left(-\frac{\delta}{\delta\phi(\mathbf{s})}\right)\left(-\frac{\delta}{\delta\phi(\mathbf{s})}\right)\left(\psi(\mathbf{s})\right)\left(\frac{1}{2}\frac{\delta}{\delta\psi^{+}(\mathbf{s})}\right)WP\nonumber \\
 &  & +\frac{g}{2}\int d\mathbf{s}\left(-\frac{\delta}{\delta\phi(\mathbf{s})}\right)\left(-\frac{\delta}{\delta\phi(\mathbf{s})}\right)\left(\frac{1}{2}\frac{\delta}{\delta\psi^{+}(\mathbf{s})}\right)\left(\psi(\mathbf{s})\right)WP\nonumber \\
 &  & +\frac{g}{2}\int d\mathbf{s}\left(-\frac{\delta}{\delta\phi(\mathbf{s})}\right)\left(-\frac{\delta}{\delta\phi(\mathbf{s})}\right)\left(\frac{1}{2}\frac{\delta}{\delta\psi^{+}(\mathbf{s})}\right)\left(\frac{1}{2}\frac{\delta}{\delta\psi^{+}(\mathbf{s})}\right)WP\nonumber \\
 &  & \,\end{eqnarray}
\begin{eqnarray}
 &  & WP[\psi(\mathbf{r}),\psi^{+}(\mathbf{r}),\phi(\mathbf{r}),\phi^{+}(\mathbf{r})]_{17-20}\nonumber \\
 & = & +\frac{g}{2}\int d\mathbf{s}\left(\psi^{+}(\mathbf{s})\right)\left(\psi^{+}(\mathbf{s})\right)\left(\phi(\mathbf{s})\right)\left(\phi(\mathbf{s})\right)WP\nonumber \\
 &  & +\frac{g}{2}\int d\mathbf{s}\left(\psi^{+}(\mathbf{s})\right)\left(-\frac{1}{2}\frac{\delta}{\delta\psi(\mathbf{s})}\right)\left(\phi(\mathbf{s})\right)\left(\phi(\mathbf{s})\right)WP\nonumber \\
 &  & +\frac{g}{2}\int d\mathbf{s}\left(-\frac{1}{2}\frac{\delta}{\delta\psi(\mathbf{s})}\right)\left(\psi^{+}(\mathbf{s})\right)\left(\phi(\mathbf{s})\right)\left(\phi(\mathbf{s})\right)WP\nonumber \\
 &  & +\frac{g}{2}\int d\mathbf{s}\left(-\frac{1}{2}\frac{\delta}{\delta\psi(\mathbf{s})}\right)\left(-\frac{1}{2}\frac{\delta}{\delta\psi(\mathbf{s})}\right)\left(\phi(\mathbf{s})\right)\left(\phi(\mathbf{s})\right)WP\nonumber \\
 &  & \,\end{eqnarray}
\begin{eqnarray}
 &  & WP[\psi(\mathbf{r}),\psi^{+}(\mathbf{r}),\phi(\mathbf{r}),\phi^{+}(\mathbf{r})]_{21-28}\nonumber \\
 & = & +2g\int d\mathbf{s}\left(\phi^{+}(\mathbf{s})\right)\left(\psi^{+}(\mathbf{s})\right)\left(\phi(\mathbf{s})\right)\left(\psi(\mathbf{s})\right)WP\nonumber \\
 &  & +2g\int d\mathbf{s}\left(\phi^{+}(\mathbf{s})\right)\left(\psi^{+}(\mathbf{s})\right)\left(\phi(\mathbf{s})\right)\left(\frac{1}{2}\frac{\delta}{\delta\psi^{+}(\mathbf{s})}\right)WP\nonumber \\
 &  & +2g\int d\mathbf{s}\left(\phi^{+}(\mathbf{s})\right)\left(-\frac{1}{2}\frac{\delta}{\delta\psi(\mathbf{s})}\right)\left(\phi(\mathbf{s})\right)\left(\psi(\mathbf{s})\right)WP\nonumber \\
 &  & +2g\int d\mathbf{s}\left(\phi^{+}(\mathbf{s})\right)\left(-\frac{1}{2}\frac{\delta}{\delta\psi(\mathbf{s})}\right)\left(\phi(\mathbf{s})\right)\left(\frac{1}{2}\frac{\delta}{\delta\psi^{+}(\mathbf{s})}\right)WP\nonumber \\
 &  & +2g\int d\mathbf{s}\left(-\frac{\delta}{\delta\phi(\mathbf{s})}\right)\left(\psi^{+}(\mathbf{s})\right)\left(\phi(\mathbf{s})\right)\left(\psi(\mathbf{s})\right)WP\nonumber \\
 &  & +2g\int d\mathbf{s}\left(-\frac{\delta}{\delta\phi(\mathbf{s})}\right)\left(\psi^{+}(\mathbf{s})\right)\left(\phi(\mathbf{s})\right)\left(\frac{1}{2}\frac{\delta}{\delta\psi^{+}(\mathbf{s})}\right)WP\nonumber \\
 &  & +2g\int d\mathbf{s}\left(-\frac{\delta}{\delta\phi(\mathbf{s})}\right)\left(-\frac{1}{2}\frac{\delta}{\delta\psi(\mathbf{s})}\right)\left(\phi(\mathbf{s})\right)\left(\psi(\mathbf{s})\right)WP\nonumber \\
 &  & +2g\int d\mathbf{s}\left(-\frac{\delta}{\delta\phi(\mathbf{s})}\right)\left(-\frac{1}{2}\frac{\delta}{\delta\psi(\mathbf{s})}\right)\left(\phi(\mathbf{s})\right)\left(\frac{1}{2}\frac{\delta}{\delta\psi^{+}(\mathbf{s})}\right)WP\nonumber \\
 &  & \,\end{eqnarray}
The functional derivatives are now placed on the left using results
in which the functional derivatives of differing fields are zero (see
(\ref{Eq.FuncDerivativeRule1-1}) and (\ref{Eq.FuncDerivativeRule2-1}))
giving \begin{eqnarray}
 &  & WP[\psi(\mathbf{r}),\psi^{+}(\mathbf{r}),\phi(\mathbf{r}),\phi^{+}(\mathbf{r})]_{1-16}\nonumber \\
 & = & g\int d\mathbf{s\{}\frac{1}{2}\phi^{+}(\mathbf{s})\phi^{+}(\mathbf{s})\psi(\mathbf{s})\psi(\mathbf{s})\}WP\emph{\qquad T1}\nonumber \\
 &  & +g\int d\mathbf{s}\left(\frac{\delta}{\delta\psi^{+}(\mathbf{s})}\right)\{\frac{1}{4}\phi^{+}(\mathbf{s})\phi^{+}(\mathbf{s})\psi(\mathbf{s})\}WP\emph{\qquad T2}\nonumber \\
 &  & +g\int d\mathbf{s}\left(\frac{\delta}{\delta\psi^{+}(\mathbf{s})}\right)\{\frac{1}{4}\phi^{+}(\mathbf{s})\phi^{+}(\mathbf{s})\psi(\mathbf{s})\}WP\emph{\qquad T3}\nonumber \\
 &  & +g\int d\mathbf{s}\left(\frac{\delta}{\delta\psi^{+}(\mathbf{s})}\right)\left(\frac{\delta}{\delta\psi^{+}(\mathbf{s})}\right)\{\frac{1}{8}\phi^{+}(\mathbf{s})\phi^{+}(\mathbf{s})\}WP\emph{\qquad T4}\nonumber \\
 &  & -g\int d\mathbf{s}\left(\frac{\delta}{\delta\phi(\mathbf{s})}\right)\{\frac{1}{2}\phi^{+}(\mathbf{s})\psi(\mathbf{s})\psi(\mathbf{s})\}WP\emph{\qquad T5}\nonumber \\
 &  & -g\int d\mathbf{s}\left(\frac{\delta}{\delta\phi(\mathbf{s})}\right)\left(\frac{\delta}{\delta\psi^{+}(\mathbf{s})}\right)\{\frac{1}{4}\phi^{+}(\mathbf{s})\psi(\mathbf{s})\}WP\emph{\qquad T6}\nonumber \\
 &  & -g\int d\mathbf{s}\left(\frac{\delta}{\delta\phi(\mathbf{s})}\right)\left(\frac{\delta}{\delta\psi^{+}(\mathbf{s})}\right)\{\frac{1}{4}\phi^{+}(\mathbf{s})\psi(\mathbf{s})\}WP\emph{\qquad T7}\nonumber \\
 &  & -g\int d\mathbf{s}\left(\frac{\delta}{\delta\phi(\mathbf{s})}\right)\left(\frac{\delta}{\delta\psi^{+}(\mathbf{s})}\right)\left(\frac{\delta}{\delta\psi^{+}(\mathbf{s})}\right)\{\frac{1}{8}\phi^{+}(\mathbf{s})\}WP\emph{\qquad T8}\nonumber \\
 &  & -g\int d\mathbf{s}\left(\frac{\delta}{\delta\phi(\mathbf{s})}\right)\{\frac{1}{2}\phi^{+}(\mathbf{s})\psi(\mathbf{s})\psi(\mathbf{s})\}WP\emph{\qquad T9}\nonumber \\
 &  & -g\int d\mathbf{s}\left(\frac{\delta}{\delta\phi(\mathbf{s})}\right)\left(\frac{\delta}{\delta\psi^{+}(\mathbf{s})}\right)\{\frac{1}{4}\phi^{+}(\mathbf{s})\psi(\mathbf{s})\}WP\emph{\qquad T10}\nonumber \\
 &  & -g\int d\mathbf{s}\left(\frac{\delta}{\delta\phi(\mathbf{s})}\right)\left(\frac{\delta}{\delta\psi^{+}(\mathbf{s})}\right)\{\frac{1}{4}\phi^{+}(\mathbf{s})\psi(\mathbf{s})\}WP\emph{\qquad T11}\nonumber \\
 &  & -g\int d\mathbf{s}\left(\frac{\delta}{\delta\phi(\mathbf{s})}\right)\left(\frac{\delta}{\delta\psi^{+}(\mathbf{s})}\right)\left(\frac{\delta}{\delta\psi^{+}(\mathbf{s})}\right)\{\frac{1}{8}\phi^{+}(\mathbf{s})\}WP\emph{\qquad T12}\nonumber \\
 &  & +g\int d\mathbf{s}\left(\frac{\delta}{\delta\phi(\mathbf{s})}\right)\left(\frac{\delta}{\delta\phi(\mathbf{s})}\right)\{\frac{1}{2}\psi(\mathbf{s})\psi(\mathbf{s})\}WP\emph{\qquad T13}\nonumber \\
 &  & +g\int d\mathbf{s}\left(\frac{\delta}{\delta\phi(\mathbf{s})}\right)\left(\frac{\delta}{\delta\phi(\mathbf{s})}\right)\left(\frac{\delta}{\delta\psi^{+}(\mathbf{s})}\right)\{\frac{1}{4}\psi(\mathbf{s})\}WP\emph{\qquad T14}\nonumber \\
 &  & +g\int d\mathbf{s}\left(\frac{\delta}{\delta\phi(\mathbf{s})}\right)\left(\frac{\delta}{\delta\phi(\mathbf{s})}\right)\left(\frac{\delta}{\delta\psi^{+}(\mathbf{s})}\right)\{\frac{1}{4}\psi(\mathbf{s})\}WP\emph{\qquad T15}\nonumber \\
 &  & +g\int d\mathbf{s}\left(\frac{\delta}{\delta\phi(\mathbf{s})}\right)\left(\frac{\delta}{\delta\phi(\mathbf{s})}\right)\left(\frac{\delta}{\delta\psi^{+}(\mathbf{s})}\right)\left(\frac{\delta}{\delta\psi^{+}(\mathbf{s})}\right)\{\frac{1}{8}\}WP\emph{\qquad T16}\nonumber \\
 &  & \,\end{eqnarray}
\begin{eqnarray}
 &  & WP[\psi(\mathbf{r}),\psi^{+}(\mathbf{r}),\phi(\mathbf{r}),\phi^{+}(\mathbf{r})]_{17-20}\nonumber \\
 & = & +g\int d\mathbf{s\{}\frac{1}{2}\psi^{+}(\mathbf{s})\psi^{+}(\mathbf{s})\phi(\mathbf{s})\phi(\mathbf{s})\}WP\emph{\qquad T17}\nonumber \\
 &  & -g\int d\mathbf{s}\left(\frac{\delta}{\delta\psi(\mathbf{s})}\right)\{\frac{1}{4}\psi^{+}(\mathbf{s})\phi(\mathbf{s})\phi(\mathbf{s})\}WP\emph{\qquad T18}\nonumber \\
 &  & -g\int d\mathbf{s}\left(\frac{\delta}{\delta\psi(\mathbf{s})}\right)\{\frac{1}{4}\psi^{+}(\mathbf{s})\phi(\mathbf{s})\phi(\mathbf{s})\}WP\emph{\qquad T19}\nonumber \\
 &  & +g\int d\mathbf{s}\left(\frac{\delta}{\delta\psi(\mathbf{s})}\right)\left(\frac{\delta}{\delta\psi(\mathbf{s})}\right)\{\frac{1}{8}\phi(\mathbf{s})\phi(\mathbf{s})\}WP\qquad T\mathit{20}\nonumber \\
 &  & \emph{\,}\end{eqnarray}
\begin{eqnarray}
 &  & WP[\psi(\mathbf{r}),\psi^{+}(\mathbf{r}),\phi(\mathbf{r}),\phi^{+}(\mathbf{r})]_{21-28}\nonumber \\
 & = & +g\int d\mathbf{s\{}2\phi^{+}(\mathbf{s})\psi^{+}(\mathbf{s})\phi(\mathbf{s})\psi(\mathbf{s})\}WP\emph{\qquad T21}\nonumber \\
 &  & +g\int d\mathbf{s}\left\{ \left(\frac{\delta}{\delta\psi^{+}(\mathbf{s})}\right)\{\phi^{+}(\mathbf{s})\psi^{+}(\mathbf{s})\phi(\mathbf{s})\}\right\} WP\qquad T\mathit{22.1}\nonumber \\
 &  & -g\int d\mathbf{s}\left\{ \phi^{+}(\mathbf{s})\delta_{C}(\mathbf{s,s})\phi(\mathbf{s})\right\} WP\emph{\qquad T22.2}\nonumber \\
 &  & -g\int d\mathbf{s}\left(\frac{\delta}{\delta\psi(\mathbf{s})}\right)\{\phi^{+}(\mathbf{s})\phi(\mathbf{s})\psi(\mathbf{s})\}WP\emph{\qquad T23}\nonumber \\
 &  & -g\int d\mathbf{s}\left(\frac{\delta}{\delta\psi(\mathbf{s})}\right)\left(\frac{\delta}{\delta\psi^{+}(\mathbf{s})}\right)\{\frac{1}{2}\phi^{+}(\mathbf{s})\phi(\mathbf{s})\}WP\emph{\qquad T24}\nonumber \\
 &  & -g\int d\mathbf{s}\left(\frac{\delta}{\delta\phi(\mathbf{s})}\right)\{2\psi^{+}(\mathbf{s})\phi(\mathbf{s})\psi(\mathbf{s})\}WP\emph{\qquad T25}\nonumber \\
 &  & -g\int d\mathbf{s}\left\{ \left(\frac{\delta}{\delta\phi(\mathbf{s})}\right)\left(\frac{\delta}{\delta\psi^{+}(\mathbf{s})}\right)\{\psi^{+}(\mathbf{s})\phi(\mathbf{s})\}\right\} WP\emph{\qquad T26.1}\nonumber \\
 &  & +g\int d\mathbf{s}\left\{ \left(\frac{\delta}{\delta\phi(\mathbf{s})}\right)\{\delta_{C}(\mathbf{s,s})\phi(\mathbf{s})\}\right\} WP\emph{\qquad T26.2}\nonumber \\
 &  & +g\int d\mathbf{s}\left(\frac{\delta}{\delta\phi(\mathbf{s})}\right)\left(\frac{\delta}{\delta\psi(\mathbf{s})}\right)\{\phi(\mathbf{s})\psi(\mathbf{s})\}WP\emph{\qquad T27}\nonumber \\
 &  & +g\int d\mathbf{s}\left(\frac{\delta}{\delta\phi(\mathbf{s})}\right)\left(\frac{\delta}{\delta\psi(\mathbf{s})}\right)\left(\frac{\delta}{\delta\psi^{+}(\mathbf{s})}\right)\{\textrm{\ensuremath{\frac{\textrm{1}}{2}}}\phi(\mathbf{s})\}WP\emph{\qquad T28}\nonumber \\
 &  & \,\end{eqnarray}
The two terms that needed extra treatment are\begin{eqnarray*}
 &  & 2g\int d\mathbf{s}\left(\phi^{+}(\mathbf{s})\right)\left(\psi^{+}(\mathbf{s})\right)\left(\phi(\mathbf{s})\right)\left(\frac{1}{2}\frac{\delta}{\delta\psi^{+}(\mathbf{s})}\right)WP\emph{\qquad T22}\\
 & = & g\int d\mathbf{s}\left\{ \left(\phi^{+}(\mathbf{s})\right)\left[\left(\frac{\delta}{\delta\psi^{+}(\mathbf{s})}\right)\left(\psi^{+}(\mathbf{s})\right)\left(\phi(\mathbf{s})\right)-\left(\delta_{C}(\mathbf{s,s})\phi(\mathbf{s})\right)\right]\right\} WP\\
 & = & g\int d\mathbf{s}\left\{ \left(\frac{\delta}{\delta\psi^{+}(\mathbf{s})}\right)\{\phi^{+}(\mathbf{s})\psi^{+}(\mathbf{s})\phi(\mathbf{s})\}\right\} WP\\
 &  & -g\int d\mathbf{s}\left\{ \phi^{+}(\mathbf{s})\delta_{C}(\mathbf{s,s})\phi(\mathbf{s})\right\} WP\end{eqnarray*}
and\begin{eqnarray*}
 &  & -2g\int d\mathbf{s}\left(\frac{\delta}{\delta\phi(\mathbf{s})}\right)\left(\psi^{+}(\mathbf{s})\right)\left(\phi(\mathbf{s})\right)\left(\frac{1}{2}\frac{\delta}{\delta\psi^{+}(\mathbf{s})}\right)WP\emph{\qquad T26}\\
 & = & -g\int d\mathbf{s}\left\{ \left(\frac{\delta}{\delta\phi(\mathbf{s})}\right)\left[\left(\frac{\delta}{\delta\psi^{+}(\mathbf{s})}\right)\left(\psi^{+}(\mathbf{s})\right)\left(\phi(\mathbf{s})\right)-\delta_{C}(\mathbf{s,s})\phi(\mathbf{s})\right]\right\} WP\\
 & = & -g\int d\mathbf{s}\left\{ \left(\frac{\delta}{\delta\phi(\mathbf{s})}\right)\left(\frac{\delta}{\delta\psi^{+}(\mathbf{s})}\right)\{\psi^{+}(\mathbf{s})\phi(\mathbf{s})\}\right\} WP\\
 &  & +g\int d\mathbf{s}\left\{ \left(\frac{\delta}{\delta\phi(\mathbf{s})}\right)\{\delta_{C}(\mathbf{s,s})\phi(\mathbf{s})\}\right\} WP\end{eqnarray*}

Collecting terms with the same order of functional derivatives we
have\begin{eqnarray}
 &  & WP[\psi(\mathbf{r}),\psi^{+}(\mathbf{r}),\phi(\mathbf{r}),\phi^{+}(\mathbf{r})]\nonumber \\
 & \rightarrow & WP^{0}+WP^{1}+WP^{2}+WP^{3}+WP^{4}\label{Eq.V2RhoResult1}\end{eqnarray}
where we have used upper subscripts for the $\,\widehat{V}_{2}\widehat{\rho}$
contributions and\begin{eqnarray}
 &  & WP^{0}\nonumber \\
 & = & g\int d\mathbf{s\{}\frac{1}{2}\phi^{+}(\mathbf{s})\phi^{+}(\mathbf{s})\psi(\mathbf{s})\psi(\mathbf{s})\}WP\emph{\qquad T1}\nonumber \\
 &  & +g\int d\mathbf{s\{}\frac{1}{2}\psi^{+}(\mathbf{s})\psi^{+}(\mathbf{s})\phi(\mathbf{s})\phi(\mathbf{s})\}WP\emph{\qquad T17}\nonumber \\
 &  & +g\int d\mathbf{s\{}2\phi^{+}(\mathbf{s})\psi^{+}(\mathbf{s})\phi(\mathbf{s})\psi(\mathbf{s})\}WP\emph{\qquad T21}\nonumber \\
 &  & -g\int d\mathbf{s}\left\{ \phi^{+}(\mathbf{s})\delta_{C}(\mathbf{s,s})\phi(\mathbf{s})\right\} WP\emph{\qquad T22.2}\nonumber \\
 &  & \,\label{eq:V2RhoResult2}\end{eqnarray}

\begin{eqnarray}
 &  & WP^{1}\nonumber \\
 & = & +g\int d\mathbf{s}\left(\frac{\delta}{\delta\psi^{+}(\mathbf{s})}\right)\{\frac{1}{4}\phi^{+}(\mathbf{s})\phi^{+}(\mathbf{s})\psi(\mathbf{s})\}WP\emph{\qquad T2}\nonumber \\
 &  & +g\int d\mathbf{s}\left(\frac{\delta}{\delta\psi^{+}(\mathbf{s})}\right)\{\frac{1}{4}\phi^{+}(\mathbf{s})\phi^{+}(\mathbf{s})\psi(\mathbf{s})\}WP\emph{\qquad T3}\nonumber \\
 &  & -g\int d\mathbf{s}\left(\frac{\delta}{\delta\phi(\mathbf{s})}\right)\{\frac{1}{2}\phi^{+}(\mathbf{s})\psi(\mathbf{s})\psi(\mathbf{s})\}WP\emph{\qquad T5}\nonumber \\
 &  & -g\int d\mathbf{s}\left(\frac{\delta}{\delta\phi(\mathbf{s})}\right)\{\frac{1}{2}\phi^{+}(\mathbf{s})\psi(\mathbf{s})\psi(\mathbf{s})\}WP\emph{\qquad T9}\nonumber \\
 &  & -g\int d\mathbf{s}\left(\frac{\delta}{\delta\psi(\mathbf{s})}\right)\{\frac{1}{4}\psi^{+}(\mathbf{s})\phi(\mathbf{s})\phi(\mathbf{s})\}WP\emph{\qquad T18}\nonumber \\
 &  & -g\int d\mathbf{s}\left(\frac{\delta}{\delta\psi(\mathbf{s})}\right)\{\frac{1}{4}\psi^{+}(\mathbf{s})\phi(\mathbf{s})\phi(\mathbf{s})\}WP\emph{\qquad T19}\nonumber \\
 &  & +g\int d\mathbf{s}\left\{ \left(\frac{\delta}{\delta\psi^{+}(\mathbf{s})}\right)\{\phi^{+}(\mathbf{s})\psi^{+}(\mathbf{s})\phi(\mathbf{s})\}\right\} WP\emph{\qquad T22.1}\nonumber \\
 &  & -g\int d\mathbf{s}\left(\frac{\delta}{\delta\psi(\mathbf{s})}\right)\{\phi^{+}(\mathbf{s})\phi(\mathbf{s})\psi(\mathbf{s})\}WP\emph{\qquad T23}\nonumber \\
 &  & -g\int d\mathbf{s}\left(\frac{\delta}{\delta\phi(\mathbf{s})}\right)\{2\psi^{+}(\mathbf{s})\phi(\mathbf{s})\psi(\mathbf{s})\}WP\emph{\qquad T25}\nonumber \\
 &  & +g\int d\mathbf{s}\left\{ \left(\frac{\delta}{\delta\phi(\mathbf{s})}\right)\{\delta_{C}(\mathbf{s,s})\phi(\mathbf{s})\}\right\} WP\emph{\qquad T26.2}\nonumber \\
 & = & +g\int d\mathbf{s}\left(\frac{\delta}{\delta\psi^{+}(\mathbf{s})}\right)\{\frac{1}{2}\phi^{+}(\mathbf{s})\phi^{+}(\mathbf{s})\psi(\mathbf{s})\}WP\emph{\qquad T2,T3}\nonumber \\
 &  & +g\int d\mathbf{s}\left\{ \left(\frac{\delta}{\delta\psi^{+}(\mathbf{s})}\right)\{\phi^{+}(\mathbf{s})\psi^{+}(\mathbf{s})\phi(\mathbf{s})\}\right\} WP\emph{\qquad T22.1}\nonumber \\
 &  & -g\int d\mathbf{s}\left(\frac{\delta}{\delta\psi(\mathbf{s})}\right)\{\frac{1}{2}\psi^{+}(\mathbf{s})\phi(\mathbf{s})\phi(\mathbf{s})\}WP\emph{\qquad T18,T19}\nonumber \\
 &  & -g\int d\mathbf{s}\left(\frac{\delta}{\delta\psi(\mathbf{s})}\right)\{\phi^{+}(\mathbf{s})\phi(\mathbf{s})\psi(\mathbf{s})\}WP\emph{\qquad T23}\nonumber \\
 &  & -g\int d\mathbf{s}\left(\frac{\delta}{\delta\phi(\mathbf{s})}\right)\{2\psi^{+}(\mathbf{s})\phi(\mathbf{s})\psi(\mathbf{s})\}WP\emph{\qquad T25}\nonumber \\
 &  & -g\int d\mathbf{s}\left(\frac{\delta}{\delta\phi(\mathbf{s})}\right)\{\phi^{+}(\mathbf{s})\psi(\mathbf{s})\psi(\mathbf{s})\}WP\emph{\qquad T5,T9}\nonumber \\
 &  & +g\int d\mathbf{s}\left\{ \left(\frac{\delta}{\delta\phi(\mathbf{s})}\right)\{\delta_{C}(\mathbf{s,s})\phi(\mathbf{s})\}\right\} WP\qquad T\mathit{26.2}\nonumber \\
 &  & \emph{\,}\label{eq:V2RhoResult3B}\end{eqnarray}
\begin{eqnarray}
 &  & WP^{2}\nonumber \\
 & = & +g\int d\mathbf{s}\left(\frac{\delta}{\delta\psi^{+}(\mathbf{s})}\right)\left(\frac{\delta}{\delta\psi^{+}(\mathbf{s})}\right)\{\frac{1}{8}\phi^{+}(\mathbf{s})\phi^{+}(\mathbf{s})\}WP\emph{\qquad T4}\nonumber \\
 &  & -g\int d\mathbf{s}\left(\frac{\delta}{\delta\phi(\mathbf{s})}\right)\left(\frac{\delta}{\delta\psi^{+}(\mathbf{s})}\right)\{\frac{1}{4}\phi^{+}(\mathbf{s})\psi(\mathbf{s})\}WP\emph{\qquad T6}\nonumber \\
 &  & -g\int d\mathbf{s}\left(\frac{\delta}{\delta\phi(\mathbf{s})}\right)\left(\frac{\delta}{\delta\psi^{+}(\mathbf{s})}\right)\{\frac{1}{4}\phi^{+}(\mathbf{s})\psi(\mathbf{s})\}WP\emph{\qquad T7}\nonumber \\
 &  & -g\int d\mathbf{s}\left(\frac{\delta}{\delta\phi(\mathbf{s})}\right)\left(\frac{\delta}{\delta\psi^{+}(\mathbf{s})}\right)\{\frac{1}{4}\phi^{+}(\mathbf{s})\psi(\mathbf{s})\}WP\emph{\qquad T10}\nonumber \\
 &  & -g\int d\mathbf{s}\left(\frac{\delta}{\delta\phi(\mathbf{s})}\right)\left(\frac{\delta}{\delta\psi^{+}(\mathbf{s})}\right)\{\frac{1}{4}\phi^{+}(\mathbf{s})\psi(\mathbf{s})\}WP\emph{\qquad T11}\nonumber \\
 &  & +g\int d\mathbf{s}\left(\frac{\delta}{\delta\phi(\mathbf{s})}\right)\left(\frac{\delta}{\delta\phi(\mathbf{s})}\right)\{\frac{1}{2}\psi(\mathbf{s})\psi(\mathbf{s})\}WP\emph{\qquad T13}\nonumber \\
 &  & +g\int d\mathbf{s}\left(\frac{\delta}{\delta\psi(\mathbf{s})}\right)\left(\frac{\delta}{\delta\psi(\mathbf{s})}\right)\{\frac{1}{8}\phi(\mathbf{s})\phi(\mathbf{s})\}WP\emph{\qquad T20}\nonumber \\
 &  & -g\int d\mathbf{s}\left(\frac{\delta}{\delta\psi(\mathbf{s})}\right)\left(\frac{\delta}{\delta\psi^{+}(\mathbf{s})}\right)\{\frac{1}{2}\phi^{+}(\mathbf{s})\phi(\mathbf{s})\}WP\emph{\qquad T24}\nonumber \\
 &  & -g\int d\mathbf{s}\left\{ \left(\frac{\delta}{\delta\phi(\mathbf{s})}\right)\left(\frac{\delta}{\delta\psi^{+}(\mathbf{s})}\right)\{\psi^{+}(\mathbf{s})\phi(\mathbf{s})\}\right\} WP\emph{\qquad T26.1}\nonumber \\
 &  & +g\int d\mathbf{s}\left(\frac{\delta}{\delta\phi(\mathbf{s})}\right)\left(\frac{\delta}{\delta\psi(\mathbf{s})}\right)\{\phi(\mathbf{s})\psi(\mathbf{s})\}WP\emph{\qquad T27}\nonumber \\
 & = & +g\int d\mathbf{s}\left(\frac{\delta}{\delta\psi^{+}(\mathbf{s})}\right)\left(\frac{\delta}{\delta\psi^{+}(\mathbf{s})}\right)\{\frac{1}{8}\phi^{+}(\mathbf{s})\phi^{+}(\mathbf{s})\}WP\emph{\qquad T4\qquad A}\nonumber \\
 &  & -g\int d\mathbf{s}\left(\frac{\delta}{\delta\psi(\mathbf{s})}\right)\left(\frac{\delta}{\delta\psi^{+}(\mathbf{s})}\right)\{\frac{1}{2}\phi^{+}(\mathbf{s})\phi(\mathbf{s})\}WP\emph{\qquad T24\qquad H}\nonumber \\
 &  & +g\int d\mathbf{s}\left(\frac{\delta}{\delta\psi(\mathbf{s})}\right)\left(\frac{\delta}{\delta\psi(\mathbf{s})}\right)\{\frac{1}{8}\phi(\mathbf{s})\phi(\mathbf{s})\}WP\emph{\qquad T20\qquad G}\nonumber \\
 &  & -g\int d\mathbf{s}\left(\frac{\delta}{\delta\psi^{+}(\mathbf{s})}\right)\left(\frac{\delta}{\delta\phi(\mathbf{s})}\right)\{\phi^{+}(\mathbf{s})\psi(\mathbf{s})+\psi^{+}(\mathbf{s})\phi(\mathbf{s})\}WP\nonumber \\
 &  & \emph{\qquad T6,T7,T10,T11,T26.1\qquad BCDEI}\nonumber \\
 &  & +g\int d\mathbf{s}\left(\frac{\delta}{\delta\psi(\mathbf{s})}\right)\left(\frac{\delta}{\delta\phi(\mathbf{s})}\right)\{\phi(\mathbf{s})\psi(\mathbf{s})\}WP\emph{\qquad T27\qquad J}\nonumber \\
 &  & +g\int d\mathbf{s}\left(\frac{\delta}{\delta\phi(\mathbf{s})}\right)\left(\frac{\delta}{\delta\phi(\mathbf{s})}\right)\{\frac{1}{2}\psi(\mathbf{s})\psi(\mathbf{s})\}WP\emph{\qquad T13\qquad F}\nonumber \\
 &  & \,\label{eq:V2RhoResult4}\end{eqnarray}
\begin{eqnarray}
 &  & WP^{3}\nonumber \\
 & = & -g\int d\mathbf{s}\left(\frac{\delta}{\delta\phi(\mathbf{s})}\right)\left(\frac{\delta}{\delta\psi^{+}(\mathbf{s})}\right)\left(\frac{\delta}{\delta\psi^{+}(\mathbf{s})}\right)\{\frac{1}{8}\phi^{+}(\mathbf{s})\}WP\emph{\qquad T8}\nonumber \\
 &  & -g\int d\mathbf{s}\left(\frac{\delta}{\delta\phi(\mathbf{s})}\right)\left(\frac{\delta}{\delta\psi^{+}(\mathbf{s})}\right)\left(\frac{\delta}{\delta\psi^{+}(\mathbf{s})}\right)\{\frac{1}{8}\phi^{+}(\mathbf{s})\}WP\emph{\qquad T12}\nonumber \\
 &  & +g\int d\mathbf{s}\left(\frac{\delta}{\delta\phi(\mathbf{s})}\right)\left(\frac{\delta}{\delta\phi(\mathbf{s})}\right)\left(\frac{\delta}{\delta\psi^{+}(\mathbf{s})}\right)\{\frac{1}{4}\psi(\mathbf{s})\}WP\emph{\qquad T14}\nonumber \\
 &  & +g\int d\mathbf{s}\left(\frac{\delta}{\delta\phi(\mathbf{s})}\right)\left(\frac{\delta}{\delta\phi(\mathbf{s})}\right)\left(\frac{\delta}{\delta\psi^{+}(\mathbf{s})}\right)\{\frac{1}{4}\psi(\mathbf{s})\}WP\emph{\qquad T15}\nonumber \\
 &  & +g\int d\mathbf{s}\left(\frac{\delta}{\delta\phi(\mathbf{s})}\right)\left(\frac{\delta}{\delta\psi(\mathbf{s})}\right)\left(\frac{\delta}{\delta\psi^{+}(\mathbf{s})}\right)\{\frac{1}{2}\phi(\mathbf{s})\}WP\emph{\qquad T28}\nonumber \\
 & = & -g\int d\mathbf{s}\left(\frac{\delta}{\delta\psi^{+}(\mathbf{s})}\right)\left(\frac{\delta}{\delta\psi^{+}(\mathbf{s})}\right)\left(\frac{\delta}{\delta\phi(\mathbf{s})}\right)\{\frac{1}{4}\phi^{+}(\mathbf{s})\}WP\emph{\qquad T8,T12}\nonumber \\
 &  & +g\int d\mathbf{s}\left(\frac{\delta}{\delta\psi(\mathbf{s})}\right)\left(\frac{\delta}{\delta\psi^{+}(\mathbf{s})}\right)\left(\frac{\delta}{\delta\phi(\mathbf{s})}\right)\{\frac{1}{2}\phi(\mathbf{s})\}WP\emph{\qquad T28}\nonumber \\
 &  & +g\int d\mathbf{s}\left(\frac{\delta}{\delta\psi^{+}(\mathbf{s})}\right)\left(\frac{\delta}{\delta\phi(\mathbf{s})}\right)\left(\frac{\delta}{\delta\phi(\mathbf{s})}\right)\{\frac{1}{2}\psi(\mathbf{s})\}WP\qquad T\mathit{14,}T\mathit{15}\nonumber \\
 &  & \emph{\,}\label{eq:V2RhoResult5}\end{eqnarray}
\begin{eqnarray}
 &  & WP^{4}\nonumber \\
 & = & +g\int d\mathbf{s}\left(\frac{\delta}{\delta\psi^{+}(\mathbf{s})}\right)\left(\frac{\delta}{\delta\psi^{+}(\mathbf{s})}\right)\left(\frac{\delta}{\delta\phi(\mathbf{s})}\right)\left(\frac{\delta}{\delta\phi(\mathbf{s})}\right)\{\frac{1}{8}\}WP\emph{\qquad T16}\nonumber \\
 &  & \,\label{eq:V2RhoResult6}\end{eqnarray}

Now if \begin{eqnarray}
\widehat{\rho} & \rightarrow & \widehat{\rho}\,\widehat{V}_{2}\nonumber \\
 & = & \frac{g}{2}\int d\mathbf{s\widehat{\rho}\,(}\widehat{\Psi}_{NC}(\mathbf{s})^{\dagger}\widehat{\Psi}_{NC}(\mathbf{s})^{\dagger}\widehat{\Psi}_{C}(\mathbf{s})\widehat{\Psi}_{C}(\mathbf{s})+\widehat{\Psi}_{C}(\mathbf{s})^{\dagger}\widehat{\Psi}_{C}(\mathbf{s})^{\dagger}\widehat{\Psi}_{NC}(\mathbf{s})\widehat{\Psi}_{NC}(\mathbf{s}))\nonumber \\
 &  & +\frac{g}{2}\int d\mathbf{s\widehat{\rho}\,(}4\widehat{\Psi}_{NC}(\mathbf{s})^{\dagger}\widehat{\Psi}_{C}(\mathbf{s})^{\dagger}\widehat{\Psi}_{NC}(\mathbf{s})\widehat{\Psi}_{C}(\mathbf{s}))\end{eqnarray}
then\begin{eqnarray}
 &  & WP[\psi(\mathbf{r}),\psi^{+}(\mathbf{r}),\phi(\mathbf{r}),\phi^{+}(\mathbf{r})]\nonumber \\
 & \rightarrow & \frac{g}{2}\int d\mathbf{s}\left(\psi(\mathbf{s})-\frac{1}{2}\frac{\delta}{\delta\psi^{+}(\mathbf{s})}\right)\left(\psi(\mathbf{s})-\frac{1}{2}\frac{\delta}{\delta\psi^{+}(\mathbf{s})}\right)\left(\phi^{+}(\mathbf{s})\right)\nonumber \\
 &  & \times\left(\phi^{+}(\mathbf{s})\right)WP\nonumber \\
 &  & +\frac{g}{2}\int d\mathbf{s}\left(\phi(\mathbf{s})-\frac{\delta}{\delta\phi^{+}(\mathbf{s})}\right)\left(\phi(\mathbf{s})-\frac{\delta}{\delta\phi^{+}(\mathbf{s})}\right)\left(\psi^{+}(\mathbf{s})+\frac{1}{2}\frac{\delta}{\delta\psi(\mathbf{s})}\right)\nonumber \\
 &  & \times\left(\psi^{+}(\mathbf{s})+\frac{1}{2}\frac{\delta}{\delta\psi(\mathbf{s})}\right)WP\nonumber \\
 &  & +2g\int d\mathbf{s}\left(\psi(\mathbf{s})-\frac{1}{2}\frac{\delta}{\delta\psi^{+}(\mathbf{s})}\right)\left(\phi(\mathbf{s})-\frac{\delta}{\delta\phi^{+}(\mathbf{s})}\right)\left(\psi^{+}(\mathbf{s})+\frac{1}{2}\frac{\delta}{\delta\psi(\mathbf{s})}\right)\nonumber \\
 &  & \times\left(\phi^{+}(\mathbf{s})\right)WP[\psi,\psi^{+},\phi,\phi^{+}]\end{eqnarray}

Expanding this result gives\begin{eqnarray}
 &  & WP[\psi(\mathbf{r}),\psi^{+}(\mathbf{r}),\phi(\mathbf{r}),\phi^{+}(\mathbf{r})]\nonumber \\
 & = & WP[\psi,\psi^{+},\phi,\phi^{+}]_{1-4}+WP[\psi,\psi^{+},\phi,\phi^{+}]_{5-20}+WP[\psi,\psi^{+},\phi,\phi^{+}]_{21-28}\nonumber \\
 &  & \,\end{eqnarray}
where\begin{eqnarray}
 &  & WP[\psi,\psi^{+},\phi,\phi^{+}]_{1-4}\nonumber \\
 & = & \frac{g}{2}\int d\mathbf{s}\left(\psi(\mathbf{s})\right)\left(\psi(\mathbf{s})\right)\left(\phi^{+}(\mathbf{s})\right)\left(\phi^{+}(\mathbf{s})\right)WP\nonumber \\
 &  & +\frac{g}{2}\int d\mathbf{s}\left(\psi(\mathbf{s})\right)\left(-\frac{1}{2}\frac{\delta}{\delta\psi^{+}(\mathbf{s})}\right)\left(\phi^{+}(\mathbf{s})\right)\left(\phi^{+}(\mathbf{s})\right)WP\nonumber \\
 &  & +\frac{g}{2}\int d\mathbf{s}\left(-\frac{1}{2}\frac{\delta}{\delta\psi^{+}(\mathbf{s})}\right)\left(\psi(\mathbf{s})\right)\left(\phi^{+}(\mathbf{s})\right)\left(\phi^{+}(\mathbf{s})\right)WP\nonumber \\
 &  & +\frac{g}{2}\int d\mathbf{s}\left(-\frac{1}{2}\frac{\delta}{\delta\psi^{+}(\mathbf{s})}\right)\left(-\frac{1}{2}\frac{\delta}{\delta\psi^{+}(\mathbf{s})}\right)\left(\phi^{+}(\mathbf{s})\right)\left(\phi^{+}(\mathbf{s})\right)WP\nonumber \\
 &  & \,\end{eqnarray}
\begin{eqnarray}
 &  & WP[\psi,\psi^{+},\phi,\phi^{+}]_{5-20}\nonumber \\
 & = & +\frac{g}{2}\int d\mathbf{s}\left(\phi(\mathbf{s})\right)\left(\phi(\mathbf{s})\right)\left(\psi^{+}(\mathbf{s})\right)\left(\psi^{+}(\mathbf{s})\right)WP\nonumber \\
 &  & +\frac{g}{2}\int d\mathbf{s}\left(\phi(\mathbf{s})\right)\left(\phi(\mathbf{s})\right)\left(\psi^{+}(\mathbf{s})\right)\left(\frac{1}{2}\frac{\delta}{\delta\psi(\mathbf{s})}\right)WP\nonumber \\
 &  & +\frac{g}{2}\int d\mathbf{s}\left(\phi(\mathbf{s})\right)\left(\phi(\mathbf{s})\right)\left(\frac{1}{2}\frac{\delta}{\delta\psi(\mathbf{s})}\right)\left(\psi^{+}(\mathbf{s})\right)WP\nonumber \\
 &  & +\frac{g}{2}\int d\mathbf{s}\left(\phi(\mathbf{s})\right)\left(\phi(\mathbf{s})\right)\left(\frac{1}{2}\frac{\delta}{\delta\psi(\mathbf{s})}\right)\left(\frac{1}{2}\frac{\delta}{\delta\psi(\mathbf{s})}\right)WP\nonumber \\
 &  & +\frac{g}{2}\int d\mathbf{s}\left(\phi(\mathbf{s})\right)\left(-\frac{\delta}{\delta\phi^{+}(\mathbf{s})}\right)\left(\psi^{+}(\mathbf{s})\right)\left(\psi^{+}(\mathbf{s})\right)WP\nonumber \\
 &  & +\frac{g}{2}\int d\mathbf{s}\left(\phi(\mathbf{s})\right)\left(-\frac{\delta}{\delta\phi^{+}(\mathbf{s})}\right)\left(\psi^{+}(\mathbf{s})\right)\left(\frac{1}{2}\frac{\delta}{\delta\psi(\mathbf{s})}\right)WP\nonumber \\
 &  & +\frac{g}{2}\int d\mathbf{s}\left(\phi(\mathbf{s})\right)\left(-\frac{\delta}{\delta\phi^{+}(\mathbf{s})}\right)\left(\frac{1}{2}\frac{\delta}{\delta\psi(\mathbf{s})}\right)\left(\psi^{+}(\mathbf{s})\right)WP\nonumber \\
 &  & +\frac{g}{2}\int d\mathbf{s}\left(\phi(\mathbf{s})\right)\left(-\frac{\delta}{\delta\phi^{+}(\mathbf{s})}\right)\left(\frac{1}{2}\frac{\delta}{\delta\psi(\mathbf{s})}\right)\left(\frac{1}{2}\frac{\delta}{\delta\psi(\mathbf{s})}\right)WP\nonumber \\
 &  & +\frac{g}{2}\int d\mathbf{s}\left(-\frac{\delta}{\delta\phi^{+}(\mathbf{s})}\right)\left(\phi(\mathbf{s})\right)\left(\psi^{+}(\mathbf{s})\right)\left(\psi^{+}(\mathbf{s})\right)WP\nonumber \\
 &  & +\frac{g}{2}\int d\mathbf{s}\left(-\frac{\delta}{\delta\phi^{+}(\mathbf{s})}\right)\left(\phi(\mathbf{s})\right)\left(\psi^{+}(\mathbf{s})\right)\left(\frac{1}{2}\frac{\delta}{\delta\psi(\mathbf{s})}\right)WP\nonumber \\
 &  & +\frac{g}{2}\int d\mathbf{s}\left(-\frac{\delta}{\delta\phi^{+}(\mathbf{s})}\right)\left(\phi(\mathbf{s})\right)\left(\frac{1}{2}\frac{\delta}{\delta\psi(\mathbf{s})}\right)\left(\psi^{+}(\mathbf{s})\right)WP\nonumber \\
 &  & +\frac{g}{2}\int d\mathbf{s}\left(-\frac{\delta}{\delta\phi^{+}(\mathbf{s})}\right)\left(\phi(\mathbf{s})\right)\left(\frac{1}{2}\frac{\delta}{\delta\psi(\mathbf{s})}\right)\left(\frac{1}{2}\frac{\delta}{\delta\psi(\mathbf{s})}\right)WP\nonumber \\
 &  & +\frac{g}{2}\int d\mathbf{s}\left(-\frac{\delta}{\delta\phi^{+}(\mathbf{s})}\right)\left(-\frac{\delta}{\delta\phi^{+}(\mathbf{s})}\right)\left(\psi^{+}(\mathbf{s})\right)\left(\psi^{+}(\mathbf{s})\right)WP\nonumber \\
 &  & +\frac{g}{2}\int d\mathbf{s}\left(-\frac{\delta}{\delta\phi^{+}(\mathbf{s})}\right)\left(-\frac{\delta}{\delta\phi^{+}(\mathbf{s})}\right)\left(\psi^{+}(\mathbf{s})\right)\left(\frac{1}{2}\frac{\delta}{\delta\psi(\mathbf{s})}\right)WP\nonumber \\
 &  & +\frac{g}{2}\int d\mathbf{s}\left(-\frac{\delta}{\delta\phi^{+}(\mathbf{s})}\right)\left(-\frac{\delta}{\delta\phi^{+}(\mathbf{s})}\right)\left(\frac{1}{2}\frac{\delta}{\delta\psi(\mathbf{s})}\right)\left(\psi^{+}(\mathbf{s})\right)WP\nonumber \\
 &  & +\frac{g}{2}\int d\mathbf{s}\left(-\frac{\delta}{\delta\phi^{+}(\mathbf{s})}\right)\left(-\frac{\delta}{\delta\phi^{+}(\mathbf{s})}\right)\left(\frac{1}{2}\frac{\delta}{\delta\psi(\mathbf{s})}\right)\left(\frac{1}{2}\frac{\delta}{\delta\psi(\mathbf{s})}\right)WP\nonumber \\
 &  & \,\end{eqnarray}
\begin{eqnarray}
 &  & WP[\psi,\psi^{+},\phi,\phi^{+}]_{21-28}\nonumber \\
 & = & +2g\int d\mathbf{s}\left(\psi(\mathbf{s})\right)\left(\phi(\mathbf{s})\right)\left(\psi^{+}(\mathbf{s})\right)\left(\phi^{+}(\mathbf{s})\right)WP\nonumber \\
 &  & +2g\int d\mathbf{s}\left(\psi(\mathbf{s})\right)\left(\phi(\mathbf{s})\right)\left(\frac{1}{2}\frac{\delta}{\delta\psi(\mathbf{s})}\right)\left(\phi^{+}(\mathbf{s})\right)WP\nonumber \\
 &  & +2g\int d\mathbf{s}\left(\psi(\mathbf{s})\right)\left(-\frac{\delta}{\delta\phi^{+}(\mathbf{s})}\right)\left(\psi^{+}(\mathbf{s})\right)\left(\phi^{+}(\mathbf{s})\right)WP\nonumber \\
 &  & +2g\int d\mathbf{s}\left(\psi(\mathbf{s})\right)\left(-\frac{\delta}{\delta\phi^{+}(\mathbf{s})}\right)\left(\frac{1}{2}\frac{\delta}{\delta\psi(\mathbf{s})}\right)\left(\phi^{+}(\mathbf{s})\right)WP\nonumber \\
 &  & +2g\int d\mathbf{s}\left(-\frac{1}{2}\frac{\delta}{\delta\psi^{+}(\mathbf{s})}\right)\left(\phi(\mathbf{s})\right)\left(\psi^{+}(\mathbf{s})\right)\left(\phi^{+}(\mathbf{s})\right)WP\nonumber \\
 &  & +2g\int d\mathbf{s}\left(-\frac{1}{2}\frac{\delta}{\delta\psi^{+}(\mathbf{s})}\right)\left(\phi(\mathbf{s})\right)\left(\frac{1}{2}\frac{\delta}{\delta\psi(\mathbf{s})}\right)\left(\phi^{+}(\mathbf{s})\right)WP\nonumber \\
 &  & +2g\int d\mathbf{s}\left(-\frac{1}{2}\frac{\delta}{\delta\psi^{+}(\mathbf{s})}\right)\left(-\frac{\delta}{\delta\phi^{+}(\mathbf{s})}\right)\left(\psi^{+}(\mathbf{s})\right)\left(\phi^{+}(\mathbf{s})\right)WP\nonumber \\
 &  & +2g\int d\mathbf{s}\left(-\frac{1}{2}\frac{\delta}{\delta\psi^{+}(\mathbf{s})}\right)\left(-\frac{\delta}{\delta\phi^{+}(\mathbf{s})}\right)\left(\frac{1}{2}\frac{\delta}{\delta\psi(\mathbf{s})}\right)\left(\phi^{+}(\mathbf{s})\right)WP\nonumber \\
 &  & \,\end{eqnarray}
The functional derivatives are now placed on the left using results
in which the functional derivatives of differing fields are zero (see
(\ref{Eq.FuncDerivativeRule1-1}) and (\ref{Eq.FuncDerivativeRule2-1}))
giving\begin{eqnarray}
 &  & WP[\psi,\psi^{+},\phi,\phi^{+}]_{1-4}\nonumber \\
 & = & g\int d\mathbf{s\{}\frac{1}{2}\psi(\mathbf{s})\psi(\mathbf{s})\phi^{+}(\mathbf{s})\phi^{+}(\mathbf{s})\}WP\emph{\qquad U1}\nonumber \\
 &  & -g\int d\mathbf{s}\left(\frac{\delta}{\delta\psi^{+}(\mathbf{s})}\right)\{\frac{1}{4}\psi(\mathbf{s})\phi^{+}(\mathbf{s})\phi^{+}(\mathbf{s})\}WP\emph{\qquad U2}\nonumber \\
 &  & -g\int d\mathbf{s}\left(\frac{\delta}{\delta\psi^{+}(\mathbf{s})}\right)\{\frac{1}{4}\psi(\mathbf{s})\phi^{+}(\mathbf{s})\phi^{+}(\mathbf{s})\}WP\emph{\qquad U3}\nonumber \\
 &  & +g\int d\mathbf{s}\left(\frac{\delta}{\delta\psi^{+}(\mathbf{s})}\right)\left(\frac{\delta}{\delta\psi^{+}(\mathbf{s})}\right)\{\frac{1}{8}\phi^{+}(\mathbf{s})\phi^{+}(\mathbf{s})\}WP\emph{\qquad U4}\nonumber \\
 &  & \,\end{eqnarray}
\begin{eqnarray}
 &  & WP[\psi,\psi^{+},\phi,\phi^{+}]_{5-20}\nonumber \\
 & = & +g\int d\mathbf{s\{}\frac{1}{2}\phi(\mathbf{s})\phi(\mathbf{s})\psi^{+}(\mathbf{s})\psi^{+}(\mathbf{s})\}WP\emph{\qquad U5}\nonumber \\
 &  & +g\int d\mathbf{s}\left(\frac{\delta}{\delta\psi(\mathbf{s})}\right)\{\frac{1}{4}\phi(\mathbf{s})\phi(\mathbf{s})\psi^{+}(\mathbf{s})\}WP\emph{\qquad U6}\nonumber \\
 &  & +g\int d\mathbf{s}\left(\frac{\delta}{\delta\psi(\mathbf{s})}\right)\{\frac{1}{4}\phi(\mathbf{s})\phi(\mathbf{s})\psi^{+}(\mathbf{s})\}WP\emph{\qquad U7}\nonumber \\
 &  & +g\int d\mathbf{s}\left(\frac{\delta}{\delta\psi(\mathbf{s})}\right)\left(\frac{\delta}{\delta\psi(\mathbf{s})}\right)\{\frac{1}{8}\phi(\mathbf{s})\phi(\mathbf{s})\}WP\emph{\qquad U8}\nonumber \\
 &  & -g\int d\mathbf{s}\left(\frac{\delta}{\delta\phi^{+}(\mathbf{s})}\right)\{\frac{1}{2}\phi(\mathbf{s})\psi^{+}(\mathbf{s})\psi^{+}(\mathbf{s})\}WP\emph{\qquad U9}\nonumber \\
 &  & -g\int d\mathbf{s}\left(\frac{\delta}{\delta\phi^{+}(\mathbf{s})}\right)\left(\frac{\delta}{\delta\psi(\mathbf{s})}\right)\{\frac{1}{4}\phi(\mathbf{s})\psi^{+}(\mathbf{s})\}WP\emph{\qquad U10}\nonumber \\
 &  & -g\int d\mathbf{s}\left(\frac{\delta}{\delta\phi^{+}(\mathbf{s})}\right)\left(\frac{\delta}{\delta\psi(\mathbf{s})}\right)\{\frac{1}{4}\phi(\mathbf{s})\psi^{+}(\mathbf{s})\}WP\emph{\qquad U11}\nonumber \\
 &  & -g\int d\mathbf{s}\left(\frac{\delta}{\delta\phi^{+}(\mathbf{s})}\right)\left(\frac{\delta}{\delta\psi(\mathbf{s})}\right)\left(\frac{\delta}{\delta\psi(\mathbf{s})}\right)\{\frac{1}{8}\phi(\mathbf{s})\}WP\emph{\qquad U12}\nonumber \\
 &  & -g\int d\mathbf{s}\left(\frac{\delta}{\delta\phi^{+}(\mathbf{s})}\right)\{\frac{1}{2}\phi(\mathbf{s})\psi^{+}(\mathbf{s})\psi^{+}(\mathbf{s})\}WP\emph{\qquad U13}\nonumber \\
 &  & -g\int d\mathbf{s}\left(\frac{\delta}{\delta\phi^{+}(\mathbf{s})}\right)\left(\frac{\delta}{\delta\psi(\mathbf{s})}\right)\{\frac{1}{4}\phi(\mathbf{s})\psi^{+}(\mathbf{s})\}WP\emph{\qquad U14}\nonumber \\
 &  & -g\int d\mathbf{s}\left(\frac{\delta}{\delta\phi^{+}(\mathbf{s})}\right)\left(\frac{\delta}{\delta\psi(\mathbf{s})}\right)\{\frac{1}{4}\phi(\mathbf{s})\psi^{+}(\mathbf{s})\}WP\emph{\qquad U15}\nonumber \\
 &  & -g\int d\mathbf{s}\left(\frac{\delta}{\delta\phi^{+}(\mathbf{s})}\right)\left(\frac{\delta}{\delta\psi(\mathbf{s})}\right)\left(\frac{\delta}{\delta\psi(\mathbf{s})}\right)\{\frac{1}{8}\phi(\mathbf{s})\}WP\emph{\qquad U16}\nonumber \\
 &  & +g\int d\mathbf{s}\left(\frac{\delta}{\delta\phi^{+}(\mathbf{s})}\right)\left(\frac{\delta}{\delta\phi^{+}(\mathbf{s})}\right)\{\frac{1}{2}\psi^{+}(\mathbf{s})\psi^{+}(\mathbf{s})\}WP\emph{\qquad U17}\nonumber \\
 &  & +g\int d\mathbf{s}\left(\frac{\delta}{\delta\phi^{+}(\mathbf{s})}\right)\left(\frac{\delta}{\delta\phi^{+}(\mathbf{s})}\right)\left(\frac{\delta}{\delta\psi(\mathbf{s})}\right)\{\frac{1}{4}\psi^{+}(\mathbf{s})\}WP\emph{\qquad U18}\nonumber \\
 &  & +g\int d\mathbf{s}\left(\frac{\delta}{\delta\phi^{+}(\mathbf{s})}\right)\left(\frac{\delta}{\delta\phi^{+}(\mathbf{s})}\right)\left(\frac{\delta}{\delta\psi(\mathbf{s})}\right)\{\frac{1}{4}\psi^{+}(\mathbf{s})\}WP\emph{\qquad U19}\nonumber \\
 &  & +g\int d\mathbf{s}\left(\frac{\delta}{\delta\phi^{+}(\mathbf{s})}\right)\left(\frac{\delta}{\delta\phi^{+}(\mathbf{s})}\right)\left(\frac{\delta}{\delta\psi(\mathbf{s})}\right)\left(\frac{\delta}{\delta\psi(\mathbf{s})}\right)\{\frac{1}{8}\}WP\emph{\qquad U20}\nonumber \\
 &  & \,\end{eqnarray}
\begin{eqnarray}
 &  & WP[\psi,\psi^{+},\phi,\phi^{+}]_{21-28}\nonumber \\
 & = & +g\int d\mathbf{s\{}2\psi(\mathbf{s})\phi(\mathbf{s})\psi^{+}(\mathbf{s})\phi^{+}(\mathbf{s})\}WP\emph{\qquad U21}\nonumber \\
 &  & +g\int d\mathbf{s}\left(\frac{\delta}{\delta\psi(\mathbf{s})}\right)\{\psi(\mathbf{s})\phi(\mathbf{s})\phi^{+}(\mathbf{s})\}WP\qquad U\mathit{22.1}\nonumber \\
 &  & -g\int d\mathbf{s\{}\delta_{C}(\mathbf{s,s})\phi(\mathbf{s})\phi^{+}(\mathbf{s})\}WP\emph{\qquad U22.2}\nonumber \\
 &  & -g\int d\mathbf{s}\left(\frac{\delta}{\delta\phi^{+}(\mathbf{s})}\right)\{2\psi(\mathbf{s})\psi^{+}(\mathbf{s})\phi^{+}(\mathbf{s})\}WP\emph{\qquad U23}\nonumber \\
 &  & -g\int d\mathbf{s}\left(\frac{\delta}{\delta\phi^{+}(\mathbf{s})}\right)\left(\frac{\delta}{\delta\psi(\mathbf{s})}\right)\{\psi(\mathbf{s})\phi^{+}(\mathbf{s})\}WP\qquad U\mathit{24.1}\nonumber \\
 &  & +g\int d\mathbf{s}\left(\frac{\delta}{\delta\phi^{+}(\mathbf{s})}\right)\{\delta_{C}(\mathbf{s,s})\phi^{+}(\mathbf{s})\}WP\emph{\qquad U24.2}\nonumber \\
 &  & -g\int d\mathbf{s}\left(\frac{\delta}{\delta\psi^{+}(\mathbf{s})}\right)\{\phi(\mathbf{s})\psi^{+}(\mathbf{s})\phi^{+}(\mathbf{s})\}WP\emph{\qquad U25}\nonumber \\
 &  & -g\int d\mathbf{s}\left(\frac{\delta}{\delta\psi^{+}(\mathbf{s})}\right)\left(\frac{\delta}{\delta\psi(\mathbf{s})}\right)\{\frac{1}{2}\phi(\mathbf{s})\phi^{+}(\mathbf{s})\}WP\emph{\qquad U26}\nonumber \\
 &  & +g\int d\mathbf{s}\left(\frac{\delta}{\delta\psi^{+}(\mathbf{s})}\right)\left(\frac{\delta}{\delta\phi^{+}(\mathbf{s})}\right)\{\psi^{+}(\mathbf{s})\phi^{+}(\mathbf{s})\}WP\emph{\qquad U27}\nonumber \\
 &  & +g\int d\mathbf{s}\left(\frac{\delta}{\delta\psi^{+}(\mathbf{s})}\right)\left(\frac{\delta}{\delta\phi^{+}(\mathbf{s})}\right)\left(\frac{\delta}{\delta\psi(\mathbf{s})}\right)\{\frac{1}{2}\phi^{+}(\mathbf{s})\}WP\emph{\qquad U28}\nonumber \\
 &  & \,\end{eqnarray}
The two terms that needed extra treatment are\begin{eqnarray*}
 &  & 2g\int d\mathbf{s}\left(\psi(\mathbf{s})\right)\left(\phi(\mathbf{s})\right)\left(\frac{1}{2}\frac{\delta}{\delta\psi(\mathbf{s})}\right)\left(\phi^{+}(\mathbf{s})\right)WP\emph{\qquad U22}\\
 & = & g\int d\mathbf{s}\left\{ \left(\frac{\delta}{\delta\psi(\mathbf{s})}\right)\left[\psi(\mathbf{s})\phi(\mathbf{s})\phi^{+}(\mathbf{s})\right]-\delta_{C}(\mathbf{s,s})\left[\phi(\mathbf{s})\phi^{+}(\mathbf{s})\right]\right\} WP\\
 & = & g\int d\mathbf{s}\left(\frac{\delta}{\delta\psi(\mathbf{s})}\right)\{\psi(\mathbf{s})\phi(\mathbf{s})\phi^{+}(\mathbf{s})\}WP-g\int d\mathbf{s\{}\delta_{C}(\mathbf{s,s})\phi(\mathbf{s})\phi^{+}(\mathbf{s})\}WP\end{eqnarray*}
and\begin{eqnarray*}
 &  & 2g\int d\mathbf{s}\left(\psi(\mathbf{s})\right)\left(-\frac{\delta}{\delta\phi^{+}(\mathbf{s})}\right)\left(\frac{1}{2}\frac{\delta}{\delta\psi(\mathbf{s})}\right)\left(\phi^{+}(\mathbf{s})\right)WP\emph{\qquad U24}\\
 & = & -g\int d\mathbf{s}\left(\frac{\delta}{\delta\phi^{+}(\mathbf{s})}\right)\left\{ \left(\frac{\delta}{\delta\psi(\mathbf{s})}\right)[\psi(\mathbf{s})\phi^{+}(\mathbf{s})]-\delta_{C}(\mathbf{s,s})[\phi^{+}(\mathbf{s})]\right\} WP\\
 & = & -g\int d\mathbf{s}\left(\frac{\delta}{\delta\phi^{+}(\mathbf{s})}\right)\left(\frac{\delta}{\delta\psi(\mathbf{s})}\right)\{\psi(\mathbf{s})\phi^{+}(\mathbf{s})\}WP+g\int d\mathbf{s}\left(\frac{\delta}{\delta\phi^{+}(\mathbf{s})}\right)\{\delta_{C}(\mathbf{s,s})\phi^{+}(\mathbf{s})\}WP\end{eqnarray*}
Collecting terms with the same order of functional derivatives we
have\begin{eqnarray}
 &  & WP[\psi(\mathbf{r}),\psi^{+}(\mathbf{r}),\phi(\mathbf{r}),\phi^{+}(\mathbf{r})]\nonumber \\
 & \rightarrow & WP_{0}+WP_{1}+WP_{2}+WP_{3}+WP_{4}\label{Eq.RhoV2Result1}\end{eqnarray}
where we have used lower subscripts for the $\widehat{\rho}\,\widehat{V}_{2}$contributions
and\begin{eqnarray}
 &  & WP_{0}\nonumber \\
 & = & g\int d\mathbf{s\{}\frac{1}{2}\psi(\mathbf{s})\psi(\mathbf{s})\phi^{+}(\mathbf{s})\phi^{+}(\mathbf{s})\}WP\emph{\qquad U1}\nonumber \\
 &  & +g\int d\mathbf{s\{}\frac{1}{2}\phi(\mathbf{s})\phi(\mathbf{s})\psi^{+}(\mathbf{s})\psi^{+}(\mathbf{s})\}WP\emph{\qquad U5}\nonumber \\
 &  & +g\int d\mathbf{s\{}2\psi(\mathbf{s})\phi(\mathbf{s})\psi^{+}(\mathbf{s})\phi^{+}(\mathbf{s})\}WP\emph{\qquad U21}\nonumber \\
 &  & -g\int d\mathbf{s\{}\delta_{C}(\mathbf{s,s})\phi(\mathbf{s})\phi^{+}(\mathbf{s})\}WP\emph{\qquad U22.2}\nonumber \\
 &  & \,\label{eq:RhoV2Result2}\end{eqnarray}
\begin{eqnarray}
 &  & WP_{1}\nonumber \\
 & = & -g\int d\mathbf{s}\left(\frac{\delta}{\delta\psi^{+}(\mathbf{s})}\right)\{\frac{1}{4}\psi(\mathbf{s})\phi^{+}(\mathbf{s})\phi^{+}(\mathbf{s})\}WP\emph{\qquad U2}\nonumber \\
 &  & -g\int d\mathbf{s}\left(\frac{\delta}{\delta\psi^{+}(\mathbf{s})}\right)\{\frac{1}{4}\psi(\mathbf{s})\phi^{+}(\mathbf{s})\phi^{+}(\mathbf{s})\}WP\emph{\qquad U3}\nonumber \\
 &  & +g\int d\mathbf{s}\left(\frac{\delta}{\delta\psi(\mathbf{s})}\right)\{\frac{1}{4}\phi(\mathbf{s})\phi(\mathbf{s})\psi^{+}(\mathbf{s})\}WP\emph{\qquad U6}\nonumber \\
 &  & +g\int d\mathbf{s}\left(\frac{\delta}{\delta\psi(\mathbf{s})}\right)\{\frac{1}{4}\phi(\mathbf{s})\phi(\mathbf{s})\psi^{+}(\mathbf{s})\}WP\emph{\qquad U7}\nonumber \\
 &  & -g\int d\mathbf{s}\left(\frac{\delta}{\delta\phi^{+}(\mathbf{s})}\right)\{\frac{1}{2}\phi(\mathbf{s})\psi^{+}(\mathbf{s})\psi^{+}(\mathbf{s})\}WP\emph{\qquad U9}\nonumber \\
 &  & -g\int d\mathbf{s}\left(\frac{\delta}{\delta\phi^{+}(\mathbf{s})}\right)\{\frac{1}{2}\phi(\mathbf{s})\psi^{+}(\mathbf{s})\psi^{+}(\mathbf{s})\}WP\emph{\qquad U13}\nonumber \\
 &  & +g\int d\mathbf{s}\left(\frac{\delta}{\delta\psi(\mathbf{s})}\right)\{\psi(\mathbf{s})\phi(\mathbf{s})\phi^{+}(\mathbf{s})\}WP\emph{\qquad U22.1}\nonumber \\
 &  & -g\int d\mathbf{s}\left(\frac{\delta}{\delta\phi^{+}(\mathbf{s})}\right)\{2\psi(\mathbf{s})\psi^{+}(\mathbf{s})\phi^{+}(\mathbf{s})\}WP\emph{\qquad U23}\nonumber \\
 &  & +g\int d\mathbf{s}\left(\frac{\delta}{\delta\phi^{+}(\mathbf{s})}\right)\{\delta_{C}(\mathbf{s,s})\phi^{+}(\mathbf{s})\}WP\emph{\qquad U24.2}\nonumber \\
 &  & -g\int d\mathbf{s}\left(\frac{\delta}{\delta\psi^{+}(\mathbf{s})}\right)\{\phi(\mathbf{s})\psi^{+}(\mathbf{s})\phi^{+}(\mathbf{s})\}WP\emph{\qquad U25}\nonumber \\
 & = & -g\int d\mathbf{s}\left(\frac{\delta}{\delta\psi^{+}(\mathbf{s})}\right)\{\frac{1}{2}\psi(\mathbf{s})\phi^{+}(\mathbf{s})\phi^{+}(\mathbf{s})\}WP\emph{\qquad U2,U3}\nonumber \\
 &  & -g\int d\mathbf{s}\left(\frac{\delta}{\delta\psi^{+}(\mathbf{s})}\right)\{\phi(\mathbf{s})\psi^{+}(\mathbf{s})\phi^{+}(\mathbf{s})\}WP\emph{\qquad U25}\nonumber \\
 &  & +g\int d\mathbf{s}\left(\frac{\delta}{\delta\psi(\mathbf{s})}\right)\{\frac{1}{2}\phi(\mathbf{s})\phi(\mathbf{s})\psi^{+}(\mathbf{s})\}WP\emph{\qquad U6,U7}\nonumber \\
 &  & +g\int d\mathbf{s}\left(\frac{\delta}{\delta\psi(\mathbf{s})}\right)\{\psi(\mathbf{s})\phi(\mathbf{s})\phi^{+}(\mathbf{s})\}WP\emph{\qquad U22.1}\nonumber \\
 &  & -g\int d\mathbf{s}\left(\frac{\delta}{\delta\phi^{+}(\mathbf{s})}\right)\{2\psi(\mathbf{s})\psi^{+}(\mathbf{s})\phi^{+}(\mathbf{s})\}WP\emph{\qquad U23}\nonumber \\
 &  & -g\int d\mathbf{s}\left(\frac{\delta}{\delta\phi^{+}(\mathbf{s})}\right)\{\phi(\mathbf{s})\psi^{+}(\mathbf{s})\psi^{+}(\mathbf{s})\}WP\emph{\qquad U9,U13}\nonumber \\
 &  & +g\int d\mathbf{s}\left(\frac{\delta}{\delta\phi^{+}(\mathbf{s})}\right)\{\delta_{C}(\mathbf{s,s})\phi^{+}(\mathbf{s})\}WP\emph{\qquad U24.2}\nonumber \\
 &  & \,\label{eq:RhoV2Result3}\end{eqnarray}
\begin{eqnarray}
 &  & WP_{2}\nonumber \\
 & = & +g\int d\mathbf{s}\left(\frac{\delta}{\delta\psi^{+}(\mathbf{s})}\right)\left(\frac{\delta}{\delta\psi^{+}(\mathbf{s})}\right)\{\frac{1}{8}\phi^{+}(\mathbf{s})\phi^{+}(\mathbf{s})\}WP\emph{\qquad U4}\nonumber \\
 &  & +g\int d\mathbf{s}\left(\frac{\delta}{\delta\psi(\mathbf{s})}\right)\left(\frac{\delta}{\delta\psi(\mathbf{s})}\right)\{\frac{1}{8}\phi(\mathbf{s})\phi(\mathbf{s})\}WP\emph{\qquad U8}\nonumber \\
 &  & -g\int d\mathbf{s}\left(\frac{\delta}{\delta\phi^{+}(\mathbf{s})}\right)\left(\frac{\delta}{\delta\psi(\mathbf{s})}\right)\{\frac{1}{4}\phi(\mathbf{s})\psi^{+}(\mathbf{s})\}WP\emph{\qquad U10}\nonumber \\
 &  & -g\int d\mathbf{s}\left(\frac{\delta}{\delta\phi^{+}(\mathbf{s})}\right)\left(\frac{\delta}{\delta\psi(\mathbf{s})}\right)\{\frac{1}{4}\phi(\mathbf{s})\psi^{+}(\mathbf{s})\}WP\emph{\qquad U11}\nonumber \\
 &  & -g\int d\mathbf{s}\left(\frac{\delta}{\delta\phi^{+}(\mathbf{s})}\right)\left(\frac{\delta}{\delta\psi(\mathbf{s})}\right)\{\frac{1}{4}\phi(\mathbf{s})\psi^{+}(\mathbf{s})\}WP\emph{\qquad U14}\nonumber \\
 &  & -g\int d\mathbf{s}\left(\frac{\delta}{\delta\phi^{+}(\mathbf{s})}\right)\left(\frac{\delta}{\delta\psi(\mathbf{s})}\right)\{\frac{1}{4}\phi(\mathbf{s})\psi^{+}(\mathbf{s})\}WP\emph{\qquad U15}\nonumber \\
 &  & +g\int d\mathbf{s}\left(\frac{\delta}{\delta\phi^{+}(\mathbf{s})}\right)\left(\frac{\delta}{\delta\phi^{+}(\mathbf{s})}\right)\{\frac{1}{2}\psi^{+}(\mathbf{s})\psi^{+}(\mathbf{s})\}WP\emph{\qquad U17}\nonumber \\
 &  & -g\int d\mathbf{s}\left(\frac{\delta}{\delta\phi^{+}(\mathbf{s})}\right)\left(\frac{\delta}{\delta\psi(\mathbf{s})}\right)\{\psi(\mathbf{s})\phi^{+}(\mathbf{s})\}WP\emph{\qquad U24.1}\nonumber \\
 &  & -g\int d\mathbf{s}\left(\frac{\delta}{\delta\psi^{+}(\mathbf{s})}\right)\left(\frac{\delta}{\delta\psi(\mathbf{s})}\right)\{\frac{1}{2}\phi(\mathbf{s})\phi^{+}(\mathbf{s})\}WP\emph{\qquad U26}\nonumber \\
 &  & +g\int d\mathbf{s}\left(\frac{\delta}{\delta\psi^{+}(\mathbf{s})}\right)\left(\frac{\delta}{\delta\phi^{+}(\mathbf{s})}\right)\{\psi^{+}(\mathbf{s})\phi^{+}(\mathbf{s})\}WP\emph{\qquad U27}\nonumber \\
 & = & +g\int d\mathbf{s}\left(\frac{\delta}{\delta\psi^{+}(\mathbf{s})}\right)\left(\frac{\delta}{\delta\psi^{+}(\mathbf{s})}\right)\{\frac{1}{8}\phi^{+}(\mathbf{s})\phi^{+}(\mathbf{s})\}WP\emph{\qquad U4}\nonumber \\
 &  & -g\int d\mathbf{s}\left(\frac{\delta}{\delta\psi(\mathbf{s})}\right)\left(\frac{\delta}{\delta\psi^{+}(\mathbf{s})}\right)\{\frac{1}{2}\phi(\mathbf{s})\phi^{+}(\mathbf{s})\}WP\emph{\qquad U26}\nonumber \\
 &  & +g\int d\mathbf{s}\left(\frac{\delta}{\delta\psi(\mathbf{s})}\right)\left(\frac{\delta}{\delta\psi(\mathbf{s})}\right)\{\frac{1}{8}\phi(\mathbf{s})\phi(\mathbf{s})\}WP\emph{\qquad U8}\nonumber \\
 &  & -g\int d\mathbf{s}\left(\frac{\delta}{\delta\psi(\mathbf{s})}\right)\left(\frac{\delta}{\delta\phi^{+}(\mathbf{s})}\right)\{\phi(\mathbf{s})\psi^{+}(\mathbf{s})+\psi(\mathbf{s})\phi^{+}(\mathbf{s})\}WP\nonumber \\
 &  & \emph{\qquad U10,U11,U14,U15,U24.1}\nonumber \\
 &  & +g\int d\mathbf{s}\left(\frac{\delta}{\delta\psi^{+}(\mathbf{s})}\right)\left(\frac{\delta}{\delta\phi^{+}(\mathbf{s})}\right)\{\psi^{+}(\mathbf{s})\phi^{+}(\mathbf{s})\}WP\emph{\qquad U27}\nonumber \\
 &  & +g\int d\mathbf{s}\left(\frac{\delta}{\delta\phi^{+}(\mathbf{s})}\right)\left(\frac{\delta}{\delta\phi^{+}(\mathbf{s})}\right)\{\frac{1}{2}\psi^{+}(\mathbf{s})\psi^{+}(\mathbf{s})\}WP\emph{\qquad U17}\nonumber \\
 &  & \,\label{eq:RhoV2Result4}\end{eqnarray}
\begin{eqnarray}
 &  & WP_{3}\nonumber \\
 & = & -g\int d\mathbf{s}\left(\frac{\delta}{\delta\phi^{+}(\mathbf{s})}\right)\left(\frac{\delta}{\delta\psi(\mathbf{s})}\right)\left(\frac{\delta}{\delta\psi(\mathbf{s})}\right)\{\frac{1}{8}\phi(\mathbf{s})\}WP\emph{\qquad U12}\nonumber \\
 &  & -g\int d\mathbf{s}\left(\frac{\delta}{\delta\phi^{+}(\mathbf{s})}\right)\left(\frac{\delta}{\delta\psi(\mathbf{s})}\right)\left(\frac{\delta}{\delta\psi(\mathbf{s})}\right)\{\frac{1}{8}\phi(\mathbf{s})\}WP\emph{\qquad U16}\nonumber \\
 &  & +g\int d\mathbf{s}\left(\frac{\delta}{\delta\phi^{+}(\mathbf{s})}\right)\left(\frac{\delta}{\delta\phi^{+}(\mathbf{s})}\right)\left(\frac{\delta}{\delta\psi(\mathbf{s})}\right)\{\frac{1}{4}\psi^{+}(\mathbf{s})\}WP\emph{\qquad U18}\nonumber \\
 &  & +g\int d\mathbf{s}\left(\frac{\delta}{\delta\phi^{+}(\mathbf{s})}\right)\left(\frac{\delta}{\delta\phi^{+}(\mathbf{s})}\right)\left(\frac{\delta}{\delta\psi(\mathbf{s})}\right)\{\frac{1}{4}\psi^{+}(\mathbf{s})\}WP\emph{\qquad U19}\nonumber \\
 &  & +g\int d\mathbf{s}\left(\frac{\delta}{\delta\psi^{+}(\mathbf{s})}\right)\left(\frac{\delta}{\delta\phi^{+}(\mathbf{s})}\right)\left(\frac{\delta}{\delta\psi(\mathbf{s})}\right)\{\frac{1}{2}\phi^{+}(\mathbf{s})\}WP\emph{\qquad U28}\nonumber \\
 & = & -g\int d\mathbf{s}\left(\frac{\delta}{\delta\psi(\mathbf{s})}\right)\left(\frac{\delta}{\delta\psi(\mathbf{s})}\right)\left(\frac{\delta}{\delta\phi^{+}(\mathbf{s})}\right)\{\frac{1}{4}\phi(\mathbf{s})\}WP\emph{\qquad U12,U16}\nonumber \\
 &  & +g\int d\mathbf{s}\left(\frac{\delta}{\delta\psi(\mathbf{s})}\right)\left(\frac{\delta}{\delta\psi^{+}(\mathbf{s})}\right)\left(\frac{\delta}{\delta\phi^{+}(\mathbf{s})}\right)\{\frac{1}{2}\phi^{+}(\mathbf{s})\}WP\emph{\qquad U28}\nonumber \\
 &  & +g\int d\mathbf{s}\left(\frac{\delta}{\delta\psi(\mathbf{s})}\right)\left(\frac{\delta}{\delta\phi^{+}(\mathbf{s})}\right)\left(\frac{\delta}{\delta\phi^{+}(\mathbf{s})}\right)\{\frac{1}{2}\psi^{+}(\mathbf{s})\}WP\emph{\qquad U18,U19}\nonumber \\
 &  & \,\label{eq:RhoV2Result5}\end{eqnarray}
\begin{eqnarray}
 &  & WP_{4}\nonumber \\
 & = & +g\int d\mathbf{s}\left(\frac{\delta}{\delta\psi(\mathbf{s})}\right)\left(\frac{\delta}{\delta\psi(\mathbf{s})}\right)\left(\frac{\delta}{\delta\phi^{+}(\mathbf{s})}\right)\left(\frac{\delta}{\delta\phi^{+}(\mathbf{s})}\right)\{\frac{1}{8}\}WP\emph{\qquad U20}\nonumber \\
 &  & \,\label{eq:RhoV2Result6}\end{eqnarray}

Now if \begin{eqnarray}
\widehat{\rho} & \rightarrow & [\widehat{V}_{2},\,\widehat{\rho}]\nonumber \\
 & = & [\frac{g}{2}\int d\mathbf{s(}\widehat{\Psi}_{NC}(\mathbf{s})^{\dagger}\widehat{\Psi}_{NC}(\mathbf{s})^{\dagger}\widehat{\Psi}_{C}(\mathbf{s})\widehat{\Psi}_{C}(\mathbf{s})+\widehat{\Psi}_{C}(\mathbf{s})^{\dagger}\widehat{\Psi}_{C}(\mathbf{s})^{\dagger}\widehat{\Psi}_{NC}(\mathbf{s})\widehat{\Psi}_{NC}(\mathbf{s})),\widehat{\rho}]\nonumber \\
 &  & +[\frac{g}{2}\int d\mathbf{s(}4\widehat{\Psi}_{NC}(\mathbf{s})^{\dagger}\widehat{\Psi}_{C}(\mathbf{s})^{\dagger}\widehat{\Psi}_{NC}(\mathbf{s})\widehat{\Psi}_{C}(\mathbf{s})),\widehat{\rho}]\nonumber \\
 &  & \,\end{eqnarray}
then \begin{eqnarray}
 &  & WP[\psi(\mathbf{r}),\psi^{+}(\mathbf{r}),\phi(\mathbf{r}),\phi^{+}(\mathbf{r})]\nonumber \\
 & \rightarrow & WP_{T}^{0}+WP_{T}^{1}+WP_{T}^{2}+WP_{T}^{3}+WP_{T}^{4}\label{Eq.V2CommRhoResult1}\end{eqnarray}
where the $WP_{T}^{n}$ are obtained by subtracting the results for
$\widehat{\rho}\widehat{\, V}_{2}$ from those for $\widehat{V}_{2}\,\widehat{\rho}.$
We find that\begin{eqnarray}
 &  & WP_{T}^{0}\nonumber \\
 & = & g\int d\mathbf{s\{}\frac{1}{2}\phi^{+}(\mathbf{s})\phi^{+}(\mathbf{s})\psi(\mathbf{s})\psi(\mathbf{s})\}WP\emph{\qquad T1}\nonumber \\
 &  & -g\int d\mathbf{s\{}\frac{1}{2}\psi(\mathbf{s})\psi(\mathbf{s})\phi^{+}(\mathbf{s})\phi^{+}(\mathbf{s})\}WP\emph{\qquad U1}\nonumber \\
 &  & +g\int d\mathbf{s\{}\frac{1}{2}\psi^{+}(\mathbf{s})\psi^{+}(\mathbf{s})\phi(\mathbf{s})\phi(\mathbf{s})\}WP\emph{\qquad T17}\nonumber \\
 &  & -g\int d\mathbf{s\{}\frac{1}{2}\phi(\mathbf{s})\phi(\mathbf{s})\psi^{+}(\mathbf{s})\psi^{+}(\mathbf{s})\}WP\emph{\qquad U5}\nonumber \\
 &  & +g\int d\mathbf{s\{}2\phi^{+}(\mathbf{s})\psi^{+}(\mathbf{s})\phi(\mathbf{s})\psi(\mathbf{s})\}WP\emph{\qquad T21}\nonumber \\
 &  & -g\int d\mathbf{s\{}2\psi(\mathbf{s})\phi(\mathbf{s})\psi^{+}(\mathbf{s})\phi^{+}(\mathbf{s})\}WP\emph{\qquad U21}\nonumber \\
 &  & -g\int d\mathbf{s}\left\{ \phi^{+}(\mathbf{s})\delta_{C}(\mathbf{s,s})\phi(\mathbf{s})\right\} WP\emph{\qquad T22.2}\nonumber \\
 &  & +g\int d\mathbf{s\{}\delta_{C}(\mathbf{s,s})\phi(\mathbf{s})\phi^{+}(\mathbf{s})\}WP\emph{\qquad U22.2}\nonumber \\
 & = & 0\label{Eq.V2CommRhoResult2}\end{eqnarray}
\begin{eqnarray*}
 &  & WP_{T}^{1}\\
 & = & +g\int d\mathbf{s}\left(\frac{\delta}{\delta\psi^{+}(\mathbf{s})}\right)\{\frac{1}{2}\phi^{+}(\mathbf{s})\phi^{+}(\mathbf{s})\psi(\mathbf{s})\}WP\emph{\qquad T2,T3}\\
 &  & +g\int d\mathbf{s}\left(\frac{\delta}{\delta\psi^{+}(\mathbf{s})}\right)\{\frac{1}{2}\psi(\mathbf{s})\phi^{+}(\mathbf{s})\phi^{+}(\mathbf{s})\}WP\emph{\qquad U2,U3}\\
 &  & +g\int d\mathbf{s}\left\{ \left(\frac{\delta}{\delta\psi^{+}(\mathbf{s})}\right)\{\phi^{+}(\mathbf{s})\psi^{+}(\mathbf{s})\phi(\mathbf{s})\}\right\} WP\emph{\qquad T22.1}\\
 &  & +g\int d\mathbf{s}\left(\frac{\delta}{\delta\psi^{+}(\mathbf{s})}\right)\{\phi(\mathbf{s})\psi^{+}(\mathbf{s})\phi^{+}(\mathbf{s})\}WP\emph{\qquad U25}\\
 &  & -g\int d\mathbf{s}\left(\frac{\delta}{\delta\psi(\mathbf{s})}\right)\{\frac{1}{2}\psi^{+}(\mathbf{s})\phi(\mathbf{s})\phi(\mathbf{s})\}WP\emph{\qquad T18,T19}\\
 &  & -g\int d\mathbf{s}\left(\frac{\delta}{\delta\psi(\mathbf{s})}\right)\{\frac{1}{2}\phi(\mathbf{s})\phi(\mathbf{s})\psi^{+}(\mathbf{s})\}WP\emph{\qquad U6,U7}\\
 &  & -g\int d\mathbf{s}\left(\frac{\delta}{\delta\psi(\mathbf{s})}\right)\{\phi^{+}(\mathbf{s})\phi(\mathbf{s})\psi(\mathbf{s})\}WP\emph{\qquad T23}\\
 &  & -g\int d\mathbf{s}\left(\frac{\delta}{\delta\psi(\mathbf{s})}\right)\{\psi(\mathbf{s})\phi(\mathbf{s})\phi^{+}(\mathbf{s})\}WP\emph{\qquad U22.1}\\
 &  & -g\int d\mathbf{s}\left(\frac{\delta}{\delta\phi(\mathbf{s})}\right)\{2\psi^{+}(\mathbf{s})\phi(\mathbf{s})\psi(\mathbf{s})\}WP\emph{\qquad T25}\\
 &  & +g\int d\mathbf{s}\left(\frac{\delta}{\delta\phi^{+}(\mathbf{s})}\right)\{2\psi(\mathbf{s})\psi^{+}(\mathbf{s})\phi^{+}(\mathbf{s})\}WP\emph{\qquad U23}\\
 &  & -g\int d\mathbf{s}\left(\frac{\delta}{\delta\phi(\mathbf{s})}\right)\{\phi^{+}(\mathbf{s})\psi(\mathbf{s})\psi(\mathbf{s})\}WP\emph{\qquad T5,T9}\\
 &  & +g\int d\mathbf{s}\left(\frac{\delta}{\delta\phi^{+}(\mathbf{s})}\right)\{\phi(\mathbf{s})\psi^{+}(\mathbf{s})\psi^{+}(\mathbf{s})\}WP\emph{\qquad U9,U13}\\
 &  & +g\int d\mathbf{s}\left\{ \left(\frac{\delta}{\delta\phi(\mathbf{s})}\right)\{\delta_{C}(\mathbf{s,s})\phi(\mathbf{s})\}\right\} WP\emph{\qquad T26.2}\\
 &  & -g\int d\mathbf{s}\left(\frac{\delta}{\delta\phi^{+}(\mathbf{s})}\right)\{\delta_{C}(\mathbf{s,s})\phi^{+}(\mathbf{s})\}WP\emph{\qquad U24.2}\end{eqnarray*}
After collecting together similar terms we get\begin{eqnarray}
 &  & WP_{T}^{1}\nonumber \\
 & = & +g\int d\mathbf{s}\left(\frac{\delta}{\delta\psi^{+}(\mathbf{s})}\right)\{\phi^{+}(\mathbf{s})\phi^{+}(\mathbf{s})\psi(\mathbf{s})\}WP\emph{\qquad T2,T3,U2,U3}\nonumber \\
 &  & +g\int d\mathbf{s}\left(\frac{\delta}{\delta\psi^{+}(\mathbf{s})}\right)\{2\phi^{+}(\mathbf{s})\psi^{+}(\mathbf{s})\phi(\mathbf{s})\}WP\emph{\qquad T22.1,U25}\nonumber \\
 &  & -g\int d\mathbf{s}\left(\frac{\delta}{\delta\psi(\mathbf{s})}\right)\{\psi^{+}(\mathbf{s})\phi(\mathbf{s})\phi(\mathbf{s})\}WP\emph{\qquad T18,T19,U6,U7}\nonumber \\
 &  & -g\int d\mathbf{s}\left(\frac{\delta}{\delta\psi(\mathbf{s})}\right)\{2\phi^{+}(\mathbf{s})\phi(\mathbf{s})\psi(\mathbf{s})\}WP\emph{\qquad T23,U22.1}\nonumber \\
 &  & +g\int d\mathbf{s}\left(\frac{\delta}{\delta\phi^{+}(\mathbf{s})}\right)\{2\psi(\mathbf{s})\psi^{+}(\mathbf{s})\phi^{+}(\mathbf{s})\}WP\emph{\qquad U23}\nonumber \\
 &  & +g\int d\mathbf{s}\left(\frac{\delta}{\delta\phi^{+}(\mathbf{s})}\right)\{\phi(\mathbf{s})\psi^{+}(\mathbf{s})\psi^{+}(\mathbf{s})\}WP\emph{\qquad U9,U13}\nonumber \\
 &  & -g\int d\mathbf{s}\left(\frac{\delta}{\delta\phi^{+}(\mathbf{s})}\right)\{\delta_{C}(\mathbf{s,s})\phi^{+}(\mathbf{s})\}WP\emph{\qquad U24.2}\nonumber \\
 &  & -g\int d\mathbf{s}\left(\frac{\delta}{\delta\phi(\mathbf{s})}\right)\{2\psi^{+}(\mathbf{s})\phi(\mathbf{s})\psi(\mathbf{s})\}WP\emph{\qquad T25}\nonumber \\
 &  & -g\int d\mathbf{s}\left(\frac{\delta}{\delta\phi(\mathbf{s})}\right)\{\phi^{+}(\mathbf{s})\psi(\mathbf{s})\psi(\mathbf{s})\}WP\emph{\qquad T5,T9}\nonumber \\
 &  & +g\int d\mathbf{s}\left(\frac{\delta}{\delta\phi(\mathbf{s})}\right)\{\delta_{C}(\mathbf{s,s})\phi(\mathbf{s})\}WP\emph{\qquad T26.2}\nonumber \\
 & = & +g\int d\mathbf{s}\left(\frac{\delta}{\delta\psi^{+}(\mathbf{s})}\right)\{[\phi^{+}(\mathbf{s})\psi(\mathbf{s})+2\psi^{+}(\mathbf{s})\phi(\mathbf{s})]\phi^{+}(\mathbf{s})\}WP\nonumber \\
 &  & -g\int d\mathbf{s}\left(\frac{\delta}{\delta\psi(\mathbf{s})}\right)\{[\phi(\mathbf{s})\psi^{+}(\mathbf{s})+2\psi(\mathbf{s})\phi^{+}(\mathbf{s})]\phi(\mathbf{s})\}WP\nonumber \\
 &  & +g\int d\mathbf{s}\left(\frac{\delta}{\delta\phi^{+}(\mathbf{s})}\right)\{[2\psi(\mathbf{s})\psi^{+}(\mathbf{s})-\delta_{C}(\mathbf{s,s})]\phi^{+}(\mathbf{s})+[\psi^{+}(\mathbf{s})\psi^{+}(\mathbf{s})]\phi(\mathbf{s})\}WP\nonumber \\
 &  & -g\int d\mathbf{s}\left(\frac{\delta}{\delta\phi(\mathbf{s})}\right)\{[2\psi^{+}(\mathbf{s})\psi(\mathbf{s})-\delta_{C}(\mathbf{s,s})]\phi(\mathbf{s})+[\psi(\mathbf{s})\psi(\mathbf{s})]\phi^{+}(\mathbf{s})\}WP\nonumber \\
 &  & \,\label{eq:V2CommRhoResult3}\end{eqnarray}

\begin{eqnarray}
 &  & WP_{T}^{2}\nonumber \\
 & = & +g\int d\mathbf{s}\left(\frac{\delta}{\delta\psi^{+}(\mathbf{s})}\right)\left(\frac{\delta}{\delta\psi^{+}(\mathbf{s})}\right)\{\frac{1}{8}\phi^{+}(\mathbf{s})\phi^{+}(\mathbf{s})\}WP\emph{\qquad T4Can1}\nonumber \\
 &  & -g\int d\mathbf{s}\left(\frac{\delta}{\delta\psi^{+}(\mathbf{s})}\right)\left(\frac{\delta}{\delta\psi^{+}(\mathbf{s})}\right)\{\frac{1}{8}\phi^{+}(\mathbf{s})\phi^{+}(\mathbf{s})\}WP\emph{\qquad U4Can1}\nonumber \\
 &  & -g\int d\mathbf{s}\left(\frac{\delta}{\delta\psi(\mathbf{s})}\right)\left(\frac{\delta}{\delta\psi^{+}(\mathbf{s})}\right)\{\frac{1}{2}\phi^{+}(\mathbf{s})\phi(\mathbf{s})\}WP\emph{\qquad T24Can2}\nonumber \\
 &  & +g\int d\mathbf{s}\left(\frac{\delta}{\delta\psi(\mathbf{s})}\right)\left(\frac{\delta}{\delta\psi^{+}(\mathbf{s})}\right)\{\frac{1}{2}\phi(\mathbf{s})\phi^{+}(\mathbf{s})\}WP\emph{\qquad U26Can2}\nonumber \\
 &  & +g\int d\mathbf{s}\left(\frac{\delta}{\delta\psi(\mathbf{s})}\right)\left(\frac{\delta}{\delta\psi(\mathbf{s})}\right)\{\frac{1}{8}\phi(\mathbf{s})\phi(\mathbf{s})\}WP\emph{\qquad T20Can3}\nonumber \\
 &  & -g\int d\mathbf{s}\left(\frac{\delta}{\delta\psi(\mathbf{s})}\right)\left(\frac{\delta}{\delta\psi(\mathbf{s})}\right)\{\frac{1}{8}\phi(\mathbf{s})\phi(\mathbf{s})\}WP\emph{\qquad U8Can3}\nonumber \\
 &  & -g\int d\mathbf{s}\left(\frac{\delta}{\delta\psi^{+}(\mathbf{s})}\right)\left(\frac{\delta}{\delta\phi(\mathbf{s})}\right)\{\phi^{+}(\mathbf{s})\psi(\mathbf{s})+\psi^{+}(\mathbf{s})\phi(\mathbf{s})\}WP\nonumber \\
 &  & \emph{\qquad T6,T7,T10,T11,T26.1}\nonumber \\
 &  & +g\int d\mathbf{s}\left(\frac{\delta}{\delta\psi(\mathbf{s})}\right)\left(\frac{\delta}{\delta\phi^{+}(\mathbf{s})}\right)\{\phi(\mathbf{s})\psi^{+}(\mathbf{s})+\psi(\mathbf{s})\phi^{+}(\mathbf{s})\}WP\nonumber \\
 &  & \emph{\qquad U10,U11,U14,U15,U24.1}\nonumber \\
 &  & +g\int d\mathbf{s}\left(\frac{\delta}{\delta\psi(\mathbf{s})}\right)\left(\frac{\delta}{\delta\phi(\mathbf{s})}\right)\{\phi(\mathbf{s})\psi(\mathbf{s})\}WP\emph{\qquad T27}\nonumber \\
 &  & -g\int d\mathbf{s}\left(\frac{\delta}{\delta\psi^{+}(\mathbf{s})}\right)\left(\frac{\delta}{\delta\phi^{+}(\mathbf{s})}\right)\{\psi^{+}(\mathbf{s})\phi^{+}(\mathbf{s})\}WP\emph{\qquad U27}\nonumber \\
 &  & +g\int d\mathbf{s}\left(\frac{\delta}{\delta\phi(\mathbf{s})}\right)\left(\frac{\delta}{\delta\phi(\mathbf{s})}\right)\{\frac{1}{2}\psi(\mathbf{s})\psi(\mathbf{s})\}WP\emph{\qquad T13}\nonumber \\
 &  & -g\int d\mathbf{s}\left(\frac{\delta}{\delta\phi^{+}(\mathbf{s})}\right)\left(\frac{\delta}{\delta\phi^{+}(\mathbf{s})}\right)\{\frac{1}{2}\psi^{+}(\mathbf{s})\psi^{+}(\mathbf{s})\}WP\emph{\qquad U17}\nonumber \\
 & = & -g\int d\mathbf{s}\left(\frac{\delta}{\delta\psi^{+}(\mathbf{s})}\right)\left(\frac{\delta}{\delta\phi(\mathbf{s})}\right)\{\phi^{+}(\mathbf{s})\psi(\mathbf{s})+\psi^{+}(\mathbf{s})\phi(\mathbf{s})\}WP\nonumber \\
 &  & +g\int d\mathbf{s}\left(\frac{\delta}{\delta\psi(\mathbf{s})}\right)\left(\frac{\delta}{\delta\phi^{+}(\mathbf{s})}\right)\{\phi(\mathbf{s})\psi^{+}(\mathbf{s})+\psi(\mathbf{s})\phi^{+}(\mathbf{s})\}WP\nonumber \\
 &  & +g\int d\mathbf{s}\left(\frac{\delta}{\delta\psi(\mathbf{s})}\right)\left(\frac{\delta}{\delta\phi(\mathbf{s})}\right)\{\phi(\mathbf{s})\psi(\mathbf{s})\}WP\nonumber \\
 &  & -g\int d\mathbf{s}\left(\frac{\delta}{\delta\psi^{+}(\mathbf{s})}\right)\left(\frac{\delta}{\delta\phi^{+}(\mathbf{s})}\right)\{\phi^{+}(\mathbf{s})\psi^{+}(\mathbf{s})\}WP\nonumber \\
 &  & +g\int d\mathbf{s}\left(\frac{\delta}{\delta\phi(\mathbf{s})}\right)\left(\frac{\delta}{\delta\phi(\mathbf{s})}\right)\{\frac{1}{2}\psi(\mathbf{s})\psi(\mathbf{s})\}WP\nonumber \\
 &  & -g\int d\mathbf{s}\left(\frac{\delta}{\delta\phi^{+}(\mathbf{s})}\right)\left(\frac{\delta}{\delta\phi^{+}(\mathbf{s})}\right)\{\frac{1}{2}\psi^{+}(\mathbf{s})\psi^{+}(\mathbf{s})\}WP\nonumber \\
 &  & \,\label{eq:V2CommRhoResult4}\end{eqnarray}
\begin{eqnarray}
 &  & WP_{T}^{3}\nonumber \\
 & = & -g\int d\mathbf{s}\left(\frac{\delta}{\delta\psi^{+}(\mathbf{s})}\right)\left(\frac{\delta}{\delta\psi^{+}(\mathbf{s})}\right)\left(\frac{\delta}{\delta\phi(\mathbf{s})}\right)\{\frac{1}{4}\phi^{+}(\mathbf{s})\}WP\emph{\qquad T8,T12}\nonumber \\
 &  & +g\int d\mathbf{s}\left(\frac{\delta}{\delta\psi(\mathbf{s})}\right)\left(\frac{\delta}{\delta\psi(\mathbf{s})}\right)\left(\frac{\delta}{\delta\phi^{+}(\mathbf{s})}\right)\{\frac{1}{4}\phi(\mathbf{s})\}WP\emph{\qquad U12,U16}\nonumber \\
 &  & +g\int d\mathbf{s}\left(\frac{\delta}{\delta\psi(\mathbf{s})}\right)\left(\frac{\delta}{\delta\psi^{+}(\mathbf{s})}\right)\left(\frac{\delta}{\delta\phi(\mathbf{s})}\right)\{\frac{1}{2}\phi(\mathbf{s})\}WP\emph{\qquad T28}\nonumber \\
 &  & -g\int d\mathbf{s}\left(\frac{\delta}{\delta\psi(\mathbf{s})}\right)\left(\frac{\delta}{\delta\psi^{+}(\mathbf{s})}\right)\left(\frac{\delta}{\delta\phi^{+}(\mathbf{s})}\right)\{\frac{1}{2}\phi^{+}(\mathbf{s})\}WP\emph{\qquad U28}\nonumber \\
 &  & +g\int d\mathbf{s}\left(\frac{\delta}{\delta\psi^{+}(\mathbf{s})}\right)\left(\frac{\delta}{\delta\phi(\mathbf{s})}\right)\left(\frac{\delta}{\delta\phi(\mathbf{s})}\right)\{\frac{1}{2}\psi(\mathbf{s})\}WP\emph{\qquad T14,T15}\nonumber \\
 &  & -g\int d\mathbf{s}\left(\frac{\delta}{\delta\psi(\mathbf{s})}\right)\left(\frac{\delta}{\delta\phi^{+}(\mathbf{s})}\right)\left(\frac{\delta}{\delta\phi^{+}(\mathbf{s})}\right)\{\frac{1}{2}\psi^{+}(\mathbf{s})\}WP\emph{\qquad U18,U19}\nonumber \\
 &  & \,\label{eq:V2CommRhoResult5}\end{eqnarray}
\begin{eqnarray}
 &  & WP_{T}^{4}\nonumber \\
 & = & +g\int d\mathbf{s}\left(\frac{\delta}{\delta\psi^{+}(\mathbf{s})}\right)\left(\frac{\delta}{\delta\psi^{+}(\mathbf{s})}\right)\left(\frac{\delta}{\delta\phi(\mathbf{s})}\right)\left(\frac{\delta}{\delta\phi(\mathbf{s})}\right)\{\frac{1}{8}\}WP\emph{\qquad T16}\nonumber \\
 &  & -g\int d\mathbf{s}\left(\frac{\delta}{\delta\psi(\mathbf{s})}\right)\left(\frac{\delta}{\delta\psi(\mathbf{s})}\right)\left(\frac{\delta}{\delta\phi^{+}(\mathbf{s})}\right)\left(\frac{\delta}{\delta\phi^{+}(\mathbf{s})}\right)\{\frac{1}{8}\}WP\emph{\qquad U20}\nonumber \\
 &  & \,\label{eq:V2CommRhoResult6}\end{eqnarray}

Thus we see that the $\widehat{\, V}_{2}$ term produces functional
derivatives of orders one, two, three and four. We may write the contributions
to the functional Fokker-Planck equation in the form\begin{eqnarray}
 &  & \left(\frac{\partial}{\partial t}WP[\psi,\psi^{+},\phi,\phi^{+}]\right)_{V2}\nonumber \\
 & = & \left(\frac{\partial}{\partial t}WP[\psi,\psi^{+},\phi,\phi^{+}]\right)_{V2}^{1}+\left(\frac{\partial}{\partial t}WP[\psi,\psi^{+},\phi,\phi^{+}]\right)_{V2}^{2}\nonumber \\
 &  & +\left(\frac{\partial}{\partial t}WP[\psi,\psi^{+},\phi,\phi^{+}]\right)_{V2}^{3}+\left(\frac{\partial}{\partial t}WP[\psi,\psi^{+},\phi,\phi^{+}]\right)_{V2}^{4}\label{Eq.FFPEConNonCondV2Intn}\end{eqnarray}
where on reverting to the original notation\begin{eqnarray}
 &  & \left(\frac{\partial}{\partial t}P[\underrightarrow{\psi}(\mathbf{r}),\underrightarrow{\psi^{\ast}}(\mathbf{r})]\right)_{V2}^{1}\nonumber \\
 & = & \frac{-i}{\hbar}\left\{ +g\int d\mathbf{s}\left(\frac{\delta}{\delta\psi_{C}^{+}(\mathbf{s})}\right)\{[\psi_{NC}^{+}(\mathbf{s})\psi_{C}(\mathbf{s})+2\psi_{C}^{+}(\mathbf{s})\psi_{NC}(\mathbf{s})]\psi_{NC}^{+}(\mathbf{s})\}P[\underrightarrow{\psi}(\mathbf{r}),\underrightarrow{\psi^{\ast}}(\mathbf{r})]\right\} \nonumber \\
 &  & \frac{-i}{\hbar}\left\{ -g\int d\mathbf{s}\left(\frac{\delta}{\delta\psi_{C}(\mathbf{s})}\right)\{[\psi_{NC}(\mathbf{s})\psi_{C}^{+}(\mathbf{s})+2\psi_{C}(\mathbf{s})\psi_{NC}^{+}(\mathbf{s})]\psi_{NC}(\mathbf{s})\}P[\underrightarrow{\psi}(\mathbf{r}),\underrightarrow{\psi^{\ast}}(\mathbf{r})]\right\} \nonumber \\
 &  & \frac{-i}{\hbar}\left\{ +g\int d\mathbf{s}\left\{ \left(\frac{\delta}{\delta\psi_{NC}^{+}(\mathbf{s})}\right)\{\psi_{NC}(\mathbf{s})\psi_{C}^{+}(\mathbf{s})\psi_{C}^{+}(\mathbf{s})\}\right\} P[\underrightarrow{\psi}(\mathbf{r}),\underrightarrow{\psi^{\ast}}(\mathbf{r})]\right\} \nonumber \\
 &  & \frac{-i}{\hbar}\left\{ +g\int d\mathbf{s}\left\{ \left(\frac{\delta}{\delta\psi_{NC}^{+}(\mathbf{s})}\right)\{[2\psi_{C}(\mathbf{s})\psi_{C}^{+}(\mathbf{s})-\delta_{C}(\mathbf{s},\mathbf{s})]\psi_{NC}^{+}(\mathbf{s})\}\right\} P[\underrightarrow{\psi}(\mathbf{r}),\underrightarrow{\psi^{\ast}}(\mathbf{r})]\right\} \nonumber \\
 &  & \frac{-i}{\hbar}\left\{ -g\int d\mathbf{s}\left\{ \left(\frac{\delta}{\delta\psi_{NC}(\mathbf{s})}\right)\{\psi_{NC}^{+}(\mathbf{s})\psi_{C}(\mathbf{s})\psi_{C}(\mathbf{s})\}\right\} P[\underrightarrow{\psi}(\mathbf{r}),\underrightarrow{\psi^{\ast}}(\mathbf{r})]\right\} \nonumber \\
 &  & \frac{-i}{\hbar}\left\{ -g\int d\mathbf{s}\left\{ \left(\frac{\delta}{\delta\psi_{NC}(\mathbf{s})}\right)\{[2\psi_{C}^{+}(\mathbf{s})\psi_{C}(\mathbf{s})-\delta_{C}(\mathbf{s,s})]\psi_{NC}(\mathbf{s})\}\right\} P[\underrightarrow{\psi}(\mathbf{r}),\underrightarrow{\psi^{\ast}}(\mathbf{r})]\right\} \nonumber \\
 &  & \,\label{eq:FFPEConNonCondV2Linear}\end{eqnarray}
\begin{eqnarray}
 &  & \left(\frac{\partial}{\partial t}P[\underrightarrow{\psi}(\mathbf{r}),\underrightarrow{\psi^{\ast}}(\mathbf{r})]\right)_{V2}^{2}\nonumber \\
 & = & \frac{-i}{\hbar}\left\{ -g\int d\mathbf{s}\left\{ \left(\frac{\delta}{\delta\psi_{C}^{+}(\mathbf{s})}\right)\left(\frac{\delta}{\delta\psi_{NC}(\mathbf{s})}\right)\{\psi_{NC}^{+}(\mathbf{s})\psi_{C}(\mathbf{s})+\psi_{C}^{+}(\mathbf{s})\psi_{NC}(\mathbf{s})\}\right\} P[\underrightarrow{\psi}(\mathbf{r}),\underrightarrow{\psi^{\ast}}(\mathbf{r})]\right\} \nonumber \\
 &  & \frac{-i}{\hbar}\left\{ +g\int d\mathbf{s}\left\{ \left(\frac{\delta}{\delta\psi_{C}(\mathbf{s})}\right)\left(\frac{\delta}{\delta\psi_{NC}^{+}(\mathbf{s})}\right)\{\psi_{NC}(\mathbf{s})\psi_{C}^{+}(\mathbf{s})+\psi_{C}(\mathbf{s})\psi_{NC}^{+}(\mathbf{s})\}\right\} P[\underrightarrow{\psi}(\mathbf{r}),\underrightarrow{\psi^{\ast}}(\mathbf{r})]\right\} \nonumber \\
 &  & \frac{-i}{\hbar}\left\{ +g\int d\mathbf{s}\left\{ \left(\frac{\delta}{\delta\psi_{C}(\mathbf{s})}\right)\left(\frac{\delta}{\delta\psi_{NC}(\mathbf{s})}\right)\{\psi_{NC}(\mathbf{s})\psi_{C}(\mathbf{s})\}\right\} P[\underrightarrow{\psi}(\mathbf{r}),\underrightarrow{\psi^{\ast}}(\mathbf{r})]\right\} \nonumber \\
 &  & \frac{-i}{\hbar}\left\{ -g\int d\mathbf{s}\left\{ \left(\frac{\delta}{\delta\psi_{C}^{+}(\mathbf{s})}\right)\left(\frac{\delta}{\delta\psi_{NC}^{+}(\mathbf{s})}\right)\{\psi_{C}^{+}(\mathbf{s})\psi_{NC}^{+}(\mathbf{s})\}\right\} P[\underrightarrow{\psi}(\mathbf{r}),\underrightarrow{\psi^{\ast}}(\mathbf{r})]\right\} \nonumber \\
 &  & \frac{-i}{\hbar}\left\{ +g\int d\mathbf{s}\left\{ \left(\frac{\delta}{\delta\psi_{NC}(\mathbf{s})}\right)\left(\frac{\delta}{\delta\psi_{NC}(\mathbf{s})}\right)\{\frac{1}{2}\psi_{C}(\mathbf{s})\psi_{C}(\mathbf{s})\}\right\} P[\underrightarrow{\psi}(\mathbf{r}),\underrightarrow{\psi^{\ast}}(\mathbf{r})]\right\} \nonumber \\
 &  & \frac{-i}{\hbar}\left\{ -g\int d\mathbf{s}\left\{ \left(\frac{\delta}{\delta\psi_{NC}^{+}(\mathbf{s})}\right)\left(\frac{\delta}{\delta\psi_{NC}^{+}(\mathbf{s})}\right)\{\frac{1}{2}\psi_{C}^{+}(\mathbf{s})\psi_{C}^{+}(\mathbf{s})\}\right\} P[\underrightarrow{\psi}(\mathbf{r}),\underrightarrow{\psi^{\ast}}(\mathbf{r})]\right\} \nonumber \\
 &  & \,\label{eq:FFPEConNonCondV2Quadratic}\end{eqnarray}
\begin{eqnarray}
 &  & \left(\frac{\partial}{\partial t}P[\underrightarrow{\psi}(\mathbf{r}),\underrightarrow{\psi^{\ast}}(\mathbf{r})]\right)_{V2}^{3}\nonumber \\
 & = & \frac{-i}{\hbar}\left\{ -g\int d\mathbf{s}\left\{ \left(\frac{\delta}{\delta\psi_{C}^{+}(\mathbf{s})}\right)\left(\frac{\delta}{\delta\psi_{C}^{+}(\mathbf{s})}\right)\left(\frac{\delta}{\delta\psi_{NC}(\mathbf{s})}\right)\{\frac{1}{4}\psi_{NC}^{+}(\mathbf{s})\}\right\} P[\underrightarrow{\psi}(\mathbf{r}),\underrightarrow{\psi^{\ast}}(\mathbf{r})]\right\} \nonumber \\
 &  & \frac{-i}{\hbar}\left\{ +g\int d\mathbf{s}\left\{ \left(\frac{\delta}{\delta\psi_{C}(\mathbf{s})}\right)\left(\frac{\delta}{\delta\psi_{C}(\mathbf{s})}\right)\left(\frac{\delta}{\delta\psi_{NC}^{+}(\mathbf{s})}\right)\{\frac{1}{4}\psi_{NC}(\mathbf{s})\}\right\} P[\underrightarrow{\psi}(\mathbf{r}),\underrightarrow{\psi^{\ast}}(\mathbf{r})]\right\} \nonumber \\
 &  & \frac{-i}{\hbar}\left\{ +g\int d\mathbf{s}\left\{ \left(\frac{\delta}{\delta\psi_{C}^{+}(\mathbf{s})}\right)\left(\frac{\delta}{\delta\psi_{NC}(\mathbf{s})}\right)\left(\frac{\delta}{\delta\psi_{NC}(\mathbf{s})}\right)\{\frac{1}{2}\psi_{C}(\mathbf{s})\}\right\} P[\underrightarrow{\psi}(\mathbf{r}),\underrightarrow{\psi^{\ast}}(\mathbf{r})]\right\} \nonumber \\
 &  & \frac{-i}{\hbar}\left\{ -g\int d\mathbf{s}\left\{ \left(\frac{\delta}{\delta\psi_{C}(\mathbf{s})}\right)\left(\frac{\delta}{\delta\psi_{NC}^{+}(\mathbf{s})}\right)\left(\frac{\delta}{\delta\psi_{NC}^{+}(\mathbf{s})}\right)\{\frac{1}{2}\psi_{C}^{+}(\mathbf{s})\}\right\} P[\underrightarrow{\psi}(\mathbf{r}),\underrightarrow{\psi^{\ast}}(\mathbf{r})]\right\} \nonumber \\
 &  & \frac{-i}{\hbar}\left\{ +g\int d\mathbf{s}\left\{ \left(\frac{\delta}{\delta\psi_{C}(\mathbf{s})}\right)\left(\frac{\delta}{\delta\psi_{C}^{+}(\mathbf{s})}\right)\left(\frac{\delta}{\delta\psi_{NC}(\mathbf{s})}\right)\{\frac{1}{2}\psi_{NC}(\mathbf{s})\}\right\} P[\underrightarrow{\psi}(\mathbf{r}),\underrightarrow{\psi^{\ast}}(\mathbf{r})]\right\} \nonumber \\
 &  & \frac{-i}{\hbar}\left\{ -g\int d\mathbf{s}\left\{ \left(\frac{\delta}{\delta\psi_{C}^{+}(\mathbf{s})}\right)\left(\frac{\delta}{\delta\psi_{C}(\mathbf{s})}\right)\left(\frac{\delta}{\delta\psi_{NC}^{+}(\mathbf{s})}\right)\{\frac{1}{2}\psi_{NC}^{+}(\mathbf{s})\}\right\} P[\underrightarrow{\psi}(\mathbf{r}),\underrightarrow{\psi^{\ast}}(\mathbf{r})]\right\} \nonumber \\
 &  & \,\label{eq:FFPEConNonCondV2Cubic}\end{eqnarray}
\begin{eqnarray}
 &  & \left(\frac{\partial}{\partial t}P[\underrightarrow{\psi}(\mathbf{r}),\underrightarrow{\psi^{\ast}}(\mathbf{r})]\right)_{V2}^{4}\nonumber \\
 & = & \frac{-i}{\hbar}\left\{ g\int d\mathbf{s}\left\{ \left(\frac{\delta}{\delta\psi_{C}^{+}(\mathbf{s})}\right)\left(\frac{\delta}{\delta\psi_{C}^{+}(\mathbf{s})}\right)\left(\frac{\delta}{\delta\psi_{NC}(\mathbf{s})}\right)\left(\frac{\delta}{\delta\psi_{NC}(\mathbf{s})}\right)\{\frac{1}{8}\}\right\} P[\underrightarrow{\psi}(\mathbf{r}),\underrightarrow{\psi^{\ast}}(\mathbf{r})]\right\} \nonumber \\
 &  & \frac{-i}{\hbar}\left\{ -g\int d\mathbf{s}\left\{ \left(\frac{\delta}{\delta\psi_{C}(\mathbf{s})}\right)\left(\frac{\delta}{\delta\psi_{C}(\mathbf{s})}\right)\left(\frac{\delta}{\delta\psi_{NC}^{+}(\mathbf{s})}\right)\left(\frac{\delta}{\delta\psi_{NC}^{+}(\mathbf{s})}\right)\{\frac{1}{8}\}\right\} P[\underrightarrow{\psi}(\mathbf{r}),\underrightarrow{\psi^{\ast}}(\mathbf{r})]\right\} \nonumber \\
 &  & \,\label{eq:FFPEConNonCondV2Quartic}\end{eqnarray}

\subsection{Condensate - Non-Condensate Interaction - Third Order in Non-Condendate}

For the Bogoliubov Hamiltonian for which we derive the functional
Fokker-Planck equation the term $\widehat{V}_{3}$ is discarded, but
for completeness we treat it here. The third order term in the interaction
between the condensate and the non-condensate is\begin{align}
\widehat{V}_{3} & =g\int d\mathbf{s(}\widehat{\Psi}_{NC}(\mathbf{s})^{\dagger}\widehat{\Psi}_{NC}(\mathbf{s})^{\dagger}\widehat{\Psi}_{NC}(\mathbf{s})\widehat{\Psi}_{C}(\mathbf{s}))+\widehat{\Psi}_{C}(\mathbf{s})^{\dagger}\widehat{\Psi}_{NC}(\mathbf{s})^{\dagger}\widehat{\Psi}_{NC}(\mathbf{s})\widehat{\Psi}_{NC}(\mathbf{s}))\nonumber \\
 & \,\end{align}
This term is due to the boson-boson interaction.

Now if \begin{eqnarray}
\widehat{\rho} & \rightarrow & \widehat{V}_{3}\,\widehat{\rho}\nonumber \\
 & = & g\int d\mathbf{s(}\widehat{\Psi}_{NC}(\mathbf{s})^{\dagger}\widehat{\Psi}_{NC}(\mathbf{s})^{\dagger}\widehat{\Psi}_{NC}(\mathbf{s})\widehat{\Psi}_{C}(\mathbf{s})+\widehat{\Psi}_{C}(\mathbf{s})^{\dagger}\widehat{\Psi}_{NC}(\mathbf{s})^{\dagger}\widehat{\Psi}_{NC}(\mathbf{s})\widehat{\Psi}_{NC}(\mathbf{s}))\widehat{\rho}\nonumber \\
 &  & \,\end{eqnarray}
then\begin{eqnarray}
 &  & WP[\psi(\mathbf{r}),\psi^{+}(\mathbf{r}),\phi(\mathbf{r}),\phi^{+}(\mathbf{r})]\nonumber \\
 & \rightarrow & g\int d\mathbf{s}\left(\phi^{+}(\mathbf{s})-\frac{\delta}{\delta\phi(\mathbf{s})}\right)\left(\phi^{+}(\mathbf{s})-\frac{\delta}{\delta\phi(\mathbf{s})}\right)\left(\phi(\mathbf{s})\right)\nonumber \\
 &  & \times\left(\psi(\mathbf{s})+\frac{1}{2}\frac{\delta}{\delta\psi^{+}(\mathbf{s})}\right)WP[\psi,\psi^{+},\phi,\phi^{+}]\nonumber \\
 &  & +g\int d\mathbf{s}\left(\psi^{+}(\mathbf{s})-\frac{1}{2}\frac{\delta}{\delta\psi(\mathbf{s})}\right)\left(\phi^{+}(\mathbf{s})-\frac{\delta}{\delta\phi(\mathbf{s})}\right)\left(\phi(\mathbf{s})\right)\nonumber \\
 &  & \times\left(\phi(\mathbf{s})\right)WP[\psi,\psi^{+},\phi,\phi^{+}]\nonumber \\
 &  & \,\end{eqnarray}

Expanding gives\begin{eqnarray}
 &  & WP[\psi(\mathbf{r}),\psi^{+}(\mathbf{r}),\phi(\mathbf{r}),\phi^{+}(\mathbf{r})]\nonumber \\
 & = & WP[\psi,\psi^{+},\phi,\phi^{+}]_{1-8}+WP[\psi,\psi^{+},\phi,\phi^{+}]_{9-12}\end{eqnarray}

where\begin{eqnarray}
 &  & WP[\psi(\mathbf{r}),\psi^{+}(\mathbf{r}),\phi(\mathbf{r}),\phi^{+}(\mathbf{r})]_{1-8}\nonumber \\
 & = & g\int d\mathbf{s}\left(\phi^{+}(\mathbf{s})\right)\left(\phi^{+}(\mathbf{s})\right)\left(\phi(\mathbf{s})\right)\left(\psi(\mathbf{s})\right)WP\nonumber \\
 &  & +g\int d\mathbf{s}\left(\phi^{+}(\mathbf{s})\right)\left(\phi^{+}(\mathbf{s})\right)\left(\phi(\mathbf{s})\right)\left(\frac{1}{2}\frac{\delta}{\delta\psi^{+}(\mathbf{s})}\right)WP\nonumber \\
 &  & +g\int d\mathbf{s}\left(\phi^{+}(\mathbf{s})\right)\left(-\frac{\delta}{\delta\phi(\mathbf{s})}\right)\left(\phi(\mathbf{s})\right)\left(\psi(\mathbf{s})\right)WP\nonumber \\
 &  & +g\int d\mathbf{s}\left(\phi^{+}(\mathbf{s})\right)\left(-\frac{\delta}{\delta\phi(\mathbf{s})}\right)\left(\phi(\mathbf{s})\right)\left(\frac{1}{2}\frac{\delta}{\delta\psi^{+}(\mathbf{s})}\right)WP\nonumber \\
 &  & +g\int d\mathbf{s}\left(-\frac{\delta}{\delta\phi(\mathbf{s})}\right)\left(\phi^{+}(\mathbf{s})\right)\left(\phi(\mathbf{s})\right)\left(\psi(\mathbf{s})\right)WP\nonumber \\
 &  & +g\int d\mathbf{s}\left(-\frac{\delta}{\delta\phi(\mathbf{s})}\right)\left(\phi^{+}(\mathbf{s})\right)\left(\phi(\mathbf{s})\right)\left(\frac{1}{2}\frac{\delta}{\delta\psi^{+}(\mathbf{s})}\right)WP\nonumber \\
 &  & +g\int d\mathbf{s}\left(-\frac{\delta}{\delta\phi(\mathbf{s})}\right)\left(-\frac{\delta}{\delta\phi(\mathbf{s})}\right)\left(\phi(\mathbf{s})\right)\left(\psi(\mathbf{s})\right)WP\nonumber \\
 &  & +g\int d\mathbf{s}\left(-\frac{\delta}{\delta\phi(\mathbf{s})}\right)\left(-\frac{\delta}{\delta\phi(\mathbf{s})}\right)\left(\phi(\mathbf{s})\right)\left(\frac{1}{2}\frac{\delta}{\delta\psi^{+}(\mathbf{s})}\right)WP\nonumber \\
 &  & \,\end{eqnarray}
\begin{eqnarray}
 &  & WP[\psi(\mathbf{r}),\psi^{+}(\mathbf{r}),\phi(\mathbf{r}),\phi^{+}(\mathbf{r})]_{9-12}\nonumber \\
 & = & +g\int d\mathbf{s}\left(\psi^{+}(\mathbf{s})\right)\left(\phi^{+}(\mathbf{s})\right)\left(\phi(\mathbf{s})\right)\left(\phi(\mathbf{s})\right)WP\nonumber \\
 &  & +g\int d\mathbf{s}\left(\psi^{+}(\mathbf{s})\right)\left(-\frac{\delta}{\delta\phi(\mathbf{s})}\right)\left(\phi(\mathbf{s})\right)\left(\phi(\mathbf{s})\right)WP\nonumber \\
 &  & +g\int d\mathbf{s}\left(-\frac{1}{2}\frac{\delta}{\delta\psi(\mathbf{s})}\right)\left(\phi^{+}(\mathbf{s})\right)\left(\phi(\mathbf{s})\right)\left(\phi(\mathbf{s})\right)WP\nonumber \\
 &  & +g\int d\mathbf{s}\left(-\frac{1}{2}\frac{\delta}{\delta\psi(\mathbf{s})}\right)\left(-\frac{\delta}{\delta\phi(\mathbf{s})}\right)\left(\phi(\mathbf{s})\right)\left(\phi(\mathbf{s})\right)WP\nonumber \\
 &  & \,\end{eqnarray}
The functional derivatives are now placed on the left using results
in which the functional derivatives of differing fields are zero (see
(\ref{Eq.FuncDerivativeRule1-1}) and (\ref{Eq.FuncDerivativeRule2-1}))
giving\begin{eqnarray}
 &  & WP[\psi(\mathbf{r}),\psi^{+}(\mathbf{r}),\phi(\mathbf{r}),\phi^{+}(\mathbf{r})]_{1-8}\nonumber \\
 & = & g\int d\mathbf{s\{}\phi^{+}(\mathbf{s})\phi^{+}(\mathbf{s})\phi(\mathbf{s})\psi(\mathbf{s})\}WP\qquad\emph{V1}\nonumber \\
 &  & +g\int d\mathbf{s}\left(\frac{\delta}{\delta\psi^{+}(\mathbf{s})}\right)\{\frac{1}{2}\phi^{+}(\mathbf{s})\phi^{+}(\mathbf{s})\phi(\mathbf{s})\}WP\qquad\emph{V2}\nonumber \\
 &  & -g\int d\mathbf{s}\left(\frac{\delta}{\delta\phi(\mathbf{s})}\right)\{\phi^{+}(\mathbf{s})\phi(\mathbf{s})\psi(\mathbf{s})\}WP\qquad\emph{V3}\nonumber \\
 &  & -g\int d\mathbf{s}\left(\frac{\delta}{\delta\phi(\mathbf{s})}\right)\left(\frac{\delta}{\delta\psi^{+}(\mathbf{s})}\right)\{\frac{1}{2}\phi^{+}(\mathbf{s})\phi(\mathbf{s})\}WP\qquad\emph{V4}\nonumber \\
 &  & -g\int d\mathbf{s}\left(\frac{\delta}{\delta\phi(\mathbf{s})}\right)\{\phi^{+}(\mathbf{s})\phi(\mathbf{s})\psi(\mathbf{s})\}WP\qquad\emph{V5}\nonumber \\
 &  & -g\int d\mathbf{s}\left(\frac{\delta}{\delta\phi(\mathbf{s})}\right)\left(\frac{\delta}{\delta\psi^{+}(\mathbf{s})}\right)\{\frac{1}{2}\phi^{+}(\mathbf{s})\phi(\mathbf{s})\}WP\qquad\emph{V6}\nonumber \\
 &  & +g\int d\mathbf{s}\left(\frac{\delta}{\delta\phi(\mathbf{s})}\right)\left(\frac{\delta}{\delta\phi(\mathbf{s})}\right)\{\phi(\mathbf{s})\psi(\mathbf{s})\}WP\qquad\emph{V7}\nonumber \\
 &  & +g\int d\mathbf{s}\left(\frac{\delta}{\delta\phi(\mathbf{s})}\right)\left(\frac{\delta}{\delta\phi(\mathbf{s})}\right)\left(\frac{\delta}{\delta\psi^{+}(\mathbf{s})}\right)\{\frac{1}{2}\phi(\mathbf{s})\}WP\qquad\emph{V8}\nonumber \\
 &  & \,\end{eqnarray}
\begin{eqnarray}
 &  & WP[\psi(\mathbf{r}),\psi^{+}(\mathbf{r}),\phi(\mathbf{r}),\phi^{+}(\mathbf{r})]_{9-12}\nonumber \\
 & = & +g\int d\mathbf{s\{}\psi^{+}(\mathbf{s})\phi^{+}(\mathbf{s})\phi(\mathbf{s})\phi(\mathbf{s})\}WP\qquad\emph{V9}\nonumber \\
 &  & -g\int d\mathbf{s}\left(\frac{\delta}{\delta\phi(\mathbf{s})}\right)\{\psi^{+}(\mathbf{s})\phi(\mathbf{s})\phi(\mathbf{s})\}WP\qquad\emph{V10}\nonumber \\
 &  & -g\int d\mathbf{s}\left(\frac{\delta}{\delta\psi(\mathbf{s})}\right)\{\frac{1}{2}\phi^{+}(\mathbf{s})\phi(\mathbf{s})\phi(\mathbf{s})\}WP\qquad\emph{V11}\nonumber \\
 &  & +g\int d\mathbf{s}\left(\frac{\delta}{\delta\psi(\mathbf{s})}\right)\left(\frac{\delta}{\delta\phi(\mathbf{s})}\right)\{\frac{1}{2}\phi(\mathbf{s})\phi(\mathbf{s})\}WP\qquad\emph{V12}\nonumber \\
 &  & \,\end{eqnarray}
Collecting terms with the same order of functional derivatives we
have\begin{eqnarray}
 &  & WP[\psi(\mathbf{r}),\psi^{+}(\mathbf{r}),\phi(\mathbf{r}),\phi^{+}(\mathbf{r})]\nonumber \\
 & \rightarrow & WP^{0}+WP^{1}+WP^{2}+WP^{3}\label{Eq.V3RhoResult1}\end{eqnarray}
where we have used upper subscripts for the $\widehat{V}_{3}\widehat{\rho}\,$contributions
and\begin{eqnarray}
 &  & WP^{0}\nonumber \\
 & = & g\int d\mathbf{s\{}\phi^{+}(\mathbf{s})\phi^{+}(\mathbf{s})\phi(\mathbf{s})\psi(\mathbf{s})\}WP\qquad\emph{V1}\nonumber \\
 &  & +g\int d\mathbf{s\{}\psi^{+}(\mathbf{s})\phi^{+}(\mathbf{s})\phi(\mathbf{s})\phi(\mathbf{s})\}WP\qquad\emph{V9}\nonumber \\
 &  & \,\label{eq:V3RhoResult2}\end{eqnarray}
\begin{eqnarray}
 &  & WP^{1}\nonumber \\
 & = & +g\int d\mathbf{s}\left(\frac{\delta}{\delta\psi^{+}(\mathbf{s})}\right)\{\frac{1}{2}\phi^{+}(\mathbf{s})\phi^{+}(\mathbf{s})\phi(\mathbf{s})\}WP\qquad\emph{V2}\nonumber \\
 &  & -g\int d\mathbf{s}\left(\frac{\delta}{\delta\phi(\mathbf{s})}\right)\{\phi^{+}(\mathbf{s})\phi(\mathbf{s})\psi(\mathbf{s})\}WP\qquad\emph{V3}\nonumber \\
 &  & -g\int d\mathbf{s}\left(\frac{\delta}{\delta\phi(\mathbf{s})}\right)\{\phi^{+}(\mathbf{s})\phi(\mathbf{s})\psi(\mathbf{s})\}WP\qquad\emph{V5}\nonumber \\
 &  & -g\int d\mathbf{s}\left(\frac{\delta}{\delta\phi(\mathbf{s})}\right)\{\psi^{+}(\mathbf{s})\phi(\mathbf{s})\phi(\mathbf{s})\}WP\qquad\emph{V10}\nonumber \\
 &  & -g\int d\mathbf{s}\left(\frac{\delta}{\delta\psi(\mathbf{s})}\right)\{\frac{1}{2}\phi^{+}(\mathbf{s})\phi(\mathbf{s})\phi(\mathbf{s})\}WP\qquad\emph{V11}\nonumber \\
 & = & +g\int d\mathbf{s}\left(\frac{\delta}{\delta\psi^{+}(\mathbf{s})}\right)\{\frac{1}{2}\phi^{+}(\mathbf{s})\phi^{+}(\mathbf{s})\phi(\mathbf{s})\}WP\qquad\emph{V2}\nonumber \\
 &  & -g\int d\mathbf{s}\left(\frac{\delta}{\delta\psi(\mathbf{s})}\right)\{\frac{1}{2}\phi^{+}(\mathbf{s})\phi(\mathbf{s})\phi(\mathbf{s})\}WP\qquad\emph{V11}\nonumber \\
 &  & -g\int d\mathbf{s}\left(\frac{\delta}{\delta\phi(\mathbf{s})}\right)\{[2\phi^{+}(\mathbf{s})\psi(\mathbf{s})+\psi^{+}(\mathbf{s})\phi(\mathbf{s})]\phi(\mathbf{s})\}WP\qquad\emph{V3,V5,V10}\nonumber \\
 &  & \,\label{eq:V3RhoResult3}\end{eqnarray}
\begin{eqnarray}
 &  & WP^{2}\nonumber \\
 & = & -g\int d\mathbf{s}\left(\frac{\delta}{\delta\phi(\mathbf{s})}\right)\left(\frac{\delta}{\delta\psi^{+}(\mathbf{s})}\right)\{\frac{1}{2}\phi^{+}(\mathbf{s})\phi(\mathbf{s})\}WP\qquad\emph{V4}\nonumber \\
 &  & -g\int d\mathbf{s}\left(\frac{\delta}{\delta\phi(\mathbf{s})}\right)\left(\frac{\delta}{\delta\psi^{+}(\mathbf{s})}\right)\{\frac{1}{2}\phi^{+}(\mathbf{s})\phi(\mathbf{s})\}WP\qquad\emph{V6}\nonumber \\
 &  & +g\int d\mathbf{s}\left(\frac{\delta}{\delta\phi(\mathbf{s})}\right)\left(\frac{\delta}{\delta\phi(\mathbf{s})}\right)\{\phi(\mathbf{s})\psi(\mathbf{s})\}WP\qquad\emph{V7}\nonumber \\
 &  & +g\int d\mathbf{s}\left(\frac{\delta}{\delta\psi(\mathbf{s})}\right)\left(\frac{\delta}{\delta\phi(\mathbf{s})}\right)\{\frac{1}{2}\phi(\mathbf{s})\phi(\mathbf{s})\}WP\qquad\emph{V12}\nonumber \\
 & = & -g\int d\mathbf{s}\left(\frac{\delta}{\delta\psi^{+}(\mathbf{s})}\right)\left(\frac{\delta}{\delta\phi(\mathbf{s})}\right)\{\phi^{+}(\mathbf{s})\phi(\mathbf{s})\}WP\qquad\emph{V4,V6}\nonumber \\
 &  & +g\int d\mathbf{s}\left(\frac{\delta}{\delta\psi(\mathbf{s})}\right)\left(\frac{\delta}{\delta\phi(\mathbf{s})}\right)\{\frac{1}{2}\phi(\mathbf{s})\phi(\mathbf{s})\}WP\qquad\emph{V12}\nonumber \\
 &  & +g\int d\mathbf{s}\left(\frac{\delta}{\delta\phi(\mathbf{s})}\right)\left(\frac{\delta}{\delta\phi(\mathbf{s})}\right)\{\phi(\mathbf{s})\psi(\mathbf{s})\}WP\qquad\emph{V7}\nonumber \\
 &  & \,\label{eq:V3RhoResult4}\end{eqnarray}
\begin{eqnarray}
 &  & WP^{3}\nonumber \\
 & = & +g\int d\mathbf{s}\left(\frac{\delta}{\delta\psi^{+}(\mathbf{s})}\right)\left(\frac{\delta}{\delta\phi(\mathbf{s})}\right)\left(\frac{\delta}{\delta\phi(\mathbf{s})}\right)\{\frac{1}{2}\phi(\mathbf{s})\}WP\qquad V8\nonumber \\
 &  & \emph{\,}\label{eq:V3RhoResult5}\end{eqnarray}

Now if \begin{eqnarray}
\widehat{\rho} & \rightarrow & \,\widehat{\rho}\,\widehat{V}_{3}\nonumber \\
 & = & g\int d\mathbf{s\,\widehat{\rho}\,(}\widehat{\Psi}_{NC}(\mathbf{s})^{\dagger}\widehat{\Psi}_{NC}(\mathbf{s})^{\dagger}\widehat{\Psi}_{NC}(\mathbf{s})\widehat{\Psi}_{C}(\mathbf{s})+\widehat{\Psi}_{C}(\mathbf{s})^{\dagger}\widehat{\Psi}_{NC}(\mathbf{s})^{\dagger}\widehat{\Psi}_{NC}(\mathbf{s})\widehat{\Psi}_{NC}(\mathbf{s}))\nonumber \\
 &  & \,\end{eqnarray}
then\begin{eqnarray}
 &  & WP[\psi(\mathbf{r}),\psi^{+}(\mathbf{r}),\phi(\mathbf{r}),\phi^{+}(\mathbf{r})]\nonumber \\
 & \rightarrow & g\int d\mathbf{s}\left(\psi(\mathbf{s})-\frac{1}{2}\frac{\delta}{\delta\psi^{+}(\mathbf{s})}\right)\left(\phi(\mathbf{s})-\frac{\delta}{\delta\phi^{+}(\mathbf{s})}\right)\left(\phi^{+}(\mathbf{s})\right)\nonumber \\
 &  & \times\left(\phi^{+}(\mathbf{s})\right)WP[\psi,\psi^{+},\phi,\phi^{+}]\nonumber \\
 &  & +g\int d\mathbf{s}\left(\phi(\mathbf{s})-\frac{\delta}{\delta\phi^{+}(\mathbf{s})}\right)\left(\phi(\mathbf{s})-\frac{\delta}{\delta\phi^{+}(\mathbf{s})}\right)\left(\phi^{+}(\mathbf{s})\right)\nonumber \\
 &  & \times\left(\psi^{+}(\mathbf{s})+\frac{1}{2}\frac{\delta}{\delta\psi(\mathbf{s})}\right)WP[\psi,\psi^{+},\phi,\phi^{+}]\nonumber \\
 &  & \,\end{eqnarray}

Expanding gives\begin{eqnarray}
 &  & WP[\psi(\mathbf{r}),\psi^{+}(\mathbf{r}),\phi(\mathbf{r}),\phi^{+}(\mathbf{r})]\nonumber \\
 & = & WP[\psi,\psi^{+},\phi,\phi^{+}]_{1-4}+WP[\psi,\psi^{+},\phi,\phi^{+}]_{5-12}\end{eqnarray}
where\begin{eqnarray}
 &  & WP[\psi,\psi^{+},\phi,\phi^{+}]_{1-4}\nonumber \\
 & = & g\int d\mathbf{s}\left(\psi(\mathbf{s})\right)\left(\phi(\mathbf{s})\right)\left(\phi^{+}(\mathbf{s})\right)\left(\phi^{+}(\mathbf{s})\right)WP\nonumber \\
 &  & +g\int d\mathbf{s}\left(\psi(\mathbf{s})\right)\left(-\frac{\delta}{\delta\phi^{+}(\mathbf{s})}\right)\left(\phi^{+}(\mathbf{s})\right)\left(\phi^{+}(\mathbf{s})\right)WP\nonumber \\
 &  & +g\int d\mathbf{s}\left(-\frac{1}{2}\frac{\delta}{\delta\psi^{+}(\mathbf{s})}\right)\left(\phi(\mathbf{s})\right)\left(\phi^{+}(\mathbf{s})\right)\left(\phi^{+}(\mathbf{s})\right)WP\nonumber \\
 &  & +g\int d\mathbf{s}\left(-\frac{1}{2}\frac{\delta}{\delta\psi^{+}(\mathbf{s})}\right)\left(-\frac{\delta}{\delta\phi^{+}(\mathbf{s})}\right)\left(\phi^{+}(\mathbf{s})\right)\left(\phi^{+}(\mathbf{s})\right)WP\nonumber \\
 &  & \,\end{eqnarray}
\begin{eqnarray}
 &  & WP[\psi,\psi^{+},\phi,\phi^{+}]_{5-12}\nonumber \\
 & = & +g\int d\mathbf{s}\left(\phi(\mathbf{s})\right)\left(\phi(\mathbf{s})\right)\left(\phi^{+}(\mathbf{s})\right)\left(\psi^{+}(\mathbf{s})\right)WP\nonumber \\
 &  & +g\int d\mathbf{s}\left(\phi(\mathbf{s})\right)\left(\phi(\mathbf{s})\right)\left(\phi^{+}(\mathbf{s})\right)\left(\frac{1}{2}\frac{\delta}{\delta\psi(\mathbf{s})}\right)WP\nonumber \\
 &  & +g\int d\mathbf{s}\left(\phi(\mathbf{s})\right)\left(-\frac{\delta}{\delta\phi^{+}(\mathbf{s})}\right)\left(\phi^{+}(\mathbf{s})\right)\left(\psi^{+}(\mathbf{s})\right)WP\nonumber \\
 &  & +g\int d\mathbf{s}\left(\phi(\mathbf{s})\right)\left(-\frac{\delta}{\delta\phi^{+}(\mathbf{s})}\right)\left(\phi^{+}(\mathbf{s})\right)\left(\frac{1}{2}\frac{\delta}{\delta\psi(\mathbf{s})}\right)WP\nonumber \\
 &  & +g\int d\mathbf{s}\left(-\frac{\delta}{\delta\phi^{+}(\mathbf{s})}\right)\left(\phi(\mathbf{s})\right)\left(\phi^{+}(\mathbf{s})\right)\left(\psi^{+}(\mathbf{s})\right)WP\nonumber \\
 &  & +g\int d\mathbf{s}\left(-\frac{\delta}{\delta\phi^{+}(\mathbf{s})}\right)\left(\phi(\mathbf{s})\right)\left(\phi^{+}(\mathbf{s})\right)\left(\frac{1}{2}\frac{\delta}{\delta\psi(\mathbf{s})}\right)WP\nonumber \\
 &  & +g\int d\mathbf{s}\left(-\frac{\delta}{\delta\phi^{+}(\mathbf{s})}\right)\left(-\frac{\delta}{\delta\phi^{+}(\mathbf{s})}\right)\left(\phi^{+}(\mathbf{s})\right)\left(\psi^{+}(\mathbf{s})\right)WP\nonumber \\
 &  & +g\int d\mathbf{s}\left(-\frac{\delta}{\delta\phi^{+}(\mathbf{s})}\right)\left(-\frac{\delta}{\delta\phi^{+}(\mathbf{s})}\right)\left(\phi^{+}(\mathbf{s})\right)\left(\frac{1}{2}\frac{\delta}{\delta\psi(\mathbf{s})}\right)WP\nonumber \\
 &  & \,\end{eqnarray}
The functional derivatives are now placed on the left using results
in which the functional derivatives of differing fields are zero (see
(\ref{Eq.FuncDerivativeRule1-1}) and (\ref{Eq.FuncDerivativeRule2-1}))
giving\begin{eqnarray}
 &  & WP[\psi,\psi^{+},\phi,\phi^{+}]_{1-4}\nonumber \\
 & = & g\int d\mathbf{s\{}\psi(\mathbf{s})\phi(\mathbf{s})\phi^{+}(\mathbf{s})\phi^{+}(\mathbf{s})\}WP\qquad\emph{W1}\nonumber \\
 &  & -g\int d\mathbf{s}\left(\frac{\delta}{\delta\phi^{+}(\mathbf{s})}\right)\{\psi(\mathbf{s})\phi^{+}(\mathbf{s})\phi^{+}(\mathbf{s})\}WP\qquad\emph{W2}\nonumber \\
 &  & -g\int d\mathbf{s}\left(\frac{\delta}{\delta\psi^{+}(\mathbf{s})}\right)\{\frac{1}{2}\phi(\mathbf{s})\phi^{+}(\mathbf{s})\phi^{+}(\mathbf{s})\}WP\qquad\emph{W3}\nonumber \\
 &  & +g\int d\mathbf{s}\left(\frac{\delta}{\delta\psi^{+}(\mathbf{s})}\right)\left(\frac{\delta}{\delta\phi^{+}(\mathbf{s})}\right)\{\frac{1}{2}\phi^{+}(\mathbf{s})\phi^{+}(\mathbf{s})\}WP\qquad\emph{W4}\nonumber \\
 &  & \,\end{eqnarray}
and\begin{eqnarray}
 &  & WP[\psi,\psi^{+},\phi,\phi^{+}]_{5-12}\nonumber \\
 & = & +g\int d\mathbf{s\{}\phi(\mathbf{s})\phi(\mathbf{s})\phi^{+}(\mathbf{s})\psi^{+}(\mathbf{s})\}WP\qquad\emph{W5}\nonumber \\
 &  & +g\int d\mathbf{s}\left(\frac{\delta}{\delta\psi(\mathbf{s})}\right)\{\frac{1}{2}\phi(\mathbf{s})\phi(\mathbf{s})\phi^{+}(\mathbf{s})\}WP\qquad\emph{W6}\nonumber \\
 &  & -g\int d\mathbf{s}\left(\frac{\delta}{\delta\phi^{+}(\mathbf{s})}\right)\{\phi(\mathbf{s})\phi^{+}(\mathbf{s})\psi^{+}(\mathbf{s})\}WP\qquad\emph{W7}\nonumber \\
 &  & -g\int d\mathbf{s}\left(\frac{\delta}{\delta\phi^{+}(\mathbf{s})}\right)\left(\frac{\delta}{\delta\psi(\mathbf{s})}\right)\{\frac{1}{2}\phi(\mathbf{s})\phi^{+}(\mathbf{s})\}WP\qquad\emph{W8}\nonumber \\
 &  & -g\int d\mathbf{s}\left(\frac{\delta}{\delta\phi^{+}(\mathbf{s})}\right)\{\phi(\mathbf{s})\phi^{+}(\mathbf{s})\psi^{+}(\mathbf{s})\}WP\qquad\emph{W9}\nonumber \\
 &  & -g\int d\mathbf{s}\left(\frac{\delta}{\delta\phi^{+}(\mathbf{s})}\right)\left(\frac{\delta}{\delta\psi(\mathbf{s})}\right)\{\frac{1}{2}\phi(\mathbf{s})\phi^{+}(\mathbf{s})\}WP\qquad\emph{W10}\nonumber \\
 &  & +g\int d\mathbf{s}\left(\frac{\delta}{\delta\phi^{+}(\mathbf{s})}\right)\left(\frac{\delta}{\delta\phi^{+}(\mathbf{s})}\right)\{\phi^{+}(\mathbf{s})\psi^{+}(\mathbf{s})\}WP\qquad\emph{W11}\nonumber \\
 &  & +g\int d\mathbf{s}\left(\frac{\delta}{\delta\phi^{+}(\mathbf{s})}\right)\left(\frac{\delta}{\delta\phi^{+}(\mathbf{s})}\right)\left(\frac{\delta}{\delta\psi(\mathbf{s})}\right)\{\frac{1}{2}\phi^{+}(\mathbf{s})\}WP\qquad\emph{W12}\nonumber \\
 &  & \,\end{eqnarray}

Collecting terms with the same order of functional derivatives we
have\begin{eqnarray}
 &  & WP[\psi(\mathbf{r}),\psi^{+}(\mathbf{r}),\phi(\mathbf{r}),\phi^{+}(\mathbf{r})]\nonumber \\
 & \rightarrow & WP_{0}+WP_{1}+WP_{2}+WP_{3}\label{Eq.RhoV3Result1}\end{eqnarray}
where we have used lower subscripts for the $\widehat{\rho}\,\widehat{V}_{3}$
contributions and\begin{eqnarray}
 &  & WP_{0}\nonumber \\
 & = & g\int d\mathbf{s\{}\psi(\mathbf{s})\phi(\mathbf{s})\phi^{+}(\mathbf{s})\phi^{+}(\mathbf{s})\}WP\qquad\emph{W1}\nonumber \\
 &  & +g\int d\mathbf{s\{}\phi(\mathbf{s})\phi(\mathbf{s})\phi^{+}(\mathbf{s})\psi^{+}(\mathbf{s})\}WP\qquad W5\nonumber \\
 &  & \emph{\,}\label{eq:RhoV3Result2}\end{eqnarray}
\begin{eqnarray}
 &  & WP_{1}\nonumber \\
 & = & -g\int d\mathbf{s}\left(\frac{\delta}{\delta\phi^{+}(\mathbf{s})}\right)\{\psi(\mathbf{s})\phi^{+}(\mathbf{s})\phi^{+}(\mathbf{s})\}WP\qquad\emph{W2}\nonumber \\
 &  & -g\int d\mathbf{s}\left(\frac{\delta}{\delta\psi^{+}(\mathbf{s})}\right)\{\frac{1}{2}\phi(\mathbf{s})\phi^{+}(\mathbf{s})\phi^{+}(\mathbf{s})\}WP\qquad\emph{W3}\nonumber \\
 &  & +g\int d\mathbf{s}\left(\frac{\delta}{\delta\psi(\mathbf{s})}\right)\{\frac{1}{2}\phi(\mathbf{s})\phi(\mathbf{s})\phi^{+}(\mathbf{s})\}WP\qquad\emph{W6}\nonumber \\
 &  & -g\int d\mathbf{s}\left(\frac{\delta}{\delta\phi^{+}(\mathbf{s})}\right)\{\phi(\mathbf{s})\phi^{+}(\mathbf{s})\psi^{+}(\mathbf{s})\}WP\qquad\emph{W7}\nonumber \\
 &  & -g\int d\mathbf{s}\left(\frac{\delta}{\delta\phi^{+}(\mathbf{s})}\right)\{\phi(\mathbf{s})\phi^{+}(\mathbf{s})\psi^{+}(\mathbf{s})\}WP\qquad\emph{W9}\nonumber \\
 & = & -g\int d\mathbf{s}\left(\frac{\delta}{\delta\psi^{+}(\mathbf{s})}\right)\{\frac{1}{2}\phi(\mathbf{s})\phi^{+}(\mathbf{s})\phi^{+}(\mathbf{s})\}WP\qquad\emph{W3}\nonumber \\
 &  & +g\int d\mathbf{s}\left(\frac{\delta}{\delta\psi(\mathbf{s})}\right)\{\frac{1}{2}\phi(\mathbf{s})\phi(\mathbf{s})\phi^{+}(\mathbf{s})\}WP\qquad\emph{W6}\nonumber \\
 &  & -g\int d\mathbf{s}\left(\frac{\delta}{\delta\phi^{+}(\mathbf{s})}\right)\{[2\phi(\mathbf{s})\psi^{+}(\mathbf{s})+\psi(\mathbf{s})\phi^{+}(\mathbf{s})]\phi^{+}(\mathbf{s})\}WP\qquad\emph{W7,W9,W2}\nonumber \\
 &  & \,\label{eq:RhoV3Result3}\end{eqnarray}
\begin{eqnarray}
 &  & WP_{2}\nonumber \\
 & = & +g\int d\mathbf{s}\left(\frac{\delta}{\delta\psi^{+}(\mathbf{s})}\right)\left(\frac{\delta}{\delta\phi^{+}(\mathbf{s})}\right)\{\frac{1}{2}\phi^{+}(\mathbf{s})\phi^{+}(\mathbf{s})\}WP\qquad\emph{W4}\nonumber \\
 &  & -g\int d\mathbf{s}\left(\frac{\delta}{\delta\phi^{+}(\mathbf{s})}\right)\left(\frac{\delta}{\delta\psi(\mathbf{s})}\right)\{\frac{1}{2}\phi(\mathbf{s})\phi^{+}(\mathbf{s})\}WP\qquad\emph{W8}\nonumber \\
 &  & -g\int d\mathbf{s}\left(\frac{\delta}{\delta\phi^{+}(\mathbf{s})}\right)\left(\frac{\delta}{\delta\psi(\mathbf{s})}\right)\{\frac{1}{2}\phi(\mathbf{s})\phi^{+}(\mathbf{s})\}WP\qquad\emph{W10}\nonumber \\
 &  & +g\int d\mathbf{s}\left(\frac{\delta}{\delta\phi^{+}(\mathbf{s})}\right)\left(\frac{\delta}{\delta\phi^{+}(\mathbf{s})}\right)\{\phi^{+}(\mathbf{s})\psi^{+}(\mathbf{s})\}WP\qquad\emph{W11}\nonumber \\
 & = & +g\int d\mathbf{s}\left(\frac{\delta}{\delta\psi^{+}(\mathbf{s})}\right)\left(\frac{\delta}{\delta\phi^{+}(\mathbf{s})}\right)\{\frac{1}{2}\phi^{+}(\mathbf{s})\phi^{+}(\mathbf{s})\}WP\qquad\emph{W4}\nonumber \\
 &  & -g\int d\mathbf{s}\left(\frac{\delta}{\delta\psi(\mathbf{s})}\right)\left(\frac{\delta}{\delta\phi^{+}(\mathbf{s})}\right)\{\phi(\mathbf{s})\phi^{+}(\mathbf{s})\}WP\qquad\emph{W8,W10}\nonumber \\
 &  & +g\int d\mathbf{s}\left(\frac{\delta}{\delta\phi^{+}(\mathbf{s})}\right)\left(\frac{\delta}{\delta\phi^{+}(\mathbf{s})}\right)\{\phi^{+}(\mathbf{s})\psi^{+}(\mathbf{s})\}WP\qquad\emph{W11}\nonumber \\
 &  & \,\label{eq:RhoV3Result4}\end{eqnarray}
\begin{eqnarray}
 &  & WP_{3}\nonumber \\
 & = & +g\int d\mathbf{s}\left(\frac{\delta}{\delta\psi(\mathbf{s})}\right)\left(\frac{\delta}{\delta\phi^{+}(\mathbf{s})}\right)\left(\frac{\delta}{\delta\phi^{+}(\mathbf{s})}\right)\{\frac{1}{2}\phi^{+}(\mathbf{s})\}WP\qquad\emph{W12}\nonumber \\
 &  & \,\label{eq:RhoV3Result5}\end{eqnarray}

Now if \begin{eqnarray}
\widehat{\rho} & \rightarrow & [\widehat{V}_{3},\,\widehat{\rho}]\nonumber \\
 & = & [g\int d\mathbf{s(}\widehat{\Psi}_{NC}(\mathbf{s})^{\dagger}\widehat{\Psi}_{NC}(\mathbf{s})^{\dagger}\widehat{\Psi}_{NC}(\mathbf{s})\widehat{\Psi}_{C}(\mathbf{s})+\widehat{\Psi}_{C}(\mathbf{s})^{\dagger}\widehat{\Psi}_{NC}(\mathbf{s})^{\dagger}\widehat{\Psi}_{NC}(\mathbf{s})\widehat{\Psi}_{NC}(\mathbf{s})),\widehat{\rho}]\nonumber \\
 &  & \,\end{eqnarray}
then \begin{eqnarray}
 &  & WP[\psi(\mathbf{r}),\psi^{+}(\mathbf{r}),\phi(\mathbf{r}),\phi^{+}(\mathbf{r})]\nonumber \\
 & \rightarrow & WP_{T}^{0}+WP_{T}^{1}+WP_{T}^{2}+WP_{T}^{3}\label{Eq.V3CommRhoResult1}\end{eqnarray}
where the $WP_{T}^{n}$ are obtained by subtracting the results for
$\widehat{\rho}\widehat{\, V}_{3}$ from those for $\widehat{V}_{3}\,\widehat{\rho}.$
We find that\begin{eqnarray}
 &  & WP_{T}^{0}\nonumber \\
 & = & g\int d\mathbf{s\{}\phi^{+}(\mathbf{s})\phi^{+}(\mathbf{s})\phi(\mathbf{s})\psi(\mathbf{s})\}WP\qquad\emph{V1}\nonumber \\
 &  & -g\int d\mathbf{s\{}\psi(\mathbf{s})\phi(\mathbf{s})\phi^{+}(\mathbf{s})\phi^{+}(\mathbf{s})\}WP\qquad\emph{W1}\nonumber \\
 &  & +g\int d\mathbf{s\{}\psi^{+}(\mathbf{s})\phi^{+}(\mathbf{s})\phi(\mathbf{s})\phi(\mathbf{s})\}WP\qquad\emph{V9}\nonumber \\
 &  & -g\int d\mathbf{s\{}\phi(\mathbf{s})\phi(\mathbf{s})\phi^{+}(\mathbf{s})\psi^{+}(\mathbf{s})\}WP\qquad\emph{W5}\nonumber \\
 & = & 0\label{Eq.V3CommRhoResult2}\end{eqnarray}
\begin{eqnarray}
 &  & WP_{T}^{1}\nonumber \\
 & = & +g\int d\mathbf{s}\left(\frac{\delta}{\delta\psi^{+}(\mathbf{s})}\right)\{\frac{1}{2}\phi^{+}(\mathbf{s})\phi^{+}(\mathbf{s})\phi(\mathbf{s})\}WP\qquad\emph{V2}\nonumber \\
 &  & +g\int d\mathbf{s}\left(\frac{\delta}{\delta\psi^{+}(\mathbf{s})}\right)\{\frac{1}{2}\phi(\mathbf{s})\phi^{+}(\mathbf{s})\phi^{+}(\mathbf{s})\}WP\qquad\emph{W3}\nonumber \\
 &  & -g\int d\mathbf{s}\left(\frac{\delta}{\delta\psi(\mathbf{s})}\right)\{\frac{1}{2}\phi^{+}(\mathbf{s})\phi(\mathbf{s})\phi(\mathbf{s})\}WP\qquad\emph{V11}\nonumber \\
 &  & -g\int d\mathbf{s}\left(\frac{\delta}{\delta\psi(\mathbf{s})}\right)\{\frac{1}{2}\phi(\mathbf{s})\phi(\mathbf{s})\phi^{+}(\mathbf{s})\}WP\qquad\emph{W6}\nonumber \\
 &  & -g\int d\mathbf{s}\left(\frac{\delta}{\delta\phi(\mathbf{s})}\right)\{[2\phi^{+}(\mathbf{s})\psi(\mathbf{s})+\psi^{+}(\mathbf{s})\phi(\mathbf{s})]\phi(\mathbf{s})\}WP\qquad\emph{V3,V5,V10}\nonumber \\
 &  & +g\int d\mathbf{s}\left(\frac{\delta}{\delta\phi^{+}(\mathbf{s})}\right)\{[2\phi(\mathbf{s})\psi^{+}(\mathbf{s})+\psi(\mathbf{s})\phi^{+}(\mathbf{s})]\phi^{+}(\mathbf{s})\}WP\qquad\emph{W7,W9,W2}\nonumber \\
 & = & +g\int d\mathbf{s}\left(\frac{\delta}{\delta\psi^{+}(\mathbf{s})}\right)\{\phi^{+}(\mathbf{s})\phi^{+}(\mathbf{s})\phi(\mathbf{s})\}WP\qquad\emph{V2,W3}\nonumber \\
 &  & -g\int d\mathbf{s}\left(\frac{\delta}{\delta\psi(\mathbf{s})}\right)\{\phi(\mathbf{s})\phi(\mathbf{s})\phi^{+}(\mathbf{s})\}WP\qquad\emph{V11,W6}\nonumber \\
 &  & +g\int d\mathbf{s}\left(\frac{\delta}{\delta\phi^{+}(\mathbf{s})}\right)\{[2\phi(\mathbf{s})\psi^{+}(\mathbf{s})+\psi(\mathbf{s})\phi^{+}(\mathbf{s})]\phi^{+}(\mathbf{s})\}WP\qquad\emph{W7,W9,W2}\nonumber \\
 &  & -g\int d\mathbf{s}\left(\frac{\delta}{\delta\phi(\mathbf{s})}\right)\{[2\phi^{+}(\mathbf{s})\psi(\mathbf{s})+\psi^{+}(\mathbf{s})\phi(\mathbf{s})]\phi(\mathbf{s})\}WP\qquad\emph{V3,V5,V10}\nonumber \\
 &  & \,\label{eq:V3CommRhoResult3}\end{eqnarray}
\begin{eqnarray}
 &  & WP_{T}^{2}\nonumber \\
 & = & -g\int d\mathbf{s}\left(\frac{\delta}{\delta\psi^{+}(\mathbf{s})}\right)\left(\frac{\delta}{\delta\phi^{+}(\mathbf{s})}\right)\{\frac{1}{2}\phi^{+}(\mathbf{s})\phi^{+}(\mathbf{s})\}WP\qquad\emph{W4}\nonumber \\
 &  & +g\int d\mathbf{s}\left(\frac{\delta}{\delta\psi(\mathbf{s})}\right)\left(\frac{\delta}{\delta\phi(\mathbf{s})}\right)\{\frac{1}{2}\phi(\mathbf{s})\phi(\mathbf{s})\}WP\qquad\emph{V12}\nonumber \\
 &  & -g\int d\mathbf{s}\left(\frac{\delta}{\delta\psi^{+}(\mathbf{s})}\right)\left(\frac{\delta}{\delta\phi(\mathbf{s})}\right)\{\phi^{+}(\mathbf{s})\phi(\mathbf{s})\}WP\qquad\emph{V4,V6}\nonumber \\
 &  & +g\int d\mathbf{s}\left(\frac{\delta}{\delta\psi(\mathbf{s})}\right)\left(\frac{\delta}{\delta\phi^{+}(\mathbf{s})}\right)\{\phi(\mathbf{s})\phi^{+}(\mathbf{s})\}WP\qquad\emph{W8,W10}\nonumber \\
 &  & -g\int d\mathbf{s}\left(\frac{\delta}{\delta\phi^{+}(\mathbf{s})}\right)\left(\frac{\delta}{\delta\phi^{+}(\mathbf{s})}\right)\{\phi^{+}(\mathbf{s})\psi^{+}(\mathbf{s})\}WP\qquad\emph{W11}\nonumber \\
 &  & +g\int d\mathbf{s}\left(\frac{\delta}{\delta\phi(\mathbf{s})}\right)\left(\frac{\delta}{\delta\phi(\mathbf{s})}\right)\{\phi(\mathbf{s})\psi(\mathbf{s})\}WP\qquad\emph{V7}\nonumber \\
 &  & \,\label{eq:V3CommRhoResult4}\end{eqnarray}
\begin{eqnarray}
 &  & WP_{T}^{3}\nonumber \\
 & = & +g\int d\mathbf{s}\left(\frac{\delta}{\delta\psi^{+}(\mathbf{s})}\right)\left(\frac{\delta}{\delta\phi(\mathbf{s})}\right)\left(\frac{\delta}{\delta\phi(\mathbf{s})}\right)\{\frac{1}{2}\phi(\mathbf{s})\}WP\qquad\emph{V8}\nonumber \\
 &  & -g\int d\mathbf{s}\left(\frac{\delta}{\delta\psi(\mathbf{s})}\right)\left(\frac{\delta}{\delta\phi^{+}(\mathbf{s})}\right)\left(\frac{\delta}{\delta\phi^{+}(\mathbf{s})}\right)\{\frac{1}{2}\phi^{+}(\mathbf{s})\}WP\qquad\emph{W12}\nonumber \\
 &  & \,\label{eq:V3CommRhoResult5}\end{eqnarray}

Thus we see that the $\widehat{\, V}_{3}$ term produces functional
derivatives of orders one, two and three. We may write the contributions
to the functional Fokker-Planck equation in the form\begin{eqnarray}
 &  & \left(\frac{\partial}{\partial t}WP[\psi,\psi^{+},\phi,\phi^{+}]\right)_{V3}\nonumber \\
 & = & \left(\frac{\partial}{\partial t}WP[\psi,\psi^{+},\phi,\phi^{+}]\right)_{V3}^{1}+\left(\frac{\partial}{\partial t}WP[\psi,\psi^{+},\phi,\phi^{+}]\right)_{V3}^{2}+\left(\frac{\partial}{\partial t}WP[\psi,\psi^{+},\phi,\phi^{+}]\right)_{V3}^{3}\nonumber \\
 &  & \,\label{eq:FnalFPConNonConV3Intn}\end{eqnarray}
where on reverting to the original notation\begin{eqnarray}
 &  & \left(\frac{\partial}{\partial t}P[\underrightarrow{\psi}(\mathbf{r}),\underrightarrow{\psi^{\ast}}(\mathbf{r})]\right)_{V3}^{1}\nonumber \\
 & = & \frac{-i}{\hbar}\left\{ +g\int d\mathbf{s}\left\{ \left(\frac{\delta}{\delta\psi_{C}^{+}(\mathbf{s})}\right)\{\psi_{NC}^{+}(\mathbf{s})\psi_{NC}^{+}(\mathbf{s})\psi_{NC}(\mathbf{s})\}\right\} P[\underrightarrow{\psi}(\mathbf{r}),\underrightarrow{\psi^{\ast}}(\mathbf{r})]\right\} \nonumber \\
 &  & \frac{-i}{\hbar}\left\{ -g\int d\mathbf{s}\left\{ \left(\frac{\delta}{\delta\psi_{C}(\mathbf{s})}\right)\{\psi_{NC}^{+}(\mathbf{s})\psi_{NC}(\mathbf{s})\psi_{NC}(\mathbf{s})\}\right\} P[\underrightarrow{\psi}(\mathbf{r}),\underrightarrow{\psi^{\ast}}(\mathbf{r})]\right\} \nonumber \\
 &  & \frac{-i}{\hbar}\left\{ +g\int d\mathbf{s}\left\{ \left(\frac{\delta}{\delta\psi_{NC}^{+}(\mathbf{s})}\right)\{[2\psi_{NC}(\mathbf{s})\psi_{C}^{+}(\mathbf{s)+}\psi_{C}\mathbf{(\mathbf{s})}\psi_{NC}^{+}\mathbf{(\mathbf{s})]}\psi_{NC}^{+}(\mathbf{s})\}\right\} P[\underrightarrow{\psi}(\mathbf{r}),\underrightarrow{\psi^{\ast}}(\mathbf{r})]\right\} \nonumber \\
 &  & \frac{-i}{\hbar}\left\{ -g\int d\mathbf{s}\left\{ \left(\frac{\delta}{\delta\psi_{NC}(\mathbf{s})}\right)\{[2\psi_{NC}^{+}(\mathbf{s})\psi_{C}(\mathbf{s})+\psi_{C}^{+}(\mathbf{s})\psi_{NC}(\mathbf{s})]\psi_{NC}(\mathbf{s})\}\right\} P[\underrightarrow{\psi}(\mathbf{r}),\underrightarrow{\psi^{\ast}}(\mathbf{r})]\right\} \nonumber \\
 &  & \,\label{eq:FnalFPEV3InteractionLinear}\end{eqnarray}
\begin{eqnarray}
 &  & \left(\frac{\partial}{\partial t}P[\underrightarrow{\psi}(\mathbf{r}),\underrightarrow{\psi^{\ast}}(\mathbf{r})]\right)_{V3}^{2}\nonumber \\
 & = & \frac{-i}{\hbar}\left\{ -g\int d\mathbf{s}\left\{ \left(\frac{\delta}{\delta\psi_{C}^{+}(\mathbf{s})}\right)\left(\frac{\delta}{\delta\psi_{NC}^{+}(\mathbf{s})}\right)\left(\frac{1}{2}\psi_{NC}^{+}(\mathbf{s})\right)\left(\psi_{NC}^{+}(\mathbf{s})\right)\right\} P[\underrightarrow{\psi}(\mathbf{r}),\underrightarrow{\psi^{\ast}}(\mathbf{r})]\right\} \nonumber \\
 &  & \frac{-i}{\hbar}\left\{ +g\int d\mathbf{s}\left\{ \left(\frac{\delta}{\delta\psi_{C}(\mathbf{s})}\right)\left(\frac{\delta}{\delta\psi_{NC}(\mathbf{s})}\right)\{\frac{1}{2}\psi_{NC}(\mathbf{s})\psi_{NC}(\mathbf{s})\}\right\} P[\underrightarrow{\psi}(\mathbf{r}),\underrightarrow{\psi^{\ast}}(\mathbf{r})]\right\} \nonumber \\
 &  & \frac{-i}{\hbar}\left\{ -g\int d\mathbf{s}\left\{ \left(\frac{\delta}{\delta\psi_{C}^{+}(\mathbf{s})}\right)\left(\frac{\delta}{\delta\psi_{NC}(\mathbf{s})}\right)\{\psi_{NC}^{+}(\mathbf{s})\psi_{NC}(\mathbf{s})\}\right\} P[\underrightarrow{\psi}(\mathbf{r}),\underrightarrow{\psi^{\ast}}(\mathbf{r})]\right\} \nonumber \\
 &  & \frac{-i}{\hbar}\left\{ +g\int d\mathbf{s}\left\{ \left(\frac{\delta}{\delta\psi_{C}(\mathbf{s})}\right)\left(\frac{\delta}{\delta\psi_{NC}^{+}(\mathbf{s})}\right)\{\psi_{NC}(\mathbf{s})\psi_{NC}^{+}(\mathbf{s})\}\right\} P[\underrightarrow{\psi}(\mathbf{r}),\underrightarrow{\psi^{\ast}}(\mathbf{r})]\right\} \nonumber \\
 &  & \frac{-i}{\hbar}\left\{ -g\int d\mathbf{s}\left\{ \left(\frac{\delta}{\delta\psi_{NC}^{+}(\mathbf{s})}\right)\left(\frac{\delta}{\delta\psi_{NC}^{+}(\mathbf{s})}\right)\{\psi_{NC}^{+}(\mathbf{s})\psi_{C}^{+}(\mathbf{s})\}\right\} P[\underrightarrow{\psi}(\mathbf{r}),\underrightarrow{\psi^{\ast}}(\mathbf{r})]\right\} \nonumber \\
 &  & \frac{-i}{\hbar}\left\{ +g\int d\mathbf{s}\left\{ \left(\frac{\delta}{\delta\psi_{NC}(\mathbf{s})}\right)\left(\frac{\delta}{\delta\psi_{NC}(\mathbf{s})}\right)\{\psi_{NC}(\mathbf{s})\psi_{C}(\mathbf{s})\}\right\} P[\underrightarrow{\psi}(\mathbf{r}),\underrightarrow{\psi^{\ast}}(\mathbf{r})]\right\} \nonumber \\
 &  & \,\label{eq:FnalFPEV3InteractionQuadratic}\end{eqnarray}
\begin{eqnarray}
 &  & \left(\frac{\partial}{\partial t}P[\underrightarrow{\psi}(\mathbf{r}),\underrightarrow{\psi^{\ast}}(\mathbf{r})]\right)_{V3}^{3}\nonumber \\
 & = & \frac{-i}{\hbar}\left\{ +g\int d\mathbf{s}\left\{ \left(\frac{\delta}{\delta\psi_{C}^{+}(\mathbf{s})}\right)\left(\frac{\delta}{\delta\psi_{NC}(\mathbf{s})}\right)\left(\frac{\delta}{\delta\psi_{NC}(\mathbf{s})}\right)\{\frac{1}{2}\psi_{NC}(\mathbf{s})\}\right\} P[\underrightarrow{\psi}(\mathbf{r}),\underrightarrow{\psi^{\ast}}(\mathbf{r})]\right\} \nonumber \\
 &  & \frac{-i}{\hbar}\left\{ -g\int d\mathbf{s}\left\{ \left(\frac{\delta}{\delta\psi_{C}(\mathbf{s})}\right)\left(\frac{\delta}{\delta\psi_{NC}^{+}(\mathbf{s})}\right)\left(\frac{\delta}{\delta\psi_{NC}^{+}(\mathbf{s})}\right)\{\frac{1}{2}\psi_{NC}^{+}(\mathbf{s})\}\right\} P[\underrightarrow{\psi}(\mathbf{r}),\underrightarrow{\psi^{\ast}}(\mathbf{r})]\right\} \nonumber \\
 &  & \,\label{eq:FnalFPEV3InteractionCubic}\end{eqnarray}
Note that these third order terms are not included in the functional
Fokker-Planck equation for the Bogoliubov Hamiltonian.

\subsection{Summary of Results}

The \emph{functional Fokker-Planck equation} may be written in the
form

\begin{eqnarray}
 &  & \left(\frac{\partial}{\partial t}P[\underrightarrow{\psi}(\mathbf{r}),\underrightarrow{\psi^{\ast}}(\mathbf{r})]\right)\nonumber \\
 & = & \left(\frac{\partial}{\partial t}P[\underrightarrow{\psi}(\mathbf{r}),\underrightarrow{\psi^{\ast}}(\mathbf{r})]\right)_{C}+\left(\frac{\partial}{\partial t}P[\underrightarrow{\psi}(\mathbf{r}),\underrightarrow{\psi^{\ast}}(\mathbf{r})]\right)_{NC}\nonumber \\
 &  & +\left(\frac{\partial}{\partial t}P[\underrightarrow{\psi}(\mathbf{r}),\underrightarrow{\psi^{\ast}}(\mathbf{r})]\right)_{V}\label{Eq.FokkerPlanck2}\end{eqnarray}
of the sum of terms from the condensate, non-condensate and interaction
terms in the Hamiltonian.

\subsubsection{Condensate Hamiltonian Terms}

The contributions to the \emph{functional Fokker-Planck equation}
from the \emph{condensate Hamiltonian} may be written in the form

\begin{eqnarray}
 &  & \left(\frac{\partial}{\partial t}P[\underrightarrow{\psi}(\mathbf{r}),\underrightarrow{\psi^{\ast}}(\mathbf{r})]\right)_{C}\nonumber \\
 & = & \left(\frac{\partial}{\partial t}P[\underrightarrow{\psi}(\mathbf{r}),\underrightarrow{\psi^{\ast}}(\mathbf{r})]\right)_{K}+\left(\frac{\partial}{\partial t}P[\underrightarrow{\psi}(\mathbf{r}),\underrightarrow{\psi^{\ast}}(\mathbf{r})]\right)_{V}\nonumber \\
 &  & +\left(\frac{\partial}{\partial t}P[\underrightarrow{\psi}(\mathbf{r}),\underrightarrow{\psi^{\ast}}(\mathbf{r})]\right)_{U}\label{Eq.FnalFPCondensate}\end{eqnarray}
of the sum of terms from the kinetic energy, the trap potential and
the boson-boson interaction. Derivations of the form for each term
are given in \ref{App 5}. Here and elsewhere $\partial_{\mu}$ is
short for $\frac{{\LARGE\partial}}{{\LARGE\partial s}_{\mu}}$.

The contribution to the functional Fokker-Planck equation from the
\emph{kinetic energy} is given by\begin{eqnarray}
 &  & \left(\frac{\partial}{\partial t}P[\underrightarrow{\psi}(\mathbf{r}),\underrightarrow{\psi^{\ast}}(\mathbf{r})]\right)_{K}\nonumber \\
 & = & \frac{-i}{\hbar}\left\{ -\int d\mathbf{s}\left\{ \frac{\delta}{\delta\psi_{C}^{+}(\mathbf{s})}\left(\sum\limits _{\mu}\frac{\hbar^{2}}{2m}\partial_{\mu}^{2}\psi_{C}^{+}(\mathbf{s})\right)P[\underrightarrow{\psi}(\mathbf{r}),\underrightarrow{\psi^{\ast}}(\mathbf{r})]\right\} \right\} \nonumber \\
 &  & +\frac{-i}{\hbar}\left\{ +\int d\mathbf{s}\left\{ \frac{\delta}{\delta\psi_{C}(\mathbf{s})}\left(\sum\limits _{\mu}\frac{\hbar^{2}}{2m}\partial_{\mu}^{2}\psi_{C}(\mathbf{s})\right)P[\underrightarrow{\psi}(\mathbf{r}),\underrightarrow{\psi^{\ast}}(\mathbf{r})]\right\} \right\} \nonumber \\
 &  & \,\label{eq:FnalFPCondensateKinetic}\end{eqnarray}

The contribution to the functional Fokker-Planck equation from the
\emph{trap potential} is given by\begin{eqnarray}
 &  & \left(\frac{\partial}{\partial t}P[\underrightarrow{\psi}(\mathbf{r}),\underrightarrow{\psi^{\ast}}(\mathbf{r})]\right)_{V}\nonumber \\
 & = & \frac{-i}{\hbar}\left\{ -\int d\mathbf{s}\left\{ \frac{\delta}{\delta\psi_{C}(\mathbf{s})}\{V(\mathbf{s})\psi_{C}(\mathbf{s})\}\right\} P[\underrightarrow{\psi}(\mathbf{r}),\underrightarrow{\psi^{\ast}}(\mathbf{r})]\right\} \nonumber \\
 &  & \frac{-i}{\hbar}\left\{ +\int d\mathbf{s}\left\{ \frac{\delta}{\delta\psi_{C}^{+}(\mathbf{s})}\{V(\mathbf{s})\psi_{C}^{+}(\mathbf{s})\}\right\} P[\underrightarrow{\psi}(\mathbf{r}),\underrightarrow{\psi^{\ast}}(\mathbf{r})]\right\} \nonumber \\
 &  & \,\label{eq:FnalFPCondensateTrap}\end{eqnarray}

The contribution to the functional Fokker-Planck equation from the
\emph{boson-boson interaction} is given by\begin{eqnarray}
 &  & \left(\frac{\partial}{\partial t}P[\underrightarrow{\psi}(\mathbf{r}),\underrightarrow{\psi^{\ast}}(\mathbf{r})]\right)_{U}\nonumber \\
 & = & \frac{-i}{\hbar}\left\{ -g\int d\mathbf{s}\frac{\delta}{\delta\psi_{C}(\mathbf{s})}\left\{ [\psi_{C}^{+}(\mathbf{s})\psi_{C}(\mathbf{s})-\delta_{C}(\mathbf{s},\mathbf{s})]\psi_{C}(\mathbf{s})\right\} P[\underrightarrow{\psi}(\mathbf{r}),\underrightarrow{\psi^{\ast}}(\mathbf{r})]\right\} \nonumber \\
 &  & \frac{-i}{\hbar}\left\{ +g\int d\mathbf{s}\frac{\delta}{\delta\psi_{C}^{+}(\mathbf{s})}\left\{ [\psi_{C}^{+}(\mathbf{s})\psi_{C}(\mathbf{s})-\delta_{C}(\mathbf{s},\mathbf{s})]\psi_{C}^{+}(\mathbf{s})\right\} P[\underrightarrow{\psi}(\mathbf{r}),\underrightarrow{\psi^{\ast}}(\mathbf{r})]\right\} \nonumber \\
 &  & \frac{-i}{\hbar}\left\{ g\int d\mathbf{s}\frac{\delta}{\delta\psi_{C}(\mathbf{s})}\frac{\delta}{\delta\psi_{C}(\mathbf{s})}\frac{\delta}{\delta\psi_{C}^{+}(\mathbf{s})}\{\frac{1}{4}\psi_{C}(\mathbf{s})\}P[\underrightarrow{\psi}(\mathbf{r}),\underrightarrow{\psi^{\ast}}(\mathbf{r})]\right\} \nonumber \\
 &  & \frac{-i}{\hbar}\left\{ -g\int d\mathbf{s}\frac{\delta}{\delta\psi_{C}^{+}(\mathbf{s})}\frac{\delta}{\delta\psi_{C}^{+}(\mathbf{s})}\frac{\delta}{\delta\psi_{C}(\mathbf{s})}\{\frac{1}{4}\psi_{C}^{+}(\mathbf{s})\}P[\underrightarrow{\psi}(\mathbf{r}),\underrightarrow{\psi^{\ast}}(\mathbf{r})]\right\} \nonumber \\
 &  & \,\label{eq:FnalFPCondensateBosonIntn}\end{eqnarray}
which involves first order and third order functional derivatives.
The quantity $\delta_{C}(\mathbf{s},\mathbf{s})$ is a diagonal element
of the restricted delta function for condensate modes. We note that\begin{equation}
\int d\mathbf{s\,}\delta_{C}(\mathbf{s},\mathbf{s})=1\label{Eq.NormCondDeltaFn}\end{equation}
corresponding to there being a single occupied condensate mode in
this treatment. The total condensate number given by\begin{align}
N_{C} & =\iiiint D^{2}\psi_{C}\, D^{2}\psi_{C}^{+}\, D^{2}\psi_{NC}\, D^{2}\psi_{NC}^{+}\int d\mathbf{s}(\psi_{C}^{+}(\mathbf{s})\psi_{C}(\mathbf{s}))P[\underrightarrow{\psi}(\mathbf{r}),\underrightarrow{\psi^{\ast}}(\mathbf{r})]\nonumber \\
 & \,\label{eq:TotalCondNumber-1}\end{align}
is depleted by one.

\subsubsection{Non-Condensate Hamiltonian Terms}

The contributions to the \emph{functional Fokker-Planck equation}
from the \emph{non-condensate Hamiltonian} may be written in the form

\begin{eqnarray}
 &  & \left(\frac{\partial}{\partial t}P[\underrightarrow{\psi}(\mathbf{r}),\underrightarrow{\psi^{\ast}}(\mathbf{r})]\right)_{NC}\nonumber \\
 & = & \left(\frac{\partial}{\partial t}P[\underrightarrow{\psi}(\mathbf{r}),\underrightarrow{\psi^{\ast}}(\mathbf{r})]\right)_{K}+\left(\frac{\partial}{\partial t}P[\underrightarrow{\psi}(\mathbf{r}),\underrightarrow{\psi^{\ast}}(\mathbf{r})]\right)_{V}\nonumber \\
 &  & +\left(\frac{\partial}{\partial t}P[\underrightarrow{\psi}(\mathbf{r}),\underrightarrow{\psi^{\ast}}(\mathbf{r})]\right)_{U}\label{Eq.FnalFPNonCondensate}\end{eqnarray}
of the sum of terms from the kinetic energy, the trap potential and
the boson-boson interaction. Derivations of the form for each term
are given in \ref{App 5}.

The contribution to the functional Fokker-Planck equation from the
\emph{kinetic energy} is given by\begin{eqnarray}
 &  & \left(\frac{\partial}{\partial t}P[\underrightarrow{\psi}(\mathbf{r}),\underrightarrow{\psi^{\ast}}(\mathbf{r})]\right)_{K}\nonumber \\
 & = & \frac{-i}{\hbar}\left\{ -\int d\mathbf{s}\left\{ \frac{\delta}{\delta\psi_{NC}^{+}(\mathbf{s})}\left(\sum\limits _{\mu}\frac{\hbar^{2}}{2m}\partial_{\mu}^{2}\psi_{NC}^{+}(\mathbf{s})\right)P[\underrightarrow{\psi}(\mathbf{r}),\underrightarrow{\psi^{\ast}}(\mathbf{r})]\right\} \right\} \nonumber \\
 &  & +\frac{-i}{\hbar}\left\{ +\int d\mathbf{s}\left\{ \frac{\delta}{\delta\psi_{NC}(\mathbf{s})}\left(\sum\limits _{\mu}\frac{\hbar^{2}}{2m}\partial_{\mu}^{2}\psi_{NC}(\mathbf{s})\right)P[\underrightarrow{\psi}(\mathbf{r}),\underrightarrow{\psi^{\ast}}(\mathbf{r})]\right\} \right\} \nonumber \\
 &  & \,\label{eq:FnalFPNonCondensateKinetic}\end{eqnarray}

The contribution to the functional Fokker-Planck equation from the
\emph{trap potential} is given by\begin{eqnarray}
 &  & \left(\frac{\partial}{\partial t}P[\underrightarrow{\psi}(\mathbf{r}),\underrightarrow{\psi^{\ast}}(\mathbf{r})]\right)_{V}\nonumber \\
 & = & \frac{-i}{\hbar}\left\{ -\int d\mathbf{s}\left\{ \frac{\delta}{\delta\psi_{NC}(\mathbf{s})}\{V(\mathbf{s})\psi_{NC}(\mathbf{s})\}\right\} P[\underrightarrow{\psi}(\mathbf{r}),\underrightarrow{\psi^{\ast}}(\mathbf{r})]\right\} \nonumber \\
 &  & \frac{-i}{\hbar}\left\{ +\int d\mathbf{s}\left\{ \frac{\delta}{\delta\psi_{NC}^{+}(\mathbf{s})}V(\mathbf{s})\psi_{NC}^{+}(\mathbf{s})\right\} P[\underrightarrow{\psi}(\mathbf{r}),\underrightarrow{\psi^{\ast}}(\mathbf{r})]\right\} \nonumber \\
 &  & \,\label{eq:FnalFPNonCondensateTrap}\end{eqnarray}

The contribution to the functional Fokker-Planck equation from the
\emph{boson-boson interaction} is given by \begin{eqnarray}
 &  & \left(\frac{\partial}{\partial t}P[\underrightarrow{\psi}(\mathbf{r}),\underrightarrow{\psi^{\ast}}(\mathbf{r})]\right)_{U}\nonumber \\
 & = & \frac{-i}{\hbar}\left\{ -g\int d\mathbf{s}\frac{\delta}{\delta\psi_{NC}(\mathbf{s})}\left\{ [\psi_{NC}^{+}(\mathbf{s})\psi_{NC}(\mathbf{s})]\psi_{NC}(\mathbf{s})\right\} P[\underrightarrow{\psi}(\mathbf{r}),\underrightarrow{\psi^{\ast}}(\mathbf{r})]\right\} \nonumber \\
 &  & \frac{-i}{\hbar}\left\{ +g\int d\mathbf{s}\frac{\delta}{\delta\psi_{NC}^{+}(\mathbf{s})}\left\{ [\psi_{NC}^{+}(\mathbf{s})\psi_{NC}(\mathbf{s})]\psi_{NC}^{+}(\mathbf{s})\right\} P[\underrightarrow{\psi}(\mathbf{r}),\underrightarrow{\psi^{\ast}}(\mathbf{r})]\right\} \nonumber \\
 &  & \frac{-i}{\hbar}\left\{ +g\int d\mathbf{s}\frac{\delta}{\delta\psi_{NC}(\mathbf{s})}\frac{\delta}{\delta\psi_{NC}(\mathbf{s})}\{\frac{1}{2}\psi_{NC}(\mathbf{s})\psi_{NC}(\mathbf{s})\}P[\underrightarrow{\psi}(\mathbf{r}),\underrightarrow{\psi^{\ast}}(\mathbf{r})]\right\} \nonumber \\
 &  & \frac{-i}{\hbar}\left\{ -g\int d\mathbf{s}\frac{\delta}{\delta\psi_{NC}^{+}(\mathbf{s})}\frac{\delta}{\delta\psi_{NC}^{+}(\mathbf{s})}\{\frac{1}{2}\psi_{NC}^{+}(\mathbf{s})\psi_{NC}^{+}(\mathbf{s})\}P[\underrightarrow{\psi}(\mathbf{r}),\underrightarrow{\psi^{\ast}}(\mathbf{r})]\right\} \nonumber \\
 &  & \,\label{eq:FnalFPNonCondensateBosonBoson}\end{eqnarray}
This term is part of the interaction term $\widehat{H}_{5}$ and its
contribution to the functional Fokker-Planck equation will be ignored.

\subsubsection{Interaction between Condensate and Non-Condensate Terms}

The contributions to the \emph{functional Fokker-Planck equation}
from the \emph{interaction Hamiltonian} between the condensate and
non-condensate may be written in the form

\begin{eqnarray}
 &  & \left(\frac{\partial}{\partial t}P[\underrightarrow{\psi}(\mathbf{r}),\underrightarrow{\psi^{\ast}}(\mathbf{r})]\right)_{V}\nonumber \\
 & = & \left(\frac{\partial}{\partial t}P[\underrightarrow{\psi}(\mathbf{r}),\underrightarrow{\psi^{\ast}}(\mathbf{r})]\right)_{V1}+\left(\frac{\partial}{\partial t}P[\underrightarrow{\psi}(\mathbf{r}),\underrightarrow{\psi^{\ast}}(\mathbf{r})]\right)_{V2}\nonumber \\
 &  & +\left(\frac{\partial}{\partial t}P[\underrightarrow{\psi}(\mathbf{r}),\underrightarrow{\psi^{\ast}}(\mathbf{r})]\right)_{V3}\label{Eq.FnalFPInteraction}\end{eqnarray}
of the sum of first, second and third order terms in the non-condensate
field operators. Derivations of the form for each term are given in
\ref{App 5}.

\subparagraph{First Order Terms}

The contribution to the \emph{functional Fokker-Planck equation} from
the \emph{first order term }in the \emph{interaction Hamiltonian}
between the condensate and non-condensate may be written in the form\begin{eqnarray}
 &  & \left(\frac{\partial}{\partial t}P[\underrightarrow{\psi}(\mathbf{r}),\underrightarrow{\psi^{\ast}}(\mathbf{r})]\right)_{V1}\nonumber \\
 & = & \left(\frac{\partial}{\partial t}P[\underrightarrow{\psi}(\mathbf{r}),\underrightarrow{\psi^{\ast}}(\mathbf{r})]\right)_{V14}+\left(\frac{\partial}{\partial t}P[\underrightarrow{\psi}(\mathbf{r}),\underrightarrow{\psi^{\ast}}(\mathbf{r})]\right)_{V12}\label{Eq.FnalFPFirstOrderInteraction}\end{eqnarray}

These two contributions may be written as the sum of terms which are
linear, quadratic, cubic and quartic in the number of functional derivatives.
For the $\widehat{V}_{14}$ term\begin{eqnarray}
 &  & \left(\frac{\partial}{\partial t}P[\underrightarrow{\psi}(\mathbf{r}),\underrightarrow{\psi^{\ast}}(\mathbf{r})]\right)_{V14}\nonumber \\
 & = & \left(\frac{\partial}{\partial t}P[\underrightarrow{\psi}(\mathbf{r}),\underrightarrow{\psi^{\ast}}(\mathbf{r})]\right)_{V14}^{1}+\left(\frac{\partial}{\partial t}P[\underrightarrow{\psi}(\mathbf{r}),\underrightarrow{\psi^{\ast}}(\mathbf{r})]\right)_{V14}^{2}\nonumber \\
 &  & +\left(\frac{\partial}{\partial t}P[\underrightarrow{\psi}(\mathbf{r}),\underrightarrow{\psi^{\ast}}(\mathbf{r})]\right)_{V14}^{3}+\left(\frac{\partial}{\partial t}P[\underrightarrow{\psi}(\mathbf{r}),\underrightarrow{\psi^{\ast}}(\mathbf{r})]\right)_{V14}^{4}\label{Eq.FnalFPFirstOrderV14Interaction}\end{eqnarray}

where\begin{eqnarray}
 &  & \left(\frac{\partial}{\partial t}P[\underrightarrow{\psi}(\mathbf{r}),\underrightarrow{\psi^{\ast}}(\mathbf{r})]\right)_{V14}^{1}\nonumber \\
 & = & \frac{-i}{\hbar}\left\{ +g\int d\mathbf{s\,}\left\{ \left(\frac{\delta}{\delta\psi_{C}^{+}(\mathbf{s})}\right)\{[2\psi_{C}^{+}(\mathbf{s})\psi_{C}(\mathbf{s})-\delta_{C}(\mathbf{s},\mathbf{s})]\psi_{NC}^{+}(\mathbf{s})\}\right\} P[\underrightarrow{\psi}(\mathbf{r}),\underrightarrow{\psi^{\ast}}(\mathbf{r})]\right\} \nonumber \\
 &  & \frac{-i}{\hbar}\left\{ +g\int d\mathbf{s\,}\left\{ \mathbf{\left(\frac{\delta}{\delta\psi_{C}^{+}(\mathbf{s})}\right)\{[}\psi_{C}^{+}\mathbf{(\mathbf{s})}\psi_{C}^{+}\mathbf{(\mathbf{s})]}\psi_{NC}\mathbf{(\mathbf{s})\}}\right\} P[\underrightarrow{\psi}(\mathbf{r}),\underrightarrow{\psi^{\ast}}(\mathbf{r})]\right\} \nonumber \\
 &  & \frac{-i}{\hbar}\left\{ -g\int d\mathbf{s\,}\left\{ \mathbf{\left(\frac{\delta}{\delta\psi_{C}(\mathbf{s})}\right)\{[}2\psi\mathbf{_{C}(\mathbf{s})}\psi_{C}^{+}\mathbf{(\mathbf{s})-\delta_{C}(\mathbf{s},\mathbf{s})]\psi}_{NC}\mathbf{(\mathbf{s})\}}\right\} P[\underrightarrow{\psi}(\mathbf{r}),\underrightarrow{\psi^{\ast}}(\mathbf{r})]\right\} \nonumber \\
 &  & \frac{-i}{\hbar}\left\{ -g\int d\mathbf{s\,}\left\{ \left(\frac{\delta}{\delta\psi_{C}(\mathbf{s})}\right)\{[\psi_{C}(\mathbf{s})\psi_{C}(\mathbf{s})]\psi_{NC}^{+}(\mathbf{s})\}\right\} P[\underrightarrow{\psi}(\mathbf{r}),\underrightarrow{\psi^{\ast}}(\mathbf{r})]\right\} \nonumber \\
 &  & \frac{-i}{\hbar}\left\{ -g\int d\mathbf{s\,}\left\{ \left(\frac{\delta}{\delta\psi_{NC}(\mathbf{s})}\right)\{[\psi_{C}^{+}(\mathbf{s})\psi_{C}(\mathbf{s})-\delta_{C}(\mathbf{s},\mathbf{s)}]\psi_{C}(\mathbf{s})\}\right\} P[\underrightarrow{\psi}(\mathbf{r}),\underrightarrow{\psi^{\ast}}(\mathbf{r})]\right\} \nonumber \\
 &  & \frac{-i}{\hbar}\left\{ +g\int d\mathbf{s\,}\left\{ \left(\frac{\delta}{\delta\psi_{NC}^{+}(\mathbf{s})}\right)\{[\psi_{C}(\mathbf{s})\psi_{C}^{+}(\mathbf{s})-\delta_{C}(\mathbf{s},\mathbf{s})]\psi_{C}^{+}(\mathbf{s})\}\right\} P[\underrightarrow{\psi}(\mathbf{r}),\underrightarrow{\psi^{\ast}}(\mathbf{r})]\right\} \nonumber \\
 &  & \,\label{eq:FnalFPFirstOrderV14InteractionLinear}\end{eqnarray}
\begin{eqnarray}
 &  & \left(\frac{\partial}{\partial t}P[\underrightarrow{\psi}(\mathbf{r}),\underrightarrow{\psi^{\ast}}(\mathbf{r})]\right)_{V14}^{2}\nonumber \\
 & = & \frac{-i}{\hbar}\left\{ -g\int d\mathbf{s\,}\left\{ \left(\frac{\delta}{\delta\psi_{C}^{+}(\mathbf{s})}\right)\left(\frac{\delta}{\delta\psi_{NC}(\mathbf{s})}\right)\{\psi_{C}^{+}(\mathbf{s})\psi_{C}(\mathbf{s})\}\right\} P[\underrightarrow{\psi}(\mathbf{r}),\underrightarrow{\psi^{\ast}}(\mathbf{r})]\right\} \nonumber \\
 &  & \frac{-i}{\hbar}\left\{ +g\int d\mathbf{s\,}\left\{ \left(\frac{\delta}{\delta\psi_{C}(\mathbf{s})}\right)\left(\frac{\delta}{\delta\psi_{NC}^{+}(\mathbf{s})}\right)\{\psi_{C}(\mathbf{s})\psi_{C}^{+}(\mathbf{s})\}\right\} P[\underrightarrow{\psi}(\mathbf{r}),\underrightarrow{\psi^{\ast}}(\mathbf{r})]\right\} \nonumber \\
 &  & \frac{-i}{\hbar}\left\{ +g\int d\mathbf{s\,}\left\{ \left(\frac{\delta}{\delta\psi_{C}^{+}(\mathbf{s})}\right)\left(\frac{\delta}{\delta\psi_{NC}(\mathbf{s})}\right)\left\{ \frac{1}{2}\delta_{C}(\mathbf{s},\mathbf{s})\right\} \right\} P[\underrightarrow{\psi}(\mathbf{r}),\underrightarrow{\psi^{\ast}}(\mathbf{r})]\right\} \nonumber \\
 &  & \frac{-i}{\hbar}\left\{ -g\int d\mathbf{s\,}\left\{ \left(\frac{\delta}{\delta\psi_{C}(\mathbf{s})}\right)\left(\frac{\delta}{\delta\psi_{NC}^{+}(\mathbf{s})}\right)\left\{ \frac{1}{2}\delta_{C}(\mathbf{s},\mathbf{s})\right\} \right\} P[\underrightarrow{\psi}(\mathbf{r}),\underrightarrow{\psi^{\ast}}(\mathbf{r})]\right\} \nonumber \\
 &  & \frac{-i}{\hbar}\left\{ +g\int d\mathbf{s\,}\left\{ \left(\frac{\delta}{\delta\psi_{C}(\mathbf{s})}\right)\left(\frac{\delta}{\delta\psi_{NC}(\mathbf{s})}\right)\{\frac{1}{2}\psi_{C}(\mathbf{s})\psi_{C}(\mathbf{s})\}\right\} P[\underrightarrow{\psi}(\mathbf{r}),\underrightarrow{\psi^{\ast}}(\mathbf{r})]\right\} \nonumber \\
 &  & \frac{-i}{\hbar}\left\{ -g\int d\mathbf{s\,}\left\{ \left(\frac{\delta}{\delta\psi_{C}^{+}(\mathbf{s})}\right)\left(\frac{\delta}{\delta\psi_{NC}^{+}(\mathbf{s})}\right)\{\frac{1}{2}\psi_{C}^{+}(\mathbf{s})\psi_{C}^{+}(\mathbf{s})\}\right\} P[\underrightarrow{\psi}(\mathbf{r}),\underrightarrow{\psi^{\ast}}(\mathbf{r})]\right\} \nonumber \\
 &  & \,\label{eq:FnalFPFirstOrderV14InteractionQuadratic}\end{eqnarray}

\begin{eqnarray}
 &  & \left(\frac{\partial}{\partial t}P[\underrightarrow{\psi}(\mathbf{r}),\underrightarrow{\psi^{\ast}}(\mathbf{r})]\right)_{V14}^{3}\nonumber \\
 & = & \frac{-i}{\hbar}\left\{ -g\int d\mathbf{s\,}\left\{ \left(\frac{\delta}{\delta\psi_{C}^{+}(\mathbf{s})}\right)\left(\frac{\delta}{\delta\psi_{C}^{+}(\mathbf{s})}\right)\left(\frac{\delta}{\delta\psi_{C}(\mathbf{s})}\right)\{\frac{1}{4}\psi_{NC}^{+}(\mathbf{s})\}\right\} P[\underrightarrow{\psi}(\mathbf{r}),\underrightarrow{\psi^{\ast}}(\mathbf{r})]\right\} \nonumber \\
 &  & \frac{-i}{\hbar}\left\{ +g\int d\mathbf{s\,}\left\{ \left(\frac{\delta}{\delta\psi_{C}(\mathbf{s})}\right)\left(\frac{\delta}{\delta\psi_{C}(\mathbf{s})}\right)\left(\frac{\delta}{\delta\psi_{C}^{+}(\mathbf{s})}\right)\{\frac{1}{4}\psi_{NC}(\mathbf{s})\}\right\} P[\underrightarrow{\psi}(\mathbf{r}),\underrightarrow{\psi^{\ast}}(\mathbf{r})]\right\} \nonumber \\
 &  & \frac{-i}{\hbar}\left\{ -g\int d\mathbf{s\,}\left\{ \left(\frac{\delta}{\delta\psi_{C}^{+}(\mathbf{s})}\right)\left(\frac{\delta}{\delta\psi_{C}^{+}(\mathbf{s})}\right)\left(\frac{\delta}{\delta\psi_{NC}(\mathbf{s})}\right)\{\frac{1}{4}\psi_{C}^{+}(\mathbf{s})\}\right\} P[\underrightarrow{\psi}(\mathbf{r}),\underrightarrow{\psi^{\ast}}(\mathbf{r})]\right\} \nonumber \\
 &  & \frac{-i}{\hbar}\left\{ +g\int d\mathbf{s\,}\left\{ \left(\frac{\delta}{\delta\psi_{C}(\mathbf{s})}\right)\left(\frac{\delta}{\delta\psi_{C}(\mathbf{s})}\right)\left(\frac{\delta}{\delta\psi_{NC}^{+}(\mathbf{s})}\right)\{\frac{1}{4}\psi_{C}(\mathbf{s})\}\right\} P[\underrightarrow{\psi}(\mathbf{r}),\underrightarrow{\psi^{\ast}}(\mathbf{r})]\right\} \nonumber \\
 &  & \frac{-i}{\hbar}\left\{ +g\int d\mathbf{s\,}\left\{ \left(\frac{\delta}{\delta\psi_{C}^{+}(\mathbf{s})}\right)\left(\frac{\delta}{\delta\psi_{C}(\mathbf{s})}\right)\left(\frac{\delta}{\delta\psi_{NC}(\mathbf{s})}\right)\{\frac{1}{2}\psi_{C}(\mathbf{s})\}\right\} P[\underrightarrow{\psi}(\mathbf{r}),\underrightarrow{\psi^{\ast}}(\mathbf{r})]\right\} \nonumber \\
 &  & \frac{-i}{\hbar}\left\{ -g\int d\mathbf{s\,}\left\{ \left(\frac{\delta}{\delta\psi_{C}(\mathbf{s})}\right)\left(\frac{\delta}{\delta\psi_{C}^{+}(\mathbf{s})}\right)\left(\frac{\delta}{\delta\psi_{NC}^{+}(\mathbf{s})}\right)\{\frac{1}{2}\psi_{C}^{+}(\mathbf{s})\}\right\} P[\underrightarrow{\psi}(\mathbf{r}),\underrightarrow{\psi^{\ast}}(\mathbf{r})]\right\} \nonumber \\
 &  & \,\label{eq:FnalFPFirstOrderV14InteractionCubic}\end{eqnarray}

\begin{eqnarray}
 &  & \left(\frac{\partial}{\partial t}P[\underrightarrow{\psi}(\mathbf{r}),\underrightarrow{\psi^{\ast}}(\mathbf{r})]\right)_{V14}^{4}\nonumber \\
 & = & \frac{-i}{\hbar}\left\{ g\int d\mathbf{s\,}\left\{ \left(\frac{\delta}{\delta\psi_{C}^{+}(\mathbf{s})}\right)\left(\frac{\delta}{\delta\psi_{C}^{+}(\mathbf{s})}\right)\left(\frac{\delta}{\delta\psi_{C}(\mathbf{s})}\right)\left(\frac{\delta}{\delta\psi_{NC}(\mathbf{s})}\right)\{\frac{1}{8}\}\right\} P[\underrightarrow{\psi}(\mathbf{r}),\underrightarrow{\psi^{\ast}}(\mathbf{r})]\right\} \nonumber \\
 &  & \frac{-i}{\hbar}\left\{ -g\int d\mathbf{s}\left\{ \left(\frac{\delta}{\delta\psi_{C}(\mathbf{s})}\right)\left(\frac{\delta}{\delta\psi_{C}(\mathbf{s})}\right)\left(\frac{\delta}{\delta\psi_{C}^{+}(\mathbf{s})}\right)\left(\frac{\delta}{\delta\psi_{NC}^{+}(\mathbf{s})}\right)\{\frac{1}{8}\}\right\} P[\underrightarrow{\psi}(\mathbf{r}),\underrightarrow{\psi^{\ast}}(\mathbf{r})]\right\} \nonumber \\
 &  & \,\label{eq:FnalFPFirstOrderV14InteractionQuartic}\end{eqnarray}

For the $\widehat{V}_{12}$ term

\begin{eqnarray}
 &  & \left(\frac{\partial}{\partial t}P[\underrightarrow{\psi}(\mathbf{r}),\underrightarrow{\psi^{\ast}}(\mathbf{r})]\right)_{V12}\nonumber \\
 & = & \left(\frac{\partial}{\partial t}P[\underrightarrow{\psi}(\mathbf{r}),\underrightarrow{\psi^{\ast}}(\mathbf{r})]\right)_{V12}^{1}+\left(\frac{\partial}{\partial t}P[\underrightarrow{\psi}(\mathbf{r}),\underrightarrow{\psi^{\ast}}(\mathbf{r})]\right)_{V12}^{2}\label{Eq.FnalFPFirstOrderV12Interaction}\end{eqnarray}
where for the two mode condensate case \begin{eqnarray}
 &  & \left(\frac{\partial}{\partial t}P[\underrightarrow{\psi}(\mathbf{r}),\underrightarrow{\psi^{\ast}}(\mathbf{r})]\right)_{V12}^{1}\nonumber \\
 & = & \frac{-i}{\hbar}\left\{ -g\int\int d\mathbf{s\,}d\mathbf{u}\left\{ \left(\frac{\delta}{\delta\psi_{C}^{+}(\mathbf{u})}\right)\{F(\mathbf{s},\mathbf{u})\,\psi_{NC}^{+}(\mathbf{s})\}\right\} \, P[\underrightarrow{\psi}(\mathbf{r}),\underrightarrow{\psi^{\ast}}(\mathbf{r})]\right\} \nonumber \\
 &  & +\frac{-i}{\hbar}\left\{ +g\int\int d\mathbf{s\,}d\mathbf{u}\left\{ \left(\frac{\delta}{\delta\psi_{C}(\mathbf{s})}\right)\{F(\mathbf{u},\mathbf{s})^{\ast}\,\psi_{NC}(\mathbf{u})\}\right\} \, P[\underrightarrow{\psi}(\mathbf{r}),\underrightarrow{\psi^{\ast}}(\mathbf{r})]\right\} \nonumber \\
 &  & +\frac{-i}{\hbar}\left\{ -g\int\int d\mathbf{s\, d\mathbf{u\,}}\left\{ \left(\frac{\delta}{\delta\psi_{NC}^{+}(\mathbf{u})}\right)\{F\mathbf{(\mathbf{u},\mathbf{s})^{\ast}}\,\psi_{C}^{+}(\mathbf{s})\}\right\} \, P[\underrightarrow{\psi}(\mathbf{r}),\underrightarrow{\psi^{\ast}}(\mathbf{r})]\right\} \nonumber \\
 &  & +\frac{-i}{\hbar}\left\{ +g\int\int d\mathbf{s\,}d\mathbf{u}\left\{ \left(\frac{\delta}{\delta\psi_{NC}(\mathbf{s})}\right)\{F(\mathbf{s},\mathbf{u})\,\psi_{C}(\mathbf{u})\}\right\} \, P[\underrightarrow{\psi}(\mathbf{r}),\underrightarrow{\psi^{\ast}}(\mathbf{r})]\right\} \nonumber \\
 &  & \,\label{eq:NewFnalFPV12Linear3}\end{eqnarray}
\begin{eqnarray}
 &  & \left(\frac{\partial}{\partial t}P[\underrightarrow{\psi}(\mathbf{r}),\underrightarrow{\psi^{\ast}}(\mathbf{r})]\right)_{V12}^{2}\nonumber \\
 & = & \frac{-i}{\hbar}\left\{ +g\int\int d\mathbf{s\,}d\mathbf{u}\left\{ \left(\frac{\delta}{\delta\psi_{C}^{+}(\mathbf{u})}\right)\left(\frac{\delta}{\delta\psi_{NC}(\mathbf{s})}\right)\{\frac{1}{2}F(\mathbf{s},\mathbf{u})\}\right\} \, P[\underrightarrow{\psi}(\mathbf{r}),\underrightarrow{\psi^{\ast}}(\mathbf{r})]\right\} \nonumber \\
 &  & +\frac{-i}{\hbar}\left\{ -g\int\int d\mathbf{s\, d\mathbf{u\,}}\left\{ \left(\frac{\delta}{\delta\psi_{C}(\mathbf{s})}\right)\left(\frac{\delta}{\delta\psi_{NC}^{+}(\mathbf{u})}\right)\{\frac{1}{2}\mathbf{F(\mathbf{u},\mathbf{s})^{\ast}}\}\right\} \, P[\underrightarrow{\psi}(\mathbf{r}),\underrightarrow{\psi^{\ast}}(\mathbf{r})]\right\} \nonumber \\
 &  & \,\label{eq:NewFnalFPV12Quadratic3}\end{eqnarray}
These results may also be written as \begin{eqnarray}
 &  & \left(\frac{\partial}{\partial t}P[\underrightarrow{\psi}(\mathbf{r}),\underrightarrow{\psi^{\ast}}(\mathbf{r})]\right)_{V12}^{1}\nonumber \\
 & = & \frac{-i}{\hbar}\left\{ -g\int d\mathbf{s\,}\left\{ \left(\frac{\delta}{\delta\psi_{C}^{+}(\mathbf{s})}\right)\{\int d\mathbf{u\,}F(\mathbf{u},\mathbf{s})\,\psi_{NC}^{+}(\mathbf{u})\}\right\} \, P[\underrightarrow{\psi}(\mathbf{r}),\underrightarrow{\psi^{\ast}}(\mathbf{r})]\right\} \nonumber \\
 &  & +\frac{-i}{\hbar}\left\{ +g\int d\mathbf{s\,}\left\{ \left(\frac{\delta}{\delta\psi_{C}(\mathbf{s})}\right)\{\int d\mathbf{u\,}F(\mathbf{u},\mathbf{s})^{\ast}\,\psi_{NC}(\mathbf{u})\}\right\} \, P[\underrightarrow{\psi}(\mathbf{r}),\underrightarrow{\psi^{\ast}}(\mathbf{r})]\right\} \nonumber \\
 &  & +\frac{-i}{\hbar}\left\{ -g\int d\mathbf{s\,}\left\{ \left(\frac{\delta}{\delta\psi_{NC}^{+}(\mathbf{s})}\right)\{\int d\mathbf{u\,}F\mathbf{(s,u)^{\ast}}\,\psi_{C}^{+}(\mathbf{u})\}\right\} \, P[\underrightarrow{\psi}(\mathbf{r}),\underrightarrow{\psi^{\ast}}(\mathbf{r})]\right\} \nonumber \\
 &  & +\frac{-i}{\hbar}\left\{ +g\int d\mathbf{s\,}\left\{ \left(\frac{\delta}{\delta\psi_{NC}(\mathbf{s})}\right)\{\int d\mathbf{u\,}F(\mathbf{s},\mathbf{u})\,\psi_{C}(\mathbf{u})\}\right\} \, P[\underrightarrow{\psi}(\mathbf{r}),\underrightarrow{\psi^{\ast}}(\mathbf{r})]\right\} \nonumber \\
 &  & \,\end{eqnarray}
so the quantity inside the inner brackets is just another functional.
The quadratic term is left unchanged except for interchanging positions
to make the expression more symmetrical\begin{eqnarray}
 &  & \left(\frac{\partial}{\partial t}P[\underrightarrow{\psi}(\mathbf{r}),\underrightarrow{\psi^{\ast}}(\mathbf{r})]\right)_{V12}^{2}\nonumber \\
 & = & \frac{-i}{\hbar}\left\{ +g\int\int d\mathbf{s\,}d\mathbf{u}\left\{ \left(\frac{\delta}{\delta\psi_{C}^{+}(\mathbf{s})}\right)\left(\frac{\delta}{\delta\psi_{NC}(\mathbf{u})}\right)\{\frac{1}{2}F(\mathbf{u},\mathbf{s})\}\right\} \, P[\underrightarrow{\psi}(\mathbf{r}),\underrightarrow{\psi^{\ast}}(\mathbf{r})]\right\} \nonumber \\
 &  & +\frac{-i}{\hbar}\left\{ -g\int\int d\mathbf{s\, d\mathbf{u\,}}\left\{ \left(\frac{\delta}{\delta\psi_{C}(\mathbf{s})}\right)\left(\frac{\delta}{\delta\psi_{NC}^{+}(\mathbf{u})}\right)\{\frac{1}{2}\mathbf{F(\mathbf{u},\mathbf{s})^{\ast}}\}\right\} \, P[\underrightarrow{\psi}(\mathbf{r}),\underrightarrow{\psi^{\ast}}(\mathbf{r})]\right\} \nonumber \\
 &  & \,\end{eqnarray}

For the single mode condensate case \begin{eqnarray}
 &  & \left(\frac{\partial}{\partial t}P[\underrightarrow{\psi}(\mathbf{r}),\underrightarrow{\psi^{\ast}}(\mathbf{r})]\right)_{V12}^{1}\nonumber \\
 & = & \frac{-i}{\hbar}\left\{ -g\int d\mathbf{s\,}\left\{ \left(\frac{\delta}{\delta\psi_{C}^{+}(\mathbf{s})}\right)\{\left\langle \widehat{\Psi}_{C}(\mathbf{s})^{\dagger}\widehat{\Psi}_{C}(\mathbf{s})\right\rangle \,\psi_{NC}^{+}(\mathbf{s})\}\right\} \, P[\underrightarrow{\psi}(\mathbf{r}),\underrightarrow{\psi^{\ast}}(\mathbf{r})]\right\} \nonumber \\
 &  & \frac{-i}{\hbar}\left\{ +g\int d\mathbf{s\,}\left\{ \left(\frac{\delta}{\delta\psi_{C}(\mathbf{s})}\right)\{\left\langle \widehat{\Psi}_{C}(\mathbf{s})^{\dagger}\widehat{\Psi}_{C}(\mathbf{s})\right\rangle \,\psi_{NC}(\mathbf{s})\}\right\} \, P[\underrightarrow{\psi}(\mathbf{r}),\underrightarrow{\psi^{\ast}}(\mathbf{r})]\right\} \nonumber \\
 &  & \frac{-i}{\hbar}\left\{ -g\int d\mathbf{s\,}\left\{ \left(\frac{\delta}{\delta\psi_{NC}^{+}(\mathbf{s})}\right)\{\left\langle \widehat{\Psi}_{C}(\mathbf{s})^{\dagger}\widehat{\Psi}_{C}(\mathbf{s})\right\rangle \,\psi_{C}^{+}(\mathbf{s})\}\right\} \, P[\underrightarrow{\psi}(\mathbf{r}),\underrightarrow{\psi^{\ast}}(\mathbf{r})]\right\} \nonumber \\
 &  & \frac{-i}{\hbar}\left\{ +g\int d\mathbf{s\,}\left\{ \left(\frac{\delta}{\delta\psi_{NC}(\mathbf{s})}\right)\{\left\langle \widehat{\Psi}_{C}(\mathbf{s})^{\dagger}\widehat{\Psi}_{C}(\mathbf{s})\right\rangle \,\psi_{C}(\mathbf{s})\}\right\} \, P[\underrightarrow{\psi}(\mathbf{r}),\underrightarrow{\psi^{\ast}}(\mathbf{r})]\right\} \nonumber \\
 &  & \,\end{eqnarray}
\begin{eqnarray}
 &  & \left(\frac{\partial}{\partial t}P[\underrightarrow{\psi}(\mathbf{r}),\underrightarrow{\psi^{\ast}}(\mathbf{r})]\right)_{V12}^{2}\nonumber \\
 & = & \frac{-i}{\hbar}\left\{ +g\int d\mathbf{s\,}\left\{ \left(\frac{\delta}{\delta\psi_{C}^{+}(\mathbf{s})}\right)\left(\frac{\delta}{\delta\psi_{NC}(\mathbf{s})}\right)\{\frac{1}{2}\left\langle \widehat{\Psi}_{C}(\mathbf{r})^{\dagger}\widehat{\Psi}_{C}(\mathbf{r})\right\rangle \}\right\} \, P[\underrightarrow{\psi}(\mathbf{r}),\underrightarrow{\psi^{\ast}}(\mathbf{r})]\right\} \nonumber \\
 &  & \frac{-i}{\hbar}\left\{ -g\int d\mathbf{s\,}\left\{ \left(\frac{\delta}{\delta\psi_{C}(\mathbf{s})}\right)\left(\frac{\delta}{\delta\psi_{NC}^{+}(\mathbf{s})}\right)\{\frac{1}{2}\left\langle \widehat{\Psi}_{C}(\mathbf{r})^{\dagger}\widehat{\Psi}_{C}(\mathbf{r})\right\rangle \}\right\} \, P[\underrightarrow{\psi}(\mathbf{r}),\underrightarrow{\psi^{\ast}}(\mathbf{r})]\right\} \nonumber \\
 &  & \,\end{eqnarray}

\subparagraph{Second Order Terms}

The contribution to the \emph{functional Fokker-Planck equation} from
the \emph{second order term }in the \emph{interaction Hamiltonian}
between the condensate and non-condensate may be written as the sum
of terms which are linear, quadratic, cubic and quartic in the number
of functional derivatives\begin{eqnarray}
 &  & \left(\frac{\partial}{\partial t}P[\underrightarrow{\psi}(\mathbf{r}),\underrightarrow{\psi^{\ast}}(\mathbf{r})]\right)_{V2}\nonumber \\
 & = & \left(\frac{\partial}{\partial t}P[\underrightarrow{\psi}(\mathbf{r}),\underrightarrow{\psi^{\ast}}(\mathbf{r})]\right)_{V2}^{1}+\left(\frac{\partial}{\partial t}P[\underrightarrow{\psi}(\mathbf{r}),\underrightarrow{\psi^{\ast}}(\mathbf{r})]\right)_{V2}^{2}\nonumber \\
 &  & +\left(\frac{\partial}{\partial t}P[\underrightarrow{\psi}(\mathbf{r}),\underrightarrow{\psi^{\ast}}(\mathbf{r})]\right)_{V2}^{3}+\left(\frac{\partial}{\partial t}P[\underrightarrow{\psi}(\mathbf{r}),\underrightarrow{\psi^{\ast}}(\mathbf{r})]\right)_{V2}^{4}\label{Eq.FnalFPSecondOrderV2Interaction}\end{eqnarray}
where \begin{eqnarray}
 &  & \left(\frac{\partial}{\partial t}P[\underrightarrow{\psi}(\mathbf{r}),\underrightarrow{\psi^{\ast}}(\mathbf{r})]\right)_{V2}^{1}\nonumber \\
 & = & \frac{-i}{\hbar}\left\{ +g\int d\mathbf{s}\left(\frac{\delta}{\delta\psi_{C}^{+}(\mathbf{s})}\right)\{[\psi_{NC}^{+}(\mathbf{s})\psi_{C}(\mathbf{s})+2\psi_{C}^{+}(\mathbf{s})\psi_{NC}(\mathbf{s})]\psi_{NC}^{+}(\mathbf{s})\}P[\underrightarrow{\psi}(\mathbf{r}),\underrightarrow{\psi^{\ast}}(\mathbf{r})]\right\} \nonumber \\
 &  & \frac{-i}{\hbar}\left\{ -g\int d\mathbf{s}\left(\frac{\delta}{\delta\psi_{C}(\mathbf{s})}\right)\{[\psi_{NC}(\mathbf{s})\psi_{C}^{+}(\mathbf{s})+2\psi_{C}(\mathbf{s})\psi_{NC}^{+}(\mathbf{s})]\psi_{NC}(\mathbf{s})\}P[\underrightarrow{\psi}(\mathbf{r}),\underrightarrow{\psi^{\ast}}(\mathbf{r})]\right\} \nonumber \\
 &  & \frac{-i}{\hbar}\left\{ +g\int d\mathbf{s}\left\{ \left(\frac{\delta}{\delta\psi_{NC}^{+}(\mathbf{s})}\right)\{\psi_{NC}(\mathbf{s})\psi_{C}^{+}(\mathbf{s})\psi_{C}^{+}(\mathbf{s})\}\right\} P[\underrightarrow{\psi}(\mathbf{r}),\underrightarrow{\psi^{\ast}}(\mathbf{r})]\right\} \nonumber \\
 &  & \frac{-i}{\hbar}\left\{ +g\int d\mathbf{s}\left\{ \left(\frac{\delta}{\delta\psi_{NC}^{+}(\mathbf{s})}\right)\{[2\psi_{C}(\mathbf{s})\psi_{C}^{+}(\mathbf{s})-\delta_{C}(\mathbf{s},\mathbf{s})]\psi_{NC}^{+}(\mathbf{s})\}\right\} P[\underrightarrow{\psi}(\mathbf{r}),\underrightarrow{\psi^{\ast}}(\mathbf{r})]\right\} \nonumber \\
 &  & \frac{-i}{\hbar}\left\{ -g\int d\mathbf{s}\left\{ \left(\frac{\delta}{\delta\psi_{NC}(\mathbf{s})}\right)\{\psi_{NC}^{+}(\mathbf{s})\psi_{C}(\mathbf{s})\psi_{C}(\mathbf{s})\}\right\} P[\underrightarrow{\psi}(\mathbf{r}),\underrightarrow{\psi^{\ast}}(\mathbf{r})]\right\} \nonumber \\
 &  & \frac{-i}{\hbar}\left\{ -g\int d\mathbf{s}\left\{ \left(\frac{\delta}{\delta\psi_{NC}(\mathbf{s})}\right)\{[2\psi_{C}^{+}(\mathbf{s})\psi_{C}(\mathbf{s})-\delta_{C}(\mathbf{s,s})]\psi_{NC}(\mathbf{s})\}\right\} P[\underrightarrow{\psi}(\mathbf{r}),\underrightarrow{\psi^{\ast}}(\mathbf{r})]\right\} \nonumber \\
 &  & \,\label{eq:FnalFPSecondOrderV2InteractionLinear}\end{eqnarray}
\begin{eqnarray}
 &  & \left(\frac{\partial}{\partial t}P[\underrightarrow{\psi}(\mathbf{r}),\underrightarrow{\psi^{\ast}}(\mathbf{r})]\right)_{V2}^{2}\nonumber \\
 & = & \frac{-i}{\hbar}\left\{ -g\int d\mathbf{s}\left\{ \left(\frac{\delta}{\delta\psi_{C}^{+}(\mathbf{s})}\right)\left(\frac{\delta}{\delta\psi_{NC}(\mathbf{s})}\right)\{\psi_{NC}^{+}(\mathbf{s})\psi_{C}(\mathbf{s})+\psi_{C}^{+}(\mathbf{s})\psi_{NC}(\mathbf{s})\}\right\} P[\underrightarrow{\psi}(\mathbf{r}),\underrightarrow{\psi^{\ast}}(\mathbf{r})]\right\} \nonumber \\
 &  & \frac{-i}{\hbar}\left\{ +g\int d\mathbf{s}\left\{ \left(\frac{\delta}{\delta\psi_{C}(\mathbf{s})}\right)\left(\frac{\delta}{\delta\psi_{NC}^{+}(\mathbf{s})}\right)\{\psi_{NC}(\mathbf{s})\psi_{C}^{+}(\mathbf{s})+\psi_{C}(\mathbf{s})\psi_{NC}^{+}(\mathbf{s})\}\right\} P[\underrightarrow{\psi}(\mathbf{r}),\underrightarrow{\psi^{\ast}}(\mathbf{r})]\right\} \nonumber \\
 &  & \frac{-i}{\hbar}\left\{ +g\int d\mathbf{s}\left\{ \left(\frac{\delta}{\delta\psi_{C}(\mathbf{s})}\right)\left(\frac{\delta}{\delta\psi_{NC}(\mathbf{s})}\right)\{\psi_{NC}(\mathbf{s})\psi_{C}(\mathbf{s})\}\right\} P[\underrightarrow{\psi}(\mathbf{r}),\underrightarrow{\psi^{\ast}}(\mathbf{r})]\right\} \nonumber \\
 &  & \frac{-i}{\hbar}\left\{ -g\int d\mathbf{s}\left\{ \left(\frac{\delta}{\delta\psi_{C}^{+}(\mathbf{s})}\right)\left(\frac{\delta}{\delta\psi_{NC}^{+}(\mathbf{s})}\right)\{\psi_{C}^{+}(\mathbf{s})\psi_{NC}^{+}(\mathbf{s})\}\right\} P[\underrightarrow{\psi}(\mathbf{r}),\underrightarrow{\psi^{\ast}}(\mathbf{r})]\right\} \nonumber \\
 &  & \frac{-i}{\hbar}\left\{ +g\int d\mathbf{s}\left\{ \left(\frac{\delta}{\delta\psi_{NC}(\mathbf{s})}\right)\left(\frac{\delta}{\delta\psi_{NC}(\mathbf{s})}\right)\{\frac{1}{2}\psi_{C}(\mathbf{s})\psi_{C}(\mathbf{s})\}\right\} P[\underrightarrow{\psi}(\mathbf{r}),\underrightarrow{\psi^{\ast}}(\mathbf{r})]\right\} \nonumber \\
 &  & \frac{-i}{\hbar}\left\{ -g\int d\mathbf{s}\left\{ \left(\frac{\delta}{\delta\psi_{NC}^{+}(\mathbf{s})}\right)\left(\frac{\delta}{\delta\psi_{NC}^{+}(\mathbf{s})}\right)\{\frac{1}{2}\psi_{C}^{+}(\mathbf{s})\psi_{C}^{+}(\mathbf{s})\}\right\} P[\underrightarrow{\psi}(\mathbf{r}),\underrightarrow{\psi^{\ast}}(\mathbf{r})]\right\} \nonumber \\
 &  & \,\label{eq:FnalFPSecondOrderV2InteractionQuadratic}\end{eqnarray}
\begin{eqnarray}
 &  & \left(\frac{\partial}{\partial t}P[\underrightarrow{\psi}(\mathbf{r}),\underrightarrow{\psi^{\ast}}(\mathbf{r})]\right)_{V2}^{3}\nonumber \\
 & = & \frac{-i}{\hbar}\left\{ -g\int d\mathbf{s}\left\{ \left(\frac{\delta}{\delta\psi_{C}^{+}(\mathbf{s})}\right)\left(\frac{\delta}{\delta\psi_{C}^{+}(\mathbf{s})}\right)\left(\frac{\delta}{\delta\psi_{NC}(\mathbf{s})}\right)\{\frac{1}{4}\psi_{NC}^{+}(\mathbf{s})\}\right\} P[\underrightarrow{\psi}(\mathbf{r}),\underrightarrow{\psi^{\ast}}(\mathbf{r})]\right\} \nonumber \\
 &  & \frac{-i}{\hbar}\left\{ +g\int d\mathbf{s}\left\{ \left(\frac{\delta}{\delta\psi_{C}(\mathbf{s})}\right)\left(\frac{\delta}{\delta\psi_{C}(\mathbf{s})}\right)\left(\frac{\delta}{\delta\psi_{NC}^{+}(\mathbf{s})}\right)\{\frac{1}{4}\psi_{NC}(\mathbf{s})\}\right\} P[\underrightarrow{\psi}(\mathbf{r}),\underrightarrow{\psi^{\ast}}(\mathbf{r})]\right\} \nonumber \\
 &  & \frac{-i}{\hbar}\left\{ +g\int d\mathbf{s}\left\{ \left(\frac{\delta}{\delta\psi_{C}^{+}(\mathbf{s})}\right)\left(\frac{\delta}{\delta\psi_{NC}(\mathbf{s})}\right)\left(\frac{\delta}{\delta\psi_{NC}(\mathbf{s})}\right)\{\frac{1}{2}\psi_{C}(\mathbf{s})\}\right\} P[\underrightarrow{\psi}(\mathbf{r}),\underrightarrow{\psi^{\ast}}(\mathbf{r})]\right\} \nonumber \\
 &  & \frac{-i}{\hbar}\left\{ -g\int d\mathbf{s}\left\{ \left(\frac{\delta}{\delta\psi_{C}(\mathbf{s})}\right)\left(\frac{\delta}{\delta\psi_{NC}^{+}(\mathbf{s})}\right)\left(\frac{\delta}{\delta\psi_{NC}^{+}(\mathbf{s})}\right)\{\frac{1}{2}\psi_{C}^{+}(\mathbf{s})\}\right\} P[\underrightarrow{\psi}(\mathbf{r}),\underrightarrow{\psi^{\ast}}(\mathbf{r})]\right\} \nonumber \\
 &  & \frac{-i}{\hbar}\left\{ +g\int d\mathbf{s}\left\{ \left(\frac{\delta}{\delta\psi_{C}(\mathbf{s})}\right)\left(\frac{\delta}{\delta\psi_{C}^{+}(\mathbf{s})}\right)\left(\frac{\delta}{\delta\psi_{NC}(\mathbf{s})}\right)\{\frac{1}{2}\psi_{NC}(\mathbf{s})\}\right\} P[\underrightarrow{\psi}(\mathbf{r}),\underrightarrow{\psi^{\ast}}(\mathbf{r})]\right\} \nonumber \\
 &  & \frac{-i}{\hbar}\left\{ -g\int d\mathbf{s}\left\{ \left(\frac{\delta}{\delta\psi_{C}^{+}(\mathbf{s})}\right)\left(\frac{\delta}{\delta\psi_{C}(\mathbf{s})}\right)\left(\frac{\delta}{\delta\psi_{NC}^{+}(\mathbf{s})}\right)\{\frac{1}{2}\psi_{NC}^{+}(\mathbf{s})\}\right\} P[\underrightarrow{\psi}(\mathbf{r}),\underrightarrow{\psi^{\ast}}(\mathbf{r})]\right\} \nonumber \\
 &  & \,\label{eq:FnalFPSecondOrderV2InteractionCubic}\end{eqnarray}
\begin{eqnarray}
 &  & \left(\frac{\partial}{\partial t}P[\underrightarrow{\psi}(\mathbf{r}),\underrightarrow{\psi^{\ast}}(\mathbf{r})]\right)_{V2}^{4}\nonumber \\
 & = & \frac{-i}{\hbar}\left\{ g\int d\mathbf{s}\left\{ \left(\frac{\delta}{\delta\psi_{C}^{+}(\mathbf{s})}\right)\left(\frac{\delta}{\delta\psi_{C}^{+}(\mathbf{s})}\right)\left(\frac{\delta}{\delta\psi_{NC}(\mathbf{s})}\right)\left(\frac{\delta}{\delta\psi_{NC}(\mathbf{s})}\right)\{\frac{1}{8}\}\right\} P[\underrightarrow{\psi}(\mathbf{r}),\underrightarrow{\psi^{\ast}}(\mathbf{r})]\right\} \nonumber \\
 &  & \frac{-i}{\hbar}\left\{ -g\int d\mathbf{s}\left\{ \left(\frac{\delta}{\delta\psi_{C}(\mathbf{s})}\right)\left(\frac{\delta}{\delta\psi_{C}(\mathbf{s})}\right)\left(\frac{\delta}{\delta\psi_{NC}^{+}(\mathbf{s})}\right)\left(\frac{\delta}{\delta\psi_{NC}^{+}(\mathbf{s})}\right)\{\frac{1}{8}\}\right\} P[\underrightarrow{\psi}(\mathbf{r}),\underrightarrow{\psi^{\ast}}(\mathbf{r})]\right\} \nonumber \\
 &  & \,\label{eq:FnalFPSecondOrderV2InteractionQuartic}\end{eqnarray}

\subparagraph{Third Order Terms}

The contribution to the \emph{functional Fokker-Planck equation} from
the \emph{third order term }in the \emph{interaction Hamiltonian}
between the condensate and non-condensate may be written as the sum
of terms which are linear, quadratic and cubic in the number of functional
derivatives\begin{eqnarray}
 &  & \left(\frac{\partial}{\partial t}P[\underrightarrow{\psi}(\mathbf{r}),\underrightarrow{\psi^{\ast}}(\mathbf{r})]\right)_{V3}\nonumber \\
 & = & \left(\frac{\partial}{\partial t}P[\underrightarrow{\psi}(\mathbf{r}),\underrightarrow{\psi^{\ast}}(\mathbf{r})]\right)_{V3}^{1}+\left(\frac{\partial}{\partial t}P[\underrightarrow{\psi}(\mathbf{r}),\underrightarrow{\psi^{\ast}}(\mathbf{r})]\right)_{V3}^{2}\nonumber \\
 &  & +\left(\frac{\partial}{\partial t}P[\underrightarrow{\psi}(\mathbf{r}),\underrightarrow{\psi^{\ast}}(\mathbf{r})]\right)_{V3}^{3}\label{Eq.FnalFPThirdOrderV3Interaction}\end{eqnarray}
where\begin{eqnarray}
 &  & \left(\frac{\partial}{\partial t}P[\underrightarrow{\psi}(\mathbf{r}),\underrightarrow{\psi^{\ast}}(\mathbf{r})]\right)_{V3}^{1}\nonumber \\
 & = & \frac{-i}{\hbar}\left\{ +g\int d\mathbf{s}\left\{ \left(\frac{\delta}{\delta\psi_{C}^{+}(\mathbf{s})}\right)\{\psi_{NC}^{+}(\mathbf{s})\psi_{NC}^{+}(\mathbf{s})\psi_{NC}(\mathbf{s})\}\right\} P[\underrightarrow{\psi}(\mathbf{r}),\underrightarrow{\psi^{\ast}}(\mathbf{r})]\right\} \nonumber \\
 &  & \frac{-i}{\hbar}\left\{ -g\int d\mathbf{s}\left\{ \left(\frac{\delta}{\delta\psi_{C}(\mathbf{s})}\right)\{\psi_{NC}^{+}(\mathbf{s})\psi_{NC}(\mathbf{s})\psi_{NC}(\mathbf{s})\}\right\} P[\underrightarrow{\psi}(\mathbf{r}),\underrightarrow{\psi^{\ast}}(\mathbf{r})]\right\} \nonumber \\
 &  & \frac{-i}{\hbar}\left\{ +g\int d\mathbf{s}\left\{ \left(\frac{\delta}{\delta\psi_{NC}^{+}(\mathbf{s})}\right)\{[2\psi_{NC}(\mathbf{s})\psi_{C}^{+}(\mathbf{s)+}\psi_{C}\mathbf{(\mathbf{s})}\psi_{NC}^{+}\mathbf{(\mathbf{s})]}\psi_{NC}^{+}(\mathbf{s})\}\right\} P[\underrightarrow{\psi}(\mathbf{r}),\underrightarrow{\psi^{\ast}}(\mathbf{r})]\right\} \nonumber \\
 &  & \frac{-i}{\hbar}\left\{ -g\int d\mathbf{s}\left\{ \left(\frac{\delta}{\delta\psi_{NC}(\mathbf{s})}\right)\{[2\psi_{NC}^{+}(\mathbf{s})\psi_{C}(\mathbf{s})+\psi_{C}^{+}(\mathbf{s})\psi_{NC}(\mathbf{s})]\psi_{NC}(\mathbf{s})\}\right\} P[\underrightarrow{\psi}(\mathbf{r}),\underrightarrow{\psi^{\ast}}(\mathbf{r})]\right\} \nonumber \\
 &  & \,\label{eq:FnalFPThirdOrderV3InteractionLinear}\end{eqnarray}
\begin{eqnarray}
 &  & \left(\frac{\partial}{\partial t}P[\underrightarrow{\psi}(\mathbf{r}),\underrightarrow{\psi^{\ast}}(\mathbf{r})]\right)_{V3}^{2}\nonumber \\
 & = & \frac{-i}{\hbar}\left\{ -g\int d\mathbf{s}\left\{ \left(\frac{\delta}{\delta\psi_{C}^{+}(\mathbf{s})}\right)\left(\frac{\delta}{\delta\psi_{NC}^{+}(\mathbf{s})}\right)\{\frac{1}{2}\psi_{NC}^{+}(\mathbf{s})\psi_{NC}^{+}(\mathbf{s})\}\right\} P[\underrightarrow{\psi}(\mathbf{r}),\underrightarrow{\psi^{\ast}}(\mathbf{r})]\right\} \nonumber \\
 &  & \frac{-i}{\hbar}\left\{ +g\int d\mathbf{s}\left\{ \left(\frac{\delta}{\delta\psi_{C}(\mathbf{s})}\right)\left(\frac{\delta}{\delta\psi_{NC}(\mathbf{s})}\right)\{\frac{1}{2}\psi_{NC}(\mathbf{s})\psi_{NC}(\mathbf{s})\}\right\} P[\underrightarrow{\psi}(\mathbf{r}),\underrightarrow{\psi^{\ast}}(\mathbf{r})]\right\} \nonumber \\
 &  & \frac{-i}{\hbar}\left\{ -g\int d\mathbf{s}\left\{ \left(\frac{\delta}{\delta\psi_{C}^{+}(\mathbf{s})}\right)\left(\frac{\delta}{\delta\psi_{NC}(\mathbf{s})}\right)\{\psi_{NC}^{+}(\mathbf{s})\psi_{NC}(\mathbf{s})\}\right\} P[\underrightarrow{\psi}(\mathbf{r}),\underrightarrow{\psi^{\ast}}(\mathbf{r})]\right\} \nonumber \\
 &  & \frac{-i}{\hbar}\left\{ +g\int d\mathbf{s}\left\{ \left(\frac{\delta}{\delta\psi_{C}(\mathbf{s})}\right)\left(\frac{\delta}{\delta\psi_{NC}^{+}(\mathbf{s})}\right)\{\psi_{NC}(\mathbf{s})\psi_{NC}^{+}(\mathbf{s})\}\right\} P[\underrightarrow{\psi}(\mathbf{r}),\underrightarrow{\psi^{\ast}}(\mathbf{r})]\right\} \nonumber \\
 &  & \frac{-i}{\hbar}\left\{ -g\int d\mathbf{s}\left\{ \left(\frac{\delta}{\delta\psi_{NC}^{+}(\mathbf{s})}\right)\left(\frac{\delta}{\delta\psi_{NC}^{+}(\mathbf{s})}\right)\{\psi_{NC}^{+}(\mathbf{s})\psi_{C}^{+}(\mathbf{s})\}\right\} P[\underrightarrow{\psi}(\mathbf{r}),\underrightarrow{\psi^{\ast}}(\mathbf{r})]\right\} \nonumber \\
 &  & \frac{-i}{\hbar}\left\{ +g\int d\mathbf{s}\left\{ \left(\frac{\delta}{\delta\psi_{NC}(\mathbf{s})}\right)\left(\frac{\delta}{\delta\psi_{NC}(\mathbf{s})}\right)\{\psi_{NC}(\mathbf{s})\psi_{C}(\mathbf{s})\}\right\} P[\underrightarrow{\psi}(\mathbf{r}),\underrightarrow{\psi^{\ast}}(\mathbf{r})]\right\} \nonumber \\
 &  & \,\label{eq:FnalFPThirdOrderV3InteractionQuadratic}\end{eqnarray}
\begin{eqnarray}
 &  & \left(\frac{\partial}{\partial t}P[\underrightarrow{\psi}(\mathbf{r}),\underrightarrow{\psi^{\ast}}(\mathbf{r})]\right)_{V3}^{3}\nonumber \\
 & = & \frac{-i}{\hbar}\left\{ +g\int d\mathbf{s}\left\{ \left(\frac{\delta}{\delta\psi_{C}^{+}(\mathbf{s})}\right)\left(\frac{\delta}{\delta\psi_{NC}(\mathbf{s})}\right)\left(\frac{\delta}{\delta\psi_{NC}(\mathbf{s})}\right)\{\frac{1}{2}\psi_{NC}(\mathbf{s})\}\right\} P[\underrightarrow{\psi}(\mathbf{r}),\underrightarrow{\psi^{\ast}}(\mathbf{r})]\right\} \nonumber \\
 &  & \frac{-i}{\hbar}\left\{ -g\int d\mathbf{s}\left\{ \left(\frac{\delta}{\delta\psi_{C}(\mathbf{s})}\right)\left(\frac{\delta}{\delta\psi_{NC}^{+}(\mathbf{s})}\right)\left(\frac{\delta}{\delta\psi_{NC}^{+}(\mathbf{s})}\right)\{\frac{1}{2}\psi_{NC}^{+}(\mathbf{s})\}\right\} P[\underrightarrow{\psi}(\mathbf{r}),\underrightarrow{\psi^{\ast}}(\mathbf{r})]\right\} \nonumber \\
 &  & \,\label{eq:FnalFPThirdOrderV3InteractionCubic}\end{eqnarray}
This term is part of the interaction term $\widehat{H}_{4}$ and its
contribution to the functional Fokker-Planck equation will be ignored.

\subsection{Supplementary Equations}

~

Bogoliubov Hamiltonian \begin{equation}
\widehat{H}_{B}=\widehat{H}_{1}+\widehat{H}_{2}+\widehat{H}_{3}\label{Eq.BogolHamiltonian-1}\end{equation}

Operator identities for various functional derivatives 

\begin{eqnarray}
\left(\frac{\delta}{\delta\psi_{C}(\mathbf{s})}\right)_{\mathbf{s}} & \equiv & \sum\limits _{k=1,2}\phi_{k}^{\ast}(\mathbf{s})\,\frac{\partial}{\partial\alpha_{k}}\qquad\left(\frac{\delta}{\delta\psi_{NC}(\mathbf{s})}\right)_{\mathbf{s}}\equiv\sum\limits _{k\neq1,2}^{K}\phi_{k}^{\ast}(\mathbf{s})\,\frac{\partial}{\partial\alpha_{k}}\nonumber \\
\left(\frac{\delta}{\delta\psi_{C}^{+}(\mathbf{s})}\right)_{\mathbf{s}} & \equiv & \sum\limits _{k=1,2}\phi_{k}(\mathbf{s})\,\frac{\partial}{\partial\alpha_{k}^{+}}\qquad\left(\frac{\delta}{\delta\psi_{NC}^{+}(\mathbf{s})}\right)_{\mathbf{s}}\equiv\sum\limits _{k\neq1,2}^{K}\phi_{k}(\mathbf{s})\,\frac{\partial}{\partial\alpha_{k}^{+}}\nonumber \\
 &  & \,\label{eq:RestrictFnalDerivIdent1-1}\end{eqnarray}

Field Functions

\begin{eqnarray}
\psi_{C}(\mathbf{r}) & = & \alpha_{1}\phi_{1}(\mathbf{r})+\alpha_{2}\phi_{2}(\mathbf{r})\qquad\psi_{C}^{+}(\mathbf{r})=\phi_{1}^{\ast}(\mathbf{r})\alpha_{1}^{+}+\phi_{2}^{\ast}(\mathbf{r})\alpha_{2}^{+}\label{eq:CondFieldFn-1}\\
\psi_{NC}(\mathbf{r}) & = & \sum\limits _{k\neq1,2}\alpha_{k}\phi_{k}(\mathbf{r})\qquad\psi_{NC}^{+}(\mathbf{r})=\sum\limits _{k\neq1,2}\phi_{k}^{\ast}(\mathbf{r})\alpha_{k}^{+}\label{eq:NonCondFieldFn-1}\\
\psi_{C}(\mathbf{r}) & = & \int d\mathbf{r}^{\prime}\mathbf{\,}\delta_{C}(\mathbf{r,r}^{\prime})\psi_{C}(\mathbf{r}^{\prime})\qquad\psi_{C}^{+}(\mathbf{r})=\int d\mathbf{r}^{\prime}\mathbf{\,}\psi_{C}^{+}(\mathbf{r}^{\prime})\delta_{C}(\mathbf{r}^{\prime}\mathbf{,r})\nonumber \\
\psi_{NC}(\mathbf{r}) & = & \int d\mathbf{r}^{\prime}\mathbf{\,}\delta_{NC}(\mathbf{r,r}^{\prime})\psi_{NC}(\mathbf{r}^{\prime})\,\,\,\,\,\,\psi(\mathbf{r})=\int d\mathbf{r}^{\prime}\mathbf{\,}\psi_{C}^{+}(\mathbf{r}^{\prime})\delta_{C}(\mathbf{r}^{\prime}\mathbf{,r})\label{Eq.RestrictedDeltaFns-1}\end{eqnarray}

Product rule for functional derivatives\begin{eqnarray}
 &  & \frac{\delta}{\delta\psi(\mathbf{s})}(F[\psi(\mathbf{r}),\psi^{+}(\mathbf{r})]G[\psi(\mathbf{r}),\psi^{+}(\mathbf{r})])\nonumber \\
 & = & (\frac{\delta}{\delta\psi(\mathbf{s})}F[\psi(\mathbf{r}),\psi^{+}(\mathbf{r})])G[\psi(\mathbf{r}),\psi^{+}(\mathbf{r})]+F[\psi(\mathbf{r}),\psi^{+}(\mathbf{r})](\frac{\delta}{\delta\psi(\mathbf{s})}G[\psi(\mathbf{r}),\psi^{+}(\mathbf{r})])\nonumber \\
 &  & \frac{\delta}{\delta\psi^{+}(\mathbf{s})}(F[\psi(\mathbf{r}),\psi^{+}(\mathbf{r})]G[\psi(\mathbf{r}),\psi^{+}(\mathbf{r})])\nonumber \\
 & = & (\frac{\delta}{\delta\psi^{+}(\mathbf{s})}F[\psi(\mathbf{r}),\psi^{+}(\mathbf{r})])G[\psi(\mathbf{r}),\psi^{+}(\mathbf{r})]+F[\psi(\mathbf{r}),\psi^{+}(\mathbf{r})](\frac{\delta}{\delta\psi^{+}(\mathbf{s})}G[\psi(\mathbf{r}),\psi^{+}(\mathbf{r})])\nonumber \\
 &  & \,\label{eq:ProdRuleFnalDeriv-1}\end{eqnarray}

Functional Derivative Results\begin{eqnarray}
\frac{\delta}{\delta\psi_{C}(\mathbf{s})}\psi_{C}(\mathbf{r}) & = & \delta_{C}(\mathbf{r,s})\nonumber \\
\frac{\delta}{\delta\psi_{C}^{+}(\mathbf{s})}\psi_{C}^{+}(\mathbf{r}) & = & \delta_{C+}(\mathbf{r,s})=\delta_{C}(\mathbf{s,r})\nonumber \\
\frac{\delta}{\delta\psi_{C}(\mathbf{s})}\psi_{C}^{+}(\mathbf{r}) & = & 0\qquad\frac{\delta}{\delta\psi_{C}^{+}(\mathbf{s})}\psi_{C}(\mathbf{r})=0\label{Eq.FuncDerivativeRule1-1}\end{eqnarray}
\begin{eqnarray}
\frac{\delta}{\delta\psi_{C}(\mathbf{s})}\psi_{NC}(\mathbf{r}) & = & 0\qquad\frac{\delta}{\delta\psi_{C}^{+}(\mathbf{s})}\psi_{NC}^{+}(\mathbf{r})=0\nonumber \\
\frac{\delta}{\delta\psi_{C}(\mathbf{s})}\psi_{NC}^{+}(\mathbf{r}) & = & 0\qquad\frac{\delta}{\delta\psi_{C}^{+}(\mathbf{s})}\psi_{NC}(\mathbf{r})=0\label{Eq.FuncDerivativeRule2-1}\end{eqnarray}
\medskip{}
 \pagebreak{}

\section{- Ito Stochastic Equations}

\label{Appendix Ito Equations}

The Ito stochastic equations are obtained after neglecting \emph{third},
\emph{fourth order} functional derivatives in the functional Fokker-Planck
equation. The drift and diffusion terms are then identified from the
remaining first and second order functional derivative terms that
are left and the Ito stochastic equations for the stochastic fields
can then be written down.\medskip{}

\subsection{Symmetric Forms of Functional Fokker-Planck Equation}

For the two mode case the diffusion term in (\ref{Eq.FFPETwoStart-1})
becomes \begin{eqnarray}
T_{Diff} & = & \sum_{A\leq\, B}\int\int dx\, dy\frac{\delta}{\delta\psi_{A}(x)}\frac{\delta}{\delta\psi_{B}(y)}H_{AB}(\underrightarrow{\psi}(x),x,\underrightarrow{\psi}(y),y)\, P\nonumber \\
 & = & \frac{1}{2}\sum_{A<\, B}\int\int dx\, dy\frac{\delta}{\delta\psi_{A}(x)}\frac{\delta}{\delta\psi_{B}(y)}H_{AB}(\underrightarrow{\psi}(x),x,\underrightarrow{\psi}(y),y)\, P\nonumber \\
 &  & +\frac{1}{2}\sum_{A<B}\int\int dx\, dy\frac{\delta}{\delta\psi_{A}(x)}\frac{\delta}{\delta\psi_{B}(y)}H_{AB}(\underrightarrow{\psi}(x),x,\underrightarrow{\psi}(y),y)\, P\nonumber \\
 &  & +\frac{1}{2}\sum_{A}\int\int dx\, dy\frac{\delta}{\delta\psi_{A}(x)}\frac{\delta}{\delta\psi_{A}(y)}H_{AA}(\underrightarrow{\psi}(x),x,\underrightarrow{\psi}(y),y)\, P\nonumber \\
 &  & +\frac{1}{2}\sum_{A}\int\int dx\, dy\frac{\delta}{\delta\psi_{A}(x)}\frac{\delta}{\delta\psi_{A}(y)}H_{AA}(\underrightarrow{\psi}(x),x,\underrightarrow{\psi}(y),y)\, P\nonumber \\
 &  & \,\end{eqnarray}
If we interchange $A,B$ and $x$, $y$ in the second term and just
$x$, $y$ in the fourth term, we find on using the result that double
functional differentiation can be carried out in either order that
\begin{eqnarray}
T_{Diff} & = & \frac{1}{2}\sum_{A<\, B}\int\int dx\, dy\frac{\delta}{\delta\psi_{A}(x)}\frac{\delta}{\delta\psi_{B}(y)}H_{AB}(\underrightarrow{\psi}(x),x,\underrightarrow{\psi}(y),y)\, P\nonumber \\
 &  & +\frac{1}{2}\sum_{B<A}\int\int dx\, dy\frac{\delta}{\delta\psi_{A}(x)}\frac{\delta}{\delta\psi_{B}(y)}H_{BA}(\underrightarrow{\psi}(y),y,\underrightarrow{\psi}(x),x)\, P\nonumber \\
 &  & +\frac{1}{2}\sum_{A}\int\int dx\, dy\frac{\delta}{\delta\psi_{A}(x)}\frac{\delta}{\delta\psi_{A}(y)}H_{AA}(\underrightarrow{\psi}(x),x,\underrightarrow{\psi}(y),y)\, P\nonumber \\
 &  & +\frac{1}{2}\sum_{A}\int\int dx\, dy\frac{\delta}{\delta\psi_{A}(x)}\frac{\delta}{\delta\psi_{A}(y)}H_{AA}(\underrightarrow{\psi}(y),y,\underrightarrow{\psi}(x),x)\, P\nonumber \\
 & = & \frac{1}{2}\sum_{A<\, B}\int\int dx\, dy\frac{\delta}{\delta\psi_{A}(x)}\frac{\delta}{\delta\psi_{B}(y)}H_{AB}(\underrightarrow{\psi}(x),x,\underrightarrow{\psi}(y),y)\, P\nonumber \\
 &  & +\frac{1}{2}\sum_{A>B}\int\int dx\, dy\frac{\delta}{\delta\psi_{A}(x)}\frac{\delta}{\delta\psi_{B}(y)}H_{BA}(\underrightarrow{\psi}(y),y,\underrightarrow{\psi}(x),x)\, P\nonumber \\
 &  & +\frac{1}{2}\sum_{A}\int\int dx\, dy\frac{\delta}{\delta\psi_{A}(x)}\frac{\delta}{\delta\psi_{A}(y)}(H_{AA}(\underrightarrow{\psi}(x),x,\underrightarrow{\psi}(y),y)+H_{AA}(\underrightarrow{\psi}(y),y,\underrightarrow{\psi}(x),x))\, P\nonumber \\
 &  & \,\end{eqnarray}
If we now define a new diffusion matrix such that\begin{eqnarray}
D_{AB}(\underrightarrow{\psi}(x),x,\underrightarrow{\psi}(y),y) & = & H_{AB}(\underrightarrow{\psi}(x),x,\underrightarrow{\psi}(y),y)\qquad A<\, B\nonumber \\
D_{AB}(\underrightarrow{\psi}(x),x,\underrightarrow{\psi}(y),y) & = & H_{BA}(\underrightarrow{\psi}(y),y,\underrightarrow{\psi}(x),x)\qquad A>\, B\nonumber \\
D_{AA}(\underrightarrow{\psi}(x),x,\underrightarrow{\psi}(y),y) & = & H_{AA}(\underrightarrow{\psi}(x),x,\underrightarrow{\psi}(y),y)+H_{AA}(\underrightarrow{\psi}(y),y,\underrightarrow{\psi}(x),x)\qquad A=B\nonumber \\
 &  & \,\end{eqnarray}
we see that the functional Fokker-Planck equation for the two mode
case becomes\begin{eqnarray}
\frac{\partial P}{\partial t} & = & \sum_{A}\int dx\,\frac{\delta}{\delta\psi_{A}(x)}A_{A}(\underrightarrow{\psi}(x),x)\, P\nonumber \\
 &  & +\frac{1}{2}\sum_{A,B}\int\int dx\, dy\frac{\delta}{\delta\psi_{A}(x)}\frac{\delta}{\delta\psi_{B}(y)}D_{AB}(\underrightarrow{\psi}(x),x,\underrightarrow{\psi}(y),y)\, P\nonumber \\
 &  & \,\end{eqnarray}
The expressions have been defined so that $D_{AB}$ is symmetric.
For the two mode condensate case\begin{equation}
D_{AB}(\underrightarrow{\psi}(x),x,\underrightarrow{\psi}(y),y)=D_{BA}(\underrightarrow{\psi}(y),y,\underrightarrow{\psi}(x),x)\end{equation}

\subsection{Complex Symmetric Matrices}

We present a proof that any $n\times n$ complex symmetric matrix
$F$ can also be written in the form $F=\mathcal{B}\,\mathcal{B}^{T}$,
where $\mathcal{B}$ is also complex and has dimension $n\times2n$.
The proof is adapted from material in {[}\citep{Drummond80a}{]} and
{[}\citep{Gardiner91a}{]} (see Sect. 6.4.7). This result is less
useful than the Takagi factorisation, where $\mathcal{B}$ has dimension
$n\times n$, the same as $F$.

The matrix $F$ is $n\times n$ and we have $F_{pq}=F_{qp}$.

We first write \begin{equation}
F=F^{x}+iF^{y}\end{equation}
where $F^{x}$ and $F^{y}$ are real symmetric matrices, both $n\times n$
in size.

We then construct a $2n\times2n$ matrix $D$ using $F^{x}$ and $F^{y}$
as sub-matrices\begin{equation}
D=\left[\begin{tabular}{cc}
 \ensuremath{\frac{1}{2}F^{x}} &  \ensuremath{\frac{1}{2}F^{y}}\\
\ensuremath{\frac{1}{2}F^{y}} &  \ensuremath{-\frac{1}{2}F^{x}}\end{tabular}\right]=\left[\begin{tabular}{cc}
 \ensuremath{D^{xx}} &  \ensuremath{D^{xy}}\\
\ensuremath{D^{yx}} &  \ensuremath{D^{yy}}\end{tabular}\right]\end{equation}
Clearly $D$ is both symmetric and real. We use $D^{xx},..,D^{yy}$
as an alternative notation for the $n\times n$ submatrices of $D$.

Hence we can find a real $2n\times2n$ matrix $B$ such that \begin{equation}
D=B\, B^{T}\end{equation}
Such a matrix can be obtained by construction using the real eigenvalues
$\lambda$ and real, orthogonal eigenvectors $X_{\lambda}$ of $D$.
Thus with \begin{eqnarray}
D\, X_{\lambda} & = & \lambda\, X_{\lambda}\qquad X_{\lambda}^{T}\, X_{\mu}=\delta_{\lambda\,\mu}\nonumber \\
D & = & \sum_{\lambda}\lambda\, X_{\lambda}X_{\lambda}^{T}\end{eqnarray}
we can choose \begin{equation}
B=\sum_{\lambda}\sqrt{\lambda}\, X_{\lambda}X_{\lambda}^{T}\end{equation}
from which it is easy to show that $D=B\, B^{T}$. Note that $B$
is complex unless $D$ is positive semi-definite.

We now divide $B$ into two $n\times2n$ submatrices as \begin{equation}
B=\left[\begin{tabular}{c}
 \ensuremath{B^{x}}\\
\ensuremath{B^{y}}\end{tabular}\right]\end{equation}
Clearly as\begin{eqnarray}
B\, B^{T} & = & D\nonumber \\
 & = & \left[\begin{tabular}{cc}
 \ensuremath{B^{x}\, B^{xT}=D^{xx}} &  \ensuremath{B^{x}\, B^{yT}=D^{xy}}\\
\ensuremath{B^{y}\, B^{xT}=D^{yx}} &  \ensuremath{B^{y}\, B^{yT}=D^{yy}}\end{tabular}\right]\end{eqnarray}
we can express the submatrices of $D$ in terms of $B^{x}$ and $B^{y}$.

Now define the $n\times2n$ complex matrix $\mathcal{B}$ as\begin{equation}
\mathcal{B=}B^{x}+iB^{y}\end{equation}
Then\begin{eqnarray}
\mathcal{B\, B}^{T} & = & (B^{x}+iB^{y})(B^{xT}+iB^{yT})\nonumber \\
 & = & B^{x}\, B^{xT}+iB^{x}\, B^{yT}+iB^{y}\, B^{xT}-B^{y}\, B^{yT}\nonumber \\
 & = & D^{xx}-D^{yy}+i(D^{xy}+D^{yx})\nonumber \\
 & = & \frac{1}{2}F^{x}-(-\frac{1}{2}F^{x})+i(\frac{1}{2}F^{y}+\frac{1}{2}F^{y})\nonumber \\
 & = & F^{x}+iF^{y}\nonumber \\
 & = & F\end{eqnarray}
showing that a $n\times2n$ complex matrix $\mathcal{B}$ can be found
such that $\mathcal{B\, B}^{T}=F$, as required.

\subsection{Properties of Noise Fields - Two Mode Case}

We can use the results in (\ref{eq:SumEtasTwoMode-1}) relating the
$\eta_{k}^{A;D}(\underrightarrow{\widetilde{\psi}}(x,t))$ to the
non-local diffusion terms $D_{AB}(\underrightarrow{\widetilde{\psi}}(x_{1},t_{1}),x_{1},\underrightarrow{\widetilde{\psi}}(x_{2},t_{2}),x_{2})$
and the fundamental noise properties of the Gaussian-Markov noise
variables $\Gamma_{k}^{D}$ in (\ref{Eq.GaussMarkGeneral-1}), together
with (\ref{Eq.StochAverProduct-1}) to determine the stochastic properties
of the noise fields. For a single noise field \begin{eqnarray}
 &  & \overline{\{\left(\frac{\partial}{\partial t}\widetilde{G}_{A}(\underrightarrow{\widetilde{\psi}}(x_{1},t_{1}),\underrightarrow{\Gamma}(t_{1+}))\right)\}}\nonumber \\
 & = & \sum_{Dk}\overline{\eta_{k}^{A;D}(\underrightarrow{\widetilde{\psi}}(x_{1},t_{1}))}\;\overline{\Gamma_{k}^{D}(t_{1})}=0\end{eqnarray}
and for two noise fields.\begin{eqnarray}
 &  & \overline{\{\left(\frac{\partial}{\partial t}\widetilde{G}_{A}(\underrightarrow{\widetilde{\psi}}(x_{1},t_{1}),\underrightarrow{\Gamma}(t_{1+}))\right)\left(\frac{\partial}{\partial t}\widetilde{G}_{B}(\underrightarrow{\widetilde{\psi}}(x_{2},t_{2}),\underrightarrow{\Gamma}(t_{2+}))\right)\}}\nonumber \\
 & = & \overline{\sum_{Dk}\eta_{k}^{A;D}(\underrightarrow{\widetilde{\psi}}(x_{1},t_{1}))\Gamma_{k}^{D}(t_{1+})\sum_{El}\eta_{l}^{B;E}(\underrightarrow{\widetilde{\psi}}(x_{2},t_{2}))\Gamma_{l}^{E}(t_{2+})}\;\,\nonumber \\
 & = & \sum_{Dk}\sum_{El}\overline{\eta_{k}^{A;D}(\underrightarrow{\widetilde{\psi}}(x_{1},t_{1}))\Gamma_{k}^{D}(t_{1+})\eta_{l}^{B;E}(\underrightarrow{\widetilde{\psi}}(x_{2},t_{2}))\Gamma_{l}^{E}(t_{2+})}\;\nonumber \\
 & = & \sum_{Dk}\sum_{El}\overline{\eta_{k}^{A;D}(\underrightarrow{\widetilde{\psi}}(x_{1},t_{1}))\eta_{l}^{B;E}(\underrightarrow{\widetilde{\psi}}(x_{2},t_{2}))}\;\,\overline{\Gamma_{k}^{D}(t_{1+})\Gamma_{l}^{E}(t_{2+})}\nonumber \\
 & = & \sum_{Dk}\sum_{El}\overline{\eta_{k}^{A;D}(\underrightarrow{\widetilde{\psi}}(x_{1},t_{1}))\eta_{l}^{B;E}(\underrightarrow{\widetilde{\psi}}(x_{2},t_{2}))}\;\,\delta_{kl}\delta_{DE}\delta(t_{1}-t_{2})\,\nonumber \\
 & = & \overline{\sum_{Dk}\eta_{k}^{A;D}(\underrightarrow{\widetilde{\psi}}(x_{1},t_{1,2}))\eta_{k}^{B;D}(\underrightarrow{\widetilde{\psi}}(x_{2},t_{1,2}))}\;\delta(t_{1}-t_{2})\,\nonumber \\
 & = & \overline{D_{AB}(\underrightarrow{\widetilde{\psi}}(x_{1},t_{1,2}),x_{1},\underrightarrow{\widetilde{\psi}}(x_{2},t_{1,2}),x_{2})}\,\;\delta(t_{1}-t_{2})\end{eqnarray}
Thus the stochastic average of the linear noise term is zero, and
the stochastic average of the product of two linear noise terms is
\emph{delta function correlated in time}, but is not delta function
correlated in space. Instead the spatial correlation is given by the
\emph{non-local diffusion term} in the original functional Fokker-Planck
equation!

The noise terms do however satisfy the Gaussian-Markoff conditions
that averages of products of odd numbers of noise terms are zero,
however averages of products of even numbers of noise terms can be
written as sums of stochastic averages of products of pairs of non-local
diffusion terms, rather than pairs of noise terms. Thus\begin{eqnarray}
 &  & \overline{\{\left(\frac{\partial}{\partial t}\widetilde{G}_{A}(\underrightarrow{\widetilde{\psi}}(x_{1},t_{1}),\underrightarrow{\Gamma}(t_{1+}))\right)\left(\frac{\partial}{\partial t}\widetilde{G}_{B}(\underrightarrow{\widetilde{\psi}}(x_{2},t_{2}),\underrightarrow{\Gamma}(t_{2+}))\right)\left(\frac{\partial}{\partial t}\widetilde{G}_{C}(\underrightarrow{\widetilde{\psi}}(x_{3},t_{3}),\underrightarrow{\Gamma}(t_{3+}))\right)\}}\nonumber \\
 & = & \sum_{Dk}\sum_{El}\sum_{Fm}\overline{\eta_{k}^{A;D}(\underrightarrow{\widetilde{\psi}}(x_{1},t_{1}))\eta_{l}^{B;E}(\underrightarrow{\widetilde{\psi}}(x_{2},t_{2}))\eta_{m}^{C;F}(\underrightarrow{\widetilde{\psi}}(x_{3},t_{3}))}\;\,\overline{\Gamma_{k}^{D}(t_{1+})\Gamma_{l}^{E}(t_{2+})\Gamma_{m}^{F}(t_{3+})}\nonumber \\
 & = & 0\label{Eq.StochAverThreeNoise2Mode}\end{eqnarray}
and\begin{eqnarray}
 &  & \overline{\begin{array}{c}
\{\left(\frac{{\LARGE\partial}}{{\LARGE\partial t}}\widetilde{G}_{A}(\underrightarrow{\widetilde{\psi}}(x_{1},t_{1}),\underrightarrow{\Gamma}(t_{1+}))\right)\left(\frac{{\LARGE\partial}}{{\LARGE\partial t}}\widetilde{G}_{B}(\underrightarrow{\widetilde{\psi}}(x_{2},t_{2}),\underrightarrow{\Gamma}(t_{2+}))\right)\\
\times\left(\frac{{\LARGE\partial}}{{\LARGE\partial t}}\widetilde{G}_{C}(\underrightarrow{\widetilde{\psi}}(x_{3},t_{3}),\underrightarrow{\Gamma}(t_{3+}))\right)\left(\frac{{\LARGE\partial}}{{\LARGE\partial t}}\widetilde{G}_{D}(\underrightarrow{\widetilde{\psi}}(x_{4},t_{4}),\underrightarrow{\Gamma}(t_{4+}))\right)\}\end{array}}\nonumber \\
 & = & \sum_{Hk}\sum_{El}\sum_{Fm}\sum_{Gn}\overline{\eta_{k}^{A;H}(\underrightarrow{\widetilde{\psi}}(x_{1},t_{1}))\eta_{l}^{B;E}(\underrightarrow{\widetilde{\psi}}(x_{2},t_{2}))\eta_{m}^{C;F}(\underrightarrow{\widetilde{\psi}}(x_{3},t_{3}))\eta_{n}^{D;G}(\underrightarrow{\widetilde{\psi}}(x_{4},t_{4}))}\;\nonumber \\
 &  & \times\overline{\Gamma_{k}^{H}(t_{1+})\Gamma_{l}^{E}(t_{2+})\Gamma_{m}^{F}(t_{3+})\Gamma_{n}^{G}(t_{4+})}\nonumber \\
 & = & \sum_{Hk}\sum_{El}\sum_{Fm}\sum_{Gn}\overline{\eta_{k}^{A;H}(\underrightarrow{\widetilde{\psi}}(x_{1},t_{1}))\eta_{l}^{B;E}(\underrightarrow{\widetilde{\psi}}(x_{2},t_{2}))\eta_{m}^{C;F}(\underrightarrow{\widetilde{\psi}}(x_{3},t_{3}))\eta_{n}^{D;G}(\underrightarrow{\widetilde{\psi}}(x_{4},t_{4}))}\nonumber \\
 &  & \times\left\{ \begin{array}{c}
(\delta_{kl}\delta_{HE}\delta(t_{1}-t_{2}))(\delta_{mn}\delta_{FG}\delta(t_{3}-t_{4}))+(\delta_{km}\delta_{HF}\delta(t_{1}-t_{3}))(\delta_{nl}\delta_{EG}\delta(t_{2}-t_{4}))\\
+(\delta_{kn}\delta_{HG}\delta(t_{1}-t_{4}))(\delta_{ml}\delta_{EF}\delta(t_{2}-t_{3}))\end{array}\right\} \nonumber \\
 & = & \overline{\begin{array}{c}
\left[\sum_{Hk}\eta_{k}^{A;H}(\underrightarrow{\widetilde{\psi}}(x_{1},t_{1}))\sum_{El}\eta_{l}^{B;E}(\underrightarrow{\widetilde{\psi}}(x_{2},t_{2}))\delta_{kl}\delta_{HE}\delta(t_{1}-t_{2})\right]\\
\times\left[\sum_{Fm}\eta_{m}^{C;F}(\underrightarrow{\widetilde{\psi}}(x_{3},t_{3}))\sum_{Gn}\eta_{n}^{D;G}(\underrightarrow{\widetilde{\psi}}(x_{4},t_{4}))\delta_{mn}\delta_{FG}\delta(t_{3}-t_{4})\right]\end{array}}\nonumber \\
 &  & +\overline{\begin{array}{c}
\left[\sum_{Hk}\eta_{k}^{A;H}(\underrightarrow{\widetilde{\psi}}(x_{1},t_{1}))\sum_{Fm}\eta_{m}^{C;F}(\underrightarrow{\widetilde{\psi}}(x_{3},t_{3}))\delta_{km}\delta_{HF}\delta(t_{1}-t_{3})\right]\\
\times\left[\sum_{El}\eta_{l}^{B;E}(\underrightarrow{\widetilde{\psi}}(x_{2},t_{2}))\sum_{Gn}\eta_{n}^{D;G}(\underrightarrow{\widetilde{\psi}}(x_{4},t_{4}))\delta_{nl}\delta_{EG}\delta(t_{2}-t_{4})\right]\end{array}}\nonumber \\
 &  & +\overline{\begin{array}{c}
\left[\sum_{Hk}\eta_{k}^{A;H}(\underrightarrow{\widetilde{\psi}}(x_{1},t_{1}))\sum_{Gn}\eta_{n}^{D;G}(\underrightarrow{\widetilde{\psi}}(x_{4},t_{4}))\delta_{kn}\delta_{HG}\delta(t_{1}-t_{4})\right]\\
\times\left[\sum_{El}\eta_{l}^{B;E}(\underrightarrow{\widetilde{\psi}}(x_{2},t_{2}))\sum_{Fm}\eta_{m}^{C;F}(\underrightarrow{\widetilde{\psi}}(x_{3},t_{3}))\delta_{lm}\delta_{EF}\delta(t_{2}-t_{3})\right]\end{array}}\nonumber \\
 & = & \overline{\left[D_{AB}(\underrightarrow{\widetilde{\psi}}(x_{1},t_{1,2}),x_{1},\underrightarrow{\widetilde{\psi}}(x_{2},t_{1,2}),x_{2})\right]\left[D_{CD}(\underrightarrow{\widetilde{\psi}}(x_{3},t_{3,4}),x_{3},\underrightarrow{\widetilde{\psi}}(x_{4},t_{3,4}),x_{4})\right]}\;\delta(t_{1}-t_{2})\delta(t_{3}-t_{4})\nonumber \\
 &  & +\overline{\left[D_{AC}(\underrightarrow{\widetilde{\psi}}(x_{1},t_{1,3}),x_{1},\underrightarrow{\widetilde{\psi}}(x_{3},t_{1,3}),x_{3})\right]\left[D_{BD}(\underrightarrow{\widetilde{\psi}}(x_{2},t_{2,4}),x_{2},\underrightarrow{\widetilde{\psi}}(x_{4},t_{2,4}),x_{4})\right]}\;\delta(t_{1}-t_{3})\delta(t_{2}-t_{4})\nonumber \\
 &  & +\overline{\left[D_{AD}(\underrightarrow{\widetilde{\psi}}(x_{1},t_{1,4}),x_{1},\underrightarrow{\widetilde{\psi}}(x_{4},t_{1,4}),x_{2})\right]\left[D_{BC}(\underrightarrow{\widetilde{\psi}}(x_{2},t_{2,3}),x_{2},\underrightarrow{\widetilde{\psi}}(x_{3},t_{2,3}),x_{3})\right]}\;\delta(t_{1}-t_{4})\delta(t_{2}-t_{3})\nonumber \\
 &  & \,\label{eq:StochAverFourNoise2Mode}\end{eqnarray}
using the results (\ref{eq:SumEtasTwoMode-1}). This is not quite
the same as\begin{eqnarray}
 &  & \overline{\{\left(\frac{\partial}{\partial t}\widetilde{G}_{A}(\underrightarrow{\widetilde{\psi}}(x_{1},t_{1}),\underrightarrow{\Gamma}(t_{1+}))\right)\left(\frac{\partial}{\partial t}\widetilde{G}_{B}(\underrightarrow{\widetilde{\psi}}(x_{2},t_{2}),\underrightarrow{\Gamma}(t_{2+}))\right)\}}\nonumber \\
 &  & \times\overline{\{\left(\frac{\partial}{\partial t}\widetilde{G}_{C}(\underrightarrow{\widetilde{\psi}}(x_{3},t_{3}),\underrightarrow{\Gamma}(t_{3+}))\right)\left(\frac{\partial}{\partial t}\widetilde{G}_{D}(\underrightarrow{\widetilde{\psi}}(x_{4},t_{4}),\underrightarrow{\Gamma}(t_{4+}))\right)\}}\nonumber \\
 &  & +\overline{\{\left(\frac{\partial}{\partial t}\widetilde{G}_{A}(\underrightarrow{\widetilde{\psi}}(x_{1},t_{1}),\underrightarrow{\Gamma}(t_{1+}))\right)\left(\frac{\partial}{\partial t}\widetilde{G}_{C}(\underrightarrow{\widetilde{\psi}}(x_{3},t_{3}),\underrightarrow{\Gamma}(t_{3+}))\right)\}}\nonumber \\
 &  & \times\overline{\{\left(\frac{\partial}{\partial t}\widetilde{G}_{B}(\underrightarrow{\widetilde{\psi}}(x_{2},t_{2}),\underrightarrow{\Gamma}(t_{2+}))\right)\left(\frac{\partial}{\partial t}\widetilde{G}_{D}(\underrightarrow{\widetilde{\psi}}(x_{4},t_{4}),\underrightarrow{\Gamma}(t_{4+}))\right)\}}\nonumber \\
 &  & +\overline{\{\left(\frac{\partial}{\partial t}\widetilde{G}_{A}(\underrightarrow{\widetilde{\psi}}(x_{1},t_{1}),\underrightarrow{\Gamma}(t_{1+}))\right)\left(\frac{\partial}{\partial t}\widetilde{G}_{D}(\underrightarrow{\widetilde{\psi}}(x_{4},t_{4}),\underrightarrow{\Gamma}(t_{4+}))\right)\}}\nonumber \\
 &  & \times\overline{\{\left(\frac{\partial}{\partial t}\widetilde{G}_{B}(\underrightarrow{\widetilde{\psi}}(x_{2},t_{2}),\underrightarrow{\Gamma}(t_{2+}))\right)\left(\frac{\partial}{\partial t}\widetilde{G}_{C}(\underrightarrow{\widetilde{\psi}}(x_{3},t_{3}),\underrightarrow{\Gamma}(t_{3+}))\right)\}}\nonumber \\
 &  & \,\end{eqnarray}
because in general \begin{eqnarray}
 &  & \overline{\left[D_{AB}(\underrightarrow{\widetilde{\psi}}(x_{1},t_{1,2}),x_{1},\underrightarrow{\widetilde{\psi}}(x_{2},t_{1,2}),x_{2})\right]\left[D_{CD}(\underrightarrow{\widetilde{\psi}}(x_{3,4},t_{3}),x_{3},\underrightarrow{\widetilde{\psi}}(x_{4},t_{3,4}),x_{4})\right]}\nonumber \\
 & \neq & \overline{D_{AB}(\underrightarrow{\widetilde{\psi}}(x_{1},t_{1,2}),x_{1},\underrightarrow{\widetilde{\psi}}(x_{2},t_{1,2}),x_{2})}\times\overline{D_{CD}(\underrightarrow{\widetilde{\psi}}(x_{3},t_{3,4}),x_{3},\underrightarrow{\widetilde{\psi}}(x_{4},t_{3,4}),x_{4})}\nonumber \\
 &  & \,\end{eqnarray}
etc., so the noise terms are not themselves Gaussian-Markov processes,
though there is some similarity.

\subsection{Properties of Noise Fields - Single Mode Case}

We can use the results in (\ref{eq:SumEtasOneMode-1}) relating the
$\eta_{k}^{A;D}(\underrightarrow{\widetilde{\psi}}(x,t))$ to the
local diffusion terms $D_{AB}(\underrightarrow{\widetilde{\psi}}(x,t),x)$
and the fundamental noise properties of the Gaussian-Markov noise
variables $\Gamma_{k}^{D}$ in (\ref{Eq.GaussMarkGeneral-1}), together
with (\ref{Eq.StochAverProduct-1}) to determine the stochastic properties
of the noise fields. For a single noise field\begin{eqnarray}
 &  & \overline{\{\left(\frac{\partial}{\partial t}\widetilde{G}_{A}(\underrightarrow{\widetilde{\psi}}(x_{1},t_{1}),\underrightarrow{\Gamma}(t_{1+}))\right)\}}\nonumber \\
 & = & \sum_{Dk}\overline{\eta_{k}^{A;D}(\underrightarrow{\widetilde{\psi}}(x_{1},t_{1}))}\;\overline{\Gamma_{k}^{D}(t_{1})}=0\end{eqnarray}
and for two noise fields\begin{eqnarray}
 &  & \overline{\{\left(\frac{\partial}{\partial t}\widetilde{G}_{A}(\underrightarrow{\widetilde{\psi}}(x_{1},t_{1}),\underrightarrow{\Gamma}(t_{1+}))\right)\left(\frac{\partial}{\partial t}\widetilde{G}_{B}(\underrightarrow{\widetilde{\psi}}(x_{2},t_{2}),\underrightarrow{\Gamma}(t_{2+}))\right)\}}\nonumber \\
 & = & \overline{\sum_{Dk}\eta_{k}^{A;D}(\underrightarrow{\widetilde{\psi}}(x_{1},t_{1}))\Gamma_{k}^{D}(t_{1+})\sum_{El}\eta_{l}^{B;E}(\underrightarrow{\widetilde{\psi}}(x_{2},t_{2}))\Gamma_{l}^{E}(t_{2+})}\;\nonumber \\
 & = & \sum_{Dk}\sum_{El}\overline{\eta_{k}^{A;D}(\underrightarrow{\widetilde{\psi}}(x_{1},t_{1}))\Gamma_{k}^{D}(t_{1+})\eta_{l}^{B;E}(\underrightarrow{\widetilde{\psi}}(x_{2},t_{2}))\Gamma_{l}^{E}(t_{2+})}\;\nonumber \\
 & = & \sum_{Dk}\sum_{El}\overline{\eta_{k}^{A;D}(\underrightarrow{\widetilde{\psi}}(x_{1},t_{1}))\eta_{l}^{B;E}(\underrightarrow{\widetilde{\psi}}(x_{2},t_{2}))}\;\,\overline{\Gamma_{k}^{D}(t_{1+})\Gamma_{l}^{E}(t_{2+})}\nonumber \\
 & = & \sum_{Dk}\sum_{El}\overline{\eta_{k}^{A;D}(\underrightarrow{\widetilde{\psi}}(x_{1},t_{1}))\eta_{l}^{B;E}(\underrightarrow{\widetilde{\psi}}(x_{2},t_{2}))}\;\,\delta_{kl}\delta_{DE}\delta(t_{1}-t_{2})\,\nonumber \\
 & = & \overline{\sum_{Dk}\eta_{k}^{A;D}(\underrightarrow{\widetilde{\psi}}(x_{1},t_{1,2}))\eta_{k}^{B;D}(\underrightarrow{\widetilde{\psi}}(x_{2},t_{1,2}))}\;\delta(t_{1}-t_{2})\,\nonumber \\
 & = & \overline{D_{AB}(\underrightarrow{\widetilde{\psi}}(x_{1,2},t_{1,2}),x_{1,2})}\,\;\delta(x_{1}-x_{2})\delta(t_{1}-t_{2})\end{eqnarray}
Thus the stochastic average of the linear noise term is zero, and
the stochastic average of the product of two linear noise terms is
\emph{delta function correlated in time}, and is also delta function
correlated in space. The spatial correlation is given by the \emph{local
diffusion term} in the original functional Fokker-Planck equation!

The noise terms do however satisfy the Gaussian-Markoff conditions
that averages of products of odd numbers of noise terms are zero,
but averages of products of even numbers of noise terms can be written
as sums of stochastic averages of products of pairs of non-local diffusion
terms, rather than pairs of noise terms. Thus\begin{eqnarray}
 &  & \overline{\{\left(\frac{\partial}{\partial t}\widetilde{G}_{A}(\underrightarrow{\widetilde{\psi}}(x_{1},t_{1}),\underrightarrow{\Gamma}(t_{1+}))\right)\left(\frac{\partial}{\partial t}\widetilde{G}_{B}(\underrightarrow{\widetilde{\psi}}(x_{2},t_{2}),\underrightarrow{\Gamma}(t_{2+}))\right)\left(\frac{\partial}{\partial t}\widetilde{G}_{C}(\underrightarrow{\widetilde{\psi}}(x_{3},t_{3}),\underrightarrow{\Gamma}(t_{3+}))\right)\}}\nonumber \\
 & = & \sum_{Dk}\sum_{El}\sum_{Fm}\overline{\eta_{k}^{A;D}(\underrightarrow{\widetilde{\psi}}(x_{1},t_{1}))\eta_{l}^{B;E}(\underrightarrow{\widetilde{\psi}}(x_{2},t_{2}))\eta_{m}^{C;F}(\underrightarrow{\widetilde{\psi}}(x_{3},t_{3}))}\;\,\overline{\Gamma_{k}^{D}(t_{1+})\Gamma_{l}^{E}(t_{2+})\Gamma_{m}^{F}(t_{3+})}\nonumber \\
 & = & 0\label{eq:StochAverThreeNoise1Mode}\end{eqnarray}
and\begin{eqnarray}
 &  & \overline{\begin{array}{c}
\{\left(\frac{{\LARGE\partial}}{{\LARGE\partial t}}\widetilde{G}_{A}(\underrightarrow{\widetilde{\psi}}(x_{1},t_{1}),\underrightarrow{\Gamma}(t_{1+}))\right)\left(\frac{{\LARGE\partial}}{{\LARGE\partial t}}\widetilde{G}_{B}(\underrightarrow{\widetilde{\psi}}(x_{2},t_{2}),\underrightarrow{\Gamma}(t_{2+}))\right)\\
\times\left(\frac{{\LARGE\partial}}{{\LARGE\partial t}}\widetilde{G}_{C}(\underrightarrow{\widetilde{\psi}}(x_{3},t_{3}),\underrightarrow{\Gamma}(t_{3+}))\right)\left(\frac{{\LARGE\partial}}{{\LARGE\partial t}}\widetilde{G}_{D}(\underrightarrow{\widetilde{\psi}}(x_{4},t_{4}),\underrightarrow{\Gamma}(t_{4+}))\right)\}\end{array}}\nonumber \\
 & = & \sum_{Hk}\sum_{El}\sum_{Fm}\sum_{Gn}\overline{\eta_{k}^{A;H}(\underrightarrow{\widetilde{\psi}}(x_{1},t_{1}))\eta_{l}^{B;E}(\underrightarrow{\widetilde{\psi}}(x_{2},t_{2}))\eta_{m}^{C;F}(\underrightarrow{\widetilde{\psi}}(x_{3},t_{3}))\eta_{n}^{D;G}(\underrightarrow{\widetilde{\psi}}(x_{4},t_{4}))}\;\nonumber \\
 &  & \times\overline{\Gamma_{k}^{H}(t_{1+})\Gamma_{l}^{E}(t_{2+})\Gamma_{m}^{F}(t_{3+})\Gamma_{n}^{G}(t_{4+})}\nonumber \\
 & = & \sum_{Hk}\sum_{El}\sum_{Fm}\sum_{Gn}\overline{\eta_{k}^{A;H}(\underrightarrow{\widetilde{\psi}}(x_{1},t_{1}))\eta_{l}^{B;E}(\underrightarrow{\widetilde{\psi}}(x_{2},t_{2}))\eta_{m}^{C;F}(\underrightarrow{\widetilde{\psi}}(x_{3},t_{3}))\eta_{n}^{D;G}(\underrightarrow{\widetilde{\psi}}(x_{4},t_{4}))}\nonumber \\
 &  & \times\left\{ \begin{array}{c}
(\delta_{kl}\delta_{HE}\delta(t_{1}-t_{2}))(\delta_{mn}\delta_{FG}\delta(t_{3}-t_{4}))+(\delta_{km}\delta_{HF}\delta(t_{1}-t_{3}))(\delta_{nl}\delta_{EG}\delta(t_{2}-t_{4}))\\
+(\delta_{kn}\delta_{HG}\delta(t_{1}-t_{4}))(\delta_{ml}\delta_{EF}\delta(t_{2}-t_{3}))\end{array}\right\} \nonumber \\
 & = & \overline{\begin{array}{c}
\left[\sum_{Hk}\eta_{k}^{A;H}(\underrightarrow{\widetilde{\psi}}(x_{1},t_{1}))\sum_{El}\eta_{l}^{B;E}(\underrightarrow{\widetilde{\psi}}(x_{2},t_{2}))\delta_{kl}\delta_{HE}\delta(t_{1}-t_{2})\right]\\
\times\left[\sum_{Fm}\eta_{m}^{C;F}(\underrightarrow{\widetilde{\psi}}(x_{3},t_{3}))\sum_{Gn}\eta_{n}^{D;G}(\underrightarrow{\widetilde{\psi}}(x_{4},t_{4}))\delta_{mn}\delta_{FG}\delta(t_{3}-t_{4})\right]\end{array}}\nonumber \\
 &  & +\overline{\begin{array}{c}
\left[\sum_{Hk}\eta_{k}^{A;H}(\underrightarrow{\widetilde{\psi}}(x_{1},t_{1}))\sum_{Fm}\eta_{m}^{C;F}(\underrightarrow{\widetilde{\psi}}(x_{3},t_{3}))\delta_{km}\delta_{HF}\delta(t_{1}-t_{3})\right]\\
\times\left[\sum_{El}\eta_{l}^{B;E}(\underrightarrow{\widetilde{\psi}}(x_{2},t_{2}))\sum_{Gn}\eta_{n}^{D;G}(\underrightarrow{\widetilde{\psi}}(x_{4},t_{4}))\delta_{nl}\delta_{EG}\delta(t_{2}-t_{4})\right]\end{array}}\nonumber \\
 &  & +\overline{\begin{array}{c}
\left[\sum_{Hk}\eta_{k}^{A;H}(\underrightarrow{\widetilde{\psi}}(x_{1},t_{1}))\sum_{Gn}\eta_{n}^{D;G}(\underrightarrow{\widetilde{\psi}}(x_{4},t_{4}))\delta_{kn}\delta_{HG}\delta(t_{1}-t_{4})\right]\\
\times\left[\sum_{El}\eta_{l}^{B;E}(\underrightarrow{\widetilde{\psi}}(x_{2},t_{2}))\sum_{Fm}\eta_{m}^{C;F}(\underrightarrow{\widetilde{\psi}}(x_{3},t_{3}))\delta_{lm}\delta_{EF}\delta(t_{2}-t_{3})\right]\end{array}}\nonumber \\
 & = & \overline{\left[D_{AB}(\underrightarrow{\widetilde{\psi}}(x_{1,2},t_{1,2}),x_{1,2})\right]\left[D_{CD}(\underrightarrow{\widetilde{\psi}}(x_{3,4},t_{3,4}),x_{3,4})\right]}\;\nonumber \\
 &  & \times\delta(x_{1}-x_{2})\delta(x_{3}-x_{4})\delta(t_{1}-t_{2})\delta(t_{3}-t_{4})\nonumber \\
 &  & +\overline{\left[D_{AC}(\underrightarrow{\widetilde{\psi}}(x_{1,3},t_{1,3}),x_{1,3})\right]\left[D_{BD}(\underrightarrow{\widetilde{\psi}}(x_{2,4},t_{2,4}),x_{2,4})\right]}\;\nonumber \\
 &  & \times\delta(x_{1}-x_{3})\delta(x_{2}-x_{4})\delta(t_{1}-t_{3})\delta(t_{2}-t_{4})\nonumber \\
 &  & +\overline{\left[D_{AD}(\underrightarrow{\widetilde{\psi}}(x_{1,4},t_{1,4}),x_{1,4})\right]\left[D_{BC}(\underrightarrow{\widetilde{\psi}}(x_{2,3},t_{2,3}),x_{2,3})\right]}\;\nonumber \\
 &  & \times\delta(x_{1}-x_{4})\delta(x_{2}-x_{3})\delta(t_{1}-t_{4})\delta(t_{2}-t_{3})\nonumber \\
 &  & \,\label{eq:StochAverFourNoise1Mode}\end{eqnarray}
using the results in (\ref{eq:SumEtasOneMode-1}). This is not quite
the same as\begin{eqnarray}
 &  & \overline{\{\left(\frac{\partial}{\partial t}\widetilde{G}_{A}(\underrightarrow{\widetilde{\psi}}(x_{1},t_{1}),\underrightarrow{\Gamma}(t_{1+}))\right)\left(\frac{\partial}{\partial t}\widetilde{G}_{B}(\underrightarrow{\widetilde{\psi}}(x_{2},t_{2}),\underrightarrow{\Gamma}(t_{2+}))\right)\}}\nonumber \\
 &  & \times\overline{\{\left(\frac{\partial}{\partial t}\widetilde{G}_{C}(\underrightarrow{\widetilde{\psi}}(x_{3},t_{3}),\underrightarrow{\Gamma}(t_{3+}))\right)\left(\frac{\partial}{\partial t}\widetilde{G}_{D}(\underrightarrow{\widetilde{\psi}}(x_{4},t_{4}),\underrightarrow{\Gamma}(t_{4+}))\right)\}}\nonumber \\
 &  & +\overline{\{\left(\frac{\partial}{\partial t}\widetilde{G}_{A}(\underrightarrow{\widetilde{\psi}}(x_{1},t_{1}),\underrightarrow{\Gamma}(t_{1+}))\right)\left(\frac{\partial}{\partial t}\widetilde{G}_{C}(\underrightarrow{\widetilde{\psi}}(x_{3},t_{3}),\underrightarrow{\Gamma}(t_{3+}))\right)\}}\nonumber \\
 &  & \times\overline{\{\left(\frac{\partial}{\partial t}\widetilde{G}_{B}(\underrightarrow{\widetilde{\psi}}(x_{2},t_{2}),\underrightarrow{\Gamma}(t_{2+}))\right)\left(\frac{\partial}{\partial t}\widetilde{G}_{D}(\underrightarrow{\widetilde{\psi}}(x_{4},t_{4}),\underrightarrow{\Gamma}(t_{4+}))\right)\}}\nonumber \\
 &  & +\overline{\{\left(\frac{\partial}{\partial t}\widetilde{G}_{A}(\underrightarrow{\widetilde{\psi}}(x_{1},t_{1}),\underrightarrow{\Gamma}(t_{1+}))\right)\left(\frac{\partial}{\partial t}\widetilde{G}_{D}(\underrightarrow{\widetilde{\psi}}(x_{4},t_{4}),\underrightarrow{\Gamma}(t_{4+}))\right)\}}\nonumber \\
 &  & \times\overline{\{\left(\frac{\partial}{\partial t}\widetilde{G}_{B}(\underrightarrow{\widetilde{\psi}}(x_{2},t_{2}),\underrightarrow{\Gamma}(t_{2+}))\right)\left(\frac{\partial}{\partial t}\widetilde{G}_{C}(\underrightarrow{\widetilde{\psi}}(x_{3},t_{3}),\underrightarrow{\Gamma}(t_{3+}))\right)\}}\nonumber \\
 &  & \,\end{eqnarray}
because in general \begin{eqnarray}
 &  & \overline{\left[D_{AB}(\underrightarrow{\widetilde{\psi}}(x_{1,2},t_{1,2}),x_{1,2})\right]\left[D_{CD}(\underrightarrow{\widetilde{\psi}}(x_{3,4},t_{3,4}),x_{3,4})\right]}\nonumber \\
 & \neq & \overline{D_{AB}(\underrightarrow{\widetilde{\psi}}(x_{1,2},t_{1,2}),x_{1,2})}\times\overline{D_{CD}(\underrightarrow{\widetilde{\psi}}(x_{3,4},t_{3,4}),x_{3,4})}\nonumber \\
 &  & \,\end{eqnarray}
etc., so the noise terms are not themselves Gaussian-Markov processes,
though there is some similarity.

\subsection{Supplementary Equations}

~

Functional Fokker-Planck equation for two mode case\begin{eqnarray}
\frac{\partial P}{\partial t} & = & \sum_{A}\int dx\,\frac{\delta}{\delta\psi_{A}(x)}A_{A}(\underrightarrow{\psi}(x),x)\, P\nonumber \\
 &  & +\sum_{A\leq\, B}\int\int dx\, dy\frac{\delta}{\delta\psi_{A}(x)}\frac{\delta}{\delta\psi_{B}(y)}H_{AB}(\underrightarrow{\psi}(x),x,\underrightarrow{\psi}(y),y)\, P\label{Eq.FFPETwoStart-1}\end{eqnarray}

Summation results

\begin{eqnarray}
 &  & \sum_{Dk}\eta_{k}^{A;D}(\underrightarrow{\widetilde{\psi}}(x_{1},t))\eta_{k}^{B;D}(\underrightarrow{\widetilde{\psi}}(x_{2},t))\nonumber \\
 & = & D_{AB}(\underrightarrow{\widetilde{\psi}}(x_{1},t),x_{1},\underrightarrow{\widetilde{\psi}}(x_{2},t),x_{2})\qquad Two\; Mode\nonumber \\
\label{eq:SumEtasTwoMode-1}\\ & = & D_{AB}(\underrightarrow{\widetilde{\psi}}(x_{1,2},t),x_{1,2})\,\delta(x_{1}-x_{2})\qquad One\; Mode\nonumber \\
 &  & \,\label{eq:SumEtasOneMode-1}\end{eqnarray}

Gaussian-Markoff rules\begin{eqnarray}
\overline{\Gamma_{k}^{D}(t_{1})} & = & 0\nonumber \\
\overline{\{\Gamma_{k}^{D}(t_{1})\Gamma_{l}^{E}(t_{2})\}} & = & \delta_{DE}\delta_{kl}\delta(t_{1}-t_{2})\nonumber \\
\overline{\{\Gamma_{k}^{D}(t_{1})\Gamma_{l}^{E}(t_{2})\Gamma_{m}^{F}(t_{3})\}} & = & 0\nonumber \\
\overline{\{\Gamma_{k}^{D}(t_{1})\Gamma_{l}^{E}(t_{2})\Gamma_{m}^{F}(t_{3})\Gamma_{n}^{G}(t_{4})\}} & = & \overline{\{\Gamma_{k}^{D}(t_{1})\Gamma_{l}^{E}(t_{2})\}}\;\overline{\{\Gamma_{m}^{F}(t_{3})\Gamma_{n}^{G}(t_{4})\}}\nonumber \\
 &  & +\overline{\{\Gamma_{k}^{E}(t_{1})\Gamma_{m}^{F}(t_{3})\}}\;\overline{\{\Gamma_{l}^{E}(t_{2})\Gamma_{n}^{G}(t_{4})\}}\nonumber \\
 &  & +\overline{\{\Gamma_{k}^{D}(t_{1})\Gamma_{n}^{G}(t_{4})\}}\;\overline{\{\Gamma_{l}^{E}(t_{2})\Gamma_{m}^{F}(t_{3})\}}\nonumber \\
 &  & ...\label{Eq.GaussMarkGeneral-1}\end{eqnarray}

Decorrelation Rule\begin{eqnarray}
 &  & \overline{F(\underrightarrow{\widetilde{\alpha}}(t_{1}))\{\Gamma_{k}^{D}(t_{2})\Gamma_{l}^{E}(t_{3})\Gamma_{m}^{F}(t_{4})...\Gamma_{a}^{X}(t_{l})\}}\nonumber \\
 & = & \overline{F(\underrightarrow{\widetilde{\alpha}}(t_{1}))}\,\overline{\{\Gamma_{k}^{D}(t_{2})\Gamma_{l}^{E}(t_{3})\Gamma_{m}^{F}(t_{4})...\Gamma_{a}^{X}(t_{l})\}}\qquad t_{1}<t_{2},t_{3},..,t_{l}\label{Eq.StochAverProduct-1}\end{eqnarray}

\end{document}